\documentclass[logos,parttoc,mainlanguage=english,morelanguage=french]{orsay-thesis}

\pdfoutput=1

\usepackage[latin1]{inputenc}
\usepackage[T1]{fontenc}
 \usepackage{ae}
 \usepackage{aecompl}
\usepackage{subeqnarray}
\usepackage{theorem}
\usepackage{appendix} 
\usepackage{graphics}
\usepackage{graphicx}
\usepackage{enumerate}
\usepackage{verbatim}
\usepackage{color}
\usepackage{pstricks}
\usepackage{bezier}
\usepackage[hang,footnotesize,bf]{caption}
 \usepackage{hyperref}
\hypersetup{ 
pdfstartview=FitV,
colorlinks=true, 
linkcolor=blue, 
citecolor=red, 
filecolor=black, 
urlcolor=magenta,
pdfauthor={Blaise Gout\'eraux},
pdfsubject={PhD Thesis},
pdfkeywords={Black Holes, Modifications of General Relativity, Extra Dimensions, Holography},
pdftitle={Black-Hole Solutions to Einstein's Equations in the Presence of Matter and Modifications of Gravitation in Extra Dimensions}}

\newcommand{\be} {\begin{equation}}
\newcommand{\ee} {\end{equation}}
\newcommand{\bea} {\begin{eqnarray}}
\newcommand{\eea} {\end{eqnarray}}
\newcommand{\bsea} {\begin{subeqnarray}}
\newcommand{\esea} {\end{subeqnarray}}
\newcommand{\nn}  {\nonumber}
\newcommand{\ga}{\gamma}
\newcommand{\ba}{\beta}
\newcommand{\da}{\delta}
\newcommand{\al}{\alpha}
\newcommand{\la}{\lambda}
\newcommand{\La}{\Lambda}
\newcommand{\eps}{\epsilon}
\newcommand{\ka}{\kappa}
\newcommand{\sig}{\sigma}
\newcommand{\U}{\hat U}
\renewcommand{\l}[1]{\left#1}
\renewcommand{\r}[1]{\right#1}
\renewcommand{\d}[1]{\displaystyle#1}
\newcommand{\e}{\mathrm{e}}
\newcommand{\ud}{\mathrm{d}}
\newcommand{\si}{ \mathrm{si}\,}

\newcommand{\half}{\frac{1}{2}}

\newcommand{\G}{\mathcal G}

\newcommand{\J}{\mathcal J}
\newcommand{\M}{\mathcal M}
\newcommand{\E}{\mathcal E}
\newcommand{\qed}{\nobreak \ifvmode \relax \else
      \ifdim\lastskip<1.5em \hskip-\lastskip
      \hskip1.5em plus0em minus0.5em \fi \nobreak
      \vrule height0.75em width0.5em depth0.25em\fi}

\newcommand{\Figref}[1]{Fig.\ref{#1}}
\newcommand{\Tableref}[1]{Table \ref{#1}}

\makeatletter
\DeclareRobustCommand*{\bfseries}{%
  \not@math@alphabet\bfseries\mathbf
  \fontseries\bfdefault\selectfont
  \boldmath
}
\makeatother

\numberwithin{equation}{section}

\author{Blaise \textsc{Gout\'eraux}}

\title[french]{Solutions de Trou Noir aux \'Equations d'Einstein en Pr\'esence de Mati\`ere et Modifications de la Gravitation en Dimensions Suppl\'ementaires.}
\title[english]{Black-Hole Solutions to Einstein's Equations in the Presence of Matter and Modifications of Gravitation in Extra Dimensions.}

\keywords[french]{Trous Noirs, Modifications de la Relativit\'e G\'en\'erale, Dimensions Suppl\'ementaires, Holographie}
\keywords[english]{Black Holes, Modifications of General Relativity, Extra Dimensions, Holography}

\ordernumber{9935}

\date{Lundi 27 septembre 2010}

\addcommissionmember[Directeur de th\`ese]{Pr.}{Christos}{Charmousis}
\addcommissionmember[Examinateur]{Dr.}{C\'edric}{Deffayet}
\addcommissionmember[Rapporteur]{Pr.}{Roberto}{Emparan}
\addcommissionmember[Examinateur]{Pr.}{Elias}{Kiritsis}
\addcommissionmember[Rapporteur]{Pr.}{David}{Langlois}
\addcommissionmember[Examinateur]{Pr.}{Renaud}{Parentani}

\begin{document}

\maketitle%

\pagestyle{empty}
\pagenumbering{roman}




\begin{abstract}[english]
In this thesis, we wish to examine the black-hole solutions of modified gravity theories inspired by String Theory or Cosmology. Namely, these modifications will take the guise of additional gauge and scalar fields for the so-called Einstein-Maxwell-Dilaton theories with an exponential Liouville potential; and of extra spatial dimensions for Einstein-Gauss-Bonnet theories. The black-hole solutions of EMD theories as well as their integrability are reviewed. One of the main results is that a master equation is obtained in the case of planar horizon topology, which allows to completely integrate the problem for s special relationship between the couplings. We also classify existing solutions. We move on to the study of Gauss-Bonnet black holes, focusing on the six-dimensional case. It is found that the Gauss-Bonnet coupling exposes the Weyl tensor of the horizon to the dynamics, severely restricting the Einstein spaces admissible and effectively lifting some of the degeneracy on the horizon topology. We then turn to the study of the thermodynamic properties of black holes, in General Relativity as well as in EMD theories. For the latter, phase transitions may be found in the canonical ensemble, which resemble the phase transitions for Reissner-Nordstr\"om black holes. Generically, we find that the thermodynamic properties (stability, order of phase transitions) depend crucially on the values of the EMD coupling constants. Finally, we interpret our planar EMD solutions holographically as Infra-Red geometries through the AdS/CFT correspondence, taking into account various validity constraints. We also compute AC and DC conductivities as applications to Condensed Matter Systems, and find some properties characteristic of strange metal behaviour.
\end{abstract}
\noindent\hspace*{0.35\textwidth}\hrulefill\hspace*{0.35\textwidth}\\[-\bigskipamount]



\selectlanguage{english}

\section*{Acknowledgments}
\vfill
These last three years as a PhD student have been intense, both in scientific learning and encounters. This would have been impossible without the constant support, encouragement and scientific advice of Pr. Christos Charmousis, my advisor. I hope he got as much pleasure from guiding my first steps in research as I did from working and learning with him.

I would also like to thank Pr. Jiro Soda, from Kyoto University, as well as Pr. Elias Kiritsis, from the University of Crete, for all they brought me through our collaborations, and also for their support in continuing my career in research. I am also thankful to Drs. R. Zegers and C. Bogdanos from the Laboratoire de Physique Th\'eorique at Paris-Sud University for our work and discussions.

I am very grateful to all the members in the jury for my defense, for the time they took and the useful comments; especially to Prs. Emparan and Langlois for accepting to review my manuscript. I am also indebted to the latter for his ongoing support since my early days as a young Special and General Relativity student.

The LPT at Paris-Sud University has proven to be a very warm and nurturing environement to foster the beginning of my scientific career: I wish to thank all of its members for various discussions and encouragements, and in particular the administrative staff for their help in all matters of day-to-day life. Thanks also to my SINJE fellow students, from the past and the present, for invigorating debates which I shall fondly recall. Those of them who bore with me over these last few months deserve special credit.

Finally, I wish to express my gratitude and love to my family, partner and close ones, for their love and support during all of these years, and without whom my life would undoubtedly be quite different.

\vfill

\newpage


\strut\newpage

\begingroup
\hypersetup{linkcolor=black}
\tableofcontents
\vfill
\pagebreak
\listoffigures
\vfill
\pagebreak
\listoftables
\vfill 
\endgroup

\pagebreak\strut\newpage
\selectlanguage{english}

\pagestyle{fancy}




\pagenumbering{arabic}
\selectlanguage{english}
\section*{Introduction\markboth{Introduction}{Introduction}}
\addcontentsline{toc}{section}{\protect\numberline{}Introduction}

In this thesis, we wish to review black-hole solutions in some gravitation theories modified with respect to Einstein's theory. These modifications will take the guise of additional matter fields with non-trivial couplings, as well as extra spatial dimensions. We shall have the opportunity to describe them in much greater detail in Part \ref{part:two}.

At the turn of the twentieth century, inertial motion was described by the quite new theory of Special Relativity\footnote{It was formalised in 1905 by Einstein, though there is considerable debate as to the paternity of the theory, since both Lorenz and Poincar\'e had also made extensive and essential contributions.}, while gravitation made use of concepts introduced by Newton  in the seventeenth century. Physicists like anecdotes very much: it is claimed by Newton himself that the ideas behind his universal law of gravitation came to him as he was walking his mother's garden and watched an apple fall to the ground. This inspired him the idea that gravity was an infinite range force and that it should be expressed as an inverse squared-distance law. Similarly,  it is claimed that the inspiration for Einstein's theory came from considerations on the nature of free fall, and how acceleration may counterbalance the effect of gravitational fields.

Special Relativity in itself had already revolutionised the separate concepts of space and time, by unifying them in a single entity, space-time (to stress the conceptual jump that must be made, we shall cross out the dash, and write it \emph{spacetime}). From there on, time was to be treated as just another coordinate. The only concession to previous physics was that the speed of light was promoted to a fundamental and universal status: its value should be the same for all inertial observers; and that the laws of physics should be the same in all inertial frames. Numerous, counter-intuitive consequences were derived, such as the contraction of lengths and the dilation of time: the clocks and rods used by different observers in different reference frames to measure time and distances would not agree! Moreover the notion of simultaneity in the usual sense became ill-defined: implicitly, simultaneity assumes that observers can agree on an absolute time, but how can they do this when each has its own definition and none of them coincide? This is the so-called relativity of motion in spacetime.

Shortly after publishing the theory of Special Relativity\footnote{The term Special Relativity refers to the fact that only inertial motion is considered in this version of the theory.}, Einstein started to wonder whether it was fully consistent with Newton's theory of gravitation. Indeed, Lorentz invariance is central to the new theory of relativistic motion, and the speed of light is a fundamental constant, of equal value to all inertial observers and defining the maximal speed one can attain. Yet, it did not enter anywhere in Newton's theory, and the idea that gravitation could propagate instantaneously seemed a little preposterous in the new frame of Special Relativity. For ten years, Einstein struggled laboriously, but his efforts finally came to a close in 1915. Unexpectedly, his research revealed a profound link between gravity and accelerated observers. Gravitation (matter) curves spacetime, which induces motion that can locally be mimicked by considering an observer in acceleration in flat spacetime. Special Relativity was to be made General. We shall go on in greater detail on the founding principles of General Relativity in Section \ref{section:GRPrinciples}.


Einstein's equations provided a way of quantifying (in the sense of ascribing a quantitative value) the interplay between matter and geometry. And at first, it seemed like everything fit together very well indeed! The long-lasting puzzle of the advance of the perihelion of Mercury was solved, and the missing $43$ seconds of arc computed very precisely from the General Relativity correction to the Newtonian (Keplerian) motion of the orbit of the planet. 
Einstein's original motivations had little to do with what seemed at the time a minor experimental discrepancy; with hindsight, one can but marvel that only a true revolution of the concepts of space and time could account for it.
Of course, the twentieth century saw many more verifications of General Relativity's predictions, which relied on measurements unknown before Einstein work was published. They fall under two great categories. First, the weak field predictions compute small deviations from the Newtonian theory, such as the deflection of light as it passes nearby a source of matter (exemplified in the famous measurement of the total solar eclipse of 1919 by Sir Arthur Eddington) or the gravitational redshift of light (the reddening of light passing by gravitational sources). Time delay (Shapiro effect) is another important prediction, and the latest measurements (on the Cassini spacecraft returning from Saturn) place the experimental value at the predicted one with a $10^{-5}$ precision. The light deflection measurement in the solar system is nearly as good, with an agreement with the predicted value at $10^{-4}$. A second important category of tests concerns strong gravitational fields. They are emitted by sources such as pulsars, which are rapidly-revolving neutron stars. Their period of revolution can be measured very precisely, and there again agreement with General Relativity predictions was found to be excellent. 

Two puzzles remain. General Relativity predicts gravitational waves, which are fluctuations of spacetime itself. They are expected to be significantly produced during the merger of two neutron stars, of two black holes or of a neutron star and a black hole. So far, none of the many experiments designed to detect these waves has been succesful, but the search is very active and new experimental (non-)results are expected in the near future. We wish to stress the importance of this detection: gravitational waves are a fundamental prediction of General Relativity, and what gives true legitimity to a theory is its ability to predict new, unforeseen results. The second puzzle is cosmological: the last two decades have seen numerous experiments confirming what is called the Standard Model of Cosmology, which is based on General Relativity. Yet, measurements indicating that the acceleration of the Universe was increasing caused a great surprise and puzzlement in the community. Though this can be accomodated by introducing a constant in the equations, this does not feel very natural and may result in a reconsideration of General Relativity's paradigm. There may be need of large (cosmological) distance modifications: on top of the above-mentioned addition of a \emph{cosmological constant}, another popular approach was to examine so-called braneworlds scenarii, where a four-dimensional ``brane'' (surface) moves around in a higher-dimensional ``bulk'' (surrounding background spacetime); more on this in Section \ref{section:IRModificationsGR}.

As much as we would like to go on with these very interesting topics, the tests of General Relativity are not the main subject of this thesis. We shall now turn to the heart of the matter. The reader may have noticed the occurrence of the words ``black holes'' in the previous paragraph. These are the truly fundamental ``objects'' of General Relativity. It is usual for a theory to describe a truly intrinsic class of objects. For instance, Quantum Field Theory is formulated in terms of fields, but predicts particles, which make up the elementary constituents of matter. This analogy can be carried over to General Relativity: the fundamental field in terms of which General Relativity is formulated is the metric field, which serves to measure spacetime distances; but black holes are solutions of the theory (particular metric fields solving Einstein's equations) which seem to offer profound insight into the nature of the theory itself. Their importance was not acknowledged at first, and indeed their physical properties were not decyphered before a long time after their discovery, see Section \ref{section:GenesisBH}: not up till the turn of the 1950s did they become fully-fledged, unequivocal solutions of the theory. One has to admit that the potion was a bitter one to swallow for the relativists of the first generation: black holes were regions where the trajectories of light simply seemed to stop, taking an infinite time to reach. The surface where this happened was called the horizon of the black hole. It took quite long before the following question was seriously asked: which time? And Special Relativity had been known since the beginning of the century\dots Black holes turned out to be the grounds where the relativistic revolution was truly put to the test, where relativity of motion was truly implemented, where the unification of time and space was really necessary and took meaning. 

We will attempt to account for all of this in Part \ref{part:one} of this work, first by a brief exposition of the founding principles followed by Einstein to construct his theory and a presentation of Einstein's equations in Section \ref{section:GRIntro}; then with a historical review of the birth of the concept of black hole and a survey of some of the most famous solutions in Section \ref{section:BH}. We will conclude Part \ref{part:one} with some motivations to consider modifications of General Relativity in Section \ref{section:ModificationsGR}. We have already mentioned the need for this because of cosmological reasons. Another incentive has a radically opposite origin. Indeed, one does not expect that General Relativity should remain valid up to arbitrarily high energies, or equivalently to arbitrarily small distances. This  can be related to the fact that graviton\footnote{The massless particles which mediate gravity. They are the dual description of gravitational waves, same as light can both be described by photons or electromagnetic waves.} interactions amplitudes are expected to diverge at ultra-high energies and some sort of Ultra-Violet cutoff has then to be imposed. An intuitive way to rephrase this is that, as one probes smaller and smaller distances with higher and higher energies, the point-like nature of interactions ceases to make sense. This points out the need for a quantum theory of gravity, the elusive Holy Grail of contemporary theoretical physics. 

One promising candidate is String Theory. The basic idea behind it is effectively to smear out the interaction at very small distances, where the string-like nature of elementary particles is revealed. This regularises the divergences of interaction amplitudes and provides a natural cutoff for the theory. These matters will be explained in a little bit more detail in Section \ref{section:UVModificationsGR}. Among many other properties, the most surprising one is certainly that String Theory lives in ten dimensions: six spatial dimensions are compactified small enough that they are ``hidden from view''; four large dimensions remain and correspond to our usual world. Although General Relativity is recovered in low-energy approximations to String Theory, it usually comes coupled with matter fields.

This will be the topic of Part \ref{part:two}. First, in Section \ref{section:EMDBH}, we shall examine the black-hole solutions of a class of theories called Einstein-Maxwell-Dilaton, which contain an electromagnetic and a scalar field, coupled to gravity and between themselves. These theories will be seen to derive from higher-dimensional theories of gravity and/or simply as theories of (Einstein) gravity coupled to matter. We shall review existing solutions in the literature and exhibit new ones. We will also comment upon the modification of the properties of the black holes of the theories because of the presence of the scalar field. We will move on in Section \ref{section:EGBBH} to a second set of theories, called Einstein-Gauss-Bonnet theories. These are truly higher-dimensional theories, as they only display properties different from General Relativity  in dimensions higher than four. They can be argued to be the most general theories of gravity in higher dimensions retaining the properties of General Relativity. Their black-holes solutions share some similarities with General Relativity's, but also differ in very interesting ways. An emphasis will be placed on the allowed geometries for the horizon of the black hole, compared to General Relativity both in four and higher dimensions.

In Part \ref{part:three}, we shall turn to the analysis of thermodynamics of black holes. Indeed, it was realised in the seventies that the laws of black-hole mechanics bore an uncanny resemblance to the laws of thermodynamics. It seemed that the law ruling the evolution of the area of the horizon as the mass of the black hole varied was similar to the first law of thermodynamics between the evolution of the entropy and internal energy of a system. The analogy was pushed further as the area of a black hole was proven to be only increasing by physical processes, just as the total entropy of a system can only grow. Finally, the surface gravity, which is the conjugate quantity to the area in the first law of black-hole mechanics was shown to be a constant on the horizon, and suggested it could be identified as the ``temperature'' of the black hole. This correspondence was put on firm footing when the quantum radiation of black holes was discovered by Hawking, with a black body spectrum at a temperature related to the surface gravity. This allowed to identify formally the entropy of the black hole as the quarter of its horizon area. The details will be given in Section \ref{section:BHThermo}. We will also explain how to define thermodynamic ensembles and partition functions, and then examine the thermodynamic stability of black holes in General Relativity. Section \ref{section:ThermoEMD} will see the application of these principles to the black holes of Einstein-Maxwell-Dilaton theories, and there again we shall comment on the differences due to the introduction of the scalar field.

Finally, we give an outlook and some conclusions in Part \ref{part:four}. We shall present some perspectives for future work in Section \ref{section:Holography}. These rest on recent developments motivated by String Theory. An identification is made between the physics of weakly-coupled gravitational theories living in the bulk of spacetime and those of strongly-coupled gauge (particle) theories living on its boundary. This so-called AdS/CFT correspondence establishes a precise dictionnary between quantities computed on both sides, gravitational and gauge, and gives access to results usually out of computational range. We shall see how Einstein-Maxwell-Dilaton theories can be interpreted as describing the properties of the bulk Infra-Red region, very far from the boundary. We shall also present some recent results where applications of this correspondence are carried over to Condensed Matter systems such as ``strange metals''.

\vfill


\part{Black Holes in General Relativity}
\label{part:one}

These sections owe much to lectures given by Nathalie Deruelle at the Institut Henri Poincar\'e, Paris in Autmun 2006, \cite{deruelle2006}, and at the Institut de Physique Th\'eorique at the CEA, Saclay in Winter 2009, \cite{deruelle2009}. I have also used material from \cite{Eisenstaedt:2007,Inverno:1992,Hawking:1973uf,Charmousis:2009}.

\section{Introducing Einstein's General Relativity}

\label{section:GRIntro}

\subsection{Principles in General Relativity}
\label{section:GRPrinciples}

\paragraph{Mach's principle}

One of the founding principles that guided Einstein as he was constructing his general theory of Relativity was Mach's principle. Indeed, Newton's laws hold in what are called \emph{inertial frames}, which are in uniform velocity compared to \emph{absolute space}. One means to determine such frames uses the bucket experiment: if an observer carrying a bucket of water is inertial, then the surface of the water should remain flat; on the other hand, if he is in accelerated motion or in rotation compared to absolute space, the water surface will be respectively inclined or concave.

This state of affairs was unacceptable for Mach, since it postulated some kind of absolute reference, \emph{absolute space}, the existence of which did not depend on the matter content of the Universe. Inertial motion existed independently of gravitational motion, and the equality of the inertial and gravitational masses was just coincidental. Mach tried to remedy this by introducing ``fixed stars'', compared to which all motion is relative. Thus, an inertial observer would now be in uniform motion \emph{relative} to the fixed stars, the Universe indistinguishable from the fixed stars, and there would actually be no Universe should the fixed stars be removed. Motion in an empty Universe would be devoid of meaning. 

This can be summarised by the three following statements:
\begin{itemize}
 \item The matter content determines the geometry of spacetime;
 \item Without matter there is no geometry (not even flat Minkowski spacetime);
 \item In an empty Universe, a test-body\footnote{By definition,  a test-body can only be passive even if it has a mass, it does not act as a source of geometry. The equivalent concept of test-charge is also used.} has no inertia.
\end{itemize}

\paragraph{Equivalence principle}

The motion of a gravitational test-mass in a gravitational field does not depend upon its mass. This simple statement, already well-known from the Pisa experiments at the time of Einstein, was promoted to the rank of principle by him. Indeed, the equality between the inertial mass (entering Newton's first law, $\sum\mathbf{F_{ext}}=m_i\mathbf{a} $), the passive gravitational mass (entering the expression of the force felt by a test-body in a gravitational potential $\phi$, $\mathbf{F_G}=m_p\mathbf{\nabla}\phi$) and the active gravitational mass (entering the expression of the gravitational potential created by a source, $\phi=-G_N\frac{m_a}r$), was coincidental in Newton's theory, yet by no means necessary to its internal consistency.

This is not the case for Einstein's theory. It is often written that ``gravity sees all'', meaning that it interacts, however weakly, with all kinds of matter. Thus, matter does not simply respond to geometry, it also creates it by acting as a gravitational source. Moreover, locally, one cannot distinguish between irrotational free fall in some gravitational field and uniform motion in flat space with no gravitational field. This allows to recover the equivalent of the Newtonian inertial frames, without contradicting Mach's principle, as Newtonian gravity did. The matter content of the Universe (the ``fixed stars'') explains the origin of inertial forces. Furthermore, gravitational fields and accelerated motion are the same: a linearly-accelerated observer can cancel the effect of the inertial forces he is feeling by turning on a gravitational field of the same intensity but opposite direction.

\paragraph{General covariance}

The principle of general covariance follows from that of general relativity: all observers are equivalent. This means that any given observer, independently of its properties, should be able to determine the laws of physics. If not, how can we explain that we on Earth can do so, since we are not even inertial observers\footnote{The motion of the Earth is certainly not inertial.}? Thus, the equations of physics should be in tensorial form, and any coordinate system (or equivalently, any observer) should be acceptable. This does not mean that any coordinate system can be used, but that the theory is invariant under a change of coordinate. Thus, one should be wary of seemingly physical effects due to the choice of a particular set, and use this liberty to extract what is physically meaningful. Although technical mastery of general covariance came very early, it was not understood physically until much later, and led to many misinterpretations of Schwarzschild's solution as we will shortly describe in Section \ref{section:HardBeginnings}.

Coupled with the principle of equivalence, it also implies that, locally\footnote{That is in a region where the gravitational field does not vary to leading order.}, one can always find a set of coordinates where spacetime is flat.

\subsection{Einstein's equations}

Let us start from the Einstein-Hilbert action with a cosmological constant
\be
	S_{EH} = \frac1{16\pi G_N}\int \ud^4x\sqrt{-g}\l(R-2\Lambda + \mathcal{L}_m\l[\Psi\r]\r)\,,
	\label{EinsteinHilbertAction}
\ee
in units with $c=1$, where the Ricci scalar, $R=R_{\mu\nu}g^{\mu\nu}$ is the trace of the Ricci tensor, itself the trace of the Riemann tensor, $R_{\mu\nu}=R_{\mu\rho\nu\sigma}g^{\rho\sigma}$. The Riemann tensor is a measure of the curvature of spacetime. Indeed, it appears through the geodesics\footnote{Trajectories of free particles in General Relativity: in Euclidean three-dimensional space, these would be straight lines, and on a sphere, circles.} deviation equation
\be
	\nabla_v\nabla_v\xi^\mu-R^\mu_{\nu\rho\sigma}v^\nu v^\rho\xi^\sigma=0\,, \qquad \nabla_v=v^\mu\nabla_\mu\,,
\ee
with $v^\mu$ the vector tangent along and $\xi^\mu$ the vector normal to the geodesics. It possesses a number of useful properties, such as the skew and interchange symmetries
\bsea
	&&R_{\mu\nu\rho\sigma}=-R_{\nu\mu\rho\sigma}=R_{\nu\mu\sig\rho}\,,\slabel{RiemannSkewSymmetry}\\
	&&R_{\mu\nu\rho\sigma}=R_{\rho\sigma\mu\nu}\,.\slabel{RiemannInterchangeSymmetry}
	\label{RiemannSymmetries}
\esea
It also verifies the quite useful first and second Bianchi identities
\bsea
	R_{\mu\l[\nu\rho\sigma\r]}&=&0\,,\slabel{FirstBianchiRiemann}\\
	\nabla_{\l[\la\r.}R_{\l.\rho\sigma\r]\mu\nu}&=&0\,,\slabel{SecondBianchiRiemann}
\esea
where as is customary the brackets denote total antisymmetrisation over the indices enclosed.

Coming back to \eqref{EinsteinHilbertAction}, $ \mathcal{L}_m$ is the matter Lagrangian, with $\Psi$ denoting collectively the various matter fields. In this thesis, we will mostly concern ourselves with scalar fields, $\phi$, and vector fields, $A_\mu$.
Einstein's equations, derived from this action, are
\be
	G_{\mu\nu}=R_{\mu\nu}-\half R g_{\mu\nu}= 8\pi G_NT_{\mu\nu}\,, \label{EinsteinEq}
\ee
where $G_{\mu\nu}$ is the Einstein tensor. By contracting twice the second Bianchi identity \eqref{SecondBianchiRiemann}, it can be checked to be divergenceless:
\be
	\nabla^\mu G_{\mu\nu}=0\,, \label{BianchiEq}
\ee
and provides an extra constraining equation to Einstein's equations. This geometrical property has actually a very important physical meaning, which is related to the right-hand side of \eqref{EinsteinEq}.

$T_{\mu\nu}$ is the stress-energy tensor containing the matter fields present in the theory, and can be derived from the matter Lagrangian by the following formula
\be
	T_{\mu\nu}=\frac1{8\pi G_N}\frac{\da\mathcal L_m}{\da g^{\mu\nu}}\,.
	\label{StressEnergyGeneral}
\ee
Most importantly, it needs to be divergence-free to ensure conservation of energy. This highlights the non-trivial choice of the Einstein tensor as the left-hand side of \eqref{EinsteinEq}. Indeed, had Einstein kept to the Ricci tensor only, this crucial property of sensible physical theories would not have been recovered.

In the case of a perfect fluid, this tensor is written as
\be
	T_{\mu\nu}=(p+\rho)u_\mu u_\nu + pg_{\mu\nu}\,,
	\label{StressEnergyPerfectFluid}
\ee
where $p$ is the pressure of the fluid, $\rho$ its energy density and $u_\mu$ its four-velocity. One can note that if this fluid obeys an equation of state, $p=w\rho$, with $w=-1$, this produces a term in Einstein's equations \eqref{EinsteinEq} which is proportional to the metric. We shall see in  sections \ref{section:SalvationCosmology} and \ref{section:IRModificationsGR} that this seemingly innocent remark has important consequences in what follows.

\vfill
\pagebreak

\section{Black holes}
\label{section:BH}

\subsection{The genesis of the concept of black holes}

\label{section:GenesisBH}

\subsubsection{Hard beginnings}

\label{section:HardBeginnings}

The history behind the genesis of the concept of black holes began very shortly after Einstein 
published his first articles on General Relativity in November 1915, \cite{Einstein:1915by,Einstein:1915ca,Einstein:1916vd}, after a series of lectures at the University of G\"ottingen in June. Indeed, Schwarzschild 
, then serving in the German artillery on the Russian front, found the following solution to Einstein's equations in vacuum, \cite{Schwarzschild:1916uq}:
\be
	\ud s^2 = -\l(1-\frac{2m}{r}\r)\ud t^2 +\frac{\ud r^2}{\l(1-\frac{2m}{r}\r)}+r^2\l(\ud\theta^2+\sin^2\theta\ud\varphi^2\r).
	\label{Schwarzschild}
\ee
This solution would thereafter bear his name. This is an immediate confirmation that one should be wary of Newtonian intuition while dealing with General Relativity: even without matter, Einstein's equations admit non-trivial solutions, whose properties are quite different from Minkowski spacetime:
\be
	\ud s^2 = -\ud t^2 +\ud r^2+r^2\l(\ud\theta^2+\sin^2\theta\ud\varphi^2\r),
	\label{Minkowski}
\ee
here in spherical coordinates. Schwarzschild died shortly after that, in May 1916, from a disease contracted in the trenches, but not before succeeding in matching this solution \eqref{Schwarzschild} with the interior of a star of constant energy density and pressure, \cite{Schwarzschild:1916ae}. As we shall see in the remainder of this section, this was to be the source of a lasting misinterpretation of the solution.

Coming back to the solution \eqref{Schwarzschild}, several properties immediately attract our attention. The metric coefficients are ill-behaved both at $r=0$ and at $r=2m$, but scalar invariants, like for example the Kretschmann invariant, diverge only at $r=0$:
\be
	K \doteq R_{\la\mu\nu\rho}R^{\la\mu\nu\rho} =  \frac{48m^2}{r^6}\,,
	\label{KretschmannSchwarzschild}
\ee
which signals the presence of a true curvature singularity. The nature of the so-called ``Schwarzschild singularity'' at $r_S=2m$ was however much more troubling. Were one to cross it somehow, time and space would be reversed. In pretty much the same way as an observer \emph{outside} Schwarzschild radius could follow a time-like wordline while sitting in the same point in space, an equivalent observer \emph{inside} would see space flow by irrevocably while time could be kept frozen! Namely, the metric signature changes from $\l(-,+,+,+\r)$ to $\l(+,-,+,+\r)$. Several people (Schwarzschild himself, Droste in 1916, \cite{Droste:1916}, von Laue in 1921, de Jans in 1923\dots) also studied particle trajectories in Schwarzschild spacetime and found that geodesics seemed to stop on the Schwarzschild sphere. From this, they naturally inferred (in a pure Newtonian frame of mind) that the interior of the Schwarzschild solution could never be reached and only the exterior region should be considered. These pecularities were dismissed at first, since no one seriously expected that such astrophysical bodies could exist. Indeed, a quick calculation shows that, for the Sun, $r_S=3km$ while $R_{Sun}=7.10^5km$. Thus, it was hastily concluded that this Schwarzschild radius was unphysical and would always fall well inside realistic stars. As Eddington 
put it, \cite{Eddington:1987tk},
\begin{quotation}
	``There is a magic circle $r=2m$ which no measurement can bring us inside. It is not unnatural that we should picture something obstructing our closer approach and say that a particle of matter fills the interior [of Schwarzschild's solution].''
\end{quotation}
Eisenstaedt calls this the \emph{neo-Newtonian bias}, \cite{Eisenstaedt:2007}.

Moreover, although general covariance was already established as one of the founding principles of General Relativity, Hilbert 
wrote in 1917, \cite{Hilbert:1917}, that 
\begin{quotation}  
``A line element or a gravitational field $g_{ij}$ is regular at a point if it is possible to introduce by a reversible, one-to-one transformation a coordinate system, such that in this system the corresponding functions $g_{ij}'$ are regular at that point, i.e. they are continuous and arbitrarily differentiable at the point and at the neighbourhood of the point and the determinant $g'$ is different from zero.''
\end{quotation}
Thus, although Painlev\'e
-Gullstrand 
coordinates were introduced as soon as 1921, \cite{Painleve:1921,Gullstrand:1922}, 
\bea
	\ud s^2 &=& -\l(1-\frac{2m}{r}\r)\ud\tilde{t}^2 +2\sqrt{\frac{2m}r}\ud r\ud\tilde{t} + \ud r^2+r^2\l(\ud\theta^2+\sin^2\theta\ud\varphi^2\r),	\label{PainleveGullstrand}\\
	\tilde t&=& t+4m\l[\sqrt{\frac{2m}r}+\ln\sqrt{\frac{r-2m}{r+2m}}\r],\nn
\eea
followed by Eddington's in 1924, \cite{Eddington:1924},
\bea
	\ud s^2 &=& -\l(1-\frac{2m}{r}\r)\ud\bar{t}^2 +\frac{4m}r\ud r\ud\bar{t} +\l(1+\frac{2m}r\r) \ud r^2+r^2\l(\ud\theta^2+\sin^2\theta\ud\varphi^2\r),	\label{Eddington}\\
	\bar t&=& t+2m\ln\l|\frac{r}{2m}-1\r|,\nn
\eea
nobody remarked on the fact that the Schwarzschild solution is regular at the Schwarzschild radius in these sets, and they were dismissed on account of the transformation being singular at this point. This constituted the second mental block, and is dubbed \emph{geometrical}.

The last nail on the coffin came from works by Flamm
, \cite{Flamm:1917}, and Weyl 
, \cite{Weyl:1919a}. Taking the $t=constant$ and $\theta=\frac\pi2$ slices of \eqref{Schwarzschild}, one finds this describes the induced metric on the paraboloid $z^2=8m(r-2m)$ for the exterior solution $r>2m$, embedded in three-dimensional Euclidean space. Explicitly, taking for $\mathcal{E}_3$
\be
	\ud s^2 = \ud r^2 + r^2\ud\varphi^2 + \ud z^2\,,
\ee
and substituting $\ud z ^2 = \frac{2m\ud r^2}{r-2m}$, we get
\be
	\ud s^2 = \frac{\ud r^2}{1-\frac{2m}r} + r^2\ud\varphi^2\,,
\ee
as advocated. Nothing then prevents us from extending the paraboloid to the lower half-plane $z<0$ and this led Weyl to write that
\begin{quotation}
	``The complete realisation of this solution would imply that space is doubly connected, that is, contains not one but two boundaries accessible at infinity.''
\end{quotation}
The third mental block, the \emph{topological} one, is in place.


\subsubsection{Salvation from Cosmology and Nuclear Physics}
\label{section:SalvationCosmology}
\begin{table}
\centering
	\caption{Mach's principle and Einstein's ``biggest blunder''.}
\begin{tabular}{|cp{0.9\textwidth}c|}
\hline&&\\
&\setlength{\parindent}{10pt}
Einstein tried to follow Mach's steps, though he never quite succeeded in incorporating Mach's principle in his theory. In particular, this was one of the main reasons why he resisted so much the interpretation of the Schwarzschild solution as a black hole. How could a \emph{vacuum} solution of his equations define the entirety of spacetime? Indeed, Einstein refused on principle the recovery of the Minkowski spacetime at the asymptotic infinity of Schwarzschild's. This was unacceptable as it seemed to him an implicit comeback to Newton's idea of absolute space and time, and he firmly believed that, without matter, the Universe should not ``exist'', or at least in more modern language, that the background spacetime should emerge from the interplay between geometry and matter and not have an existence of its own in the theory. 

Moreover, he believed the Universe to be static. Applying his newfound theory to Cosmology, he realised he could not find any solution both static and containing matter. To remedy this, he introduced a constant in the equations of motion of General Relativity, in an attempt to safeguard his precious Mach's principle, \cite{Einstein:1917ce}. Then, one solution existed with uniform matter density, globally static, and last but not least, positively curved: the Einstein static Universe. Mach's principle was saved! Soon after that, exhibition by de Sitter of a curved vacuum solution to the new equations of motion turned his hopes to dust. General Relativity did not embody Mach's principle, geometry existed independently from matter. 

The coup de grace came as Hubble put forward the proof of the expansion of the Universe by measuring the peculiar velocity of receding galaxies. Einstein revised his views, barred the constant from his equations and called it ``his biggest blunder''. In the end, Mach's principle was never fully implemented in Einstein's theory. However, the cosmological constant would reappear, nearly one century later.&\\&&\\
\hline
\end{tabular}
	\label{Table:EinsteinLambda}
\end{table}

Progress came in the 1930s from fields exterior to relativistic circles proper, namely Cosmology and Nuclear Physics (a perfect example of interplay between different fields). To witness this, we need to go back in time, to the early years of General Relativity. Dissatisfied that his theory did not obey Mach's principle, see \Tableref{Table:EinsteinLambda}, Einstein introduced the celebrated cosmological constant in his equations of motion, \cite{Einstein:1917ce},
\be
	G_{\mu\nu} +\Lambda g_{\mu\nu}=8\pi G_N T_{\mu\nu}\,, \label{LambdaEinsteinEq}
\ee
and very soon after that, de Sitter 
presented the following vacuum solution to this new set of equations, \cite{deSitter:1917a,deSitter:1917b,deSitter:1917c,deSitter:1918},
\be
	\ud s^2 = -\l(-\frac{\Lambda}{3}r^2+1\r)\ud t^2 + \frac{\ud r^2}{\l(-\frac{\Lambda}{3}r^2+1\r)} +r^2\l(\ud\theta^2+\sin^2\ud\varphi^2\r),
	\label{deSitter}
\ee
where one should note that this metric is static \emph{inside} the de Sitter radius, $r=\sqrt{\frac3{\Lambda}}$, for positive $\Lambda$. New asymptotic conditions had to be imposed, which were not flat, and so the quest for implementing Mach's principle (see \Tableref{Table:EinsteinLambda}) in General Relativity was back to square one. Klein 
as well as de Sitter remarked that this solution  could be embedded in five-dimensional Minkowski space as a hyperboloid, see  \Tableref{Table:deSitter}, thus removing the singularity at $r=\sqrt{\frac3{\Lambda}}$.

\begin{table}
\centering
	\caption{De Sitter space as a hyperboloid in Minkowski space.}
\begin{tabular}{|cp{0.9\textwidth}c|}
\hline&&\\
&
Consider five-dimensional Minkowski space, $\ud s^2 = -\ud X_0^2 + \sum^{4}_{i=1}\ud X_i^2$
and define the hyperboloid $-X_0^2+\sum^{4}_{i=1}X_i^2=a^2$. Then, set 
\be
	X_0 = \sqrt{a^2-r^2}\sinh\frac ta\,,\quad X_1 = \sqrt{a^2-r^2}\cosh\frac ta\,,\quad\sum^{4}_{i=2}X_i^2 = r^2\,,
\ee
so that the induced metric on the hyperboloid is
\be
	\ud s^2 = -\l(-\frac{r^2}{a^2}+1\r)\ud t^2 + \frac{\ud r^2}{\l(-\frac{r^2}{a^2}+1\r)} +r^2\l(\ud\theta^2+\sin^2\ud\varphi^2\r),
	\label{deSitter2}
\ee
and $a$ is called the de Sitter radius. Alternatively, the change of coordinates
\be
	X_0 = a\sinh\frac Ta\,,\quad\sum^{4}_{i=1}X_i^2 = a^2\cosh^2\frac Ta = R^2\,,
\ee
 yields the following form of de Sitter solution \eqref{deSitter}:
\be
	\ud s^2 = -\ud T^2+a^2\cosh^2\frac Ta\l[\ud R^2 + \sin^2R\l(\ud\theta^2+\sin^2\ud\varphi^2\r)\r].
		\label{deSitter3}
\ee
This metric is manifestly regular everywhere, there is no singularity at $r=a$. 
&\\
&&\\
\hline
\end{tabular}
	\label{Table:deSitter}
\end{table}

However, Einstein dimissed this as the metric was no longer static, and the coordinate transformation to the regular form was singular at the de Sitter radius. Resolution would wait until 1925, as Lema\^itre 
was studying the motion of nebul\ae, and described it by the de Sitter metric in another coordinate system
\be
	\ud s^2 = -\ud \tilde{T}^2+e^{2\frac Ta}\l[\ud r^2 +r^2\l(\ud\theta^2+\sin^2\ud\varphi^2\r)\r],
	\label{Lemaitre1}
\ee
which was manifestly regular at the de Sitter radius. It was then understood that, inside the de Sitter radius, this metric described a Universe in expansion and the $r$-coordinate of \eqref{deSitter} was \emph{time-like}. Later, in 1932, the same attempted to describe the collapse of a nebula in an expanding Universe. He found the following exterior solution
\be
	\ud s^2 = -\ud\tau^2 + \frac{2m\ud \rho^2}{\l[\frac32\sqrt{2m}\l(\rho-\tau\r)\r]^{\frac23}}+\l[\frac32\sqrt{2m}\l(\rho-\tau\r)\r]^{\frac43}\l(\ud\theta^2+\sin^2\ud\varphi^2\r).
	\label{Lemaitre2}
\ee

This is a spherically symmetric solution of Einstein's equations in vacuum, and so Lema\^itre understood that it had to be related to Schwarzschild solution by some change of coordinate, after Birkhoff's theorem\footnote{This theorem was proven by Birkhoff in 1923, \cite{Birkhoff:1923}. It states that a spherically symmetric solution to Einstein's equations in the vacuum must necessarily be static. If asymptotically flat boundary conditions are imposed (no cosmological constant term), then the exterior solution must be described by Schwarzschild solution. It was pointed out by Deser recently that a Norwegian physicist, Jebsen, had also given a demonstration two years earlier in 1921, \cite{Jebsen:1921}.}. Moreover, one now has $r=\l[\frac32\sqrt{2m}\l(\rho-\tau\r)\r]^{\frac23}$, and the locus $r_S=2m$ is perfectly regular. However, as Lema\^itre published his results in French and in a Belgian journal, \cite{Lemaitre:1933}, they were not advertised in the relativistic community until much later.

From there on, it was clear, at least in the (small) Cosmology community, that Einstein's theory should be approached in a radically different frame of mind from Newton's theory. On top of allowing the existence of non-trivial spacetime topologies in absence of any matter, the fact that spacetime could be curved \emph{globally} as in de Sitter space, be closed, flat or open, and not just \emph{locally} by a distribution of matter (just as waves ripple on the surface of a lake when a stone is thrown in), was a revolution in itself, which would start the study of dynamic Cosmology. Yet, this idea would not prevail in relativistic circles before some time yet, as the concept of black hole was not yet ready to be birthed. A further step towards this was taken by the quantum and nuclear physics community.

Indeed, from the 1930s on, studies were carried on to understand what happened during gravitational collapse. Combining gravitational results with the Pauli exclusion principle from quantum mechanics, it was found out by Chandrasekhar as soon as September 1930, \cite{Chandrasekhar:1930}, that white dwarfs have a critical mass of around $1.4$ solar masses beyond which the electron pressure could not counterbalance the gravitational pressure and the star would at some point collapse on itself. Later, after Rutherford had postulated the existence of the neutron in 1931 and Chadwick had discovered it in 1932, \cite{Chadwick:1932ma}, Oppenheimer and Volkov reiterated Chandrasekhar's calculations on maximal critical masses, but applied them to neutron stars (understood as possible endstates of stellar evolution), \cite{Oppenheimer:1939}. They found that a neutron star would collapse on itself, should it weigh more than approximately six solar masses. It was then that Oppenheimer and Snyder wrote about the possibility for an astrophysical body to actually cross its Schwarzschild radius, \cite{Oppenheimer:1939ue}:
\begin{quotation}
	``When all thermonuclear sources of energy are exhausted a sufficiently heavy star will collapse\dots the radius of
the star approaches asymptotically its gravitational radius; light from the surface of the star is progressively
reddened, and can escape over a progressively narrower range of angles\dots The total time of collapse for
an observer comoving with the stellar matter is finite\dots an external observer sees the star asymptotically
shrinking to its gravitational radius.''
\end{quotation}
The way was now clear for a new generation of relativists, who, in the years 1950-1970, would bring about a second relativistic revolution, the black-hole revolution.


\subsubsection{Crossing the horizon}

In 1958, Finkelstein reintroduced Eddington's coordinates and showed that they described ingoing light rays, \cite{Finkelstein:1958zz}, after emphasis had been put by Synge in 1950 that, in order to study such spacetimes, one should learn the fate of lightcones, \cite{Synge:1950gk}.  The main point is that, in Schwarzschild coordinates, null geodesics cannot cross the horizon: the closer one gets to this null surface, the more the redshift increases, up to the point where it diverges. 
However, reexpressing Eddington's coordinates \eqref{Eddington} in light-cone coordinates,
\be
	v=\bar t + r = t + r^\star\,, \quad r^\star = r+2m\ln\l|\frac r{2m}-1\r|\,,
	\label{Tortoise}
\ee
the metric becomes
\be
	\ud s^2 = -\l(1-\frac{2m}r\r)\ud v^2+2\ud v\ud r+r^2\l(\ud\theta^2+\sin^2\theta\ud\varphi^2\r),
	\label{EddingtonFinkelstein}
\ee
and this describes ingoing light rays. 
The ingoing light rays are diagonal lines penetrating straight through the Schwarzschild radius, while outgoing light rays cannot cross it. The same transformation may be applied to outgoing light rays, with the opposite results. Thus, it was finally understood in relativistic circles that the Schwarzschild singularity was no true singularity at all, but simply an artifact of the set of coordinates used to describe this spacetime. As the star collapses, it passes at some point through its Schwarzschild radius, and then none of the lightrays it emits can ever reach an observer sitting in the outside region. The horizon is formed as a perfect \emph{mathematical} surface, independently of the fate of the star in the inside radius.

Yet, one issue remained: it was related to the concept of what is now called \emph{maximal extension} and had to do with the fact that it seemed impossible to describe the whole of Schwarzschild spacetime using only one set of coordinates. Indeed, one had to use two separate sets, one for ingoing lightrays, the other for outgoing ones: both could not be straightened out using Eddington-Finkelstein advanced or retarded null coordinates. That is, a solution describing both regions, the inside and the outside of Schwarzschild radius, was still lacking. Moreover, an explanation to Flamm and Weyl's double asymptotic structure had yet to be provided. Both answers to these came from Kruskal in 1959, \cite{Kruskal:1959vx}, and, independently, from Szekeres in 1960, \cite{Szekeres:1960gm}.

They both proposed a change of coordinates, see \Tableref{Table:SchwMaximal}, which implemented the concept of maximal extension. Indeed, in Schwarzschild coordinates \eqref{Schwarzschild}, it is easily seen that null geodesics stop at $r_S=2m$ and thus are not complete. Again, in Eddington-Finkelstein coordinates \eqref{EddingtonFinkelstein}, they cover the coordinate range $0<r<+\infty$, but only for either the ingoing or outgoing lightrays. The set of coordinates introduced by Kruskal and Szekeres solved this issue and provided an extension of any null geodesic over the whole range of coordinates. Such an extension is called \emph{maximal}. It has to be distinguished from \emph{geodesically complete} spacetimes for which geodesics are extended to infinite values of their affine (intrinsic) parameter: this is not the case for Schwarzschild's spacetime as one end of radial geodesics will always terminate on a true, curvature singularity. 

\begin{figure}[t]
\begin{center}
	 \includegraphics[width=0.45\textwidth]{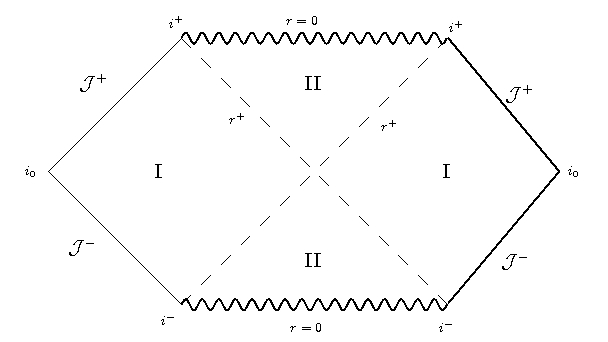}
	 \caption[Penrose-Carter diagram of Schwarzschild black hole]{Penrose-Carter conformal diagram of Schwarzschild's spacetime.}
\label{Fig:SchwMaximal}
\end{center}
\end{figure}

A further step to facilitate the treatment of asymptotics is to make use of the fact that conformal transformations such as $\tilde{g}_{\mu\nu}=\Omega^2g_{\mu\nu}$ do not change the causal structure of spacetime, as they leave the null geodesics invariant. This was introduced by Penrose, \cite{Penrose:1962ij,Penrose:1964ge}, so as to be able to carry out computations at asymptotic infinity. Some of the original motivation came from the definition of energy in General Relativity: since general covariance holds, it is impossible to give a local definition of energy. The metric carries both a background and a dynamical part, which cannot be disentangled. As energy is usually associated with dynamics, one can intuitively understand the issue. However, it turns out to be possible to define globally conserved charges, and in particular energy. Isolated systems suggested to compute such quantities far away, and thus the necessity of precising  what ``being at infinity'' meant arose. The method crafted by Penrose consisted in using conformal transformations to ``bring infinity'' to a finite locus. Using the following coordinate transformation:
\bea
	T\pm X&=&\half\tan\l(\psi\pm\xi\r)\,,\quad \Omega^{-2} =\frac{32m^3}{r}\frac{\e^{-\frac{r}{2m}}}{\cos^2\l[\half\l(\psi+\xi\r)\r]\cos^2\l[\half\l(\psi+\xi\r)\r]}\,,  \label{SchwPenroseCoords}\\
	\ud\tilde{s}^2& =& -\ud\psi^2+\ud\xi^2 + r^2\Omega^{2}\l(\ud\theta^2+\sin^2\theta\ud\varphi^2\r)\,,\quad \l(\psi,\xi\r)\in\l[-\frac\pi2,+\frac\pi2\r]^2, \label{SchwPenrose}
\eea
the unphysical metric $\tilde{g}_{\mu\nu}$ has now a finite locus corresponding to asymptotic infinity of the physical metric $g_{\mu\nu}$. Looking at the right panel of \Figref{Fig:SchwMaximal}, where such a Penrose-Carter diagram for Schwarzschild spacetime is displayed, \cite{Carter:1966zz}, one can now define several kinds of asymptotics: 
\begin{itemize}
 \item $\J^\pm$ are hypersurfaces representing asymptotic future/past null infinity;
 \item $i^\pm$ are points\footnote{Two-spheres in reality, since two spatial dimensions are suppressed in Penrose-Carter diagrams.} standing for future/past time-like infinity, $i^0$ space-like infinity.
\end{itemize}
This can be readily seen by determining the $(T,X)$ locus corresponding to $\psi\pm\xi=\pm\frac\pi2$ for instance. Note that, on the one hand, all time-like trajectories in region $I$ start at past time-like infinity $i^-$ while they end on the future time-like infinity $i^+$. In region $II$, they start and end at the two copies of $i^+$, though now they are space-like, time and space have been reversed. On the other hand, space-like trajectories all end at $i^0$. Ingoing and outgoing null geodesics are both straight diagonal lines, and the horizon (past and future) at $r_S=2m$ is manifestly a null hypersurface. Finally, there is both a space-like past and future singularity. Thus, having exchanged the role of time and space inside the Schwarzschild radius, space (which is now time) can but flow, and so one falls inexorably in the singularity: the singularity is the endpoint, past or future, of all time-like trajectories, in region $II$ or $II'$. 

This also provides a global definition for black holes, instead of depending upon a local system of coordinates. A spacetime will contain a black hole if there is a region of spacetime causally disconnected from future infinity. The horizon of the black hole is the boundary of such a region.

To conclude, although Kruskal-Szekeres' maximal extension has allowed to cross the horizon both for ingoing and outgoing lightrays, or in other words to extend geodesics from $r=0$ to $r\to+\infty$, Schwarzschild spacetime may not be analytically continued across the singularity at $r=0$: it is geodesically incomplete. 

\begin{table}
\centering
	\caption{Maximal extension of Schwarzschild spacetime.}
\begin{tabular}{|cp{0.9\textwidth}c|}
\hline&&\\
&
Consider the $\theta=\frac\pi2$, $\varphi=ct$ of Schwarzschild spacetime \eqref{Schwarzschild} in lightcone coordinates, introducing both of Eddington-Finkelstein advanced and retarded coordinates $u,v$:
\bsea
	\ud s^2 &=& -\frac{2m}r\e^{-\frac r{2m}}\e^{\frac{v-u}{4m}}\ud u\ud v\,,\\
	&=& -\l(1-\frac{2m}r\r)\ud u\ud v\,, \quad u,v=t\mp r^\star\,,
\esea
where $r^\star$ is Wheeler's ``tortoise'' coordinate \eqref{Tortoise}. Although the $u,v$ coordinates now both go from minus to plus infinity, the metric itself is still singular across $r=2m$ since it changes sign. Then, setting $U=-\e^{-\frac u{4m}}\in\l[-\infty,0\r]$, $V=\e^{\frac v{4m}}\in\l[0,+\infty\r]$, the metric is
\be
	\ud s^2 = -\frac{32m^3}r\ud U\ud V\,, 
\ee
which is regular everywhere except at $r=0$, and one can \emph{extend} the $U,V$ coordinates to the whole plane $U,V\in\l[-\infty,+\infty\r]$, notwithstanding the fact that the coordinate transformations we used were not necessarily defined there (General Covariance). Going back to Minkowski-like coordinates $T=\frac{U+V}2,X=\frac{V-U}2$, the metric reads in \emph{maximally extended} Kruskal-Szekeres coordinates
\be
	\ud s^2 = -\frac{32m^3}r\e^{-\frac r{2m}}\l(-\ud T^2+\ud X^2\r) + r^2\l(ud\theta^2+\sin^2\theta\ud\varphi^2\r).
	\label{SchwKruskal} 
\ee
Then, the spacetime diagram of the solution in this form is shown in the left panel of \Figref{Fig:SchwMaximal}.
&\\
&&\\
\hline
\end{tabular}
	\label{Table:SchwMaximal}
\end{table}

\subsubsection{The power of a name}

Now that the object described by Schwarzschild solution is well understood, it is worth our while to linger a little more before introducing the whole crowd of subsequent generalisations that followed, and to tell the story of a name. The name ``black hole'' did not come about before 1967, when it was devised by Wheeler. Such a late occurence can seem surprising nowadays but reflects the lack of understanding of the solution until that time. This is even more obvious when one studies the various names by which it went about: to describe Schwarzschild radius, people spoke of a ``singularity'', a ``catastrophy'' (Hadamard), a ``magic circle'' (Eddington)... All of these denominations suggested that something terrible was happening coming upon this locus, and were above all \emph{lapsus}, since, as pointed out by Eisenstaedt, \cite{Eisenstaedt:2007}, they betrayed how one thought of this place, and thus contributed to fix its meaning in the community's minds.

As the topological signification of Schwarzschild radius evolved and then was understood at the turn of the fifties, expressions like ``photon well'', ``wormhole'' or  ``matter horn'' popped up. But they still failed to encompass all that the solution represented. ``Fixed star'', or ``collapsed star'' were not more satisfactory to Wheeler, who devised the name ``black hole'' in 1967, as is related by Thorne. He would thereafter use no other name, and he popularised it in such a way that it is nowadays unanimously accepted, all traces of its controversial history erased.

Names have power, but this power is intimately linked to the understanding we have of the object they describe. So, from now on and without further reservations, black holes!

\subsection{Charged black holes}

\begin{figure}[t]
\begin{center}
\begin{tabular}{cc}
	 \includegraphics[width=0.45\textwidth]{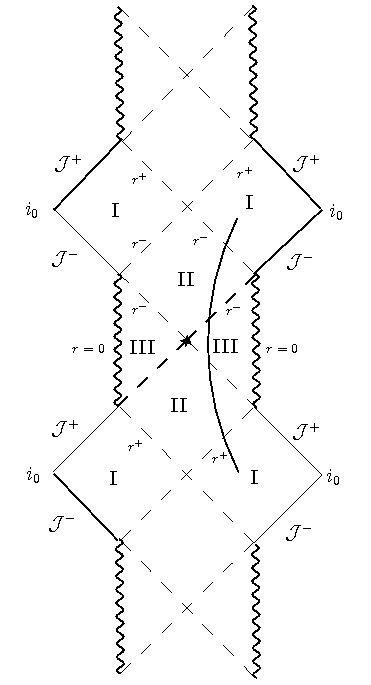}&
	 \includegraphics[height=0.45\textheight]{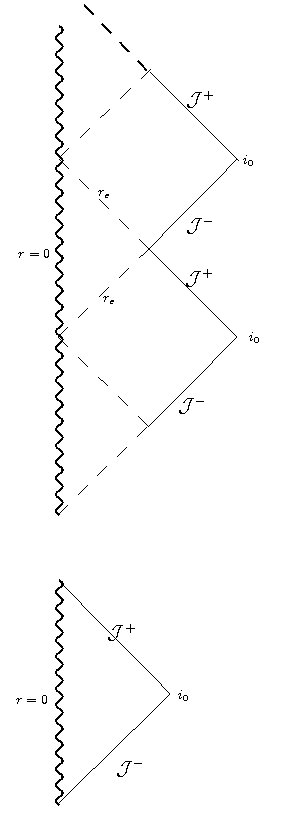}
\end{tabular}
\caption[Penrose-Carter diagrams for Reissner-Nordstr\"om spacetime]{Penrose-Carter diagrams for Reissner-Nordstr\"om spacetime, for the case with two horizons on the left, and for the extremal black hole and naked singularity on the right.}
\label{Fig:RNPenrose}
\end{center}
\end{figure}

There is an obvious and easy generalisation of Schwarzschild spacetime, including Maxwell's electromagnetism and describing the electric field of a charged point-mass. One has to modify Einstein's equations \eqref{EinsteinEq} with a Maxwell contribution in the stress-energy tensor
\be
	T_{\mu\nu}^M = \half F^{\;\rho}_\mu F_{\nu\rho}-\frac{g_{\mu\nu}}8 F^2\,,
	\label{MaxwellStressEnergy}
\ee
and one also has to add Maxwell equation,
\be
	0=\nabla_\mu\l(\sqrt{-g} F^{\mu\nu}\r), \label{MaxwellEq} 
\ee
which now yields the Reissner-Nordstr\"om black hole, \cite{Reissner:1916,Nordstrom:1916},
\bsea
	\ud s^2 &=& -\l(1-\frac{2m}{r}+\frac{q^2}{r^2}\r)\ud t^2 +\frac{\ud r^2}{\l(1-\frac{2m}{r}+\frac{q^2}{r^2}\r)}+r^2\l(\ud\theta^2+\sin^2\theta\ud\varphi^2\r), \slabel{RNmetric} \\
	A&=&\l(\Phi-\frac{2q}r\r)\ud t\,. \slabel{RNA}
	\label{RN}
\esea
It has a curvature singularity, and the number of event horizons depends on the number of roots of the $g_{tt}$ metric element (to which we will also refer in the remainder of this text as the black hole/blackness potential/function). Namely, if 
\begin{itemize}
 \item $m^2>q^2$, there are two roots at $r^\pm = m\pm\sqrt{m^2-q^2}$. The outer root at $r^+$ is an event horizon, and the region outside of it is static and asymptotically flat. The region between the inner and the outer horizon is time-like, and so an observer falling into the hole has to cross to the interior region, which is space-like again. Thus, although the curvature singularity sits there at $r=0$, it is time-like and can be avoided by an observer following a time-like worldline.
 \item $m=\pm q$, there is a single double root, $r_e=m_e=q$. The black hole is called extremal, but there is no event horizon as the $tt$- and $rr$-metric elements cannot change sign any longer. Yet, this spacetime can still be interpreted as a black hole, since the causal past of any given geodesic at null future infinity is bounded by a null surface, which is now called a \emph{Killing horizon}. This is a good opportunity to stress out the difference between an event horizon and a Killing horizon: the latter does not involve a change of nature of two of the coordinates, that is a reversion of time and space.
 \item $m^2<q^2$, there is no root, and this spacetime is a naked singularity.
\end{itemize}
The maximal extension and causal structure of the various cases of the Reissner-Nordstr\"om solution can be found in \Figref{Fig:RNPenrose}, and was presented in 1960 by Brill and Graves for the non-extremal case, \cite{Graves:1960zz}, and later by Carter for the extremal case, \cite{Carter1966423}.

This solution also has a non-trivial gauge field, with a constant limit at infinity, its electric potential. This constant is usually arbitrary and is part of the gauge invariance of $U(1)$ Maxwell theory. However, contrarily to vacuum spacetime, it cannot be set to zero. Indeed, the following quantity would then be singular on the outer horizon of the hole, \cite{Gibbons:1976ue},
\be
	A_\mu A_\nu g^{\mu\nu} = \frac{\l(\Phi-\frac{2q}r\r)^2}{\l(1-\frac{2m}{r}+\frac{q^2}{r^2}\r)}\underset{r\to r^+}{\longrightarrow}+\infty\,.
\ee
This can be remedied by generically taking the gauge field to be zero on the horizon
\be
	\Phi = \frac{2q}{r^+}\,.
	\label{RNPotential}
\ee
This is what we will systematically do in all charged solutions considered subsequently.

\subsection{Black holes embedded in constant curvature spacetimes}

Using Einstein's equations supplemented by a cosmological constant \eqref{LambdaEinsteinEq}, there are two different backgrounds one can consider, depending on the sign of $\Lambda$, but always with the same metric \eqref{deSitter}:
\begin{itemize}
 \item if $\Lambda>0$, then this is de Sitter spacetime, whose symmetry group is no longer Poincar\'e, $SO(3,1)$, but $SO(4,1)$. We have already seen that this is regular and can be embedded into five-dimensional Minkowski spacetime, \Tableref{Table:deSitter}.
 \item if $\Lambda<0$, then this is Anti-de Sitter spacetime, whose symmetry group is $SO(3,2)$. It has a boundary.
\end{itemize}
As already stated, both solutions are written the same, whatever the sign of $\La$, and can be generalised to an $m\neq0$ black hole spacetime, \cite{Kottler:1918},
\be
	\ud s^2 = -\l(-\frac{\Lambda}{3}r^2+1-\frac{2m}r\r)\ud t^2 + \frac{\ud r^2}{\l(-\frac{\Lambda}{3}r^2+1-\frac{2m}r\r)} +r^2\l(\ud\theta^2+\sin^2\theta\ud\varphi^2\r),
	\label{Kottler}
\ee
with $\La>0$ ($<0$) for de Sitter (Anti-de Sitter). We will define and use throughout the rest of the manuscript the de Sitter and Anti-de Sitter radii, as follows,
\bsea
	\La &=& \frac3{a^2}\,,\qquad \La>0\,,\\
	-\La &=& \frac3{\ell^2}\,,\qquad \La<0\,,
\esea
and denote for short dS (de Sitter) and AdS (Anti-de Sitter).


	\subsubsection{Positively curved backgrounds and de Sitter black holes}
	
We have already seen in \Tableref{Table:deSitter} how de Sitter spacetime \eqref{deSitter} could be embedded in five-dimensional Minkowski and thus was perfectly regular across the horizon $r=a$. From the form of the metric in $(T,R)$ coordinates \eqref{deSitter3}, its topology is $\mathbf{R}\times\mathbf{S}^3$. For completeness, we will quickly go over its Penrose-Carter diagram and point out the main differences with Minkowski, and then go on to the Schwarzschild-de Sitter black hole living in this background.

\begin{figure}[t]
\begin{center}
\begin{tabular}{cc}
	 \includegraphics[width=0.35\textwidth]{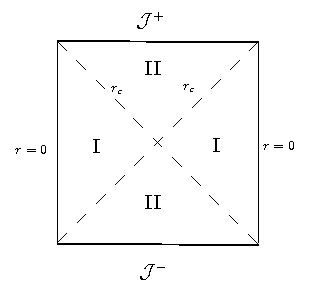}&
	 \includegraphics[width=0.55\textwidth]{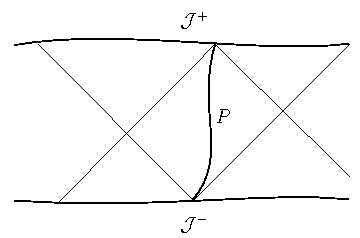}
\end{tabular}
\caption[Conformal structure and particle horizons in de Sitter space]{Left panel, Penrose-Carter diagram for de Sitter space; right panel, future and past particle horizons in de Sitter.}
\label{Fig:DeSitter}
\end{center}
\end{figure}

In order to study dS infinity, the following coordinates are introduced:
\bsea
	T'&=&2\arctan\l[\exp\l(\frac Ta\r)\r]-\frac\pi2\,,\qquad -\frac\pi2<T'<\frac\pi2\,,\\
	\ud s^2&=& a^2\cosh^2\l(\frac{T'}a\r)\l[-\ud T'^2 + \ud R^2 +\sin^2R\l(\ud\theta^2+\sin^2\theta\ud\varphi^2\r)\r].
	\label{deSitter4}
\esea
This shows that there is no global time-like Killing vector in dS, and that dS is conformally equivalent to the region $ -\frac\pi2<T'<\frac\pi2$ of Einstein's static universe\footnote{Remember that Minkowski is a diamond embedded in the same cylinder, with $i^\pm$ at $T'=\pm\pi,R=0$ and $i^0$ is at $T'=0,R=\pi$.}.
De Sitter spacetime's Penrose-Carter diagram is drawn in \Figref{Fig:DeSitter}, and takes the form of a square, with horizontal lines depicting constant $T$ lines and vertical ones constant $R$ lines. Null lines are not straight lines inclined at $45^\circ$ degrees as for Minkowski, but hyperboloids. Time-like and null lines have a space-like infinity, both future (top horizontal line,  $T=+\infty$) and past (bottom horizontal line,  $T=-\infty$). As we will shortly see this allows particle (or cosmological) horizons on top of event horizons. This is quite different from Minkowski, where all time-like geodesics started from $i^-$ and ended at $i^+$, space-like geodesics started and ended at $i_0$, while $\J^\pm$ were null surfaces. 

Particle horizons arise in the following sense: consider a family of particle timelines, following time-like geodesics. They originate on $\J^-$ and end on $\J^+$. Given an observer $O$ sitting at some point $p$ along one of these lines, he will only be able to observe a fraction of the other particle timelines, those originating in the projection of its past null cone on $\J^-$. All the other particle timelines originating somewhere on $\J^-$ but outside this projection will be invisible to him. By taking the intersection $p^+$ of $O$'s worldline with future space-like infinity, $\J^+$, one can define a \emph{future event horizon} for this worldline: this will be the boundary of the causal past of $O$'s worldline, that is the region in dS spacetime outside the past null cone drawn from $p^+$, see \Figref{Fig:DeSitter}. In the same way, one can also define \emph{past event horizons}. This differs greatly from the situation in Minkowski's spacetime: since there $\J^-$ is null, an observer on a time-like geodesic will always see the whole spacetime in its past lightcone. However, accelerated observers in Minkowski (Rindler observers) will experience the same phenomenon, although no black hole is present.

\begin{figure}[t]
\begin{center}
\begin{tabular}{c}
	 \includegraphics[width=0.85\textwidth]{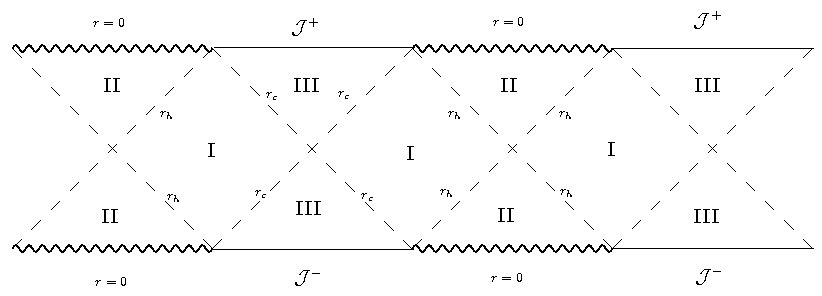}
\end{tabular}
\caption[Penrose-Carter diagram for Schwarzschild-de Sitter black hole]{Penrose-Carter diagram for Schwarzschild-de Sitter black hole, \cite{Gibbons:1977mu}.}
\label{Fig:SchwDS}
\end{center}
\end{figure}

De Sitter metric is easily generalised to a black hole metric, \eqref{Kottler}, and there is a curvature singularity at $r=0$ while putative horizons will be given by the zeros of the black-hole potential
\be
	V(r_h) =-\l(\frac{r_h}a\r)^2+1-\frac{2m}{r_h} = 0\,.
\ee
Though no simple analytic expression can be obtained\footnote{It amounts to solving a third-order polynomial, for which explicit expressions are not very enlightening.}, studying the location of the minimum of this function yields their number. Indeed, if $a^2>27m^2$, or equivalently $9\La m^2<1$, there are two positive non-degenerate zeros at $r_h$ and $r_{c}$. $V(r)$ is positive for $r_h<r<r_c$, negative otherwise, and so the Killing vector $\frac{\partial}{\partial t}$ is time-like only for $r_h<r<r_c$. Said otherwise, the latter is the only region where spacetime is static. There is an event horizon at $r_h$, and a cosmological horizon at $r_c$. The position of the black-hole horizon increases as the parameter $m$ increases, while $r_c$ decreases; conversely, if  $\La$ increases, it is the cosmological horizon that increases while the black-hole one decreases. The Penrose-Carter diagram for this spacetime is drawn in \Figref{Fig:SchwDS}, \cite{Gibbons:1977mu}, and shows a succession of future and past space-like infinities, intersped with curvature singularities for the top and bottom horizontal lines. Diagonal null lines inside define black-hole and cosmological horizons for time-like observers moving in the intermediary region $r_h<r<r_c$.

If $a^2=27m^2$, then there is a single degenerate zero, delimiting two space-like regions of spacetime, one containing the singularity, the other the space-like infinity. Then, an observer moving along some timeline may either
\begin{itemize}
 \item go to future space-like infinity $\J^+$ if he is in a region where the Killing vector is ``outside''-directed;
 \item fall into the singularity at $r=0$  if he is in a region where the Killing vector is ``inside''-directed;
\end{itemize}
this may involve crossing from one causal triangle to another, or bouncing back against the cosmological degenerate horizon.

Finally, if $a^2<27m^2$, then there is a naked future space-like singularity, e.g. a Big Crunch, or a past space-like singularity, e.g. a Big Bang.

	\subsubsection{Negatively curved backgrounds and Anti-de Sitter black holes}
	
	\label{section:AdSspace}

We will start by showing how AdS can be embedded into Minkowski in one dimension higher, same as for dS. From \eqref{Kottler} with $m=0$, one goes to the coordinates
\bsea
	X_0&=&\sqrt{\ell^2+r^2}\cos\frac t\ell\,,\qquad 	X_1=\sqrt{\ell^2+r^2}\sin\frac t\ell\,,\qquad 	\sum_{i=2}^4X_i^2=r^2\,, \\
	-\ell^2&=&- X_0^2 - X_1^2 + \sum_{i=2}^4 X_i^2\,,\slabel{AdSHyperboloid}\\
	\ud s^2 &=& -\ud X_0^2 -\ud X_1^2 + \sum_{i=2}^4\ud X_i^2\,.
	\label{AdS5DEmbedding}
\esea
This shows the anounced properties. Let us continue with the proof of one of AdS space most interesting properties, which has generated a flurry of activity over the last decade\footnote{Over ten thousand citations for the founding papers of the AdS/CFT correspondence, \cite{Maldacena:1997re,Witten:1998qj,Witten:1998zw}\dots But we shall come back to this in Section \ref{section:Holography}.}: AdS space has a boundary which is Minkowski with one less dimension. The structure of the boundary may be exposed by rescaling $X_0\dots X_4\to \la X_0\dots \la X_4$, and then by sending $\la\to+\infty$, so that the boundary verifies
\be
	- X_1^2 + \sum_{i=2}^4 X_i^2=X_0^2\,.
\ee
Two cases arise: either $X_0=0$, and then the boundary is simply the two-sphere $\mathbf S^2$ $\sum_{i=2}^4 X_i^2=X_1^2$, times the point $X_0=+\infty$; or $X_0\neq0$ (and finite), in which case we use it to rescale the other coordinates and get the unit four-dimensional hyperboloid, that is three-dimensional de Sitter space. The topology is then $\mathbf R\times\mathbf S^2$. Adding these two spaces, we have an $\mathbf S^2$ multiplied by a straight line plus a point at infinity, and this yields a circle, so that the topology of the full boundary is $\mathbf S^1\times\mathbf S^2$.

This highlights another characteristic: there can be closed time-like curves $X_0^2+X_1^2=1$. This $\mathbf S^1$ is present explicitly in the metric using the coordinates
\bsea
	X_0&=&\sin\tau\,,\qquad -X_1^2+\sum_{i=2}^4X_i^2=\cos^2\tau\,,\\
	\ud s^2&=&-\ud\tau^2 +\cos^2\tau\l[\ud\chi^2+\sinh^2\chi\l(\ud\theta^2+\sin^2\theta\ud\varphi^2\r)\r].
	\label{AdS2}
\esea
This coordinate set covers only half the space, with $-\frac\pi2<\tau<\frac\pi2$ compactified on a circle and the bounds of the $\tau$-range are coordinate singularities. However, there is no prescription to accept this \emph{possibility} offered by the equations of motion, so we can unwrap the circle $\mathbf S^1$ to a straight line $\mathbf R^1$, its universal covering, and avoid entirely this issue of closed time-like curves in AdS. We shall assume we have done so from now on, and the topology of AdS is then $\mathbf R^4$ instead of $\mathbf S^1\times\mathbf R^3$.

\begin{figure}
\begin{tabular}{cc}
 	 \includegraphics[height=0.4\textheight]{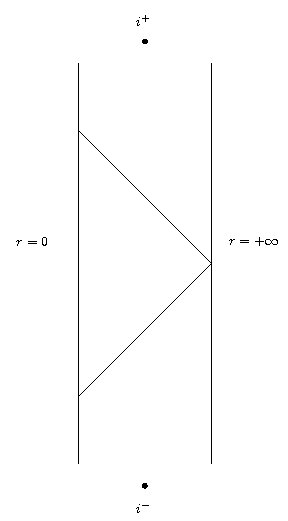}&
 	 \includegraphics[width=0.45\textwidth]{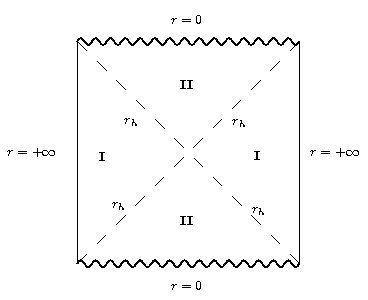}
\end{tabular}

\caption[Penrose diagram of AdS and Schwarzschild-AdS space-time]{Left panel, Penrose diagram of AdS spacetime,\cite{Hawking:1973uf}; right panel, Penrose diagram for Schwarzschild-AdS, \cite{Brill:1997mf}.}
\label{Fig:SchwAdS}
\end{figure}

Another useful set of coordinates is the \emph{Poincar\'e} set, defined from the higher-dimensional hyperboloid as
\bsea
	r&=&X_1+X_2\,,\qquad t=\frac{X_0}{X_1+X_2}\,,\qquad x_{3,4} = \frac{X_{3,4}}{X_1+X_2}\,,\nn\\
	\ud s^2 &=& r^2\l(-\ud t^2 + \ud x_i\ud x^i\r)+\frac{\ell^2}{r^2}\ud r^2\,,
	\label{AdSPoincare}
\esea
which covers only half of the hyperboloid, $r=X_1+X_2>0$. To extend it, one should also consider the lower half-plane $r<0$. This set of coordinates shows a degenerate\footnote{The factor of $r^2$ shows that this is a double zero, thus degenerate.} Killing horizon at $r=0$. This opens the way for \emph{conformal} coordinates:
\be
	r=\frac1z\,, \qquad \ud s^2 = \frac1{z^2}\l(-\ud t^2 + \ud x_i\ud x^i+\ud z^2\r),
	\label{AdSConformal}
\ee
where AdS is manifestly conformally flat at the $z=0$ spatial infinity, but still does not encompass the $z<0$ half-plane. To cover the whole space, we go to the coordinates
\bsea
	X_0^2+X_1^2&=&\cosh^2r\,,\qquad \sum_{i=2}^4=\sinh^2r\,,\\
	\ud s^2&=&-\cosh^2r\,\ud t'^2 +\ud r^2 +\sinh^2r\l(\ud\theta^2+\sin^2\theta\ud\varphi^2\r).
	\label{AdS3}
\esea
The surfaces $t'=constant$ cover the whole space with space-like hypersurfaces. Now that AdS has been maximally extended, we can study its conformal infinity by defining
\bsea
	r'&=&2\arctan\l(\exp r\r)-\frac\pi2\,,\\
	\ud s^2&=&\frac1{\cos^2r'}\l[-\ud t'^2+\ud r'^2 +\sin^2r'\l(\ud\theta^2+\sin^2\theta\ud\varphi^2\r)\r],
	\label{AdS4}
\esea
which is again conformally equivalent to the region $0\leq r'<\frac\pi2$ of the Einstein cylinder. Having unwrapped the circle $t'$, this shows that the boundary at spatial infinity is time-like and has topology $\mathbf  R\times\mathbf S^2$, with an infinite series of the $\mathbf S^1$ contained on the real axis. AdS Penrose-Carter diagram is shown in \Figref{Fig:SchwAdS}, displaying the time-like boundary at null and spatial infinity. Time infinity is displayed as two points, $i^\pm$, but cannot be compactified without destroying the space-like surfaces.

Now that we have unravelled the main properties of the background geometry, let us go over to the black hole case $m\neq0$, \eqref{Kottler}. The behaviour of the solution is controlled by the sign and zeros of the black-hole potential
\be
	V(r_h)  = \l(\frac r\ell\r)^2 +1 -\frac{2m}{r}\,.
	\label{AdSPotential}
\ee
Here again, no enlightening expression can be obtained for the zeros of the potential. However, it is readily seen that it will be space-like for large enough $r$ and time-like for small enough $r$, the same as Schwarzschild solution. Thus, for positive mass parameter $m$, an event horizon is always present, cloaking a singularity at $r=0$. The Penrose-Carter diagram for Schwarzschild-AdS is show in \Figref{Fig:SchwAdS}, and quite resembles Schwarzschild solution's diagram, \Figref{Fig:SchwMaximal}. There are four regions, a future and past event horizon cloaking the space-like curvature singularity. However, the global shape is not a losange as for Schwarzschild but a square, as in the AdS background, and the straight vertical lines represent the conformal boundary at infinity.

\vfill
\pagebreak

\subsection{Topological black holes}

\subsubsection{Spherical topology of the horizon in General Relativity}

We have seen that in the case of the Schwarzschild solution, the horizon metric is the round two-sphere. One could ask the question: how general is this result? Could the horizon be something other than the round two-sphere, and even have a completely different topology? This question was answered for stationary spacetimes by a combination of theorems established by Israel and Hawking. One the one hand, Hawking showed that, for asymptotically flat stationary spacetimes, the horizon of any black hole should have the topology of a two-sphere, \cite{Hawking:1971vc}. On the other hand, Israel showed that for static spacetimes, uncharged (charged) black holes should be isometric to Schwarzschild (Reissner-Nordstr\"om) solution, \cite{Israel:1967wq,Israel:1967za}. This suggests that the only black-hole solution of General Relativity is Schwarzschild (Reissner-Nordstr\"om) solution and that its horizon can only be the round two-sphere. 

However, for non-stationary spacetimes, the theorems are relaxed, and the topology can also be a two-torus, keeping the asymptotic flatness condition, \cite{Gannon:1976}. The topological censorship theorem seemed to put some tension on this result: it states that, in a globally hyperbolic and asymptotically flat spacetime, two causal curves extending from past to future null infinity must be homotopic, \cite{Friedman:1993ty}. Were the topology of the horizon toroidal for instance, a light ray coming from past null infinity, through the hole of the doughnut and then out to future null infinity, could not be homotopic to another light ray going straight from past to future null infinity without passing through the hole, \cite{Jacobson:1994hs}. Moreover, using numerics, black holes were found that had a toroidal horizon during the collapse, before settling into the expected two-sphere, \cite{Hughes:1994ea}. Upon investigation, it was then shown that the doughnut hole closed up faster than light after it was formed, so that topological censorship is preserved, \cite{Shapiro:1995rr}.

We will see in Section \ref{section:EGBBH} that, once extra dimensions are included, the topology of the horizon in Einstein gravity is greatly relaxed. 

More immediately, we will examine the possibility of topological black holes once a cosmological constant is added, and the asymptotic flatness condition is abandonned.


\subsubsection{Topological black holes in Anti-de Sitter space}
\label{section:TopologicalAdSBH}

It is possible to generalise Kottler's solution \eqref{Kottler} to the case where the horizon is not a two-sphere but simply a constant curvature space $\Sigma_\ka$. We will denote its metric by $\ud\sigma_\ka^2=\ud\theta^2+\si^2\theta\ud\varphi^2$, with
\be
	 \si\theta = \begin{cases}
\sin\theta \, ,&\ka=1\,;\\
\theta\ ,&\ka=0\,;\\
\sinh\theta\ ,&\ka=-1\,.\\
\end{cases}
\label{HorizonKappa}
\ee
so that the constant curvature space topology is respectively a two-sphere, a two-torus and a two-hyperbolic space with curvature $\ka=+1,0,-1$ and its isometry group is $SO(3)$, $E^2$ and $SO_c(2,1)$ (the connected component of $SO(2,1)$). Then, it is a matter of calculation to show that the solution \eqref{Kottler} is generalised to\footnote{Including electric charge.}
\be
	V(r) = -\frac{\La}{3}r^2 + \ka -\frac{2m}{r}+\frac{q^2}{r^2}\,.
	\label{BHPotKappa}
\ee
One sees quickly that
\begin{itemize}
 \item if $\La=0$ and $\ka=0,-1$, then $V(r)$ has no real zeros and so in General Relativity, for static spacetimes, no other black-hole solution than Schwarzschild is permitted.
 \item if $\La>0$ and $\ka=0,-1$, the same happens and one only has Schwarzschild-de Sitter solution with $\ka=1$; or one has to consider $m<0$, which does not make much physical sense.
 \item if $\La<0$ and $\ka=0,-1$, horizons are still allowed and we will thereafter focus on this case.
\end{itemize}
We will add charge, keeping an Anti-de Sitter background since this will be the most relevant to later considerations. Let us start by noting that these black holes will not quite be AdS asymptotically, but only \emph{locally} so. Indeed, the asymptotic metric associated to \eqref{BHPotKappa} is
\be
	\ud s^2 = -\l(\frac{r}\ell\r)^2\ud t^2 + \l(\frac\ell{r}\r)^2\ud r^2 + r^2\ud\sigma_\ka^2\,,
	\label{TopologicalBackground}
\ee
which, for $\ka=1$, is the background metric in which the Kottler solution \eqref{Kottler} is embedded, and can be brought to the five-dimensional hyperboloid through \eqref{AdS5DEmbedding}. Note also that for $\ka=0$ planar horizons, this looks exactly like the \emph{Poincar\'e} patch of AdS \eqref{AdSPoincare}, but the topology of the slicing will not be the same ($\mathbf R\times \mathbf R^2$ here). Let us make these notions a little bit more precise by giving a more rigorous definition of what an asymptotically AdS space is, quite elegantly formalised by Skenderis in \cite{Skenderis:2002wp}.

\paragraph{Asymptotically AdS spaces}

Up till now, we have mosly defined AdS spaces through metric representations. But AdS space can also be defined as the hyperbolic (negative) constant curvature space solution to Einstein's equations with a cosmological constant, \eqref{LambdaEinsteinEq}. It is conformally flat, so its Weyl tensor\footnote{The traceless part of the Riemann tensor.} vanishes and its Riemann tensor can be shown to be proportional to the metric using \eqref{LambdaEinsteinEq}
\be
	R_{\mu\nu\rho\sigma} = \frac2{\ell^2}g_{\mu\l[\rho\right.}g_{\left.\sigma\right]\nu}\,,
\ee
where the brackets denote antisymmetrisation with respect to the indices enclosed. Taking a look at AdS space in the coordinates of \eqref{AdS4}, the bulk metric yields the conformal structure of AdS instead of a given boundary metric at $r'=\frac\pi2$: the bulk metric is undefined there and has a double pole. So, let us call a \emph{defining} function $\Omega(x)$, which is positive in the interior of AdS and has a single pole at the boundary. This gives the definition for an equivalence class of conformal metrics:
\be
	\bar g = \Omega^2(x) g\,,
\ee
where $\Omega(r')=\cos r'$ is an example of defining function, but can also be multiplied by any positive-definite function without poles at the boundary.

An asymptotically AdS will be any space that 
\begin{itemize}
 \item is asymptotically a solution of Einstein's equations with a negative cosmological constant which asymptotically has constant (negative) curvature;
 \item has an asymptotically flat conformal structure with topology $\mathbf R\times \mathbf S^2$.
\end{itemize}
We can now turn to the definition of asymptotically \emph{locally} AdS spaces.

\paragraph{Asymptotically locally AdS spaces}

We can generalise the previous definitions to englobe the case of \emph{conformally compact manifolds}. Let $\M$ be a manifold with a boundary $\partial\M$. A metric $g$ defined on $\M$ will be conformally compact if it has a double pole on its boundary and there exists a defining function $r(\M)$
\be
	r(\partial\M)=0\,,\qquad \ud r(\partial\M)  \neq 0\,, \qquad r\l(\M\r)>0\,,
\ee
such that
\be
	\bar g = r\l(\partial\M\r) g
	\label{ConformalClass}
\ee
smoothly extends to $\partial\M$ (thus note that $\bar g$ is a bulk metric and not the induced metric on the boundary). Another quantity that can be defined is
\be
	|\ud r|^2_{\bar g} = \bar g^{\mu\nu}\partial_\mu r\partial_\nu r\,,
	\label{ConformalClassInvariant}
\ee
which has the following two properties: it can be extended smoothly on $\partial\M$ and its value there is independent of the choice of the function $r\l(\partial\M\r)$. To prove the first property, it suffices to note that by definition, $\bar g^{\mu\nu}$ has no pole on the boundary, and $ \ud r\l(\partial\M\r)  \neq 0\Rightarrow \partial_\mu r  \neq 0$. The second property follows from the following remark: \eqref{ConformalClassInvariant} is manifestly reparameterisation-invariant under a change of defining function $r\l(\partial\M\r)\to\tilde{r}\l(\partial\M\r)$. But such a reparameterisation can also be interpreted as having a different bulk metric $\tilde g$, with the same boundary $\bar g$ as previously, but of course a different defining function $\tilde r$,
\be
	\bar g_{\mu\nu} = \tilde r \tilde g_{\mu\nu}\,,
\ee
which effectively defines a conformal equivalent to \eqref{ConformalClass}. Then, using the properties of the Riemann tensor under conformal transformations, one can show that
\be
	R_{\mu\nu\rho\sigma} \l[g\r]= 2|\ud r|^2_{\bar g} g_{\rho\l[\mu\r.} g_{\l.\nu\r]\sigma} + O\l(r^{-3}\r)\,,
\ee
where the leading term is of order $r^{-4}$ near the boundary $r=0$. Inserting this in Einstein's equations, it is straightforward to show that $|\ud r|^2_{\bar g}=
\ell^{-2}$, so that the Riemann tensor of the bulk metric coincides with that of AdS space near the boundary and the bulk metric is Einstein (e.g., satisfies Einstein's equations). The following definition holds: \emph{an asymptotically locally AdS space is a conformally compact Einstein space}. However, the topology of the boundary is left completely unconstrained and can differ greatly from that of AdS.

\paragraph{Playing around with topology}

Now that we have at our disposal a working definition of what an asymptotically locally AdS space is, we can play around with the horizon's topology. We will focus on $3+1$-dimensional black holes, leaving aside for example the B(H)TZ solution, \cite{Banados:1992wn,Banados:1992gq}. For more details, we refer to \cite{Mann:1997iz,Brill:1997mf} and references therein, where such solutions are reviewed (see also \cite{Vanzo:1997gw} for the uncharged case). In our case, the starting point are equations \eqref{HorizonKappa} and \eqref{BHPotKappa}, defining the horizon of the black hole and its topology, and its blackness function with possible horizons sitting where it cancels out. Let us distinguish the three cases,
\begin{itemize}
 \item $\ka=1$: the topology is that of the two-sphere, and the horizon is either the round two-sphere, $\mathbf S^2$,  for the simply connected case or the two-dimensional real projective space, $\mathbf{RP}^2=\mathbf{S}^2/\mathbf{Z}^2$, for the multiply connected case.
 \item $\ka=0$: the topology is that of the plane $\mathbf R^2$, and the multiply connnected cases are the cylinder, the torus, the M\"obius strip and the Klein bottle.
 \item $\ka=-1$: the topology is hyperbolic $\mathbf R^2$, and the space must contain the proper identifications so that there are no conical singularities left. Closed surfaces are Riemann surfaces with genus $\geq2$, the non-closed cases are the cylinder and the M\"obius band, for instance.
\end{itemize}
Asymptotically, these black holes are locally AdS in the sense defined above, except for the $\ka=1$ round two-sphere case which is exactly asymptocally AdS. The spatial infinity is both space-like and null, as stated before for AdS, and the causal structure depends upon the number and nature of horizons. Let us examine the $q\neq0$ case first.

\begin{figure}[ht]
\begin{tabular}{ccc}
\begin{minipage}{0.3\linewidth}
\includegraphics[width=0.95\textwidth]{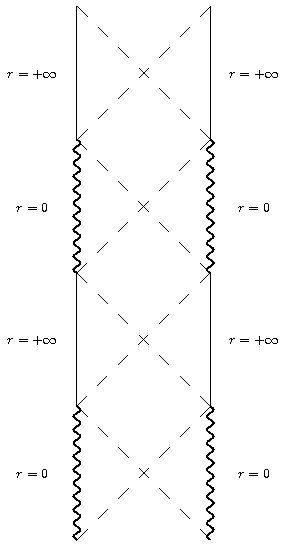}
\end{minipage}&
\begin{minipage}{0.3\linewidth}
\includegraphics[width=0.95\textwidth]{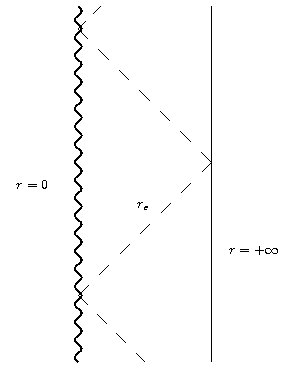}
\end{minipage}&
\begin{minipage}{0.3\linewidth}
 \includegraphics[width=0.95\textwidth]{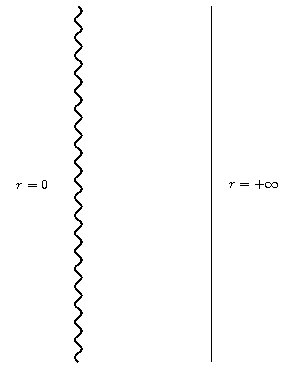}
\end{minipage}
\end{tabular}

\caption[Penrose diagrams of topological Reissner-Nordstr\"om-AdS space-time]{Penrose diagrams for Reissner-Nordstr\"om-AdS ($q\neq0$, $\ka=\pm1,0$) space-time: left pannel, two non-degenerate horizons ($m>m_e$); center, one degenerate ``horizon''  (extremal case, $m=m_e$); right pannel, naked singularity ($m<m_e$), \cite{Brill:1997mf}.}
\label{Fig:RNAdS}
\end{figure}

\begin{itemize}
 \item Two non-degenerate horizons: $m>m_e$ (with $V'(r_e)=0$), independently from the topology ($\ka=\pm1,0$). This is the same as the RN black hole, there is an inner and an outer horizon, and the infinity is doubly connected for each cell, but here it is both space-like and null. The Penrose-Carter diagram in this case is presented in the left panel of \Figref{Fig:RNAdS}.
 \item One degenerate horizon: extremal case, $m=m_e$, $\ka=\pm1,0$. Here, the null and space-like infinity is simply connected, but the spacetime cannot be interpreted as a black hole. Indeed, the Penrose-Carter diagram (\Figref{Fig:RNAdS}, center) shows that the past of the future infinity consists of the entire spacetime, and so there cannot be any event horizon (compare with the RN case, \Figref{Fig:RNPenrose}, upper right panel).
 \item No horizon: $m<m_e$, $\ka=\pm1,0$. There is a curvature singularity (vertical wavy line) and a singly connected null-space-like infinity, see the right panel of \Figref{Fig:RNAdS}.
\end{itemize}

\begin{figure}[ht]
\centering
\includegraphics[width=0.3\textwidth]{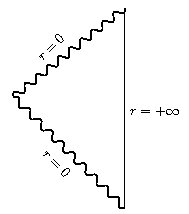}
\caption[Penrose diagram for planar AdS ($q=m=\ka=0$) space-time]{Penrose diagram for ``planar'' AdS ($q=m=\ka=0$), \cite{Brill:1997mf}.}
\label{Fig:PlanarAdS}
\end{figure}

The uncharged case $q=0$ is quite interesting, since it will allow topological effects in full. As the $\ka=1$ case has already been analysed, we will concentrate on the planar $\ka=0$ and hyperbolic case $\ka=-1$.
\begin{itemize}
 \item Planar case: if $m>0$, the structure is the same as Schwarzschild-Ads, \Figref{Fig:SchwAdS}; if $m=0$, there can be no horizons, see \Figref{Fig:PlanarAdS}, but the singularity at $r=0$ is always a coordinate one if $\Sigma_0$ is simply connected\footnote{If not, it depends on the identifications.}.
 \item Hyperbolic case: if $m<m_e$ or $m=m_e$, there is a naked singularity or a single degenerate horizon, described by the same Penrose-Carter diagrams as before; if $m_e<m<0$, there are two non-degenerate horizons  (again, see \Figref{Fig:RNAdS}); if $m\geq0$, the diagram is Schwarzschild-AdS, \Figref{Fig:SchwAdS}, though the singularity is a coordinate one when the inequality is saturated and $\Sigma_1$ is simply connected.
\end{itemize}

\vfill
\pagebreak

\section{Modifications of General Relativity}
\label{section:ModificationsGR}
In the next part, we will study black-hole solutions in various gravity theories. On one hand, they may be interpreted as theories with matter. But on the other hand, we may also consider more profound modifications of General Relativity at both ends of the energy spectrum, either in the Ultra-Violet (high energy, small length scales) or in the Infra-Red (low energy, large length scales). In this sense, General Relativity is really a theory valid at scales not too small but not too large either, stuck in-between.

\subsection{Ultra-Violet divergences in General Relativity}

\label{section:UVModificationsGR}

The topic of UV divergences in General Relativity is a very large and complex one, and would easily fill the contents of several PhDs. So we will just restrict ourselves to a few simple arguments as to what the problem is and what can be (and in some case, has been) done to remedy it. Einstein's equations can be linearised around Minkowski space, and made to display a wave equation for a spin-two, massless particle, aptly named the graviton as it is believed to mediate gravitation, just the same as light can be thought of as both a wave or propagating photons. Since it is massless, gravitation is an infinite-range interaction, contrarily to the weak interaction for instance, which has massive gauge bosons.

\begin{figure}[h]
\begin{center}
\begin{tabular}{lr}
	 \includegraphics[width=0.472\textwidth]{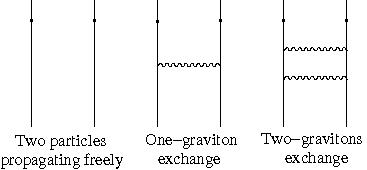}&
	 \includegraphics[width=0.472\textwidth]{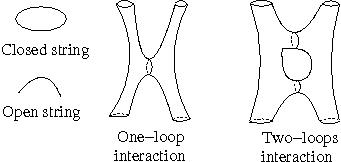}
\end{tabular}
\caption[Graviton exchanges in General Relativity; strings interactions in String Theory]{On the left panel, graviton exchanges in General Relativity; on the right panel, strings interactions.}
\label{Fig:GravitonExchange}
\end{center}
\end{figure}

Now, let us consider a process in which two particles, propagating freely in spacetime, exchange a graviton, \Figref{Fig:GravitonExchange}. Then, each vertex contributes a factor $G_N^{-\half}$, the Newton constant. So each graviton exchange contributes a factor of $G_N^{-1}$ to the amplitude of the interaction. Comparing the ratio between the one-graviton exchange amplitude and the zero-graviton one, at some energy scale $E$, one finds that it must be controlled by the ratio $\l(E/M_P\r)^2$, since the Planck mass $M_P^2=G_N^{-1}=1.22\times10^{19}GeV$ is the only independent energy scale one can define (in units where $\hbar=c=1$). So this one-loop correction will be irrelevant at low energies compared to the Planck scale, particularly at the particle scales (around the $TeV$). However, at scales comparable to or greater than the Planck scale, one clearly needs to take into account the one-loop correction, whose coupling constant diverges as the energy. More generally, as pointed out by Weinberg in \cite{Weinberg:1980gg}, a process of order N with a coupling constant of dimension $\l[mass\r]^d$ will have its amplitude proportional to $\int p^{A-dN}\ud p$, with $A$ a number characterising the interaction. Clearly, since $d=-2$ for gravitation, the high energy interactions are dangerous because their amplitudes can grow without bound if no cut-off is imposed. Carrying out detailled loop-calculations showed that the UV-divergences took the form of arbitrarily high powers of the curvature invariants, signalling the non-renormalisability of the theory, or at least its need for a UV-completion. We will leave aside the issue of renormalisability of Einstein's theory and refer to \cite{Weinberg:1980gg} and focus on the second possibility.

The diverging of N$^{\mathrm{th}}$-order graviton-exchange Feynamn diagrams for high energies in momentum space can be translated in position space: at arbitrarily high energies, the graviton vortices all become coincident. One way to cure this is to smear out the interaction in the UV, that is impose some kind of cut-off at, equivalently, small lengths. This is the idea behind String Theory, \cite{Polchinski:1998rq,Polchinski:1998rr}, where the cut-off is implemented in a natural way: as we look at higher and higher energy, or alternatively at smaller and smaller length, substructures become visible, and point-like particles turn into strings. Interactions now take place over a finite region in position space, see \Figref{Fig:GravitonExchange}, and String Theory constitutes a possible UV-completion of General Relativity, which is recovered at low energies. However, it comes with accessories: there exist at least five versions of String Theory (interrelated by dualities), which live in ten dimensions and reduce to supergravity at low energies. Six of these dimensions are supposed to be compactified, and upon compactification, string theory low-energy actions include the Einstein-Hilbert lagrangian, as a first-order term, but also scalar and gauge matter fields. We will elaborate a little more on these actions in the next section, before moving on to analyse their equations of motion and examining the existence of black-hole solutions.

\subsection{Infra-Red modifications of General Relativity}
\label{section:IRModificationsGR}
\paragraph{The expansion of the Universe is accelerating}

The UV-problem we described in the previous subsection mainly has a theoretical origin, since it does not prevent one in any way from doing valid experiments at ``day-to-day'' energy scales, much below the Planck scale. The IR-problem has, on the contrary, an experimental origin and came as quite a surprise. Riess \emph{et al.}, \cite{Riess:1998cb}, and also Perlmutter \emph{et al.}, \cite{Perlmutter:1998np}, published measurements of SN Ia luminosity that could only be accommodated in the frame of the Standard Cosmological Model if about seventy percent of the total energy contained in the Universe was under the form of some mysterious Dark Energy, modelised as a homogeneous, negative pressure and repulsive fluid driving a late phase of acceleration of the expansion of the Universe.

 These measurements were then confirmed by other sources, such as the Baryonic Acoustic Oscillations, \cite{Eisenstein:2005su}, or the position of the Acoustic Peaks in the Cosmic Microwave Background, \cite{Spergel:2003cb}. 

\paragraph{Cosmological constant problem}

The state of the problem is the following: experiments reveal that a positive and very small extra contribution to the energy content of the Universe exists, $\rho_{DE}=10^{-47}GeV^4$, and that it is best modelised as a perfect fluid, homogeneous and isotropic. Moreover, its equation of state, $p_{DE}=w_{DE}\rho_{DE}$, is measured to be very close to $-1$.

The easiest way to explain this extra energy density is simply to plug a bare cosmological constant, $\La$, in Einstein's equations. This seems the most economical approach, entailing no extra matter content or modifications of General Relativity (understood in the loose sense of Einstein's theory plus a cosmological constant). The late phase of acceleration can then be modelised by a de Sitter Universe and the cosmological constant's energy density and pressure are $\rho_\La=-P_\La=\La/8\pi G_N$. A first issue with this modelisation is the following: though this might appear as a theorist's fancy, it is not quite satisfying to introduce in the theory an \emph{a priori} free parameter, to be fixed solely by experiment. There are very few instances where such a procedure is tolerated, like for instance the value of the electric charge of the electron. In this case, one would rather try and find some justification for this constant. 

A second and more pressing concern comes from quantum corrections: the vacuum is expected to be the scene of unceasing spontaneous creation and annihiliation of electrons and positrons, predicted by Heisenberg's uncertainty principle. However, upon evaluating this quantity, it was quite soon realised that this yielded a huge discrepancy with the expected value from Cosmology, of more than $120$ orders of magnitude. The introduction of Supersymmetry in the game reduced this somewhat, but some $60$ orders of magnitude difference still remains.  A simple argument to highlight the depth of the problem is due to Weinberg, \cite{Weinberg:1988cp}. Let us sum the zero-point energies, $\hbar\omega/2$, of a scalar field of some mass $m$, up to some cut-off $\La\gg m$. Then, this gives a total vacuum energy density
\be
	<\rho> = \int_0^\La\frac{4\pi k^2\ud k}{\l(2\pi\r)^3}\half\sqrt{m^2+k^2} \simeq \frac{\La^4}{16\pi^2}\,,
\ee	
which will obviously be much greater than the measured value for any reasonable cut-off, be it the GUT scale ($E_{GUT}=10^{16}GeV$) or even worse, the Plank scale. 

On top of that, particle physics adds an extra contribution: as the Universe cools down, the symmetries of the Standard Model\footnote{$SU_c(3)\times SU(2)\times U(1)$.} or Supersymmetry\footnote{If it is indeed a symmetry of Nature.}, are broken spontaneously in turn, at various scales (GUT, electroweak and strong interaction scale). This implies that some Goldstone boson settles in a minimum of its potential, effectively breaking the symmetry and acquiring a vacuum expectation value. This vev contributes to the matter stress-energy tensor exactly as a cosmological constant or a vacuum energy would.

In the end, we get an effective cosmological constant, made from adding up the various components: bare cosmological constant, vacuum energy from quantum fluctuations and from spontaneous symmetry-breaking cosmological phase transitions. 

The original problem was a little bit different from its modern counterpart. At first, there only existed an upper bound on the value of the effective cosmological constant, so one could entertain the hope that it was actually zero. The question raised was how one would fit various contributions, and make them all cancel out. Though this promised to prove quite challenging, some symmetry or other might exist which would enforce the cancellation.

However, matters were made quite worse when experiments confirmed that the value of the effective cosmological constant was not zero but positive and very small. An adjustment of this scope does not suggest the existence of a symmetry, and leaves one at a loss.  Even if a suitably \emph{fine-tuning mechanism} was devised, it would seem contrived, unnatural, and lacking the elegance that theorists often look for in their research (which might, admittedly, be a somewhat challengeable attitude\ldots). For a much more detailled account of these matters, we refer to the classic text by Weinberg \cite{Weinberg:1988cp}, or to \cite{Carroll:2000fy} by Carroll for a more recent review.

A second, related problem is that of \emph{coincidence} and is two-fold: how can one explain the recent (in cosmological time) onset of the phase of acceleration, precisely at a time where we (Humanity) are here to measure it? And how come the measured value is of the same order as the matter energy density?

\paragraph{Dark energy? Right-hand side modifications}

A very popular, particle physics-oriented approach to the acceleration problem involves extensive use of scalar fields. The principle is that of a heavy scalar, gently rolling down its potential and simulating the role of a cosmological constant, with a $w_\phi=-1$ equation of state verified in contemporary times. This is different from a cosmological constant where $w_\La=-1$ at all times, and the onset of acceleration is generated by the dilution of matter in the Universe due to expansion. For the slow-rolling scalar field, acceleration is achieved as the field enters a plateau of its potential flat enough so that the above equation of state holds. A plethora of such models exist, and might be traced back to \cite{Caldwell:1997ii}. We will not dwell any longer on these \emph{quintessence} models and instead take a more gravitationally-inclined approach.

\paragraph{Acceleration from geometry. Left-hand side modifications}

The previous paragraph described essentially right-hand side modifications of Einstein's equations, that is modifications of the matter content of the Universe. An alternate view relies upon the liberty to add a cosmological constant on the left-hand side as well, that is considering that the acceleration has a geometrical origin. The game then is either to devise some geometrical mechanism to fix the value of $\La$ or to simulate its effect. We will focus on braneworlds approaches, the heart of which are extra dimensions and which can achieve both of the previous effects.

\subparagraph{Braneworlds: Randall-Sundrum and Dvali-Gabadadze-Porrati scenarii}

There exists models where matter is localised on a four-dimensional hypersurface, named \emph{brane}, while gravity propagates in the entire background spacetime, named \emph{bulk}. The bulk can be five-dimensional, and then the brane is a true hypersurface: these are called codimension-one models. In codimension-two models, the bulk is six-dimensional. Loosely, the codimension is the number of independent vectors normal to the brane. These approaches, thought not uncorrelated, are different from String Theory since the extra dimensions in the former are large and uncompactified.

The first example of codimension-one braneworld theory is the Randall-Sundrum theory, \cite{Randall:1999vf}, where the fundamental idea is to use a warped product between the extra-dimension and the brane to trap the zero-mode of the gravitational fluctuations around the vacuum. The volume of the extra-dimension is finite in this setup, and so the zero-mode is normalisable and can represent a bound-state localised on the brane, that is the graviton. Then, gravity is effectively four-dimensional on the brane, the tower of Kaluza-Klein modes generating a negligible correction to the usual Newtonian potential. Another realisation of these ideas is the Dvali-Gabdadze-Porrati scenario, \cite{Dvali:2000hr}, but here an infinite bulk is implemented, and there is a crossover scale for gravity: at small distances, gravity behaves four-dimensionally, while at large distances, it can be made five-dimensional, that is, weaker. The mechanism does not rely on a normalisable zero-mode dominating the low-energy physics, but rather on resonance of massive gravitons, \cite{Dvali:2000rv}. This scenario was then extended to general codimension setups, \cite{Dvali:2000xg}.

\subparagraph{From General Relativity to Lovelock gravity}

We will conclude this section by motivating the second class of actions we will examine in the second part of this work. The previous scenarii made use of gravity in higher dimensions extensively, based on General Relativity. However, in dimensions higher than four, there is no good reason for restricting the modelisation of gravity to the Einstein-Hilbert action. It was proved that this was the unique theory with a symmetric metric two-tensor, second-order equations of motion and general covariance \emph{in four dimensions}. In more than four dimensions, \emph{this unicity property is lost}! Lovelock proved in the seventies that this desirable trait could be recovered, at the cost of adding in the Lagrangian specific combinations of higher powers of the Riemann tensor and its associated scalar invariants, \cite{Lovelock:1971yv,Lovelock:1972vz}.

\vfill


\part{Black holes in String Theory and Cosmology Inspired Theories}
\label{part:two}
\section{Einstein-Maxwell-Dilaton black holes}

\label{section:EMDBH}

\subsection{Gravity coupled to matter and no-hair theorems}

In Part \ref{part:one} of this work, we have reviewed a collection of black-hole solutions of General Relativity either pure or with the inclusion of a cosmological constant and an abelian gauge field. For a long time these solutions were deemed unphysical, and this was described in detail in the previous part. In a gist, they were seen by Einstein as an unacceptable contradiction to Mach's principle, since they were the undeniable proof that spacetimes with a non-trivial topology could exist independently of any interplay with matter. Moreover, misinterpretation of the coordinate singularity at the Schwarzschild radius prevented the unravelment of its physical meaning, that is the trapping of light inside the event horizon. After black holes were proven to be the endstates of the gravitational collapse of heavy enough stars, and the Schwarzschild solution to be stable against perturbations, \cite{Regge:1957td}, they were finally seen as proper physical solutions representing either an astrophysical body of their own or the exterior solution to a star. Their study became then a fully-fledged field of gravitational research.

Black holes in a sense are similar to solitons in the theory of gravitation: they have a mass and a charge, in the same way as atoms in quantum theories of matter have an atomic mass and number. Moreover, generalisations to spinning uncharged or charged ``point-masses'' were quickly discovered, \cite{Kerr:1963ud,Newman:1965my}. So the question arose as to how many ``hairs'' a black hole could actually have, that is how many independent parameters could characterise the endpoint of stellar evolution. Inspired by unicity theorems for the Schwarzschild, Reissner-Nordstr\"om, Kerr and Kerr-Newman solutions in electrovacuum, \cite{Israel:1967wq,Israel:1967za,Carter:1971zc,Wald:1971iw,Robinson:1975bv,Mazur:1982db,Mazur:1984wz}, Wheeler put forward what came to be known as the ``no-hair conjecture''\footnote{Another proof of Wheeler's ability to come up with names that st(r)uck.}, \cite{Ruffini:1973}, hypothesising that as the star collapsed down to its black-hole endstate, it lost all its hair, that is all information about its constituents apart from its mass, electro-magnetic charge and angular momentum. The particularity of these numbers is that they are all conserved charges which can be computed by a Gaussian-type flux integral, and so be measured from afar (for instance at infinity). Other information, such as baryon number or other kinds of quantum numbers, usually cannot. To pursue our atomic analogy, other kinds of hair would somehow correspond to excited states of the solitons. On the other hand, restriction to a small number of conserved charges would support the thermodynamic interpretation of black holes: black holes hide in their interior a large number of invisible degrees of freedom (hairs that they have shed during the collapse), which can be evaluated by computing a finite, large entropy for the black hole. This was part of Bekenstein's original argument supporting a thermodynamical interpretation of black holes, \cite{Bekenstein:1973ur,Bekenstein:1980}. But this will be the concern of Part \ref{part:three}.

Making crucial use of the hypothesis of asymptotic flatness, a number of no-hair theorems were proven, including the case of massless scalars by Chase \cite{Chase:1970}, massive scalar, vectors and spin-2 matter fields by Bekenstein for non-decreasing positive definite potentials, \cite{Bekenstein:1972ny,Bekenstein:1972ky}, as well as neutral meson, electromagnetic and neutrino sectors coupled to Schwarzschild or Kerr black holes, \cite{Hartle:1971qq,Teitelboim:1972qx}. By hair, we now mean a non-trivial matter field coupled to a black-hole spacetime, abiding boundary conditions of specific interest. Following the course of history, we will focus first on asymptotically flat boundary conditions.

Let us outline the argument for a real minimally-coupled static scalar field due to Bekenstein, as this has particular relevance to what follows, \cite{Bekenstein:1972ny}. The action for such a scalar is written
\be
	S_\phi = -\half\int_\M \sqrt{-g}\l[\phi_{,\mu}\phi^{,\mu}+V(\phi^2)\r],
\ee
and the Klein-Gordon equation derived from it is
\be
	\square\phi-\phi V'(\phi^2) = 0\,.
\ee
Multiplying the previous equation by $\phi$ and integrating over spatial coordinates at a given time-like point $x^0$ of spacetime (so that the $x^i$ are free), one is left after integrating by parts with a three-dimensional integral over spatial directions $\mathcal V$ and a boundary term evaluated on the spatial boundary $\partial\mathcal V$ of $\mathcal V$
\be
	\int_{\mathcal V}\sqrt{-g}\ud^3x\l[g^{ab}\partial_a\phi\partial_b\phi+\phi^2V'(\phi^2)\r]+ \int_{\partial\mathcal V}\phi\partial_a\phi\ud\Sigma^a=0\,,
	\label{NoHairEquation}
\ee
where $\ud\Sigma_a$ is the metric line element of $\partial\mathcal V$. The boundary has two components, an outer one at spatial infinity, and an inner one at the horizon. On the outer boundary, imposing asymptotically flat boundary conditions $\phi=O\l(r^{-1}\r)$ and $g^{rr} = 1+O\l(r^{-1}\r)$, the surface term is seen to vanish. On the horizon, Schwarz inequality states that $|\phi\phi^{,a}\ud\Sigma_a|\leq\sqrt{\phi^2\phi_{,a}\phi^{,a}\ud\Sigma^b\ud\Sigma_b}$, so that since the horizon is a null surface and thus $\ud\Sigma^b\ud\Sigma_b=0$, the inner boundary does not contribute anything either. So, the whole boundary term cancels, as long as the scalar is bounded on the horizon (which is a physically reasonable assumption).

The three-dimensional term remains, and can be seen to cancel for non-negative $V'\l(\phi^2\r)$ if the scalar field is constant or null outside the black hole where the spatial metric components are positive-definite.

These early-years results were subsequently extended to the case of arbitrary positive potentials $V>0$ (instead of $V'>0$), \cite{Heusler:1992ss,Bekenstein:1995un,Sudarsky:1995zg}, for the spherical, static and neutral case, and then to the charged and the non-minimally coupled cases, \cite{Mayo:1996mv}.

Up till now, we have only mentioned scalar hair, and seen that its existence in asymptotically flat situations was quite restricted in a large class of potentials. The reason why most of the attention on no-hair theorems has focused on the scalar case is that for other kinds of matter fields, a number of solutions were obtained early on which quite obviously rendered these theorems obsolete. This was the case with Abelian, \cite{Nordstrom:1916,Reissner:1916}, and non-Abelian gauge fields for instance, \cite{Greene:1992fw}. A first simple and intuitive argument to justify this has to do with gauge invariance, \cite{Bekenstein:1971hc}: indeed, in the massless case, the fundamental vector field is subject to gauge transformations and cannot \emph{a priori} be bounded on the horizon. In the massive Proca case, Bekenstein's argument holds as the mass term breaks gauge invariance, \cite{Bekenstein:1971hc}.

One can also ask what happens with non-asymptotically flat boundary conditions. This topic was spurred on a few years back by the advent of the AdS/CFT correspondence. Without entering (yet) in too many details, since a finite gauge theory on the boundary of AdS is associated with a bulk spacetime of non-zero temperature, e.g. a black hole, it makes sense to consider bulk matter fields which will be related \emph{via} the correspondence to relevant deformations of the CFT on the boundary. Although most of the motivation comes from a negative cosmological constant, we shall first review the positive case. When the scalar field is minimally coupled and for static, spherically symmetric spacetimes, it was proven that no scalar hair could exist in the massless or convex potential case, \cite{Cai:1997ij,Torii:1998ir}. Scalar hair could not be excluded in the general positive semi-definite potential case: numerical regular solutions were found for a double-well potential but turned out to be unstable against linear perturbations. So one may conclude that for asymptotically de Sitter spacetimes some version of the no-hair theorem still holds.

Let us turn now to the case with negative cosmological constant, that is AdS boundary conditions. Numerous papers have been published about it, \cite{Torii:2001pg,Sudarsky:2002mk,Martinez:2004nb,Henneaux:2004zi,Henneaux:2006hk,Hertog:2004ns,Hertog:2004bb,Hertog:2004dr,Hertog:2005hm,Martinez:2006an,Hertog:2006rr,Hertog:2006wj,Amsel:2007im}, exhibiting numerical and analytical hairy solutions, albeit with unsual boundary conditions. Namely, the fall-off towards the asymptotically locally AdS background is slower than usual and results in non-conventional energy definition requiring the inclusion of a scalar contribution, which vanishes in the usual situation\footnote{We shall come back to this issue in greater detail in Section \ref{section:ThermoEMD}. Let us simply state for now that the extra contribution is related to the non-vanishing of boundary terms of the type that is present in \eqref{NoHairEquation}.}. Large classes of such boundary conditions have been studied and have been dubbed ``designer gravity'', in keeping with the fact that their properties depend significantly on the choice of boundary conditions. They are conjectured to be stable, and thus seem in violation of no-hair theorems for asymptotically AdS spacetimes. However, it was conjectured by Hertog that if one adds the requirement that the Positive Energy Theorem (PET) holds, then so do the no-hair theorems as all hairy solutions just mentioned are in violation of the PET.

We will not linger on the non-minimally coupled case, although it has been the subject of continued scrutiny since the 70s, see for instance \cite{Bocharova:1970,Bekenstein:1974sf,Bekenstein:1975ts,Martinez:2002ru,Winstanley:2002jt,Harper:2003wt,Winstanley:2005fu,Martinez:2005di,Dotti:2007cp}.

\pagebreak

\subsection{Einstein-Maxwell-Dilaton theories}

We will now shift focus to a class of theories containing scalar and Abelian gauge fields coupled minimally to gravity, the so-called Einstein-Maxwell-Dilaton (EMD) theories:
\be
	\label{EMDaction}
	S = \int \ud^dx\sqrt{-g}\l[R-\frac12(\partial \phi)^2-\frac14\e^{\ga\phi}F^2-V(\phi)\r],
\ee
where $\ga$, $\da$ are coupling constants, $F$ is the two-form field strength of the Maxwell one-form $A$, and $\phi$ is the scalar field. It is canonically coupled to gravity, and has a potential $V(\phi)$. We will mainly concentrate on Liouville potentials, which have an exponential shape:
\be 
	V(\phi)=2\La\e^{-\da\phi}\,.
	\label{LiouvillePotential}
\ee 
They reduce to a constant in the case when $\da=0$ and are zero when the ``cosmological constant'' $\La=0$. Similarly, for $\ga=0$, the gauge coupling between the scalar and the gauge field is constant. Thus, for $\ga=\da=0$, we may expect to recover the Reissner-Nordstr\"om black hole in Minkowski, dS or AdS spacetime for $\La$ null, positive or negative respectively. We shall give several motivations for studying such theories shortly.

The classical covariant equations of motion derived from this action by varying the matter fields are the following:
\bsea
	0&=&\partial_\mu\l(\sqrt{-g}\e^{\ga\phi}F^{\mu\nu}\r), \slabel{MaxwellEqEMD} \\
	\square{\phi}&=&\frac{\ga}4\e^{\ga\phi}F^2-2\da\La\e^{-\da\phi}\,, \slabel{DilatonEqEMD} \\
	G_{\mu\nu}&=& \half\partial_\mu\phi\partial_\nu\phi-\frac{g_{\mu\nu}}4\l(\partial\phi\r)^2+\half \e^{\ga\phi}F^{\;\rho}_\mu F_{\nu\rho}-\frac{g_{\mu\nu}}8\e^{\ga\phi}F^2-\La\e^{-\da\phi}g_{\mu\nu}\,, \slabel{EinsteinEqEMD}
	\label{EOMEMD}
\esea
with $\square$ the $d$-dimensional d'Alembertian operator.

This allows us to write the Ricci scalar on-shell:
\be
	R = \frac{2}{2-d}T=\half\l(\partial\phi\r)^2+\frac{4-d}{4(2-d)}\e^{\ga\phi}F^2-\frac{2d}{2-d}\Lambda \e^{-\da\phi}.
	\label{RicciScalarOnShell}
\ee
The electromagnetic contribution vanishes as expected for $d=4$ (the Maxwell stress-energy tensor is traceless in four dimensions). Then, the only matter singular points of spacetime will be those present in the scalar field. However, for higher dimensions, there might be a richer variety of singular points, though, in all the solutions we show in the next subsections, all singular points of the Maxwell field are always contained in the dilaton field.

The next question we ask is: Can we relate to the no-hair theorems of the previous section?

In the zero-potential case, they are quite obviously contradicted since an asymptotically flat charged solution exists for all values of the coupling $\ga$ and non-trivial scalar and gauge fields, both of which are regular and bounded on the horizon, see \cite{Gibbons:1987ps} (and also \cite{Garfinkle:1990qj} for the $\ga=1$ case). How is Bekenstein's proof circumvented? Equation \eqref{NoHairEquation} is no longer valid: the potential term is absent and replaced by an effective potential term from the gauge field, due to the non-trivial coupling with the dilaton. This term is \emph{a priori} not positive-definite, so that the three-dimensional integral \eqref{NoHairEquation} can be satisfied with a non-trivial scalar profile. This of course carries over to the non-zero potential case. Furthermore, these solutions were proven to be stable against linear perturbations, \cite{Holzhey:1991bx}, and so may legitimately be considered as hairy black holes. However, we need to moderate this statement as the hair is not of ``primary'' type, that is the Gaussian flux integral built out of the gradient of the dilaton field is not independent from the mass and charge of the black hole. This is referred to as a ``secondary'' hair and does not constitute so serious a violation of no-hair theorems as an independent scalar hair would.

When the potential is not zero but instead as in \eqref{LiouvillePotential}, $V(\phi)=2\La\e^{-\da\phi}$, the previous no-hair theorems are again evaded. So as to connect with previous literature, the potential should be written $\tilde V(\phi)=2\La\l(-1+\e^{-\da\phi}\r)$, once we have subtracted a constant piece and explicitly displayed a cosmological constant in the action. Then, the potentials are never positive-definite, nor do they have local minima. Yet, some version of a no-hair theorem was provided by Wiltshire \emph{et al.}, \cite{Poletti:1994ww,Wiltshire:1994de}.

We introduce \emph{black hole coordinates}, in terms of which the metric is written
\be
	\ud s^2 = -f\l(r\r)\ud t^2 + \frac{\ud r^2}{f\l(r\r)} + R\l(r\r)^2\l(\frac{\ud z^2}{1-\ka z^2}+z^2\ud\Omega_{d-3}^2\r),
	\label{BHcoordEMD}
\ee
which is spherically symmetric but where the horizon can \emph{a priori} have the topology of the sphere, the plane or the hyperbolic plane for $\ka=+1,0,-1$ respectively ($\ud\Omega_{d-3}^2$ is the metric of the round $d-3$-dimensional sphere). In these coordinates, suitable combinations of the equations of motion \eqref{EOMEMD} go as
\bsea
	\frac{R''}{R}&=&-\frac{\l(\phi'\r)^2}{2\l(d-2\r)}\,,\slabel{Ein1EqBHcoordEMD}\\
	\frac1{R^{d-2}}\l[\l(R^{d-2}\r)'f\r]'&=&\l(d-2\r)\l(d-3\r)\frac{\ka}{R^2}-2\La\e^{-\da\phi}-\frac{q^2}{2R^{2\l(d-2\r)}}\e^{-\ga\phi}\,,\slabel{Ein2EqBHcoordEMD}\\
	\frac1{R^{d-2}}\l(R^{d-2}f\phi'\r)'&=&-2\da\La\e^{-\da\phi}-\frac{\ga q^2}{2R^{2\l(d-2\r)}}\e^{-\ga\phi}\,,\slabel{DilEqBHcoordEMD}
	\label{EOMBHcoordEMD}
\esea
where primes denote derivatives with respect to $r$ and we have plugged in an electric Ansatz
\be
	F = \frac{q}{R^{d-2}}e^{-\ga\phi}\ud t\wedge\ud r\,.
	\label{ElectricalAnsatzEMD}
\ee
One more equation can be deduced from \eqref{EOMBHcoordEMD} by a Bianchi identity. Let us first deal with the case of a constant potential, $\da=0$, which is closest to our intuition. We will show, after Wiltshire \emph{et al.}, \cite{Poletti:1994ww,Wiltshire:1994de}, that a black-hole solution with a constant potential cannot have more than one regular horizon.

Suppose first that the Killing vector $\partial/\partial t$ is space-like in the outer region, and so that there exist two horizons. Then, we label them $r_-<r_+$; we also suppose that $r_+$ is non-degenerate, so that $f\l(r\r)\sim r-r_+$ near the outer horizon, and also that it is regular, so that $\phi\l(r_+\r)$ and $A\l(r_+\r)$ are bounded and $R\l(r_+\r)$ is non-zero. Then, evaluating \eqref{DilEqBHcoordEMD} at both horizons, we get
\be
	\l.\l(f'\phi'\r)\r|_{\pm} = -\l.\frac{\ga q^2}{2R^{2\l(d-2\r)}}\e^{-\ga\phi}\r|_{\pm}\,.
\ee
Suppose now that $\ga<0$. Then, since the outer region is space-like, $f\l(r\r)$ must be positive in-between the two horizons, with $f'_->0$ and $f'_+<0$. This implies from the above equation that $\phi'_->0$ and $\phi'_+<0$, so that, $\phi$ being smooth in this region, there must exist an $r_0$ such that $\phi_0'=0$ and $\phi_0''<0$. Coming back to \eqref{DilEqBHcoordEMD} and evaluating it at $r_0$, we find $sign\l(\phi_0''\r)=-sign\l(\ga\r)>0$ by assumption. So we end up with a contradiction, and the argument can also be seen to hold if we suppose in turn that $\ga>0$\footnote{Though this should also be obvious by symmetry arguments on the sign of $\ga$, $\da$ and $\phi$.}. Thus, there cannot exist a solution with more than one regular, non-degenerate horizon, which effectively rules out dS asymptotics. In the case of AdS asymptotics, the outer region is time-like, and a similar argument rules out the existence of more than one horizon.

What is the status for a proper Liouville potential, in the form displayed in \eqref{LiouvillePotential}? This question has been addressed by Wiltshire \emph{et al.}, \cite{Poletti:1994ff}, who carried out a general analysis of the global properties of the phase space for spherically symmetric solutions with a horizon of undetermined topology. Let us use the metric function $R(r)$ in \eqref{BHcoordEMD} as a coordinate, so that the metric is now
\be
	\ud s^2 = -f(R)\ud t^2 + h(R)\ud R^2+ R^2\l(\frac{\ud z^2}{1-\ka z^2}+z^2\ud\Omega_{d-3}^2\r).
\ee
Concentrating on critical points representing the asymptotic spatial infinity $R\to\infty$, one may classify the solutions as in the \Tableref{TableWilt}.
\begin{table}[htb]
\centering
\begin{tabular}{|c|c|c|c|c|}
	\hline
	&  $|f|$ & $|h|$ & $\e^{\phi}$&Solutions\\
	\hline
	$K_{1,2}$ &  $R^{\frac{4(d-3)^2}{(d-2)\ga^2}}$ & $1$ & $R^{-\frac{2(d-3)}\ga}$& \eqref{CHMii}
	\\
	\hline
	$M_{1,2}$ & $1$ & $1$ & $1$ &\eqref{SolZeroLa}\\
	\hline 
	$N_{1,2}$ & $R^2$ & $R^{2(\frac{d-2}{2}\da^2-1)}$ & $R^{(d-2)\da}$&\eqref{solution:2}, \eqref{solution:4}, \eqref{solution:5},\\
															&&&&\eqref{Sol2}, \eqref{solution:10}, \eqref{CHMiii} 
	\\
	\hline
	$P_{1,2}$ & $R^{\frac{4}{(d-2)\da^2}}$ & $1$ & $R^{\frac2\da}$ &\eqref{CHMi}
	 		\\
	\hline
	$T_{1,2}$ &  $R^{2\frac{\ga^2-(d-3)\da^2+(d-4)\ga\da+2(d-2)}{(\ga-\da)^2}}$ & $R^{-2\frac{(d-3)\da+\ga}{\ga-\da}}$ & $R^{\frac{2(d-2)}{\da-\ga}}$&	\eqref{solution:7} 
	\\
	\hline
\end{tabular}
\caption[Asymptotics of black-holes solutions in Einstein-Maxwell-Dilaton theories]{Asymptotic form of solutions for trajectories approaching critical points at phase space infinity from within the sphere at infinity, in the case $R\to\infty$. In some cases, conditions exist on the values of $\La$, $\ga$, $\da$ and $\ka$ in order to have regular black-hole solutions, see \cite{Poletti:1994ff}.}
\label{TableWilt}
\end{table}

One readily sees that only the families of solutions ending on the points $M_{1,2}$ are asymptotically flat, and these families are entirely contained in the $\La=0$ (zero potential) plane. In particular, these contain the black hole solutions found in \cite{Gibbons:1987ps,Garfinkle:1990qj}. Moreover, it is obvious that the asymptopia of the solutions are irregular,  except for the family with endpoints $N_{1,2}$, and then only for $\da=0$. This is precisely the constant potential case we have just studied. As soon as $\La\neq0$ and $\da\neq0$, no realistic asymptotics can be found in these theories. However, we will see in Section \ref{section:EHT} how to give them a physical meaning through holography.

So, it seems that the no-hair theorem holds in the dS case: combining the impossibility of having regular dS asymptotics in the Liouville case and that of having two regular horizons in the cosmological constant case seems to forbid any asymptotically-dS black hole ever occuring in these EMD theories. This certainly agrees with the fact that dS hairy black holes are not allowed for convex potentials, \cite{Torii:1998ir}.

In the case of AdS hairs, the existing theorems \cite{Sudarsky:2002mk,Hertog:2006rr} rely on the assumption that the scalar field settles at spatial infinity in a local or global finite extremum of the potential: in our case, intuitively, the scalar field can but roll down its potential and will diverge asymptotically. This is apparent from the results displayed in \Tableref{TableWilt}, which confirm that no solution can have AdS boundary conditions, except for a flat $\da=0$ potential. However, such a solution in closed analytical form has yet to be put forward.

We close this section by writing the ``background'' of the theory with non-zero potential, which corresponds to the asymptotics of the family $N_{1,2}$, as will become clear in the subsequent study of the black holes of EMD theories,
\bsea
	\ud s^2& = & r^2(-\ud t^2 + \ud\Omega_{d-2,\ka}^2)+ r^{(d-2)\da^2-2}\ud r^2\,,\\
	e^\phi &=& r^{2\da} \,,
	\label{LiouvilleBackground}
\esea
where $\ud\Omega_{d-2,\ka}^2$ is the line element for the $d-2$-dimensional maximally symmetric subspace of spherical, planar or hyperbolic topology depending on the value of $\ka$. This is not $d$-dimensional AdS and explicitly breaks its $SO(d-1,2)$ symmetry group. This breaking corresponds to a non-zero value for $\da$, and goes together with a non-trivial scalar field profile. On the other hand, setting $\da=0$ restores the $SO(d-1,2)$ invariance, yields a constant scalar field, and spacetime is now locally isometric to AdS (exactly global AdS if $\ka=1$). The potential $V(\phi)$ then is simply a constant, as expected. However, none of the analytic solutions presented below have both a non-trivial dilaton and AdS$_d$ asymptotics in the case of a pure cosmological constant.

By going through a conformal transformation of the metric $g_{\mu\nu}\to e^{-\frac\phi\da}g_{\mu\nu}$ in appropriate coordinates, one can show that this spacetime is conformally flat, recovering Poincar\'e invariance on the boundary. In the string case ($\ga=\da=1$, see Section \ref{section:StringEffectiveActions}), the coordinate transformation induces a logarithmic branch and then the background in the string frame is simply Minkowski spacetime. For a more detailed discussion on (non-SUSY) String Theory dilatonic backgrounds, see \cite{Dudas:2000ff}.

\subsubsection{Kaluza-Klein reductions of \texorpdfstring{$d+1$}{d+1}-dimensional theories}

In this section, we show how to obtain $d$-dimensional Einstein-Maxwell-Dilaton theories from $D=(d+1)$-dimensional Einstein theories with a cosmological constant. Indeed the uplifted theory is just
\be
  S= \int d^{(D)} x \sqrt{-g^{(D)}} \l[ R^{(D)} -2 \La \r],
\ee
where $g^{(D)}$, $R^{(D)}$ and $\La$ are the determinants of the $D$-dimensional metric, the $D$-dimensional scalar curvature and the cosmological constant, respectively. The argument is a standard one: by taking the metric ansatz
\be
\ud s^2_{D}=e^{-\da\phi}\ud s^2_{d} + \e^{(d-2)\da\phi}\l(\ud w + A_\nu\ud x^\nu\r)^2 ,
\ee
one can reduce the $D$-dimensional theories to the $d$-dimensional Einstein-Maxwell-Dilaton action \eqref{EMDaction}, where we have the relations
\be
	\ga=\pm\sqrt{2\frac{d-1}{d-2}}\,,\qquad \da=\pm\sqrt{\frac2{(d-1)(d-2)}}\,,\qquad \gamma\delta = \frac2{d-2}\,.
	\label{KK}
\ee
As an illustration, let us uplift a four-dimensional metric to five dimensions. This can be done only in the two cases $\ga=\pm\sqrt{3}$ and $\da=\pm\frac1{\sqrt3}$, which satisfies the relation $\ga\da=1$. The general way to uplift the metric is to use the relation
\be
	\ud s^2_{5}=e^{\mp\frac\phi{\sqrt3}}\ud s^2_{4} + e^{\pm2\frac\phi{\sqrt3}}\l(\ud w + A_\nu\ud x^\nu\r)^2,
\ee
where the four-dimensional metric $\ud s^2_{4}$ can be obtained from the results in the following sections.

\subsubsection{String theory effective actions}
\label{section:StringEffectiveActions}

Once one has subscribed to the necessity of dealing with the problem of quantum gravity (UV-divergences, quantisation of the gravitational field\ldots) by adopting a string-theoretical approach, one is faced with the daunting task of actually solving the String Theory equations of motion. Leaving aside ambiguities related to the choice of a particular realisation of String Theory (type I, type IIa and IIb, heterotic on $O(32)$ or $E8\times E8$) and the dualities linking them all, or to the mysterious ``mother of all'' string theories (the so-called M-theory), the theory to be solved is one of many-dimensional extended objects, whose equations of motion are functionals and not easily dealt with. In the meantime, one may resort to more familiar, perturbative approaches based on field theory tools, which amount to formulate the problem in an effective field theory perspective.

One such possibility is to consider a finite number of massless modes of the string evolving in some background, after the massive modes have been integrated out. The massless modes left should then acquire a vev, which should derive from the appropriate equations of motion. This should be consistent as long as String Theory is weakly-coupled, so that the perturbative expansion in the $\al'$ parameter (the string loop-expansion parameter, inverse to the string tension) makes sense. 

The bosonic string can be modelised as a nonlinear sigma model and propagates on a two-dimensional spacetime (the world-sheet), coupled to a number of massless background fields. The minimal set required by the $N=1$ supergravity bosonic sector is a symmetric two-tensor $g_{\mu\nu}$ (the graviton), an antisymmetric two-tensor $B_{\mu\nu}$ and a scalar field $\phi$ (the dilaton) in the case of type II theory (for heterotic one should add a $U(1)$ gauge field $A_\mu$). It is well-known that two-dimensional spacetime is conformally flat, and invariance under conformal transformations is a subgroup of two-dimensional reparameterisation invariance. Imposing local scale invariance on the sigma model requires the beta functions of the various background fields to vanish, and in fact yields the equations of motion of the model, which in turn provide the expectation values of the background fields. Explicitly, to first order in $\al'$, one finds, \cite{Callan:1985ia,Callan:1986jb},
\bsea
	\frac{\ba^\phi}{\al'} &=& \frac{d-d_{crit}}{48\pi^2\al'}+\frac1{16\pi}\l[4\l(\nabla\phi\r)^2-4\nabla^2\phi-R+\frac1{12}\mathcal H^2\r], \slabel{BetaFunctionDilaton}\\
	\ba^g_{\mu\nu}  &=& R_{\mu\nu}-\frac14\mathcal H_\mu^{\la\sig}\mathcal H_{\nu\la\sig}+2\nabla_\mu\nabla_\nu\phi+O\l(\al'\r),\slabel{BetaFunctionMetric}\\
	\ba^B_{\mu\nu}&=& \nabla_\la\mathcal H^\la_{\mu\nu}-2\l(\nabla_\la\phi\r)\mathcal H^\la_{\mu\nu}+ O\l(\al'\r),\slabel{BetaFunctionTwoForm}
	\label{FirstOrderBetaFunctions}
\esea
where $\mathcal H_{\la\mu\nu} = 3\nabla_{\l[\mu\r.}B_{\l.\nu\la\r]}$ is the field strength associated to $B_{\mu\nu}$. The number $d_{crit}$ appearing in \eqref{BetaFunctionDilaton} determines whether the string theory under scrutiny is critical and possesses conformal invariance (no conformal anomaly), while if not, it is non-critical and the non-linear Liouville sigma model with a conformal anomaly should be quantised. This was emphasised by Polyakov, both in the bosonic case, \cite{Polyakov:1981rd}, where $d_{crit}=26$, and in the fermionic case, \cite{Polyakov:1981re}, where the critical dimension is $d_{crit}=10$.

Requesting that the beta functions cancel gives equations of motion, which can be shown to derive from the following effective action:
\be
	S=\int \ud^{d}x\sqrt{-g^{(d)}}\e^{\phi}\l[R+\l(\nabla\phi\r)^2-\frac1{12}\mathcal H^2-\frac{d-d_{crit}}{3\al'}\r].
	\label{TypeIIEffActionStringFrame}
\ee
The role of the dilaton as a loop-expansion parameter is apparent from the conformal factor $\e^\phi$. Now, a conformal transformation $g_{\mu\nu}\to\e^{-\sqrt{2/(d-2)}\phi}g_{\mu\nu}$ allows to go to the \emph{Einstein frame}
\be
	S=\int \ud^{d}x\sqrt{-g^{(d)}}\l[R-\frac12\l(\nabla\phi\r)^2-\frac1{12}\e^{2\sqrt{\frac2{d-2}}\phi}\mathcal H^2-\frac{d-d_{crit}}{3\al'}\e^{-\sqrt{\frac2{d-2}}\phi}\r],
	\label{TypeIIEffActionEinsteinFrame}
\ee
which has a more familiar form. In non-critical theories, the dilaton will have a Liouville potential, $V(\phi)=2\La\e^{-\sqrt{\frac2{d-2}}\phi}$, with $2\La=(d-d_{crit})/3\al'$. For the heterotic string, one must add a field strength squared term in the effective action to ensure the vanishing of the corresponding beta-function,
\be
	S=\int \ud^{d}x\sqrt{-g^{(d)}}\l[R-\frac1{2}\l(\nabla\phi\r)^2-\frac1{12}\e^{2\sqrt{\frac2{d-2}}\phi}\mathcal H^2-\frac14\e^{\sqrt{\frac2{d-2}}\phi}F^2-\frac{d-d_{crit}}{3\al'}\e^{-\sqrt{\frac2{d-2}}\phi}\r].
	\label{HeteroticEffActionEinsteinFrame}
\ee
Then, generalising to the EMD action \eqref{EMDaction} (with $\mathcal H=0$ for simplicity), one finds that the ``string case'' corresponds to $\ga=\da=1$. 

In a different physical setting and taking $F=0$ in $d=10$ dimensions, the action \eqref{EMDaction} describes  tachyon-free non-supersymmetric String Theory, \cite{Dixon:1986iz,AlvarezGaume:1986jb,Sagnotti:1995ga,Sagnotti:1996qj,Sugimoto:1999tx,Dudas:2000ff}. The Liouville coupling $\gamma$ plays the role of the leading string surface  ($g_s$) correction  in the Liouville term which appears due to the breaking of supersymmetry. For example  we have $\ga=3/2$ for the type I string and $\ga=5/2$ for the closed heterotic string.  As mentioned in the introduction, the characteristic of these string theories is that they do not have maximally symmetric backgrounds and as a result, the solutions of maximal possible symmetry are $SO(9)$-symmetric backgrounds \cite{Dudas:2000ff}.

\subsubsection{Field redefinitions}

In this work we will consider a $d$-dimensional metric of the form (see also \cite{Charmousis:2003wm,Charmousis:2006fx}):
\be
\label{cylindrical_metric}
	\ud s^2 = \e^{2\chi}\al^{-\frac{d-3}{d-2}}(\ud \rho^2+ \ud \theta^2)+\al^\frac{2}{d-2}\l(-\e^{2 U_t}dt^2+\e^{2 U_\varphi} \si(\theta) ^2 \ud \varphi^2 +\sum_{i=1}^{i=d-4}\e^{2U_i}\ud x_i^2\r),
\ee
where the Maxwell field will be restricted to be either electric, $A=A\l(\rho,\theta\r)\ud t$ or magnetic $A=A\l(\rho,\theta\r)\ud\varphi$. Dyonic solutions have been studied: with zero potential and asymptotically flat boundary conditions, \cite{Gibbons:1987ps,Shapere:1991ta,Kallosh:1992ii}, or not \cite{Clement:2005vn}; with non-zero potential and (A)dS asymptotics, \cite{Poletti:1995yq}, or irregular asymptotics, \cite{Yazadjiev:2005du,Yazadjiev:2005pf}. The function $\si(\theta)$ denotes $\sin\l(\theta\r)$, $\sinh\l(\theta\r)$ and unity for $\kappa=1,-1,0$ respectively. We can also choose the potentials $U_i$ so that  they sum to zero
\be
\label{sum}
\sum_{i=1}^{i=d-4}U_i+U_t+U_\varphi=0
\ee
without any loss of generality.

When $\kappa=0$ and all metric components are locally only $\rho$-dependent we have cylindrical symmetry ($\rho$ is not the normal coordinate). For $d=4$, $\kappa=\pm1$ will correspond  to a spherically symmetric and hyperbolic two-dimensional space-like sections respectively{\footnote{There is no particular reason in choosing two-dimensional sections for a $d$-dimensional spacetime except that in the present analysis we will specialise later on to four-dimensional spacetimes. This can be easily generalised, \cite{Charmousis:2003wm}.}}.
It is rather useful now to go to a different set of variables, \cite{Charmousis:2006fx}, for which the field equations will take a simpler form,
\bea
	\psi_\star& = & \sqrt\frac{d-2}{d-3}\l[\frac{d-3}{d-2}(\phi -\da\ln \al)+ \ga U_\star\r],	\slabel{chgt1}\\
	\psi_i& =& U_i+\frac{1}{d-3}U_\star,\qquad i=1,\ldots,d-4\,,\slabel{chgt12}\\	
	\Omega & = & \ga(\phi-\da\ln\al)-2U_\star\,,	\slabel{chgt2}\\
	2\nu & = & 2\chi -\da \phi +\frac{\da^2}2 \ln\al\,,	\slabel{chgt3}
\eea
where $\epsilon=-1$ corresponds to an electric potential and $\epsilon=1$ to a magnetic one. The $\star$ symbol denotes  $t$ for the electric case and $\varphi$ for the magnetic case respectively.
These technicalities put aside, the field equations for the electric case ($\eps=-1$) are:
\bsea
	\al'' -\kappa \al & = & -2\Lambda \al^{\frac{1}{d-2}-\frac{\da^2}{2}}\e^{2\nu}\,, \slabel{m1}\\
	0 & = & \overrightarrow{\nabla} \cdot \l(\e^{\Omega} \al^{\ga\da+\frac{d-4}{d-2}} \si(\theta)^{-\eps} \overrightarrow{\nabla} A  \r), \slabel{m2}\\
	\l(\al \Omega'\r)' +\Big(\ga\da-\frac{2}{d-2}\Big)\alpha \kappa & = & \dfrac{\eps s}2 \e^\Omega \al^{\ga\da+\frac{d-4}{d-2}} \si(\theta)^{-\eps}\l( \overrightarrow{\nabla} A\r)^2 ,	\slabel{m3}\\
	\l(\al \psi'_\star\r)' +\alpha \kappa \sqrt{\frac{d-3}{d-2}}\Big(\da+\frac{\ga}{d-3}\Big) &=& 0,	\slabel{m4}\\
	\l(\al \psi_i'\r)' & = & 0,	\qquad i=1,\ldots,d-4\,, \slabel{m5}\\
	2\nu'\dfrac{\al'}\al-\dfrac{\al''}\al -\kappa & = & \dfrac1s\l[\l(\psi_\star'\r)^2+\dfrac12\l(\Omega'\r)^2\r]+\nn\\
		&&\quad +\dfrac\eps2 \e^\Omega \al^{\ga\da -\frac{2}{d-2}} (A'^2-\dot{A}^2)+\sum_{i=1}^{d-4} 		\psi_{i}^{'2} ,	\slabel{m6} \\
	2\al\nu' \kappa -\l(\frac{d-1}{d-2}-\frac{\da^2}{2}\r)\al'\kappa &=&-\frac{2\kappa \alpha}{s}\l[\l(\frac{\ga}{d-3}+\da\r)\sqrt{\frac{d-3}{d-2}}\psi'+\r.\nn\\
	&&\qquad\qquad+\l.\dfrac12 \l(\ga\da-\frac{2}{d-2}\r)\Omega'\r]. \slabel{m7}	
\esea
All fields depend on $\rho$ (according to cylindrical symmetry), except for the electric potential for which we allow a $(\rho,\theta)$ dependence which will be useful for the extension of the electro-magnetic duality in four dimensions later on. For the same reason we keep $\epsilon$. Note equation (\ref{m7}) which is an additional equation present for $\kappa \neq 0 $ which constrains the metric elements (\ref{cylindrical_metric}) in such a way as to obtain maximally symmetric two-dimensional sections. We have also set
\be
s=\gamma^2+2\frac{d-3}{d-2}\,.
\ee
For the magnetic case ($\eps=1$) on the other hand, we have
\bsea
	\al'' -\kappa \al& = & -2\Lambda \al^{\frac{1}{d-2}-\frac{\da^2}{2}}\e^{2\nu}\,,	\slabel{m11}\\
	0 & = & \overrightarrow{\nabla} \cdot \l(\e^\Omega\si{\theta}^{-\eps} \al^{{\ga\da}+\frac{d-4}{d-2}}\overrightarrow{\nabla} A \r), \slabel{m12}\\
	\l(\al \Omega'\r)' +\l[\ga\da+\frac{2(d-3)}{d-2}\r]\alpha \kappa & = & \dfrac{\eps s}{2}\si(\theta)^{-\eps} \e^\Omega \al^{\ga\da+\frac{d-4}{d-2}} \l( \overrightarrow{\nabla} A\r)^2,	\slabel{m13}\\
	\l(\al \psi_\star'\r)' +\alpha \kappa \sqrt{\frac{d-3}{d-2}}(\da-\ga)& = & 0,	\slabel{m14}\\
	\l(\al \psi_i'\r)' & = & 0,	\qquad i=1,\ldots,d-4\slabel{m15}\\
	2\nu'\dfrac{\al'}\al-\dfrac{\al''}\al -\kappa &=& \dfrac1s\l[\l(\psi'\r)^2+\dfrac12\l(\Omega'\r)^2\r]+\sum_{i=1}^{d-4}\psi_i^{'2} + \nonumber\\
	&&+\frac{\eps}{2\si^2(\theta)}\e^\Omega \al^{\ga\da -\frac{2}{d-2}} \l(A'^2-\dot{A}^2\r),	\slabel{m16} \\
	2\al\nu' \kappa- \l(\frac{d-1}{d-2}-\frac{\da^2}{2}\r)\al'\kappa&=&\frac{2\alpha \kappa}{s}\l[(\ga-\da)\sqrt{\frac{d-3}{d-2}}\psi'-\r.\nn\\
	&&\qquad\qquad \l.-\dfrac12 \l(\ga\da+2\frac{d-3}{d-2}\r)\Omega'\r].\slabel{m17}	
\esea

The field equations written in this form are quite straightforward to reduce to one or two coupled second-order ODE's with respect to one or two  variables respectively. In reducing the system of equations, we adapt our system of coordinates accordingly. It turns out that the judicious system of coordinates differs for $\kappa=0$ (cylindrical symmetry) and for $\kappa\neq 0$. Let us reduce the system in turn now for each case, starting with $\kappa=0$. Note that (\ref{m7}) and (\ref{m17}) drop out in this case.
\vfill

\subsubsection{Electro-magnetic duality in four dimensions}

Let us consider now the symmetries of the magnetic and electric field equations \eqref{m1}-\eqref{m7} and \eqref{m11}-\eqref{m17}, following \cite{Charmousis:2006fx}. We can define a dual potential $\omega$ to $A$ by
\be
\label{dualpotential}
\l(-\partial_\theta \omega, \partial_\rho \omega\r)=\e^\Omega \al^{\ga\da} \si\l(\theta\r)^{-\epsilon}\l(\partial_\rho A, \partial_\theta A\r).
\ee
To be definite, we take $\epsilon=-1$ and apply \eqref{dualpotential}. After this, the field equations \eqref{m2}, \eqref{m3}, \eqref{m6} and \eqref{m7} take the form
\bsea
	0 & = & \overrightarrow{\nabla} \cdot \l(\e^{-\Omega} \al^{-\ga\da} \l[\si\l(\theta\r)\r]^{-1} \overrightarrow{\nabla} \omega  \r), \slabel{m2d}\\
	\l(\al \Omega'\r)' +\l(\ga\da-1\r)\alpha \kappa & = & \dfrac{\eps s}2 \e^{-\Omega} \al^{-\ga\da} \l[\si\l(\theta\r)\r]^{-1}\l( \overrightarrow{\nabla} \omega\r)^2 ,	\slabel{m3d}\\
	2\nu'\dfrac{\al'}\al-\dfrac{\al''}\al -\kappa & = & \dfrac1s\l((\psi_\star')^2+\dfrac12(\Omega')^2\r)+ \nonumber\\
	&&\qquad+\dfrac\eps2 \e^{-\Omega} \al^{-\ga\da -1} \l(\omega'^2-\dot{\omega}^2\r)+\sum_{i=1}^{d-4} \psi_{i}^{'2} \,,	 \slabel{m6d} \\
	2\al\nu' \kappa -\l(\frac{3-\da^2}{2}\r)\al'\kappa &=&\frac{2\kappa \alpha}{s}\l[-\l(\ga+\da\r)\frac{\sqrt{2}}2\psi'-\dfrac12 \l(\ga\da-1\r)\Omega'\r]. \slabel{m7d}
\esea
Now, consider the following map
\be
\label{dualitymap}
\bar{\Omega}=-\Omega\,,\qquad \bar{A}=\omega\,, \qquad \bar{\epsilon}=-\epsilon\,,\qquad \bar{\gamma}=-\ga\,,\qquad \bar{\da}=\da\,,
\ee
then \eqref{m2d}, \eqref{m3d}, \eqref{m6d} and \eqref{m7d} are exactly \eqref{m12}, \eqref{m13}, \eqref{m16} and \eqref{m17}
 for the barred variables $\bar{A}\,, \bar{\Omega}$ and constants $\bar{\ga}\,,\bar{\da}\,,\bar{\epsilon}\,$. The remaining equations \eqref{m11}, \eqref{m14}, \eqref{m15}, \eqref{m17} do not yield any additional constraint and hence the map \eqref{dualitymap} generates a novel solution. In other words, the duality is valid for any $\ga$ and $\da$. The application \eqref{dualitymap} provides a simple way to obtain a magnetic/electric solution from another given electric/magnetic solution. Although \eqref{dualitymap} is clearly an extension of the EM duality for $\Lambda=0$, it is of quite a different nature since \eqref{dualitymap} changes the coupling $\ga$, hence maps solutions belonging to \emph{different} theories. We will use this symmetry in order to construct solutions in $d=4$ dimensions for $\kappa\neq 0$. This is also particularly useful to construct solutions for the uplifted metrics.

\vfill

\subsection{Non-planar solutions in four-dimensional spacetime}

\subsubsection{Reduction of the equations of motion: non-planar case}

We now turn our attention to the case of $\kappa\neq 0$. Let us stick to the electric case here and note that the judicious choice of coordinates dictated for example from \eqref{m4} is
\be
	\al=\frac{\ud r}{\ud \rho}\,.
\ee

We also note that, for the electric case, in order for the $A$ field not to be trivial - as imposed by separability requirements -, it has to be a function of $\rho$. On the other hand, for the magnetic case, $A$ has to be a function of $\theta$.  This can easily be seen by inspecting the equations of motion in both cases. In other respects, the magnetic resolution is very similar to the electric one.
Let us denote by a dot the derivation with respect to $r$. Integrating \eqref{m2}, \eqref{m3} and then \eqref{m4}, \eqref{m5}, we obtain
\bsea
	q&=&\e^\Omega\al^{1+\ga\da}\dot A\,, \slabel{Q_max} \\
	\al^2\dot\Omega &=& \eps \frac s2qA +a +\l(\frac{2}{d-2}-\ga\da\r)\ka r\,, \slabel{A_r_max}\\
	\al^2\dot\psi &=&c- \ka x \sqrt{\frac{d-3}{d-2}}\l(\da+\frac{\ga}{d-3}\r), \slabel{c_x_max}\\
	\al^2\dot\psi_i&=& c_i  \slabel{ci_x_max}\,.
\esea
Combining \eqref{m6} and \eqref{m7} with \eqref{A_r_max}, \eqref{c_x_max}, we obtain
\bea
	2\al^2\dot\nu &=& \l[\frac{2}{(d-2)(d-3)}+\da^2\r]\ka \l(\frac{d-1}{d-2}-\frac{\delta^2}{2}\r)\al\dot\al+\nn\\
	&&\qquad\qquad +\l(\ga\da-\frac{2}{d-2}\r)\l(\frac{a}{s}-\frac{qA}{2}\r)-h\,,
	\label{h_x_max}
\eea
where we now have
\be
\label{hconstraint}
	h=\frac{2c}{s}\sqrt{\frac{d-3}{d-2}}\l(\delta+\frac{\gamma}{d-3}\r)\,,
\ee
so that  \eqref{m6} and \eqref{m7} are compatible: maximal symmetry imposes one more  relation between the integration constants $h$ and $c$. In fact, for $\kappa=0$ this means that the $(\theta,\varphi)$ plane is homogeneous in the cylindrical case \eqref{cylindrical_metric_bis}. We have now solved the system with respect to the variables $\al$ and $A$. Indeed using \eqref{A_r_max} and \eqref{m2}, we obtain
\be
\label{int1}
	\al^2 \frac{\ddot{A}}{\dot{A}}+2\dot{\al}\al-\l(\ga\da-\frac{2}{d-2}\r)\l(\ka r-\dot{\al}\al\r)+a-\frac{sq A}{2}=0\,,
\ee
and then, using \eqref{h_x_max} with \eqref{m6}, we get
\bea
\label{int2}
	&\dot{\al} \al\left[\l(\delta^2 +\frac{2}{(d-2)(d-3)}\r)\kappa r+\frac{\dot{\al}\al}{2}\l(\frac{2}{d-2}-\da^2\r)\right] -\l[\l(\ga\da-\frac{2}{d-2}\r)\l(\frac{a}{s}-\frac{qA}{2}\r)+h\r]\l(\dot{\al}\al-\kappa r\r)=&\nn\\ 
&=\ddot{\al}\al^3+\kappa \al^2+ \frac{c^2}{s}+\kappa^2r^2\l(\frac{\da^2}{2}+\frac{1}{(d-2)(d-3)}\r) + \frac{1}{2s}\l(a-\frac{sqA}{2}\r)^2-\frac{q\alpha^2\dot{A}}{2}\,.&
\eea
By solving \eqref{int1}, \eqref{int2} for $A$ and $\al$, we find a solution to the full system \eqref{m1}-\eqref{m7} by direct integration of \eqref{A_r_max}-\eqref{h_x_max}. In particular, note that for $\ga\da=\frac{2}{d-2}$, \eqref{int1} integrates out, giving
\be
q \e^{-\Omega}=\al^2 \dot{A}= k-a A-\frac{sq A^2}{4}\,,
\ee
where we have also used \eqref{Q_max} to obtain $\Omega$. This reduces the full system to the resolution of equation \eqref{int2} with respect to $A$.

This completes our analysis of the theory  in arbitrary dimension $d$. From now on we will concentrate on the case of $d=4$ and give explicit solutions.

\subsubsection{Zero potential black holes}

We start this section by very briefly considering the case $\Lambda=0$ which yields insight on our case of interest $\Lambda\neq 0$. This case was first analysed by Gibbons and Maeda, \cite{Gibbons:1987ps}, and later on revisited in the case of $\ga=\da=1$ by Horowitz \emph{et al.}, \cite{Garfinkle:1990qj}. In the coordinate system \eqref{X}, it is trivial to integrate, since from \eqref{m1},
\be
\beta(r)\equiv \alpha^2=\kappa r^2+\beta_1 r+\beta_0\,,
\ee
where $\beta_1, \beta_0$ are arbitrary constants. As in that case the coupling $\da$ can be chosen at will, we fix it to be  $\da=1/\ga$ and then \eqref{int1} is simply an identity, whereas \eqref{int2} gives $A$ by direct integration as in \eqref{cond}. The important thing to note is that the second-order coefficient of $\beta$ is directly given by $\kappa$. Whenever this is the highest-order coefficient of $\beta$ this immediately means that $\Lambda=0$. According to the prescription we described in the second section, we easily find the remaining metric components obtaining the general solution for $\Lambda=0$, \cite{Gibbons:1987ps}:
\bsea
	\ud s^2&=&-V(r)\ud t^2 + \frac{\ud r^2}{V(r)}+R(r)^2\l(\ud\theta^2+\sin^2\theta\ud\varphi \r),\\
	V(r)&=&\l(1-\frac{r_+}r\r)\l(1-\frac{r_-}r\r)^{\frac{1-\ga^2}{1+\ga^2}}\,,\slabel{PotZeroLa}\\
	R(r)&=&r\l(1-\frac{r_-}r\r)^{\frac{\ga^2}{1+\ga^2}}\,,\slabel{WarpZeroLa}\\
	\e^\phi&=&\l(1-\frac{r_-}r\r)^{-\frac{2\ga}{1+\ga^2}}\,,\slabel{DilZeroLa}\\
	A&=&\sqrt{\frac{4r_-}{(1+\ga^2)r_+}}\l(1-\frac{r_+}r\r)\ud t\,.\slabel{MaxwellZeroLa}
	\label{SolZeroLa}
\esea

Note that there are two singularities, at $r=0$ and $r=r_-$, though the former is never attained if the gauge field is turned on, that is if $r_-\neq0$: the horizon has finite size at the singularity, contrarily to the usual black holes from General Relativity; this seems to be a defining property of dilatonic black holes. On the other hand, there is an event horizon at $r=r_+$ and, remarkably, the blackness potential \eqref{PotZeroLa} is identical to Schwarzschild's. Similarly, no topological black hole can exist, the horizon must have spherical topology.

Asymptotically, Minkowski is recovered, and so this solution co-exists with Schwarzschild's solution, which is a clear violation of the no-hair theorems. However, the hair is secondary, as there is no other integration constant independent from $r_\pm$ associated to the scalar field. There is also the possibility of having a non-zero asymptotic value for the dilaton, reflecting the classical scale invariance of the action without potential. Another view on this matter is that the coupling $\ga$ classifies different EMD theories, that one should not consider that a solution for a given $\ga$ competes with the General Relativity solutions (recovered for $\ga=0$). The causal structure resembles that of Schwarzschild (and not Reissner-Nordstr\"om) , and so should the Penrose-Carter diagram.

For $\ga=0$, we recover Reissner-Nordstr\"om, while for $\ga=1$, this is the string case discussed by Garfinkle, Horowitz and Strominger, \cite{Garfinkle:1990qj}. 

There is an extremal limit where the two horizons coincide and become degenerate, $r_\pm=r_e$, which is quite different from the usual Reissner-Nordstr\"om black hole. Indeed, the extremal horizon is regular here, and the horizon size is finite and equal to $r_e$. Here, the horizon size collapses at $r_e$ and it is a singular point of spacetime. Moreover, while the distance to the extremal horizon is infinite for Reissner-Nordstr\"om\footnote{Though of course one can cross it in finite \emph{proper} time.}, this is not the case for the dilatonic black holes where the distance to the extremal singularity is finite, \cite{Holzhey:1991bx}:
\be
	\int_{r_e}^{+\infty} \frac{\ud r}{\sqrt{V(r)}} <\infty\,.
\ee
So indeed, dilatonic black holes appear to have quite different properties from the usual General Relativity ones.

The magnetic dual solution can be obtained the usual way, setting $\ga\to-\ga$ and taking for the magnetic field $A=q\cos\theta\,\ud\varphi$.

\subsubsection{\texorpdfstring{$\ga\da=1$}{gamma*delta=1} solution}

Let us now consider $\Lambda\neq 0$. We have to simultaneously solve for two coupled equations (\ref{int1}) and (\ref{int2}). For $d=4$, these read:
\be
\label{int14}
\beta \ddot{A}+\dot{\beta}\dot{A}-(\ga\da-1)(\ka r-\half \dot{\beta})\dot{A}+(a-\frac{sq A}{2})\dot{A}=0,
\ee
\bea
\label{int24}
&-\frac{\da^2+1}{2}\left[\frac{1}{2}\dot{\beta}-\kappa r+\frac{h}{\da^2+1} \right]^2-\kappa \beta -\half (\ddot{\beta}\beta-\dot{\beta^2})&\nn\\
	&=&\nn\\
	&\frac{h^2 (1-\ga \da)^2}{2(\da^2+1)(\ga+\da)^2}-\frac{(1-\ga\da)}{2s}\left(a-\frac{sq}2A\right)(\dot\ba-2\ka r)+\frac1{2s}\left(a-\frac{sq}2A\right)^2-\half q\dot A\ba.&
	\label{al_x_max0}
\eea

The case $\ga\da=1$ is special since \eqref{int14}, \eqref{int24} decouple and furthermore \eqref{int14} is integrable. For this case:
\be
	\ba\dot A = \frac{sq}4A^2-aA+k\,,
	\label{k_x_max}
\ee
\be
-\frac{1}{2(\da^2+1)}\left[\frac{\da^2+1}{2}\dot{\beta}-\kappa r (\delta^2+1)+h \right]^2=\kappa \beta +\half (\ddot{\beta}\beta-\dot{\beta^2})-\frac{qk}{2} +\frac{a^2}{2s}\,.
	\label{al_x_max}
\ee
The general solution to this equation can be obtained by numerical integration. Some explicit solutions can be obtained by supposing that $\beta$ is of polynomial form. One of them is the $\Lambda=0$ solution discussed above and the second is a black hole solution first obtained in \cite{Chan:1995fr} for $\ka=1$. The potential reads
\be
\beta(r)  =  \ka\frac{\da^2+1}{\da^2-1} r^2 -\frac{2h}{\da^2-1}r +(\da^2-1)\frac{qk}{4\da^2\ka} +\frac{h^2}{\ka(\da^4-1)}-\frac{a^2(\da^2-1)}{4\ka(\da^2+1)}\,.
\ee
The solution is not valid for $\ga=\da=1$. After a translation and some redefinitions of parameters, the solution takes the form of \cite{Chan:1995fr}
\be
	\ud s^2 = -V(r)\ud t^2 + \frac{\ud r^2}{V(r)} +R^2(r)d\Omega^2\,, \qquad V=\frac{\beta}{R^2}\,,
	\label{CHMmetric}
\ee
where a suitable change of the origin and rescaling of constants gives
\bsea
	\beta(r) & = & \ka\frac{\da^2+1}{\da^2-1}r^2-2sMr+\frac{sq^2}{4}\,\e^{-\frac{\phi_0}{\da}}\,, \\
	\e^{\phi} & = & \e^{\phi_0}r^{\frac{2\da}{1+\da^2}}\,, \\
	\dot A(r) &= & q r^{-2}\e^{-\frac{\phi_0}{\da}}\,, \\
	R^2(r)&=&r^{\frac{2\da^2}{1+\da^2}}\,, \\
	\Lambda & = & \frac{\ka}{1-\da^2}\,\e^{\da\phi_0}\,.
	\label{CHMi}
\esea

Note the absence of an extra parameter presented in \cite{Chan:1995fr} (see also \cite{Cai:1997ii} for $\kappa=-1$) which can be gauged away. This solution is clearly valid only for $\ka\neq0$. The $\kappa=0$ black holes need to be treated separately. We describe the spherical case below.

If $\da^2>1$, the solution has one singularity in $r=0$ and two horizons at the two roots of $\ba(r)$. If these two roots are degenerate, the solution is extremal but regular. However, it can also be a naked singularity if $q<q_e$.

If $\da^2<1$, there is a single positive root for $\ba(r)$ and the $r$-coordinate is space-like inside the single horizon, so we have a cosmological horizon with  a singularity at $r=0$ and a cosmological horizon cloaking it.

The dual magnetic solution is readily obtained from \eqref{CHMi}. Using the dual potential \eqref{dualpotential} and the duality map \eqref{dualitymap}, we get the magnetic solution by simply replacing the Maxwell field of \eqref{CHMi}:
\be
	A = \frac{q}{\ka}\,\textrm{co}(\theta)\,\ud\varphi ,
\ee
and setting $\da=-\frac1\ga$ in the solution (\ref{CHMi}).

This particular solution is not defined for $\ga=\da=\pm1$. If we do try to find a solution for the string case, the only permitted polynomial solution  is one of second degree verifying:
\be
	\kappa(\ba_2-\kappa)=0\,,
\ee
where $\ba_2$ is the highest-order coefficient.
Therefore we either have a toroidal black hole (and we will see explicitly that this is the case in Section \ref{section:NearExtremalPlanarEMD}) or a $\Lambda = 0$ solution, \eqref{SolZeroLa} above.

For the adequate couplings, \eqref{CHMi} could be uplifted in order to obtain a five-dimensional metric, \cite{Charmousis:2009xr}. 

\vfill 

\subsubsection{\texorpdfstring{$\ga+\da=0$}{gamma-delta=0} solution}

If $\ga\da\neq1$, we have to make some starting assumption in order to solve for  $A(r)$. A simple starting point is to assume that $A$ is a linear function and from (\ref{int14}), we get:
\be
	sq\ddot A = 0 = (1+\ga\da)\beta^{(3)}.
	\label{A_x_max2}
\ee
This last equation gives us two constraints: either $\beta(r)$ is a second-order polynomial or $\ga\da=-1$. Suppose that  $\ga\da\neq \pm 1$.
Then, solving for a second-order polynomial in (\ref{al_x_max}) gives us three distinct possibilities. First of all $\Lambda=0$ solutions, \cite{Gibbons:1987ps}, or again a subclass of Reissner-Nordstr\"om-AdS where the dilaton is trivial. The third case lies within the interest of our study and the action parameters are related \emph{via} $\ga+\da=0$. The solution reads:
\bsea
	\beta(r) & = &\ba_2r^2-2\frac{(1+\da^2)}{\da^2}Mr\,,  \\
	\e^{\phi} & = & \e^{\phi_0}r^{\frac{2\da}{1+\da^2}}, \quad \e^{\da\phi_0} = \frac{2\Big[(1-\da^2)\ba_2+\ka(1+\da^2)\Big]}{q^2(1+\da^2)}\,,\\
	\dot A(r) &= & \frac{2}{(1+\ga^2)q}\l[(1-\ga^2)\ba_2 +\ka(1+\ga^2) \r], \\
	R^2(r)&=& r^{\frac{2\da^2}{1+\da^2}}\,, \\
	\Lambda & = & \frac{\ka-\ba_2}{2}\e^{\da\phi_0}\,.
	\label{CHMii}
\esea
This is again the solution presented in \cite{Chan:1995fr} and \cite{Cai:1997ii}. In order to have the $rr$-metric element space-like outside the horizon, we need $\ba_2>0$. For $\kappa=1$, it has one singularity at $r=0$ and one horizon at $r_h = \frac{2sM}{\da^2}$. So, even though this is a charged solution, it has no extremal limit with a regular black hole, as was remarked in \cite{Chan:1995fr}. Anticipating on Section \ref{section:NearExtremalPlanarEMD}, we find that it has the same asymptotics as the planar near-extremal solution \eqref{Sol3}. Indeed, \Tableref{TableWilt} reveals that the family of solutions $K_{1,2}$ to which the $\ga+\da=0$ solution \eqref{CHMii} belongs have the same asymptotics as the family $T_{1,2}$, to which the near-extremal planar solution \eqref{Sol3} belongs, provided one sets $\da=-\ga$ in the latter.

\subsubsection{\texorpdfstring{$\ga\da=-1$}{gamma*delta=-1} solution}

As we noticed from (\ref{A_x_max2}) when $\ga \da=-1$ we can have a higher-order polynomial. Upon making this assumption for $\beta$,
\be
\beta(r) = \ba_N r^N+\ba_2r^2+\ba_1r+\ba_0\,,
\ee
where $N$ is assumed to be different from $2$, $1$ or $0$,
we obtain:
\be
\al_2=\ka \quad \mathrm{or} \quad \da^2=\frac13\,,
\ee
which both lead to the black hole solution of \cite{Chan:1995fr}
\bsea
	\beta(r) &=& \ba_Nr^{\frac4{1+\da^2}} +\ka r^2-2(1+\da^2)Mr\,,\\
	\dot A(r) & =& \frac{4\ka}{qs}, \\
	\e^{\phi(r)} & =& \e^{\phi_0}r^{\frac{2\da}{1+\da^2}}\,, \qquad \e^{\frac{\phi_0}\da}=\frac{4\ka}{q^2(1+\ga^2)},\\
	R(r)^2 &=& r^{\frac2{1+\da^2}}\,,\\
	\Lambda&=&-\ba_N\frac{(3-\da^2)\e^{\da\phi_0}}{(1+\da^2)^2}\,.
	\label{CHMiii}
\esea

For the spherical case, this solution exhibits a variety of causal structures which can be neatly summarised in the \Tableref{Table:CHM}, \cite{Chan:1995fr}.

\begin{table}[htb]
	\centering
	\begin{tabular}{|c|c|c|}
\hline
	 	&$\La<0$& $\La>0$\\
\hline
		&&({\bf O, {\bf C}}),\qquad $q>q_e$\\
	$\ga^2>1$&{\bf O}& ({\bf O}={\bf C}),\qquad $q=q_e$\\
		&& {\bf B},\qquad $q<q_e$\\
\hline
	$\frac13<\ga^2<1$	&{\bf O}&{\bf O}\\
\hline
		&&({\bf I},{\bf O}),\qquad $q>q_e$\\
	$\ga^2<\frac13$	&{\bf O}& ({\bf I}={\bf O}),\qquad $q=q_e$\\
		&& {\bf N},\qquad $q<q_e$\\
\hline
	\end{tabular}
\caption[Various causal structures for spherically symmetric black holes in Einstein-Maxwell-Dilaton theories]{Various causal structures depending upon the values of $\La$, $q$ and $\da$ for solution \eqref{CHMiii}, assuming $\ka=1$. {\bf O}=Outer horizon, {\bf I} = Inner horizon, {\bf C} = Cosmological horizon, {\bf N}= Naked singularity, {\bf B} = cosmological singularity.}
 \label{Table:CHM}
\end{table}

Let us focus on the case $\delta=1$.  It actually coincides with the previous solution \eqref{CHMii} for which $\ga+\da=0$ as can be easily checked. Setting $r=\bar r^2$ the solution reads
\be
\ud s^2 = -\bar r^2\l[\l(\ba_N+\kappa\r)\bar r^2-\frac{4M}{\bar r^2}\r]\ud t^2 + \frac{4\ud\bar  r^2}{\l(\ba_N+\kappa\r)\bar r^2-\frac{4M}{\bar r^2}} +\bar r^2d\Omega^2\,, 
\ee
with $\e^{\phi(\bar r)} = \e^{\phi_0}\bar r^{2}$. This solution is singular at $\bar r=0$ and has an event horizon at $\bar r_h=\sqrt{\frac{4M}{\beta_N+\kappa}}$. By use of the duality, the above metric is a magnetic solution with
\be
	A = \frac{q}{\ka}\,\textrm{co}(\theta)\,\ud\varphi ,
\ee
and $\ga=1$. This is the only $\kappa=1$ solution for the couplings $\ga=\da=1$ we could find (though see later for one with planar horizon $\ka=0$).

For the adequate couplings, \eqref{CHMiii} could be uplifted in order to obtain a five-dimensional metric, \cite{Charmousis:2009xr}. 

\vfill

\subsection{Planar solutions in four-dimensional spacetime}

\label{section:PlanarEMDBH}

\subsubsection{Reduction of the equations of motion: planar case}

\label{PlanarReduction}

Directly integrating \eqref{m2}, \eqref{m4}, \eqref{m5} and using \eqref{m3}, we get
\bsea
q&=&\e^\Omega\al^{\ga\da}A'\,,\slabel{Q}\\
\al \Omega' &=& \dfrac{s\eps}2 qA +a\,, \slabel{A}\\ 
c_t & = & \al \psi'_t \slabel{c_r}\,,\qquad c_i = \al \psi'_i\,, \slabel{c_i}
\esea
where $q$ is the electric charge and $c_t, c_i$ are the constant scalar charges associated to the $\psi$ fields. The constant $a$ can be gauged away but we choose to keep it and fix it later to simplify integrated quantities.
Using now \eqref{m1}, \eqref{m6} and \eqref{A}, we solve for  $\nu'$:
\be
2\nu'\al = \l( \frac{d-1}{d-2} -\frac{\da^2}{2}\r)\al' + \l(\frac{2}{d-2}-\ga\da\r)\l(\frac{\eps}{2}qA+\frac as\r)-h\,.
\label{h_r}
\ee
Here $h$ is a constant and can be related to the asymptotic mass of the solution.
At the end of the day, the whole system boils down to two coupled ordinary differential equations (ODEs):
\bsea
	 \dfrac{s\eps}2 qA + a + \l( \ga\da-\frac{2}{d-2} \r)\al' +\al \dfrac{A''}{A'}&=&0\,, \slabel{alphaA1}\\
	\al'  \l[ \l( \frac{d-1}{d-2}-\frac{\da^2}{2} \r)\al'+\l(\frac{2}{d-2}-\ga\da \r)\l(\frac as +\frac{\eps A q}{2} - h\r)\r] - \al'' \al &=& \nn\\
	=\dfrac1s \l[c_\star^2+\dfrac12\l(\dfrac{s\eps}2 qA+a\r)^2\r] +\frac{\eps q}{2} \al A' +\sum c_i^2\,,&& \slabel{alphaA2}
\esea
which once solved give a solution for the theory \eqref{EMDaction} with cylindrical symmetry \eqref{cylindrical_metric}. To go further we fix the coordinate system by setting $\al' = p$. The two coordinate systems ($p$ for $\ka=0$, $r$ for $\ka=\pm1$) are related by $2p=\dot{\al^2}$. Hence, if $\al^2$ is a second-degree polynomial in $r$, only then are $p$ and $r$ identical coordinates.
The integration of equation \eqref{alphaA1} then gives:
\be
k = \frac{s\eps q}4A^2 +aA +\al A' - \l(\frac{2}{d-2}-\ga\da\r)\l(pA-\int A\,\ud p\r).
\label{kr}
\ee
On the other hand, \eqref{alphaA2} becomes
\bea
\label{X}
X(p)&\equiv& \l( \frac{d-1}{d-2} -\frac{\da^2}{2}\r)p^2 +\l[\l(\frac{2}{d-2}-\ga\da \r)\frac as - h\r]p -\frac{\eps q}2 k-\frac{c^2}s -\frac{a^2}{2s} \nonumber \\
&=& - \frac{q\eps}2\l(  \frac{2}{d-2}-\ga\da \r)\int A\,\ud p +p\frac{\ud p}{\ud \ln\al}\,.
\eea
Let us note now that, taking $\frac{2}{d-2}-\ga\da=0$, \eqref{X}  completely decouples from $A$ and gives immediately $\alpha$. Both the Kaluza-Klein and the string case falls in this category, see \eqref{KK} and \eqref{HeteroticEffActionEinsteinFrame}. Once $\alpha$ is known, $A$ is obtained from \eqref{kr}. We will examine shortly and in detail the solutions emanating for the four-dimensional case. Lastly, by defining
\be
	B(p)=\int A(p)\ud p
	\label{DefB}
\ee
and combining \eqref{X} and \eqref{kr}, we get
\be
	k-\frac{s\eps q}4\dot B^2 -a\dot B + \l( \frac{2}{d-2}-\ga\da \r)(p\dot B -B) = \ddot B\l[X(p)+  \frac{q\eps}2\l(  \frac{2}{d-2}-\ga\da \r)B\r].
	\label{radial_eq}
\ee
This second-order non-linear and autonomous ODE with respect to $B$ is one of the  main results in this section. In this form, it is quite obvious that the case $\frac{2}{d-2}-\ga\da$ is special, we will come back to that in the next subsection for $d=4$. But let us work a little bit more on it and put it in a form which will yield another interesting case.

First, it is obvious that there are ``gauge symmetries'' hidden in $B(p)$, given its definition, \eqref{DefB}. Since $A(p)$ possesses global gauge invariance (from the symmetries of the action), we can certainly shift it by a constant. In terms of $B$, this means that $B$ can be shifted by a first-order polynomial. So let
\be
	B(p)\to \tilde B(p)-\frac{2ap}{s\eps q}+\frac{1}{\eps q\l(\frac2{d-2}-\ga\da\r)}\l(\frac{a^2}s+\eps qk\r),
\ee
and we drop the tildas in the following.

Then, \eqref{radial_eq} becomes
\be
	-\frac{s\eps q}4\dot B^2+ \l( \frac{2}{d-2}-\ga\da \r)(p\dot B -B)=  \ddot B\l[X(p)+  \frac{q\eps}2\l(  \frac{2}{d-2}-\ga\da \r)B\r],
\label{radial_eq2}
\ee	
where $X(p)$ does not contain any term proportional to $a$ or $k$ anylonger. Reshuffling some terms,
\bea
	&&\frac{\ud}{\ud p}\l[\frac{\l(\ga^2-2\ga\da+\frac{d+2}{d-2}\r)p\dot B-2\l(\frac2{d-2}-\ga\da\r)B}{p-\frac{\eps q}{2}\dot B}\r] =\qquad\qquad\qquad\qquad\qquad\qquad\qquad\nn\\
	&&\qquad\qquad\qquad\qquad = \frac{\ddot B}{\l(p+\frac{\eps q}2\dot B\r)^2}\l\{\l[\l(\ga-\da^2\r)^2-\frac{d-4}{d-2}\r]p^2+2hp+\frac{c^2}s\r\}\,. \label{radial_eq3}
\eea
It is quite straightforward from this expression to see that the case $\ga=\da$, $h=c=0$ will be integrable in four dimensions.

Once we have determined $B$ analytically or numerically from \eqref{radial_eq}, we can then find a solution of the entire system which corresponds to an exact solution of  the action \eqref{EMDaction} for a metric of cylindrical symmetry \eqref{cylindrical_metric}. We will find several solutions in the next sections for the case of four dimensions. Indeed, once $B$ is known from \eqref{X}, it is easy to see that
\be
	\ln \alpha(p)= \int \frac{p\,\ud p}{X(p)- \frac{q\eps}2(\ga\da-\frac2{d-2})B}\,.
	\label{logalpha}
\ee
Using \eqref{h_r}, we solve for $\nu$:
\bea
	2\nu&=&\l( \frac{d-1}{d-2} -\frac{\da^2}{2}\r)\ln\alpha +\nn\\
	&&+\int \frac{\ud p}{X(p)- \frac{q\eps}2(\ga\da-\frac2{d-2})B} \l[-h+\l(  \frac{2}{d-2}-\ga\da  \r)\l(\frac{\eps}{2}qA+\frac as\r)\r].
	\label{nu_eq}
\eea
Note that, alternatively, equation \eqref{m1} enables us to write:
\be
	\e^{2\nu}=-\frac{1}{2\Lambda}\l[X(p)+ \frac{q\eps}2\l(   \frac{2}{d-2}-\ga\da \r)B\r]\alpha^{-\frac{d-4}{d-2}+\frac{\da^2}2}\,.
	\label{solnu}
\ee
These two equations fix $\Lambda$ with respect to the integration constants.
We make use of \eqref{c_r} to write
\be
	\psi = c\int \frac{\ud p}{X(p)+ \frac{q\eps}2\l( \frac{2}{d-2}-\ga\da  \r)B}\,,
	\label{solpsi}
\ee
and similarly for $\psi_i$. Finally, to get $\Omega$, we use \eqref{Q}:
\be
	\e^{\Omega}= \frac{q\al^{\frac2{d-2}-\ga\da}}{\dot A(p) \l[X(p)+ \frac{q\eps}2\l( \frac{2}{d-2}-\ga\da  \r)B\r]}\,.
	\label{solOmega}
\ee
Note that, in terms of $p$, the line element becomes:
\be
	\label{cylindrical_metric_bis}
	\ud s^2 =\frac{\al^{\frac32}\e^{2\chi}\ud p^2}{\l[X(p)+\frac{q\eps}2\l(  \frac{2}{d-2}-\ga\da \r)B\r]^2} + \e^{2\chi}\al^{-\frac{1}{2}}\ud \theta^2+\al\l(-\e^{2\U}\ud t^2+\e^{-2\U}\,\ud\varphi^2\r).
\ee
It is important to note here that \eqref{solnu} and \eqref{nu_eq} provide a relation between the action parameter $\Lambda$ and the constants of integration. This is similar to the pure dilatonic case, \cite{Charmousis:2001nq}. Here \eqref{solOmega} and \eqref{Q} provide an additional relation between constants of integration.

In all generality we need only to solve (\ref{radial_eq}) which is not integrable in general. There are, however, several special cases depending on the coupling constants of our theory $\ga$ and $\da$. In fact it is easy to see that (\ref{kr}) and (\ref{X}) are decoupled when $\ga \da=\frac2{d-2}$ and in this particular case we can obtain the general solution. We deal with this case first and focus on four-dimensional spacetimes.

\subsubsection{Generic dilatonic \texorpdfstring{$\ga \da=1$}{gamma*delta=1} solutions}

	\label{subsubsection:ka0_general}
\paragraph{Solution to the equations of motion}

Combining \eqref{kr} and \eqref{X} we obtain
\be
\label{cond}
	\frac{\ud A}{-\frac{\eps sq}4A^2 -aA+1} = \frac{\ud p}{\frac{3-\da^2}2 p^2 - hp -\frac{\eps q}2 -\frac{c^2}s -\frac{a^2}{2s} }\,,
\ee
where we have rescaled $k$ and we remind the reader that the constant $a$ is arbitrary, reflecting a choice of coordinates which we now fix.
We demand the discriminants of both polynomials to be equal to each other and positive. We set
\be
	\Delta_X=\Delta_A = a^2+\eps sqk = \la^2 > 0,
\ee
where $\la$ is now arbitrary replacing $a$, hence
\be
	X(p) = \frac{3-\da^2}2 p^2 - hp -\frac{c^2}s -\frac{\la^2}{2s}.
	\label{X:1}
\ee
The discriminant $\Delta_X$ on the other hand is always positive for $\da^2<3$. When $\da^2>3$ we need to suppose additionally that $h^2>\frac{2c^2}{s} (\da^2-3)$\footnote{The case $h^2\leq \frac{2c^2}{s} (\da^2-3)$ can be dealt with in a different coordinate system but does not present interesting black-hole solutions.}. The case of $\delta^2=3$ is marginal and we relegate it to Appendix \ref{d2=3g2=1/3Solutions}. First, let us  take $\da^2<3$
and then proceed to a rescaling of coordinates
\be
	\bar h =\frac h{|\la|}\,, \qquad \bar c= \frac c{|\la|}\,, \qquad \bar p = \frac{3-\da^2}{|\la|}p -\frac h{|\la|}+1\,.
\ee
Dropping anew all the bars and comparing with (\ref{X}) we have
\be
	h^2+(3-\da^2)\frac{2c^2}s = \frac{(\da^2-1)^2}{\da^2+1}\,,
	\label{rel_h-c-lambda:1}
\ee
which imposes certain conditions on $h$ and $c$. In particular, we can see that the case $\ga=\da=1$ has to be treated separately and will be dealt with after this section.
We now integrate (\ref{cond})
\be
	\dot A(p) = 2\e^{-\frac{\phi_0}{2\da}}\sqrt{\frac{\da^2(1-\eta^2)}{1+\da^2}}\frac1{(\eta p +1 -\eta)^2}\,, \qquad X(p)=\frac{\la^2}{2(3-\da^2)}p(p-2)\,,
	\label{A:1}
\ee
where the dot denotes derivation with respect to $p$, the electric charge has been replaced by its expression in terms of $\phi_0$, an integration constant linked to the dilaton field and the integration constant $\eta$ is such that $|\eta|<1$. The zeros of $X$, $p=0$, $p=2$ and the singularity in $A$, $p_\eta=1-1/\eta$ will be possible singularities or horizon positions for the metric. 

Using the integrals obtained in the previous section we can now write down the general solution for the case of cylindrical symmetry:

\bea
	\ud s^2 &=& -(p-2)^{C(-h,-c,\eps)}p^{C(h,c,\eps)}(\eta p+1-\eta)^{\frac{2\eps\da^2}{1+\da^2}} \ud t^2 +\nonumber\\
			&&+\frac{\e^{\da\phi_0}}{-\Lambda(3-\da^2)}p^{F(h,c)}(p-2)^{F(-h,-c)}(\eta p+1-\eta)^{\frac{2\da^2}{1+\da^2}}\ud p^2 +\nonumber \\
			&& +p^{B(h,c)} (p-2)^{B(-h,-c)}(\eta p+1-\eta)^{\frac{2\da^2}{1+\da^2}} \ud z^2+ \nonumber \\
			&& + p^{C(h,c,-\eps)} (p-2)^{C(-h,-c,-\eps)} (\eta p+1-\eta)^{-\frac{2\eps\da^2}{1+\da^2}} \ud\varphi^2,
\label{metric:1}
\eea
and dilaton field
\be
	\e^{\phi} = \e^{\phi_0}(\eta p+1-\eta)^{\frac{2\da}{\da^2+1}}p^{D(h,c)}(p-2)^{D(-h,-c)}\,,
      \label{Phi:1}
\ee
where we have rescaled the $t$, $z$ and $\varphi$ coordinates to absorb constant overall factors and
where the exponents are given by:
\bsea
\label{exp}
	F(h,c) &=& -1+\frac{\da^2}{3-\da^2}(1-h)-\da^2\frac{1+\da\sqrt2c}{1+\da^2}\,, \\
	B(h,c) &=&1+\frac{\da^2-2}{3-\da^2}(1-h)-\da^2\frac{1+\da\sqrt2c}{\da^2+1}\,,\\
	C(h,c,\eps) &=& \frac{1-h}{3-\da^2}-\eps\da\frac{\da-\sqrt2c}{\da^2+1}\,,\\
	D(h,c)&=&\frac{\da}{3-\da^2}(1-h)-\da\frac{1+\da\sqrt2c}{1+\da^2}\,.
\esea
Note the symmetry upon exchanging the sign of $h$ and $c$ and interchanging $p$ and $p-2$:
\be
	\label{symmetry}
	\ud s^2(h,c,p>2)=\ud s^2(-h,-c,p<0)\,.
\ee
Therefore we only need to study the $p>2$ interval.

To determine which points of spacetime are singular, we need to calculate the Ricci scalar, which can be done in two (equivalent) ways: either from the metric (41) or from the expression of the trace of the stress-energy tensor,
\be
	R = -  T = \half\l(\partial\phi\r)^2 + 4\Lambda \e^{-\da\phi} =- T_1-T_2\,.
	\label{Ricciscalar_T}
\ee
The expressions for $T_1$ and $T_2$ are:
\bsea
	T_1 &=& \frac{2\Lambda\da^2\e^{-\da\phi_0}}{\l(\da^2-3\r)\l(1+\da^2\r)}\l(\eta p+1-\eta\r)^{-2-\frac{2\da^2}{\da^2+1}}p^{-1-\da D(h,c)}(p-2)^{-1-\da D(-h,-c)} \nn\\
&&\l\{-(1+\da^2)p^2+\l[-(1+\da^2)\eta h+\eta(\da^2-3)\da\sqrt2c+2(1+\da^2)\eta+2(1-\da^2)\r]p\r. \nn\\
&&\l.+(\eta-1)\l[(1+\da^2)h-(\da^2-3)\da\sqrt{2}c-2(1-\da^2)\r]\r\}^2\,,\\
	T_2&=&4\Lambda \e^{-\da\phi_0}\l(\eta p+1-\eta\r)^{-\frac{2\da^2}{\da^2+1}}p^{-\da D(h,c)}(p-2)^{-\da D(-h,-c)}\,,
\esea
so that formally
\be
	R = P_4\l(p,\eta,h,c,\da\r)\l(\eta p+1-\eta\r)^{-2-\frac{2\da^2}{\da^2+1}}p^{-1-\da D(h,c)}(p-2)^{-1-\da D(-h,-c)}\,.
	\label{Ricciscalar:1}
\ee
Let us first look at the asymptotics $p\to\infty$:
\be
	R \underset{\infty}{\sim} p^{\frac{2}{\da^2-3}}\,.
\ee
We conclude that the Ricci curvature will be regular as $p\to\infty$ if and only if $\da^2<3$, while it will diverge if $\da^2>3$.  

Given (\ref{Ricciscalar:1}), we see that there is always a curvature singularity at $p_\eta$, but it is more subtle  to read what happens at $p=0$ or $p=2$, which can be curvature singularities or horizon positions. Indeed, one can look at the sign of the exponent $-1-D(-h,-c)$. This is a function of two variables, $h$ and $\da$, since $c$ is constrained by \eqref{rel_h-c-lambda:1}. Plotting it in terms of $h$ and $c$ shows that it is always negative. Computing the partial derivatives of this function with respect to $h$ and $\da$, we find a single extremum at $h=\frac{1-\da^2}2$, for which $D(-h,-c)=0$. Thus, except in the case where $h=\frac{1-\da^2}2$, $p=0$ and $p=2$ will be curvature singularities.

So, using the freedom we have in $h$ and $\da$, we will try to regularise the solutions for $p=2$. The following statements are equivalent:
\begin{enumerate}
\item  The dilaton field is regular at $p=2$ $\Longrightarrow$ $D(-h,-c)=0$.
\item The Ricci scalar is regular at $p=2$ $\Longrightarrow$ $-1-D(-h,-c)=-1$ and $P_4\l(p,\eta,h,c,\da\r)=(p-2)P_3\l(p,\eta,\da\r)$.
\item $C(-h,-c,-1)=1$.
\item $F(-h,-c)=-1$.
\item $B(\pm h,\pm c)=C(\pm h, \pm c, 1)$.
\item $h=\da\sqrt2c\Longrightarrow h=\frac{1-\da^2}2$.
\end{enumerate}

\paragraph{Black-hole and other regular solutions}

Fixing $\eps=-1$ and $h$ as stated above, we obtain the following regular solution:
\bsea
	\dot A(p) &=& 2\e^{-\frac{\phi_0}{2\da}}\sqrt{\frac{\da^2(1-\eta^2)}{1+\da^2}}\frac1{(\eta p +1 -\eta)^2}\,, \slabel{A:2}\\
	\e^{\phi} &=& \e^{\phi_0}(\eta p+1-\eta)^{\frac{2\da}{\da^2+1}}p^{\frac{4\da(\da^2-1)}{(\da^2+1)(3-\da^2)}}\,, \slabel{Phi:2}\\
	\ud s^2 &=& -(p-2)\frac{p^{\frac{-\da^4+6\da^2-1}{(1+\da^2)(3-\da^2)}}}{(\eta p+1-\eta)^{\frac{2\da^2}{1+\da^2}}} \ud t^2 -\frac{\e^{\da\phi_0}}{\Lambda(3-\da^2)}\frac{p^{\frac{5\da^4-6\da^2-3}{(1+\da^2)(3-\da^2)}}}{p-2}(\eta p+1-\eta)^{\frac{2\da^2}{1+\da^2}}\ud p^2 \nonumber \\
			&& + \,p^{\frac{2(\da^2-1)^2}{(\da^2+1)(3-\da^2)}} (\eta p+1-\eta)^{\frac{2\da^2}{1+\da^2}}\l( \ud z^2 + \ud \varphi^2 \r).\slabel{metric:2}
	\label{solution:2}
\esea
$\eta$ is constrained to be $|\eta|<1$, which implies that $p_\eta<0$ ($\eta>0$) or $p_\eta>2$ ($\eta<0$). We will of course arrange for the first eventuality. The $\eta=0$ has a special interpretation, and we will focus on it in Section \ref{section:NearExtremalPlanarEMD}. 
We immediately change coordinates and rescale the metric to get the following expression, which is more palatable:
\bsea
	\ud s^2 &=& - \frac{V(r)\ud t^2 }{\l[1-\l(\frac{r_-}{r}\r)^{3-\da^2}\r]^{\frac{4(1-\da^2)}{(3-\da^2)(1+\da^2)}}}+ e^{\da\phi}\frac{\ud r^2}{V(r)}+ \nn\\
			&&\qquad\qquad\qquad+ r^2\l[1-\l(\frac{r_-}{r}\r)^{3-\da^2}\r]^{\frac{2(\da^2-1)^2}{(3-\da^2)(1+\da^2)}}\l(\ud x^2+\ud y^2\r), \slabel{Metric1} \\
	V(r) &=& \l(\frac r\ell\r)^2-2\frac{m\ell^{-\da^2}}{r^{1-\da^2}} +\frac{(1+\da^2)q^2\ell^{2-2\da^2}}{4\da^2(3-\da^2)^2r^{4-2\da^2}}\,, \slabel{Pot1} \\
	\l(r_{\pm}\r)^{3-\da^2} &=& \ell^{2-\da^2}\l[m\pm\sqrt{m^2-\frac{(1+\da^2)q^2}{4\da^2(3-\da^2)^2}}\r], \slabel{Horizon1}\\
	\e^{\phi}&=& \l(\frac r\ell\r)^{2\da}\l[1-\l(\frac{r_-}{r}\r)^{3-\da^2}\r]^{\frac{4\da(\da^2-1)}{(3-\da^2)(1+\da^2)}}\,, \slabel{Phi1}\\
	A &=&\frac{q\ell^{2-\da^2}}{(3-\da^2)} \l[1 -\l(\frac{r_+}{r}\r)^{3-\da^2}\r]\ud t\,, \slabel{A1}
	\label{Sol1}
\esea
where the parameters $m$ and $q$ are integration constants linked to the gravitational mass and the electric charge (note that the electric potential \eqref{Sol1} has been fixed accordingly). There is an overall scale $\ell$ which can be fixed freely as the metric is scale-invariant (up to redefinitions of $m$ and $q$, of course),
\be
	\ell^2=\frac{3-\da^2}{-\Lambda}\,.
\label{107}\ee
 We could easily consider dS-like solutions by setting
\be
	a^2=\frac{3-\da^2}{\Lambda}\,,
\label{108}\ee
in the usual fashion, and these solutions can be obtained from \eqref{Sol1} by making the change $\ell\to ia$. However, this forces to Wick-rotate both the time coordinate $t$ and the radial coordinate $r$, yielding time-dependent solutions rather than black-hole spacetimes. We will thus restrict our analysis to cases where $\ell^2$ is real and positive. It is worth emphasising once again that what controls the scalar curvature of the solutions, i.e. (A)dS-like, is the sign of the product $(3-\da^2)\Lambda$. We will now focus on the static case, where the radial coordinate $r$ is space-like:
\begin{enumerate}
 \item $\Lambda<0$ and $\da^2<3$, which will yield black-hole solutions;
 \item $\Lambda>0$ and $\da^2>3$, which will yield singular solutions.
\end{enumerate}
The Ricci scalar, computed with \eqref{RicciScalarOnShell} is:
\be
	R = 4\Lambda \e^{-\da\phi} -\frac{2\da^2\Lambda}{3-\da^2}\e^{-\da\phi}\frac{\l[1-\l(\frac{r_+}{r}\r)^{3-\da^2}\r]}{\l[1-
\l(\frac{r_-}{r}\r)^{3-\da^2}\r]}\l[1-\frac{3-\da^2}{1+\da^2}\l(\frac{r_-}{r}\r)^{3-\da^2}\r],
	\label{RicciScalar1}
\ee
and using \eqref{Sol1}, it is straightforward to observe that for $\da^2<3$, both $r=0$ and $r_-$ are singular. This confirms that in the generic case, the only event horizon is at $r=r_+$, and spacetime only extends up to $r=r_-$. Quite interestingly, when $\da=1$, $r_-$ ceases to be a singular point and turns into an inner horizon, and spacetime extends all the way to $r=0$.

The solution has an extremal limit when
\be	 r_e^{3-\da^2}=m_e\ell^{2-\da^2}=\frac{\sqrt{1+\da^2}q\ell^{2-\da^2}}{2\da(3-\da^2)}\,.
\label{109}\ee
The two points $r^\pm$ then merge and become singular, except again in the limit $\da=1$ where the extremal black hole is regular.

For $\da$ taken in the range $\da^2<3$, we have AdS-like solutions. The solution \eqref{Sol1} is then truly a black-hole solution with an event horizon situated at $r_+$ and a curvature singularity at $r_-$. The spacetime is static outside the horizon and the singularity is time-like. The black-hole spacetime has to be cut off at $r_-$ and does not extend all the way to $r=0$.  The background spacetime obtained either by taking $m=q=0$ or the asymptotic limit $r\to+\infty$ of \eqref{Sol1} goes all the way to $r=0$, where a curvature singularity sits. This does not come as a surprise, as the background solution still contains a non-trivial scalar field which is singular asymptotically and thus cannot be treated as a perturbation. It is also worth pointing out that the horizon collapses at $r=r_-$, that is for finite ``radius''. This kind of feature seems to be characteristic of general dilatonic solutions, see \cite{Garfinkle:1990qj} for the zero potential case in String Theory\footnote{This solution, or more precisely its generalisation to free $\ga$, was proven in \cite{Gibbons:1987ps,Charmousis:2009xr} to be the general solution of the system of equations of motion without a scalar potential.} or references \cite{Henneaux:2004zi,Henneaux:2006hk} for related work.

\paragraph{$\da^2>3$ case :}

From \eqref{RicciScalar1}, we observe that in the case where $\da^2>3$ (which is perfectly admissible from previous considerations, provided we take a positive $\Lambda$), the Ricci scalar blows up when $r\to+\infty$ and vanishes  when $r\to0$. This is a hint that our space-like coordinate is ill-chosen, and that we should change $r\to\frac1r$. However, upon doing this, we find that the new radial coordinate is actually space-like \emph{inside} the horizon radius, so in the end we obtain a spacetime describing a cosmological horizon covering a naked singularity. Although this is reminiscent of de Sitter space, the inner singularity spoils the regularity of spacetime. A more appropriate picture would rather be that this depicts the interior of the black hole.

\paragraph{Magnetic $\eps=1$ solutions :}

Setting $\eps=1$ in (\ref{metric:1}) and $h$ as for the black hole solutions above, we obtain a "solitonic" version of \eqref{solution:2}, with Wick rotated $t=i\theta$ and $\varphi=i\tau$. This solution  is of axial symmetry at the origin $p=2$ and has a conical singularity given by $(g_{\theta\theta})'$ evaluated at $p=2$. The conical singularity can be removed by adequately rescaling the $\theta$ angle's periodicity in the standard way, given that we have infinite proper distance in $p$. Whenever there is a conical singularity, metric (\ref{solution:2}) describes the gravitational field of a magnetic straight cosmic string immersed in the $(\tau, z)$ plane.
\newline

\paragraph{Magnetic dual solutions : }
As we were careful to write every quantity with respect to $\da$, in order to obtain the dual magnetic solution, we only need to replace the electric Maxwell field with its magnetic dual:
\be
	A=-qz \,\ud\varphi\,.
\ee
Then (\ref{solution:2}) is a solution for (\ref{EMDaction}) for the theory with  $\ga\da=-1$.

\paragraph{String case $\ga=\da=\pm 1$}
\label{subsubsection:ka0_string}

In this case, \eqref{rel_h-c-lambda:1} imposes very severe constraints on $h$ and $c$ : $h=c=0$ and questions our gauge choice for $a$ (\ref{X:1}). We can try a different approach by setting
\be
	\Delta_X=b^2\la^2\,,
\ee
in order to relax (\ref{X:1}). Unfortunately, although the system is still fully integrable, this does not yield any black hole solutions other than for $|b|=1$ (by imposing homogeneous two-dimensional spatial sections and regularity at $p=2$). So, setting $\da=1$ and $h=c=0$  into \eqref{metric:2}, we get the following solution:
\bsea
	\dot A(p) &=& \e^{-\frac{\phi_0}{2}}\sqrt{2(1-\eta^2)}\frac1{(\eta p +1 -\eta)^2}\,, \slabel{A:4}\\
	\e^{\phi} &=& \e^{\phi_0}(\eta p+1-\eta)\,,\slabel{Phi:4} \\
	\ud s^2 &=& -\frac{p(p-2)}{\eta p+1-\eta} \ud t^2 -\frac{\e^{\phi_0}}{2\Lambda}\frac{\eta p+1-\eta}{p(p-2)}\ud p^2 +(\eta p+1-\eta)\big( \ud z^2 + \ud \varphi^2 \big), \slabel{metric:4}
	\label{solution:4}
\esea
and Ricci scalar:
\bea
	T_1 &=& -\eta^2\Lambda \e^{-\da\phi_0}\frac{p(p-2)}{\eta p-\eta p_\eta}\,,   \nn \\
	T_2 &=&-\frac{4\Lambda \e^{-\phi_0}}{\eta p -\eta p_\eta}\,, \nn \\
	R &= & \Lambda \e^{-\phi_0}\frac{3\eta^2p^2-2\eta(3\eta-4)p+4(\eta-1)^2}{\l(\eta p-\eta p_\eta\r)^3}.
	\label{Ricciscalar:4}
\eea
This coincides in fact with the limit $\da\to1$ taken from \eqref{solution:2}, in the $\ga\da=1$ general case. What this means is that in fact the $\ga=\da=1$ solution does belong to the same family as the general $\ga\da=1$ one, and that, although the intermediary calculations differ a little bit, there are no qualitative differences between them. 

However, there is still something special happening. Indeed, as can be seen from the solution \eqref{Sol1} and its Ricci scalar \eqref{RicciScalar1}, the singularity at $r_-$ becomes regular in the string case and turns into an inner Cauchy horizon. We thus get a black-hole spacetime extending all the way to $r=0$, whose causal structure is similar to Reissner-Nordstr\"om.

In  \cite{Poletti:1994ff}, the analysis of the global properties of the system of equations of motion revealed that some of the $\ka\neq0$ families of solutions had endpoints on the $\ka=0$ plane. This is actually a perfect example of this happening. Indeed, it can be shown that the $\da\to1$ limit in the non-planar solution  \eqref{CHMi} of  \cite{Chan:1995fr} can only be taken consistently if simultaneously $\ka\to0$ while keeping $\ka/(\da^2-1)$ fixed. Effectively, we are ``blowing up'' the sphere (for $\ka=1$) on the horizon to a plane. One could be tempted to interpret this as a ``stringy'' effect, as this happens specifically in the case descending from String Theory effective actions\footnote{At least this is the only instance we have of such a phenomenon.}.

\subsubsection{``Reissner-Nordstr\"om'' \texorpdfstring{$\ga=\da$}{gamma=delta} solutions}

As mentioned in Section \ref{PlanarReduction}, from \eqref{radial_eq}, we can integrate the case $\ga=\da$, $h=c=0$. This integration yields the following type of solution for $B(p)$:
\be
	B(p)=B_Np^{2\frac{1-\da^2}{3-\da^2}} + B_1p + B_0\,,
\ee	
which after some reparameterising of constants and rescalings, as well as changing coordinates to $r=p^{\frac1{3-\da^2}}$, is written
\bsea
	\ud s^2 &=& - V(r)\ud t^2 + e^{\da\phi}\frac{\ud r^2}{V(r)} + r^2\l(\ud x^2+\ud y^2\r)\,, \slabel{Metric2} \\
	V(r) &=& \l(\frac r \ell\r)^{2}-2m\ell^{-\da^2}r^{\da^2-1} +\frac{q^2}{4(1+\da^2)r^2}\,, \slabel{Pot2} \\
	\e^{\phi}&=& \l(\frac r\ell\r)^{2\da} \slabel{Phi2}\,, \slabel{ElectricPotential2}\\
	A &=&\frac{\ell^{\da^2}q}{(1+\da^2)r_+^{1+\da^2}} \l[1 - \l(\frac{r_+}{r}\r)^{1+\da^2}\r]\ud t\,. \slabel{A2}
	\label{Sol2}
\esea
We observe that this solution reduces to the solution \eqref{Sol1} when $\da=1$. Let us also note that for $\da=0$ (and so $\ga=0$, too), we recover the planar version of the AdS-Reissner-Nordstr\"om black hole, as expected from the action.

The scale $\ell^2=\frac{3-\da^2}{-\Lambda}$ is defined as in the previous solution, and, for the same reasons, we shall restrict our attention to black-hole solutions, that is $\Lambda<0$ and $\da^2<3$.

The Ricci scalar on-shell is:
\be
	R = 4\Lambda \e^{-\da\phi}\l[ 1-\frac{2\da^2\ell^2V(r)}{(3-\da^2)^2r^2}\r],
	\label{RicciScalar2}
\ee
so there is a curvature singularity at $r=0$ for $\da^2<3$, radial infinity is regular if $\da^2<3$ and there are event horizons at the zeros of the potential \eqref{Sol2}. The size of the horizon vanishes now at $r=0$.

 We analyse the conditions for $V(r)$ to vanish. Indeed, solving simultaneously for $V(r)=0$ and $\frac{\partial V(r)}{\partial r}=0$, we find the extremal value
\be
	r_e^4=\frac{q^2\ell^2}{4(3-\da^2)}\,,\qquad m_e = \frac{2r_e^{3-\da^2}}{(1+\da^2)\ell^{2-\da^2}}\,.
\label{132}\ee
This implies that for $r_+\geq r_e$, or equivalently $m\geq m_e$, there are two event horizons, one inner and another outer, which are degenerate in the extremal case (where the bound is saturated). Below the bound, a naked singularity exists. The extremal black hole is defined only for $\da^2<3$. We call these solutions ``Reissner-Nordstr\"om''-like since they are the only ones with two regular non-degenerate horizons. Note also that they coincide with the previous string solutions for $\ga=\da=1$, \eqref{solution:4}. This suggests the following remark: although it is clear from our derivation that the above $\ga=\da$ solutions \eqref{Sol2} should not be considered as the general solution to the equations of motion for this particular relationship between the couplings of the theory\footnote{On the face of it, the two extra integration constants $h$ and $c$ have been set to zero. This however only ``counts'' once since $c$ has to be related to $h$ to have a homogeneous horizon, as has been made clear in the $\ga\da=1$ case.}, the $\ga=\da=1$ \emph{is} the generic solution in this case, since the equations of motion \emph{impose} that $h=c=0$ and it can consistently be derived as the $\da\to1$ limit of the $\ga\da=1$ general solution \eqref{Sol1}. However, no argument can yet be put forward as to why the integration procedure for $\ga=\da$ should be the general one, there may exist other solutions.

Thus, there are two independent integration constants specifying the solution, as well as an independent overall scale that we have fixed to its ``natural'' value, e.g. the maximal number allowed by the equations of motion. They are $m$, $q$ and $\ell$ as before and there is no relation between them. $m$ and $q$ can be considered as the ``reduced'' mass and charge, while $\ell$ is the ``AdS'' radius.

\subsubsection{Near-extremal, arbitrary \texorpdfstring{$\ga$}{gamma}, \texorpdfstring{$\da$}{delta} solutions}

\label{section:NearExtremalPlanarEMD}

When seeking a solution of \eqref{radial_eq} for general coupling constants $\ga$ and $\da$ one has to make some suitable ansatz. Given the form of the general solution, found in the previous section, we expect by continuity some form of polynomial solutions in the same coordinate system. As we saw earlier on, roots of the polynomial are either singularities or Killing horizons. The generic Ansatz that works for $B(p)$ for arbitrary coupling constants is a second-order polynomial in $p$. Setting
\be
	u=\ga^2-\ga\da+2\,, \qquad v=\da^2-\ga\da-2\,, \qquad w=\frac 1u\l[s(3-\da^2)+(1-\ga\da)^2)\r],
\ee
\be
	\la^2=\Delta_X = \Big(\frac{sh}u\Big)^2+\frac{wu}{s(v+u)}\Big[2c^2+\Big(\frac{sh}u\Big)^2\Big], \quad \bar p = \frac w{|\la|}p-\bar h+1, \quad \bar h=\frac{sh}{u|\la|}, \quad \bar c=\frac{c}{|\la|},
\ee
gives for $A(p)$ and $X(p)$:
\bea
	A(p) &=& a-\frac{1-\ga\da}wh-\frac{sv}{wu}(p-1)\,, \label{A:7}\\
	X(p)&=&\frac {\la^2}{2w}p(p-2)-\frac{Q\eps}2(1-\ga\da)B(p)\,, \label{X:7}
\eea
where we have dropped all bars and our coordinate $p$ is dimensionless. Notice that $uw$, $u$ and the discriminant of $X$ are necessarily positive for all $\da^2<3$, and consequently so is $w$. Thus, for $\da^2<3$, the $p$-coordinate is  space-like. If $\da^2>3$, no general arguments can easily be made. The roots of $X(p)$, namely $p=0$ and $p=2$, are again singularities. The solution we obtain for general couplings $\ga$ and $\da$ is the following:
\bsea
	\ud s^2&=&  - p^{C(  h,  c,\eps)}(p-2)^{C(-  h,-  c,\eps)}\ud t^2 +\frac{\e^{\da\phi_0}}{-w\Lambda}p^{F(  h,  c)}(p-2)^{F(-  h,-  c)}\ud p^2+ \nonumber\\
			&&+p^{B(  h,  c)}(p-2)^{B(-  h,-  c)}\ud z^2 + p^{C(  h,  c,-\eps)}(p-2)^{C(-  h,-  c,-\eps)}\ud\varphi^2\,, \slabel{metric:7}\\
	\e^{\phi} &=& \e^{\phi_0}p^{D(h,c)}(p-2)^{D(-h,-c)} \slabel{Dilaton:7}\,,
\label{solution:7}
\esea
where $\phi_0$ is an integration constant,  and the exponents are
\bea
	F(  h,  c) &=& \frac{1-  h}w\Big[\da^2+\frac{\ga\da}s(1-\ga\da)\Big]-1-\frac1s(\da\sqrt2  c+\ga\da)\,, \\
	B(  h,  c) &=&\frac{1-  h}w\Big[\da^2-2+\frac{\ga\da}s(1-\ga\da)\Big]+1-\frac1s(\da\sqrt2  c+\ga\da)\,, \\
	C(  h,  c,\eps) &=&\frac{1-  h}w\Big[1+\frac\eps s(1-\ga\da)\Big]-\frac\eps s(1-\ga\sqrt2  c)\,,\\
	D(h,c) &=& -\frac\ga s+\frac{\ga+\da}{ws}(1-h)-\frac{\sqrt2c}s.
	\label{powers}
\eea
The roots $p=0$ and $p=2$ are again interchanged under the symmetry (\ref{symmetry}) and as a result the solutions $p>2$ or $p<0$ are equivalent, up to inversing the signs of both $h$ and $c$. Inversing the sign of $\eps$ allows us to get the magnetic solutions from the electric ones, so we will consider the case $\eps=-1$ in the following without any loss of generality. As before, the sign for the cosmological constant and the nature of spacetime will depend on the couplings $\gamma$ and $\delta$. Indeed, for $w>0$, we see that $\Lambda$ is negative and the coordinate $p$ is time-like in between 0 and 2, whereas for $p>2$ the solution is static.
Given the form of the metric and its symmetry in $h$ and $c$, we can easily find the form of the metric for large $p$ by setting $h=c=0$. Given the coordinate transformation
\be
	\bar p=p^{\frac{(\ga-\da)^2}{wu}}\,,
\ee
we obtain the solution of maximal symmetry
\be
	\ud s^2 = -\bar p^{2\frac{\ga^2-\da^2+4}{(\ga-\da)^2}}\ud t^2 +\bar p^2\l(\ud z^2+\ud\varphi^2\r) +\frac{wu^2\e^{\da\phi_0}}{(-\Lambda)(\ga-\da)^4}\bar p^{2\frac{\da+\ga}{\da-\ga}}d\bar p^2\,.
	\label{metric_asymptotic:7}
\ee
This can be locally AdS if and only if we take the limit $\da=0$, $\ga\to\infty$, which is similar to the previous subsection for $\ga\da=1$. Here again, this amounts to cancelling the Maxwell term ($\dot A=0$ in this limit) and taking the dilaton to be trivial, which is in agreement with Wiltshire \emph{et al.}, \cite{Poletti:1994ff}.

Furthermore, the Ricci scalar is:
\bea
	T_1 &=& -\frac{2\Lambda}{ws^2}\e^{-\da\phi_0}\frac{\l[(\ga+\da-\ga w)p+w(\ga+\sqrt2c)+(\ga+\da)(h-1)\r]^2}{\l(p-2\r)^{1+\da D(-h,-c)}p^{1+\da D(h,c)}}\,, \nn \\
	T_2 &=& 4\Lambda \e^{-\da\phi_0}\l(p-2\r)^{-\da D(-h,-c)}p^{-\da D(h,c)}\,, \nn \\
	R &=&-T_1-T_2\sim_\infty p^{\frac{4\da(\ga-\da)}{wu}}\,, \label{Ricciscalar:7}
\eea
which, given that $wu>0$ for $\da^2<3$, yields the regular asymptotic region for $\ga<\da$, $\da>0$ and $p$ space-like infinity (if $\da<0$, remember there is a symmetry : change simultaneously $\ga\to-\ga$, $\da\to-\da$, $\phi\to-\phi$). Taking the limit $\da=0$, $\ga\to\infty$, the Ricci scalar equals $4\Lambda$ as expected.

\paragraph{Magnetic dual solutions }

Here again, we can use the same procedure as before and obtain a magnetic dual solution to this one. We just need to set $\ga\to-\ga$ in the previous metric and take
\be
	A=-qz\,\ud\varphi
\ee
as Maxwell field. This is particularly useful since in this way we can get an extra upliftable solution when $\ga\da=-1$ (giving $\ga\da=1$ after using the duality).

\paragraph{Black-hole solutions}

To get the black-hole solutions, we need to set (note that this is the equivalent of $h=\da\sqrt2c$ as in the previous subsection $\ga\da=1$, albeit for arbitrary couplings)
\be
	h=\frac{\ga+\da}{s}\sqrt2c\,,
\ee
which yields
\be
  \tilde h = \frac{\ga^2-\da^2}{2u}\,, \qquad \sqrt2\tilde c = \frac{\ga-\da}2\,.
\ee
Then, for the electric case $\eps=-1$, putting the scale back so as to make the $p$-coordinate dimensionful,
\bsea
	\ud s^2 &=& -V(p)p^{-4\frac{\ga(\ga-\da)}{wu}}\ud t^2 +\frac{\e^{\da\phi}\ud p^2}{-w\Lambda V(p)} + p^{2\frac{(\ga-\da)^2}{wu}}\l(\ud x^2+\ud y^2\r), \slabel{Metric3bis}\\
	\e^{\phi} &=& \e^{\phi_0}p^{-4\frac{(\ga-\da)}{wu}}\,, \slabel{Phi3bis}\\	
	A&=&2\sqrt{\frac{-v}{wu}}\e^{-\frac\ga2\phi_0} \l(p-2m\r)\ud t\,, \slabel{A3bis}\\
	V(r)&=& p(p-2m)\,, \slabel{Pot3bis} \\
	wu&=&3\ga^2-\da^2-2\ga\da+4\,, \qquad u=\ga^2-\ga\da+2\,, \qquad v=\da^2-\ga\da-2\,.\nn
	\label{Sol3}
\esea
The Ricci scalar is
\be
	R = 4\Lambda \e^{-\da\phi_0}p^{-\frac{3(\ga-\da)^2+4}{wu}}\l\{\l[1-\frac{2(\ga-\da)^2}{wu^2}\r]p+4m\frac{(\ga-\da)^2}{wu^2}\r\}.
	\label{RicciScalar3}
\ee

\begin{figure}
\begin{center}
\begin{tabular}{c}
	 \includegraphics[width=0.35\textwidth]{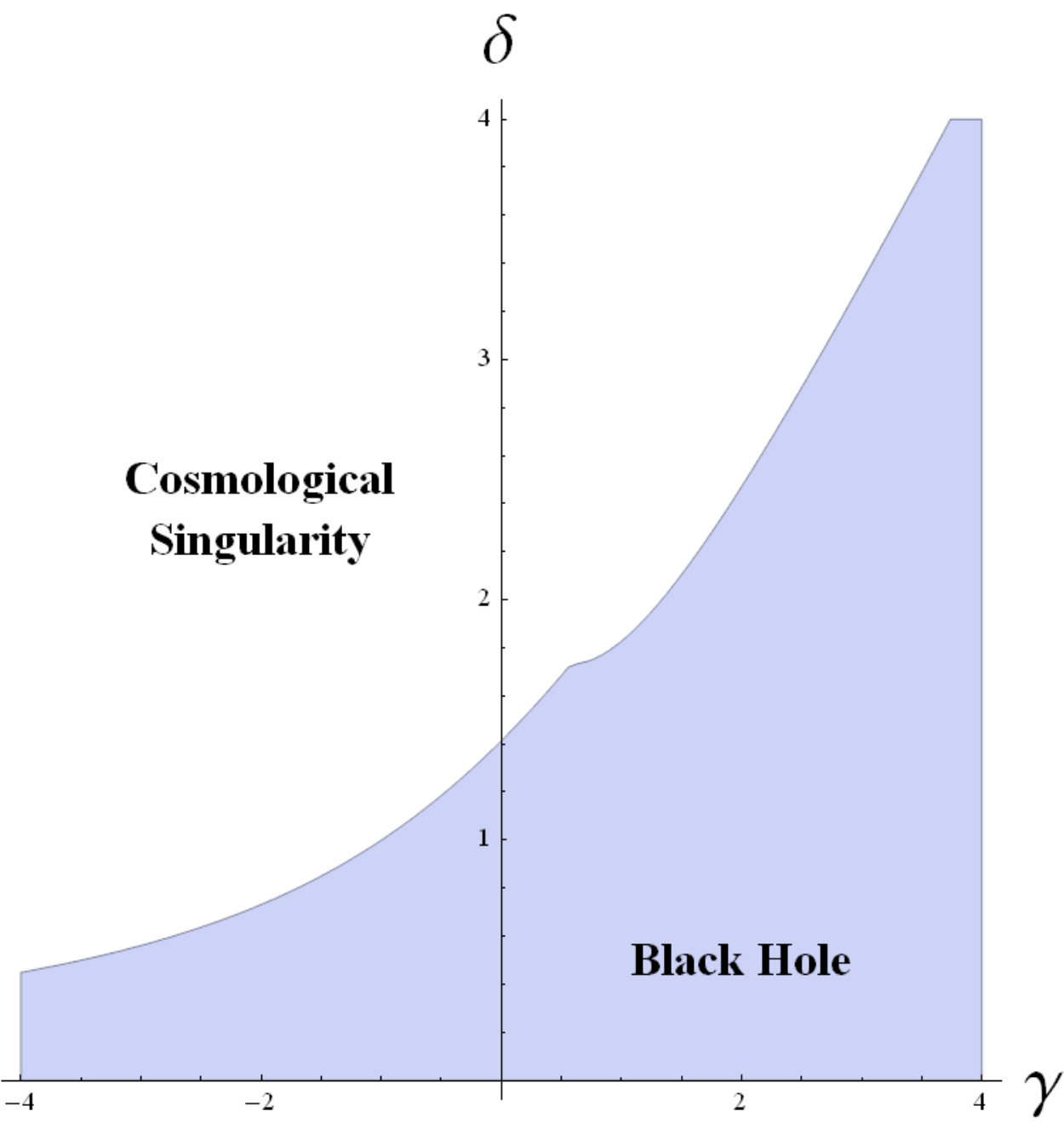}
\end{tabular}
\caption[Nature of the planar near-extremal black-hole solutions in Einstein-Maxwell-Dilaton theories]{This plot shows the nature of the near-extremal solution \eqref{Sol3} in the upper half $\l(\ga,\da\r)$ plane.}
\label{Fig:NatureSolutionNearExtremal}
\end{center}
\end{figure}

There is an event horizon at $p_h=2m$ and a curvature singularity at $p=0$ for all $\da^2<3$. Indeed, it is not hard to see that if $wu>0$ (which is true in particular for all $\da^2<3$),
\be
	wu=(3-\da^2)(1+\ga^2)+(1-\ga\da)^2>0\quad\forall \da^2<3\,, \nn
\ee
then the Ricci scalar diverges at $p=0$ and is regular at $p_h$. On the other hand, if $wu<0$, then we need to change variables $p\to\frac1p$ in order for the Ricci scalar not to diverge asymptotically, but then this changes the nature of the solution and we have a cosmological spacetime. These respective regions in the upper half $(\ga,\da)$ plane are plotted in \Figref{Fig:NatureSolutionNearExtremal}, together with the condition that the $p$-coordinate should be space-like beyond the horizon, which is simply $vu<0$ (we take $\Lambda<0$ to have a potential bounded from below). There is no need to consider the lower half-plane, as it can be obtained by a rotation of $\pi$ around the origin, reflecting that any simultaneous change of sign in $\ga$ and $\da$ can always be absorbed by a change of sign in $\phi$.

Here again, $m$ is related to the mass of the solution, and $\phi_0$ to the overall scale (and thus can be fixed at will), but we lack a second dimensionful integration constant. As such, the charge density  built through the usual Gaussian integral is finite but universal. This hints at the fact that the solution is not the generic one.

 We will now proceed and take both $\ga\da=1$ and $\ga=\da$ in \eqref{Sol3}, and compare the result with both full solutions we have at our disposal in these two cases, \eqref{Sol1} and \eqref{Sol2}, by going to the so-called \emph{near-extremal} limit in both cases.

\paragraph{Near-extremal $\ga\da=1$ solution}

Let us determine the near-extremal behaviour of the full solution \eqref{Sol1} that is, set\footnote{We could as well work with $r_+$, this is just a matter of convenience.} $r^{3-\da^2} = r_-^{3-\da^2} + p$, with $p$ small compared to $r_-$. We obtain:
\bsea
	\e^\phi &=& r_-^{2\da\frac{(3-\da^2)}{(1+\da^2)}}p^{\frac{4\da(\da^2-1)}{(3-\da^2)(1+\da^2)}}\,, \\
	A & = & \l[\phi + \sqrt{\frac{\da^2}{1+\da^2}}r_-^{-\frac{(3-\da^2)}{(1+\da^2)}}p \r]\ud t\,, \\
	\ud s^2 &=& -\l[p(p-2\tilde m)\r]p^{\frac{4(\da^2-1)}{(3-\da^2)(1+\da^2)}}  \ud t^2 +\frac{\e^{\da\phi}}{(3-\da^2)^2}\frac{\ud p^2}{p(p-2\tilde m)}+ \nn\\
					&&+p^{\frac{2(\da^2-1)^2}{(3-\da^2)(1+\da^2)}}  \l(\ud x^2+\ud y^2 \r),
	\label{NearExtremal1}
\esea
where we have set $r_-\sim r_+$ at zeroth order and $\tilde m = r_+^{3-\da^2}-r_-^{3-\da^2}$. The extremal limit is seen to be exactly $\tilde m=0 \Longleftrightarrow r_+=r_-=r_e$. Then, identifying in \eqref{Sol3}
\bea
	\e^{\phi_0} &=&  r_-^{2\da\frac{(3-\da^2)}{(1+\da^2)}}\,, \\
	\frac{-\Lambda}{3-\da^2} &=& \ell^{-2} =1\,, 
\eea
the near-extremal limit of the $\ga\da=1$ solution \eqref{Sol1} and the $\ga\da=1$ limit of \eqref{Sol3} agree. Another remarkable thing is that we are now able to give precise meaning to the integration constant $\eta$ present in the $\ga\da=1$ solution: taking this limit $\eta=0$ in \eqref{solution:2} allows to recover directly the near-extremal solution \eqref{NearExtremal1} we have just derived. So the $p$-coordinate and its related $\eta$ integration constant seem to be ``tailored'' to yield easily the near-extremal geometry of the solution.

\paragraph{Near-extremal $\ga=\da$ solution}

We now turn to the second full solution for $\ga=\da$ \eqref{Sol2}, and set this time $r^{1+\da^2} = r_-^{1+\da^2} + p$. Then, taking care to keep all the terms to high-enough order, this yields:
\bsea
	\e^\phi &=& r_-^{2\da}\,, \\
	A & = & \l[\Phi + \sqrt2r_-^{-\da^2}p \r]\ud t\,, \\
	\ud s^2 &=& -p(p-2\tilde m) \ud t^2 +\frac{r_-^{2\da^2}\ud p^2}{2(3-\da^2)p(p-2\tilde m)} + \l(\ud x^2+\ud y^2 \r).
	\label{NearExtremal2}
\esea

This is just a product of $AdS_2\times\mathbf{R}_2$, with a constant dilaton but with a non-zero gauge field. $\tilde m$ has been redefined:
\be
	\tilde m = \frac{1}2r_-\l(\frac{r_e^4}{r_-^4}-1\r),
\ee
where here again $\tilde m=0$ corresponds precisely to the extremal limit $r_-=r_+=r_e$. We then compare with \eqref{Sol3}, and find perfect agreement, provided we set in \eqref{Sol3}
\bea
	\e^{\phi_0} & = & r_-^{2\da}\,, \\
	-(3-\da^2)\Lambda &=& \ell^{-2}= 1\,.
\eea
We can also remark upon the $\eta=0$ limit of the stringy solution $\ga=\da=1$, \eqref{solution:4}: by taking this limit, we recover exactly the near-extremal geometry for the $\ga=\da$ case \eqref{NearExtremal2} we have just derived. This confirms the interpretation given to $\eta$: it is a measure of the deviation from the near-extremal to the full geometry.

\paragraph{The arbitrary $\ga$, $\da$ solution is the near-extremal solution}

Now,  we are able to ascribe a precise interpretation to the generic $\ga$, $\da$ solution \eqref{Sol3}. Indeed, it turns out that, both in the limit $\ga\da=1$ and $\ga=\da$, it coincides exactly with the near-extremal limit of the full solution. Therefore {\em we interpret \eqref{Sol3} as the near-extremal limit of the full generic solution}. Yet, such a general solution remains elusive and we shall keep a more detailed search for later work.
\vfill
\pagebreak

\subsection{To the next order: from Einstein-Maxwell-Dilaton to Gauss-Bonnet}

We have examined at length the black-hole solutions (as well as other regular solutions) of Einstein-Maxwell-Dilaton theories, with or without a potential. Some of them greatly differ from General Relativity black holes. In the no-potential case, they have an irregular extremal limit in the Einstein frame, but not in the string frame. The horizon size collapses at a singularity which is located at finite radial coordinate, and there is no inner Cauchy horizon, contrarily to Reissner-Nordstr\"om black holes. Finally, these solutions are stable linearly and have asymptotically flat boundary conditions, so they are natural competitors of the Reissner-Nordstr\"om solutions and evade the usual no-hair ``theorems''.

In the Liouville potential case and for planar horizon topology, the equations of motion are not fully integrable in the general case, which yields only a near-extremal version of the full solution for arbitrary couplings $\ga$ and $\da$. Imposing the relation $\ga\da=1$ allows to integrate the problem, and provides a black-hole solution which has a singularity at finite radial coordinate and an irregular extremal limit. Contrarily to the no-potential case, it does not have regular boundary conditions, except for the flat potential case (in which case it reduces to Schwarzschild-(A)dS) and so does not circumvent no-hair theorems. A third family of solutions exists in the $\ga=\da$ case and results from integration of the equations of motion once one of the integration constants has been set to zero. It has the same irregular asymptotics as the previous family, but its causal structure is that of Reissner-Norstr\"om with two (outer event and inner Cauchy) regular horizons cloaking a singularity at zero radial coordinate. Thus, we have dubbed this family ``Reissner-Norstr\"om''-like. The linear stability of these solutions remains to be fathomed.

Finally, in the non-planar case, no integrability statement can be made, though several families of irregular black-hole solutions can be found ($\ga\da=\pm1$, $\ga+\da=0$), and were shown to display a variety of properties similar to those mentioned above. There is one difference though: they are all scaling solutions, meaning that they never have a singularity at finite radial coordinate. Thus, the state of affairs in this case is clearly not very satisfactory, but no obvious way exists to go beyond the results known so far.

We have also stressed that these theories, on top of being studied in their own right as theories of gravity with matter, can descend from higher-dimensional gravity theories: either from $d+1$-dimensional General Relativity through Kaluza-Klein reduction, or from $d$-dimensional (bosonic or supersymmetric) string theories as effective low-energy actions at the first order of the string loop expansion. One may then ask the following question: Can we go further? The answer is most certainly yes, in both directions. First, giving a flavour of what is to come, one may generalise the Einstein-Hilbert action in higher dimensions and incorporate a Gauss-Bonnet term. As we will argue in the next section, this theory is the natural candidate for a theory of gravity in five and six dimensions, similarly as General Relativity is in four dimensions. For instance, in $D=5$, the action is written
\bea
	S_{GB} &=& \int\ud^{5}x\sqrt{-g^{(5)}}\l[-2\La + R^{(5)} + \al\hat G^{(5)} \r],\\
	\hat G &=& R_{\la\mu\nu\rho}R^{\la\mu\nu\rho}-4R_{\mu\nu}R^{\mu\nu}+R^2\,,
\eea
and we have given the Gauss-Bonnet term explicitly. Second-order equations are derived from it, which do not contain any poles or ghosts in the graviton propagator. Next, we can apply the Kaluza-Klein reduction procedure
\bsea
	\ud s^2_5& = &\e^{-2a\phi}\ud w^2+\e^{a\phi}g_{\mu\nu}^{(4)}\ud x^\mu\ud x^\nu\,,\\
	R^{(5)}&=& \e^{-a\phi}R^{(4)}-a\e^{-a\phi}\nabla^2\phi-\frac52a^2\e^{-a\phi}\l(\nabla\phi\r)^2\,,\\
	\hat G^{(5)}&=&\hat G^{(4)}-4a\G^{(4)}_{\mu\nu}\nabla^\mu\nabla^\nu\phi+14a^2\G^{(4)}_{\mu\nu}\nabla^\mu\phi\nabla\nu\phi+3a^2\l(\nabla\phi\r)^2R^{(4)}-6a^2\l(\square\phi\r)^2+\nn\\
				&&+6a^2\l(\nabla_\mu\nabla_\nu\phi\r)^2+3a^3\square\phi\l(\nabla\phi\r)^2-18a^3\l(\nabla_\mu\nabla_\nu\phi\r)+6a^4\l(\nabla\phi\r)^4,
\esea
 and find that the action boils down to (after getting rid of some higher-derivative terms by integration by parts and discarding the corresponding boundary terms)
\bea
	S_{GBD} &=& \int \ud^4x\sqrt{-g^{(4)}}\l\{R^{(4)}-\frac32a^2\l(\partial\phi\r)^2-2\La\e^{a\phi}+\r.\nn\\
				&& \l.+\e^{-a\phi}\l[\al\hat G^{(4)}+4a^2\G^{(4)}_{\mu\nu}\nabla^\mu\phi\nabla^\nu\phi+3a^3\square\phi\l(\nabla\phi\r)^2\r]\r\}.
\eea
Here, we have not taken into account the Maxwell terms, which would give many more higher-order contributions. Moreover, we have written the action in five dimensions, but for dimensions higher than five, a term remains that is proportional to $\l[\l(\nabla\phi\r)^2\r]^2$. Taking $a^2=1/3$ allows to recover canonical normalisation for the kinetic term of the scalar field, and the values of the exponents $\ga$, $\da$ of the previous section appropriate for Kaluza-Klein reduction. Even though the action of the Gauss-Bonnet term is trivial in four-dimensional General Relativity, here it contributes to the dynamics through its coupling to the scalar field. Actions of this kind have been studied in braneworld setups, \cite{Binetruy:2002ck,Charmousis:2003ke}, as well as in Cosmology, \cite{Amendola:2005cr,Amendola:2007ni,Amendola:2008vd,Nojiri:2005vv,Carter:2005fu,Koivisto:2006xf,Koivisto:2006ai}, and of course their black-hole solutions have been studied, mostly perturbatively or numerically\footnote{It seems analytical solutions even for the simplest case with only the extra Gauss-Bonnet term remain to be found.},\cite{Mignemi:1991wa,Mignemi:1992nt,Kanti:1995vq,Torii:1996yi,Kanti:1997br,Melis:2005xt,Melis:2005ji,Guo:2008hf,Guo:2008eq,Ohta:2009tb,Ohta:2009pe,Maeda:2009uy}.

On the other hand, we may also consider two-loop expansions of the string surface amplitude, calculate the beta functions, and impose that they must vanish to verify conformal invariance locally. This method already allows to compute the presence of quadratic terms in the Riemann curvature, \cite{Callan:1986jb},
\be
	S=\int \ud^{d}x\sqrt{-g^{(d)}}\e^{\phi}\l[R+\l(\nabla\phi\r)^2-\frac14F^2+\frac14\al'R_{\la\mu\nu\rho}R^{\la\mu\nu\rho}-\frac{d-d_{crit}}{3\al'}\r],
	\label{HeteroticEffActionStringFrameTwoLoop}
\ee
in the bosonic, heterotic and Type II String Theory\footnote{For simplicity, we do not consider the two-form $B_{\mu\nu}$.}, with equations of motion
\bsea
	R_{\mu\nu}+\half\nabla_\mu\nabla_\nu\phi+\la R_{\mu\rho\sigma\tau}R^{\phantom{1}\rho\sigma\tau}_{\nu}&=&O\l(\al'^2\r),\\
	\square\phi-\l(\partial\phi\r)^2+\frac14R+\frac18\la R_{\la\mu\nu\rho}R^{\la\mu\nu\rho}&=&O\l(\al'^2\r),
\esea
with $\la=\al'/2,\,\al'/4,\,0$ for bosonic, heterotic and supersymmetric string theories. Note that ``string'' effects may only be felt when the curvature is strong with respect to the string scale $1/\al'$. Else, the higher-order terms in the curvature drop out and gravity is similar to Einstein's theory (e.g., no change in the graviton propagator). It is possible to go to the Einstein frame as in \eqref{HeteroticEffActionEinsteinFrame} and get the action
\be
	S=\int \ud^{d}x\sqrt{-g^{(d)}}\l[R- \frac2{d-2}\l(\nabla\phi\r)^2+\frac\la2\al'\e^{\frac2{d-2}\phi}R_{\la\mu\nu\rho}R^{\la\mu\nu\rho}-\frac{d-d_{crit}}{3\al'}\e^{-\frac2{d-2}\phi}\r],
	\label{EffActionEinsteinFrameTwoLoop}
\ee
where higher-order derivatives of the dilaton have been eliminated through further field redefinitions, and the kinetic terms for the dilaton and graviton have been diagonalised. Of course, an immediate issue with this type of action is that the equations of motion contain four-derivatives of the metric, and the graviton propagator a pole with the wrong sign, that is, a ghost degree of freedom, \cite{Stelle:1977ry}. 

Another way to tackle this is to consider the scattering amplitudes of the massless particles in the classical approximation (small string coupling),\cite{Gross:1986mw}. Like for the sigma-model approach, the effective actions obtained from the two-, three- or four-point interactions are not unique, since they can be subjected to a field redefinition, the S-matrices being identical under such transformations so long as they are not singular. The relation between these two approaches has been considered in detail in \cite{Brustein:1987qw}. For the heterotic string, the two-loop action is given in \cite{Gross:1986mw} by
\bea
	S_{10}&=&\int\ud^{10}x\sqrt{-g}\l\{R-\nabla\phi^2+\e^{\frac{\phi}{\sqrt2}}\l[-\frac14F^2+\frac{\al'}8\hat G+\r.\r.\nn\\
		&&+\l.\l.\frac{3\al'}8G_{\mu\nu}\nabla^\mu\phi\nabla^\nu\phi+\frac{\al'}{2\sqrt2}\nabla^\mu\nabla^\nu\phi\nabla_\mu\phi\nabla_\nu\phi+\frac{\al'}{16}\l[\l(\nabla\phi\r)^2\r]^2\r]\r\}.
\eea

Neutral black holes with String Theory corrections were studied in \cite{Callan:1988hs} for bosonic and heterotic $O(\al')$ corrections, while \cite{Myers:1987qx} studied $O(\al'^3)$ for Type II superstring. The solutions are constructed as perturbations around Schwarzschild background with a constant dilaton field. They also display a generic decrease of the dilaton field as one approaches the outer event horizon. This means that, in these perturbative expansions, the stringy effects are becoming smaller as one goes near the hole (since the string coupling is proportional to $\exp\phi$). One can see the same kind of behaviour displayed by the dilatonic solutions above. While this is a trivial statement for all the scaling solutions, the exact planar $\ga\da=1$ \eqref{Sol1} provides a non-trivial verification of these results. Inspecting the dilaton field \eqref{Phi1} and in particular its derivative, one finds, assuming $\da>0$ as usual, that it decreases as one nears the black hole in the intermediate and upper ranges for $\da$; however, in the lower range, there is a turning point where the scalar field starts increasing again at
\be
	r_0^{3-\da^2}=r_-^{3-\da^2}\frac{3-\da^2}{1+\da^2}\,.
	\label{TurningPoint1}
\ee
One finds that requesting this turning point to be inside the event horizon places a lower bound on the charge parameter $q$ using \eqref{Horizon1}:
\be
	q>2\da\sqrt{3-\da^2}r_0^{3-\da^2}\,.
	\label{LowerboundQPhi1}
\ee
This differs from the analysis in \cite{Callan:1988hs,Myers:1987qx}, but they were made in the neutral case. For neutral dilatonic black holes, $r_-=0$ so there would be no turning point.

The analytic study of the black-hole solutions of such actions is a natural further step, but one that we shall not attempt to take in this work, and keep for the future. In the meantime, we will put aside the scalar field (let us say, take it to be a constant) and study the black-hole solutions of Einstein-Gauss-Bonnet theories. From a string theory point of view, this makes sense as long as the curvature of spacetime is kept small compared to the string scale, which is true far enough from curvature singularities. However, taking a constant dilaton certainly changes the physics of such solutions a lot, so that one may not necessarily expect properties of Gauss-Bonnet black holes to carry over once stringy corrections are properly taken into account.
\vfill
\pagebreak

\begin{subappendices}
	\subsection{Appendix: \texorpdfstring{$\da^2=3$}{delta2=3} planar solutions \label{d2=3Solutions}}

	\subsubsection{$\ga\da=1$ solutions\label{d2=3g2=1/3Solutions}}

Let us write directly the general electric solution here (as before, $\eps=1$ solutions are obtained by exchanging $t$ and $\varphi$-coordinates), following the method exposed in subsection \ref{subsubsection:ka0_general}:
\bsea
	 A(p) &= & \frac{3a}{2Q}+\frac{3|\la|}{2Q}\frac{\l[\eta(p-p_0) \r]^{\frac1h}-1}{\l[\eta(p-p_0) \r]^{\frac1h}+1}\,, \slabel{A:5}\\
	\e^{\phi} &=& \e^{\sqrt3\phi_0}\l\{\l[\eta(p-p_0) \r]^{\frac1h}+1\r\}^{\frac32}\e^{-\frac{\sqrt3p}h}\l(p-p_0\r)^{-\frac3{4h}+\frac9{8h^2}-\frac{3\sqrt6c}{4h}+\frac{9c^2}{4h^2}}\,,\slabel{Phi:5}\\		 
	\quad \e^{\sqrt3\phi_0}&=&\l[-\frac{Q^2}{3\la^2\eta^{\frac1h}}\r]^{\frac34}\al_0^3\e^{\frac{3\sqrt6}{4}\Psi_0}\,,\nn\\
	\ud s^2 &=&-\al_0^4\e^{-\sqrt3\phi_0+\sqrt6\Psi_0-\frac{p}h}\l\{\l[\eta(p-p_0) \r]^{\frac1h}+1\r\}^{-\frac32}\l(p-p_0\r)^{\frac3{4h}+\frac3{8h^2}-\frac{\sqrt6c}{4h}+\frac{3c^2}{4h^2}}	\ud t^2 \nn \\
	&&+\frac{\e^{\sqrt3\phi_0-\frac{3p}{h}}}{2h\Lambda}\l\{\l[\eta(p-p_0) \r]^{\frac1h}+1\r\}^{\frac32}\l(p-p_0\r)^{-1-\frac3{4h}+\frac9{8h^2}-\frac{3\sqrt6c}{4h}+\frac{9c^2}{4h^2}}\ud p^2 \nn \\	
	&&+\frac{ h\la^2\e^{\sqrt3\phi_0-\frac ph}}{2\al_0^2\Lambda}\l\{\l[\eta(p-p_0) \r]^{\frac1h}+1\r\}^{\frac32}\l(p-p_0\r)^{1-\frac3{4h}+\frac3{8h^2}-\frac{3\sqrt6c}{4h}+\frac{3c^2}{4h^2}}\ud z^2 \nn \\	 
	&&+ \al_0^{-2}\e^{\sqrt3\phi_0-\sqrt6\Psi_0-\frac ph}\l\{\l[\eta(p-p_0) \r]^{\frac1h}+1\r\}^{\frac32}\l(p-p_0\r)^{-\frac3{4h}+\frac3{8h^2}+\frac{\sqrt6c}{4h}+\frac{3c^2}{4h^2}}\ud \varphi^2\,,\nn \\
	&&\slabel{metric:5}\\
	p_0&=&-\frac{3c^2}{4h}-\frac3{8h}\,.\nn
	\label{solution:5}
\esea
From \eqref{Phi:5}, we can deduce that $h$ has to be of the form $\frac1{2n+1}$ with $n$ an integer and $\eta<0$, otherwise the sign of $\e^{\phi_0}$ is not well-defined. Also, examination of the $pp$ and $zz$-metric elements signals that
\be
	h\Lambda>0\,.
\ee
This tells us that the sign of $\Lambda$ will determine the sign of $h$, and vice-versa. Let us specialise to black-hole solutions, by the same procedure used previously to regularise the horizon and the singularity at $p_0$:
\be
	h=\sqrt6 c=1 \Rightarrow \Lambda>0, \; p_0=-\half\,.
\ee
Then, rescaling some of the overall factors and taking $p\rightarrow -p$, we get
\bsea
	 A(p) &= & \sqrt{\frac{-3\eta}4}\l[a+\frac{p+p_\eta-1}{p-p_\eta}\r], \slabel{A:6}\\
	\e^{\phi} &=& \e^{\sqrt3\phi_0}\l[-\eta(p-p_\eta) \r]^{\frac32}\e^{\sqrt3p}\,, \slabel{Phi:6}\\
	\ud s^2 &=&\e^{p}\l[-\eta(p-p_\eta) \r]^{-\frac32}\l(p-\half\r)\ud t^2 -\frac{\e^{\sqrt3\phi_0+3p}}{2\Lambda\l(p-\half\r)}\l[-\eta(p-p_\eta) \r]^{\frac32}\ud p^2 \nn \\	
	&&+\e^{p}\l[-\eta(p-p_\eta) \r]^{\frac32}\l(\ud z^2+\ud\varphi^2\r). \slabel{metric:6}
	\label{solution:6}
\esea
Thus, we have $p_\eta=\half+\frac1\eta<\half$ and computation of the Ricci scalar gives
\bea
	T_1&=& 3\Lambda(\eta p-1)^2(p-\half)\e^{-\sqrt3\phi_0-3p}\l[-\eta(p-p_\eta) \r]^{-\frac72}, \nn \\
	T_2 &=& -4\Lambda \e^{-\sqrt3\phi_0-3p}\l[-\eta(p-p_\eta) \r]^{-\frac32}, \nn \\
	R &=& T_1+T_2 = P_3(p,\eta)\e^{-\sqrt3\phi_0-3p}\l[-\eta(p-p_\eta) \r]^{-\frac72}.
\label{Ricciscalar:6}
\eea
This solution displays two curvature singularities, at $p\to-\infty$ and $p_\eta=\half+\frac1\eta<\half$ for all $\eta<0$, as is required by \eqref{Phi:5}. We thus have $-\infty<p_\eta<\half$, depending on the value of $\eta$, but both singularities are always screened by a event horizon at $p=\half$. Asymptotic infinity is regular, but the coordinate $p$ is time-like when $p>\half$ and space-like when $0<p<\half$. The solution is therefore cosmological.

Let us now look at the $h=-1$ solutions, which have $\Lambda<0$. We get:
\bsea
	 A(p) &= &  \sqrt{\frac{-3\eta}4}\l[a+\frac{\eta(p-\half)-1}{\eta(p-\half)+1}\r], \slabel{A:6bis}\\
	\e^{\phi} &=& \e^{\sqrt3\phi_0}\l[\eta(p-\half)+1 \r]^{\frac32}\e^{\sqrt3p}, \qquad \e^{\sqrt3\phi_0}=\l[-\frac{Q^2}{3\la^2\eta}\r]^{\frac34}\al_0^3\e^{\frac{3\sqrt6}{4}\Psi_0}\,, \slabel{Phi:6bis}\\
	\ud s^2 &=&-\e^{p}\l[\eta(p-\half)+1 \r]^{-\frac32}\l(p-\half\r)\ud t^2 -\frac{\e^{\sqrt3\phi_0+3p}}{2\Lambda\l(p-\half\r)}\l[\eta(p-\half)+1 \r]^{\frac32}\ud p^2 \nn \\	
	&&+\e^{p}\l[\eta(p-\half)+1 \r]^{\frac32}\l(\ud z^2+\ud\varphi^2\r). \slabel{metric:6bis}
	\label{solution:6bis}
\esea
We still need $\eta<0$, but now
\bea
	R &\sim&-4\Lambda \e^{-\sqrt3\phi_0-3p}\l[1+\eta(p-\half) \r]^{-\frac32},
\label{Ricciscalar:6bis}
\eea
so there is a curvature singularity both at $p\to-\infty$ and $p_\eta=\half-\frac1\eta$. $p=\half$ and $p\to+\infty$ are regular points, but $p=\half$ does not screen the $\eta$-singularity any longer.

	\subsubsection{$\ga=\da$ solutions}

The master equation \eqref{radial_eq} can be solved directly for certain values of the coupling constants. For example taking $\ga=\da=-\sqrt3$ we have
\be
	\eps q\ddot{\overbrace{\log B}}+2\dot{\l(\frac{p}{B}\r)}+\frac{k}{B^2}-a\frac{\dot B}{B}-\frac{\ddot B}{B}\l[-\l(\frac a2+h\r)p-\frac{c^2}4-\frac{a^2}8-\eps\frac{Qk}2B\r].
		\label{radial_eq:10}
\ee
Considering $k=c=h=a=0$, rescaling  $\bar p=\frac{\mu p}{\eps q}$ and then dropping the bars yields the solution:
\bea
	B(p)&=&\lambda e^{-p}-\frac{2\eps q}{\mu^2}(p-1)\,, \label{B:10}\\
	Y(p)&=&X(p)-\eps qB(p)=-\eps Q\lambda \e^{-p}+\frac{2Q^2}{\mu^2}(p-1)\,. \label{X:10}
\eea
Here, $\mu,\la$ are integration constants.
The resolution is then straightforward and the solution reads, for $\eps=-1$:
\bsea
	A(p)&=&\l(\lambda \e^{-p}-\frac{2q}{\lambda\mu^2}\r)\sqrt{\lambda q}\e^{\frac{\sqrt3\phi_0}2}\,, \slabel{A::10} \\
	\e^{\phi}&=&\e^{\phi_0}\e^{-\frac{\sqrt3}2p}\,, \slabel{Dilaton::10} \\
	\ud s^2&=&-Y(p)e^{\frac p2}\ud t^2-\frac{q^2\e^{-\sqrt3\phi_0}}{2\Lambda\mu^2Y(p)}\e^{\frac32p}\ud p^2+\e^{\frac p2}\l(\ud z^2+\ud\varphi^2\r)\slabel{metric::10}.
	\label{solution:10}
\esea
We have set $\eps=-1$ in the metric expression in order to get a black hole.  Staticity outside the hole imposes $\Lambda<0$, so we have an AdS-like solution. The solution verifies $\lambda q>0$ and admits several horizons. To check this, let us calculate the Ricci scalar
\be
	R = -\frac{\Lambda}{4q^2}\e^{\sqrt3\phi_0}\l(3\lambda\mu^2 \e^{-p}+6Q^2p-22q^2\r)\e^{-\frac32p}\,,
\ee
which is regular for all possible finite zeros of $X(p)$, singular as $p\to-\infty$ and cancels as $p\to\infty$. The same behaviour is exhibited by the Weyl square. There is a curvature singularity as $p\to-\infty$  screened by two horizons if and only if
\be
	\la q<2\frac{Q^2}{\mu^2}\,,
\ee
the unequality being saturated in the extremal case. Again the horizon structure is similar to planar Reissner-Nordstr\"om in AdS. Finally, to keep our radial coordinate space-like as we approach asymptotic infinity, we need
\be
\frac{\mu}{q}<0 \textrm{ if } \eps=-1\,.
\ee

\end{subappendices}
\vfill
\pagebreak

 \section{Einstein-Gauss-Bonnet black holes}

\label{section:EGBBH}

\subsection{Einstein-Gauss-Bonnet theories}

\subsubsection{Theories of higher-dimensional gravity}

Over the years, incentives to consider theories in which the dynamics of gravity is modified have accumulated. The puzzles of quantum gravity, as well as the recent discovery of the acceleration of the expansion of the Universe, showed some progress towards resolution thanks to the study of models of gravity in dimensions greater than four (String Theory, braneworld paradigm). However, it has been known for a long time that, in four dimensions, General Relativity was the unique gravity theory verifying the following statements:
\begin{enumerate}
	\item Gravity is described by a symmetric two-tensor, the metric $g_{\mu\nu}$, and its dynamics is encoded in second-order non-linear ordinary differential equations. In particular, no higher-order derivatives are contained in the equations of the theory.
	\item General Relativity is a covariant theory and verifies a Bianchi equation (which amounts to conservation of energy).
\end{enumerate}
Ideally, one would strive to keep these desirable traits in any higher-dimensional theory of gravity, and it just so happens that General Relativity fills this slot only in four dimensions. So the question is: What kind of gravity theory should be used in higher dimensions?

Fortunately, answers have been provided over the course of the twentieth century. Indeed, expanding the String Theory equations of motion beyond the linear order in the string coupling brought the inclusion of higher powers of the Riemann curvature in the action. Yet, this is not good enough: it is well-known that higher powers of the Ricci scalar in the action result directly in equations of motion that contain higher derivatives, \cite{Stelle:1977ry}. Consequently, these theories contain ghost degrees of freedom: by this, we mean massive gravitational degrees of freedom propagating with a kinetic term of the wrong sign, so that the theory is generically unstable. Nonetheless, using ``gauge freedom'' in the background fields, one may operate field redefinitions in the effective equations of motion so that the effective action displays to second order a very specific combination of quadratic terms, the so-called Gauss-Bonnet combination:
\be
	\hat G = R_{\la\mu\nu\rho}R^{\la\mu\nu\rho}-4R_{\mu\nu}R^{\mu\nu}+R^2 = \da_{\l[b_1b_2b_3b_4\r]}^{\l[a_1a_2a_3a_4\r]}R^{b_1b_2}_{a_1a_2}R^{b_3b_4}_{a_3a_4}\,,
	\label{GB}
\ee
where the brackets mean antisymmetrisation in the following sense:
$$\da_{\l[b_1\ldots b_k\r]}^{\l[a_1\ldots a_k\r]}=\frac1{k!}\l(\da^{a_1}_{b_1}\times\ldots\times\da^{a_k}_{b_k}+\ldots\r).$$
This term allows to have well-defined propagators for the graviton, which is still a massless spin-2 degree of freedom, \cite{Zwiebach:1985uq}. It  is then a small step to adopt this combination as the correct one, all the more since, at the time, String Theory had already been proven to be perturbatively renormalisable (free of UV-divergences). So, these ghost degrees of freedom were guaranteed to be absent once the expansion was completed at all orders.

It was known at the time of these expansions that in higher dimensions, General Relativity's unicity properties were lost. It was also known that there was a way to recover them: it consisted precisely in the inclusion of the Gauss-Bonnet term and its higher-order counterparts, as the dimensionality of the theory increased. Indeed, the Gauss-Bonnet combination can be related to dimensionally-extended Euler characteristics. In the theory of two-dimensional surfaces, the Euler characteristic is a topological invariant:
\be
	\chi\l(\M\r) = \frac1{4\pi}\int_{\M}R = 2-2h\,,
	\label{Euler2D}
\ee
where $R$ is the Ricci scalar and $h$ the number of holes of the surface. Thus, if $h=0,2,4$, one has respectively a two-sphere with $\chi\l(\mathbf S^2\r)=2$, a two-torus with $\chi\l(\mathbf T^2\r)=0$, or double torus (with two holes) with $\chi\l(\mathbf T_2^2\r)=-2$. The Gauss-Bonnet theorem states that this number uniquely classifies the two-dimensional surfaces with no boundary (a property which is lost for higher-dimensional surfaces). This provides another connection with String Theory, where the Euler characteristic of the worldsheet (the two-dimensional surface of spacetime spanned by the string during its motion) plays a role in the string coupling used for calculating string surface amplitudes,  $g_s=\e^{\chi\phi}$ (where, Lo!, $\phi$ is the dilaton field). That this number is a topological invariant is of course related to the fact that the Ricci scalar does not carry any dynamics in two dimensions, it is simply a total divergence. A reformulation of this in geometrical term is that any two-dimensional space(time) is conformally flat.

We may then define dimensionally-extended Euler characteristics in higher even dimensions:
\be
	\chi\l(\M\r) = \frac1{(4\pi)^{D/2}\l(D/2\r)!}\int_{\M}\mathcal L_{D/2}\,,
	\label{EulerD}
\ee
though their meaning as topological numbers classifying the surfaces does not carry over. $\mathcal L_{k}$ is the $k^{\mathrm{th}}$ appropriate Lovelock density, $k=\l[(D-1)/2\r]$ being the integer part of $(D-1)/2$:
\be
	\mathcal L_{k}=\da_{\l[b_1\ldots b_k\r]}^{\l[a_1\ldots a_k\r]}R^{b_1b_2}_{a_1a_2}\times\ldots\times R^{b_{k-1}b_k}_{a_{k-1}a_k}\,.
	\label{LovelockDensity}
\ee
The zeroth-order density is simply a constant, the first-order one is the Ricci scalar which is a topological invariant in $D=2$ but becomes dynamic in $D>2$, the second-order one is the Gauss-Bonnet term \eqref{GB} which is a total divergence in $D=4$ but becomes dynamic in $D>4$. It also turns out that in dimension $D$, the $k^{\mathrm{th}}$-order Lovelock density is precisely the term that should be added to the gravity action so that the afore-mentionned unicity properties of the theory are recovered, so that the total Lagrangian may be written:
\be
	\mathcal L = \sum_{k=0}^{\l[(D-1)/2\r]}\al_{k}\mathcal L_k= \al_0 + R + \al\hat G+\ldots 
	\label{LovelockLagrangian}
\ee
These Lovelock theories were first studied in the seventies by Lovelock, \cite{Lovelock:1971yv,Lovelock:1972vz}, and they  were proven to be the natural higher-dimensional generalisations of General Relativity\footnote{Though one should mention some even earlier work in four dimensions by Lanczos, \cite{Lanczos:1938sf}.}. For reviews of their general properties as well as their solutions, see for instance \cite{Charmousis:2008kc,Garraffo:2008hu}; for a focus on the black-hole solutions, one may consult \cite{Wheeler:1985qd,Myers:1988ze} for early results, \cite{Crisostomo:2000bb} for a focus on the unique vacuum case and \cite{Zegers:2005vx} for a generalisation of Birkhoff's theorem. We will focus in this section on the quadratic case where one adds the Gauss-Bonnet term to the Einstein-Hilbert Lagrangian, with the Gauss-Bonnet coupling constant traditionally denoted by $\al$. This yields a theory which lives naturally in five or six dimensions. Up till now, we have left aside manifolds with a boundary and we shall keep doing so until the next part. Indeed, we are interested for the moment in solutions to the classical equations of motion, and the boundary terms in the action have no influence over this. They will however play a major role when we study the thermodynamics of the solutions.

The Gauss-Bonnet action is then written as
\be
	S_{GB} = \frac1{16\pi G_D}\int_{\M}\ud^Dx\sqrt{-g}\l(R-2\La+\al\hat G\r),
	\label{GBAction}
\ee
where we have set $\al_0=-2\La$, $\al_1=1$ and $\al_2=\al$ in the Lovelock Lagrangian \eqref{LovelockLagrangian} to relate to the usual Einstein-Hilbert action, and $G_D$ is the $D$-dimensional Newton constant. The equations of motion stemming from this action are
\be
	\E_{\mu\nu} = G_{\mu\nu} + \La g_{\mu\nu} -\al H_{\mu\nu} = 8\pi G_DT_{\mu\nu}\,,
	\label{EOMGB}
\ee
where $G_{\mu\nu}$ is the usual Einstein tensor and $\La$ of course the bare cosmological constant, while $H_{\mu\nu}$ is the Lanczos tensor:
\be
	H_{\mu\nu}=\half\hat Gg_{\mu\nu}-2RR_{\mu\nu}+4R_{\mu\rho}R^{\rho}_{\phantom{1}\nu}+4R_{\rho\sigma}R^{\rho\phantom{1}\sigma}_{\phantom{1}\mu\phantom{1}\nu}-2R_{\mu}^{\phantom{1}\rho\sigma\tau}R_{\nu\rho\sigma\tau}\,.
	\label{LanczosTensor}
\ee
Note that $g^{\mu\nu}H_{\mu\nu}=(D-4)\hat G/2$, which confirms explicitly that this term does not play any role in the equations of motion in four dimensions. Moreover,  the Lanczos tensor can be rewritten as
\bea
		H_{\mu\nu} &=& \frac12\hat Gg_{\mu\nu}-2P_{\mu\rho\sigma\tau}R_{\nu}^{\phantom{1}\rho\sigma\tau}\,,\\
		P_{\mu\nu\rho\sigma}&=& R_{\mu\nu\rho\sigma}-2\l(R_{\mu\l[\rho\r.}g_{\l.\sig\r]\nu}-R_{\nu\l[\rho\r.}g_{\l.\sig\r]\mu}\r)+Rg_{\mu\l[\rho\r.}g_{\l.\sig\r]\nu}\,,
\eea
where the brackets denote total antisymmetrisation over the indices enclosed. We have introduced a four-tensor $P_{\mu\nu\rho\sig}$ which has several interesting properties: it is divergence-free since the Bianchi identities of the curvature tensor are simply $\nabla^\sig P_{\mu\nu\rho\sig}=0$\footnote{Which can be checked straightforwardly by applying a contraction of the second Bianchi identity \eqref{SecondBianchiRiemann} and the Bianchi equation  \eqref{BianchiEq}.}. It has also has the same index symmetries as the Riemann curvature tensor. Tracing two of its indices yields $P^{\rho}_{\phantom{1}\mu\nu\rho}=-(D-3)G_{\mu\nu}$, from which the divergence free property of the Einstein tensor follows. In rather loose terms, one can say that $P$ is the curvature tensor associated to the Einstein tensor, just as the Ricci tensor is associated to the Riemann tensor. In four dimensions, this statement is far more precise since $P_{\mu\nu\rho\sig}$ coincides with the double dual (i.e. for each pair of indices) of the Riemann tensor $^\star R^{\rho\sig\star}_{\phantom{2}\mu\nu} \doteq -\frac{1}{2} \epsilon^{\mu\nu\ka\la}\,R_{\ka\la}^{\phantom{2}\ga\da} \, \frac{1}{2} \epsilon_{\ga\da\rho\sig}$, where $\epsilon_{\mu\nu\rho\sig}$ is the rank four Levi-Civita tensor. In four dimensions, we have $H_{\mu\nu}=0$ thus picking up the following Lovelock identity (for extensions see \cite{Edgar:2001vv}),
\be
\label{lock}
P_{\mu\rho\sig\tau}R_\nu^{\phantom{1}\rho\sig\tau}=\frac{{g_{\mu\nu} }}{2}\hat G\,,
\ee
which will be useful to us later on.

\subsubsection{The vacua of the theory}
\label{GBVacuum}

Now that we have become moderately familiar with the theory, let us ask what the vacua of the theory are, and in particular its \emph{constant curvature}, maximally symmetric spaces. They are defined by requesting that they have the maximal number of Killing vectors allowed by the symmetries of the theory, or equivalently by imposing that the Riemann tensor is proportional to the metric:
\be
	\textrm{{\it Constant curvature space: }}R_{\la\mu\nu\rho}=\frac{2\La}{(D-1)(D-2)}\l(g_{\la\nu}g_{\mu\rho}-g_{\la\rho}g_{\mu\nu}\r).
	\label{RiemannConstantCurvatureSpace}
\ee
Of course, in four-dimensional General Relativity, they are the familiar Minkowsky, de Sitter or Anti-de Sitter spaces, depending whether $\La$ is null, positive or negative. They should be distinguished from \emph{Einstein spaces}, where only the Ricci tensor is proportional to the metric and the Riemann tensor retains an antisymmetric, traceless part, called the Weyl tensor\footnote{The Weyl tensor has the same symmetry properties as the Riemann tensor. It is trivial in $D<4$. In $D=3$, its role is played by the Cotton tensor. It is invariant under conformal transformations of the metric. Thus, a conformally flat space has zero Weyl tensor. Conversely, in $D\geq4$, a spacetime with zero Weyl tensor is conformally flat.}:
\bsea
	\textrm{{\it Einstein space: }}&R_{\la\mu\nu\rho}&=C_{\la\mu\nu\rho}+\frac{2\La}{(D-1)(D-2)}\l(g_{\la\nu}g_{\mu\rho}-g_{\la\rho}g_{\mu\nu}\r), \slabel{RiemannEinsteinSpace}\\
	&R_{\mu\nu}&=\frac{2\La}{(D-2)} g_{\mu\nu}, \slabel{RicciEinsteinSpace}
	\label{EinsteinSpace}
\esea
for a $D$-dimensional spacetime. Plugging \eqref{RiemannConstantCurvatureSpace} in the equations of motion \eqref{EOMGB}, we find that it cannot be a solution if the two $\La$s in both equations are identical. Rather, one needs to define an effective cosmological constant by replacing the bare one $\La$ in \eqref{RiemannConstantCurvatureSpace} by
\bsea
	\La_{e}^{\pm} &=& 2\La_{CS}\l[1\mp\sqrt{1-\frac{\La}{\La_{CS}}}\r],\\ 
\Leftrightarrow	\La &=& \La_e\l(1-\frac{\La_e}{4\La_{CS}}\r)\,,\\
	\La_{CS}&=&-\frac{(D-1)(D-2)}{8\al(D-3)(D-4)}\,,
	\label{GBEffectiveCC}
\esea	
where $\La=\La_{CS}$ denotes a special value for which there is actually only one branch of vacuum. If $\La\neq\La_{CS}$, there are two, a plus and a minus one. The vacua of the theory are easily found to be
\bsea
	\ud s^2&=&-V(r)\ud t^2 + \frac{\ud r^2}{V(r)} + r^2\ud\Omega^2_{D-2}\,,\\
	V_{\pm}(r)&=& 1-\frac{2\La_e^{\pm}r^2}{(D-1)(D-2)}\,.
\esea
It is instructive to examine the small $\al$ limit of these vacua. Indeed, a cursory look at the action \eqref{GBAction} would lead us to believe that Einstein gravity should be recovered as $\al\to0$. Yet,
\bsea
	V_+(r)&\underset{\al\to0}{\sim}&1-\frac{2\La r^2}{(D-1)(D-2)} + O(\al)\,,\slabel{PlusVacSmallAlpha}\\
	V_-(r)&\underset{\al\to0}{\sim}&1+\frac{r^2}{\al(D-3)(D-4)}\l[1+\frac{2\al(D-3)(D-4)\La}{(D-1)(D-2)}\r] + O(\al)\,,\slabel{MinusVacSmallAlpha}
	\label{GBVacSmallAlpha}
\esea
so that it is clear that only the plus branch \eqref{PlusVacSmallAlpha} has a sensible small $\al$ limit, recovering Einstein gravity and Minkowski or (A)dS vacua depending upon the value of $\La$. However, the minus branch \eqref{MinusVacSmallAlpha} is undefined as $\al$ goes to zero, so that it is characteristic of Gauss-Bonnet gravity and represents a family of solutions disconnected from Einstein ones in the phase space. These branches are called respectively the \emph{Einstein branch} and \emph{Gauss-Bonnet branch} for reasons that are now clear.

Is there a way to discriminate between these two branches? There are two schools of thought in this matter. One circumvents the problem by arguing that one should place oneself in the Chern-Simons limit $\La=\La_{CS}$. Indeed, only one, degenerate branch of vacuum is left, and a connection can be made with a theory that is $D+1$-dimensional, either Chern-Simons (odd-dimensional spaces) or Born-Infeld theory (even-dimensional spaces). For this particular value of $\La$, one can show that the Einstein-Gauss-Bonnet action can be rewritten under a Chern-Simons and Born-Infeld form by including one extra dimension. One then enlarges the symmetry group and the theory has enhanced integrability properties thanks to this. We will not dwell too much upon this and refer to other works, \cite{Banados:1993ur,Crisostomo:2000bb}.

Without resorting to such symmetry arguments, one can still examine both branches a little closer: in particular, one may wonder at their stability against linear perturbations of the equations of motion \eqref{EOMGB}. Let us set $g_{\mu\nu}=\bar g_{\mu\nu}+h_{\mu\nu}$, so that $\da g_{\mu\nu}=h_{\mu\nu}$ and expand the Gauss-Bonnet equations at linear order in $\da g$, \cite{Charmousis:2008ce}. To this effect, one can use the formul\ae\vphantom{1} in the appendix of \cite{Deser:2002jk}, which give the linear part of the variations of all necessary quadratic combinations of the Riemann tensor. Varying the Einstein part of the equations of motion is easy and yields
\be
	\da\l(G_{\mu\nu}+\La g_{\mu\nu}\r) = \da G_{\mu\nu} + \La_e\l(1-\frac{\La_e}{4\La_{CS}}\r)\da g_{\mu\nu}\,.
\ee
$\da\l(G_{\mu\nu}\r)$ is the graviton operator, proportional to $\square h_{\mu\nu}$ in a transverse, traceless gauge: this shows that in pure $\La=0$ Einstein theory, gravity is mediated by a massless spin-2 particle, the graviton. The variation of the Gauss-Bonnet part requires a little more care:
\be
	\da\l(\al H_{\mu\nu}\r) = \frac{\La_e}{2\La_{CS}}\da\l(G_{\mu\nu}\r)+\frac{\La_e^2}{4\La_{CS}}h_{\mu\nu}\,,
\ee
so that the whole put together gives for the linear perturbations of the Gauss-Bonnet equations of motion around the vacuum
\be
	\da\l(\E_{\mu\nu}\r)=\l(1-\frac{\La_e}{2\La_{CS}}\r)\l[\da\l(G_{\mu\nu}\r) + \La_eh_{\mu\nu}\r]=8\pi G_D\,\da\l(T_{\mu\nu}\r).
	\label{GBLinearPert}
\ee
This result is extremely interesting for several reasons. First, it shows that the Einstein branch is stable at the linear level and couples with matter with the correct sign, since the left-hand side factor multiplying the tensorial variation of the Einstein tensor is positive. The tensorial structure is unaltered, we still have a spin-2 particle, with the stability controlled as for Einstein gravity by the value of the \emph{effective} cosmological constant. At the linear level, this is the sole effect of the Gauss-Bonnet term, it merely changes the value of the effective cosmological constant. However, the situation is very different for the minus, Gauss-Bonnet branch: in this case, the overall multiplying factor on the left-hand side is negative, so that gravitation couples to matter with the wrong sign, rendering the whole branch unstable. This is equivalent to Einstein gravity but with a negative Newton's constant, \cite{Boulware:1985wk,Deser:2002jk,Charmousis:2008ce}. 

Thus, one may be tempted to discard entirely this branch of solution on the very physical ground that it contains a ghost - although it does not manifest itself as a wrong-sign pole in the propagator as in higher-derivative theories of gravity, \cite{Stelle:1977ry}, but in the coupling of gravity to matter.

Finally, another remark must be made: should $\La_e$ have the value $2\La_{CS}$, the theory would become strongly-coupled to matter, which entails a whole host of complications and makes the previous computation ill-defined. One has to resort to non-perturbative techniques: for instance, one can study transitions between the vacua of the theory as in \cite{Charmousis:2008ce}. Such an analysis reveals that there is very strong mixing between the vacua of the two branches, with bubbles of one vacuum nucleating into the other. The true, quantum, vacuum has certainly ingredients of both classical vacua but no precise statement can be made.

\subsection{Six-dimensional Gauss-Bonnet black holes}

We now turn to the study of the spherically-symmetric, possibly time-dependent solutions of Einstein-Gauss-Bonnet theory.  These solutions have first been exhibited by Boulware and Deser, \cite{Boulware:1985wk}, and independently by Wheeler, \cite{Wheeler:1985nh}. Wiltshire then provided the generalisation to the Maxwell case and an equivalent of Birkhoff's theorem, \cite{Wiltshire:1985us}, as well as the Dirac-Born-Infeld electromagnetic theory\footnote{These theories can be thought of as non-linear versions of Maxwell theory for electromagnetism.}, \cite{Wiltshire:1988uq}. A more modern and quite elegant version of Birkhoff's theorem was proven in \cite{Charmousis:2002rc}. We will set the spacetime dimension to six, $D=6$, for simplicity's sake: the physics of the five-dimensional case are entirely contained in the six-dimensional one, which, as we will see, has very interesting properties not shared by its lower-dimensional counterpart. We will also deliberately leave aside the Maxwell version of the solution, as its extension should be quite straightforward and does not significantly influence the point we wish to make. The following discussion heavily draws on reference \cite{Bogdanos:2009pc}.

\subsubsection{Symmetries of the metric and equations in a lightcone gauge}

In order to proceed with the solution of the equations, we are now going to choose an appropriate symmetry for the metric. We distinguish between the transverse two-space, which also carries the time-like coordinate $t$, and the internal four-space, which is going to represent the possible horizon line element of the six-dimensional black hole. The metric of the internal space $h_{\mu\nu}$ is an arbitrary metric of the internal coordinates $x^\mu, \mu=0,1,2,3$ but we are imposing that the internal and transverse spaces are orthogonal to each other. This is an additional hypothesis we have to make since $h_{\mu\nu}$ is not a homogeneous metric and because our six-dimensional space is {\it not} an Einstein space (in General Relativity such an orthogonal foliation is possible for an Einstein metric). For loss of a better name, we will call this a warped metric Ansatz. Guided by the analogous procedure of analyzing Birkhoff's theorem in \cite{Charmousis:2002rc}, we write the metric as
\be
\label{GBmetric}
	\ud s^2  = e^{2\nu \left( {t,z} \right)} B\left( {t,z} \right)^{ - 3/4} \left( { - \ud t^2  + \ud z^2 } \right) + B\left({t,z}\right)^{1/2} h^{\left( 4 \right)} _{\mu \nu } \left( x \right)\ud x^\mu  \ud x^\nu  \,.
\ee
Lowercase Greek indices correspond to internal coordinates of the four-space. We then switch the coordinates of the transverse space to light-cone coordinates,
\be
	u = \frac{{t - z}}{{\sqrt 2 }}\,,\qquad v = \frac{{t + z}}{{\sqrt 2 }}\,,
\ee
in terms of which the metric reads
\be
	\ud s^2  =  - 2e^{2\nu \left( {u,v} \right)} B\left( {u,v} \right)^{ - 3/4} \ud u\ud v + B\left( {u,v} \right)^{1/2} h^{\left( 4 \right)} _{\mu \nu } \left( x \right)\ud x^\mu  \ud x^\nu  \,.
	\label{GBmetricLC}
\ee
Using the above prescription, we are now able to write down the equations of motion. The $u u $- and $vv$-equations yield
\be
	\label{GBequu}
	{\cal E}_{uu}=\frac{2 \nu_{,u} B_{,u}- B_{,uu}}{B} \left[ 1+\alpha \left( B^{-1/2} R^{(4)}+\frac{3}{2} e^{-2\nu} B^{-5/4} B_{,u} B_{,v}  \right) \right]\,,
\ee
\be
	\label{GBeqvv}
	{\cal E}_{vv}=\frac{2 \nu_{,v} B_{,v}- B_{,vv}}{B} \left[ 1+\alpha \left( B^{-1/2} R^{(4)}+\frac{3}{2} e^{-2\nu} B^{-5/4} B_{,u} B_{,v}  \right) \right].
\ee
The off-diagonal equation reads
\bea
	\label{GBequv}
	{\cal E}_{uv}  &=&\frac{{B_{,uv} }}{B} - \Lambda e^{2\nu } B^{ - 3/4}  + \frac{\alpha }{2}e^{2\nu } B^{ - 7/4} \hat G^{(4)}+ \nonumber\\
									&&\qquad + R^{(4)} \left[ {\frac{1}{2}e^{2\nu } B^{ - 5/4}  - \alpha B^{ - 3/2} \left( {\frac{1}{2}\frac{{B_{,u} B_{,v} }}{B} - B_{,uv} } \right)} \right]+ \nonumber\\
									&&\qquad +  \alpha e^{ - 2\nu } B^{ - 5/4} \left[ { - \frac{{15}}{{16}}\left( {\frac{{B_{,u} B_{,v} }}{B}} \right)^2  + \frac{3}{2}\frac{{B_{,u} B_{,v} }}{B}B_{,uv} } \right] \,.
\eea
We also have the $\mu \nu$-equations, which can be brought into the form
\bea
	\label{GBeqmunu}
	{\cal E}_{\mu \nu}&=& G_{\mu \nu }^{(4)}  - e^{ - 2\nu } B^{1/4} \left( {\frac{3}{4}B_{,uv}  + 2B\nu _{,uv} } \right)h^{\left( 4 \right)} _{\mu \nu }  + \Lambda B^{1/2} h^{\left( 4 \right)} _{\mu \nu } +\nonumber\\
										&&\qquad +\frac{3}{2}\alpha e^{ - 4\nu } \left( {B_{,uu}  - 2\nu _{,u} B_{,u} } \right)\left( {B_{,vv}  - 2\nu _{,v} B_{,v} } \right)h^{\left( 4 \right)} _{\mu \nu }- \nonumber \\
										&&\qquad-\alpha e^{ - 4\nu } \left[ {\frac{{45}}{{32}}\left( {\frac{{B_{,u} B_{,v} }}{B}} \right)^2  - \frac{{21}}{8}\frac{{B_{,u} B_{,v} }}{B}B_{,uv}  + \frac{3}{2}B_{,uv} ^2  + 3B_{,u} B_{,v} \nu _{,uv} } \right]h^{\left( 4 \right)} _{\mu \nu }-\nonumber\\
										&&\qquad-\alpha e^{ - 2\nu } B^{ - 1/4} \left( {\frac{3}{4}\frac{{B_{,u} B_{,v} }}{B} - \frac{1}{2}B_{,uv}  + 4B\nu _{,uv} } \right)\left( {R^{(4)} h^{\left( 4 \right)} _{\mu \nu }  - 2R_{\mu \nu }^{(4)} } \right)
\,\,.
\eea

In this way, we have decomposed the gravitational equations into expressions depending on either transverse space quantities or internal coordinates. The integrability conditions, \cite{Bowcock:2000cq}, are unchanged compared to the original version of the theorem, \cite{Charmousis:2002rc}, and this will permit us to obtain the staticity conditions.  Furthermore, the internal geometry of the horizon only enters these equations through expressions involving the four-dimensional Gauss-Bonnet scalar density, the Ricci tensor and scalar of the internal metric $h_{\mu\nu}$. Note the absence of $H^{(4)}_{\mu\nu}$ terms due to the fact that the internal space is four-dimensional. Also note that terms proportional to the Gauss-Bonnet coupling constant are responsible for the appearance of $R^{(4)}$ and $R^{(4)}_{\mu \nu}$ and in this way, the Gauss-Bonnet term exposes the internal geometry to the transverse space dynamics in a non-trivial way, something which would obviously not occur in ordinary General Relativity. As we will see, this decomposition imposes severe constraints on the allowed form of the horizon geometry in order to get a spacetime solution.

The $uu$- and $vv$-equations \eqref{GBequu}, \eqref{GBeqvv} can lead to three different classes of solutions, depending on whether the first or second factor is zero (an additional class will emerge for constant $B$). The corresponding solutions have distinct characteristics and are thus treated separately in what follows. Class-I and II are both warped solutions whereas for Class-III we have $B=constant$.

\subsubsection{Class-I solutions}

This class corresponds to solutions which can generally be time-dependent and, hence, for which a Birkhoff-type theorem does not hold. As we shall soon see, all of them imply $5+12 \alpha \Lambda=0$. The latter corresponds to the so-called \emph{Born-Infeld} limit, an even-dimensional counterpart of the well-known odd-dimensional Chern-Simons limit in which the Lovelock action can be written as a Chern-Simons action for some (A)dS connection, see e.g. \cite{Zanelli:2005sa}. In the Born-Infeld limit, the Lovelock action can be written as a Born-Infeld action for some curvature two-form, hence its name. For the class of spacetime metrics under consideration here, it typically leads to an underdetermined set of equations and the unconstrained components of the metric subsequently allow for a possible time dependence. This is reminiscent of class-I Lovelock solutions with spherical, hyperbolic or planar symmetry \cite{Charmousis:2002rc, Zegers:2005vx} and is expectedly related to perturbative strong-coupling problems as in the case of Chern-Simons gravity \cite{Charmousis:2008ce}.

Setting the second factor of the $uu$- and $vv$-equations \eqref{GBequu} equal to zero leads to the common equation
\be
	\label{classIcond}
	1 + \alpha {B^{ - 1/2} R^{(4)}  + \frac{3}{2} \alpha e^{ - 2\nu } B^{ - 5/4} B_{,u} B_{,v} }  = 0\,,
\ee
from which we can solve for the function $\nu(u,v)$ in terms of $B(u,v)$, according to
\be
	\label{int11}
	\nu \left( {u,v} \right) = \frac{1}{2}\ln \left( { - \frac{{3\alpha }}{2}\frac{{B_{,u} B_{,v} }}{{B^{5/4} \left( {1 + \alpha B^{ - 1/2} R^{\left( 4 \right)} } \right)}}} \right) \,.
\ee
Note that this equation immediately constrains the Ricci scalar $R^{(4)}$ of the internal space to be a constant. We are thus required to consider only horizon geometries of constant scalar curvature as candidate solutions. Substituting the above expression for $\nu(u,v)$ into \eqref{GBequv} yields the two additional constraints
\be
	5 + 12\alpha \Lambda  = 0,\,\,\hat G^{\left( 4 \right)}  = \frac{1}{6}{R^{\left( 4 \right)} }^2 \,.
	\label{cons}
\ee
The second of these constraints tells us that the Gauss-Bonnet scalar $\hat G$ is also constant.
By taking the trace of \eqref{GBeqmunu} with $h_{\mu\nu}$ and performing the same substitution, we end up with the equation
\be
	{\cal E} \equiv {\cal E}_\mu^{\mu} = \frac{{5 + 12\alpha \Lambda }}{{3\alpha }} = 0\,,
\ee
Finally, we can rewrite the complete  equation \eqref{GBeqmunu} in terms of the trace as
\bea
	{\cal E}_{\mu \nu} &=& \frac{1}{4}B^{1/2} {\cal E} h^{\left( 4 \right)} _{\mu \nu }+ \left( {R^{\left( 4 \right)} _{\mu \nu }  - \frac{1}{4}R^{\left( 4 \right)} h^{\left( 4 \right)} _{\mu \nu } } \right)\bigg[ 1+ \nonumber\\
										&& \qquad+\l. 2\alpha e^{ - 2\nu } B^{ - 1/4} \left( {\frac{3}{4}\frac{{B_{,u} B_{,v} }}{B} - \frac{1}{2}B_{,uv}  + 4B\nu _{,uv} } \right) \right].
	\label{mneq}
\eea
Given the above mentioned constraints, the first term vanishes because it is proportional to ${\mathcal E}$. The second term can vanish in one of two ways giving us two distinct cases of Class-I solutions both verifying (\ref{int11}) and (\ref{cons}). We can have
\be
	R^{\left( 4 \right)} _{\mu \nu }  = \frac{1}{4}R^{\left( 4 \right)} h^{\left( 4 \right)} _{\mu \nu } \,,
\ee
which is the definition of a four-dimensional \emph{Einstein space}, \eqref{RiemannEinsteinSpace}, for which the Ricci tensor (but not the Riemann one) is proportional to the metric. Coupled with the condition  $\hat G^{\left( 4 \right)}  = \frac{1}{6}\left[R^{\left( 4 \right)} \right]^2$, this leads to
\be
	C^{\left( 4 \right)} _{\alpha \beta \mu \nu } C^{\left( 4 \right)\alpha \beta \mu \nu }  = 0\,,
\ee
i.e. the square of the Weyl tensor of the internal space must be zero. We then have a \emph{constant curvature space}, \eqref{RiemannConstantCurvatureSpace}, for which the Riemann tensor is proportional to the metric.
Since \eqref{GBeqmunu} is in this way automatically satisfied, there is no dynamical equation defining the function $B(u,v)$ and thus the system of field equations becomes underdetermined. This is a typical feature of the Class-I solutions which have been discussed in \cite{Charmousis:2002rc}.

If, on the contrary, we demand the second factor in the second term of equation (\ref{mneq}) to be zero, the requirement for a four-dimensional Einstein space on the horizon of the black hole can be relaxed. Instead, we get a third-order partial differential equation for $B(u,v)$, which reads
\bea
&&\left( {1 + \alpha B^{ - 1/2} R^{\left( 4 \right)} } \right)^2 \left( {B_{,u} ^2 B_{,vv} B_{,uv}  + B_{,v} ^2 B_{,uu} B_{,uv}  - B_{,u} ^2 B_{,v} B_{,uvv}  - B_{,v} ^2 B_{,u} B_{,uuv} } \right)+ \nonumber\\
&&\qquad+ \frac{{B_{,uv} }}{B}B_{,u} ^2 B_{,v} ^2 \left[ {\frac{3}{2} + \frac{5}{2}\alpha B^{ - 1/2} R^{\left( 4 \right)}  + \left( {\alpha B^{ - 1/2} R^{\left( 4 \right)} } \right)^2 } \right]- \nonumber\\
&&\qquad- \frac{{B_{,u} ^3 B_{,v} ^3 }}{{B^2 }}\left[ {\frac{5}{4} + \frac{{17}}{8}\alpha B^{ - 1/2} R^{\left( 4 \right)}  + \frac{9}{8}\left( {\alpha B^{ - 1/2} R^{\left( 4 \right)} } \right)^2 } \right] = 0\,.\label{Beqn_class_notEinstein}
\eea
This equation can in principle be solved for $B(u,v)$, again for an internal space of constant Ricci scalar and given the constraints (\ref{cons}). Note that the horizon is not necessarily an Einstein space but instead we have the four-dimensional geometrical constraint
\be
	{C^{\left( 4 \right)}}^2+2{R^{\left( 4 \right)}_{\mu\nu}}^2=\frac12{R^{\left( 4 \right)}}^2= \mbox{constant}\,.
\ee

We now summarise the results for the Class-I solutions. We can distinguish two subclasses, both requiring the fine-tuning condition $5+12 \alpha \Lambda=0$, which is the six-dimensional version of the Born-Infeld gravity condition, and a constant Ricci scalar $R^{\left( 4 \right)}$:
\begin{itemize}
	\item Class-Ia: we have an underdetermined system for the transverse dimension geometry (free function $B$ and (\ref{int11})) and an internal space which is an Einstein space of zero Weyl squared curvature, that is a \emph{constant curvature space}.
	\item Class-Ib: A completely determined system of transverse dimensions (\ref{int11}), (\ref{Beqn_class_notEinstein}) with an internal geometry  obeying  (\ref{cons}) (non-zero Weyl curvature).
\end{itemize}
The former of the two subclasses is certainly incompatible with Birkhoff's theorem as demonstrated in \cite{Charmousis:2002rc}, whereas for the latter we could not find the general solution to \eqref{Beqn_class_notEinstein}.

\subsubsection{Class-II solutions}
Class-II solutions are obtained by imposing, instead of (\ref{classIcond}), that
\bea		
	2 \nu_{,u} B_{,u}- B_{,uu} &=& 0\,,\\
	2 \nu_{,v} B_{,v}- B_{,vv} &=& 0\,.
	\label{nuBB}
\eea
These integrability conditions are the same as in the case of ordinary General Relativity. We will again assume that $B$ is not constant.

Equation (\ref{nuBB}) implies that
\be
	e^{2\nu} = B_{,u} f(v) = B_{,v} g(u)\,, 
	\label{enuBB}
\ee
for some functions $f$ and $g$, which, in turn, yields
$B=B(U+V)$, with $U=U(u)$ and $V=V(v)$. In this way, under the change of coordinates
\be
	\label{uvUV}
	U = \frac{{\bar{z} - \bar{t}}}{{\sqrt 2 }}\,,\qquad V = \frac{{\bar{z} + \bar{t}}}{{\sqrt 2 }}\,,
\ee
the function $B$ becomes independent of time and Birkhoff's theorem holds. Additionally, rewriting (\ref{enuBB}), $\nu(u,v)$ is now defined as
\be
	e^{2\nu }  = B'U'V'\,,
\ee
where primes denote differentiation with respect to the single argument of each function. Under (\ref{uvUV}), we get $e^{2 \nu}= \partial_{\bar{z}} B$. The $uu$- and $vv$-equations thus determine the staticity of the metric, as well as the relation between $B$ and $\nu$. We can then determine $B(u,v)$, or equivalently the form of the black hole potential, from the $uv$-equation. Taking advantage of the already-deduced staticity, we can express this as
\bea
	0&=& B''+ \frac{1}{2}R^{\left( 4 \right)} B^{ - 1/4} B' - \frac{{15}}{{16}}\alpha B^{ - 9/4} B'^3  + \frac{3}{2}\alpha B^{ - 5/4} B''B'-\frac{1}{2}\alpha R^{(4)} B^{ - 3/2} B'^2+   \nonumber\\
	&& \qquad  + \alpha B^{ - 1/2} R^{\left( 4 \right)} B'' + \frac{1}{2}\alpha B^{ - 3/4} B'\hat G^{\left( 4 \right)}  - \Lambda B^{1/4} B'\,.
\eea
Inspection of the above expression leads to the conclusion that \emph{a priori} only solutions with a constant Ricci scalar and Gauss-Bonnet density for the internal space are permissible. However, this is not always the case, we have to be wary of special cases. Upon integration, this leads to a quadratic equation for $B'$. We can then solve for $B'$ and determine the black hole potential $V$
\be
	\label{Class2a}
	\ud s^2  =  - V\left( r \right)\ud t^2  + \frac{{\ud r^2 }}{{V\left( r \right)}} + r^2 h^{\left( 4 \right)} _{\mu \nu } \left( x \right)\ud x^\mu  \ud x^\nu \,,
\ee
using the change of variables $r=B^{1/4}$. The corresponding potential turns out to be
\be
	V(r) = \frac{{R^{\left( 4 \right)} }}{{12}} + \frac{{r^2 }}{{12\alpha }}\left[ {1 \pm \sqrt {1 +\frac{12\alpha \Lambda}{5}  +\frac{{\alpha ^2 \left({R^{(4)}}^2-6\hat G^{(4)} \right)}}{{r^4 }} + 24\frac{{\alpha M}}{{r^5 }}} } \right]\,,
	\label{potentialclassII}
\ee
where $M$ is an integration constant independent of $x$, related to the mass of the six-dimensional black hole \footnote{We note that the Gauss-Bonnet coupling constant has dimensions $mass^{-2}$, $k$ of $mass$ and $\kappa$ is dimensionless. The latter is justified by the fact that the internal metric $h^{(4)}_{\mu \nu} dx^{\mu} dx^{\nu}$ is multiplied by $r^{2}$, so the internal coordinates must be of an angular nature and carry no dimension. Consequently, derivatives with respect to them as well as the Riemmann, Ricci and Weyl tensors are dimensionless.}.

We now turn to the $\mu \nu$-equations \eqref{GBeqmunu}. Taking the trace with respect to the internal metric  leads to the expression
\bea
{\cal E} &=& 4\Lambda  - R^{\left( 4 \right)} B^{ - 1/2}  - B^{ - 1/4} \left( {3\frac{{B''}}{{B'}} + 4\frac{{BB''}}{{B'^2 }} - 4\frac{{BB''^2 }}{{B'^3 }}} \right)- \\
&&\qquad - \alpha B^{ - 1/2} \left( {\frac{{45}}{8}\frac{{B'^2 }}{{B^2 }} - \frac{{21}}{2}\frac{{B''}}{B} + 6\frac{{B'''}}{{B'}}} \right) -\\
&&\qquad - \alpha R^{\left( 4 \right)} B^{ - 3/4} \left( {\frac{3}{2}\frac{{B'}}{B} - \frac{{B''}}{{B'}} + 4\frac{{BB'''}}{{B'^2 }} - 4\frac{{BB''^2 }}{{B'^3 }}} \right) = 0\,.
\eea
It can be shown that this equation can be rewritten as $ - \partial _v \left( {\frac{{B^{3/4} }}{{B'}}{\mathcal E}_{uv} } \right) = 0$, which is identically satisfied as a Bianchi identity.

The $\mu \nu$-equation then gives
\be
	0=\left(R^{\left( 4 \right)}_{\mu\nu}-\frac14R^{\left( 4 \right)}h_{\mu\nu}\right)\left[1 + \alpha B^{ - 1/4} \left( {\frac{3}{2}\frac{{B'}}{B} - \frac{{B''}}{B'} + 8\frac{{BB'''}}{{B'^2 }} - 8\frac{{BB''^2 }}{{B'^3 }}} \right) \right].
	\label{munueqclassII}
\ee
Therefore, we have two distinct cases, depending on which of the two factors of \eqref{munueqclassII} cancels.

For the first case, the horizon has to  be an Einstein space with constant scalar curvature, defined by $R^{\left( 4 \right)}_{\mu \nu}= 3 \kappa h_{\mu \nu}$. This is similar to ordinary General Relativity. However, given that $\hat G^{(4)}$ is also  constant, we have that $C^{\alpha \beta \gamma \mu} C_{\alpha \beta \gamma \mu}=4\Theta $ where $\Theta$ is a positive constant. This is the solution obtained by \cite{Dotti:2005rc}. Now using the properties of the $P_{\mu\nu\alpha\beta}$ tensor and (\ref{lock}), we immediately get
\be
	C^{\alpha \beta \gamma \mu} C_{\alpha \beta \gamma \nu}=\Theta \delta^\mu_\nu\,.
\ee
This is a supplementary condition imposed on the usual Einstein space condition for the horizon. Both have a similarity in that we ask for (part of) a curvature tensor to be analogous to the spacetime metric. The main difference being that the curvature tensor in question here is the Weyl tensor and, given its symmetries, it is actually its square which is analogous to the spacetime metric. Clearly horizons with $\Theta\neq 0$ will not be homogeneous spaces and not even asymptotically so in the non-compact cases. We will see in a forthcoming section that they can be related to squashed sphere geometries. Another interesting point is that the Gauss-Bonnet scalar, whose spacetime integral is the Euler characteristic of the horizon, has to be constant. In other words the Euler Poincar\'e characteristic of the horizon is in this case simply the volume integral of the horizon. In this sense, $\Theta$ could be thought of as a topological charge, though this may be reaching a little.

The Gauss-Bonnet scalar of the internal space then reads $\hat G^{(4)}=4 \Theta+24 \kappa^2$ and the potential \cite{Dotti:2005rc}
\be
	V(r)=\kappa+\frac{r^2}{12 \alpha} \left(1\pm \sqrt{1+\frac{12}5\alpha\Lambda - 24 \frac{\alpha^2 \Theta}{r^4} + 24 \frac{\alpha M}{r^5}} \right)\,.
	\label{potentialclassIIEinstein}
\ee
For $\Theta=0$, we obtain the well-known black holes discussed by Boulware and Deser, and Wheeler, \cite{Boulware:1985wk,Wheeler:1985nh}. Note that it is quite remarkable that the internal geometry of the horizon affects the black hole potential. This is very different from General Relativity, where only the topology (spherical, planar or hyperbolic) can be distinguished from the sign of $\kappa$.

Alternatively (\ref{munueqclassII}) tells us that we can have a horizon which is potentially not Einstein, if and only if $B$ satisfies
\be
	1 + \alpha B^{ - 1/4} \left( {\frac{3}{2}\frac{{B'}}{B} - \frac{{B''}}{B'} + 8\frac{{BB'''}}{{B'^2 }} - 8\frac{{BB''^2 }}{{B'^3 }}} \right) = 0 \,.
	\label{munueqclassIInotEinstein}
\ee
Note that in this case we have two equations for $B$ and the system is overdetermined. Integrating \eqref{munueqclassIInotEinstein}, we obtain the following potential
\be
	\tilde{V}(r) = \frac{r^2}{12\alpha}+\frac{\rho}{2\alpha}  - \frac{\mu}{2\alpha r}\,,
\ee
where $\mu$ and $\rho$ are integration constants. Comparing with \eqref{potentialclassII}, we make the following identifications:
\be
	5+12\alpha \Lambda=0\,, \qquad \mu=0\,,\qquad M=0\,.
	\label{constclassIInotEinstein}
\ee
and
\be
	\label{extra}
	\rho=\frac{R^{\left( 4 \right)}}6\pm\frac16\sqrt{{R^{\left( 4 \right)}}^2-6\hat G^{\left( 4 \right)}}\,.
\ee
The potential \eqref{potentialclassII} reduces to
\be
	V(r) =  \frac{\rho}{2}+ \frac{r^2}{12\alpha}\,.
	\label{potentialclassIInotEinstein}
\ee
This corresponds to a massless solution resembling (A)dS space, with a curvature  radius dependent on both the internal geometry and the Gauss-Bonnet coupling. The solution is defined only for $\left[R^{\left( 4 \right)}\right]^2-6\hat G^{\left( 4 \right)}>0$. Equation (\ref{extra}) is now a geometric equation constraining the four-dimensional horizon geometry. Indeed $R^{\left( 4 \right)}$ and $G^{\left( 4 \right)}$ no longer have to be constant individually. In Section \ref{sec:BIBS}, by Wick rotating these solutions to Lorentzian internal sections, we shall construct Born-Infeld black string solutions.

\noindent
Thus, Class-II contains the folllowing solutions:
\begin{itemize}
	\item Class-IIa: The solution is locally static (\ref{Class2a}), and the horizon is an Einstein space with $\Theta\geq0$.
	\item Class-IIb: The solution is again locally static with potential given by (\ref{potentialclassIInotEinstein}), but the horizon is constrained by \eqref{extra} and the Born-Infeld condition is imposed.
\end{itemize}
Both subclasses of Class-II obey a local staticity theorem.

\subsubsection{Class-III solutions}

The remaining Class of solutions is given by  $B=:\beta^4= constant \neq 0$. In this case, the metric is no longer warped in the internal directions and the Einstein-Gauss-Bonnet equations (\ref{GBequv}), (\ref{GBeqmunu}) reduce to
\bea
0&=& -2 \Lambda \beta^4 + R^{(4)} \beta^2 + \alpha \hat{G}^{(4)}\,, \label{beta} \\
G^{(4)}_{\mu \nu} + \Lambda \beta^2 h^{(4)}_{\mu\nu} &=& 2 \beta^3 \nu_{,uv} e^{-2\nu} \left ( \beta^2 h^{(4)}_{\mu\nu} - 4 \alpha G^{(4)}_{\mu\nu} \right ) \,.\label{NariaiGmunu}
\eea
It follows from contracting the second of the above equations, (\ref{NariaiGmunu}), with the metric $h^{ \mu\nu}$ that
\be
	4\Lambda \beta^2 -R^{(4)} = 8 \beta^3 \nu_{,uv} e^{-2\nu} \left ( \beta^2 +  \alpha R^{(4)} \right )\,.
	\label{nunonsep}
\ee
If $R^{(4)}=-\beta^2/\alpha$, then we have the fine-tuning relation $1+ 4\Lambda \alpha =0$, (\ref{beta}) implies that $\hat{G}^{(4)} = \beta^4 /(2 \alpha^2)$ and (\ref{NariaiGmunu}) can be rewritten as
\be 
	\left (G^{(4)}_{\mu \nu} + \frac{1}{4} R^{(4)} h^{(4)}_{\mu\nu}\right ) \left ( \frac{2 \beta^3}{\Lambda} \nu_{,uv} e^{-2\nu} -1 \right )  = 0\,,
\ee
which implies that either $h^{(4)}_{\mu\nu}$ is Einstein and $\nu$ is not determined (and thus possibly time-dependent), or $h^{(4)}_{\mu\nu}$ is not necessarily Einstein and $\nu$ obeys the Liouville equation
\be 
	\nu_{,uv} = \frac{\Lambda}{2 \beta^3} e^{2\nu} \,.
\ee
The latter can be solved exactly, yielding
\be 
	e^{2\nu} = \frac{2\beta^3}{\Lambda}\,,\qquad \frac{U' V'}{(U+V)^2} \,,
\ee
for some functions $U=U(u)$ and $V=V(v)$. Now we can perform a change of coordinates of the form (\ref{uvUV}),
under which $\nu$ transforms in such a way that eventually
\be
	\label{z}
	e^{2\nu} = \frac{2\beta^3}{\Lambda} \frac{1}{\bar{z}^2} \,.
\ee
The metric obviously admits the locally time-like Killing vector $\partial_{\bar{t}}$ and Birkhoff's theorem holds in this case. If on the contrary $R^{(4)}\neq -\beta^2/\alpha$, (\ref{nunonsep}) can be rewritten in the separable form
\be
	\frac{4\Lambda \beta^2 -R^{(4)}}{\beta^2 +  \alpha R^{(4)}} = 8 \beta^3 \nu_{,uv} e^{-2\nu}  = constant\,.
	\label{nusep}
\ee
Provided that $1+ 4\Lambda \alpha \neq0$, we can have $R^{(4)} = 4\Lambda\beta^2$, which implies $\nu_{,uv}=0$ and $2\nu = \ln U' + \ln V'$ for some functions $U=U(u)$ and $V=V(v)$. We can perform a change of coordinates of the form (\ref{uvUV}) so that, in the end, $e^{2\nu}=1$ and the metric admits the Killing vector $\partial_{\bar{t}}$. It also follows from (\ref{beta}) that $\hat{G}^{(4)} = -2\Lambda \beta^4/\alpha$ and from (\ref{NariaiGmunu}) that $h^{(4)}_{\mu\nu}$ is Einstein. Otherwise, for non-vanishing values of the constant in (\ref{nusep}), say $\lambda$, $\nu$ obeys once again the Liouville equation
\be
	\nu_{,uv} = \frac{\lambda}{8\beta^3} e^{2 \nu} \,.
\ee
After a change of coordinates of the form of (\ref{uvUV}), we therefore have
\be 
	e^{2\nu} = \frac{8\beta^3}{\lambda} \frac{1}{\bar{z}^2} \,,
\ee
and the metric admits the Killing vector $\partial_{\bar{t}}$. If $\lambda=4\Lambda=-1/\alpha$, (\ref{NariaiGmunu}) is trivially satisfied and the only constraint on $h^{(4)}_{\mu\nu}$ comes from (\ref{beta}). Otherwise, it follows from (\ref{NariaiGmunu}) that $h^{(4)}_{\mu\nu}$ is Einstein and from (\ref{beta}) that $\hat{G}^{(4)}$ is a constant.

Wick-rotating the solutions obtained in the former case allows to construct axially symmetric black string-type solutions, provided we impose a certain amount of symmetry to the internal manifold. Some static examples of this subclass of solutions have already been studied (see  \cite{Maeda:2006hj,Maeda:2006iw,Molina:2008kh} and references therein). We will briefly study an example in Section \ref{sec:6dSBS}. It is worth noting that, once we allow for lesser symmetry, the scalar equation \eqref{beta} does not suffice to determine the full horizon metric.

\noindent
The solutions contained in Class-III are the following:
\begin{itemize}
	\item Class-IIIa: $1+4\al\Lambda=0$, $R^{(4)}$, $\hat{G}^{(4)}$ are constant, and the horizon is Einstein.
	\item Class-IIIb: $1+4\al\Lambda\neq 0$, the transverse space is of constant curvature, and  \eqref{nusep} is satisfied, and the horizon is Einstein.
	\item Class-IIIc: $1+4\al\Lambda=0$, the transverse space is of constant curvature, and the horizon satisfies (\ref{beta}) and does not have to be Einstein.
\end{itemize}
Birkhoff's theorem holds for two of the subclasses, Class-IIIb and Class-IIIc.

\subsubsection{A staticity theorem}

For generic Class-II and certain Class-III solutions, we have the following local staticity theorem.
\vskip.5cm

\noindent {\bf Theorem} {\it Let $({\mathcal M},g)$ be a six-dimensional pseudoriemannian spacetime whose metric $g$ satisfies the Gauss-Bonnet equations of motion (\ref{EOMGB}) and whose manifold ${\mathcal M}$ admits a foliation into two-dimensional submanifolds $\Sigma_{(x_1, \dots x_4)}^{(2)}$ and a foliation
into four-dimensional submanifolds $H_{(t_1, t_2)} ^{(4)}$ such that:
\begin{itemize}
\item the tangent bundles of the leaves $T \Sigma_{(x_1, \dots, x_4)}^{(2)}$ and $T H_{(t_1, t_2)} ^{(4)}$ are orthogonal with respect to $g$;
\item for all $(t_1, t_2)$, the four-dimensional induced metric $h_{(t_1,t_2)}^{(4)}$ on $H_{(t_1, t_2)}^{(4)}$ is conformal to a given four-dimensional metric $h^{(4)}$ with conformal factor depending only on $(t_1, t_2)$.
\end{itemize}
If in addition, either
\begin{enumerate}[i)]
	\item $1+4\Lambda \alpha \neq 0$ and $5+12 \alpha \Lambda \neq0$, or \label{casei}
	\item $1+4\Lambda \alpha = 0$ and $h^{(4)}$ is $\mathrm{not}$ an Einstein space, or \label{caseii}
	\item $5+12\al\Lambda=0$, $h^{(4)}$ is $\mathrm{not}$ an Einstein space and $R^{(4)}$ is not constant, \label{caseiii}
\end{enumerate}
then ${\mathcal M}$ admits a locally time-like Killing vector. Furthermore, in case \ref{casei}), $h^{(4)}$ is an Einstein metric with $\hat G^{(4)}= \mbox{constant}$, whereas in cases \ref{caseii}) and \ref{caseiii}), $h^{(4)}$ is not Einstein and solves respectively \eqref{beta} and \eqref{extra}.}
\vskip.5cm

\noindent This is a restatement of the properties of generic Class-II and some Class-III solutions we studied above, as these are the ones leading to necessarily static solutions. Note that the above theorem does not restrict the horizon geometry to be spherically symmetric. We can thus have horizons which are anisotropic as admissible static solutions. It should also be stressed that this is qualitatively different from the corresponding theorem in five dimensions, since there the black hole horizon is three-dimensional and its Weyl tensor is automatically zero. $D=6$ is the first case where the Weyl tensor $C_{\alpha \beta \gamma \delta}$ of the internal space plays a non-trivial role and can impose constraints. In dimensions $D>6$, one has a similar situation, \cite{Dotti:2010bw}, although one would normally be required to also consider the corresponding higher Lovelock densities in such a setup. The theorem of course makes no claims about the stability of such configurations (though see \cite{Dotti:2005sq,Gleiser:2005ra,Beroiz:2007gp}). As we see, allowed horizons are four-dimensional Einstein spaces of Euclidean signature, with an added constraint on their Weyl tensor. Since $\Theta$ is non-zero, note that in the non-compact cases these spaces are not asymptotically flat, for otherwise they should satisfy $C_{\alpha \beta \gamma \delta} \to 0$ at four-dimensional infinity.

\subsubsection{Horizon Structure}
We now focus on static Class-II solutions and elaborate on the form of the corresponding potential $V(r)$, (\ref{potentialclassIIEinstein}), which determines the occurrence of event horizons. In particular, we clarify the role of $\Theta$ in this case. There exist two branches of solutions, depending on the sign choice in (\ref{potentialclassIIEinstein}): the \emph{Einstein branch} solutions (-), which tend to Einstein solutions in the limit $\alpha\rightarrow 0$, and the \emph{Gauss-Bonnet branch} solutions (+), which have been argued to be unstable \cite{Charmousis:2008ce}. Because of the stability problems associated with the latter, we now restrict ourselves to the Einstein branch, whose potential is given by
\be
	V\left( r \right) = \kappa  + \frac{{r^2 }}{{12\alpha }}\left( {1 - \sqrt {1 + \frac{12 \alpha \Lambda}{5}  - 24\Theta \frac{{\alpha ^2 }}{{r^4 }} + 24\alpha \frac{M}{{r^5 }}} } \right)\,.
	\label{bhpotential}
\ee
We will then take $M$ to be positive, as required to have a correct definition of mass in the usual $\Theta=0$ situation \cite{Deser:2002rt,Deser:2002jk}. We should stress that once $\Theta\neq 0 $, the proper definition of  mass is no longer  clear, as the constant $\Theta$ changes the spacetime asymptotics. By continuity we take $M>0$, entrusting further study on the meaning of these charges to later work.

In the Born-Infeld limit, $5+12\alpha \Lambda=0$, the only contributions come from the $\Theta$ and mass terms. At large $r$, the $\Theta\geq 0 $ term becomes dominant, developing a branch cut-type singularity. Solutions with $1+\frac{12 \alpha \Lambda}{5} =0$ and $\Theta\neq 0$ are therefore singular. The Born-Infeld case thus falls into the second family of solutions verifying (\ref{lock}) which have to be treated separately.

From the above observation for the Born-Infeld limit, we already see that the $\Theta>0$ term will increase the possibility of a branch singularity near the Born-Infeld limit. We assume for the rest of this section that $5+12\alpha \Lambda>0$.
A branch cut occurs at $r=r_\mathrm{bc}$ whenever
\be
	Q(r_\mathrm{bc}) = (1 + \frac{12 \alpha \Lambda}{5})r_\mathrm{bc}^5  - 24\Theta \alpha ^2 r_\mathrm{bc} + 24\alpha M=0\,.
\ee
When does this actually happen? First, let us consider the simple case where $M$ is switched off. Then, provided $5+12\alpha\Lambda>0$, there is always a branch singularity at
\be
	r_\mathrm{bc}=\left(\frac{24\alpha^2\Theta}{1+\frac{12\alpha\Lambda}5}\right)^{\frac14} =: 5^{1/4} r_0
	\label{branchcut_M0}
\ee
due to the non-vanishing of $\Theta$. On the other hand, if $M$ is not switched off, there is a branch-cut if and only if
\be
	\alpha M <  \frac45\alpha^2\Theta r_0\,, \label{ineqM}
\ee
where $r_0 >0$ is the minimum of $Q(r)$. The constraint (\ref{ineqM}) is the generalisation of the $M=0$ result, the unequality on $M$ being trivially satisfied then. Generically, the effect of the $M$ term will be to decrease $r_\mathrm{bc}$, even if its exact expression cannot be computed analytically in the general case.

To go on, let us turn to the horizon analysis, first by considering the background solution, with $\Theta$ and $M$ switched off (or equivalently for $r$ large enough to make the $\Theta$ and $M$ terms negligible),
\be
	V(r) = \frac {\left(1-\sqrt{1+\frac{12}{5}\alpha\Lambda}\right)}{12\alpha}\left(r^2-r_{\mathrm{c}}^2\right) = 0, \qquad r_c^2 = -\frac{12\alpha\kappa}{1-\sqrt{1+\frac{12}{5}\alpha\Lambda}}\,,
\ee
which is defined if and only if
\be
	\kappa\Lambda>0, \quad \alpha\Lambda>-\frac5{12}\,.
	\label{condition_horizon_lambda}
\ee
We obtain
\bea
	V(r<r_c)>0 &\Longleftrightarrow& \Lambda>0\,, \\
	V(r>r_c)>0 &\Longleftrightarrow& \Lambda<0\,. 
\eea
As noticed in subsection (\ref{GBVacuum}), the solution behaves exacty like four-dimensional (A)dS space in General Relativity with effective cosmological constant
\be
	\Lambda_{e}=\frac {\left(1-\sqrt{1+\frac{12}{5}\alpha\Lambda}\right)}{12\alpha}\,.
\ee

Now, as for the existence of event horizons, following \cite{Charmousis:2008kc} and \cite{Myers:1988ze}, $r=r_h$ is a horizon if and only if
\begin{itemize}
	\item{$r_h>r_{bc}$,}
	\item{$r_h^2\geq -12\alpha \kappa$} (trivial if $\al\ka>0$),
	\item{$r=r_h$ is a root of $P\left( r \right) = -\frac{\Lambda}{10} r^5  + \kappa r^3  + \alpha \left( {\Theta  + 6\kappa ^2 } \right)r - M\,$.}
\end{itemize}
Whenever $\Theta=0$, the black holes behave similarly (modulo the branch singularity that puts some constraints on the smallness of the black hole mass) to their General Relativity black-hole counterparts. Typically, $\Lambda<0$ permits planar and hyperbolic black holes, $\Lambda>0$ an event and a cosmological horizon, and $\Lambda=0$ a unique event horizon. The key question we want to answer here is: does $\Theta\neq 0$ introduce novel horizons to the above black holes, keeping in mind that $\Theta>0$? To answer this question, we momentarily switch off the ``mass'' parameter $M$ and we note that if $\alpha<0$, the resulting black hole potential can be identified with that (tilded quantities) of the five-dimensional Boulware and Deser solution \cite{Boulware:1985wk}:
\be
	\tilde{\alpha}=3\alpha\,, \qquad \tilde \Lambda=\frac{3\Lambda}{5}\,, \qquad \Theta=\frac{-3\tilde M}{\tilde \alpha} \qquad M=0\,.
\ee
Thus, we expect that horizons will be formed even if $M$ is set to zero.
In that case, $P(r)$ is a bisquare polynomial and its zeros $P(r_h>0)=0$ are easily found:
\be
	r_h^2=-\frac5\Lambda\left[-\kappa \pm 	\sqrt{\frac{2\al\Lambda}{5}\left(\Theta+\mathrm{sign}(\al\Lambda)\Theta_\mathrm{max}\right)}\right],
\ee
where
\be
	\frac{2\al\Lambda}5\left(\Theta+\mathrm{sign}(\al\Lambda)\Theta_{\mathrm{max}}\right)>0, \quad \Theta_{\mathrm{max}} = \frac{5\kappa^2}{2|\al\Lambda|}\left(1+\frac{12\al\Lambda}{5}\right).
\ee
This unequality is always true if $\alpha\Lambda>0$, whereas when $\al\Lambda<0$ we need $\Theta<\Theta_\mathrm{max}$. These horizons, when defined, are always greater than the corresponding branch cut position $r_\mathrm{bc}$ \eqref{branchcut_M0}. When $\al\ka<0$, verifying $r_\mathrm{h}^2>-12\al\ka$ yields
\be
	\Theta>\Theta_0\,, \qquad \Theta_0=6\ka^2\left(1+\frac{12}5\al\Lambda\right).
\ee
The occurrence of horizons due to the $\Theta$-term is summarised in the following \Tableref{TableM0}, for various signs of the cosmological constant and zero mass term. In short, $\Theta$ has no effect on the advent of horizons if $\al\ka>0$, whereas it will generate a new event horizon if $\al\ka<0$: when $\al\Lambda\geq0$ for an infinite range, bounded from below ; or for a finite range if $\al\Lambda<0$. It is quite interesting to see that there is a natural separation between these two cases, specifying clearly the effect of $\Theta$, depending on the respective signs of $\al\ka$.

\begin{table}
\centering
{\footnotesize
\begin{tabular}{|c||c|c||c|c||c|c|c|c|}
	\hline
	&\multicolumn{2}{|c||}{$\Lambda=0$}&\multicolumn{2}{|c||}{$\Lambda>0$ ($\ka>0$)}&\multicolumn{4}{|c|}{$\Lambda<0$}\\
	\hline
	 $\Theta$& $\al\ka>0$ & $\al\ka<0$&$\al >0$ & $\al<0$&$\ka<0$, $\al>0$&$\ka,\al>0$&$\ka>0$, $\al<0$&$\ka,\al<0$\\
	\hline
	 0 & $\varnothing$ &$\varnothing$&  $\mathbf C$ &  $\mathbf C$& $\mathbf K$&$\varnothing$&$\varnothing$& $\mathbf K$\\
	\hline
	$\neq0$ & $\varnothing$ &  $\mathbf E$ & $\mathbf C$& $\mathbf C\,  + \mathbf E$&$\mathbf E+\mathbf K$ &$\varnothing$&$\mathbf E$ & $\mathbf K$ \\
	&& iff $\Theta_0<\Theta$&&\multicolumn{2}{|c|}{iff $\Theta_0<\Theta<\Theta_{\textrm{max}}$ }&&iff $\Theta_0<\Theta$&\\
	\hline
\end{tabular}}
\caption[Occurrence of black-hole horizons in six-dimensional Einstein-Gauss-Bonnet theory]{Occurrence of horizons, for parameter $M=0$, depending on the respective signs of $\kappa$ and $\alpha$. $\varnothing$ = no horizons, $\mathbf E\, =$ Event horizon, $\mathbf C\, =$ Cosmological horizon and $\mathbf K\,=$ Killing horizon.  $\Theta_0=6\ka^2(1+\frac{12}5\al\Lambda)$, $\Theta_{\mathrm{max}}=\frac{5\Theta_0}{12|\al\Lambda|}$.}
\label{TableM0}
\end{table}

Let us now examine the special case of planar horizons ($\ka=0$):
\begin{itemize}
	\item Usually, if $\Lambda=0$, no planar horizons are allowed. Here, there is one at $r_{\mathrm h}=\frac{M}{\al\Theta}$ provided $\al M>0$.
	\item For $\Lambda>0$, $M=0$, there is a cosmological horizon ($V(r>r_\mathrm{c})<0$) at $r_\mathrm{c}=10\frac{\al\Theta}\Lambda$ provided $\al>0$ (quite differently from the usual General Relativity case).
	\item For $\Lambda<0$, $M=0$, there is an event horizon ($V(r>r_\mathrm{h})>0$) at $r_\mathrm{h}=10\frac{(-\al)\Theta}{(-\Lambda)}$ provided $\al<0$.
\end{itemize}

If $M$ is not taken to be zero, it is difficult to evaluate quantitatively the impact of $\Theta$, and, apparently, little interesting information can be gained without resorting to a numerical study.

\subsection{Horizon Geometries in the Static Case}

After providing the general discussion of the theorem and the allowed static solutions, we proceed to give some concrete examples. As already mentioned, the geometry of the internal space on the horizon cannot be asymptotically flat due to the non-vanishing Weyl tensor. Candidate solutions are consequently not going to approximate flat space at infinity. Two simple examples of such configurations include an $\mathbf S^{2} \times\mathbf S^{2}$ geometry, as well as a variation of the Taub-NUT space, known as Bergman space. Finally, we will consider solutions that may have some interest for codimension-two setups.

\subsubsection{\texorpdfstring{$\mathbf S^{2} \times\mathbf S^{2}$}{S^2xS^2}}
This four-dimensional space is the product of two two-spheres, with Euclidean signature and the metric
\be
	\ud s^2  = \rho _1 ^2 \left( {\ud \theta _1 ^2  + \sin ^2 \theta _1 \ud \phi _1 ^2 } \right) + \rho _2 ^2 \left( {\ud \theta _2 ^2  + \sin ^2 \theta _2 \ud \phi _2 ^2 } \right)\,,
\ee
where we take the (dimensionless) radii $\rho_{1}$ and $\rho_{2}$ of the spheres to be constant. The entire six-dimensional space has the form
\be
	\ud s^2  =  - V\left( r \right)\ud t^2  + \frac{{\ud r^2 }}{{V\left( r \right)}} + r^2 \rho _1 ^2 \left( {\ud \theta _1 ^2  + \sin ^2 \theta _1 \ud \phi _1 ^2 } \right) + r^2 \rho _2 ^2 \left( {\ud \theta _2 ^2  + \sin ^2 \theta _2 \ud \phi _2 ^2 } \right),
	\label{2sphere}
\ee
with the potential
\be
	V\left( r \right) = \frac{{R^{\left( 4 \right)} }}{{12}} + \frac{{r^2 }}{{12\alpha}}\left( {1 \pm \sqrt {1 - 24k^2 \alpha - 24\Theta \frac{{\alpha ^2 }}{{r^4 }} + 24 \alpha \frac{M}{{r^5 }}} } \right).
\ee
In order for (\ref{2sphere}) to be a solution to the Gauss-Bonnet equations of motion, we are led to the condition of equal sphere radii, $\rho_{1}=\rho_{2}$.
In that case, we have $\ka=\frac1{3\rho_1^2}>0$, $\Theta=\frac4{3\rho_1^4}$.
Since we want to look at the possible creation of an event horizon by
$\Theta$ if $M=0$, it suffices to check the case $\al<0$ for all values and
signs of the cosmological constant: \Tableref{TableM0} clearly shows that
such a creation only occurs as $\al\ka<0$, that is $\al<0$ in our case.
If $\Lambda=0$ or $\Lambda<0$, the constraint $\Theta_0<\Theta$ implies
\be
   0\leq\al\Lambda<\frac5{12}\,,
\ee
which is trivially satisfied if $\Lambda=0$ and yields a \emph{minimum}
value for negative cosmological constant,
$\Lambda_{\textrm{min}}=\frac{5}{12\al}<0$.
On the other hand, if $\Lambda>0$, the constraint
$\Theta<\Theta_{\textrm{max}}$ (necessary to have any horizon at all)
implies
\be
  -\frac5{36}<\al\Lambda<0\,,
\ee
This gives this time a \emph{maximum} value for $\Lambda$,
$\Lambda_{\textrm{max}}=-\frac5{36\al}>0$, which is a more stringent constraint
than the one imposed to have a properly-defined background,
$5+12\al\Lambda>0$.

\subsubsection{Bergman Space}

The Bergman space is a homogenous but non-isotropic space which can be derived as a special case of the AdS Taub-NUT vacuum\cite{Taub:1950ez,Newman:1963yy,Hunter:1998qe,Hawking:1998jf,Chamblin:1998pz,Hawking:1998ct,Zoubos:2002cw}. The ordinary Taub-NUT metric\footnote{Since we consider the horizon geometry to carry a Euclidean signature, in this section all references to known metrics implicitly or explicitly assume a Euclidean version of them. These metrics are usually referred to in literature as gravitational instantons, since they represent solutions to Einstein's equations in Euclidean space with finite actions.} can be written as
\be
	\ud s^2  = W\left( \rho \right)\left( {\ud\tau ^2  + 2n\cos \theta \ud\phi } \right)^2  + \frac{{\ud\rho^2 }}{{W\left( \rho \right)}} + \left( {\rho^2  - n^2 } \right)\left( {\ud\theta ^2  + \sin ^2 \theta \ud\phi ^2 } \right),
	\label{TaubNUT1}
\ee
with the potential $W(\rho)=\frac{\rho-n}{\rho+n}$. The Euclidean time coordinate has a period of $8\pi n$. Here, $n$ is what is usually called the ``nut'' parameter. It has dimensions of $mass^{-1}$. Mathematically, we define a nut as a zero-dimensional (point-like) space where the Killing vector generating the $U(1)$ Euclidean time isometry\footnote{The presence of this isometry is just a mathematical restatement of the property of the Taub-NUT solution being a static spacetime. In the case of Lorentzian Taub-NUT, the Killing vector shows the direction in spacetime (meaning, time $t$) towards which the metric remains unchanged. The isometry generated is thus a non-compact, one-parameter group of translations, while the parameter manifold is isomorphic to $\mathbf R^{1}$. Once we Wick-rotate to imaginary time, $t \to i\tau$, Euclidean time $\tau$ becomes periodic and the parameter manifold is now $\mathbf S^{1}$. The isometry, now generating rotations on the circle charactering the $\tau$ dimension turns into a $U(1)$.} vanishes. The nut is thus a fixed-point of the Euclidean time isometry. The Killing vector generating the isometry is in the case of Taub-NUT $K = \frac{\partial }{{\partial \tau }}$. A fixed-point occurs where $K=0$, or equivalently, $\left| K \right|^2  = g_{\mu \nu } K^\mu  K^\nu   = W\left( \rho \right) = 0$. Zeros of the Taub-NUT potential are then identified as positions of nuts. For the given potential, this occurs at $\rho=n$. We see that, at this position, the factor $\rho^{2}-n^{2}$ in front of the 2-sphere part of the metric is also zero, so the fixed-point set is really zero-dimensional as we would expect from the definition of a nut. This should be juxtaposed with the related concept of a ``bolt'', as a two-dimensional fixed-point {\it set}. We encounter such sets if the potential vanishes at some position different than $\rho=n$, which signifies the position of a two-dimensional sphere. In that sense, bolts are similar to black-hole horizons, since they too are examples of such two-dimensional fixed-point sets for the Euclidean time isometry, although without a nut parameter. To have a regular solution for (\ref{TaubNUT1}), we only consider the range $\rho \ge n$.

In order to make contact with the parametrisations used for the description of the Bergman metric, we introduced the $SU(2)$ one-forms to parametrise the three-sphere
\bea
	\sigma _1  &=& \frac{1}{2}\left( {\cos \psi d\theta  + \sin \psi \sin \theta d\phi } \right),\\
	\sigma _2  &=& \frac{1}{2}\left( { - \sin \psi d\theta  + \cos \psi \sin \theta d\phi } \right),\\
	\sigma _3  &=& \frac{1}{2}\left( {d\psi  + \cos \theta d\phi } \right).
\eea
These satisfy the cyclic relations $d\sigma _1  =  - 2\sigma _2  \wedge \sigma _3 $ etc. The angles $\theta, \phi, \psi$ vary in the ranges $0 \le \theta \le \pi$, $0 \le \phi \le 2\pi$, $0 \le \psi \le 4\pi$. The choice of parameters has to do with the asymptotic behavior of metric at infinity ($r \to 0$). There, the metric three remaining coordinates (angular and time) are combined to give a three-sphere, which we parametrise using $\theta$, $\phi$ and $\psi$. We say that the metric is asymptotically locally flat. This should be contrasted with the usual asymptotically flat metrics, where the corresponding boundary geometry at infinity is a direct product space $\mathbf S^{1} \times\mathbf S^{2}$, instead of $\mathbf S^{3}$. For the Taub-NUT space, the time coordinate indices a non-trivial fibration of $\mathbf S^{3}$.

Using the $SU(2)$ one-forms, and setting $\tau=2 n \psi$, we can eliminate the angular and time coordinates of the metric (\ref{TaubNUT1}) in favor of the one-forms. For the radial coordinate, we make the successive redefinitions $\rho \to \rho+n$, (so that $\rho$ starts at $\rho=0$) and then $\rho \to \frac{\rho^{2}}{2 n}$. The Taub-NUT metric can thus be rewritten as
\be
	\ud s^2  = 4\left( {1 - \mu ^2 \rho^2 } \right)\left[ {\ud\rho^2  + \rho^2 \left( {\sigma _1 ^2  + \sigma _2 ^2 } \right)} \right] + \frac{{4\rho^2 }}{{1 - \mu ^2 \rho^2 }}\sigma _3 ^2  \,,
	\label{TaubNUT}
\ee
where $\mu^{2}=\frac{1}{4n^{2}}$. The metric (\ref{TaubNUT}) can be considered to be a special case of the more general AdS Taub-NUT, of the form
\be
	\ud s^2  = \frac{4}{{\left( {1 - k^2 \rho^2 } \right)^2 }}\left[ {\frac{{1 - \mu ^2 \rho^2 }}{{1 - k^2 \mu ^2 \rho^4 }}\ud\rho^2  + \rho^2 \left( {1 - \mu ^2 \rho^2 } \right)\left( {\sigma _1 ^2  + \sigma _2 ^2 } \right) + \rho^2 \frac{{1 - k^2 \mu ^2 \rho^4 }}{{1 - \mu ^2 \rho^2 }}\sigma _3 ^2 } \right].
\ee
Note that the mass parameter $\mu$ is now defined in terms of $k$ and the nut parameter by $\mu^{2}=k^{2}-\frac{1}{4n^{2}}$. This is a Taub-NUT space with a cosmological constant $-3k^{2}$. We consider the space of radial coordinates where the metric is non-singular, i.e. $0 \le \rho \le 1/k$, so that $\rho_{h}=1/k$ is the horizon of the $AdS$ space. For vanishing cosmological constant ($k=0$), this reduces to the ordinary Taub-NUT geometry of (\ref{TaubNUT}), while for $\mu=0$, the $AdS_{4}$ is recovered. $AdS$ Taub-NUT has in general an $SU(2) \times U(1)$ isometry group, which can however be enhanced for special parameter values.

None of the above mentioned spaces is a good candidate solution for the horizon, since they do not possess a constant $\Theta$. For AdS Taub-NUT, we obtain
\be
	\Theta  = 6\mu ^4 \frac{{\left( {1 - k^2 \rho^2 } \right)^6 }}{{\left( {1 - \mu^{2} \rho^2 } \right)^6 }}\,,
	\label{Theta}
\ee
which only becomes constant at radial infinity (past the $AdS$ horizon), $\Theta  \sim \frac{{6k^{12} }}{{\mu ^8 }}$. Setting $k=0$ in this relation we obtain the corresponding value for the ordinary Taub-NUT, $\Theta=\frac{6\mu^{2}}{(1-\mu^{2} \rho^{2})^{6}}$. The space is asymptotically (locally) flat, so $\Theta \sim 0$ at infinity.

Let us now consider the case where $\mu=k$. We then recover the Bergman metric
\be
	\ud s^2  = \frac{4}{{\left( {1 - k^2 \rho^2 } \right)^2 }}\left[ {\frac{1}{{1 + k^2 \rho^2 }}\ud\rho^2  + \rho^2 \left( {1 - k^2 \rho^2 } \right)\left( {\sigma _1 ^2  + \sigma _2 ^2 } \right) + \rho^2 \left( {1 + k^2 \rho^2 } \right)\sigma _3 ^2 } \right].
\label{Bergman}
\ee
It describes the coset space $SU(2,1)/U(2)$, which is a K\"ahler-Einstein manifold with K\"ahler potential
\be 
	K(z_1, \bar{z}_1, z_2, \bar{z_2} ) = 1- z_1 \bar{z}_1 - z_2 \bar{z}_2 \, , \qquad \mbox{for $z_1 \bar{z}_1 + z_2 \bar{z}_2<1$,}
\ee
and the topology of the open ball in $\mathbf C^2$. Setting $z_1= k \xi \cos (\theta /2 ) e^{i(\phi+\psi)/2}$ and $z_2= k \xi \sin (\theta /2) e^{i(\phi-\psi)/2}$ the metric $g_{\alpha \bar{\beta}} = - \partial_\alpha \partial_{\bar{\beta}} \ln K^{1/k^2}$ reproduces exactly (\ref{Bergman}) after a change of coordinate $\xi^2=2 \rho^2/(1+k^2 \rho^2)$. The Bergman metric (\ref{Bergman}) has an isometry group of $SU(2,1)$. In practice, the choice $\mu=k$ corresponds to infinite ``squashing'' of the three-sphere at the boundary $\rho \to 1/k$, such that only a one-dimensional circle remains intact at spatial infinity. By comparing the terms multiplying $\sigma_{1}^{2}+\sigma_{2}^{2}$ (two-sphere) and $\sigma_{3}^{2}$, we see that as we approach the boundary, the $\sigma_{3}^{2}$ part blows up faster and becomes dominant. The space has this circle as its conformal boundary. It is now possible to see from the expression (\ref{Theta}) for $\Theta$ in $AdS$ Taub-NUT that the Bergman space has $\Theta=6k^{4}$ and is thus a suitable horizon solution. Substituting (\ref{Bergman}) as the metric of the internal space $h^{(4)}_{\mu \nu}$, we verify that it is a solution to the equations of motion. To do so, we first rescale the radial coordinate as $\rho\to\rho/l$, with $l$ having dimensions of $mass^{-1}$ in order to make the metric dimensionless. As a result, we identify the dimensionless curvature scale $k\to kl$. The bulk potential of the solution is then given by
\be
	V\left( r \right) =  -k^{2}  + \frac{{r^2 }}{{12 \alpha}}\left( {1 \pm \sqrt {1 +\frac{12}5\alpha\Lambda - 144k^2 \frac{{\alpha ^2 }}{{r^4 }} + 24 \alpha\frac{M}{{r^5 }}} } \right).
\ee
Bergman space exists in the case $\ka=-k^2<0$,
$\Theta=6k^4$. According to Table \ref{TableM0}, when $M$ is set to zero,
the only case where a horizon may originate from the $\Theta$-term is when
$\al>0$ and $\Lambda$, the bulk cosmological constant, is negative. Then, the condition $\Theta_0<\Theta<\Theta_{\textrm{max}}$ needs
to be verified in order to have a new event horizon, on top of the
pre-existing Killing horizon.
The left part of the unequality yields $\al>0$ and is thus trivially
satisfied, and the right half gives a \emph{minimum} value for $\Lambda$,
\be
  \Lambda_\textrm{min}=-\frac5{24\al}<\Lambda<0\,.
\ee
This is a more stringent constraint than the one imposed to have a
properly-defined background, $5+12\al\Lambda>0$, which yields a lower
minimum value. If this is verified, the Bergman space with $M=0$,
$\Theta\neq0$ allows an event horizon.

We should note at this point that previous studies have shown the Bergman geometry to be unstable, both perturbatively and non-perturbatively,  in the context of ordinary General Relativity, \cite{Kleban:2004bv}. It is not known whether this property persists also in Gauss-Bonnet theory.

As we mentioned above, apart from zero-dimensional fixed-points of the Euclidean time isometry (nuts), one could also consider spaces exhibiting the two-dimensional variety (bolts). This is known and appropriately termed as the Taub-Bolt space and is very similar to the already discussed Taub-NUT. Indeed, the metric for Taub-Bolt is the same as (\ref{TaubNUT1}) and (\ref{TaubNUT}), with the only distinction that the potential is now
\be
	W\left( \rho \right) = \frac{{\rho^2  - 2m\rho + n^2  + k^2 \left( {\rho^4  - 6n^2 \rho^2  - 3n^4 } \right)}}{{\rho^2  - n^2 }}\,.
\ee
The position at which $W(\rho)=0$ is no longer $\rho=n$ and consequently the term $\rho^{2}-n^{2}$ multiplying the two-sphere does not vanish at this point, providing the two-dimensional bolt. Imposing regularity of the potential at the position of the bolt $\rho=\rho_{b}$, we end up with the following prescriptions
\bea
	m &=& \frac{{\rho_b ^2  + n^2 }}{{2\rho_b }} + \frac{{k^2 }}{2}\left( {\rho_b ^3  - 6n^2 \rho_b  - 3\frac{{n^4 }}{{\rho_b }}} \right), \\
	\rho_{b \pm }  &=& \frac{1}{{12k^2 n}}\left( {1 \pm \sqrt {1 - 48k^2 n^2  + 144k^4 n^4 } } \right).
\eea
Is it possible to take the Bergman limit for the Taub-Bolt space as we did with Taub-NUT? To do so, we should retrace our steps and first recast the metric into the Pedersen form. Unfortunately, this is now non-trivial due to the more involved potential and bolt radius. We can however consider the limit $\mu=k$ without deriving the full metric for arbitrary $\mu$. Inspecting the definition of $\mu$ for Taub-NUT, we see that $\mu=k$ corresponds to the limit $n \to \infty$. To find the form of the metric in that limit, we first make the shift $\rho \to \rho+\rho_{b}$. The potential can then be written as
\be
	W\left( \rho \right) = \frac{{\rho\left( {C_0  + C_1 \rho + C_2 \rho^2  + C_3 \rho^3 } \right)}}{{(\rho + \rho_b  + n)(\rho + \rho_b  - n)}}
\ee
with the parameters
\bea
	C_0  &=& \frac{{\left( {\rho_b ^2  - n^2 } \right)\left( {1 + 3k^2 \left( {\rho_b ^2  - n^2 } \right)} \right)}}{{\rho_b }}\mathop  \sim \limits_{n \to \infty } 0 \,,\\
	C_1  &=& 1 + 6k^2 \left( {\rho_b ^2  - n^2 } \right)\mathop  \sim \limits_{n \to \infty } 1 \,,\\
	C_2  &=& 4k^2 \rho_b \mathop  \sim \limits_{n \to \infty } 4k^2 n \,,\\
	C_3  &=& k^2 \,.
\eea
In determining the limit of parameters we used the fact that $\rho_{b} \mathop \sim \limits_{n \to \infty } n$. We then set $\rho \to \frac{\rho^{2}}{2n (1-k^{2}\rho^{2})}$ and keeping only finite terms in the metric, we recover the Bergman space (\ref{Bergman}). Taub-Bolt has thus the same limit as Taub-NUT for infinite nut parameter.

We would like to conclude this section by noting that, by taking $k$ to be purely imaginary in (\ref{Bergman}), we end up with the Fubini-Study metric on $\mathbf{CP}^2$ and that the latter also constitutes a possible horizon metric for a static Lovelock black hole.

\subsection{Six-dimensional black strings}

Let us now turn to some special solutions which resemble black string metrics. Here we assume that the ``horizon'' surface is of Lorentzian signature.
Both solutions presented in this section admit an extra axially symmetric Killing vector, see also \cite{Charmousis:2008bt,Charmousis:2009uk}.

\subsubsection{Six-dimensional warped Born-Infeld black strings}
\label{sec:BIBS}
Throughout this section, the Born-Infled limit is assumed, that is we set $5+12\Lambda \alpha=0$. In this case, we would like to discuss a particular subclass of Class-II solutions, which appears to contain black string solutions as well as solutions that may be relevant to codimension-two braneworld Cosmology. They correspond to the overdetermined solutions (\ref{constclassIInotEinstein}-\ref{potentialclassIInotEinstein}). After Wick rotation, these solutions can be rewritten as
\be 
	\ud s^2 = r^2 h^{(4)}_{\mu\nu} \ud x^\mu dx^\nu + \frac{\ud r^2}{\frac{\rho}{2} + \frac{r^2}{12\alpha}} +\left(\frac{\rho}{2} + \frac{r^2}{12\alpha} \right ) \ud \theta^2 
\ee
where the four-dimensional Lorentzian metric $h^{(4)}_{\mu\nu}$ needs not be Einstein and is only subject to equation (\ref{extra}) that we reproduce here
\be
	\rho=\frac{R^{\left( 4 \right)}}6\pm\frac16\sqrt{{R^{\left( 4 \right)}}^2-6\hat G^{\left( 4 \right)}}\,. 
	\label{extrabis}
\ee
In order to solve (\ref{extrabis}), we assume, for example, that $h^{(4)}_{\mu\nu}$ is of the form
\be
	\ud s^2_{\left( 4 \right)} = -f(\xi)\ud t^2 +\frac{\ud \xi^2}{f(\xi)} + \xi^2\ud \Omega_{2,k}^{2} \,, 
	\label{sphansatz}
\ee
where $\ud \Omega_{2,k}^{2}$ denotes the two-dimensional metric with constant curvature on the sphere, the plane or the hyperbolic space, depending on whether $k=1, 0$ or $-1$ respectively. $h^{(4)}_{\mu\nu}$ therefore has spherical, planar or hyperbolic symmetry, although it is certainly not the most general ansatz with these symmetries. Now, it follows from (\ref{extrabis}) that
\be
	f(\xi)=k-\frac{\rho}2\xi^2\left(1\pm\sqrt{\frac{c_1}{\xi^3}+\frac{c_2}{\xi^4}}\right)\, ,
\ee
where $c_1$ and $c_2$ are integration constants. The corresponding four-dimensional metric $h^{(4)}_{\mu\nu}$ is not an Einstein space and distributional sources at $r^2= -6 \alpha \rho$ are therefore expected from the matching conditions.
These four-dimensional metrics $h^{(4)}_{\mu\nu}$ do not correspond to any known General Relativity solutions at large distance. Born-Infeld theory has been shown to suffer from strong-coupling problems, \cite{Charmousis:2008ce}.

The total space is, in the end, a warped product between a constant curvature two-space and a four-dimensional lorentzian space. This particular black string solution has been first discussed in \cite{CuadrosMelgar:2008kn}.

\subsubsection{Six-dimensional straight black strings}
\label{sec:6dSBS}
We finally consider the special case of Class-III solutions, with a time-like local Killing vector and an undetermined horizon geometry:
\be
	\ud s^2 = \frac{2}{\Lambda \bar{z^2}}\left(-\ud t^2 + \ud z^2\right) + \beta^2h_{\mu\nu}\ud x^\mu\ud x^\nu\,.
\ee
The only constraint on the internal geometry comes from the scalar equation \eqref{beta}, i.e.
\be
	0=-2\Lambda\beta^4+\beta^2R^{\left( 4 \right)}+\al\hat G^{\left( 4 \right)}, \label{beta2}
\ee
where $\beta$ is a constant ``warp factor'' and $1+4\al\Lambda=0$. As in the previous section, we consider a Wick-rotated version in which the internal space is lorentzian and we assume the same particular ansatz for $h^{(4)}_{\mu\nu}$, (\ref{sphansatz}). It then follows from (\ref{beta2}) that
\bea
	\ud s^2_{\left( 4 \right)} &=& -f(\rho)\ud t^2 +\frac{\ud \rho^2}{f(\rho)} +\rho^2\ud \Omega_{2,k}^{2}\,, \\
	f(\rho)&=&k+\frac{\beta^2\rho^2}{4\al}\left[1\pm\sqrt{\frac{2}{3\beta^2}+\frac{32\al\mu}{3\beta^4\rho^3}-\frac{16\al q}{3\beta^4\rho^4}}\right],
\eea
where $\mu$ and $q$ are both integration constants. They have been rescaled so that the metric resembles the Reissner-Nordstr\"om solution far from the source in the minus branch, provided $\beta^2$ is set to two-thirds.

The six-dimensional metric finally reads
\be
	\ud s^2 = \frac{2}{\Lambda z^2}\left(\ud\theta^2 + \ud z^2\right) + \beta^2\left[-f(\rho)\ud t^2 +\frac{\ud \rho^2}{f(\rho)} +\rho^2\ud \Omega_{2,k}^{2}\right],
\ee
and is an unwarped product between a constant curvature two-dimensional space and a four-dimensional unwarped brane admitting Schwarzschild as a limit in one of the branches of solutions, with $\beta^2=\frac23$. This coincides with the Kaluza-Klein black hole reported in \cite{Maeda:2006hj}, provided $\beta^2=1$. We should emphasize here that, as an equation for $h^{(4)}_{\mu\nu}$, (\ref{beta2}) is underdetermined. In particular, had we considered a generic spherically symmetric ansatz, we would have had a free metric function appearing in the internal geometry.

\subsection{Summary of results and outlook}

We have found the general solution{\footnote{The case of Class-Ib still demands the resolution of (\ref{Beqn_class_notEinstein}).}} to the metric  (\ref{GBmetric})  and have investigated generalisations of Birkhoff's theorem in six-dimensional Einstein-Gauss-Bonnet theory. This analysis  significantly generalises previous treatments in five dimensions and six dimensions, or cases where spherical symmetry of the horizon is imposed from the beginning. Furthermore, the analysis undertaken here agrees with \cite{Dotti:2008pp} where staticity is assumed. Permitting the Weyl tensor of the internal space in the equations of motion through the combination $C^{\alpha \beta \gamma \mu} C_{\alpha \beta \gamma \nu}=\Theta \delta^{\mu}_{\nu}$ leads to severe restrictions. We analysed the way this new contribution modifies the available solutions. We distinguish three categories.

The so called Class-I leads both to an underdetermined system of equations and the application of a specific condition (Born-Infeld in even dimensions, Chern-Simons in odd ones) to the parameters of the theory. We find two possibilities:
\begin{itemize}
\item the internal space is a constant curvature space (with $\Theta=0$) and one of the metric functions in transverse space is undetermined (Ia),
\item the internal space is not necessarily Einstein (and generically $\Theta\neq0$) and all metric functions can be determined (Ib).
\end{itemize}

The possibility of an underdetermined system of equations once a particular choice of parameters is used seems to hint at the presence of an increased ``symmetry'' in such a case. Class-I solutions do not obey some variant of Birkhoff's theorem, i.e. static solutions are not unique in this context. Class-II solutions on the other hand give rise to a generalised Birkhoff's theorem; static solutions are unique, provided some conditions related to the structure of the internal space are satisfied:
\begin{itemize}
\item the internal space is Einstein with a constant four-dimensional Gauss-Bonnet charge  and constant curvature (IIa),
\item the internal space is not necessarily Einstein but is constrained by a scalar equation \eqref{extra} and the Born-Infeld condition holds (IIb).
\end{itemize}

Class-III case corresponds to unwarped metrics, and Birkhoff's theorem also holds in some specific subcases:
\begin{itemize}
\item  $1+4\al\Lambda\neq0$ and the internal space is Einstein (IIIb), or
\item $1+4\al\Lambda=0$, the internal space is not Einstein and can or not be constrained by a scalar equation \eqref{beta} (IIIc).
\end{itemize}
A third case exists where Birkhoff's theorem does not hold, when both the horizon is Einstein and the condition $1+4\al\Lambda=0$ is applied (IIIa).

\noindent We summarise our results in \Tableref{TableResults}.

\begin{table}
\centering
\begin{tabular}{|c|c|c||c|c||c|c|c|}
	\hline
		& Ia & Ib & IIa & IIb & IIIa & IIIb & IIIc \\
	\hline
	\bf{Birkhoff} & $\varnothing$& $\varnothing$ & $\surd$ & $\surd$ & $\varnothing$& $\surd$& $\surd$\\
	\hline
	\bf{Einstein} & $\surd $& $\varnothing $ &$\surd$ & $\varnothing$, \eqref{extra}&$\surd $ &$\surd$ &$\varnothing$, \eqref{beta} \\
	\hline	
	$\Theta$ & $0$ & $\geq 0$ & $\geq 0$ & $\geq 0$ & $>0$ & $\geq0$ & $\geq0$ \\
	\hline
	\bf{Fine-tuning} & \bf{BI}& \bf{BI}& $\varnothing$&  \bf{BI}& $1+4\al\Lambda=0$ & $\varnothing$&$1+4\al\Lambda=0$ \\
	\hline
\end{tabular}
\caption[Classes of solutions of six-dimensional Einstein-Gauss-Bonnet theory and their characteristics]{Classes of solutions and their characteristics. {\bf Einstein} : horizon is an Einstein space. {\bf BI} : $5+12\al\Lambda=0$. $\Theta\doteq\frac14C^{abcd}C_{abcd}$.}
\label{TableResults}
\end{table}

For the Class-II solutions, the generalised staticity theorem holds, and we studied some examples of non-trivial horizon geometries. The spaces we consider are in general anisotropic, such as the $\mathbf S^{2}\times \mathbf S^{2}$ product space and the Euclidean Bergman geometry. The latter can be considered as the appropriate limit of either an AdS Taub-NUT or Taub-Bolt space with infinite nut charge. Bergman space has the squashed three-sphere as its conformal boundary and is thus anisotropic.

It would be interesting to investigate further cases of suitable horizon geometries satisfying the requirements of Birkhoff's theorem and also to study the general conditions under which a class of such solutions may arise. A consistent generalisation to higher dimensions would require the inclusion of higher-order Lovelock densities in the action, which will be the topic of future work. 

The most interesting departure from General Relativity arises due to the non-vanishing of the constant $\Theta$. The latter appears, at the level of the static black hole potential, as a novel integration constant or ``charge'' and
is directly related to the Gauss-Bonnet scalar of the four-dimensional horizon, a quantity whose integral yields a topological invariant: the relevant Euler-Poincar\'e characteristic. It is remarkable that it appears directly in the black hole potential, allowing to distinguish the internal geometry of the horizon, and not simply its topology (spherical, planar or hyperbolic) as in General Relativity.

We saw that the presence of this constant imposes particular and non-trivial asymptotic conditions and certainly a particular topology. It probably cannot be interpreted as ``hair'', since it is not a conserved charge defined by a Gauss integral at infinity and it does not indicate an extra Killing symmetry. However, it does severely constrain the horizon geometry, and this is very good news compared to General Relativity. Indeed, the study of General Relativity in higher dimensions, on top of the higher-dimensional counterpart of Schwarzschild solution, \cite{Tangherlini:1963bw}, is the theater of a huge degeneracy: not only the horizon geometry is a lot less constrained, it is an Einstein space instead of a constant curvature one, but on top of this, numerous other solutions exist (see \cite{Emparan:2008eg} for a review), such as black strings, \cite{Gregory:1993vy}, and black rings, \cite{Emparan:2001wn}, to cite only the most famous. It is not clear if this whole zoo persists in Gauss-Bonnet gravity, and the Einstein spaces admissible are significantly constrained. Moreover, among the non-trivial examples provided, the Bergman space is believed to be unstable.

\vfill
\pagebreak



\part{Thermodynamics}
\label{part:three}

\section{Thermodynamics of black holes}

\label{section:BHThermo}

\subsection{The laws of black-hole mechanics and thermodynamics \label{section:BHThermoLaws}}

Nowadays, it is taken for granted that there is some deep connection between black holes and thermodynamics, and concepts like black-hole entropy or temperature are commonly used. Thus, it seems worthwhile to recount how such a correspondence came to be and to briefly sketch the results which set it on firm footing. For a more complete overview, the reader is referred to \cite{Wald:1999vt} or \cite{Ross:2005sc} for instance.

To this end, we shall enounce the four laws of black-hole mechanics, \cite{Bardeen:1973gs}:
\begin{enumerate}
	\item[0.] the \emph{zeroth law}, \cite{Bardeen:1973gs}, states that the surface gravity $\ka_H$ of the black-hole horizon is constant on the horizon itself. The surface gravity is calculated when there is a Killing horizon in spacetime, that is a null hypersurface generated by a time-like Killing vector\footnote{By definition, the Lie derivative of the metric is null along a Killing vector, which translates as the Killing equation: $\nabla_{\l(\mu\r.}\chi_{\l.\nu\r)}=0$.}. Loosely speaking, the surface gravity can be understood as the acceleration exerted at infinity to keep a test body on the horizon. In terms of the Killing vector $\chi^\mu$,
	\be
		\l.\chi^\nu\nabla_\mu\chi_\nu = \ka_H\chi^\mu\r|_H \Rightarrow-\ka_H^2 = \l.\l(\nabla_\mu\chi_\nu\r)^2\r|_H\,,
		\label{SurfaceGravity}
	\ee
	where both equations are evaluated on the horizon\footnote{To prove this, one can use the identity $\chi_{\l[\mu\r.}\chi_{\l.\nu ;\rho\r]}=0$ valid on the horizon.}. By commuting properties of Killing vectors and assuming the existence of a bifurcation surface\footnote{Surface where the Killing vector cancels.}, the proof of constancy follows.
	\item the \emph{first law}, \cite{Smarr:1972kt,Bekenstein:1973mi}, states that the following relation holds
	\be
		\delta M = \frac{\ka_H}{8\pi}\,\delta A+\Omega_H\delta J+\Phi_H\delta Q\,,
		\label{FirstLawVariational}
	\ee
	where $M$ is the gravitational mass of the hole, $\l(\kappa_H,A\r)$ the surface gravity and the area of its horizon, $\l(\Omega_H,J\r)$ its angular velocity and momentum\footnote{We write the version of the first law generalised to rotating stationary black holes for completeness, but in this work we shall only consider non-rotating, static black holes.} and $\l(\Phi_H,Q\r)$ its electric chemical potential and charge. The first law relates the changes in the characteristic quantities of two neighbouring black-hole solutions $\l(M,A,J,Q\r)$ and $\l(M+\da M,A+\da A,J+\da J,Q+\da Q\r)$, keeping the surface gravity, angular velocity and electric potential constant. This is the stationary comparison interpretation, \cite{Jacobson:2003wv}. A second, complementary one is the physical process interpretation, \cite{Penrose:1969pc,Penrose:1971uk,Christodoulou:1970wf,Christodoulou:1972kt}. In this version, the first law gives the change in area when a small amount of mass, angular momentum or charge is thrown into the hole.
	\item the \emph{second law}, \cite{Hawking:1971tu,Hawking:1971vc}, states that the area of a black hole can only increase:
	\be
		\delta A \geq 0\,.
		\label{AreaLaw}
	\ee
	This means that black holes can never bifurcate, and that the area of a black hole resulting from the merging of two black holes is greater than the sum of the areas of the original ones. However, this does not mean that no energy can be extracted from it. Gravitational radiation is allowed provided the area condition holds, \cite{Hawking:1971tu}. 
	\item the \emph{third law} states that no classical process can lower the surface gravity down to zero. Indeed, the rate of absorption of some small quantity of mass, angular momentum or charge, decreases exponentially as the surface gravity decreases, and so it would take an infinite time to end this process, \cite{Israel:1986}\footnote{The weak energy condition is assumed in this reference.}. Another formulation would be that the area of the event horizon goes to zero with  the surface gravity. However, the example of extremal black holes seem to contradict this version, since they retain a finite area when the extremal limit is taken (though see Section \ref{section:EntropyExtremalBlackHoles}). It is however not clear if this last law should be taken as seriously as the previous three, see for instance \cite{Wald:1997qp} for a more detailled discussion.
\end{enumerate}

These laws of course beg for an identification with usual thermodynamics, but at the turn of the seventies, it was not clear if there was any real meaning to it. Clearly, the gravitational mass and electric charge could be taken to be the analogs of the internal energy and electric charge entering in the first law of thermodynamics. But to pursue the analogy to the end, one should also identify some kind of entropy and temperature for the hole, and early proposals were put forward to do just that, \cite{Bekenstein:1973ur,Bardeen:1973gs}, with both quantities identified as multiples of the horizon area and surface gravity. The Generalised Second Law would then state that the entropy of the black hole and its surroundings could only grow, \cite{Bekenstein:1972tm,Bekenstein:1973ur,Bekenstein:1974ax}. The analogy was hampered by the thermal instability of Schwarzschild black hole. Indeed, its temperature grows as it looses mass, and so the black hole either disappears or expands indefinitely because of (classical) statistical fluctuations. This prevented from actually attributing it a proper temperature and to consider it fully as a thermodynamical system, \cite{Hawking:1976de} and Section \ref{section:ThermoSchw}.

The tables turned drastically when Hawking proved that in the semiclassical approximation, black holes emitted a small amount of radiation due to quantum fluctuations in the (curved) vacuum, \cite{Hawking:1974rv,Hawking:1974sw}. In loose terms, different observers in curved spacetimes do not agree on the quantum vacuum of the theory, and one will see particle creation and annihilation while the other sees vacuum, and vice and versa. We will not enter into details as this is well beyond the scope of this work. The spectrum of emission is thermal and has blackbody radiation
\be
	T_H = \frac{\ka_H}{2\pi\hbar}\,,
	\label{HawkingTemperature}
\ee
which means that the correct quantity to be identified with entropy is the quarter of the area of the horizon of the black hole, 
\be
	S_h = \frac{\hbar A}4\,.
	\label{BlackHoleEntropy}
\ee
The analogy between black holes and thermodynamics could then be formalised and it was found that black holes could be in equilibrium with blackbody radiation if put in a box, \cite{Hawking:1976de}.

At this point, we should mention that in fact the analogy should not quite be between black holes and thermodynamics. The temperature and the entropy are actually to be attributed to the \emph{horizon}, independently of its covering a singularity or even being an event horizon. Shortly after Hawking's discovery, Unruh showed that the same rate of emission could be measured by an accelerated Rindler observer, \cite{Unruh:1976db}, since he too would observe an horizon, albeit with a different temperature
\be
	T_R = \frac{a_R}{2\pi\hbar}\,,
	\label{UnruhTemperature}
\ee
where $a_R$ is the acceleration of the observer. Finally, Hawking and Gibbons studied a little bit later cosmological horizons such as that of de Sitter space, \cite{Gibbons:1977mu} , and found that there too a perfectly thermal spectrum of emission could be computed. So we are led to the idea that the concept of entropy has more to do with that of Killing horizon than to that of black hole properly, \cite{Jacobson:2003wv}. 

Lastly, the famous formula \eqref{BlackHoleEntropy} for the entropy of a black-hole horizon is valid \emph{a priori} only in the semi-classical approximation and is expected to change as quantum corrections to the classical Einstein-Hilbert action are taken into acount. A generic, Lagrangian-independent, formulation of entropy has been put forward by Wald, \cite{Wald:1993nt}, and one can check that the effect of higher powers of the curvature in Lovelock theory on black-hole entropy is indeed to generate corrections to the area value, \cite{Myers:1988ze,Jacobson:1993xs,Banados:1993qp}.

The rest of this section will proceed along the following lines. In subsection \ref{section:EuclideanPathIntegral}, we shall introduce the concept of Euclidean path integral, and its close relationship with thermodynamic potentials. Next, in subsection \ref{section:PartitionFunctionEM}, we shall give a prescription on how to calculate such potentials using the Hamiltonian formalismin General Relativity, focusing in particular on boundary conditions. We continue with an exposition of the link between Killing horizons (e.g., inner boundaries) and entropy in subsection \ref{section:Entropy}. Once thermodynamic quantities in General Relativity have been defined, we make a short detour by classical thermodynamics and recall how to define thermodynamic ensembles through the use of extensive or intensive variables in subsection \ref{section:ThermoStability}. Finally, subsection \ref{section:ThermoBHGR} ends this section with an overview of the thermodynamic behaviours of neutral and charged black holes in General Relativity.

\subsection{Euclidean gravitational path integral}

\label{section:EuclideanPathIntegral}

One most popular technique for computing the partition function is certainly through the Euclidean path integral method for gravity, \cite{Gibbons:1976ue,York:1972sj}. Leaving aside intrinsic problems linked to quantisation of gravity, we may consider the (Lorentzian) path integral over all possible metrics (and possibly all matter fields if needed) of the effective action for gravity
\be
	Z = \int \mathcal D\l[g\r]\mathcal D\l[\Psi\r] \e^{\frac{i}{\hbar}S\l[g,\Psi\r]}\,,
	\label{PathIntegral}
\ee
where $g$ and $\Psi$ denote respectively the metric fields and all the matter fields collectively.

This procedure should be valid at energies well below the Planck scale or at distances well over the Planck length (where quantum corrections are expected to matter), so that we do not need a theory of quantum gravity. We have included explicitly the $\hbar$ factor to underline that such a procedure is intrisically quantum mechanical. Thinking in terms of particle trajectories for definiteness, this amounts to calculating a quantum transition amplitude between two different states, taking into account \emph{all possible trajectories} between the initial and final state. Indeed, we know perfectly well that classical particles should follow classical trajectories (that is, solutions to the equations of motion derived through the variation of the action $S\l[g,\Psi\r]$). Yet, the quantum formulation of field theory only knows about \emph{probabilities}: all trajectories ought to be taken into account, affected by different probabilistic weights. The path integral is then the computation of the quantum amplitude by integrating over all possible trajectories, classical as well as virtual, each contributing to some extent to the value of the amplitude. The classical trajectories will yield the bulk of it, while virtual ``quantum'' trajectories will contribute all the less than they stray farther and farther from the classical ones. 

How can this be consistent? The reader may find a little puzzling that, knowing only the initial and final state, the system ``chooses'' the classical trajectories as the most probable ones. One can shed some light on this picture in view of optical interference theory. The classical trajectories are stationnary points of the action, and so are not subject to interferences. However, the virtual trajectories are so, and, given a long enough evolution time, the probabilistic weight assigned to them will be suppressed by interferences, leaving only the classical trajectories with significant weights.

Let us now discuss the connection with statistical physics. First, we need to Wick-rotate the Lorentzian time to render spacetime Euclidean, since the statistical physics phase space of variables is Euclidean (positive signature). Second, the Euclideanised time needs to be made periodic.
Namely, we formulate the statistical mechanics partition function $Z$ as
\be
	Z = \sum_{states} \e^{-\ba E_n} = \int \mathcal D\l[g\r]\mathcal D\l[\Psi\r] \e^{-I}\,,
	\label{EuclideanPartitionFunction}
\ee
where we have set $\hbar=1$ again and $I=iS$ is the Euclidean action. The path integral then coincides with the partition function of statistical mechanics in the canonical ensemble, provided we identify the inverse temperature $\ba$ with the periodicity of the Euclidean time\footnote{In the case of a static metric Ansatz, it is straightforward to prove that this is equivalent to the formula in terms of the surface gravity on the horizon, \eqref{HawkingTemperature}, see Appendix \ref{section:TempCalculation}.}. This is the reason why the Euclidean time should be made periodic: in the canonical ensemble, the temperature of the system is fixed (by connection to a reservoir, see Section \ref{section:ThermoStability} below) and the evolution should be made between two states with the same initial and final temperature. In field theory vocabulary,  this means that the initial and final state should be the same, and so the evolution should be periodic in time.

The next issue is how to evaluate the path integral: calculating it through direct integration is usually a daunting (if not outright impossible) task, given that it is a functional of fields. A popular technique is to evaluate it in a saddle-point approximation. In essence, most of the value of the integral will come from maxima of the integrand, which turn out to be minima of the Euclidean action. These of course are classical solutions, hence we approximate the value of the path integral to the sum of classical contributions: this is the semi-classical approximation.

The sum is realised over the various $n$ states of the system with energy $E_n$, and the inverse equilibrium temperature $\ba$ is identified with the period of the Euclidean time. The mean, expectation value for the energy of the system\footnote{E.g., the macroscopic value that would be measured by experiments.} can be calculated by the following formula:
\be
	\langle E\rangle = -\frac{\partial}{\partial\ba}\log Z\,,
	\label{EnergyExpectationValue}
\ee
while the macroscopic entropy is
\be
	\langle S\rangle = -\sum_n p_n\log p_n = -\ba\frac{\partial}{\partial\ba}\log Z+\log Z\,,
	\label{EntropyExpectationValue}
\ee
which relates the Shannon interpretation of entropy in information theory to its statistical physics by use of the probability for the system to be in the $n^{\mathrm{th}}$ state
\be
	p_n = \frac{\e^{-\ba E_n}}{Z}\,.
	\label{StateProbability}
\ee

Let us now come back to the matter at hand, that is the thermodynamic interpretation of horizons in General Relativity and their assignment of a certain quantity of entropy. We have to worry  whether the Euclidean path integral is well-defined in General Relativity. Basically, the integral needs to be convergent. This is trivially ensured if the Euclidean action is positive semi-definite, but there is no reason why the General Relativity action should be so. It is written as
\be
	I = -\frac1{16\pi}\int_\M\sqrt{g}\l(R-2\La\r) - \frac1{8\pi}\int_{\partial \M}\sqrt{h}K\,,
	\label{EuclideanActionEinstein}
\ee
where we included an extra term evaluated on the boundary $\partial\M$ of spacetime, with $g$ the four-metric relative to four-dimensional spacetime $\M$ and $h$ the induced three-metric on $\partial \M$. $K$ is the extrinsic curvature of the boundary with respect to spacetime, that is some kind of measure of how much the boundary is curved compared to the embedding spacetime\footnote{For conciseness, we postpone a more precise definition to Section \ref{section:IntegralHamiltonian}.}. This term is needed so that only the components of the induced metric on the boundary need to be specified as boundary conditions and not their derivatives. Note that the overall minus sign in the Euclidean action comes from the Wick-rotation procedure\footnote{A factor of $i$ comes from the integration measure of the time coordinate, another from the extraction of the minus sign under the square-root of the determinant of the metric which is now positive.}.

Let us address two convergence issues, very different in nature. First, the Euclidean action generically develops some conformal negative modes, which can render it arbitrarily negative. Then, it is not bounded from below, and thus seems divergent, \cite{Gibbons:1978ac,Gibbons:1978ac}. These works by Gibbons, Hawking and Perry also showed that these modes were physically irrelevant provided one chose the integration contour of the path integral carefully. For a more modern and pedagogical treatment of the question of negative modes of the Euclidean path integral, generalised to Reissner-Nordstr\"om and rotating spacetimes in generic dimensions, see \cite{Monteiro:2010cq} and references therein.

The second convergence issue is related to the fact that some saddle points of the Euclidean action may have infinite value, for non-compact spacetimes. They must be regularised, by subtracting an appropriate ``background'' contribution. By background, we mean here the asymptotic metric on the boundary of spacetime (which fortunately for black holes, our case of interest, also coincides with the spacetime obtained by switching off the back hole entirely, that is by cancelling all integration constants). Let us give a few examples. For asymptotically flat boundary conditions, one may usually use Minkowski (or its Euclidean counterpart) as a background. Of course, since in this case the Ricci scalar cancels (flat spacetime), the subtraction term is simply the boundary term evaluated on the background
\be
	\bar I = - \frac1{8\pi}\int_{\partial \M}\sqrt{h} \bar K\,,
\ee
so that the full Euclidean action is now
\be
	I- \bar I  = -\frac1{16\pi}\int_\M\sqrt{g}\l(R-2\La\r) - \frac1{8\pi}\int_{\partial \M}\sqrt{h}\l[K-\bar K\r].
	\label{SubEuclideanActionEinstein}
\ee
It is very important to note that for this procedure the (black-hole) spacetime and the background must coincide on the boundary, so that the induced three-metric $h$ \emph{is the same}. 

For Schwarzschild spacetime, \cite{Gibbons:1976ue}, the bulk term is zero, while the boundary term is
\be
	\int_{\partial \M}\sqrt{h}K = \partial_n\int_{\partial\M}\ud\Sigma = 4\pi\beta(2r-3m)\,,
\ee
where $n^\mu$ is the outward-pointing normal vector to the boundary, and $\ud\Sigma$ the coordinate measure of integration on the boundary. The background subtraction term is then
\be
	\int_{\partial \M}\sqrt{h}\bar K = \partial_n\int_{\partial\M}\ud\Sigma = 4\pi\bar\beta2r\,.
\ee
There is an extremely important step in the computation hidden here: there is no natural periodicity associated to Euclidean space. This is because it is a perfectly regular, positive-definite instanton and contains no conical singularity. So in fact any periodicity may be attributed, there is no prescription \emph{a priori}. This would be true were it not for the black-hole spacetime. Remember that the path integral implies summing over metrics obeying certain boundary conditions, in particular the induced metric elements on the boundary must be the same. In thermodynamic terms, the temperature of the black hole and of the background redshifted from the horizon to the boundary should be the same. Imagining first a boundary at finite $r=r_B$, this means the redshifted boundary periodicity for the black hole and the background are respectively
\be
	\frac{\ba\l(r_B\r)}{\sqrt{-g_{tt}\l(r_B\r)}}=\ba\,,\qquad  \frac{\bar\ba\l(r_B\r)}{\sqrt{-\bar g_{tt}\l(r_B\r)}}= \bar\ba\,. 
\ee
We can now use the freedom in choosing the background periodicity so that
\be
	\bar\ba\l(r_B\r)=\ba\l(r_B\r)\,\Rightarrow\, \bar\ba\l(r_B\r) = \sqrt{\frac{-g_{tt}\l(r_B\r)}{- \bar g_{tt}\l(r_B\r)}}\ba\l(r_B\r)\,,
\ee
and from there send the boundary to infinity (or not, if one wished to keep a regulator and define some kind of gravitational box). The total Euclidean action is then finite
\be
	I-\bar I = \half\ba m\,,
\ee
and the canonical thermodynamic potential 
\be
	W=\frac1\ba\log Z = \half m\,.
\ee
These results are easily generalised to the charged case, \cite{Gibbons:1976ue}. So let us now turn to the case with a negative cosmological constant, which is of special interest. $n+1$-dimensional AdS space has $n$-dimensional Minkowski space as a boundary, see Section \ref{section:AdSspace}. Both bulk and boundary terms diverge, but setting the zero of the action for the AdS background, the action for Schwarzschild-AdS may be regularised (so that the boundary contribution cancels), and the thermodynamics analysed, \cite{Hawking:1982dh} and Section \ref{section:ThermoRNAdS}.

Concerning de Sitter space, the same procedure carries over, but here one has to keep in mind that Euclideanised de Sitter space has no boundary, arising from considerations in \Tableref{Table:deSitter}. Euclideanised de Sitter space can be embedded in five-dimensional Euclidean space as a four-sphere of radius $a=\sqrt{3/\La}$. Since spheres have no boundary (they are themselves the boundary of balls), no boundary term can occur.  Moreover it is compact, so the action is finite. The thermodynamics of black holes in de Sitter space have been analysed in \cite{Gibbons:1977mu} and show that both the event and particle horizon can emit radiation.

One may question the naturalness of such a procedure. Indeed, it seems that the background subtraction leaves a lot of freedom to choose the background in question. Even though it appears convenient to choose it simply as the asymptotic spacetime, there are certainly many choices possible depending on the situation and, in the end, the admissible ones should be those allowing to regularise the divergences. This seems to be an impassable ambiguity of the Euclidean path integral method. We feel that the correct way of thinking about this issue should provide some sort of prescription. In a later section, we will review a method based solely on variational Hamiltonian techniques which lift this ambiguity by making full use of the conditions imposed on the boundary, see \ref{section:DefEnsemblesEM} and \cite{Regge:1957td,Henneaux:1985tv}.

Recently however, some progress has been made in the context of the AdS-CFT correspondence. Indeed, counterterms on the boundary of AdS (on which the gauge theory lives) could be devised which make the action finite without resorting to background subtraction. Of course, this is only valid for asymptotically AdS spacetimes, \cite{Balasubramanian:1999re,Emparan:1999pm}, strictly speaking (though see \cite{Cai:1999xg} for irregular asymptotics). Remarkably, these counterterms depend only on the Ricci curvature of the boundary and its derivatives, so that no background metric is needed to match the spacetime on the boundary.

In the next subsection, we show how the various contributions in the thermodynamic potential can be computed by the Hamiltonian formalism and how boundary conditions come into play.

\subsection{Calculation of the partition function}

\label{section:PartitionFunctionEM}

\subsubsection{Gravitational Hamiltonian}

\label{section:IntegralHamiltonian}

\begin{figure}
 \begin{center}
	\includegraphics[width=.5\textwidth]{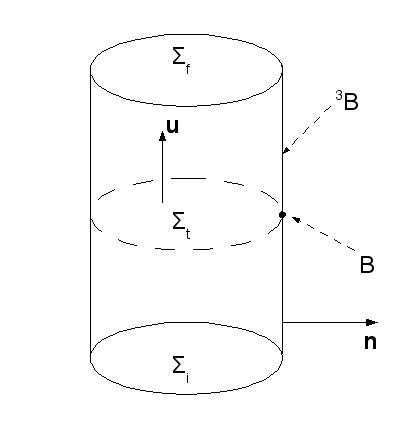}
 \end{center}
	\caption{Foliation of spacetime according to the time-coordinate.}
	\label{Fig:foliation}
\end{figure}

We recall the (unregularised) Einstein-Maxwell action supplemented by the Gibbons-Hawking three-boundary term, \cite{Gibbons:1976ue,York:1972sj},
\be
	S_{EM} = \frac1{16\pi}\int_{\M}\ud^4x\sqrt{-g}\l(R-2\La-\frac14F^2\r)\mp\frac1{8\pi}\int_{\partial\M}\ud^3x\sqrt{\pm h}K\,.
	\label{EMActionGHY}
\ee
The boundary terms are necessary so that for classical solutions obeying the equations of motion, only the components of the three-metric $h$ induced on the boundary need to be specified there and not their derivatives, \cite{York:1972sj,Regge:1974zd,Gibbons:1976ue}. The plus and minus sign allow for a space-like and time-like boundary, as the boundary may consist in several pieces. In order to define the Hamiltonian of the theory, we foliate spacetime along the time direction, and we consider only static and spherically symmetric spacetimes for simplicity. The full analysis without the staticity assumption may be found in the original work by Brown and York, \cite{Brown:1992br} (see also later work by Hawking and Horowitz, \cite{Hawking:1995fd}). We thus take as Ansatz for the metric:
\be
	\ud s^2 = -N^2(r)\ud t^2 +n^2(r)\ud r^2 + R^2(r)\ud \Omega^{2}_2\,.
	\label{ADMmetric}
\ee
With these assumptions, the time coordinate $t$ does generate a foliation of spacetime into space-like hypersurfaces $\Sigma_t$ with unit normal time-like vector $u^\mu=N^{-1}\da^\mu_0$. The space-like boundary then consists of two initial and final space-like hypersurfaces $\Sigma_{t_i}$ and $\Sigma_{t_f}$, which are the initial and final states of the time evolution. The induced three-metric on the hypersurfaces $\Sigma_t$ is called $h^{(t)}_{ij}$, $i,j=1,2,3$ and has Euclidean signature and  is related to the bulk metric by $h^{(t)}_{\mu\nu}=g_{\mu\nu}+u_\mu u_\nu$. We also define $K^{(t)}_{\mu\nu}=h^{(t)}_{\mu\al}\nabla^\al u_\nu$ as the extrinsic curvature\footnote{Note that our sign convention is the same as  \cite{Hawking:1995fd} but opposite to \cite{Brown:1992br}.} of the hypersurface, which is a measure of how much it is curved compared to the spacetime it is embedded into.

To this picture we add a time-like boundary at spatial infinity, ${}^{3}B$, with induced metric $h_{\mu\nu}=g_{\mu\nu}-n_{\mu}n_\nu$, and $n^\mu=n^{-1}\da^\mu_r$ is the unit normal space-like vector to the boundary. Again, for simplicity, we assume that $u\cdot n=0$, that is the foliation surfaces  $\Sigma_t$ are always othogonal to the time-like boundary  ${}^{3}B$ (though see \cite{Hawking:1996ww,Booth:1998eh,Booth:2000iq,Brown:2000dz} for non-orthogonal boundaries). We define its extrinsic curvature $K_{\mu\nu} =h^\al_\mu\nabla_\al n_\nu$. Finally, the two-boundary $B$ is the intersection of the slices $\Sigma_t$ foliating spacetime with the time-like boundary ${}^{3}B$. It has an induced two metric $\sigma_{\mu\nu}=g_{\mu\nu}-n_{\mu}n_\nu+u_{\mu}u_\nu$ and extrinsic curvature $k=\sig^\al_\mu D_\al n_\nu$ with $D_\mu$ the covariant derivative on the $\Sigma$-foliation. This picture is summarised in \Figref{Fig:foliation}.

In all this work, we will only be interested in asymptotic conserved quantities and asymptotic thermodynamics. So the boundary ${}^{3}B$ will be systematically sent at spatial infinity $r\to+\infty$. We note that one can also define quasilocal quantities defined on boundaries at finite distance, along the lines advocated by Brown, York and others, \cite{Brown:1992br,Brown:1992bq}. We will not deal at all with these matters here and refer the interested reader to the references just cited. However, this will not prevent us from relying on these in order to define properly the thermodynamics on the asymptotic boundary.

The canonical variables of gravitation were identified by Arnowitt, Deser and Misner, \cite{Arnowitt:1962hi}, and consist in the spatial components $h^{(t)}_{ij}$ of the induced metric on the foliation and their conjugate momenta $P^{ij}=-\frac1{16\pi G}\sqrt {h^{(t)}}\l(K^{(t)}h_{(t)}^{ij}-K_{(t)}^{ij}\r)$. The lapse $N$ (and the shift vector $N_i$ when spacetime is stationary) act as Lagrange multipliers for the Hamiltonian and momentum constraints. On the other hand, the canonical variables for Maxwell theory are the spatial components of the electric potential $A_i$ and their conjugate momenta are the electric fields $E^i=F^{i0}$. So, by definition, the Hamiltonian of the Einstein-Maxwell theory is formed by decomposing the action on the foliation and writing it in the form:
\be
	S_{EM} = \int_{t_i}^{t_f}\ud t\l\{\int_{\Sigma_t} \l[P^{ij}\dot h^{(t)}_{ij} + \l(\frac{\sqrt{-g}}{16\pi}E^i\r)\dot A_i  \r]\ud^3x-H\r\},
	\label{EMActionHamiltonian}
\ee
where $H$ is the total (off-shell) Hamiltonian of the theory. The latin indices $i,\,j$ are raised and lowered with the three-metric $h^{(t)}_{ij}$. Under the form \eqref{EMActionHamiltonian}, the action is now first-order in time-derivatives (but not in space), and treating the various pairs of canonical conjugate variables as independent allows to recover the usual equations of motion, by varying the action with respect to them and to the Lagrange multipliers.
 According to Brown and York, \cite{Brown:1992br}, and Hawking and Horowitz, \cite{Hawking:1995fd}, the Hamiltonian derived from the Einstein-Maxwell action is written as:
\bea
	H &=& \int_{\Sigma_t}\l[N\mathcal{H}+A_0 \partial_i\l(\frac{\sqrt{-g}}{16\pi} E^i\r)\r]\ud^3x  -\frac1{8\pi} \int_{B} \sqrt{-h}\l[k-\bar k\r] -\nn\\
		&&+\frac1{16\pi} \int_{B} \sqrt{-h}E^{i} n_i A_0\,,
	\label{HamiltonianEinsteinMaxwell}
\eea
where crucial use has been made of the decomposition on the foliation of the four-dimensional Ricci scalar and the three-extrinsic curvature on the spatial boundary ${}^{3}B$:
\bea
	R&=&\mathcal R+K{(t)}_{\mu\nu}K_{(t)}^{\mu\nu}-K_{(t)}^2+2\nabla_\mu\l(K^{(t)}u^\mu-a^\mu\r)\,,\label{RicciDecomposition}\\
	a^\mu&=&u^\nu\nabla_\nu u^\mu\,, \label{AccelerationNormalVector}\\
	K &=&k+n_\mu a^\mu\,. \label{ThetaDecomposition}
\eea
$\mathcal R$ is the three-dimensional Ricci scalar on the foliation $\Sigma_t$, calculated from $h^{(t)}$. $a^\mu$ is the acceleration of the hypersurface normal vector $u^\mu$. It verifies $u_\mu a^\mu=0$ as can be straightforwardly checked by taking the covariant derivative of $u^2=-1$. Using \eqref{RicciDecomposition} and comparing with \eqref{HamiltonianEinsteinMaxwell}, one infers that
\be
	\mathcal H = -\frac{\sqrt{h^{(t)}}}{16\pi}\l[\mathcal R-2\La+K^{(t)}_{\mu\nu}K_{(t)}^{\mu\nu}-K_{(t)}^2\r]+\frac{\sqrt{-g}}{32\pi}NE_iE^i\,.
	\label{HamiltonConstraintEM}
\ee
The Hamiltonian and Maxwell constraints cancel on-shell since they are just the time components of the Einstein and Maxwell equations of motion, \cite{Arnowitt:1962hi}.

Barred quantities in \eqref{HamiltonianEinsteinMaxwell} refer to the background metric taken as reference for the zero of energy, so that these quantities do not diverge (this is often the case for non-compact spaces). There is a debate as to what prescription to choose. Brown and York argue that any choice is admissible, what only matters is to obtain regular finite expressions and we should let physical sense guide us in the various situations encountered. Hawking and Horowitz take a different viewpoint, arguing that the action should be zero for some reference background, which is a static solution to the equations of motion. To achieve this, the induced metric and various matter fields should agree on the spatial boundary $\phantom{1}^{3}B$. Thus one can choose to label the static slices such that $\bar N(r)=N(r)$ on the boundary, so that both the spacetime and the background metric agree there. Most importantly, it leaves the \emph{bulk} metric elements free to fluctuate.

One may then evaluate the total Hamiltonian \eqref{HamiltonianEinsteinMaxwell} on-shell for classical solutions, so that the three-boudary term cancels and only the two-boundary terms are left, \cite{Brown:1992br,Hawking:1995fd}:
\bsea
	\l.H\r|_{cl}&=&-\frac1{8\pi} \int_{B} \sqrt{\sigma}N\l(k-\bar k\r) -\frac1{16\pi} \int_{B} \sqrt{\sigma} NF^{\mu\nu}A_\mu n_\nu\\
							&=& M_g - \Phi Q\,,
	\label{HamiltonianOnShellEM}
\esea
which is the expression we may expect from usual physics and is the total energy of the system.

The total asymptotic gravitational mass of the solution will be defined at spatial infinity $r\to\infty$ on $B$:
\be
	M_g = -\frac1{8\pi} \int_{B} \sqrt{\sigma}N\l[k-\bar k\r],
	\label{GravitationalMassHH}
\ee
and gives the gravitational mass on the asymptotic boundary

We identify the second boundary term in \eqref{HamiltonianOnShellEM} with the electric energy of the system:
\be
	\Phi Q = \frac1{16\pi} \int_{B} \sqrt{\sigma} NF^{\mu\nu} n_\nu A_\mu\,.
\ee
This suggests that the electric chemical potential $\Phi$ is the asymptotic value of the electric potential (reached in particular on the surfaces $B_\infty$), and allows to define the electric charge as
\be
	Q = \frac1{16\pi}\int_{B} \sqrt{\sigma}NF^{0\nu}n_\nu\,.
	\label{ElectricChargeEM}
\ee
We did not include any background contribution since usually the background is neutral, though no particular difficulty arises if it is not so and the above expression needs to be amended. Moreover, if there were a source term for the electric field, then the spatial components of the above definition would give rise to conserved currents and to the usual charge-current equation of conservation.

\subsubsection{Action variation and boundary conditions}

\label{section:DefEnsemblesEM}

We now tackle a subtle point, which has to do with the conditions imposed on the boundary. Indeed the Hamiltonian \eqref{HamiltonianEinsteinMaxwell} and its on-shell version \eqref{HamiltonianOnShellEM} are all well and good, but do they define a proper and consistent variational principle? This will be the case if the Hamilton-Jacobi equations with appropriate boundary conditions are truly equivalent to the Euler-Lagrange equations (the usual Einstein-Maxwell equations of motion).

So let us take the variation of \eqref{EMActionHamiltonian} with respect to the canonical variables, and evaluate it on-shell (so that bulk variations proportional to the Einstein-Maxwell equations of motion cancel out) for the static ADM metric \eqref{ADMmetric}. A computation by Brown and York, \cite{Brown:1992br}, yields
\bea
	\l.\delta S_{EM}\r|_{cl} &=&  \frac1{8\pi}\int_{{}^3B}\sqrt{\sig}\l[k\,\delta (N) +\frac N2\l(k\sig^{ab}-k^{ab}+n_\mu a^\mu\sig^{ab}\r)\,\delta(\sig_{ab})\r] -\nn\\
			&&- \frac1{16\pi}\int_{\phantom{1}^3B} \sqrt{-h}E^{i}n_i\,\delta (A_0)\,.
		\label{EMActionHamiltonianVariation}
\eea
We shall not redemonstrate this result, but simply illustrate that it is the correct one in the static ADM case. From \eqref{HamiltonConstraintEM} and plugging \eqref{ADMmetric} (and setting $f(r)=n(r)^{-1}$), we get
\bsea
	\mathcal H &=& -\frac{1}{16\pi}\l[2\La R^2f^{-1}-4RfR''-4RR'f'-2fR'^2+2f^{-1}\r]+\frac{fR^2A'^2}{32\pi N}\,,\\
	k&=&2\frac{fR'}{R}\,.
\esea
The variation of the action \eqref{EMActionHamiltonian} is, taking the on-shell limit:
\bsea
	\l. \delta (S_{EM})\r|_{cl} &=& \frac1{4\pi}\int_{{}^3B}\l[-NRf\,\da(R')-RR'N\,\da(f)+N'Rf\,\da(R)\r]+\nn\\
														&&+ \frac1{4\pi}\int_{{}^3B}\da(NRR'f)+\slabel{EMActionHamiltonianVariationSuperspace1Grav}\\
														&& +\frac1{16\pi}\int_{{}^3B}\da\l(\sqrt{-h}\,n_iE^i\r)A_0- \frac1{16\pi}\int_{{}^3B}\da\l(\sqrt{-h}\,n_iE^iA_0\r)\slabel{EMActionHamiltonianVariationSuperspace1Maxwell}\,,
	\label{EMActionHamiltonianVariationSuperspace1}
\esea
where in \eqref{EMActionHamiltonianVariationSuperspace1Grav} the first boundary term comes solely from the bulk constraints while the second boundary term comes from the Gibbons-Hawking-York boundary term. Thus, if the latter is absent, then in order to have a well-defined variational problem, one needs to fix derivatives of the metric as well as bulk metric components on the boundary. Similarly, in \eqref{EMActionHamiltonianVariationSuperspace1Maxwell}, the first term comes from varying and integrating by parts the bulk constraint, while the second term is generated by the electric boundary term present in \eqref{EMActionHamiltonianVariation}. In the electric expressions, we have kept the electric canonical decomposition intact since it is more suggestive. However, when all boundary terms in \eqref{EMActionHamiltonianVariation} are included, simplifying the expressions in \eqref{EMActionHamiltonianVariationSuperspace1} yields
\be
	\l. \delta (S_{EM})\r|_{cl} = \frac1{4\pi}\int_{{}^3B}\l[RR'f\,\da(N)+(NR)'f\,\da(R)\r]- \frac1{16\pi}\int_{{}^3B}\sqrt{-h}\,n_iE^i\da(A_0),
	\label{EMActionHamiltonianVariationSuperspace}
\ee
where indeed only boundary metric and electric field components need to be fixed. However, once this is done, that is $\l.\da(N,R,A_0)\r|_{{}^3B}=0$, an extremum of the action \eqref{EMActionGHY} is finally reached. For the electric field, this precisely means keeping the electric potential at infinity $\Phi$ fixed.

Expression \eqref{EMActionHamiltonianVariationSuperspace} is exactly what one would have obtained by substituting an ADM Ansatz \eqref{ADMmetric} in the Brown and York expression, \eqref{EMActionHamiltonianVariation}. Note also that \eqref{EMActionHamiltonianVariationSuperspace} provides a concrete realisation of the effect of the Gibbons-Hawking-York term, in not having to fix any derivative of the metric elements on the boundary.

One can also take a different view, as advocated in \cite{Regge:1974zd,Martinez:2004nb,Martinez:2006an}, keeping the boundary terms $B_g$, $B_{em}$ general \emph{a priori} in \eqref{EMActionHamiltonian} and \eqref{HamiltonianEinsteinMaxwell}, and then asking simply that once the action is varied, the variations of the boundary terms  $\da(B_g)$, $\da(B_{em})$ should be fixed so as to compensate the boundary terms generated by the variation of the bulk terms after integration by parts. Inspecting \eqref{EMActionHamiltonianVariationSuperspace1}, we identify
\bsea
	\da(B_g)&=&\frac1{4\pi}\int_{{}^3B}\l[NRf\,\da(R')+RR'N\,\da(f)-N'Rf\,\da(R)\r], \slabel{VarGravBoundaryTerm}\\
	\da(B_{em})&=&-\frac1{16\pi}\int_{{}^3B}\da\l(\sqrt{-h}n_iE^i\r)A_0 = \frac1{16\pi}\int_{\phantom{1}^3B}\da\l(\frac{R^2fA'}{N}\r)A\,,\slabel{VarElecBoundaryTerm}
	\label{VarBoundaryTerms}
\esea
where we have used $A_0=A(r)$ in the second line. Then, going over to Euclidean signature, one identifies the Euclidean action $I=iS_{EM}$
\be
	\da(I) = \da(\ba H) = -\da(B_g)-\da(B_{em})\,, \qquad \textrm{{\it on-shell.}}
	\label{EuclActionVariation}
\ee
Let us apply this to Reissner-Nordstr\"om spacetime \eqref{RN} as an example (leaving aside for now the question of the inner boundary). We impose the following boundary conditions:
\bsea
	\da(N)&=&0\,,\\
	\da(R)&=&\da(R')\,,\\
	\da(f)&=&\sqrt{1-2\frac{m+\da(m)}{r}+\frac{q^2+2q\da(q)}{r^2}}-\sqrt{1-2\frac{m}{r}+\frac{q^2}{r^2}} \nn\\
		&=& \frac{1}{\sqrt{1-\frac{2m}r+\frac{q^2}{r^2}}}\l(-\frac{\da(m)}{r}+\frac{q\da(q)}{r^2}\r)\,,\\
	\da(A_0)&=&0\,.
\esea
The boundary conditions on the lapse and on the electric potential are a choice of physics: this means that we keep the temperature and the electric chemical potential fixed on the boundary. The boundary conditions on the warp factor $R(r)$ simply reflect the properties of the Reissner-Nordstr\"om black hole: the warp factor of the horizon does not depend on the black hole parameters\footnote{In Section \ref{section:ThermoEMD}, devoted to the thermodynamics of Einstein-Maxwell-Dilaton theories, we shall see explicit examples where it is not so}. Using Maxwell's equation,
\be
	\da\l(\sqrt{-h}\,n_iE^i\r)=2\da(q)\,,
\ee
one gets
\be
	\da(I) = \da(\ba H) = -\da (B_g)-\da(B_{em}) = \ba\da(m)-\ba\Phi\da\l(\frac q2\r)\,, 
\ee
and one may integrate, keeping the Euclidean periodicity $\ba$ fixed (and undetermined for now) as well as the electric potential at infinity $\Phi$:
\bea
	H= m - \Phi\frac q2 &\Rightarrow& M = m\,,   Q=\frac q2\,,
\eea
as expected from the usual ADM or Hawking-Horowitz expressions \eqref{GravitationalMassHH} and \eqref{ElectricChargeEM}. Note then that the action \eqref{EMActionGHY} as it stands defines a \emph{grand-canonical ensemble}, where all intensive quantities ($\ba$, which is to become the temperature on the inner boundary, and $\Phi$) are fixed, while all extensive quantities are varied and can be determined in terms of them (the gravitational mass $M$ and the electric charge $Q$)\footnote{For a reminder of thermodynamic definitions of variables and ensembles, as well as their corresponding stability, see Appendix \ref{section:ThermoStability}}.

We can now easily define the \emph{canonical ensemble}, as the ensemble where the Euclidean periodicity (inverse temperature) $\ba$ and the electric charge $Q$ are kept fixed. From \eqref{EMActionHamiltonianVariationSuperspace1Maxwell}, we see that it suffices to compensate the total variation in the first term by adding to the action this exact same boundary term unvaried. \eqref{EMActionHamiltonianVariationSuperspace} becomes
\be
	\l. \delta (S^c_{EM})\r|_{cl} = \frac1{4\pi}\int_{{}^3B}\l[RR'f\,\da(N)+(NR)'f\,\da(R)\r]- \frac1{16\pi}\int_{{}^3B}\da\l(\sqrt{-h}\,n_iE^i\r)A_0,
	\label{VariationEMActionCanonical}
\ee
which does amount to keeping the electric charge \eqref{ElectricChargeEM} fixed. In the alternate procedure by Regge and Teitelboim, the variation of the electric boundary term $\da\,B_{em}$ \eqref{VarElecBoundaryTerm} remains the same, except that now it will yield zero once the fixed charge requirement is taken into account. With both methods, one thus gets a Hamiltonian which has only pure gravitational boundary term,and which once integrated will be equal to the gravitational mass of the classical solution, as befits the canonical ensemble. To summarise, the action for the canonical ensemble is:
\be
	S^c_{EM}=\frac1{16\pi}\int_{\M}\sqrt{-g}\l(R-2\La-\frac14F^2\r)\mp\frac1{8\pi}\int_{\partial\M}\sqrt{\pm h}K+\frac1{16\pi}\int_{{}^3B} \sqrt{-h} F^{\mu\nu}n_\mu A_\nu\,.
	\label{EMActionCanonical}
\ee

Up till now, we have concentrated here on the case without inner boundary, and we see that the entropy of such systems must be zero, since the Euclideanised action on-shell for a static classical solution will simply be the classical value of the Hamiltonian \eqref{HamiltonianOnShellEM}. There is no entropic contribution, which is associated with inner boundaries with Killing horizons as we shall review in the next section.

\subsection{Black hole entropy}
\label{section:Entropy}

\subsubsection{Inner boundaries and Killing horizons}
\label{section:EntropyInnerBoundaries}

Let us now assume that the topology of our space is not simply $\Sigma\times\mathbf{R^1}$ (or $\Sigma\times\mathbf{S^1}$ in the Euclidean case), but rather  $\mathbf{R^2}\times\mathbf{S^2}$ (or $\mathbf{S^1}\times\mathbf{R^1}\times\mathbf{S^2}$ in the Euclidean case), so that black-hole topologies are allowed. In this derivation, we shall follow the procedure of Brown and York, \cite{Brown:1992bq}, and then connect with other definitions. For simplicity, we restrict to the purely gravitational case. These considerations are easily generalised to the Maxwell case, since any Maxwell boundary term does not contribute on a Killing horizon. 

Start with the Euclidean Einstein action in the canonical ensemble\footnote{As there is no Maxwell term, we do not define a grand-canonical ensemble.}:
\be
	I =-\frac1{16\pi} \int_{\mathcal M} \ud^{4}x\sqrt{g}\l(R-2\La\r)-\frac1{8\pi}\int_{{}^3B_o}\sqrt{h}\, K\,.
	\label{EinsteinActionCanonical}
\ee
Note that we have deliberately restricted the spatial boundary term to its outer component. We do not take into account boundary terms over the initial and final time-like surface which do not play any role here. Fixing the lapse to cancel on the inner boundary, $N(B_i)=0$, and the surface gravity to be constant there, \cite{Brown:1992bq}, fixes the variation of all fields on the inner boundary so that \eqref{VariationEMActionCanonical} is still valid. In Hamiltonian form, \eqref{EinsteinActionCanonical} becomes for a static spacetime of the form \eqref{EuclideanADMmetric}
\be
	 I = \int_{\M} N\mathcal H-\frac1{8\pi}\int_{{}^3B_o}\sqrt{\sig}N\l[k-\bar k\r] -\frac1{8\pi}\int_{{}^3B_i}\sqrt{\sig}Nn_\mu a^{\mu}\,.
	\label{EuclideanActionInnerBoundaryCanonical}
\ee
Since the reference spacetime has no inner boundary, it does not contribute there. The term on the inner boundary is proportional to the projection of the acceleration vector along the opposite normal to the inner boundary $n^\mu$ and we have
\be
	n_\mu a^{\mu}=-n_\mu u^\nu\nabla_\nu u^\mu = \frac{\partial_r N}{Nn}\,,
\ee
which differs from the Lorentzian definition \eqref{AccelerationNormalVector} by a minus sign. This explains why we kept this boundary term, since indeed it will not cancel as $N\to0$. Imposing regularity on the inner boundary by identifying the periodicity $\ba$ of the Euclidean time as in \eqref{EuclideanPeriod}, we find that
\be
	\l.I\r|_{cl} = \ba M_g -S\,, \qquad S = \frac14A_h = \frac14\int_{B_i} \sqrt{\sigma}\,,
	\label{IntegralInnerBoundary}
\ee
where the usual area law is recovered. Identifying the above formula with the partition function $\ba F$, the correct expressions for the energy and the entropy are recovered through  \eqref{EnergyExpectationValue} and  \eqref{EntropyExpectationValue} :
\be
	\langle E \rangle = \l.H\r|_{cl}=M_g\,, \qquad \langle S \rangle = \frac14A_h\,.
\ee
Moreover, although we have focused in particular on black-hole spacetimes, this argument carries over for any kind of Killing horizon (acceleration or cosmological, for instance, and being careful with subtleties linked to the infinite areas of the former).

Are the variations of the action \eqref{EinsteinActionCanonical} consistent with the boundary conditions? Varying \eqref{EinsteinActionCanonical} and keeping only the terms on the inner boundary, one finds 
\be
	\l.\da(I)\r|_{{^3}B_i} = \frac1{4\pi}\int_{{}^3B_i}\l[-RNf\da(R')-RR'N\da(f)+RN'f\da(R)\r]-\frac1{8\pi}\int_{{}^3B_i}\da\l(R^2fN'\r),
	\label{EinsteinActionCanonicalInnerBoundaryVariation}
\ee
where the first boundary term comes from the variation of the bulk term in  \eqref{EinsteinActionCanonical}, while the second one is the total variation of the inner boundary term in \eqref{EinsteinActionCanonical}. Simplifying the above, one gets
\be
	\l.\da(I)\r|_{{^3}B_i} = \frac1{4\pi}\int_{{}^3B_i}\l[-RNf\da\, R'-RR'N\da\,f-\half R^2\da(fN')\r],
	\label{EinsteinActionCanonicalInnerBoundaryVariation2}
\ee
evaluated on the boundary. However, on the inner Killing boundary, by definition the following boundary conditions hold
\be
	\l.N\r|_{{}^3B_i}=0\,, \qquad  \l.N'f\r|_{{}^3B_i}=\frac{2\pi}{\ba}=ct\,,
	\label{InnerBoundaryConditions}
\ee
where the first condition is simply that defining a Killing horizon while the second comes from regularising the conical singularity at the origin of the polar coordinates as prescribed in Section \ref{section:EuclideanPathIntegral}, formula \eqref{EuclideanPeriod}. So no further boundary conditions than the two above need to be imposed, and the variational problem is well-defined as it stands.

Following up on the Regge-Teitelboim approach, one finds that there is a gravitational variational term generated on the inner boundary, which reads
\be
	\l.\da B_g\r|_{{}^3B_i} = \frac1{4\pi}\int_{{}^3B_i}\l[-RNf\da(R')-RR'N\da(f)+RN'f\da(R)\r],
\ee
which, once \eqref{InnerBoundaryConditions} are imposed, reduces to
\be
	\l.\da B_g\r|_{{}^3B_i} =\frac\ba{4\pi}N'f\half\da\l[R^2\l({}^3B_i\r)\r]\int_{{}^3B_i}=\frac14\da(A_h)\,,
	\label{VarGravInnerBoundaryTerm}
\ee
where $A_h=4\pi R^2\l({}^3B_i\r)$ is the area of the Killing horizon and this yields once integrated the same result as that obtained by integral methods, \eqref{IntegralInnerBoundary}.

In passing, we note that with this approach, the extremal black holes do not contribute any entropy for a very simple reason: the zero of the Killing vector is degenerate on the inner boundary, so the surface gravity and the contribution of the acceleration vector in \eqref{EuclideanActionInnerBoundaryCanonical} both cancel. In effect, we recover the formula $I_{extremal}=\ba H$, which yields zero entropy through its partition function definition \eqref{EntropyExpectationValue}. In the next section, we discuss in greater detail this surprising result, which seems to contradict the extremal limit of the area law.

What happens if we do not fix the temperature of our thermodynamical system, but rather its entropy? This means going over to the microcanonical ensemble. However, in order to carry calculations in this ensemble, we would need to be able to count the available microstates needed to fix the entropy to a particular value. Unfortunately, this requires a quantum theory of gravity: some progress in the case of extremal (or so-called BPS) black holes have been made in some specific realizations of String Theory, \cite{Strominger:1996sh}, but this is as far as our knowledge extends for the present. For a review on the counting of microstates, one may consult \cite{Sen:2007qy}. Thus, we will leave aside the microcanonical ensemble in the following.

\subsubsection{The entropy of extreme black holes}
\label{section:EntropyExtremalBlackHoles}

In a now famous paper, \cite{Hawking:1994ii}, Hawking, Horowitz and Ross claimed that the entropy of the extremal Reissner-Nordstr\"om black holes should be zero. Though this seems in flagrant contradiction with the extremal limit $r_h\to r_e$ ($m=q$) taken from the usual quarter-of-the-horizon-area formula, they make a strong case based on topology arguments, which were then confirmed by Teitelboim by using the Hamiltonian formalism, \cite{Teitelboim:1994az}. Moreover, on the face of it, this would seem to be in agreement with the usual thermodynamic formulation  of the third law, that the entropy of a zero-temperature system should also be zero.

In a nutshell, the argument goes as follows, and has to do with the topology of spacetime one integrates over during the Euclidean saddle-point approximation procedure, \cite{Hawking:1994ii,Ross:2005sc}. It requires to evaluate the value of the Euclidean action at a classical solution, both in the bulk but also on boundary terms. For a Reissner-Nordstr\"om black-hole spacetime (though the argument also applies to other kinds of Killing horizons), there is both an inner (on the horizon) and an outer boundary (at asymptotic infinity $r\to\infty$). The Euclidean spacetime thus ranges over $r_+\leq r<+\infty$. There lies the catch. Consider flat Euclidean space, whose topology is $\mathbf S^1\times\mathbf R^1\times\mathbf S^2$,
\be
	\ud s^2 = \ud t^2 + \ud r^2 + r^2\ud\Omega^2_2\,.
\ee
There is a global Killing vector $\partial_t$, which vanishes nowhere over the range where this space is defined, $0\leq+\infty$. The Killing coordinate $t$ is periodic with period $\ba=2\pi$ and the origin at $r=0$ is perfectly well-defined, except for the usual collapsing of the two-sphere. Since the Killing symmetry is global, it is also a symmetry of the Euclidean action:
\be
	I = \int\ud t\int\ud^3x\mathcal L = \ba H\,
\ee
where the Hamiltonian $H=\int\ud^3x\mathcal L$ is $t$-independent. Thus, the entropy calculated from flat Euclidean space (taken as a saddle-point of the action) through the usual formula vanishes as expected for flat space:
\be
	S = \ba\frac{\partial I}{\partial \ba}-I = 0\,.
\ee	
Technically speaking, this is simply a consequence of the fact that the Euclidean action evaluated for flat space (or, as we saw in the previous section, for any space containing no non-degenerate Killing horizon) is simply proportional to the Euclidean periodicity.
This is quite different for a space with a Killing horizon, since at that point, the $S^1$ corresponding to the Killing isometry shrinks to zero. The Killing symmetry ceases to be global, and one cannot consider that the previous integration is $\ba$-independent yielding a term linear in $\ba$. 

As we saw in the previous section in equation \eqref{EuclideanActionInnerBoundaryCanonical}, there is a contribution on the Killing surface which does not vanish when evaluated on the non-degenerate Killing horizon. The non-degeneracy (e.g. the non-extremality) is a crucial hypothesis, since this is precisely why this term is unequivocally non-zero: when reduced with the generic form of the metric \eqref{EuclideanADMmetric}, all factors of $N$ cancel out, and only factors of $\partial_r N$ are left which are non-zero on the Killing surface if it is non-degenerate. 

Let us now consider the extremal black holes in two ways. First, take the extremal limit right from the start in the metric, which is equivalent to supposing that both $N$ and $\partial_r N$ vanish on $B_i$, but not $\partial_r^2 N$. Then one would be inclined to conclude that the whole inner boundary term in \eqref{EuclideanActionInnerBoundaryCanonical} vanishes \emph{a priori}, and no entropy contribution is generated: the partition function is proportional to $\ba$ and there is no prescription to fix the periodicity. Hawking \emph{et al.}, \cite{Hawking:1994ii} noted this fact and put forward a topological explanation for this: the extremal horizon is at an infinite proper distance from any point of space outside the horizon. This means that we should remove the point $r=r_e$ from the topology of the $\mathbf R^1$ $r$-direction, thus actually turning it into an $\mathbf S^1$. The overall Euclidean topology of the extremal black holes has now gone from $\mathbf S^1\times\mathbf R^1\times\mathbf S^2$ to $\mathbf S^1\times\mathbf S^1\times\mathbf S^2$. Since we have removed the inner boundary to an internal infinity, there is no need for a prescription to close off the hole in the topology, where the Euclidean time $\mathbf S^1$ vanished and the foliation along the time direction broke down. Since this locus is now down an infinite throat, it can be removed from the topology and no inner boundary contribution needs be taken into account: the extremal black holes have zero entropy\dots Or, do they?

This proposal raised quite a lot of controversy at the time it was published, since it seemed to point out a contradiction both with the extremal semi-classical limit taken from non-extremal black holes, and also with String Theory microstates counting for extremal black holes, \cite{Strominger:1996sh}, which yielded the same finite result as the former. Some amount of understanding comes from careful analysis of the commutativity of the operations of taking the extremal limit and quantising the theory by evaluating the path integral, \cite{Ghosh:1996gp,Kiefer:1998rr}. This can readily be seen in the previous formul\ae. Indeed, taking a look at \eqref{EuclideanActionInnerBoundaryCanonical}, it is possible to first evaluate the inner boundary contribution and then take the extremal limit between the expectation values of the gravitational mass and electric charge, $\langle E \rangle = \langle Q \rangle$. Then the entropy of extremal black hole is non-zero and is simply the (finite) area of the degenerate Killing horizon.

It seems that summing over all the possible topologies in the path integral (and not simply the asymptotically flat ones) allows for a consistent picture. This means also integrating over the Bertotti-Robinson topologies, which are not asymptotically flat and have topology $\mathbf{AdS}_2\times\mathbf S^2$, \cite{Carroll:2009maa}. This implies that the extreme black holes lie separately from the non-extreme black holes. Though this may seem consistent with results from physical processes trying to lower the surface gravity of a charged black hole to zero (this happens in infinite advanced time), other results examining the Hawking radiation of both extreme and non-extreme black holes and the related energy fluxes show that the extremal limit is perfectly continuous, \cite{Balbinot:2007kr}. Taking a stand on this thorny matter lies well beyond the scope of this work. However, in the Section \ref{section:ThermoEMD}, the addition of scalars in the game shall bring about an interesting twist.

\subsection{Thermodynamics of black holes in Einstein-Maxwell theory}

\label{section:ThermoBHGR}

\subsubsection{Thermodynamics of Schwarzschild black holes}
\label{section:ThermoSchw}
Let us start by considering the situation for the Schwarzschild spacetime \eqref{Schwarzschild}, \cite{Hawking:1976de}. The temperature of the solution is readily computed to be 
\be
	T=\frac1{8\pi M}\,,
	\label{TemperatureSchwarzschild}
\ee
where $M=m$ is the gravitational mass, which can be calculated using the various formul\ae described in Section \ref{section:PartitionFunctionEM}, see  \eqref{GravitationalMassHH} or \eqref{MassCanonical}. The partition function is determined in the canonical ensemble to be
\be
	W = M-TS = \frac M2\,,
	\label{HelmholtzSchwarzschild}
\ee
and the entropy is as usual equal to the quarter of the area of the horizon. One easily calculates the heat capacity \eqref{HeatCapacity},
\be
	C = -8\pi M^2 <0\,,
\ee
so that the Schwarzschild black hole is unstable both globally and locally in the canonical ensemble. Physically, it means that if the black hole is in contact with a reservoir containing some blackbody radiation at higher temperature, then the temperature of the hole decreases as matter is absorbed by the hole. As the temperature goes down, so does the rate of Hawking emission, which worsens the imbalance between absorption and emission even further. The temperature continues to drop down and the hole grows without bounds.

On the other hand, if the temperature of the reservoir is initially lower than the hole's, then the emission rate by Hawking radiation is greater than the absorption rate. Then, as energy is being radiated out of the hole, the temperature grows, which again makes the emission rate increase, so that the hole completely evaporates.

Finally, what happens if we start at equilibrium, with a hole at the same temperature as the reservoir? Some statistical fluctuation in the emission or absorption rate is bound to occur, which then brings us back to the two previous arguments.

One way to make the hole stable is to put it in a box, \cite{Hawking:1976de}. Then, there is no infinite reservoir with which it can exchange energy. The box containing the hole contains also a finite quantity of energy, $E_{tot}=E_h+E_{outside}$, which is constant. Maximising the total entropy $S_{tot}=S_h+S_{outside}$ gives the following constraints,
\bsea
	T_h&=&T_{outside}\,,\\
	T_{outside}\frac{\partial E_{outside}}{\partial T_{outside}}&<&E_h\,.
\esea
If we suppose that the outside of the hole contains only gravitons with a blackbody thermal spectrum, then $E_{outside}\sim T_{outside}^{-4}$ by Stefan-Boltzman law, and this yields the constraint 
\be
	E_{outside}<\frac14E_h\,,
\ee
which translates as a higher bound on the volume of the box. When it is satisfied, the hole is in thermal equilibrium with the blackbody gravitons in the box.

These results can be recovered by applying a Hamiltonian-type thermodynamic analysis for the Schwarzschild blak hole, \cite{Louko:1994tv}. In a nutshell, the partition function is computed after quantising the Lorentzian Hamiltonian and including appropriate boundary terms to make the variational problem well-defined. However, one realises that without the inclusion of an Infra-Red regulator, the Euclidean action diverges and, although this prevents in no way the identification of saddle points, does make the canonical ensemble ill-defined. Joined with the thermodynamic instability of Schwarzschild solution in such an ensemble, this seals the coffin. However, putting the regulator back is equivalent physically to enclose the black hole in a box and allows to render it stable if the box is small enough (but not so small as to incur gravitational collapse.

This whole procedure is however not very well-defined, since putting the black hole in a box of constant total energy actually requires to work in the microcanonical ensemble, where one should count the black-hole microstates. In order to do this, one needs to know a consistent way to quantise gravity. Such a theory of quantum gravity is not (yet) available, though string theory does constitute a promising candidate. Progress has been made over the last decade in counting the microstates of certain classes of ``BPS'' black holes, \cite{Strominger:1996sh}, recovering the famous result that the microscopic entropy is indeed equal to the quarter of the area of the horizon (classically, since quantum corrections are expected to provide corrections to this result).

\subsubsection{Thermodynamics of Reissner-Nordstr\"om black holes}
\label{section:ThermoRN}

\begin{figure}[!ht]
\begin{center}
\begin{tabular}{c}
	 \includegraphics[width=0.45\textwidth]{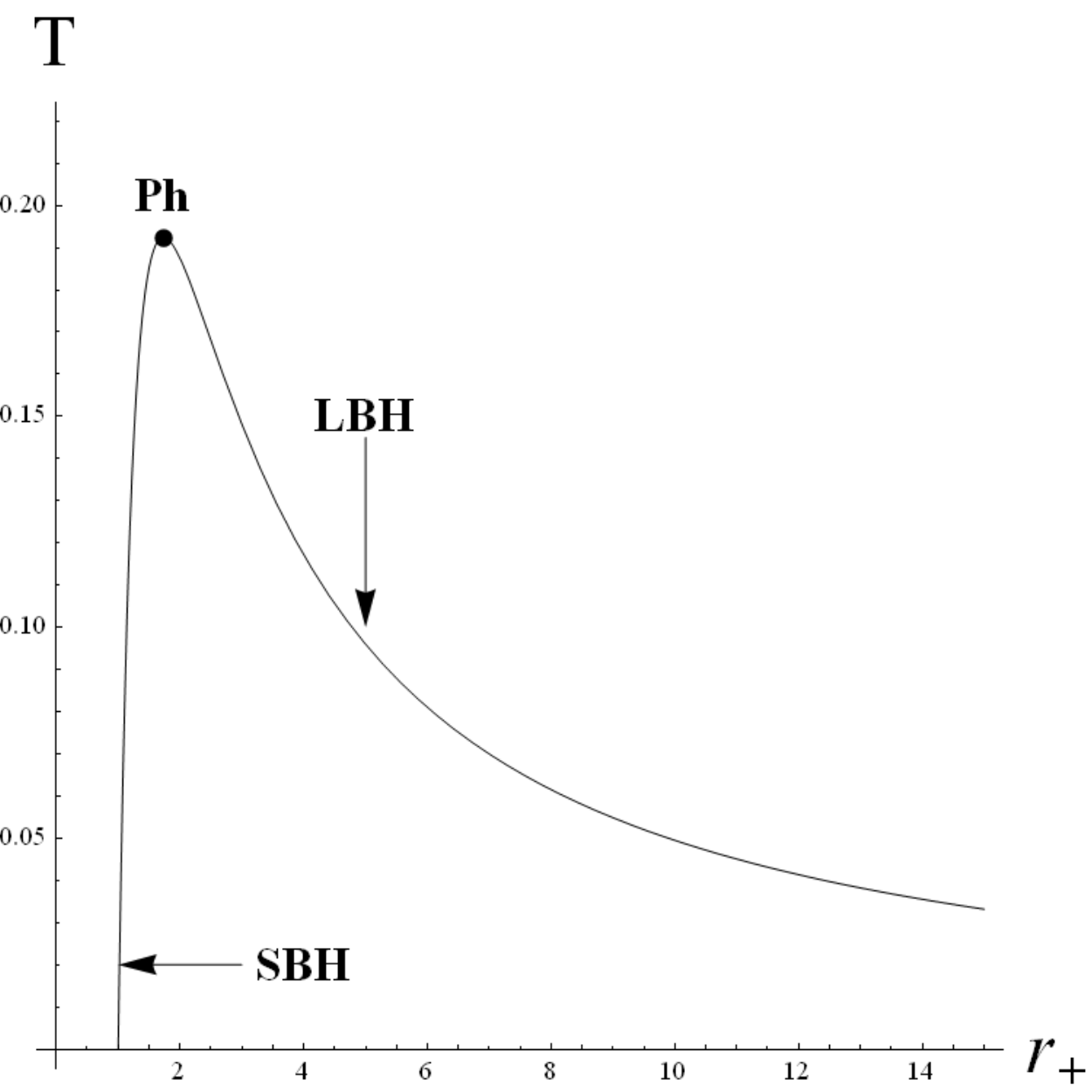}
 \end{tabular}
\caption[Equation of state at fixed charge for the Reissner-Nordstr\"om black holes]{Equation of state $T(r_+,Q)$ at fixed $Q$ for the Reissner-Nordstr\"om black holes.}
\label{Fig:EqOfStateRNCanonical}
\end{center}
\end{figure}

For conciseness, we will only examine the canonical ensemble, that is we will compare solutions with the same temperature and electric charge. For the Reissner-Nordsrt\"om solution, we find
\be
	T = \frac{\pi}{r_+}\l(1-\frac{q^2}{r_+^2}\r),
	\label{TemperatureRN}
\ee
keeping in mind that
\be
	r_+ = m+\sqrt{m^2-q^2}\,,\qquad r_+r_-=q^2\,,
\ee
where $M=m$ and $Q=q/2$ are the gravitational mass and electric charge derived from the solutions \eqref{RN}. The equation of state is an implicit equation $T(r_+,Q)$ and can be plotted, revealing two branches of black holes, small and large ones (see \Figref{Fig:EqOfStateRNCanonical}). Moreover, these black holes exist only below some maximal temperature. Beyond that, only the charged background solution exists. This is already very different from Schwarzschild solution! The small black holes end on the extremal radius at zero temperature, while the large black holes have infinite radius at zero temperature. From this, one can note that the small black holes are truly specific of the charged case, while the large ones are Schwarzschild-like. We will see by examining the free energy and the heat capacity that the comparison can be pushed further.

\begin{figure}[!ht]
\begin{center}
\begin{tabular}{cc}
	 \includegraphics[width=0.45\textwidth]{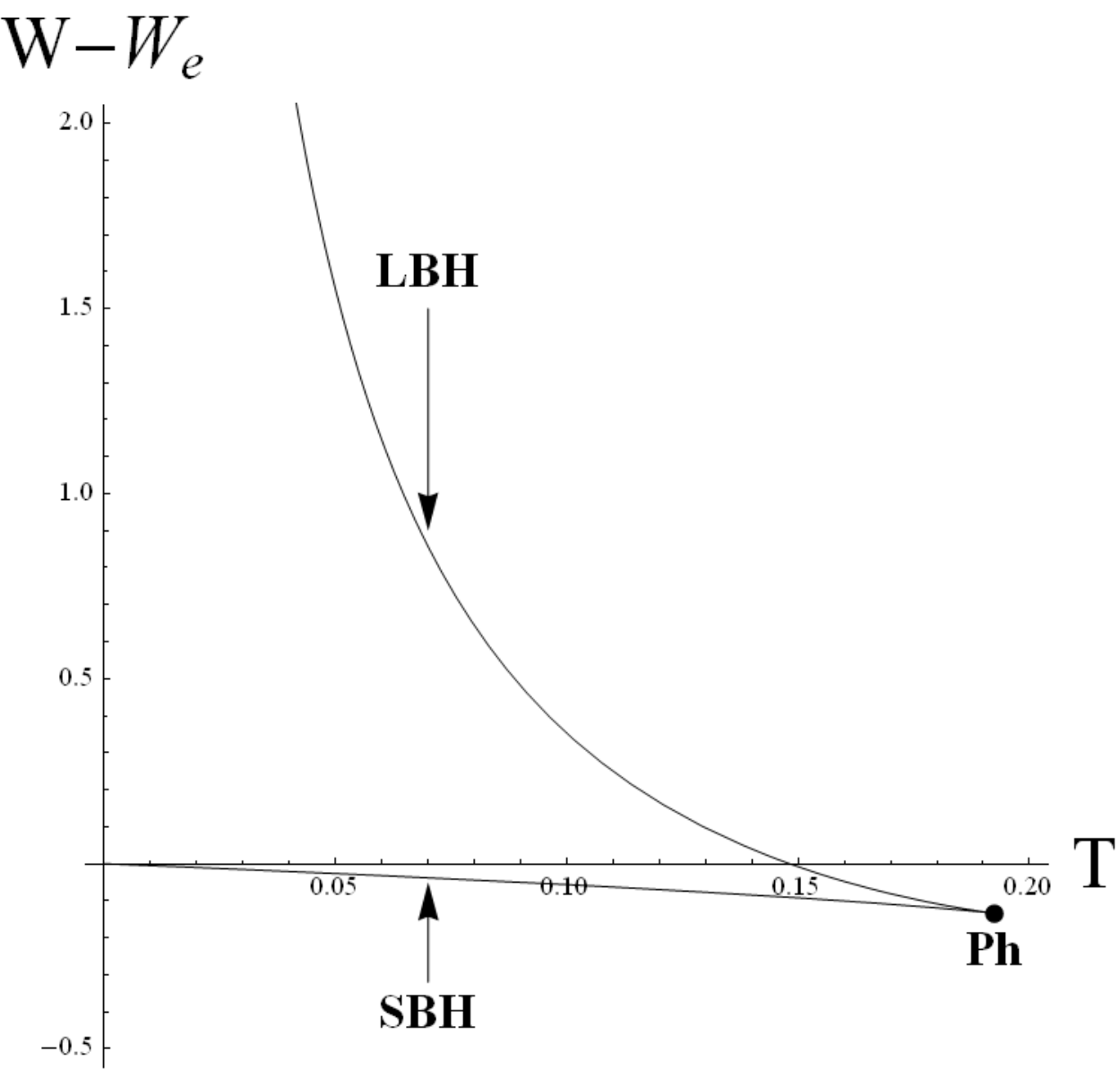}&
	 \includegraphics[width=0.45\textwidth]{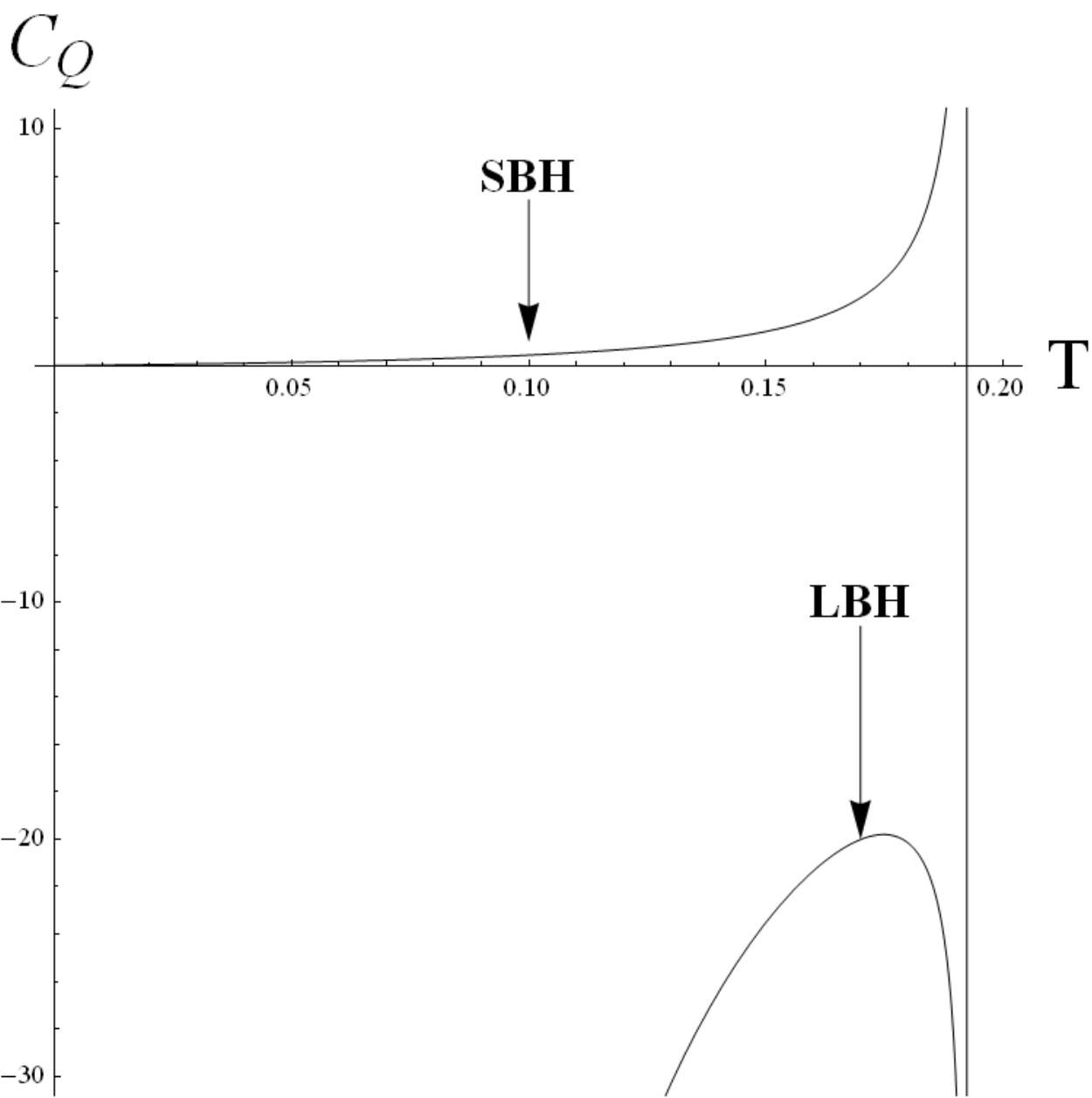}
 \end{tabular}
\caption[Thermal stability of the Reissner-Nordstr\"om black holes]{Thermal stability of the Reissner-Nordstr\"om black holes: on the left pannel, the Helmholtz free energy against the temperature, compared to the extremal charged competing background; on the right, the heat capacity.}
\label{Fig:ThermalStabilityRNCanonical}
\end{center}
\end{figure}

The Helmholtz free energy is
\be
	W = m-\half\sqrt{m^2-q^2} = \frac{r_+}4\l(1+3\frac{q^2}{r_+^2}\r)\,.
	\label{HelmholtzRNCanonical}
\ee
and shows, see the left pannel of \Figref{Fig:ThermalStabilityRNCanonical}, that the small black holes are globally stable when compared to the competing solution, that is the charged extremal black holes. The Gibbs potential in the grand-canonical ensemble is
\be
	G=\half\sqrt{m^2-q^2}\geq0\,,
\ee
so the charged Reissner-Nordst\"om black holes are never a global minimum of the partition function in this ensemble. 
For more considerations on charged black holes in the grand-canonical ensemble, see for example \cite{Braden:1990hw}.

The heat capacity at fixed charge, \eqref{HeatCapacity},
\be
	C_Q = -\pi r_+^2\frac{1-\frac{q^2}{r_+^2}}{1-3\frac{q^2}{r_+^2}}
	\label{HeatCapacityRNCanonical}
\ee
 shows that the small black hole branch is also locally stable with respect to thermal fluctuations, see the right pannel of \Figref{Fig:ThermalStabilityRNCanonical}. This is a well known result by Davies, \cite{Davies:1978mf}, and shows that in the parameter range $3m^2/4<q^2<m^2$, the Reissner-Nordstr\"om black holes are stable against decay by Hawking radiation if the canonical ensemble can be defined.
Turning to the local electric stability, we find that the electric permittivity \eqref{ElectricPermittivityCanonical} is
\be
	\eps_Q = \frac2{r_+}\frac{r_+^2-q^2}{r_+^2-3q^2}\,,
	\label{ElectricPermittivityRNCanonical}
\ee
which has opposite sign compared to the heat capacity \eqref{HeatCapacityRNCanonical} and thus shows that the large black holes are stable to electric fluctuations while the small ones are not. This result is precisely opposite to that for thermal fluctuations.

At this point, we find that there exists an ambiguity in the literature. Indeed, this analysis makes sense if and only if the canonical ensemble can be defined. On the one hand, the above analysis makes implicit use of the fact that we are using extremal black holes as a background. How is this consistent, given that they have zero temperature by taking the extremal limit from \eqref{TemperatureRN}? Well, it is argued in \cite{Hawking:1994ii} that, since the inner boundary (e.g., the extremal horizon) is at infinite proper distance from a given point of Euclidean spacetime, then there is no inner boundary in the Euclidean path integral computation of the partition function but an infinite throat in its stead and so no periodicity can be ascribed to the Euclideanised time. Thus, it can be chosen at will as is usual for a spacetime without any inner boundary, and \cite{Chamblin:1999tk,Chamblin:1999hg} then argue that this allows to use the extremal black hole as a background. All the charge is localised inside the extremal horizon, and only thermal neutral quanta are left free to fluctuate in the reservoir. The thermodynamic stability results then follow. Although these authors, \cite{Chamblin:1999tk,Chamblin:1999hg}, make their case for the AdS charged black holes (see below), we see no reason why this line of reasoning should not be valid here too. We shall come back to this issue in greater detail in Section \ref{section:EntropyExtremalBlackHoles}.

On the other hand, Hamiltonian thermodynamic analysis indicates that, just as Schwarzschild, Reissner-Nordstr\"om  black holes with asymptotically flat boundary conditions do not allow a consistent definition of a (grand-)canonical ensemble, see \cite{Braden:1990hw,Louko:1996dw}. The point in this case is even sharper than for Schwarzschild, since we did find a region of thermal stability in the canonical ensemble $3m^2/4<q^2<m^2$. However the same kind of divergence occurs as in the neutral case, and the partition function diverges as the regulator of the Euclidean action is sent to infinity. The same cure as before is possible, that is to put the hole in a box, provided one knows how to count microstates. We shall now turn to a way to evade this problem, which is to examine (charged) black holes in a setup with a negative cosmological constant and AdS boundary conditions. The negative cosmological constant generates some kind of attractive gravitational potential at large distances, so that it simulates the effect of a box confining gravitational energy and allows a consistent definition of the thermodynamic ensembles. This is actually a better solution than putting a box in flat spacetime, since this allows to confine gravitons in a virtual AdS box\footnote{This statement is not quite correct: in truth, gravitons (and other massless particles) can escape to infinity, but there is an equal ingoing flux from infinity since AdS is not globally hyperbolic, \cite{Hawking:1973uf}.}, whereas this difficulty cannot be easily resolved in flat space, in which the gravitons can \emph{a priori} escape the box.

\subsubsection{Thermodynamics of Reissner-Nordstr\"om black holes in Anti-de Sitter spacetime}
\label{section:ThermoRNAdS}

Let us review the case of Einstein-Maxwell theory supplemented by a negative cosmological constant. Let us begin by noting that the previous definitions in Sections \ref{section:PartitionFunctionEM} and \ref{section:DefEnsemblesEM} are easily generalised to the case with a cosmological constant since this addition does not entail any extra boundary term, and it should now be clear that these boundary terms are what controls how the ensembles are defined.

It is well-known that negative cosmological constant gives rise to an attractive gravitational potential at large distances, so we may expect intuitively that the stability and definiteness of the canonical ensemble will be improved with regards to the zero cosmological constant results. The effect of the negative cosmological constant should be to simulate a box and stabilise the ensemble. The thermodynamics of neutral black holes in AdS space were first studied by Hawking and Page in \cite{Hawking:1982dh}. They found that there exists a minimum temperature $T_{min}$ above which two branches of black holes, small and large, exist, the latter of which are thermodynamically locally stable. The small black holes resemble Schwarzschild black holes, while the large ones are specific to AdS. Moreover, above a given temperature $T_{HP}>T_{min}$, the large ``AdS'' black holes are found to be thermodynamically stable compared to AdS space with thermal radiation, so that there is a first-order phase transition from the latter to the former. Thus, they conclude that the canonical ensemble at fixed temperature can be defined. As we saw earlier in Section \ref{section:ThermoRN}, this is going a little too fast as thermodynamic stability does not necessarily garantee that the ensemble can be defined. As pointed by Louko and Winters-Hilt in \cite{Louko:1996dw}, one also has to make sure that the integral of the Euclidean partition function converges. This is not the case for asymptotically flat boundary conditions, \cite{Louko:1994tv,Louko:1996dw} but can be remedied by putting a regulator (a box) or imposing different boundary conditions, such as AdS boundary conditions, \cite{Hawking:1982dh,York:1986it,Brown:1989fa,Braden:1990hw,Louko:1996dw}.

When charge is added, both the grand-canonical and the canonical ensemble can also be defined consistently, \cite{Braden:1990hw,Louko:1996dw}. We begin by reviewing black holes with spherical topology on the horizon, \cite{Louko:1996dw,Cvetic:1999ne,Chamblin:1999tk,Chamblin:1999hg}. In the grand-canonical ensemble, two regimes must be distinguished, low and high chemical potential. 
\begin{itemize}
	\item For low chemical potential $\Phi<\Phi_c$, the situation is similar to the neutral limit $\Phi=0$ (which is the lower bound on this regime): on top of thermal AdS, there are two branches of black holes with small and large radius for $T>T_{min}$,  the latter stable and the former unstable. Moreover, for $T>T_{HP}>T_{min}$, the large black holes dominate the phase space.
	\\
	\item For large chemical potential $\Phi\geq\Phi_c$, there is a single black-hole branch, locally stable to thermal fluctuations, which dominate the phase space at all temperatures. It also does so at zero temperature, so that the extremal black holes are still a global minimum of the Gibbs free energy, but they turn out to be unstable. This case is quite different from the neutral case, as can be expected since it is disconnected from it by the lower bound on the chemical potential.
\end{itemize}
In both cases, the dominant black hole phase is locally stable, both thermally and electrically.

\begin{figure}[t]
\begin{center}
\begin{tabular}{cc}
	 \includegraphics[width=0.45\textwidth]{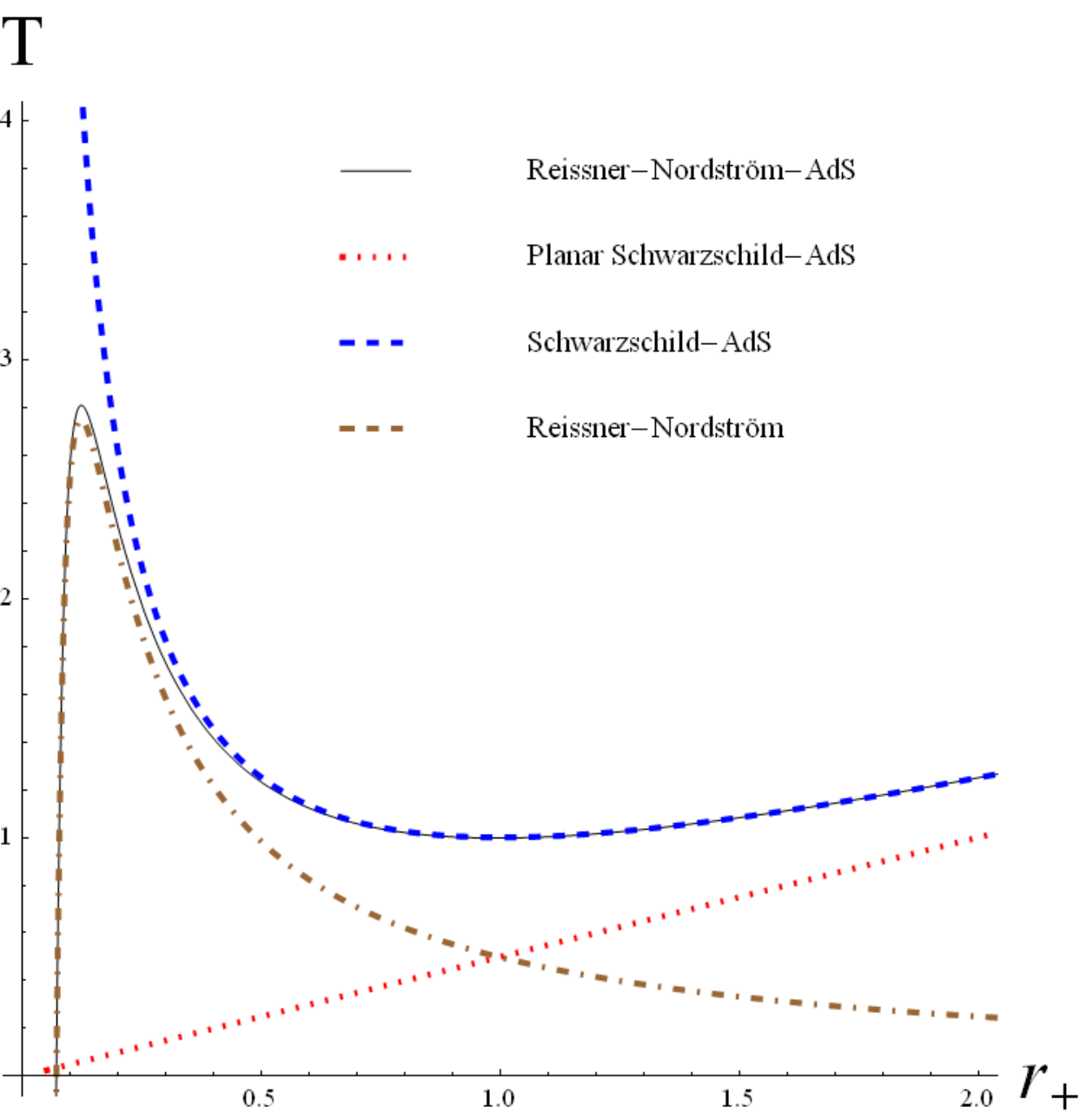}&
	 \includegraphics[width=0.45\textwidth]{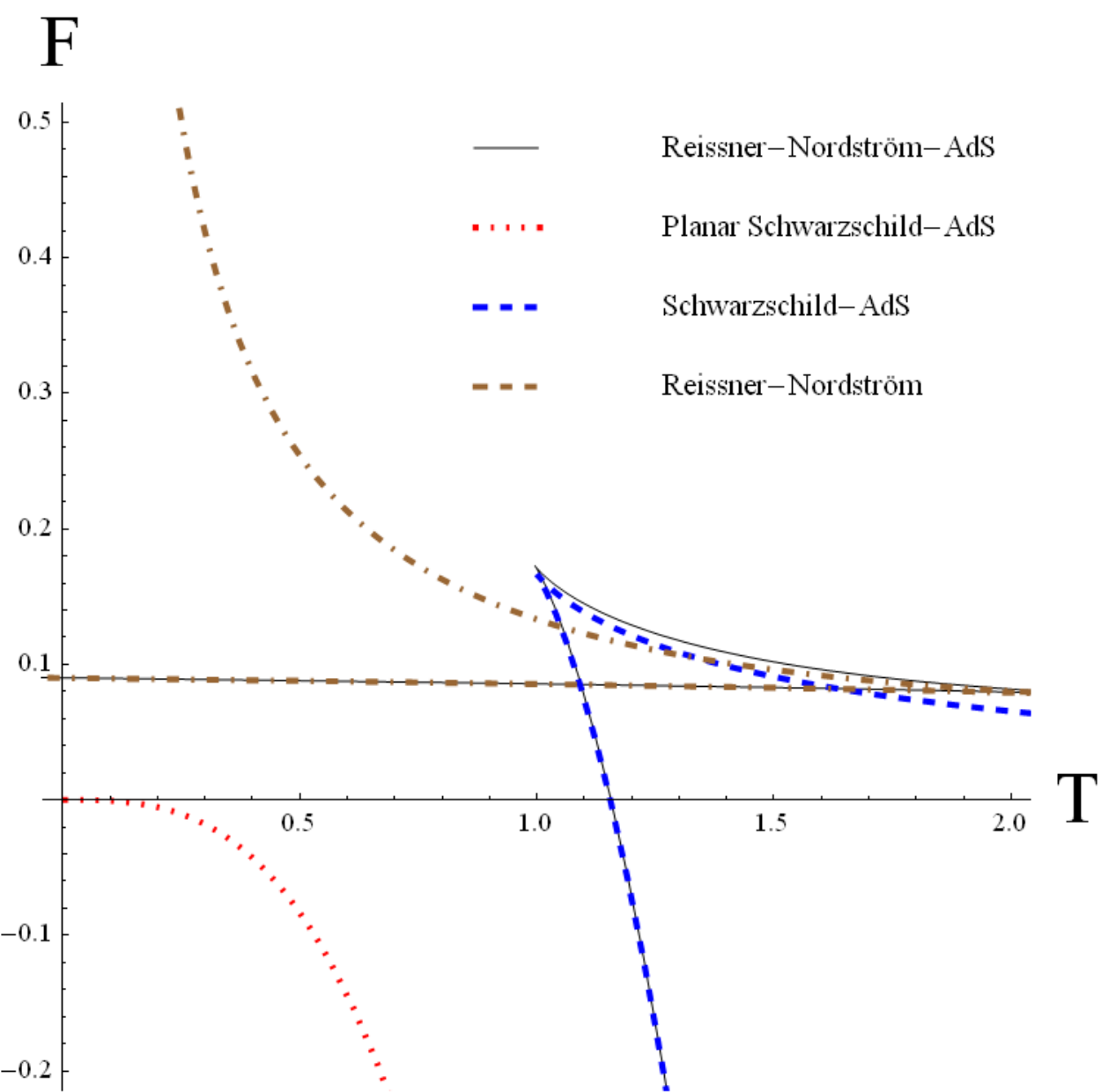}
 \end{tabular}
\caption[Thermodynamics of the Reissner-Nordstr\"om Anti-de Sitter black holes in the canonical ensemble]{These plots summarise the thermodynamics of the Reissner-Nordstr\"om Anti-de Sitter black holes in the canonical ensemble, for low charge. On the left pannel, the equation of state at fixed charge, with the spherical Reissner-Nordstr\"om (brown, dashed-dotted), Schwarzschild Anti-de Sitter (blue, dashed), Reissner-Nordstr\"om Anti-de Sitter (black, solid) and planar Schwarzschild Anti-de Sitter (red, dotted). On the right pannel, the Helmholtz free energy against the temperature at fixed charge, compared to the extremal charged competing background, with the same color and curve conventions.}
\label{Fig:ThermodynamicsRNAdSCanonical}
\end{center}
\end{figure}

We now turn to the the result for the canonical ensemble, for which two regimes can again be distinguished:
\begin{itemize}
	\item For low charge $Q<Q_c$, one finds three branches of black holes. The large and medium black holes correspond respectively to the large ``AdS'' and small ``Schwarzschild'' black holes of the neutral Schwarzschild-AdS case. Correspondingly, the large ones are locally thermally stable while the medium ones are not. On the other hand, the small and medium ones can be identified to the small ``Reissner-Nordst\"om'' and large ``Schwarzschild'' black holes of the zero cosmological constant charged black holes, and correspondingly the small ones are locally thermally stable. 
	
	The medium branch always has positive free energy and so can be dismissed, it is never a global minimum. However, remember that both the small charged flat Reissner-Nordst\"om black holes and the large neutral Schwarzschild-AdS black holes were global minima of the free energy for some range of temperature $T<T_{max}$ and $T>T_{HP}$ respectively. For charged AdS black holes, we thus have a combination of these two effects: there is a globally and locally (thermally) stable black-hole branch at all temperatures. At low temperatures, the small ``Reissner-Nordst\"om'' black holes dominate while at high temperatures, the large ``AdS'' black holes do, and the ``Schwarzschild'' black holes are never stable. There is a first-order phase transition between the small and large black holes. Regarding electric stability, the medium black holes are stable, while the small and large black holes are only stable in a given area.
	
	These results are summarised on \Figref{Fig:ThermodynamicsRNAdSCanonical}, where each of the three branches can be connected to a precise feature of the black hole: the small black holes are generated by the electric charge, the medium black holes by the non-zero positive curvature of the horizon, and the large black holes by the non-zero negative cosmological constant.
	\\
	\item For large charge $Q\geq Q_c$, there is a single locally thermally stable black-hole branch and a situation analogous to large potential, the black-hole phase dominating the phase space at all temperatures.\\
\end{itemize}

Finally, we address the thermodynamics of topological charged black holes in AdS spacetime, as discussed in Section \ref{section:TopologicalAdSBH}. The thermodynamics of such solutions were first reported by Brill, Louko and Peld\`an in \cite{Brill:1997mf} (though see \cite{Vanzo:1997gw} for the uncharged case). They found that both in the grand-canonical and in the canonical ensembles, only one black-hole branch survives, which is both a global and local (thermal as well as electric) minimum of the free energy.

\vfill
\pagebreak

\begin{subappendices}

\subsection{Appendix: Equivalent calculations of the temperature associated to a Killing horizon \label{section:TempCalculation}}

Let us show explicitly that the Hawking temperature \eqref{HawkingTemperature} calculated from the expression of the surface gravity \eqref{SurfaceGravity} in subsection \ref{section:BHThermoLaws} does coincide with the inverse periodicity of the Euclideanised time-coordinate defined in subsection \ref{section:EuclideanPathIntegral},
\be
	T_H=\ba^{-1}\,.
\ee

Let us take a $3+1$ Euclidean ADM Ansatz,
\be
	\ud s^2 = N^2(r)\ud t^2 +n^2(r)\ud r^2 + r^2\ud \Omega^{2}_2\,.
	\label{EuclideanADMmetric}
\ee
The time-like Killing vector is $\chi^\mu=\da^\mu_0$ and it should cancel at some Killing horizon\footnote{Note that we do not specify the nature of the horizon, black-hole, cosmological or acceleration.} $r_h$, 
\be
	\l.\chi^2\r|_h=N^2(r_h)=0\,.
\ee
Then, we may go over to a system of coordinates where the near-horizon $\mathbf R^2$-part of the Euclidean (black-hole) spacetime will described by polar coordinates by choosing
\be
	\eps = \int_{r_h}^r n(\rho)\ud\rho\, \Longrightarrow \ud\eps = n(r)\ud r\,,
\ee
so that the horizon is reached for the origin of the $\eps$-coordinate $\eps=0$. The $\mathbf R^2$-part of the metric then takes the form
\be
	\ud s^2 = \tilde N^2(\eps)\ud t^2 + \ud\eps^2\,, \qquad \tilde N(\eps) = N(r)\,.
\ee
We expand around the horizon position $r\to r_h$, or equivalently $\eps\to 0$,
\bea
	\tilde N(\eps) &\underset{\eps\to0}{\sim}& \tilde N(0) + \eps \l.\frac{\partial\tilde N}{\partial\eps}\r|_{\eps=0} + O(\eps^2)\nn\\
								&\sim& N(r_h) + \eps \l.\l(\frac{\partial r}{\partial\eps}\frac{\partial N}{\partial r}\r)\r|_{r=r_h} + O(\eps^2)\nn\\
								&\sim& \eps \l.\l(\frac{N'}{n}\r)\r|_{r=r_h} + O(\eps^2)\,.
\eea
and find close to the horizon
\be
	\ud s^2  \sim \l.\l(\frac{N'}{n}\r)\r|_h^2\eps^2\ud t^2+\ud\eps^2\,,
\ee
which is indeed in polar coordinates. The metric will be regular (no conical singularity) at the polar origin if the angular variable has period $2\pi$
\be
	\Delta\tau = \Delta t\l.\l(\frac{N'}{n}\r)\r|_h = 2\pi\,,
\ee
which implies that the $t$-coordinate has period 
\be
	\ba=2\pi\l.\l(\frac{n}{\partial_r N}\r)\r|_h\,.
	\label{EuclideanPeriod}
\ee
This latter expression can straightforwardly be seen to correspond to the inverse Hawking temperature \eqref{HawkingTemperature} by computing the surface gravity for the Killing horizon \eqref{SurfaceGravity}.

\subsection{Appendix: Global versus local thermodynamic stability}

\label{section:ThermoStability}

First, we recall the definition of an \emph{intensive} versus an {extensive} variable. In a homogeneous system, \emph{intensive} parameters such as the temperature, the electric chemical potential or the pressure do not vary when the volume of the system does. On the other hand, \emph{extensive} parameters such as the energy, the electric charge or the volume are all proportional to the volume of the system and vary with it. Which combination of intensive and extensive variables are fixed will then determine the kind of thermodynamic ensemble one is considering.

Let us now define the grand-canonical ensemble: we use only intensive quantities as parameters, that is the gauge potential at infinity and the temperature in the Einstein-Maxwell case. The other extensive variables $S$ and $Q$ are determined as functions of them. The Gibbs thermodynamical potential in the grand-canonical ensemble is
\be
	G[T,\Phi] =  M - \Phi Q - TS\,,
	\label{ThermPotentialGrandCanonical}
\ee
and is related for a particular solution to the Euclidean continuation of the Lorentzian action by $I-\bar I=\beta G$ where $\bar I$ is the background regularisation. The first law is expressed (in the micro-canonical ensemble) as:
\be
	\ud S =\ba \ud M- \ba\Phi\ud Q\,,
	\label{FirstLaw}
\ee
and is consistent with the expression in the grand-canonical ensemble:
\be
	\ud G = -S\,\ud T -Q\,\ud\Phi\,.
	\label{FirstLawGibbs}
\ee
This confirms that it is indeed the temperature and chemical potential which parameterise the ensemble.
We can then deduce the other thermodynamic quantities of the solution by taking the appropriate derivatives of the Gibbs potential with respect to the thermodynamic variables,
\bea
	M&=&G-T\frac{\partial G}{\partial T}\Big|_\Phi -\Phi\frac{\partial G}{\partial \Phi}\Big|_T\,, \label{MassGrandCanonical}\\
	S&=&-\frac{\partial G}{\partial T}\Big|_\Phi \label{EntropyGrandCanonical}\,,\\
	Q&=&\frac{\partial G}{\partial\Phi}\Big|_T \label{ElectricChargeGrandCanonical}\,.
\eea

There are two checks to perform in order to evaluate the stability of a given solution in the ensemble: a global and a local one. First, one has to verify if this solution corresponds to a global minimum of the phase space of solutions. Since they are classical solutions, we know they correspond to saddle points, that is local minima. But there might be several such minima, that is several such competing solutions. For instance, in Anti-de Sitter spacetime, there is a competition between the AdS-Schwarzschild black hole and the AdS background, giving rise to the famous Hawking-Page phase transition (see Section \ref{section:ThermoRNAdS}).

In order to determine this global minimum, one simply compares the value of the thermodynamic potential for each competing solution and identifies the lowest one. This saddle point is then \emph{globally thermodynamically stable}.

Let us describe a little more in detail how this happens. From the second law of thermodynamics, one will aim to maximise the entropy of the (whole) system (black hole plus reservoir, for instance). Equivalently, one will try to minimise the free energy of the system. In the grand-canonical ensemble, the thermodynamic system (black hole), with temperature $T$ and chemical potential $\Phi$, is connected to a reservoir $\l(T_{res},\Phi_{res}\r)$ with which it exchanges gravitational energy ($\Delta M\neq0$) and electric charge ($\Delta Q\neq0$). By conservation of mass and charge, one should have under some transformation 
\be
	\Delta\l(M\r)=-\Delta\l(M_{res}\r),\qquad \Delta\l(Q\r)=-\Delta\l(Q_{res}\r).
\ee
Now, recall that the second law states that the entropy of the whole system (black hole plus reservoir) should always increase, so that equilibrium will be reached for a maximum of the entropy $S\l(M,Q\r)$,
\be
	\Delta S + \Delta S_{res}\geq 0\,.
\ee
The maximisation condition implies, together with the first law $\Delta M = T\Delta S+\Phi\Delta Q$ and conservation of total mass and charge, that
\be
	\frac{\Delta S}{\Delta M} = \frac{\Delta S_{res}}{\Delta M_{res}}\,,\qquad \frac{\Delta S}{\Delta Q} = \frac{\Delta S_{res}}{\Delta Q_{res}}\,,
\ee
or equivalently,
\be
	T=T_{res}\,, \qquad \Phi=\Phi_{res}\,.
\ee
These are the usual thermodynamic equilibrium conditions.
Using these, we can express the variations in the Gibbs potential,
\bsea
	\Delta G&=&\Delta M-T\Delta S-\Phi\Delta Q\,,\\
	\Delta G_{res}&=&-\Delta M-T\Delta S_{res}+\Phi\Delta Q=0\,,
\esea
where the second equality follows from the definition of a reservoir, which by nature cannot be disturbed from equilibrium by the system connected to it. Adding these two equations and using the second law, one finds
\be
	\Delta G\leq 0\,,
\ee
which proves that global equilibrium is attained at a \emph{global minimum of the Gibbs thermodynamic potential} of the system, that is the competing spacetime solution for which it is lowest is the global minimum the phase space of solutions.

Let us now examine the canonical ensemble, where the appropriate thermodynamic potential is the Helmholtz potential,
\be
	W[T,Q] = M - TS\,,
	\label{FreeEnergyCanonical}
\ee
related to the Euclidean action by $\beta W = I_c -\bar I_c$. Defined in this way, the thermodynamic potential for the canonical ensemble is as expected the Legendre transform of the thermodynamic potential for the grand-canonical ensemble \eqref{ThermPotentialGrandCanonical}. The first law in the canonical ensemble is
\be
  \ud W = -S\,\ud T + \Phi\,\ud Q\,.
	\label{FirstLawCanonical}
\ee
We can then deduce the other thermodynamic quantities of the solution by taking the appropriate derivatives of the Helmholtz potential with respect to the temperature and electric charge,
\bea
	M_W&=&W-T\frac{\partial W}{\partial T}\Big|_Q \label{MassCanonical}\,,\\
	S_W&=&-\frac{\partial W}{\partial T}\Big|_Q \label{EntropyCanonical}\,,\\
	\Phi_W&=&\frac{\partial W}{\partial Q}\Big|_T \label{ElectricPotentialCanonical}\,.
\eea

Let us now examine the stability of a solution in this ensemble. In this case, the system can only exchange gravitational energy with the reservoir, not electric charge (which is held fixed): $\Delta M\neq0\,$, $\Delta Q=0$. The equilibrium conditions $T=T_{res}\,$, $\Phi=\Phi_{res}$ follow as above. However, variations in entropy are connected with variations of the Helmholtz potential, because there are no charge variations,
\bsea
	\Delta W&=&\Delta M-T\Delta S\,,\\
	\Delta W_{res}&=&-\Delta M-T\Delta S_{res}=0\,,
\esea
which yields that thermodynamic equilibrium is reached for a \emph{global minimum of the Helmholtz potential} in the canonical ensemble, \be
	\Delta W\leq0\,.
\ee 
In this way, we have determined that the correct energy definition to use so as to determine global thermodynamic stability are respectively the Gibbs and Helmholtz free energies in the grand-canonical and canonical ensembles.

The next question to arise is the response of the globally stable solution to thermodynamic fluctuations, that is to small perturbations in the thermodynamic variables of the ensemble (temperature $T$ and electric charge $Q$ or chemical potential $\Phi$ for the canonical or grand-canonical ensembles respectively).

The equilibrium point is reached at a maximum of the entropy: this maximum is stable under small fluctuations if and only if
\bsea
	\l.\frac{\partial^2 S}{\partial T^2}\r|_{\Phi}\geq0\,,&\qquad& \l.\frac{\partial^2 S}{\partial \Phi^2}\r|_{T}\geq0\,,\\
	\l.\frac{\partial^2 G}{\partial T^2}\r|_{\Phi}\leq0\,,&\qquad& \l.\frac{\partial^2 G}{\partial \Phi^2}\r|_{T}\leq0\,,
	\label{LocalStabilityGrandCanonical}
\esea
in the grand-canonical ensemble, or
\bsea
	\l.\frac{\partial^2 S}{\partial T^2}\r|_{Q}\geq0\,,&\qquad& \l.\frac{\partial^2 S}{\partial Q^2}\r|_{T}\geq0\,,\\
	\l.\frac{\partial^2 W}{\partial T^2}\r|_{Q}\leq0\,,&\qquad& \l.\frac{\partial^2 W}{\partial Q^2}\r|_{T}\geq0\,,
	\slabel{LocalStabilityCanonical}
\esea
in the canonical ensemble. The constraints on the second derivatives of the thermodynamic potential follow from the differential version of the first law.

Defining the heat capacity at constant electric charge or chemical potential,
\be
	C_Q = T\l(\frac{\partial S}{\partial T}\r)_{Q}\,,\qquad
	C_\Phi = T\l(\frac{\partial S}{\partial T}\r)_{\Phi}\,,
	\label{HeatCapacity}
\ee
as well as the electric permittivity,
\be
	\epsilon_T = \l(\frac{\partial Q}{\partial\Phi}\r)_{T}\,,
	\label{ElectricPermittivityCanonical}
\ee
it can be seen that the above local stability conditions amount to requiring that the heat capacity and electric permittivity are positive quantities in both ensembles. This corresponds respectively to the fact that larger black holes should heat up and radiate more, while smaller black holes should go colder and radiate less; and that the chemical potential should increase when more charge is added to black hole \cite{Chamblin:1999tk}, making it harder to move away from equilibrium, as expected from classical physics or more generally, from Le Chatellier's principle. The sign difference for the conditions of local electric stability in the grand-canonical \eqref{LocalStabilityGrandCanonical} and canonical ensembles \eqref{LocalStabilityCanonical} is linked to the sign difference when the first law is expressed for the Gibbs potential \eqref{FirstLawGibbs} or for the Helmholtz potential \eqref{FirstLawCanonical}: a small increase in the electric potential results in a small increase in the Gibbs potential, while it results in a small decrease in the Helmholtz potential.

There are other criterions that one could use to rigorously ensure the thermodynamic local stability (such as cross-derivatives of the entropy), but they are not relevant to the physics we wish to descibe in this text. We refer the reader to \cite{Monteiro:2010cq} for a more careful and general derivation of the above.

\end{subappendices}

\vfill
\pagebreak

\section{Thermodynamics of black holes in Einstein-Maxwell-Dilaton theories}

\label{section:ThermoEMD}

Let us introduce this section by collecting a few results from previous studies. We recall that the zero-potential black holes were first studied by Gibbons and Maeda in \cite{Gibbons:1987ps}, as well as their thermodynamics, in generic dimension. Restricting to four dimensions, we summarise their results. The black holes \eqref{SolZeroLa} have one event horizon and a curvature singularity at finite radius, where the scalar field diverges. Since the black holes are asymptotically flat, the conserved quantities can be defined the usual way, yielding the mass and electric charge of the black hole. Together with the temperature on the horizon and the entropy, they verify the usual first law. One has to distinguish two ranges:
\begin{itemize}
 \item a lower range $\ga<1$ where the extremal limit is well-defined (the temperature goes to zero). There is a change of sign in the heat capacity as in the Einstein-Maxwell case, so that one has a transition between stable small black holes and unstable large black holes. 
 \item an upper range $\ga>1$ where the extremal limit is ill-defined (the temperature diverges). The heat capacity is always negative and never changes sign, so that the black holes are unstable.
 \item in the limiting (String Theory) case $\ga=1$, the temperature does not depend on the charge and so is finite in the extremal limit. There is a single stable branch.
\end{itemize}
All in all, the EMD black holes in the lower range behave like Reissner-Nordstr\"om black holes, while the black holes in the upper range belong to a purely dilatonic category. These behaviours were studied more closely in \cite{Preskill:1991tb,Holzhey:1991bx}. In all cases, the entropy in zero when the extremal limit is taken in the area law, which is quite a different behaviour from the Reissner-Nordstr\"om case (although we have seen in Section \ref{section:EntropyExtremalBlackHoles} how one may argue that the entropy of Reissner-Nordstr\"om black holes can also be zero depending on the way the extremal limit is taken). The $\ga<1$ black holes are interpreted as fixed objects of finite size, capable of absorbing arbitrarily small amounts of energy. As the entropy goes down along the temperature, there is a breaking down of the statistical description in the extremal limit, since the authors of \cite{Preskill:1991tb,Holzhey:1991bx} argue that the black hole contains very few thermal states. So as $T\to0$, the absorption of some arbitrary amount of energy is liable to change significantly the underlying states, and no \emph{equilibrium} temperature can be defined.

On the other hand, for $\ga\geq1$, the black hole is prevented from absorbing too small amounts of energy by a mass gap. This gap grows with the temperature, so that it becomes infinite for $\ga>1$ in the (infinite) extremal limit, thereby isolating the black hole from the outside world. At intermediate, finite temperatures, the black holes behave like particles, which can only be probed with energies high enough.

The EMD black holes with a Liouville potential were briefly studied by Cai and Ohta using holographic renormalisation, adapted to non-AdS boundary conditions, \cite{Cai:1999xg}. They examined some of the solutions with spherical topology obtained in \cite{Chan:1995fr} and some of the solutions with planar and hyperbolic topology one of them had obtained in \cite{Cai:1997ii}, restricting to the grand-canonical ensemble.
\begin{itemize}
 \item $\ka=0$: the solutions \eqref{Sol2} are found to be thermodynamically stable for $\da<1$, unstable for $\da>1$.
 \item $\ka=-1$: they find that, for the solution \eqref{CHMi}, a Hawking-Page phase transition takes place from small unstable black holes to large stable ones.
 \item $\ka=1$ : for the solution \eqref{CHMi} and $\ga^2<1$, they find a phase transition between small stable black holes and large unstable ones, which is a Reissner-Nordstr\"om-like behaviour, although there is a negative cosmological constant.
\end{itemize}

We will now focus on the planar $\ka=0$ case, and review the charged solutions of Section \ref{section:PlanarEMDBH} in detail both in the canonical and grand-canonical ensembles. Before doing so, we will review the definition of these ensembles in EMD theories, with an emphasis on possible scalar boundary terms.
These results are drawn from our recent work \cite{Charmousis:2010zz}.

\subsection{Definition of ensembles and partition functions}

We have examined in a previous Section \ref{section:DefEnsemblesEM} how the thermodynamic ensembles should be defined in Einstein-Maxwell theory. We have also argued that generalising to a cosmological constant did not change the definitions in that it had no influence over the boundary terms. This will not however be the case as a scalar field is added. 

The action for the EMD theory in the grand-canonical ensemble differs from the Einstein-Maxwell one by the scalar part and by the modified coupling between the gauge field and the scalar field: 
\be
	\label{ActionGrandCanonicalEMD}
	S_{EMD} = \frac1{16\pi}\int_\M\l[R-\frac12(\partial \phi)^2-\frac14\e^{\ga\phi}F^2-V(\phi)\r]+\frac1{8\pi}\int_{{}^3B_o}\sqrt{-h} K\,,
\ee
where we include the usual Gibbons-Hawking-York term on the (outer spatial) boundary of spacetime, so as to have only the outer boundary metric components fixed. The Maxwell considerations of Section \ref{section:DefEnsemblesEM} are generalised straightforwardly by considering the following redefinition of the momentum conjugate to the spatial components of the electric field:
\be
	E^i \to \e^{\ga\phi} E^i\,,
\ee
so that the Maxwell conserved charge is accordingly redefined as
\be
	Q_{EMD} = \frac1{16\pi}\int_{B} \sqrt{\sigma}N\e^{\ga\phi}F^{0\nu}n_\nu\,.
	\label{ElectricChargeEMD}
\ee
On the scalar side, the canonical variables are $\phi$ and its conjugate momentum $p_\phi=-\frac{\sqrt{-g}}{16\pi}\partial^0\phi$, so that the EMD action in Hamiltonian form is written\footnote{For more detailled studies of the Hamiltonian formalism in Einstein-Maxwell-Scalar theories, we refer to \cite{Creighton:1995au,Creighton:1996st,Booth:2000iq}.}
\be
	S_{EMD} = \int_{t_i}^{t_f}\ud t\l\{\int_{\Sigma_t} \l[P^{ij}\dot h^{(t)}_{ij} + \l(\frac{\sqrt{-g}}{16\pi}E^i\r)\dot A_i+p_\phi\dot\phi  \r]\ud^3x-H\r\},
	\label{EMDActionHamiltonian}
\ee
with the Hamiltonian $H$ as in \eqref{HamiltonianEinsteinMaxwell}, except for the change in the Hamiltonian constraint $\mathcal H$, \cite{Leygnac:2004bb},
\be
	\mathcal H \to \mathcal H + \frac{8\pi}{\sqrt{h^{(t)}}}p_\phi^2+\frac{\sqrt{h^{(t)}}}{32\pi}h_{(t)}^{ij}\partial_i\phi\partial_j\phi+\frac{\sqrt{h^{(t)}}}{16\pi}\l[V(\phi)-2\La\r]\,.
	\label{HamiltonianConstraintScalar}
\ee
In particular, no scalar boundary term on ${}^3B_0$ is needed. Then the Hamiltonian on-shell has the same expression as for the Einstein-Maxwell theory\footnote{Neglecting the inner boundary contribution which is not relevant here.}:
\be
	\l.H\r|_{cl}=-\frac1{8\pi} \int_{B} \sqrt{\sigma}N\l(k-\bar k\r) -\frac1{16\pi} \int_{B} \sqrt{\sigma} N\e^{\ga\phi}F^{\mu\nu}A_\mu n_\nu\,.
	\label{HamitonianEMDOnShell}
\ee
But then, we may ask the question: Does this Hamiltonian generate the correct conserved charges on the outer boundary? To study this, we need to compute the variation of the EMD action, which will be the opposite of that of the Hamiltonian once evaluated on a static solution of the equations of motion. One finds, along the lines of Section \ref{section:DefEnsemblesEM}, that \eqref{EuclActionVariation} receives an extra contribution from the scalar field on the outer boundary:
\bea
	\da(I_{EMD}) &=& \da(\ba H_{EMD}) = -\da \,B_g-\da \,B_{em} - \da\,B_\phi\,, \qquad \textrm{{\it on-shell.}}
	\label{EuclActionVariationEMD}\\
	\da\,B_\phi&=& - \frac1{16\pi}\int_{{}^3B_o}\sqrt{\sig}Nn^\mu\partial_\mu\phi\,\da\phi\,.\label{ScalarBoundaryVariation}
\eea
The integral of this variation \eqref{EuclActionVariationEMD} will \emph{a priori} be different from the integral on-shell Hamiltonian \eqref{HamitonianEMDOnShell}, except if the scalar outer boundary contribution \eqref{ScalarBoundaryVariation} vanishes. It has been known for a long time that regular asymptotically flat, \cite{Regge:1974zd}, or AdS boundary conditions, \cite{Henneaux:1985tv}, will make this term cancel asymptotically. Yet, instances of modified AdS boundary conditions have been exhibited in the recent years where it is not so for given scalar potentials with hyperbolic sines or cosines, \cite{Henneaux:2004zi,Henneaux:2006hk}. We will see in the following that this does happen too in the case of the $\ga\da=1$ black hole, were the total mass has a scalar part. One can understand this as the fact that the conserved charges at infinity under the symmetry group there pick up extra contributions because of the modified slow-off of the matter fields towards the boundary. In the asymptotically flat or AdS case, this analysis was performed carefully in \cite{Regge:1974zd,Henneaux:1985tv} as the symmetry group and its generators are known; here, we have not done this analysis, and so we will simply adopt a pragmatic philosophy: since the conserved charges thus defined are finite, integrable and verify the first law, we shall satisfy ourselves with them.

To conclude this section, the action \eqref{ActionGrandCanonicalEMD} defines the grand-canonical ensemble, since it requires to fix as boundary conditions the horizon temperature, the asymptotic electric potential, and the asymptotic value of the scalar field. The canonical ensemble where instead the electric charge \eqref{ElectricChargeEMD} is fixed is defined as in \eqref{EMActionCanonical} by adding the relevant Maxwell term on the outer boundary:
\bea
	S_{EMD}^c &=& \frac1{16\pi}\int_\M\l[R-\frac12(\partial \phi)^2-\frac14\e^{\ga\phi}F^2-V(\phi)\r]+\frac1{8\pi}\int_{{}^3B_o}\sqrt{-h} K +\nn\\
			&&\qquad\qquad\qquad\qquad\qquad\qquad\qquad\qquad+\frac1{16\pi}\int_{{}^3B_o} \sqrt{-h}\e^{\ga\phi} F^{\mu\nu}n_\mu A_\nu\,.	\label{ActionCanonicalEMD}
\eea

\subsection{\texorpdfstring{$\ga\da=1$ charged planar black holes}{gamma*delta=1 charged planar black holes}}

\label{section:ThermoLEMDGaDa=1}

In this section, we study the thermodynamics of the dilatonic solution \eqref{Sol1} presented in Section \ref{subsubsection:ka0_general} in the grand-canonical and canonical ensembles.
	
	The temperature is
\be	 T=\frac{3-\da^2}{4\pi\ell}\l(\frac{r_+}\ell\r)^{1-\da^2}\l[1-\l(\frac{r_-}{r_+}\r)^{3-\da^2}\r]^{1-2\frac{(\da^2-1)^2}{(3-\da^2)(1+\da^2)}}\,.
	\label{Temperature1}
\ee

Let us note already that the behaviour of the temperature in the extremal limit is not uniform in the $\da^2<3$ range. Indeed, for $\da^2<1+\frac2{\sqrt3}$, the temperature vanishes in the extremal limit, whereas it diverges for $\da^2>1+\frac2{\sqrt3}$ and is finite and non-zero for $\da^2=1+\frac2{\sqrt3}$. Thus, the extremal black hole seems ill-defined in the range $1+\frac2{\sqrt3}\leq\da^2<3$, as it does not seem that they can be end states of the evaporation through thermal radiation of the black hole. This behaviour has been noted  a long time ago in String Theory black holes and is commented upon in \cite{Preskill:1991tb,Holzhey:1991bx}. From similar cases of uncharged black holes investigated in \cite{gkmn2}, this behaviour corresponds to the small unstable black hole solution, and behaves like flat space Schwarzschild black holes.

In order to determine the gravitational energy of the solution, one can integrate the contribution at infinity of the boundary term
\eqref{VarGravBoundaryTerm},
\be
	M_g = \frac{\omega_2}{8\pi}\ell^{\da^2-2}\l[(r_+)^{3-\da^2}+\frac{2\da^6-15\da^4+20\da^2-3}{(3-\da^2)(1+\da^2)}(r_-)^{3-\da^2}\r],
	\label{GravitationalMass1}
\ee
where $\omega_2$ is defined as the volume of the compact two-dimensional horizon. This reduces in particular for $\da^2=0,1$ to $M_g=\frac{\omega_2}{4\pi}m$.
However, the boundary term at infinity stemming from the scalar field \eqref{ScalarBoundaryVariation} gives a non-zero contribution,
\be
	M_\phi = \frac{\omega_2}{8\pi}\ell^{\da^2-2}\frac{4\da^2(\da^2-1)}{(3-\da^2)(1+\da^2)}r_-^{3-\da^2}\,,
	\label{ScalarMass1}
\ee
which must be added to find the total mass,
\be
	M = \frac{\omega_2}{8\pi}\ell^{\da^2-2}\l[(r_+)^{3-\da^2}-\frac{2\da^4-5\da^2+1}{(1+\da^2)}(r_-)^{3-\da^2}\r].
	\label{Mass1}
\ee
The conserved electric charge \eqref{ElectricChargeEMD} is
\be
	Q =  \frac{\omega_2}{16\pi}q\,.
	\label{ElectricCharge1}
\ee

We will make the following rescalings in order to  absorb volume factors in the thermodynamic quantities, throughout the rest of this paper:
\bea
	W[T,Q],\,G[T,\Phi] &\to& \frac{16\pi}{\omega_2}W[T,Q],\,\frac{16\pi}{\omega_2}G[T,\Phi]\nn\,,\\
	M &\to& \frac{16\pi}{\omega_2}M\,,\\
	T &\to& 4\pi T, \quad S\to\frac{4}{\omega_2}S\nn\,,\\
	Q &\to& \frac{4\pi}{\omega_2} Q, \quad \Phi\to4\Phi\nn\,,
\label{rescale}
\eea
which will not modify the expression of the first law.
	
	\subsubsection{Grand-canonical ensemble}

The grand-canonical ensemble is defined by keeping the temperature and the electric chemical potential fixed: this means that we can take as the thermal background the spacetime with the black hole switched off ($m=q=0$), which also coincides with the asymptotic limit of the solution if $\da^2<3$:
\bsea
	\ud s^2_0 &=& r^2\l(-\ud t^2 +\ud x^2 + \ud y^2 \r) + \l(\frac r\ell \r)^{2\da^2-2}\ud r^2\,, \slabel{BackgroundMetric1} \\
	A_0 &=& \Phi\ud t\,,   \slabel{BackgroundA1}  \\
	e^{\phi_0} &=& \l(\frac r\ell \r)^{2\da}\,,  \slabel{BackgroundPhi1}
\label{LinearDilaton}
\esea
For a flat potential $\da=0$, one recognizes AdS as expected. In what follows, we denote this spacetime as the neutral (dilatonic) background.

\begin{figure}[t]
\begin{center}
\begin{tabular}{cc}
	 \includegraphics[width=0.45\textwidth]{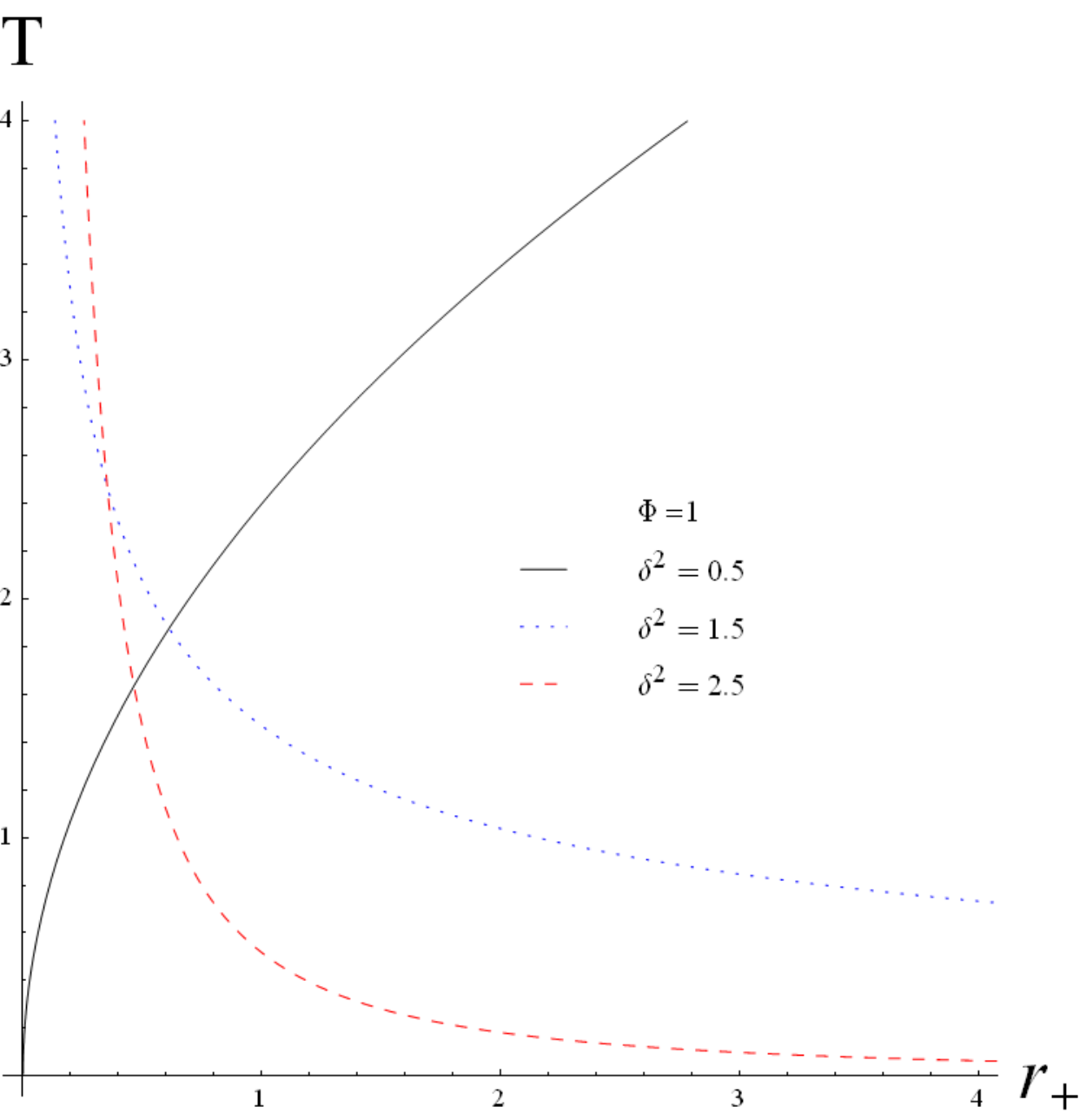}&	 
	 \includegraphics[width=0.45\textwidth]{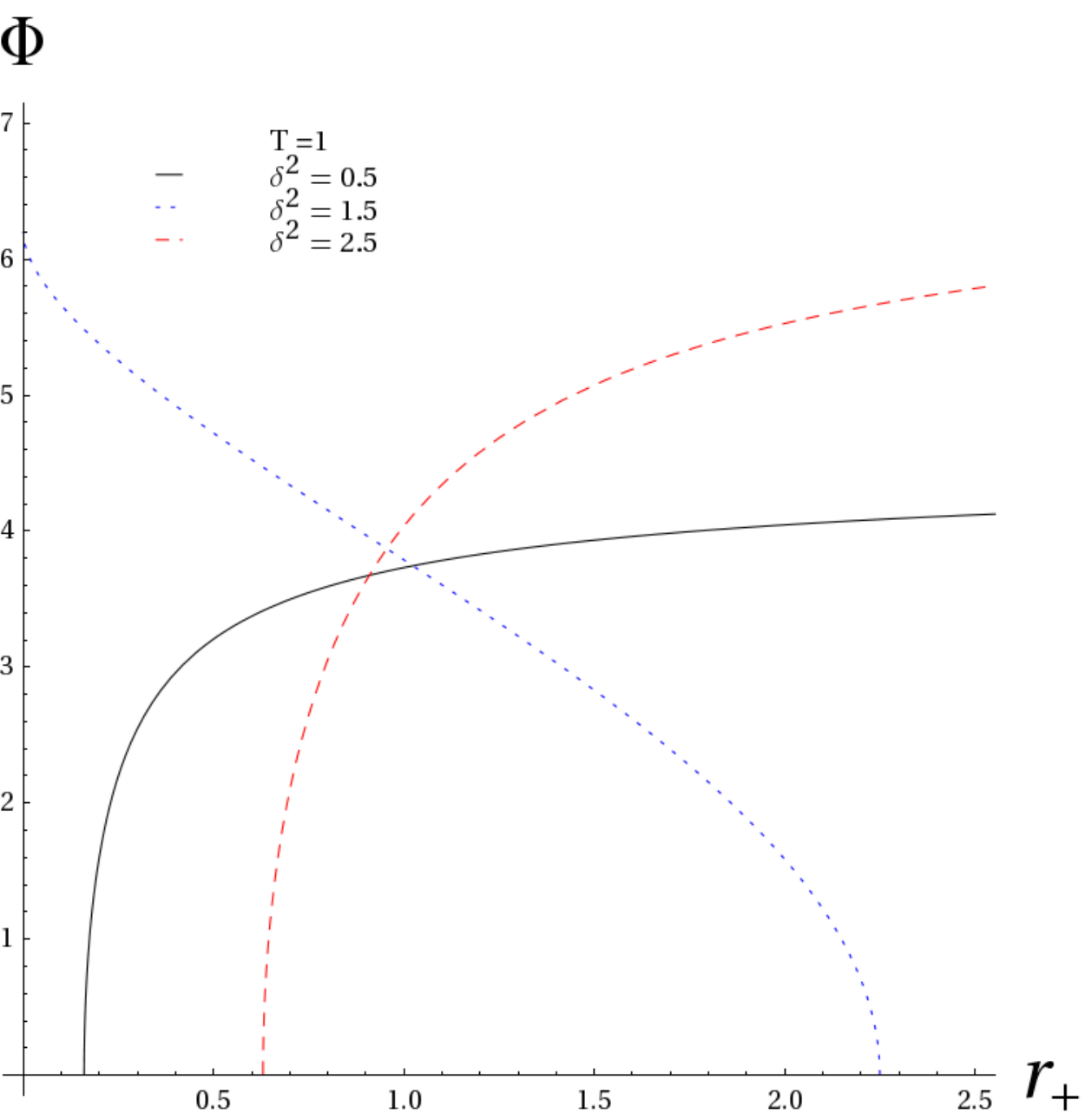}
 \end{tabular}
\caption[Equation of state for the planar $\ga\da=1$ EMD solution in the grand-canonical ensemble]{Slices of the temperature and the electric potential versus the horizon radius $r_+$ for the planar $\ga\da=1$ EMD solution \eqref{Sol1}.}
\label{Fig:HorizonSizeGaDa1GrandCanonical}
\end{center}
\end{figure}

\begin{figure}[t]
\begin{center}
\begin{tabular}{ccc}
	 \includegraphics[width=0.3\textwidth,trim=15mm 0mm 0mm 0mm,clip]{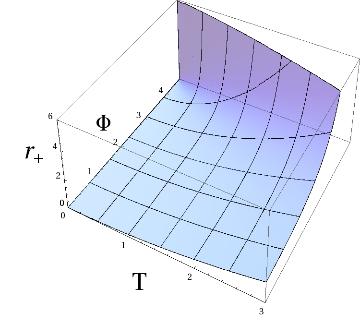} &
	 \includegraphics[width=0.3\textwidth,trim=15mm 0mm 0mm 0mm,clip]{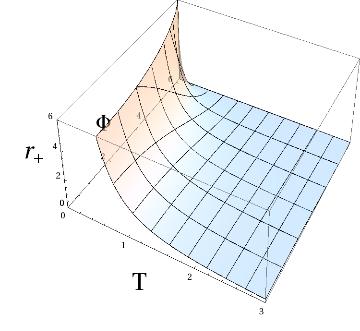} &
	 \includegraphics[width=0.3\textwidth,trim=15mm 0mm 0mm 0mm,clip]{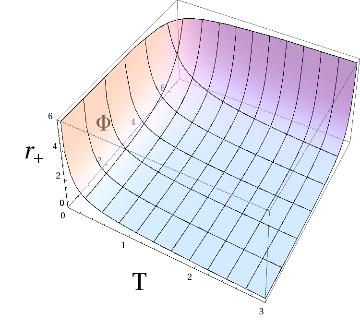}
 \end{tabular}
\caption[Three-dimensional representation of the equation of state for the planar $\ga\da=1$ EMD solution in the grand-canonical ensemble]{Three-dimensional equation of state $r_+(T,\Phi)$ for the lower ($0\leq\da^2<1$), intermediate ($1<\da^2<1+\frac2{\sqrt{3}}$) and upper ($1+\frac2{\sqrt{3}}<\da^2<3$) range from left to right for the planar $\ga\da=1$ EMD solution \eqref{Sol1}.}
\label{Fig:EquationOfState3DGaDa1GrandCanonical}
\end{center}
\end{figure}

The equation of state \eqref{Temperature1}
\be
	r_+^{1-\da^2}= \ell^{2-\da^2}\frac{T}{3-\da^2}\l[1-\frac{(1+\da^2)\Phi^2}{64\da^2}\r]^{\frac{2(\da^2-1)^2}{(1+\da^2)(3-\da^2)}-1}\,,
	\label{EqOfStateGaDa1GrandCanonical}
\ee
gives the horizon radius in terms of the grand-canonical thermodynamic variables $(T,\Phi)$. From the previous expression, the temperature and the chemical potential cannot be considered as independent variables in the $\da\to1$ limit. Thus, the grand-canonical ensemble is ill-defined for this value of $\da$ and we shall have to restrict the discussion to the canonical ensemble.

For all temperatures, there is a maximum value to the chemical potential, corresponding to
\be
	\Phi_e = \frac{64\da^2}{1+\da^2}\,,
\ee
and the region $\Phi\geq\Phi_{e}$ is forbidden because of the non-trivial exponent in \eqref{EqOfStateGaDa1GrandCanonical},
contrarily to what happens for AdS-Reissner-Nordstr\"om black holes \cite{Chamblin:1999tk}, see Section \ref{section:ThermoRNAdS}.
The extremal black holes $r_-=r_+=r_e$ are attained only for this particular value of the chemical potential,
and it is straightforward to observe that the equation of state \eqref{EqOfStateGaDa1GrandCanonical} becomes ill-defined in this limit,
as one cannot know both the temperature and the black hole radius.
Thus, one has to conclude that the extremal black holes do not exist in the grand-canonical ensemble.
However, for values of $\Phi\geq\Phi_e$, the neutral black holes can still exist and compete with the neutral background. Indeed,
there can be a non-zero chemical potential for a neutral black hole while keeping a zero charge, as only the derivative of the
vector potential has physical meaning. 

In this case, the equation of state \eqref{EqOfStateGaDa1GrandCanonical} reduces to
\be
	r_+^{1-\da^2}= \ell^{2-\da^2}\frac{T}{3-\da^2}\,,
	\label{EqOfStateNeutralGaDa1GrandCanonical}
\ee
and gives the horizon radius as a function of the temperature only.

We plot slices of the equation of state \eqref{EqOfStateGaDa1GrandCanonical} at fixed chemical potential and temperature in \Figref{Fig:HorizonSizeGaDa1GrandCanonical}, while the full three-dimensional plots are in \Figref{Fig:EquationOfState3DGaDa1GrandCanonical}. Three ranges can be distinguished,
\begin{itemize}
 \item Lower range $0<\da^2\leq1\,$:  The black holes resemble planar AdS-Reissner-Nordstr\"om and behave according to the standard physical intuition. Indeed, for finite $\Phi$, the black hole (charged or neutral) disappears (zero horizon radius) as the temperature goes to zero, and grows to cover the whole of spacetime as the temperature grows to infinity. This is to be compared with Fig.2 in \cite{Chamblin:1999tk}: the term responsible for the vanishing of the radius at non-zero finite temperature is absent because the spatial curvature of the horizon is zero. The extremal limit $\Phi=\Phi_e$ can never be attained on the left plot of \Figref{Fig:HorizonSizeGaDa1GrandCanonical} since $\Phi\neq\Phi_e$ there, so it comes as no surprise that the zero temperature limit switches off the black hole, instead of going to the extremal limit. 
 
The right plot of \Figref{Fig:HorizonSizeGaDa1GrandCanonical} is at finite temperature and so does not display extremal black holes. Instead, the extremal limit $\Phi\to\Phi_e^-$ leads to a black hole which engulfs the whole  spacetime as $r_+\to+\infty$. The horizontal axis $\Phi=0$ shows the uncharged dilatonic black holes.
 \item Intermediate range $1<\da^2<1+\frac2{\sqrt{3}}\,$: This range seems quite different from the usual behaviour as the black hole radius decreases with the temperature! The background spacetime, given by $r_+=0$ is attained for infinite temperature and in the zero temperature limit the black hole covers the whole of spacetime. The right plot of \Figref{Fig:HorizonSizeGaDa1GrandCanonical} shows a finite radius for the neutral dilatonic black holes and no black hole for the value $\Phi=\Phi_e$, all of this expected as the temperature is finite and does not allow extremal black holes to be reached.
 \item Upper range $1+\frac2{\sqrt{3}}\leq\da^2<3\,$:  The same analysis of the left plot of \Figref{Fig:HorizonSizeGaDa1GrandCanonical} as in the intermediate range applies, while the right plot corresponds to the same analysis as in the lower range.
\end{itemize}

\begin{figure}[t]
\begin{center}
\begin{tabular}{cc}
	 \includegraphics[width=0.45\textwidth]{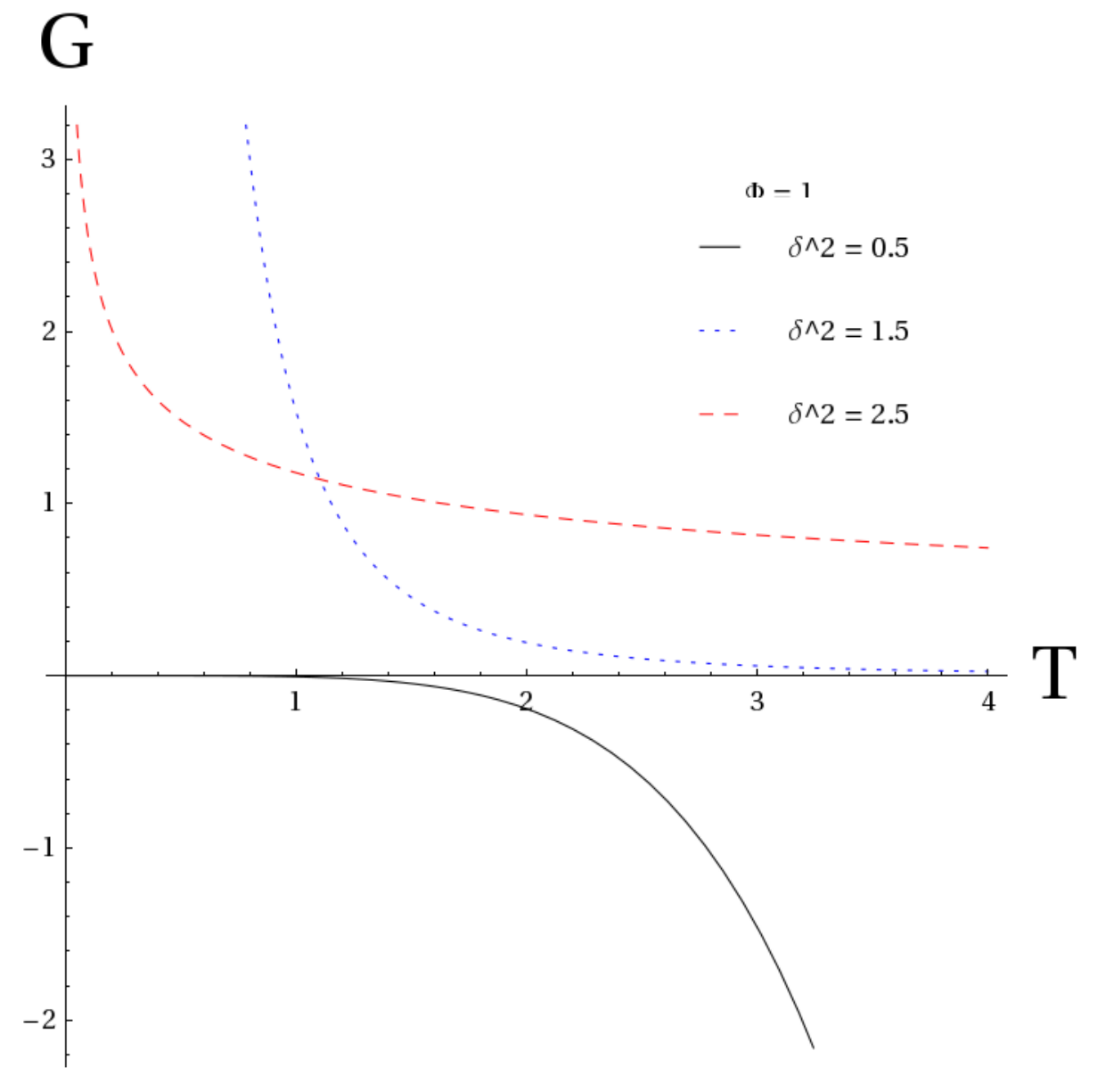}&	 
	 \includegraphics[width=0.45\textwidth]{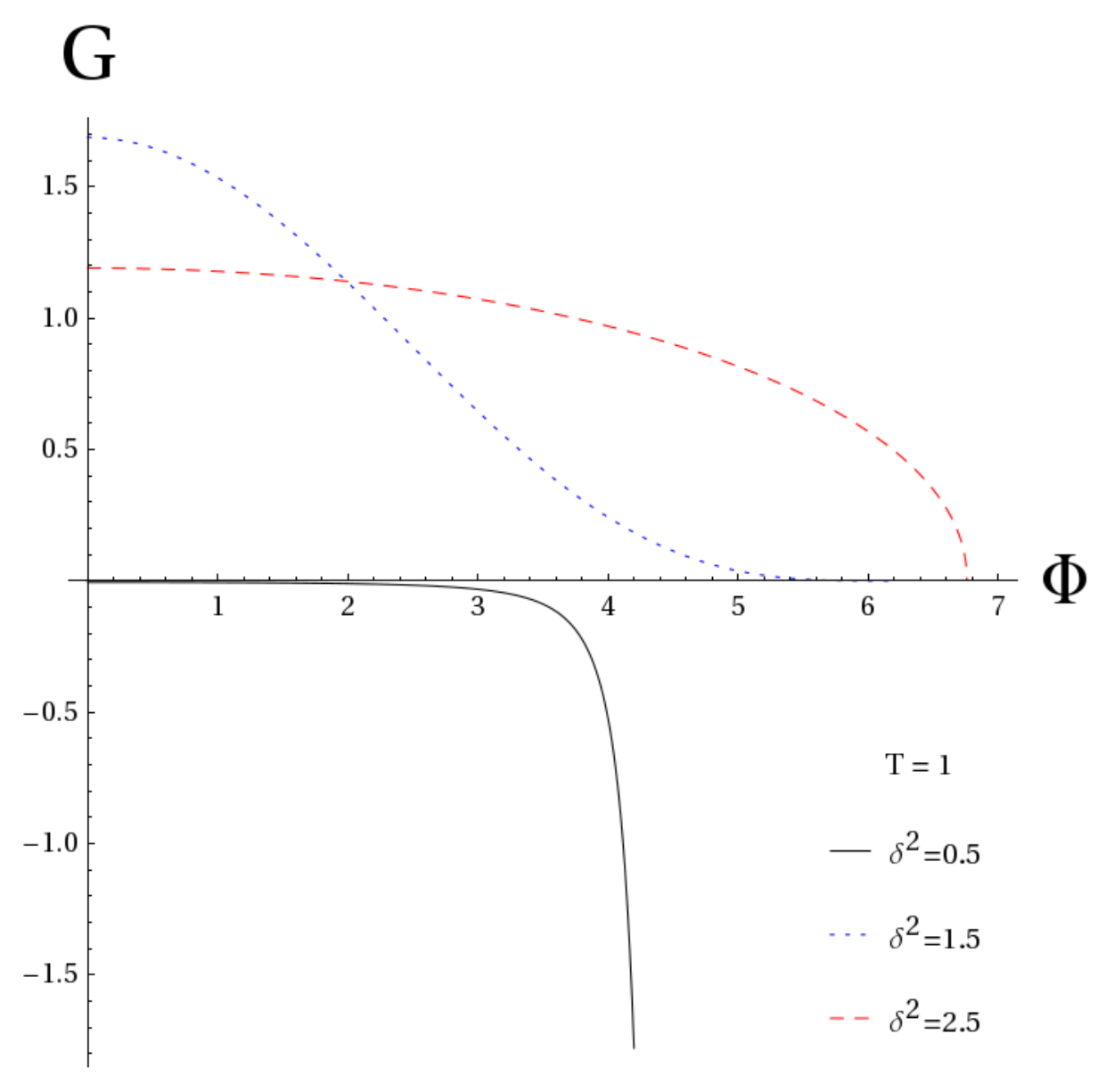}
 \end{tabular}
\caption[Free energy for the planar $\ga\da=1$ EMD solution in the grand-canonical ensemble]{Slices at constant chemical potential and constant temperature of the Gibbs potential for the planar $\ga\da=1$ EMD solution \eqref{Sol1}. The $T=2$ value for $\da^2=0.5$ allows to see graphically the non-zero limit as $\Phi\to0$}
\label{Fig:GibbsGaDa1GrandCanonical}
\end{center}
\end{figure}

\begin{figure}[t]
\begin{center}
\begin{tabular}{ccc}
	 \includegraphics[width=0.3\textwidth,trim=15mm 0mm 0mm 0mm,clip]{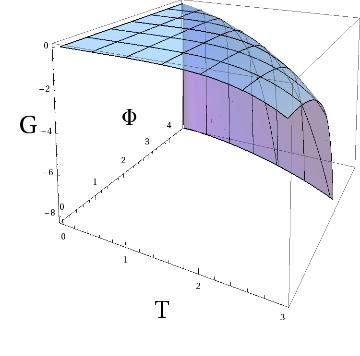} &
	 \includegraphics[width=0.3\textwidth,trim=15mm 0mm 0mm 0mm,clip]{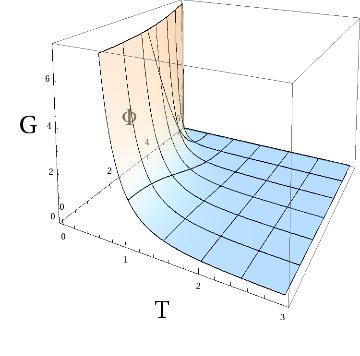} &
	 \includegraphics[width=0.3\textwidth,trim=15mm 0mm 0mm 0mm,clip]{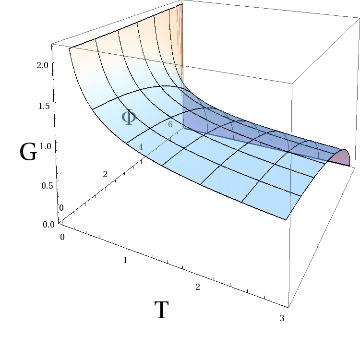}
 \end{tabular}
\caption[Three-dimensional representation of the free energy for the planar $\ga\da=1$ EMD solution in the grand-canonical ensemble]{Three-dimensional representation of the Gibbs potential for the planar $\ga\da=1$ EMD solution \eqref{Sol1} in the lower ($0\leq\da^2<1$), intermediate ($1<\da^2<1+\frac2{\sqrt{3}}$) and upper ($1+\frac2{\sqrt{3}}<\da^2<3$) range from left to right.}
\label{Fig:Gibbs3DGaDa1GrandCanonical}
\end{center}
\end{figure}

Then, calculating the value of the Euclidean action \eqref{ActionGrandCanonicalEMD} and subtracting the background contribution, we find
\be
	\beta G =  I - \bar I = \beta \ell^{\da^2-2}(\da^2-1)\l[r_+^{3-\da^2}-r_-^{3-\da^2} \r],
	\label{EuclideanActionGC1}
\ee
where we have identified the temperatures of the black hole and the thermal background on the outer boundary in order to do the subtraction. The Gibbs potential of the black hole in the thermal background \eqref{LinearDilaton} is, expressed in the thermodynamic variables $(T,\Phi)$
\be
	G[T,\Phi] = \ell^{\da^2-2}(\da^2-1)(3-\da^2)^{\frac{3-\da^2}{\da^2-1}}T^{\frac{3-\da^2}{1-\da^2}}\l[1-\frac{\Phi^2}{\Phi_e^2} \r]^{2\da^2\frac{3-\da^2}{\da^4-1}}\,,
	\label{Gibbs1}
\ee
and here we have to be careful when evaluating the Gibbs potential for extremal black holes. Inspecting the limit $\Phi\to\Phi_e^-$ (e.g. approached from below) in \eqref{Gibbs1}, we find that it yields zero in the intermediate and upper range, but minus infinity in the lower one: in this range we cannot define the extremal limit from \eqref{Gibbs1} as we have an undetermined expression.
 Thus, the proper, unambiguous way to calculate the Gibbs potential of the extremal black holes is by evaluating \eqref{EuclideanActionGC1}, from which we find it is always identically zero. We show slices at constant chemical potential and temperature of the Gibbs potential in \Figref{Fig:GibbsGaDa1GrandCanonical}, while the full three-dimensional representation is in \Figref{Fig:Gibbs3DGaDa1GrandCanonical}.

We also note that the Gibbs potential is always zero if $\da^2=1$, that is in the string case. The explanation is simple: in this limit, the equation of state \eqref{EqOfStateGaDa1GrandCanonical} becomes a relation between $T$ and $\Phi$ and thus they cannot be taken as independent variables; the grand-canonical ensemble is ill-defined in this case.

It is also easy to observe, either from \eqref{EuclideanActionGC1} or \eqref{Gibbs1}, that the neutral black holes always have greater Gibbs potential than the charged black holes, and so will not be favoured globally as long as the charges black holes exist.

We may now calculate the entropy \eqref{EntropyGrandCanonical},
\be
	 S=\ell^{\frac{4-2\da^2}{1-\da^2}}\l(\frac{T}{3-\da^2}\r)^{\frac2{1-\da^2}}\l[1-\frac{(1+\da^2)\Phi^2}{64\da^2}\r]^{2\frac{(3-\da^2)}{\da^4-1}}\,,
	\label{EntropyGrandCanonical1}
\ee
which is the quarter of the area of the horizon as expected, the electric charge \eqref{ElectricChargeGrandCanonical},
\be
	Q=\frac{(3-\da^2)}{16}\ell^{\da^2-2}\Phi r_+^{3-\da^2}\,,
\label{114}\ee
which is equal to its usual value \eqref{ElectricCharge1}, and  the energy \eqref{MassGrandCanonical},
\be
	M = 2\ell^{\da^2-2}r_+^{3-\da^2}\l[1-(2\da^4-5\da^2+1)\frac{\Phi^2}{64\da^2}\r],
	\label{MassGrandCanonical1}
\ee
which coincides with the previous expression \eqref{Mass1}

\begin{figure}[t]
\begin{center}
\begin{tabular}{cc}
	 \includegraphics[width=0.45\textwidth]{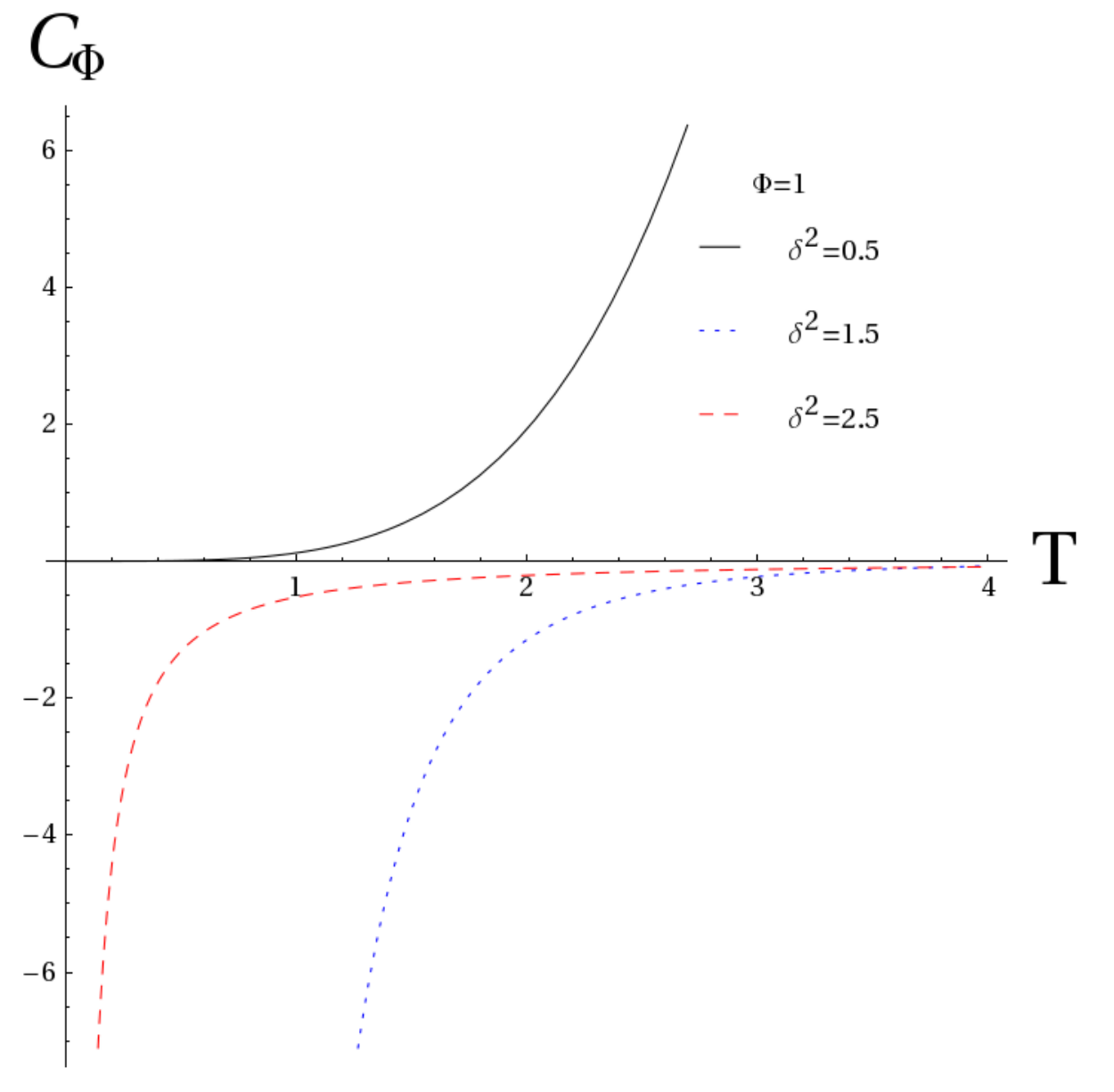}&	 
	 \includegraphics[width=0.45\textwidth]{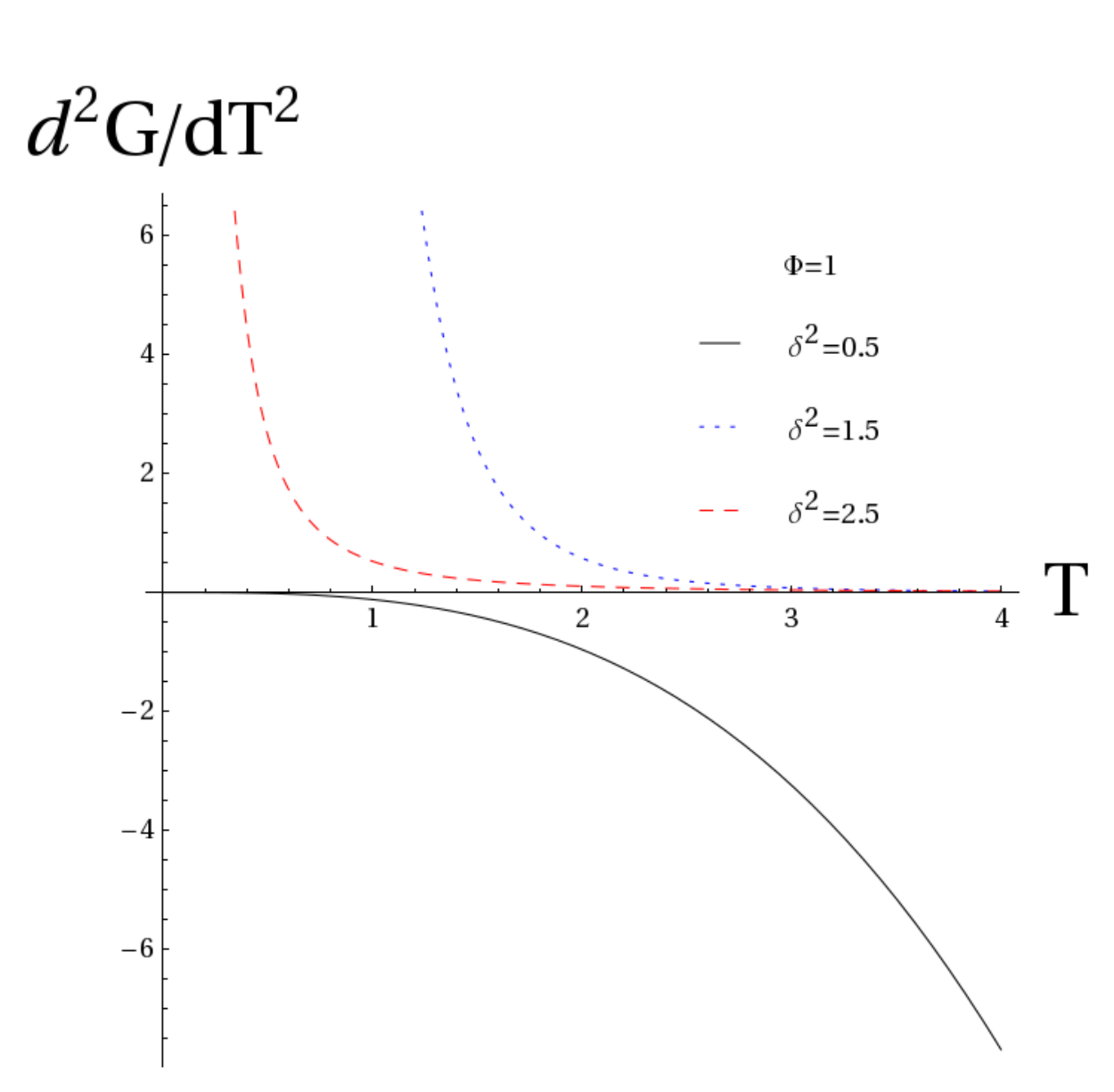}
 \end{tabular}
\caption[Thermal stability for the planar $\ga\da=1$ EMD solution in the grand-canonical ensemble]{Heat capacity and second derivative of the Gibbs potential at constant chemical potential for the planar $\ga\da=1$ EMD solution \eqref{Sol1}.}
\label{Fig:ThermalStabilityGaDa1GrandCanonical}
\end{center}
\end{figure}

Turning to the thermodynamic analysis, we find that the same three ranges as above must be distinguished.
\begin{itemize}
 \item Lower range: Similarly to the AdS-Reissner-Nordstr\"om black holes, the dilatonic black holes dominate at all values of the temperature, since their Gibbs potential is negative, see \Figref{Fig:GibbsGaDa1GrandCanonical}. The black holes are stable, as can be assessed both from the positivity of the heat capacity \eqref{HeatCapacity} and negativity of the second derivative of the Gibbs potential at constant chemical potential, see  \Figref{Fig:ThermalStabilityGaDa1GrandCanonical}. They are also stable to electric fluctuations: the slope of the Gibbs potential, as the chemical potential varies, is always negative and decreasing, which corresponds to positive electric permittivity. However, the charged black holes dominate the ensemble only for $\Phi<\Phi_e$, whereas the neutral ones do for higher values of the chemical potential.

The remarkable feature that at zero temperature and for large enough $\Phi$,  the extremal black holes were dominating \cite{Chamblin:1999tk} is no longer true, as the extremal black holes are ill-defined except if $\Phi=\Phi_e$. As the temperature approaches zero, the black hole shrinks until it disappears completely and only the uncharged background is left. This is not a proper phase transition as there are no competing solutions: only the background exists. If both $T=0$ and $\Phi=\Phi_e$, then the uncharged uncharged background and the extremal black holes are competing, but none of them dominate the other.
 \item Intermediate and upper range: The black holes, charged or uncharged, are always globally and locally unstable and decay to the uncharged background at all values of the temperature and chemical potential.
\end{itemize}

The heat capacity at constant chemical potential is given below,
\be
  C_\Phi = \frac{2}{1-\da^2}\l(\frac{T}{3-\da^2}\r)^{\frac{2}{1-\da^2}}\l(1-\frac{\Phi^2}{\Phi_e^2}\r)^{-2\da^2\frac{3-\da^2}{1-\da^4}}\,,
  \label{HeatCapacityGrandCanonical1}
\ee
and is positive for $\da^2<1$ and negative otherwise.

\begin{figure}[t]
\begin{center}
\begin{tabular}{cc}
	 \includegraphics[width=0.45\textwidth]{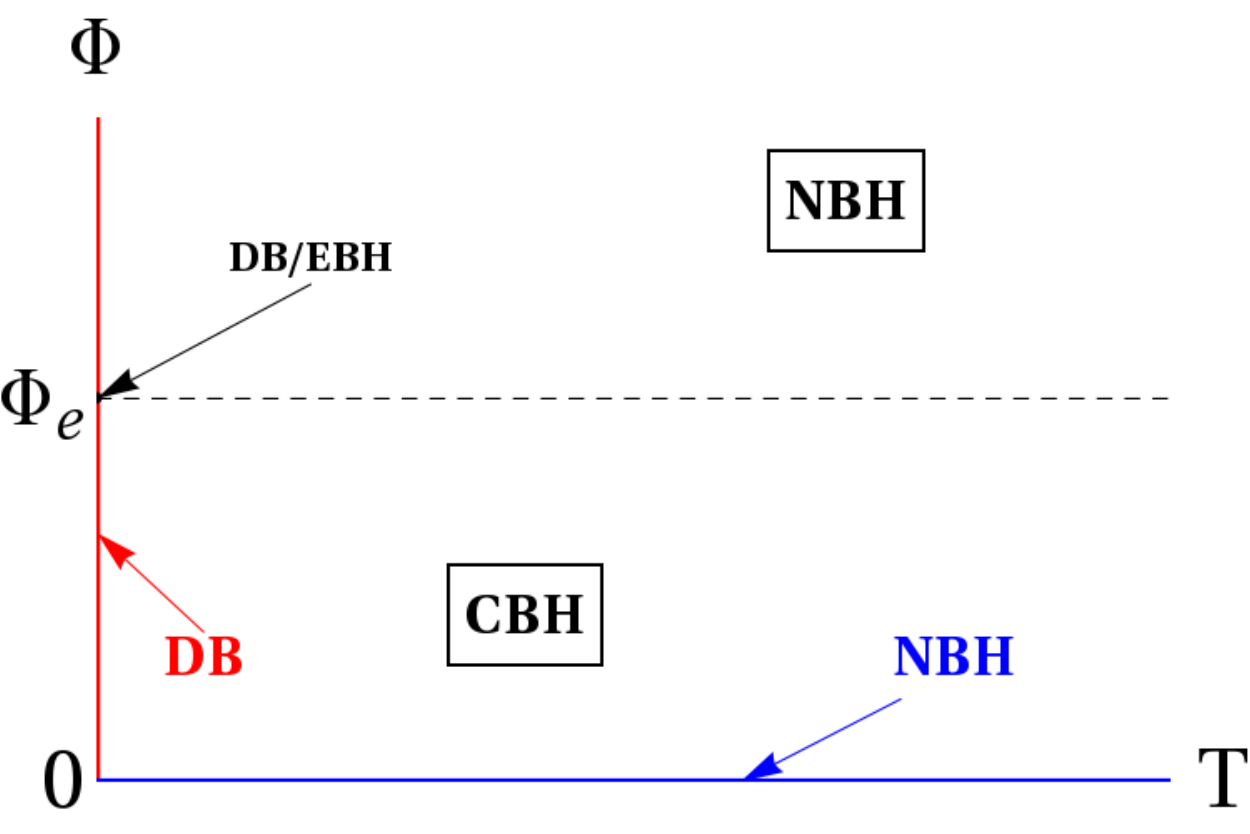}&
	 \includegraphics[height=2in]{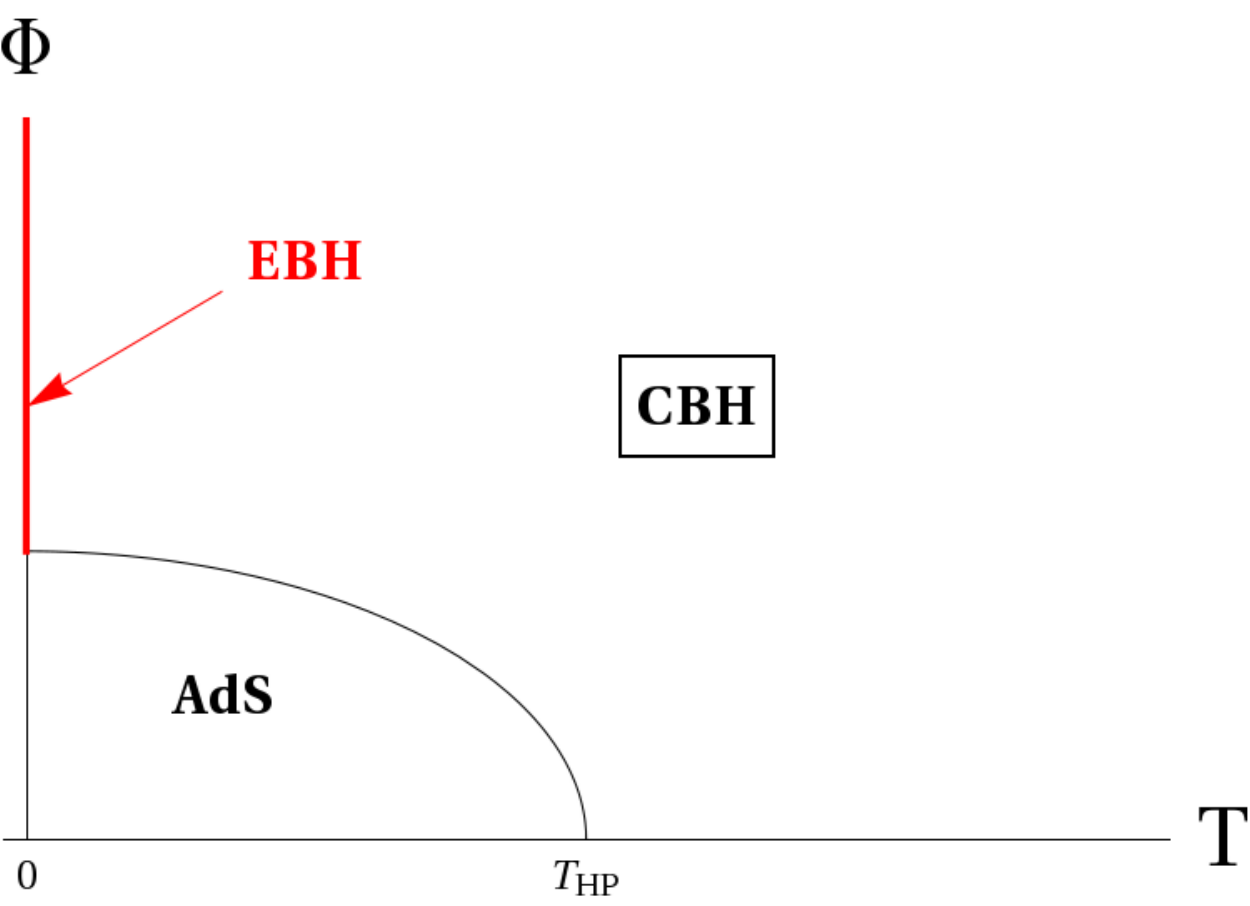}
 \end{tabular}
\caption[Phase diagrams for the planar $\ga\da=1$ EMD solution in the grand-canonical ensemble]{On the left, phase diagram in $(T,\Phi)$ space for the lower range, $\da^2 = 0.5$. CBH = Charged Black Holes, NBH = Neutral Black Holes, DB = Dilatonic Background, EBH = Extremal Black Holes. On the right, for comparison, we display the phase space for spherical AdS-RN black holes, \cite{Chamblin:1999tk}}
\label{Fig:PhaseDiagramGaDa1GrandCanonical}
\end{center}
\end{figure}

We present the (only non-trivial) phase diagram for the lower range in \Figref{Fig:PhaseDiagramGaDa1GrandCanonical}.

In the region $\Phi<\Phi_e$, the charged black holes dominate at finite temperature, and so do the neutral black holes on the vertical axis $\Phi=0$. {The transition to the neutral black holes is smooth, in particular no phase transition happens, as is obvious from \eqref{Gibbs1}.}

On the horizontal axis  $T=0$, only the neutral background exists (no black holes) as the extremal black holes can only be reached at the point $\Phi=\Phi_e$. The approach to zero temperature exhibits critical behaviour, there is an $n^\textrm{th}$-order phase transition to the uncharged background for values
\be
\frac{n-4}{n-2} < \delta^2 < \frac{n-3}{n-1}\,,\quad n=4,5,6\dots
\ee
Note that the upper bound goes to $1$ as $n\to+\infty$, allowing for abitrarily high-order phase transitions.

If $\Phi\to\Phi_e^-$ and $T$ is finite, then the Gibbs potential diverges and there is a zeroth-order phase transition to the neutral black holes. For $\Phi\geq\Phi_e$ and finite temperature, the neutral black holes dominate the ensemble over the uncharged background. It makes sense on physical grounds to expect that the naked singularity of \eqref{LinearDilaton} should preferably be cloaked by an event horizon. At zero temperature and $\Phi=\Phi_e$, the equation of state \eqref{EqOfStateGaDa1GrandCanonical} for $\delta^2<1$
can still be satisfied, but \eqref{EuclideanActionGC1} yields $G=0$ identically, and so we cannot
discriminate at this particular point between the uncharged background and
the extremal limit.

We can compare with the phase space in the case of AdS-RN spherical black holes, see Fig.6 of \cite{Chamblin:1999tk}. The effect of missing the asymptotically AdS region (and therefore the large black hole branch) is to destroy the Hawking-Page transition, and so the black holes dominate the whole phase space. Then, the effect of the scalar field is that charged black holes exist only up to some critical value for the chemical potential, and then there is a transition to the neutral black holes. Furthermore, the extremal black holes now exist only for a specific value of the chemical potential, and do not dominate the phase space anywhere. This can be seen comparing the equation of state \eqref{EqOfStateGaDa1GrandCanonical} with equation (20) of \cite{Chamblin:1999tk}, which we reproduce below (notwithstanding some numerical factors),
\be
  \Phi = \sqrt{r_+^2 -2r_+ T +\kappa}\,,
\label{116}\ee
where we have explicitly introduced the normalised spatial curvature of the horizon, $\kappa=0,\pm1$ for planar, spherical or hyperbolic geometry. Specifically, one sees that for zero temperature zero $\kappa$, one can always obtain in the planar Reissner-Nordstr\"om case an extremal black hole, whereas \eqref{EqOfStateGaDa1GrandCanonical} forces either $\Phi=\Phi_e$ (fixed value) to keep $r_+=r_e$ finite, or $r_+=0$ (no black hole).

	\subsubsection{Canonical ensemble}

The canonical ensemble is defined by keeping the temperature and the electric charge fixed. Then \eqref{Temperature1} can be considered as an implicit equation of state giving $r_+$ in terms of $T$ and $Q$:
\be
	 T=\frac{3-\da^2}{\ell}\l(\frac{r_+}{\ell}\r)^{1-\da^2}\l[1-\frac{4(1+\da^2)Q^2}{\da^2(3-\da^2)^2r_+^{6-2\da^2}} \r]^{1-2\frac{(\da^2-1)^2}{(1+\da^2)(3-\da^2)}}\,.
	\label{EquOfStateCanonical1}
\ee

\begin{figure}[t]
\begin{center}
\begin{tabular}{cc}
	 \includegraphics[width=0.45\textwidth]{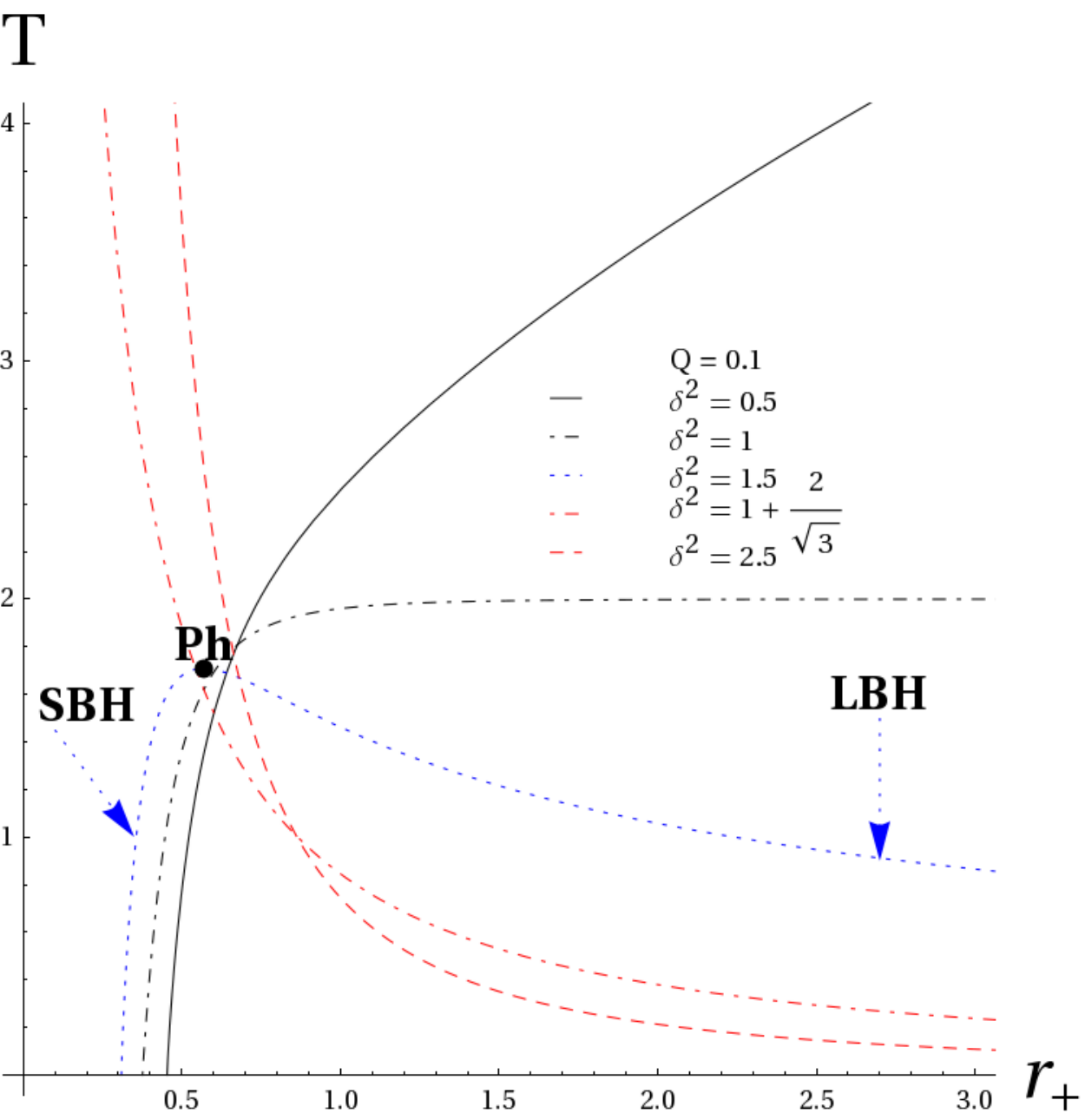}&	 
	 \includegraphics[width=0.45\textwidth]{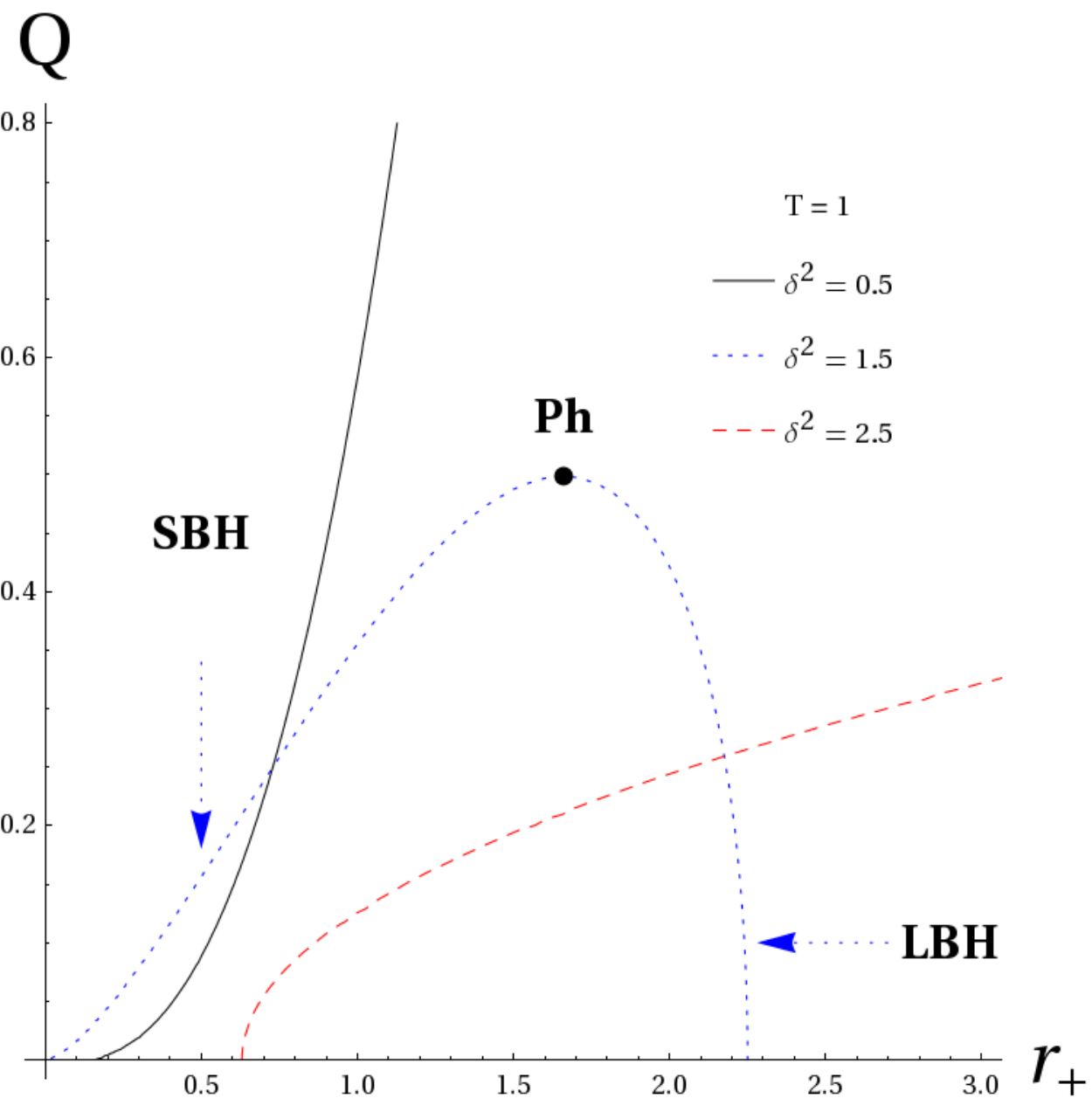}
 \end{tabular}
\caption[Equation of state for the planar $\ga\da=1$ EMD solution in the canonical ensemble]{Temperature and charge versus horizon radius $r_+$ for the planar $\ga\da=1$ EMD solution \eqref{Sol1}.}
\label{Fig:HorizonSizeGaDa1Canonical}
\end{center}
\end{figure}

\begin{figure}[t]
\begin{center}
\begin{tabular}{cc}
	 \includegraphics[width=0.45\textwidth,trim=0mm 0mm 0mm 0mm,clip]{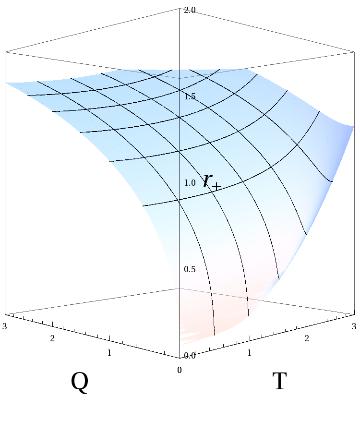}&
	 \includegraphics[width=0.45\textwidth,trim=0mm 0mm 0mm 0mm,clip]{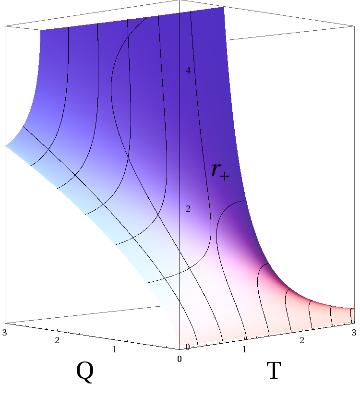}
 \end{tabular}
\caption[Three-dimensional representation of the equation of state for the planar $\ga\da=1$ EMD solution in the canonical ensemble]{Three-dimensional equation of state $r_+(T,Q)$ for the planar $\ga\da=1$ EMD solution \eqref{Sol1} in the lower ($0\leq\da^2<1$) and intermediate ($1<\da^2<1+\frac2{\sqrt{3}}$) range from left to right.}
\label{Fig:EquationOfState3DGaDa1Canonical}
\end{center}
\end{figure}

We plot it  for various values of the coupling constant $\da$ in \Figref{Fig:HorizonSizeGaDa1Canonical} and \Figref{Fig:EquationOfState3DGaDa1Canonical}. We notice three different types of behaviour:
\begin{itemize}
 \item  In the lower range, $\da^2\leq1$, there is a single black hole branch for each doublet $(T,Q)$.
This is similar to the spherical AdS-Reissner-Nordstr\"om black holes for $q>q_\mathrm{crit}$ (see \Figref{Fig:ThermodynamicsRNAdSCanonical}), or to the $\da=0$ topological black holes as described in Section \ref{section:ThermoRNAdS}. In this range, the scalar field does not spoil the effect of the negative cosmological constant.

The extremal limit $r_-\to r_+$ has zero temperature, and the radius of the black hole grows with the temperature.
At $Q=0$, the endpoint of the curve on the right in \Figref{Fig:HorizonSizeGaDa1Canonical} is the neutral black hole.
For the limiting case $\da^2=1$, there is a maximal temperature for large black holes, prefiguring the behaviour in the next range.
 \item In the intermediate range, $1<\da^2<1+\frac2{\sqrt3}$, there are two branches, small black holes (SBH) and large black holes (LBH),
which merge at a transition point
\be
	 r_\mathrm{Ph}^{6-2\da^2}=\frac{\l(-5\da^4+12\da^2+1\r)}{(\da^4-1)}r_e^{6-2\da^2}\,.
	\label{TransitionRadiusCanonical1}
\ee
This radius corresponds either to a maximal temperature at fixed electric charge, or to a maximum charge at fixed temperature,
see \Figref{Fig:HorizonSizeGaDa1Canonical}. The critical point exists only for $1<\da^2<1+\frac2{\sqrt3}$, and
the critical temperature is given by:
\be	 T_\mathrm{Ph}=(3-\da^2)\ell^{\da^2-2}r_\mathrm{Ph}^{1-\da^2}\l(\frac{-6\da^4+12\da^2+2}{-5\da^4+12\da^2+1}\r)^{\frac{-3\da^4+6\da^2+1}{(3-\da^2)(1+\da^2)}}\,.
	\label{TransitionTemperatureCanonical1}
\ee
Again, this relates to \Figref{Fig:ThermodynamicsRNAdSCanonical}, except that the third branch is gone: the scalar field has destroyed the effect of a negative comological constant. However, we recovered the two branches characteristic of charged asymptotically flat spherical black holes. To some extent, the extra scalar degree of freedom simulates a positive curvature horizon in this range.

The small black holes branch end on the other side to the extremal black holes at $T=0$ and $r_+=r_e$, while the radius of the large black holes diverges in the zero temperature limit. This alone would cast doubt on their physical relevance, see the energetic analysis below. At $Q=0$, the endpoint of the curve on the right in \Figref{Fig:HorizonSizeGaDa1Canonical} is either the neutral black hole (non-zero radius) or the background extremal black hole, which in this case has zero radius and so coincides with the uncharged background.
 \item In the upper range $1+\frac2{\sqrt3}\leq\da^2<3$, there is again a single branch. The temperature diverges in the extremal limit $r_-\to r_+$, which consitutes a lower bound for the black hole size. This is however expected for dilatonic black holes \cite{Preskill:1991tb,Holzhey:1991bx} and signals the breakdown of their statistical description. On the other side, the large black holes have arbitrarily small temperature. For the limiting case $\da^2=1+\frac2{\sqrt3}$, the horizon size is actually independent of the electric charge and therefore has no minimum : the temperature diverges for zero radius. This is similar to spherical Schwarzschild balck holes, see again \Figref{Fig:ThermodynamicsRNAdSCanonical}. In this range, the scalar field still similates a positive horizon curvature, but counters the effect of the negative cosmological constant and of the electric charge.
\end{itemize}

\begin{figure}[t]
\begin{center}
\begin{tabular}{cc}
	 \includegraphics[height=3in]{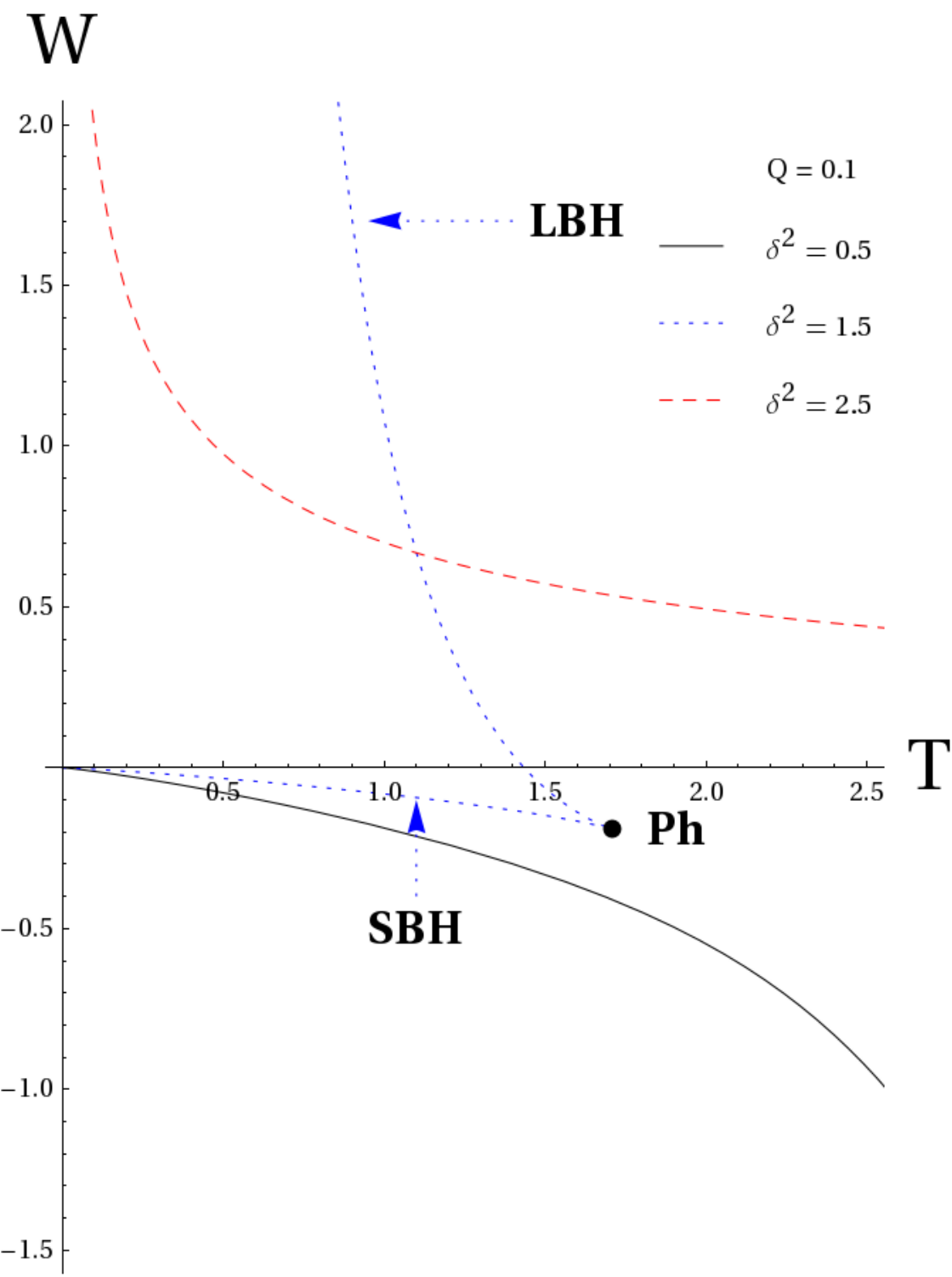}
	 &\includegraphics[height=3in]{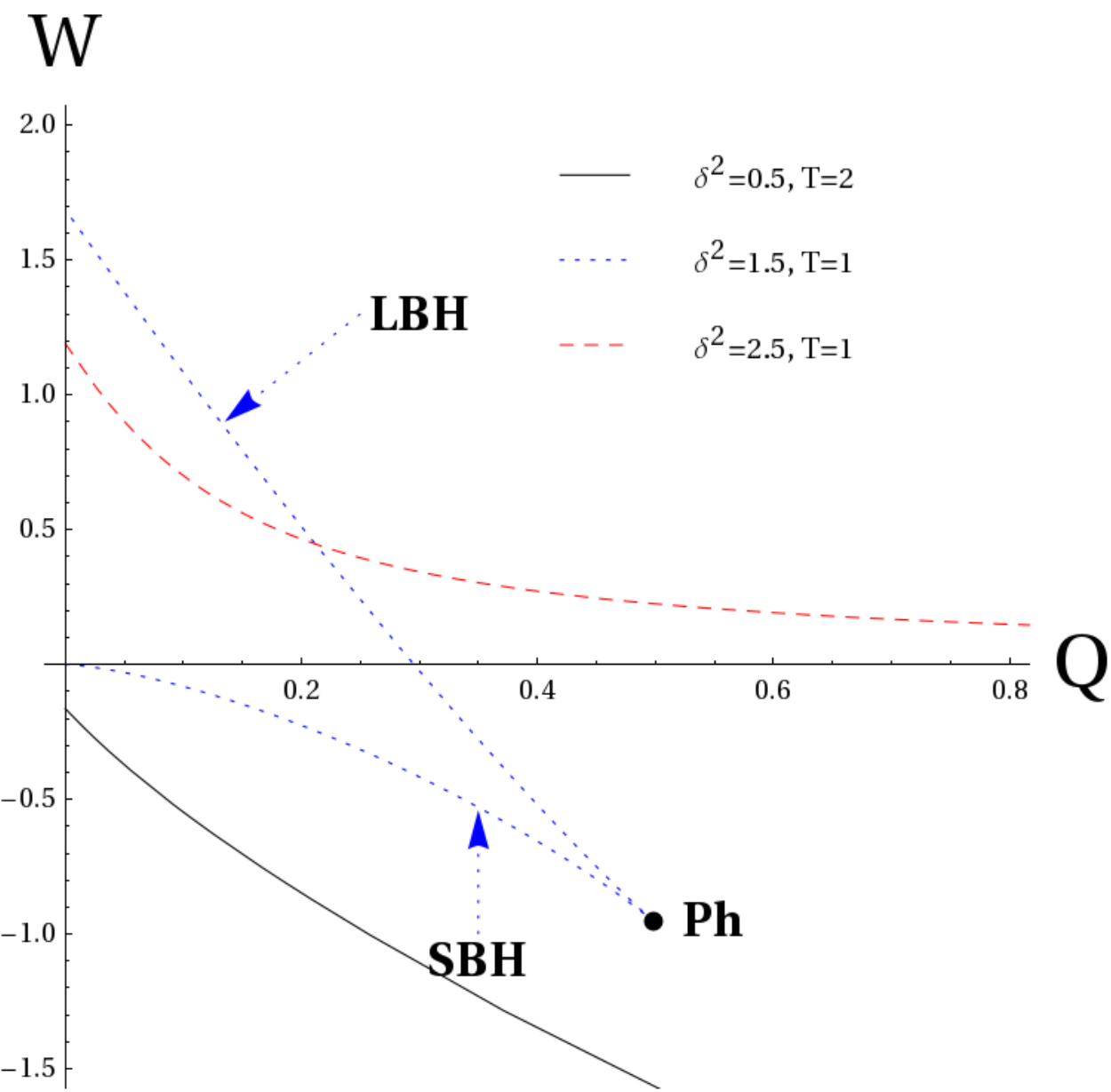}
 \end{tabular}
\caption[Free energy for the planar $\ga\da=1$ EMD solution in the canonical ensemble]{Slices of the Helmholtz potential at fixed charge and temperature for the planar $\ga\da=1$ EMD solution \eqref{Sol1}.}
\label{Fig:FreeEnergyGaDa1Canonical}
\end{center}
\end{figure}

Calculating the value of the action \eqref{ActionCanonicalEMD} in the canonical ensemble (where we have subtracted the value of the Euclidean action $I_e^c$ for the extremal black hole, which we use as the thermal background \cite{Chamblin:1999tk})
\be
	  I^c -  I^c_e = \beta \ell^{\da^2-2}\l[(\da^2-1)r_+^{3-\da^2}-\frac{5\da^4-12\da^2-1}{1+\da^2}r_-^{3-\da^2}-\frac{4\da^2(3-\da^2)}{1+\da^2}(r_e)^{3-\da^2} \r],
\ee
yields the Helmholtz potential of the black hole
\be
	W[T,Q] =\ell^{\da^2-2} \l[(\da^2-1)r_+^{3-\da^2}-\frac{4(5\da^4-12\da^2-1)Q^2}{\da^2(3-\da^2)^2r_+^{3-\da^2}} -8\sqrt{\frac{\da^2}{1+\da^2}}Q\r],
	\label{Helmholtz1}
\ee
which we plot at fixed charge density  or at fixed temperature in \Figref{Fig:FreeEnergyGaDa1Canonical} for the two ranges.

The total mass of the solution is by \eqref{MassCanonical}
\be
	M_W = M-M_e = 2\ell^{\da^2-2}\l[(r_+)^{3-\da^2}-\frac{4(2\da^4-5\da^2+1)Q^2}{\da^2(3-\da^2)^2r_+^{3-\da^2}}\r] - 8Q\sqrt{\frac{\da^2}{1+\da^2}}\,,
	\label{EnergyCanonical1}
\ee
and the chemical potential by \eqref{ElectricPotentialCanonical}
\be
	\Phi_W= \Phi-\Phi_e = \frac{16Q\ell^{2-\da^2}}{(3-\da^2)r_+^{3-\da^2}}-8\sqrt{\frac{\da^2}{1+\da^2}}\,,
	\label{ElectricPotentialCanonical1}
\ee
while the entropy \eqref{EntropyCanonical} is equal to one quarter of the area of the horizon
\be
	S_W = S-S_e = S = r_+^2\l[1-\frac{4(1+\da^2)Q^2}{\da^2(3-\da^2)^2r_+^{6-2\da^2}} \r]^{2\frac{(\da^2-1)^2}{(3-\da^2)(1+\da^2)}}\,,
	\label{EntropyCanonical1}
\ee
since the entropy of the extremal background is identically zero by our method of computation, \cite{Hawking:1994ii} and see Section \ref{section:EntropyExtremalBlackHoles}. The entropy of the non-extremal black holes actually also goes to zero at extremality, indicating the presence of a non-degenerate ground state, \cite{Preskill:1991tb,Holzhey:1991bx}. This seems to be characteristic of charged dilatonic black holes.

\begin{figure}[t]
\begin{center}
\begin{tabular}{cc}
	 \includegraphics[width=0.45\textwidth]{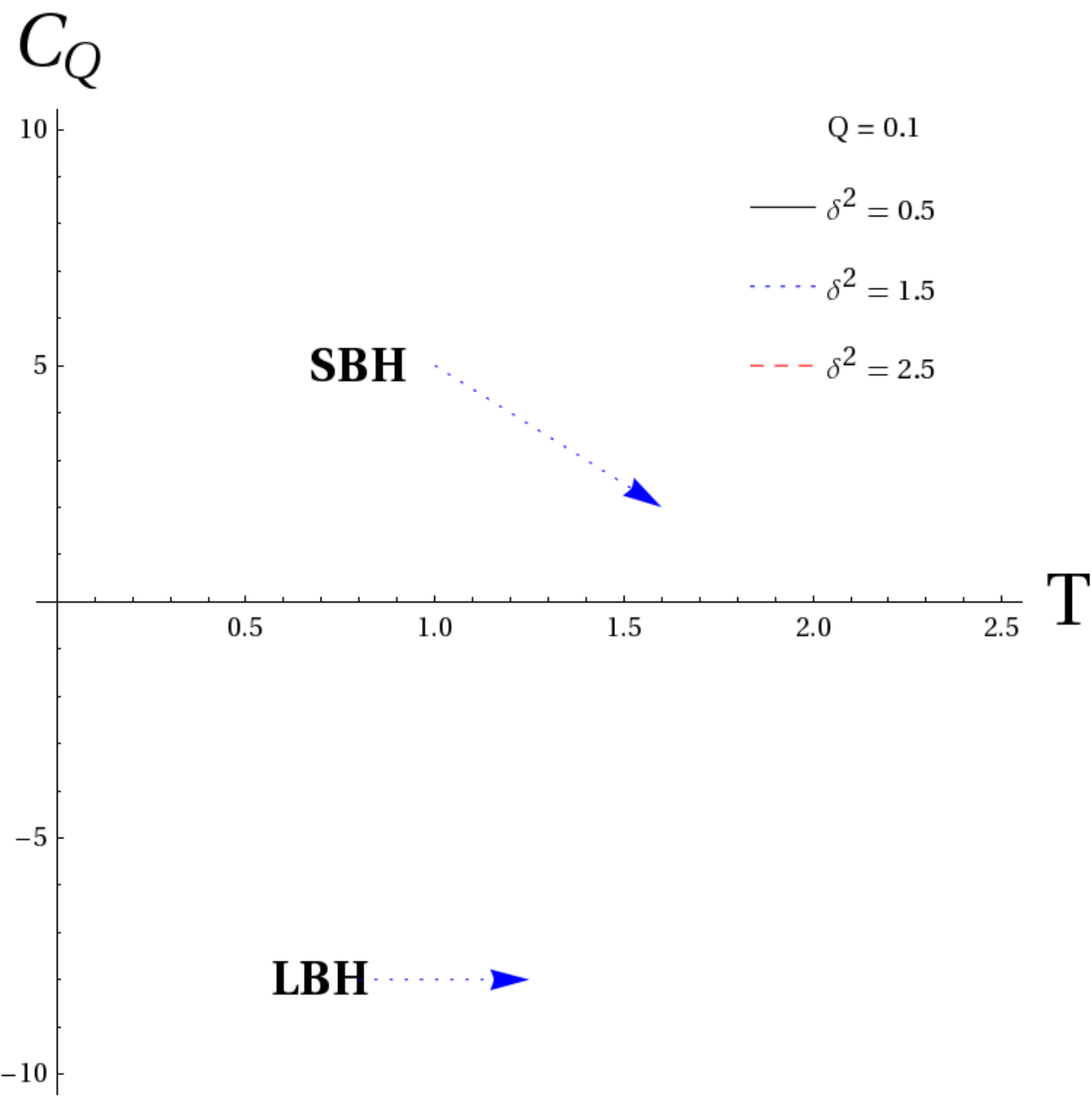}
	 &\includegraphics[width=0.45\textwidth]{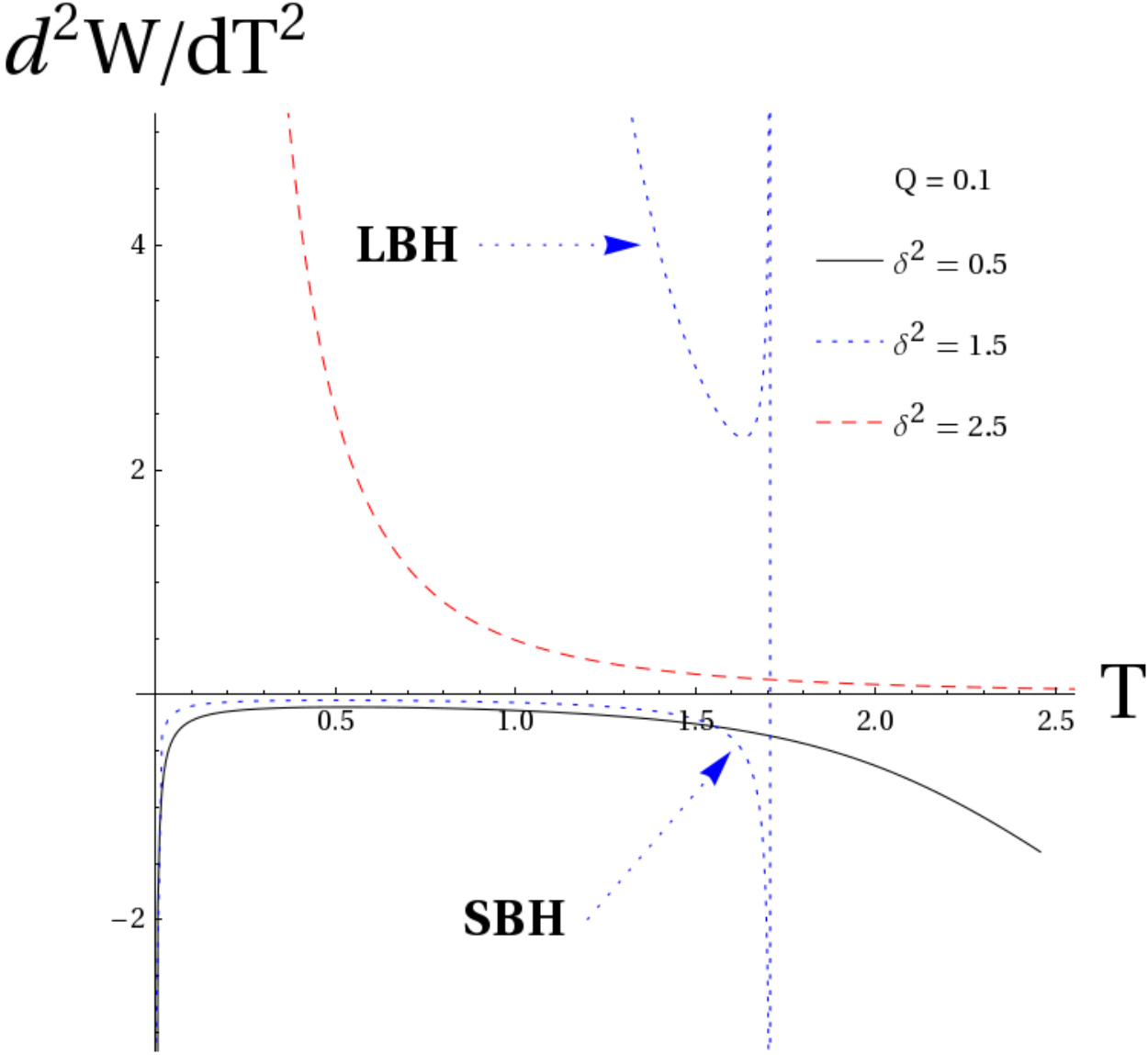}
 \end{tabular}
\caption[Local thermal stability for the planar $\ga\da=1$ EMD solution]{Heat capacity and double derivative of the Helmholtz potential against the temperature as a function of the temperature for the planar $\ga\da=1$ EMD solution \eqref{Sol1}.}
\label{Fig:HeatCapacityGaDa1Canonical}
\end{center}
\end{figure}

\begin{figure}[t]
\begin{center}
\begin{tabular}{cc}
	 \includegraphics[width=0.45\textwidth]{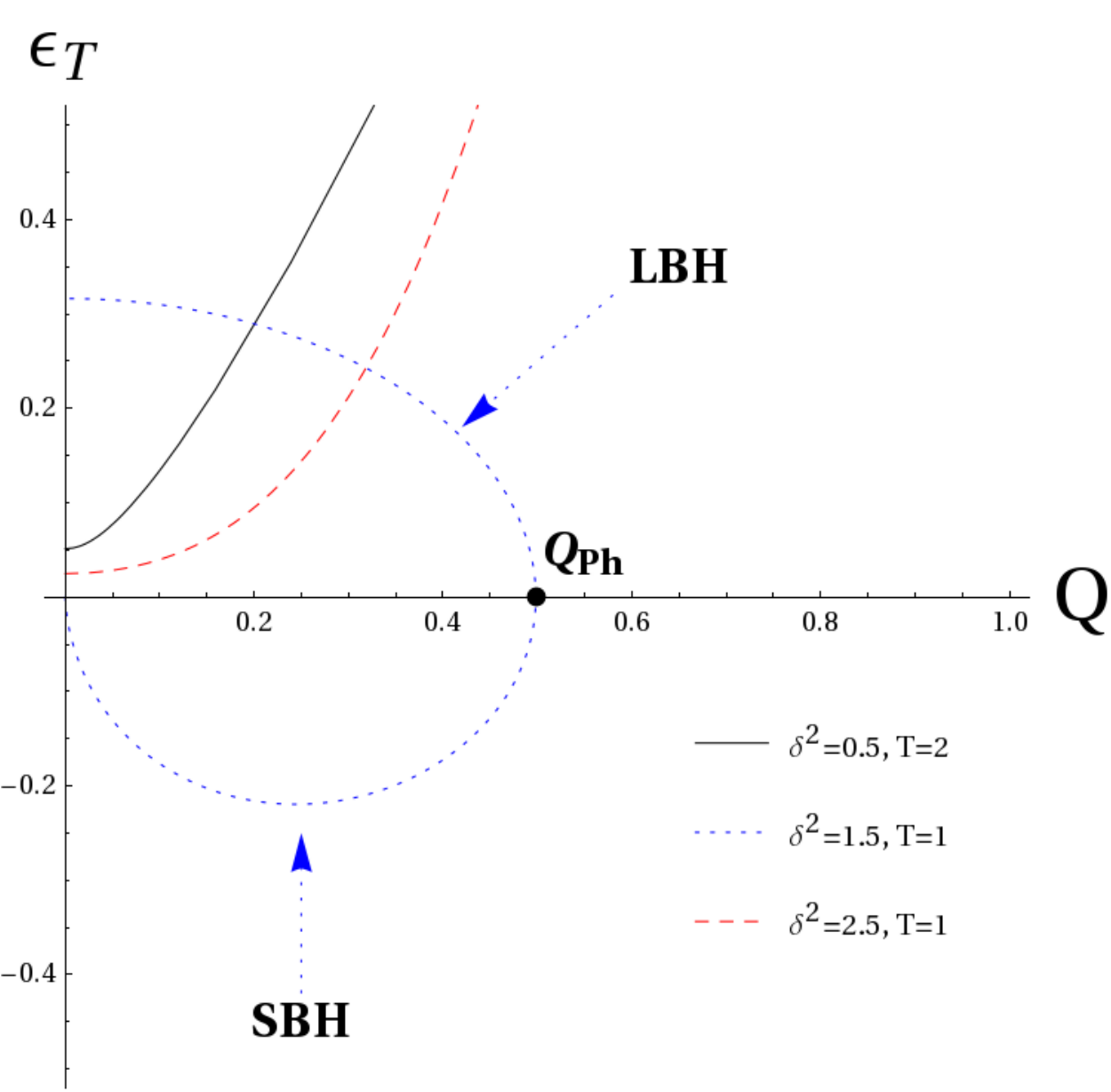}&	 
	 \includegraphics[width=0.45\textwidth]{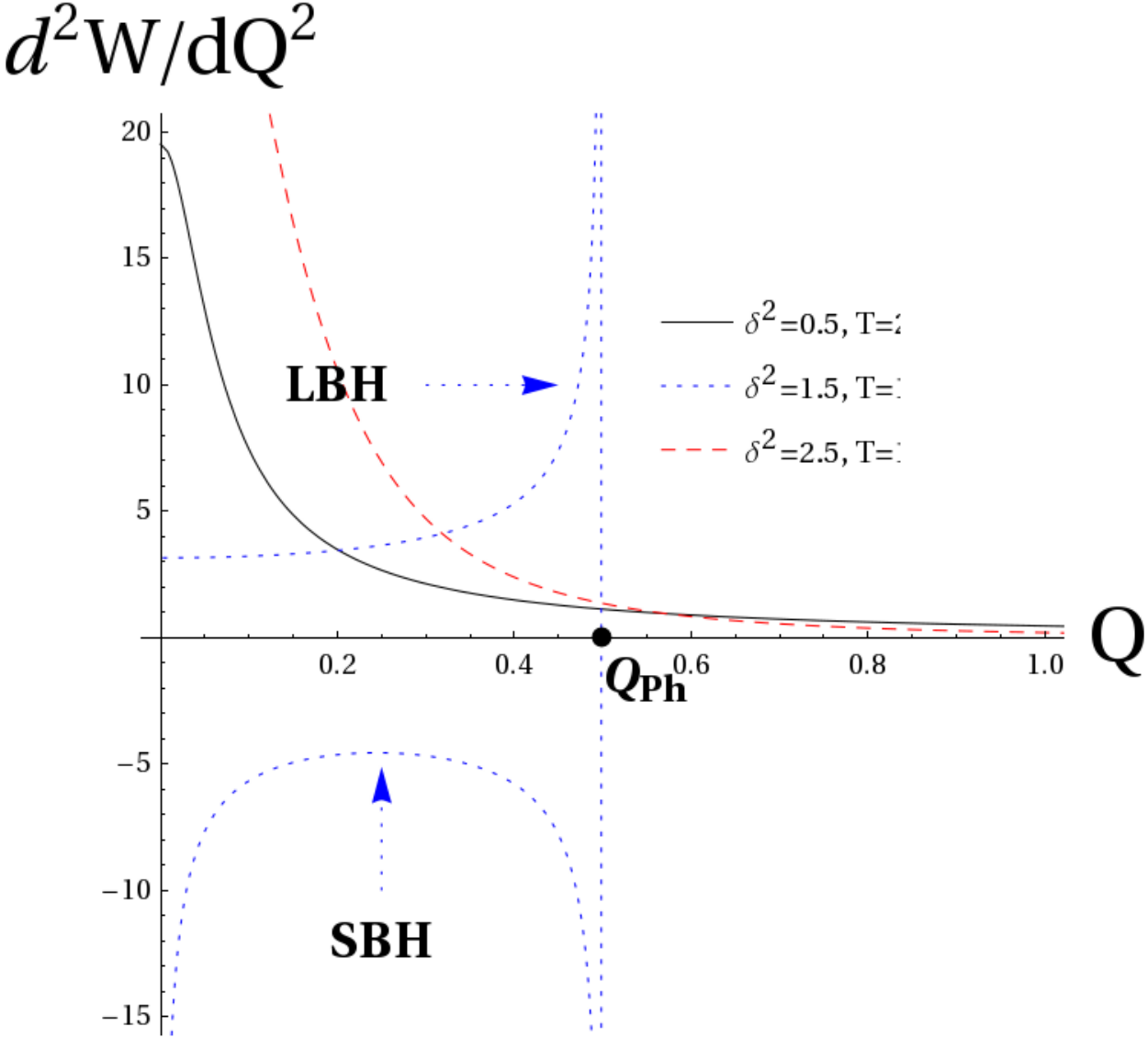}
 \end{tabular}
\caption[Local electric stability for the planar $\ga\da=1$ EMD solution]{Electric permittivity and double derivative of the Helmholtz potential against the charge for the planar $\ga\da=1$ EMD solution \eqref{Sol1}.}
\label{Fig:ElectricStabilityGaDa1Canonical}
\end{center}
\end{figure}

We also derive the heat capacity at constant electric charge, \Figref{Fig:HeatCapacityGaDa1Canonical},
\be
	C_{Q} = \frac{2S}{1-\da^2}\l[1-\frac{4(2\da^4-5\da^2-1)Q^2}{\da^2(3-\da^2)^2r_+^{6-2\da^2}}\r]\l[1-\frac{4(-5\da^4+12\da^2+1)Q^2}{\da^2(3-\da^2)^2(\da^2-1)r_+^{6-2\da^2}}\r]^{-1}\,.
	\label{HeatCapacity1}
\ee
The electric permittivity at constant temperature is by \Figref{Fig:ElectricStabilityGaDa1Canonical}
\be
	\epsilon_T = \frac{(3-\da^2)r_+^{3-\da^2}}{16\ell^{2-\da^2}}\frac{1-\frac{4(-5\da^4+12\da^2+1)Q^2}{\da^2(3-\da^2)^2(\da^2-1)r_+^{6-2\da^2}}}{1-\frac{4(1+\da^2)l^{4-2\da^2}Q^2}{\da^2(3-\da^2)^2r_+^{6-2\da^2}}}\,.
	\label{ElectricPermittivity1}
\ee

This ensures that the following first law is verified,
\bea
	\ud(M - M_e) = T\ud S + (\Phi-\Phi_e)\ud Q &\Longleftrightarrow& \ud M_W = T\ud S_W +\Phi_W\ud Q\,,
\eea
as expected since we use the extremal black hole as a thermal background.

Let us study the energetic competition between the black holes and the thermal background, e.g. the extremal limit.
\begin{itemize}
 \item Lower range: the single black hole branch is energetically favoured over the thermal background as the Helmholtz potential is always negative, all the way to the $Q=0$ axis for fixed temperature, where the neutral black holes are seen to dominate instead of the extremal background. Moreover, the black holes are stable both to thermal  \Figref{Fig:HeatCapacityGaDa1Canonical} and electric fluctuations \Figref{Fig:ElectricStabilityGaDa1Canonical}.
 \item Intermediate range : the Helmholtz potential is always more negative for the small black holes branch than for the large black holes branch. The latter crosses the $F=0$ plane at some $(T,Q)$ but this is irrelevant to the thermodynamics. Both branches exist only in a certain region of the $(T,Q)$ plane limited by the line $(T_\mathrm{Ph},Q_\mathrm{Ph})$ and beyond which only the thermal background exists. At $Q=0$ and for finite temperature, the neutral black holes have positive free energy (they are the endpoint of the large black holes branch) and so it is the extremal background that dominates (which is the endpoint of the small black holes branch). The small black holes always dominate the phase space, and are stable to thermal fluctuations \Figref{Fig:HeatCapacityGaDa1Canonical} but not to electric fluctuations  \Figref{Fig:ElectricStabilityGaDa1Canonical}. The large black holes display the opposite behaviour.
 
 Note that we recover thermodynamics typical of charged spherical Reissner-Nordstr\"om solutions as in Section \ref{section:ThermoRN}.
 \item Upper range: Since the Helmholtz potential is positive for all values $(T,Q)$, the regular black holes are always disfavoured compared to the thermal background.
\end{itemize}
In the case of the lower and intermediate range, these plots are to be compared with Fig.5 in \cite{Chamblin:1999tk} and Figs.4, 6 in \cite{Chamblin:1999hg} in which the so-called branch 3 does not exist as argued above.

\begin{figure}[t]
\begin{center}
\begin{tabular}{cc}
	 \includegraphics[width=0.45\textwidth]{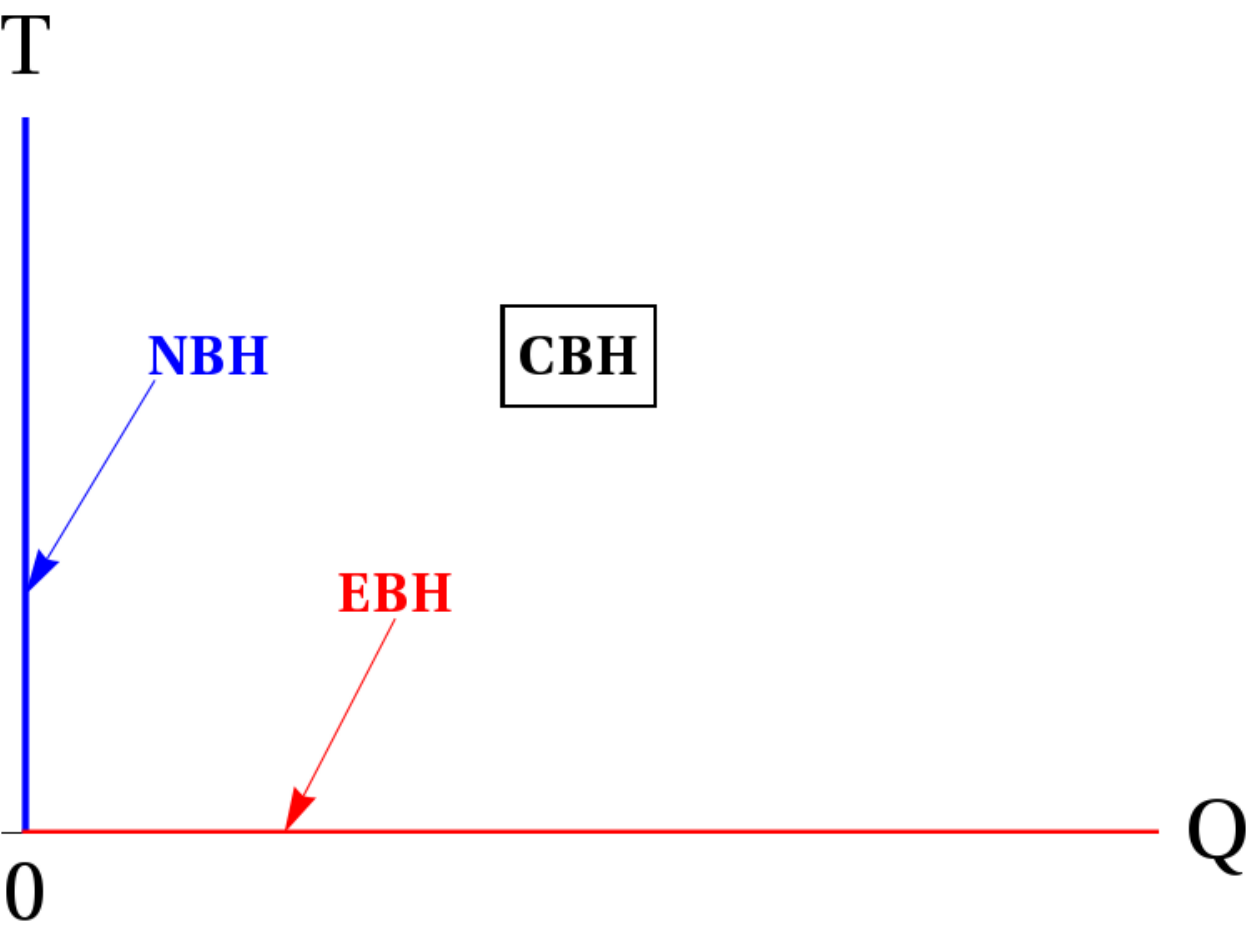}&
	 \includegraphics[width=0.45\textwidth]{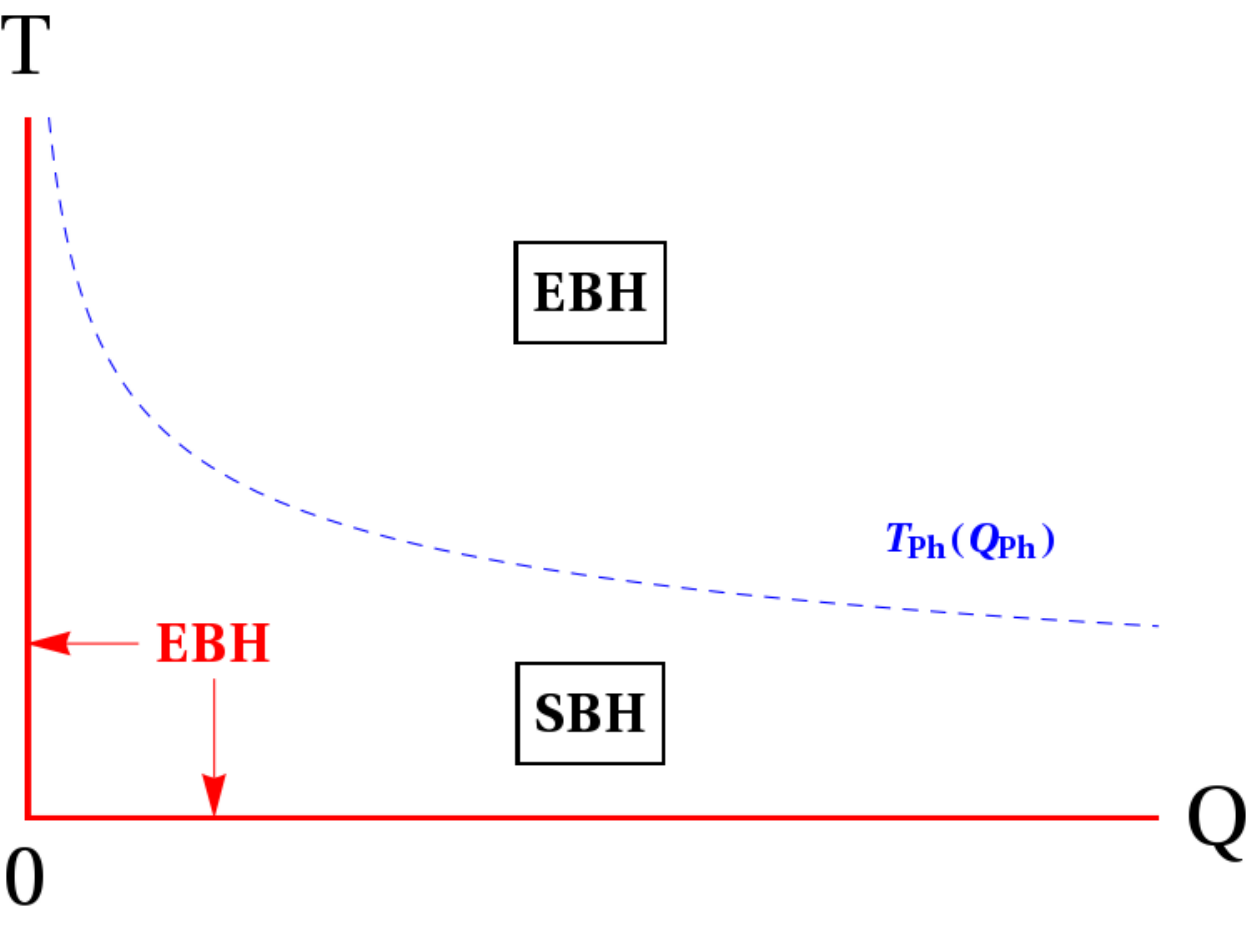}
\end{tabular}
\caption[Phase diagram for the planar $\ga\da=1$ EMD solution in the canonical ensemble]{$(T,Q)$ phase diagram for the planar $\ga\da=1$ EMD solution \eqref{Sol1} in the lower (left) and the intermediate (right) range. EBH = Extremal Black Holes, CBH = Charged Black Holes, SBH = Small Black Holes.}
\label{Fig:PhaseDiagramGaDa1Canonical}
\end{center}
\end{figure}

\begin{figure}[!ht]
\begin{center}
\begin{tabular}{c}
	 \includegraphics[width=0.45\textwidth]{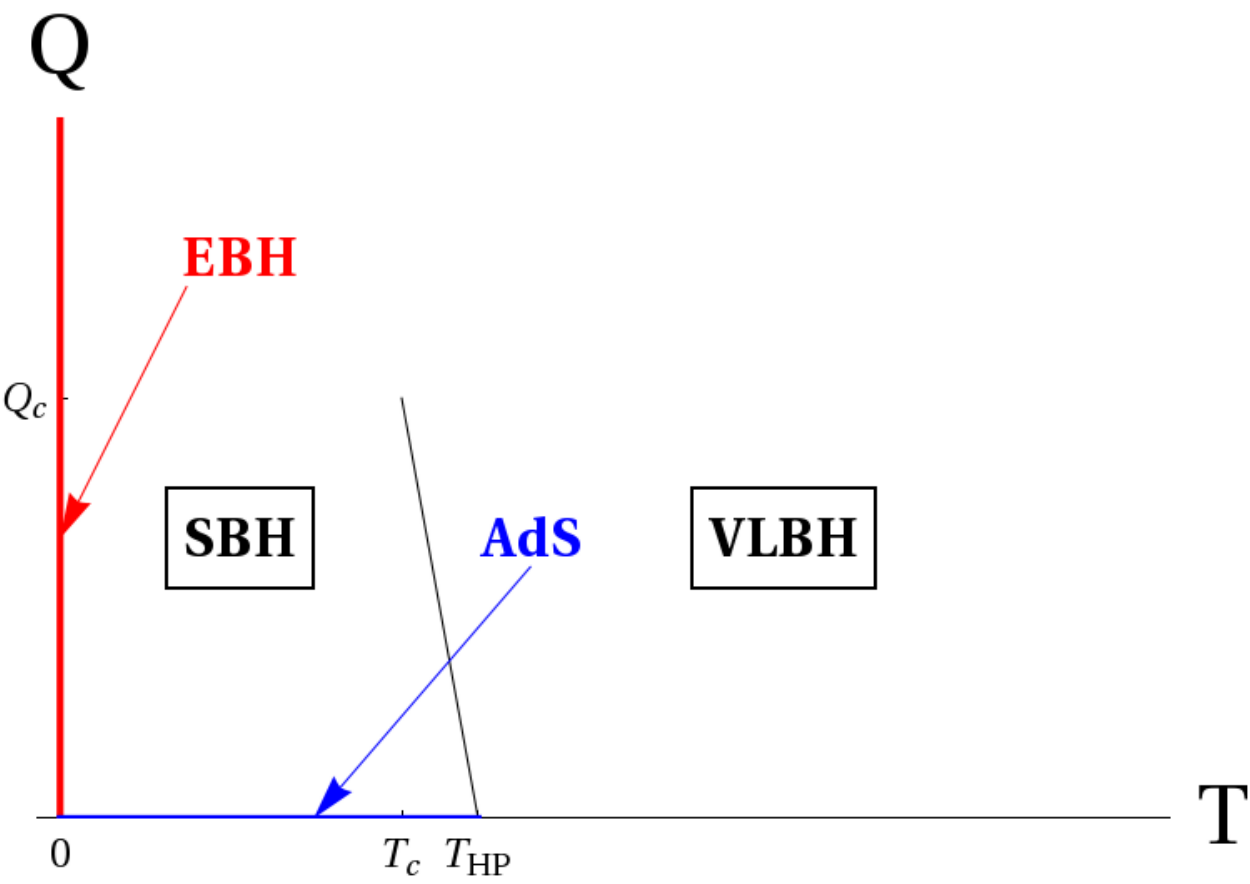}
\end{tabular}
\caption[Phase diagram for the AdS-Reissner-Nordtr\"om solution in the canonical ensemble]{$(T,Q)$ phase diagram in the case of spherical AdS-RN black holes on the bottom line, \cite{Chamblin:1999tk}.}
\label{Fig:PhaseDiagramRNAdSCanonical}
\end{center}
\end{figure}

The order of phase transitions is as follows, inspecting the Helmholtz potential \Figref{Fig:FreeEnergyGaDa1Canonical} and eq.~\eqref{Helmholtz1}.
\begin{itemize}
 \item Lower range: There is a continuous phase transition to the extremal background at finite charge in the zero-temperature limit. Inspecting \eqref{Helmholtz1}, one finds that the transition is third-order for $0 < \d^2 < 1-\frac{2}{\sqrt{5}}$ and second-order for $1-\frac{2}{\sqrt{5}} < \delta^2 < 1$. There is no discontinuity in any derivative of the free energy at the points $\delta^2=(0,1\pm \frac{2}{\sqrt{5}})$, but subleading corrections to the scalar potential might induce critical behaviour here.

 For zero charge and small temperature, the neutral black holes dominate. This is the range in which these black holes have a reasonable thermodynamic limit. The phase transition from the charged black holes is at least of fourth-order, and of $n^\textrm{th}$-order for
 \be
 \label{PTuncharged}
 \frac{4-n}{n}<\delta^2 < \frac{n-3}{n-1}\,,\quad n=4,5,6,\dots
 \ee
Note that the upper bound goes to $1$ as $n\to+\infty$, thus spanning the whole lower range and admitting phase transitions of arbitrary order. Again, there is no transition at the boundary of these intervals, but transitions can be induced through subleading terms in the scalar potential.

Finally, approaching the zero charge axis of the phase diagram at finite temperature never shows critical behaviour, since the free energy \eqref{Helmholtz1} is a series expansion in integer powers of $Q$. The charged black holes settle continuously and without transition in their stable endpoints, the neutral black holes.

 \item Intermediate range: There is a zeroth-order (discontinuous) phase transition to the thermal background at the point $(T_\mathrm{Ph},Q_\mathrm{Ph})$ since the Helmholtz potential is discontinuous there and jumps to zero.

At finite charge, in the zero-temperature limit, the Helmholtz potential and its first derivative are continuous, but higher-derivatives  diverges for the small black holes branch: there is a continuous phase transition to the thermal background of $n^\textrm{th}$-order for the parameter values
\be
1 + 2 \sqrt{\frac{n-2}{3n-4}} < \delta^2 < 1 + 2 \sqrt{\frac{n-1}{3n-1}}\,.
\ee
Note that as $n\to+\infty$, one reaches the endpoint of the Intermediate Range, $\da^2=1+2/<\sqrt3$, so phase transitions of all orders are possible. At the endpoints of these intervals there are no phase transitions with our choice of potential, but again subleading corrections to the scalar potential might change this. 

At zero charge and finite temperature, the neutral black holes are unstable and the thermal background dominates.

At finite temperature, in the zero charge limit, one follows the stable branch (the small black-holes) to its endpoint, which has zero free energy and coincides with the thermal background. This is consistent with the fact that in the neutral case,  the black holes are stable only for $\da^2<1$. There is no phase transition because the free energy is an integer-power expansion in $Q$.

 \item Upper range: The non extremal black holes are always unstable, so the background phase dominates everywhere.
\end{itemize}

We have plotted the phase diagram ($T,Q$) for the lower and intermediate ranges in \Figref{Fig:PhaseDiagramGaDa1Canonical}, and reproduced the phase diagram for spherical RNAdS black holes from Fig.6 in \cite{Chamblin:1999tk}. As branch 3 is absent in our case, there is no first-order phase transition with branch 1 in the lower and intermediate ranges : the Hawking-Page transition is absent. In the latter, this also explains the zeroth-order phase transition, which in the vocabulary of \cite{Chamblin:1999tk} corresponds to the maximal temperature where branch 1 (small black holes in our case) and branch 2 (large black holes in our case) merge. However, the fact that the extremal black hole background dominates on the $T=0$ axis is of course unchanged.

\subsection{\texorpdfstring{$\ga=\da$ charged planar black holes}{gamma=delta charged planar black holes}}

\begin{figure}[t]
\begin{center}
\begin{tabular}{cc}
	 \includegraphics[width=0.45\textwidth]{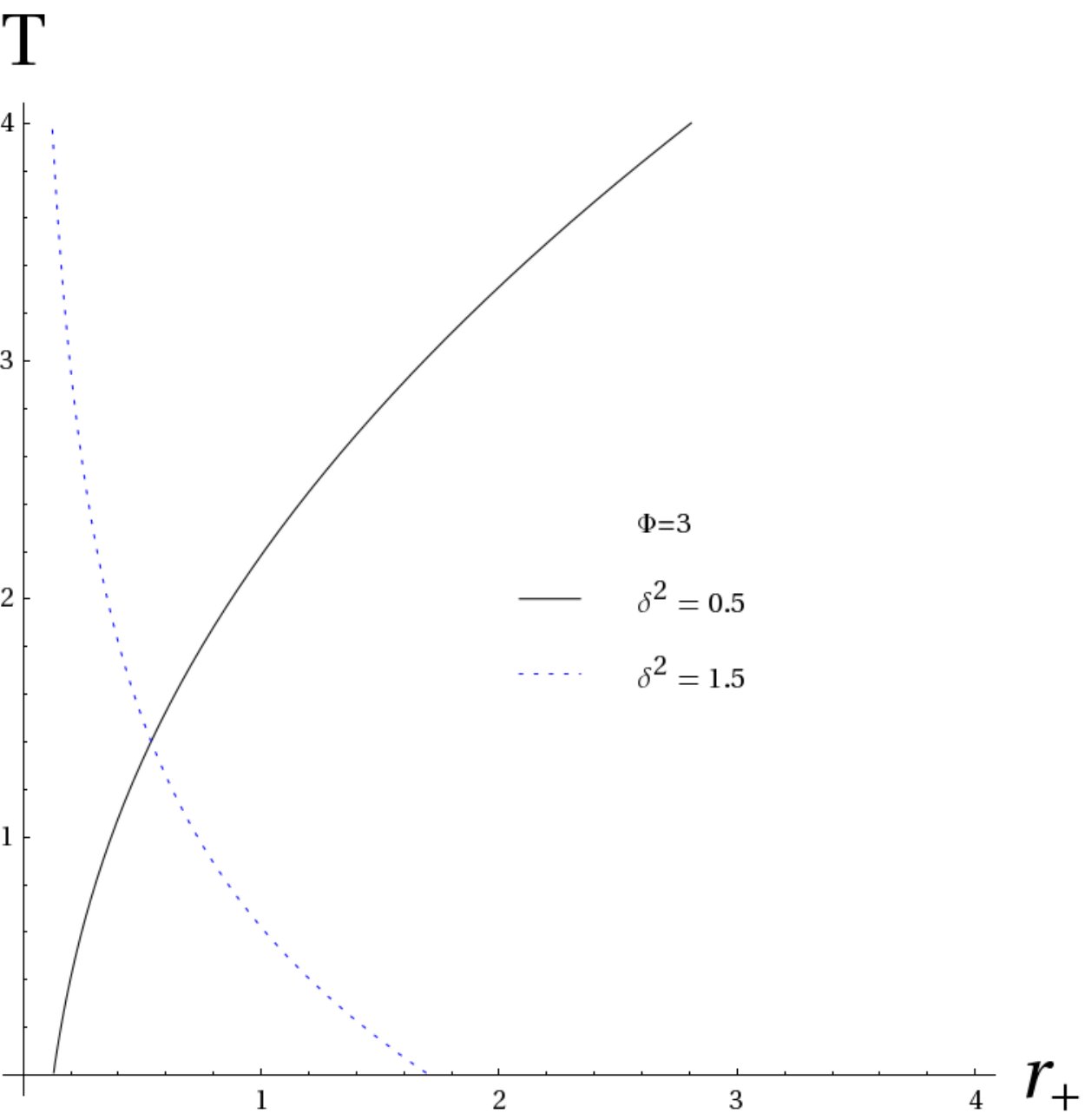}&
	 \includegraphics[width=0.45\textwidth]{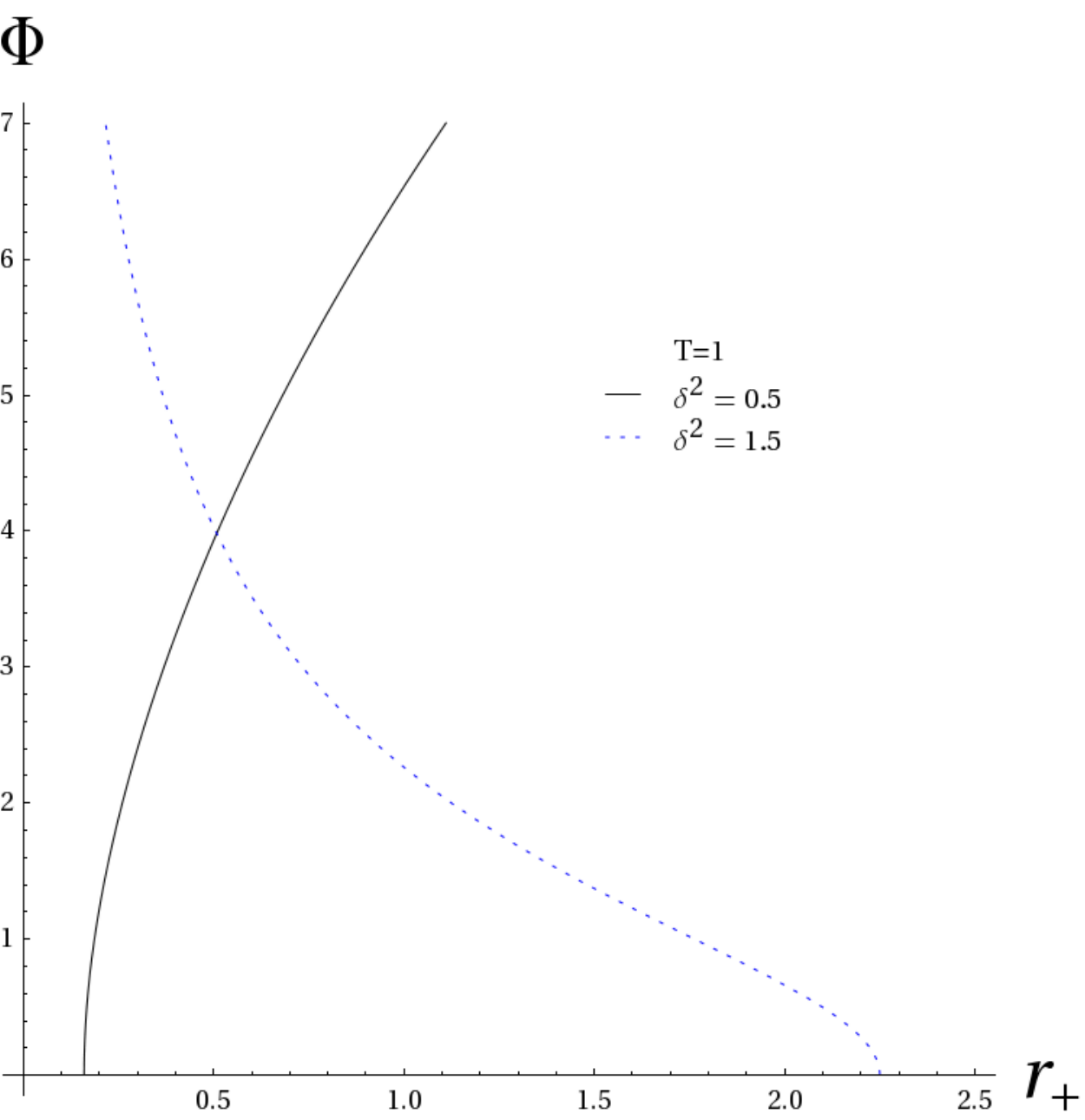}
 \end{tabular}
\caption[Equation of state for the planar $\ga=\da$ EMD solution in the grand-canonical ensemble]{Slices of the equation of state $T(r_+,\Phi)$ at fixed potential (left) and $\Phi(r_+,T)$ at fixed temperature (right) for the planar $\ga=\da$ EMD solution \eqref{Sol2}.}
\label{Fig:EqOfStateGa=DaGrandCanonical}
\end{center}
\end{figure}

\begin{figure}[t]
\begin{center}
\begin{tabular}{cc}
	 \includegraphics[width=0.45\textwidth]{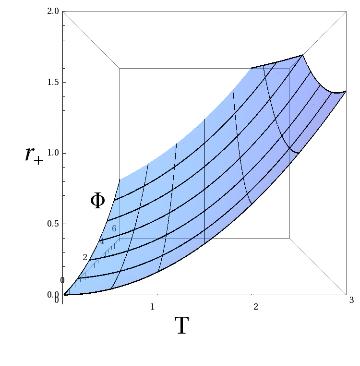}&	 
	 \includegraphics[width=0.45\textwidth]{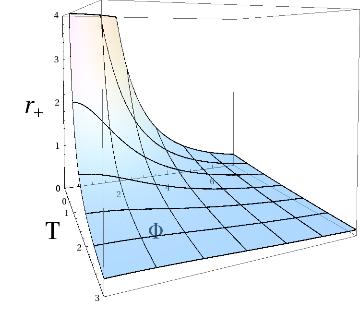}
 \end{tabular}
\caption[Three-dimensional representation of the equation of state for the planar $\ga=\da$ EMD solution in the grand-canonical ensemble]{Three-dimensional equation of state $r_+(T,\Phi)$ in the lower ($0\leq\da^2<1$) and upper ($1<\da^2<3$) range from left to right, for the planar $\ga=\da$ EMD solution \eqref{Sol2}.}
\label{Fig:EqOfState3DGa=DaGrandCanonical}
\end{center}
\end{figure}

Throughout this section and both for the grand-canonical and canonical case, several cases will be distinguished depending on the value of  $\da^2$. However, we will only plot one curve per range, representative of the behaviour in this range. Where useful, we will also display the behaviour at the limiting values for these ranges.

The temperature of the planar $\ga=\da$ solution \eqref{Sol2} is,
\be T=\frac{(3-\da^2)}{4\pi}\ell^{\da^2-2}r_+^{1-\da^2}\l[1-\l(\frac{r_e}{r_+}\r)^{4}\r],
	\label{Temperature2}
\ee
and vanishes in the extremal case. The integrated gravitational mass of the solution from \eqref{VarGravBoundaryTerm} is
\be
	M_g = \frac{\omega_2}{4\pi}m\,,
	\label{GravitationalMass2}
\ee
and in this case the scalar contribution \eqref{ScalarBoundaryVariation} is zero. This is expected since the scalar field always has its background value, it is not backreacted on by the black hole, contrarily to \eqref{Sol1}. The electric charge \eqref{ElectricChargeEMD} is
\be
	Q = \frac{\omega_2}{16\pi}q\,.
	\label{ElectricCharge2}
\ee

Thus, there are two independent integration constants specifying the solution, as well as an independent overall scale that we have fixed to its `natural' value, e.g. the maximal number allowed by the equations of motion. They are $m$, $q$ and $\ell$ as before and there is no relation between them. $m$ and $q$ can be considered as the `reduced' mass and charge, while $\ell$ is the IR radius. However, it is possible that this is not the most general solution for $\ga=\da$, because of the method by which we obtained \eqref{Sol2}.

For the remainder of this section, we shall use the same rescalings as for the previous $\ga\da=1$ solution, as in (\ref{rescale}).

	\subsubsection{Grand-canonical ensemble}

\begin{figure}[t]
\begin{center}
\begin{tabular}{cc}
	 \includegraphics[width=0.45\textwidth]{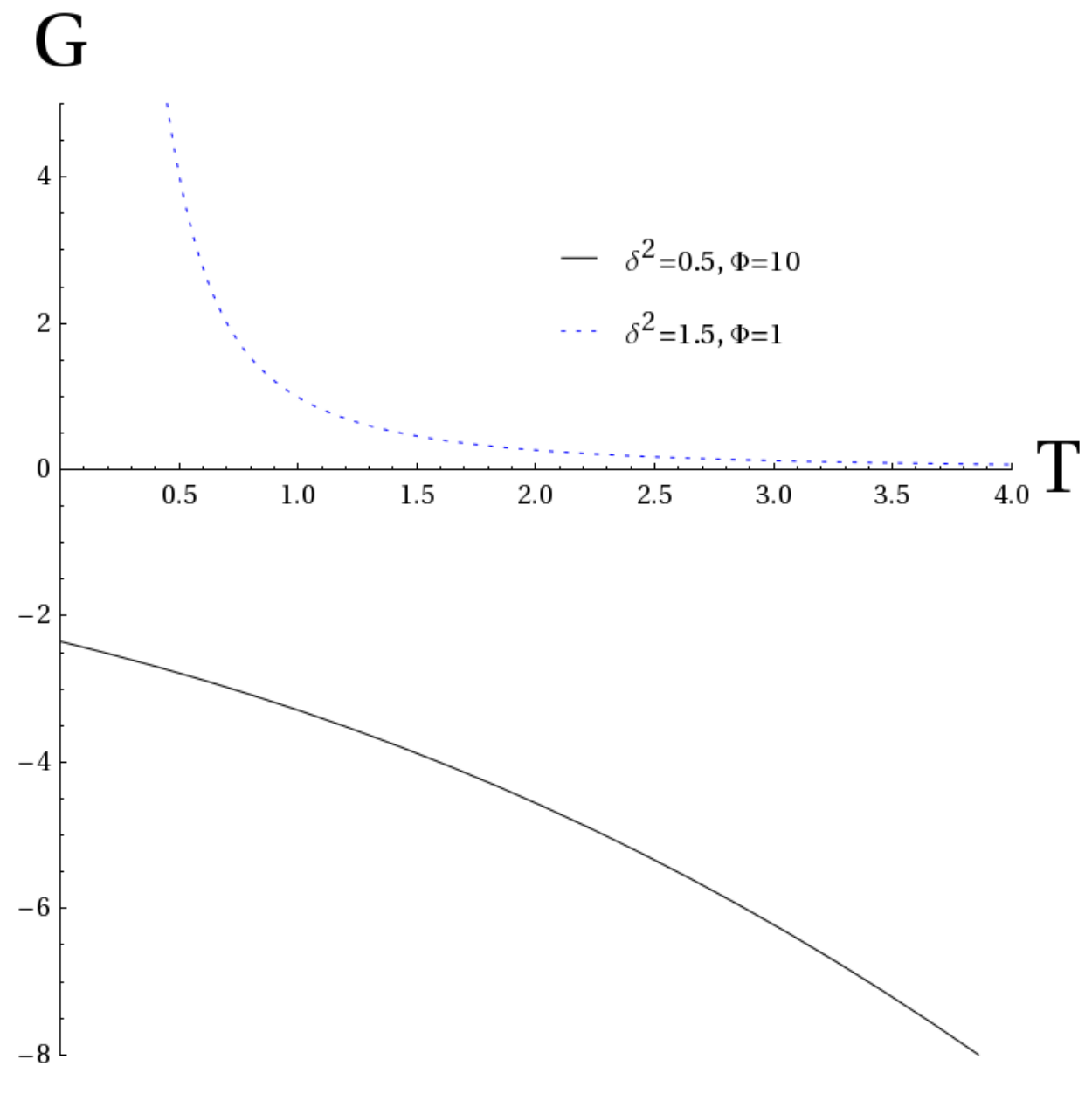}&	 \includegraphics[width=0.45\textwidth]{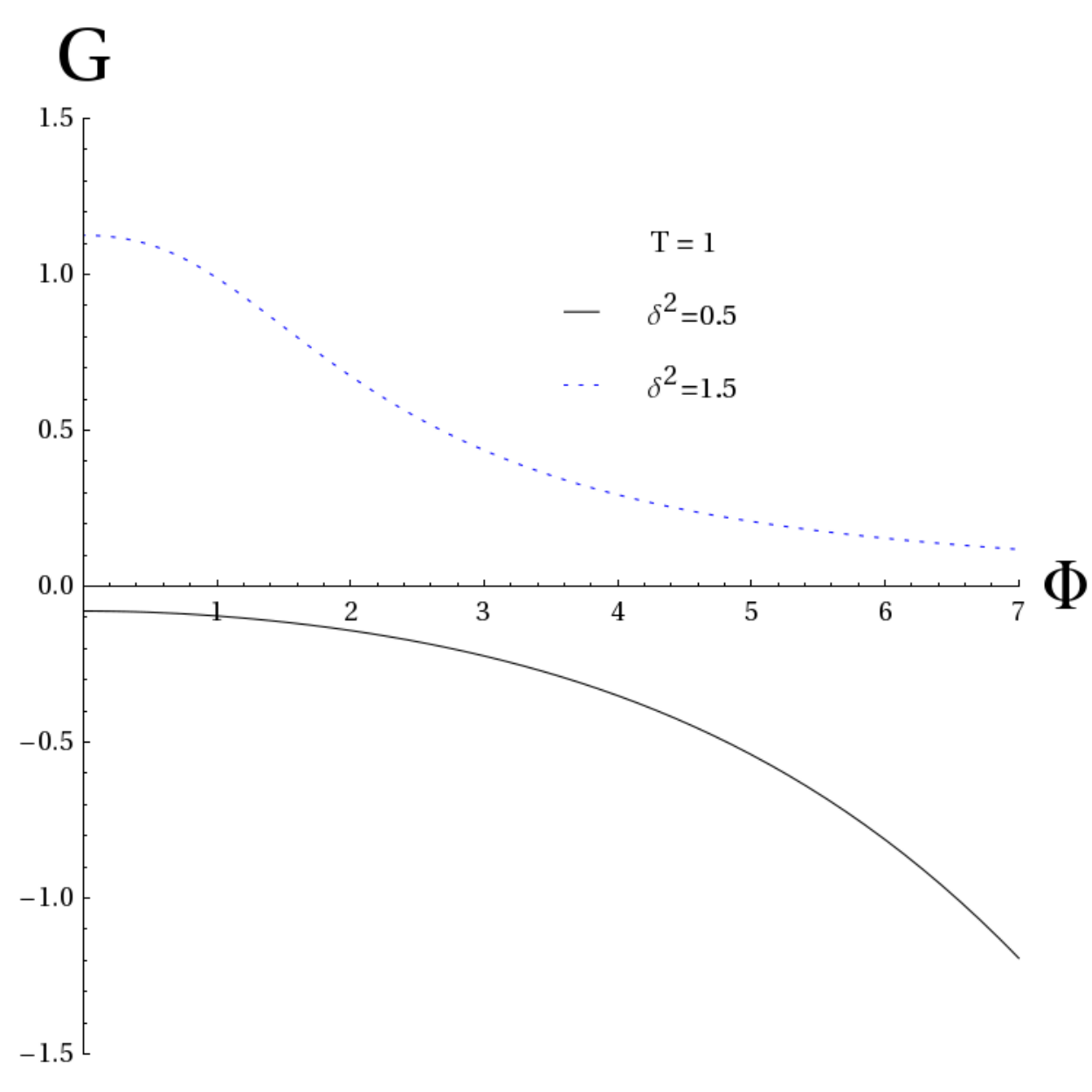}
 \end{tabular}
\caption[Free energy for the planar $\ga=\da$ EMD solution in the grand-canonical ensemble]{Slices of the Gibbs potential at fixed potential (left) and fixed temperature (right), for the planar $\ga=\da$ EMD solution \eqref{Sol2}.}
\label{Fig:GibbsGa=DaGrandCanonical}
\end{center}
\end{figure}

\begin{figure}[!ht]
\begin{center}
\begin{tabular}{cc}
	 \includegraphics[width=0.45\textwidth]{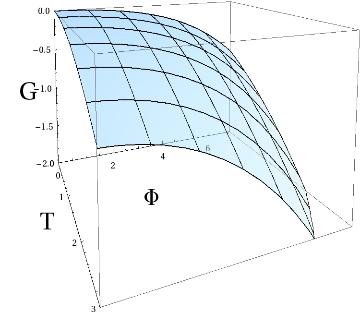}&	 
	 \includegraphics[width=0.45\textwidth]{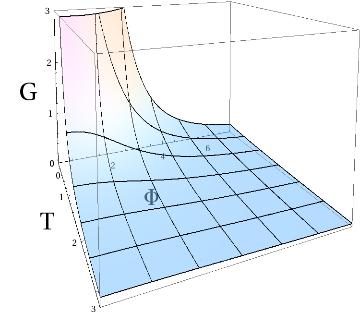}
 \end{tabular}
\caption[Three-dimensional representation of the free energy for the planar $\ga=\da$ EMD solution in the grand-canonical ensemble]{Three-dimensional representation of the Gibbs potential in the lower ($0\leq\da^2<1$) and upper ($1<\da^2<3$) range from left to right, for the planar $\ga=\da$ EMD solution \eqref{Sol2}.}
\label{Fig:Gibbs3DGa=DaGrandCanonical}
\end{center}
\end{figure}

The equation of state \eqref{Temperature2}
\be
	T= (3-\da^2)\ell^{\da^2-2}r_+^{1-\da^2}\l[1-\frac{(1+\da^2)^2\ell^{2-2\da^2}\Phi^2}{64(3-\da^2)r_+^{2-2\da^2}} \r],
\label{135}\ee
can be rewritten so as to express the horizon radius as a function of the thermodynamic variables $(T,\Phi)$
\be
	r_+^{\da^2-1} = \frac{32(3-\da^2)}{(1+\da^2)^2\Phi^2}\l[-\frac{T}{3-\da^2}+\sqrt{\frac{T^2}{(3-\da^2)^2}+\frac{(1+\da^2)^2\Phi^2}{16(3-\da^2)}} \r].
	\label{EqOfStateGa=DaGrandCanonical}
\ee
 As in the previous solution, the grand-canonical ensemble breaks down in the string limit ($\da^2= 1$) since the temperature and the chemical potential cease to be independent variables. We plot slices at fixed chemical potential and temperature of the equation of state in \Figref{Fig:EqOfStateGa=DaGrandCanonical} and the full three-dimensional representations in \Figref{Fig:EqOfState3DGa=DaGrandCanonical}.

We have to distinguish between two ranges,
\begin{itemize}
 \item Lower range $\da^2<1\,$: There is a single black hole branch. At fixed potential, the black hole grows with the temperature and the endpoint of the curve is the extremal black hole. Contrary to the previous solution, the extremal black hole can now be reached for any value of the chemical potential. At fixed temperature, the vertical axis $\Phi=0$ corresponds to the neutral black holes.
 \item Upper range $1<\da^2<3\,$: There is a single black hole branch. At fixed chemical potential, the curve starts at zero temperature at the extremal black hole, and then the radius actually decreases for the non-extremal black holes as the temperature increases. At fixed temperature, the vertical axis $\Phi=0$ displays the neutral black holes, and here also the radius diminishes as the chemical potential grows.
\end{itemize}

\begin{figure}[th]
\begin{center}
\begin{tabular}{cc}
	 \includegraphics[width=0.45\textwidth]{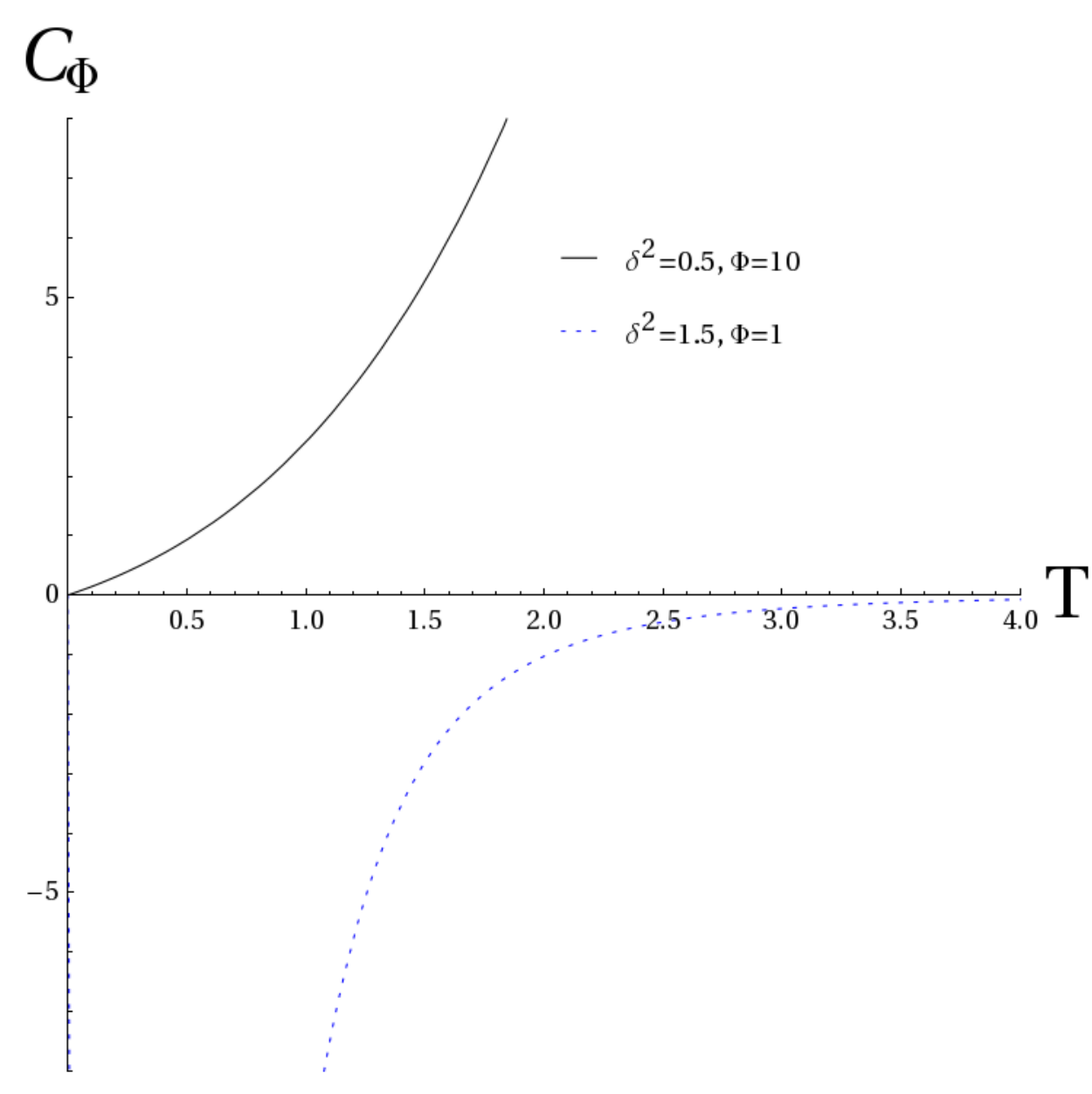}&	 \includegraphics[width=0.45\textwidth]{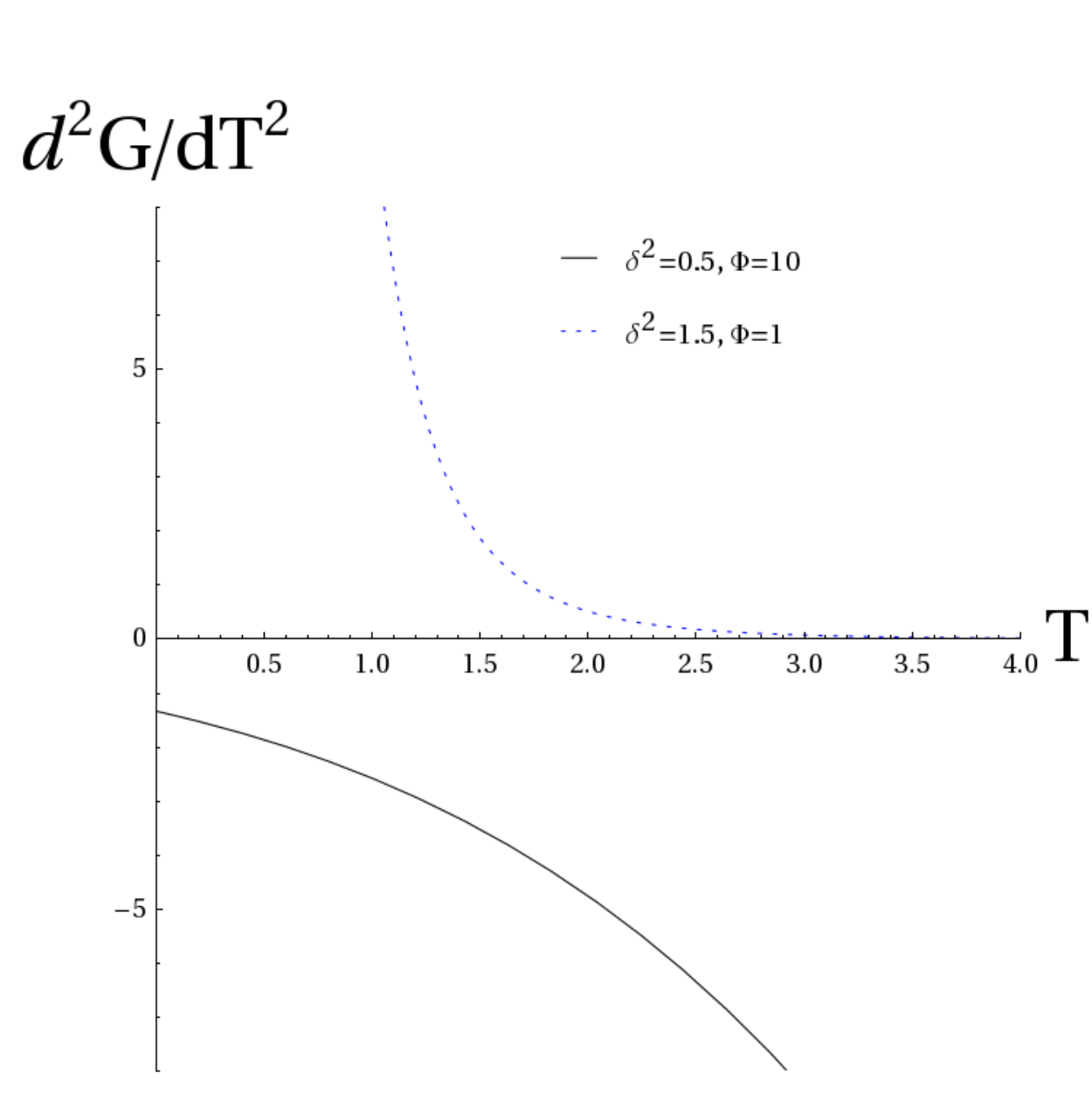}
 \end{tabular}
\caption[Local thermal stability for the planar $\ga=\da$ EMD solution in the grand-canonical ensemble]{Heat capacity (left) and second derivative of the Gibbs potential with respect to the temperature (right) at fixed potential, for the planar $\ga=\da$ EMD solution \eqref{Sol2}.}
\label{Fig:ThermalStabilityGa=DaGrandCanonical}
\end{center}
\end{figure}

We now turn to the calculation of the Gibbs thermodynamical potential. Evaluating the value of the Euclidean action \eqref{ActionGrandCanonicalEMD} and subtracting the background contribution, we find
\be
	\beta G =  I - \bar I = \beta \ell^{\da^2-2}(\da^2-1)\l[r_+^{3-\da^2}+\frac{4Q^2\ell^2}{(1+\da^2)r_+^{1+\da^2}}\r],
	\label{EuclideanActionGC2}
\ee
where we have identified the temperatures of the black hole and the thermal background on the outer boundary in order to do the subtraction. The Gibbs potential of the black hole in the thermal background \eqref{LinearDilaton} is then
\be
	G[T,\Phi] = \ell^{\da^2-2}(\da^2-1)r_+^{3-\da^2}\l[1+\frac{(1+\da^2)\Phi^2}{64}r_+^{2\da^2-2} \r],
	\label{Gibbs2}
\ee
slices of which at fixed chemical potential or fixed temperature are plotted in \Figref{Fig:GibbsGa=DaGrandCanonical}, while the three-dimensional representation is in \Figref{Fig:Gibbs3DGa=DaGrandCanonical}.

From this expression, we derive the entropy \eqref{EntropyGrandCanonical},
\be
	S=r_+^2\,,
	\label{EntropyGrandCanonical2}
\ee
which is the quarter of the area of the horizon as expected, the charge density \eqref{ElectricChargeGrandCanonical},
\be
	Q=\frac{(1+\da^2)}{16\ell^{\da^2}}r_+^{1+\da^2}\Phi\,,
\ee
which is equal to its usual value \eqref{ElectricCharge2}, and  the energy \eqref{MassGrandCanonical},
\be
	E = 2\ell^{\da^2-2}\l[r_+^{3-\da^2}+\frac{4Q^2\ell^2}{(1+\da^2)r_+^{1+\da^2}}\r]=4m = M_g\,,
	\label{EnergyGrandCanonical2}
\ee
which is equal to the gravitational mass \eqref{GravitationalMass2} as expected for a scaling dilaton. The quantities calculated by use of \eqref{MassGrandCanonical}, \eqref{EntropyGrandCanonical}, \eqref{EntropyGrandCanonical} satisfy the first law \eqref{FirstLawCanonical}.

\begin{figure}[th]
\begin{center}
\begin{tabular}{c}
	 \includegraphics[width=0.45\textwidth]{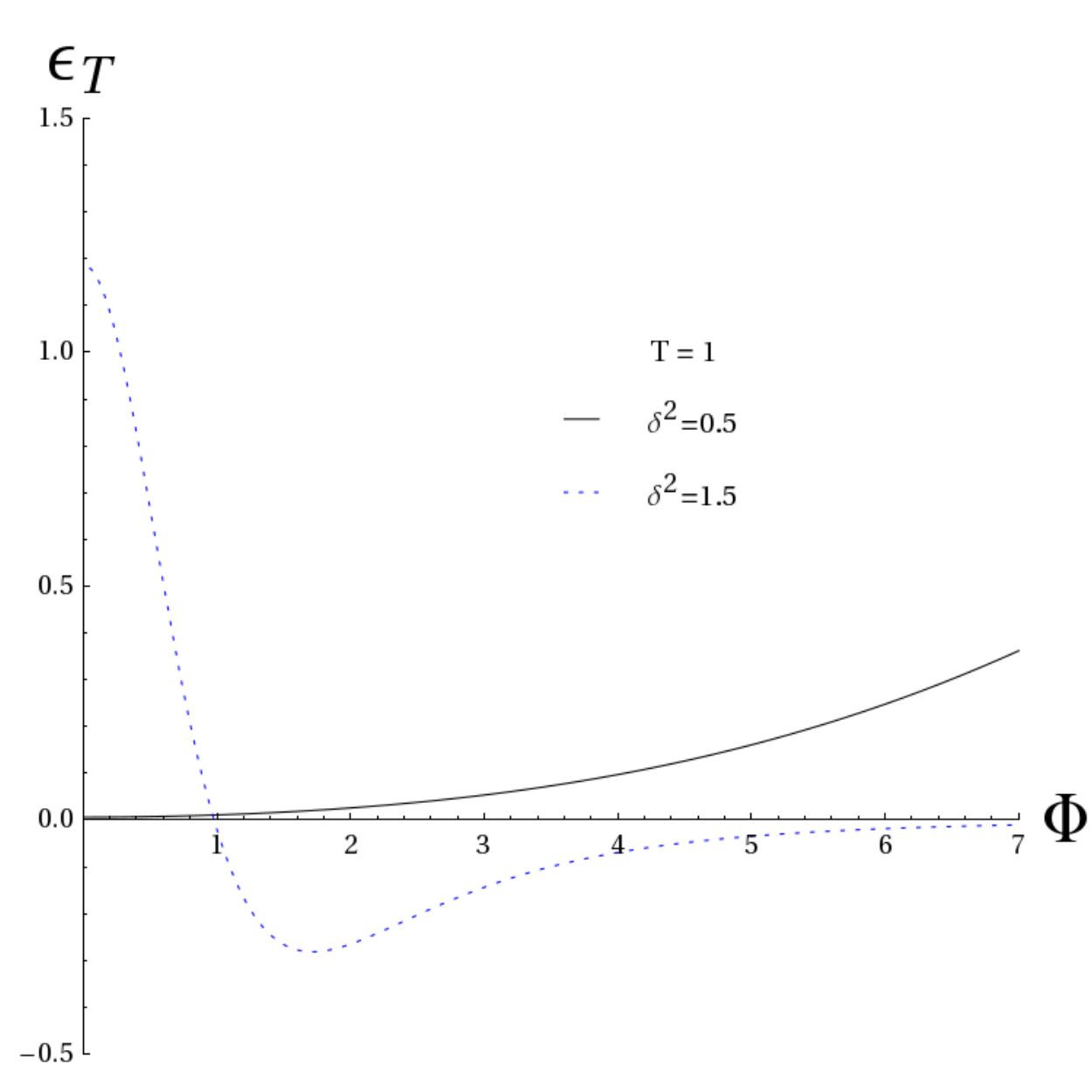}
 \end{tabular}
\caption[Local electric stability for the planar $\ga=\da$ EMD solution in the grand-canonical ensemble]{Electric permittivity at fixed temperature for the planar $\ga=\da$ EMD solution \eqref{Sol2}.}
\label{Fig:ElectricStabilityGa=DaGrandCanonical}
\end{center}
\end{figure}

Studying the thermodynamics of the solution, two behaviours have to be distinguished:
\begin{itemize}
 \item Lower range: At fixed potential, the charged black holes dominate the ensemble (\Figref{Fig:GibbsGa=DaGrandCanonical} and \Figref{Fig:Gibbs3DGa=DaGrandCanonical}) , at all temperatures, even in the extremal limit. At fixed potential, the black holes also dominate at all values of the chemical potential, including the neutral black holes at $\Phi=0$. The black holes are stable thermally and electrically, see \Figref{Fig:ThermalStabilityGa=DaGrandCanonical} and \Figref{Fig:ElectricStabilityGa=DaGrandCanonical}.
 \item Upper range: The ensemble is dominated by the dilatonic background for all $(T,\Phi)$ values.
\end{itemize}

\begin{figure}[th]
\begin{center}
\begin{tabular}{c}
	 \includegraphics[width=0.45\textwidth]{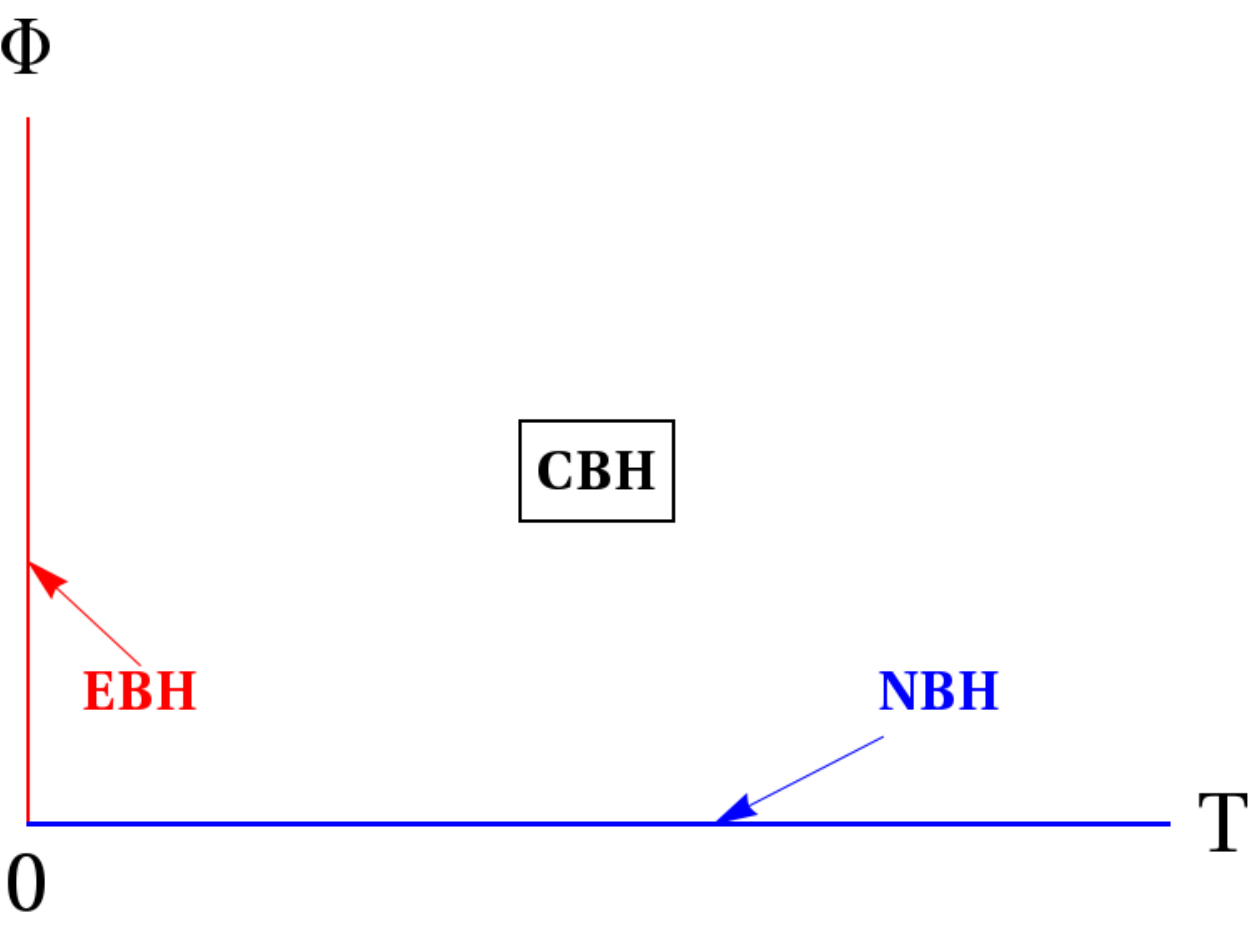}
 \end{tabular}
\caption[Phase diagram for the planar $\ga=\da$ EMD solution in the grand-canonical ensemble]{Phase diagram in the $(T,\Phi)$ phase space for the lower range $\da^2<1$ for the planar $\ga=\da$ EMD solution \eqref{Sol2}.}
\label{Fig:PhaseDiagramGa=DaGrandCanonical}
\end{center}
\end{figure}

The phase space is non-trivial only in the lower range $\da^2<1$, and is depicted in \Figref{Fig:PhaseDiagramGa=DaGrandCanonical}. In that case, the charged black holes dominate for all $(T,\Phi)$, except when $T=0$ (extremal black hole) or $\Phi=0$ (neutral black holes). There are no phase transitions to the dilatonic background in the interior of the phase diagram, the extremal black holes can exist for any values $(T,\Phi)$, contrary to the case $\ga\da=1$. These black holes behave more like AdS-RN black holes. The only critical behaviour appearing is when approaching zero temperature at $\Phi=0$, or zero chemical potential on the $T=0$ axis. In both cases there are phase transitions of $n^\textrm{th}$-order to the corresponding dominating solution in the parameter range
\be
\frac{n-4}{n-2}< \delta^2 < \frac{n-3}{n-1}\,,\quad n= 4,5,6\dots
\ee
In particular, these transitions are fourth or higher-order. In the upper range, the dilatonic background dominates everywhere and there are no phase transitions.

	\subsubsection{Canonical ensemble}

\begin{figure}[ht]
\begin{center}
\begin{tabular}{cc}
	 \includegraphics[width=0.45\textwidth]{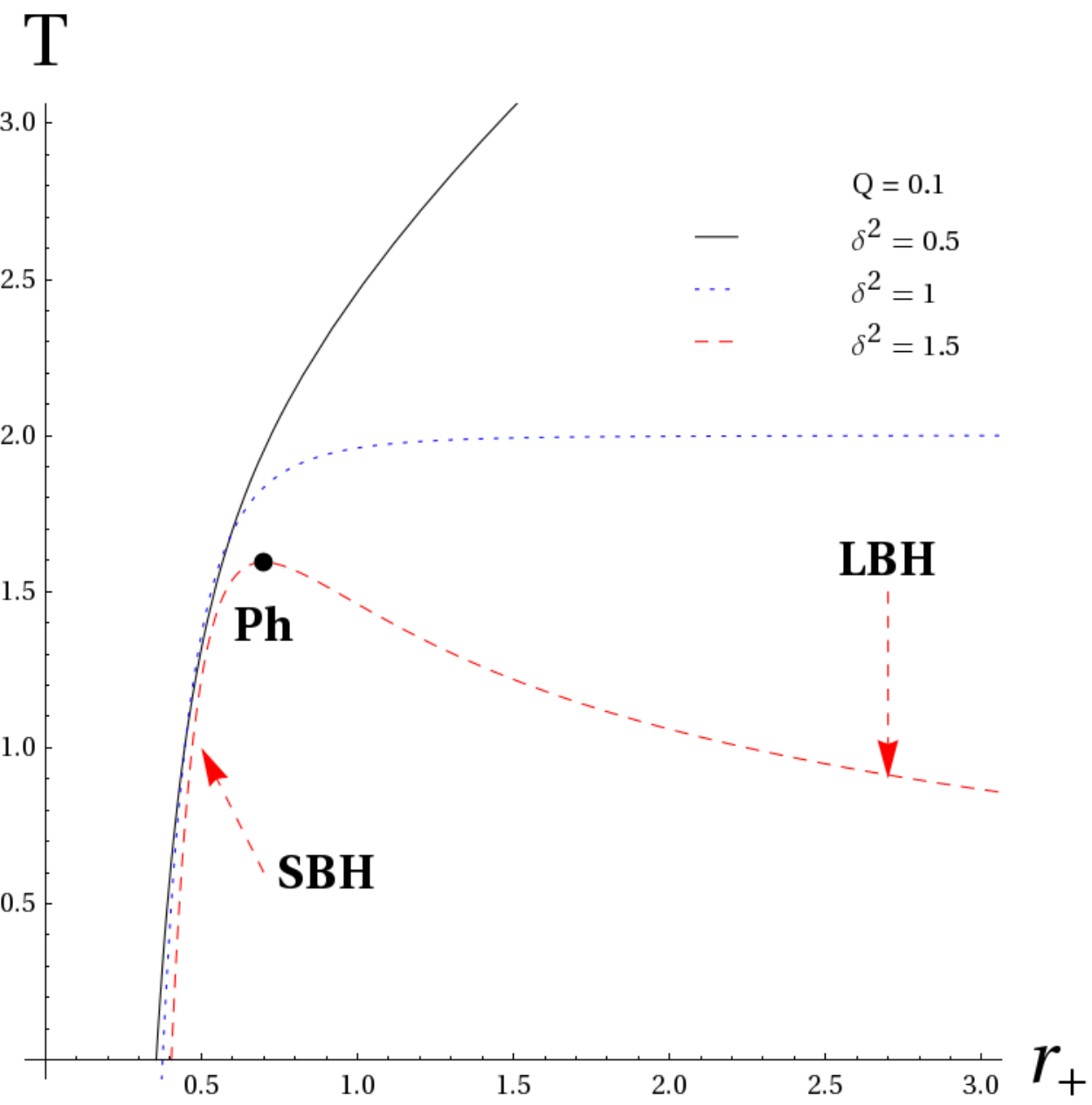}&
	 \includegraphics[width=0.45\textwidth]{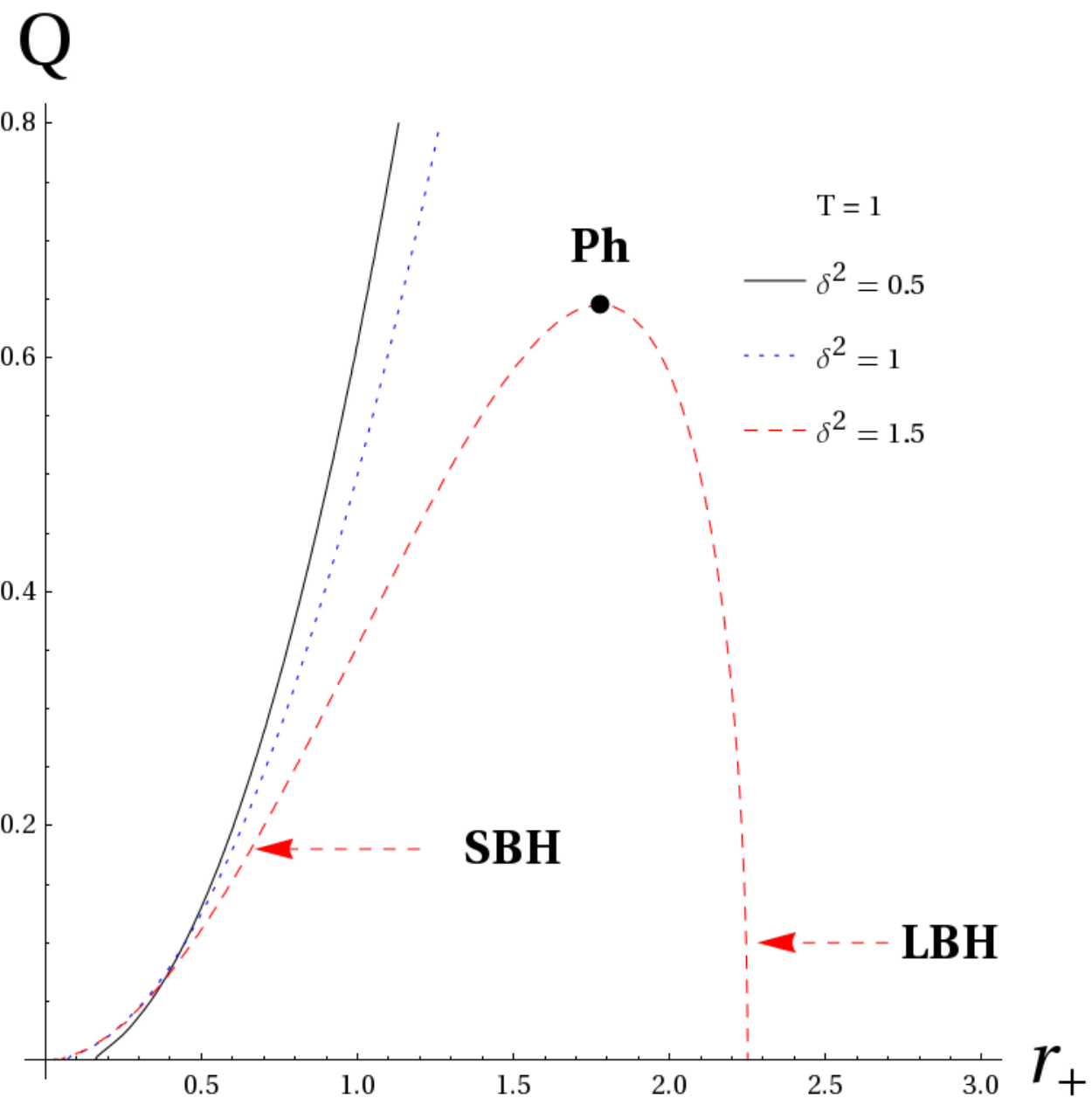}
 \end{tabular}
\caption[Equation of state for the planar $\ga=\da$ EMD solution in the canonical ensemble]{Slices of the equation of state $T(r_+,Q)$ at fixed charge (left) and $Q(r_+,T)$ at fixed temperature (right) for the planar $\ga=\da$ EMD solution \eqref{Sol2}.}
\label{Fig:HorizonSizeCanonicalGa=Da}
\end{center}
\end{figure}

\begin{figure}[!ht]
\begin{center}
\begin{tabular}{cc}
	 \includegraphics[width=0.45\textwidth]{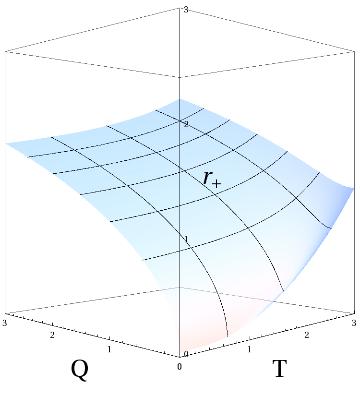}&
	 \includegraphics[width=0.45\textwidth]{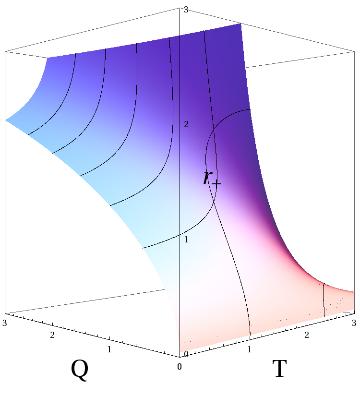}
\end{tabular}
\caption[Three-dimensional representation of the equation of state for the planar $\ga=\da$ EMD solution in the canonical ensemble]{Equation of state $r_+(T,Q)$ for the planar $\ga=\da$ EMD solution \eqref{Sol2} in the lower ($0\leq\da^2<1$) and upper ($1<\da^2<3$) range from left to right.}
\label{Fig:EqOfState3DCanonicalGa=Da}
\end{center}
\end{figure}

In the canonical ensemble, the equation of state is
\be
	 T=(3-\da^2)\ell^{\da^2-2}r_+^{1-\da^2}\l[1-\frac{4\ell^2Q^2}{(3-\da^2)r_+^{4}}\r],
	\label{EquationOfStateCanonical2}
\ee
and allows to determine implicitly the horizon radius in term of the temperature and the charge density. Slices at fixed charge density or fixed temperature are plotted in \Figref{Fig:HorizonSizeCanonicalGa=Da}, while the three-dimensional version is in \Figref{Fig:EqOfState3DCanonicalGa=Da}. We notice two behaviours depending on the value of $\da^2$:
\begin{itemize}
 \item  In the lower range, $\da^2\leq1\,$, there is a single black hole branch for each doublet $(T,Q)$, and the limiting case $\da^2=1$ has a maximal temperature at large radius, again signalling the change of behaviour in the upper range.
 \item In the upper range, $1<\da^2<3\,$, there are two branches, small black holes and large black holes, which merge at radius,
\be
	r_{Ph}^4=\frac{(3+\da^2)}{(\da^2-1)}r_e^4\,.
	\label{CriticalRadiusCanonical2}
\ee
This corresponds either to a maximal temperature at fixed charge density, or to a maximum charge at fixed temperature, see \Figref{Fig:HorizonSizeCanonicalGa=Da}. The critical point exists only for $\da^2>1$, and the line in the $(T, Q)$ phase space so defined is given by
\be T_{Ph}=4\ell^{\da^2-2}\frac{3-\da^2}{3+\da^2}\l(\frac{4(3+\da^2)\ell^2Q_{Ph}^2}{(\da^2-1)(3-\da^2)}\r)^{\frac{1-\da^2}{4}}\,.
	\label{CriticalTemperatureCanonical2}
\ee
\end{itemize}

\begin{figure}[t]
\begin{center}
\begin{tabular}{cc}
	 \includegraphics[width=0.45\textwidth]{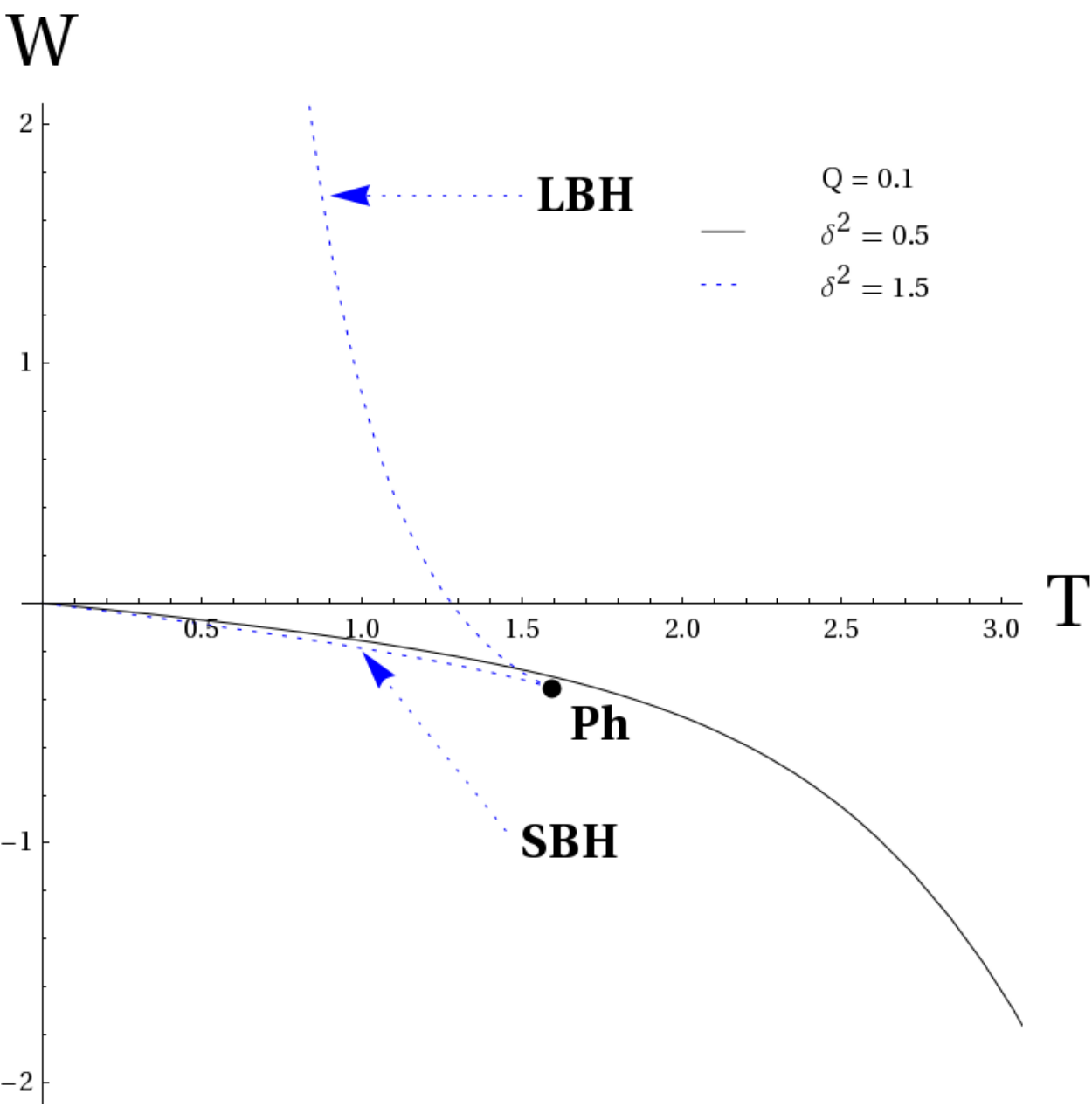}&
	 \includegraphics[width=0.45\textwidth]{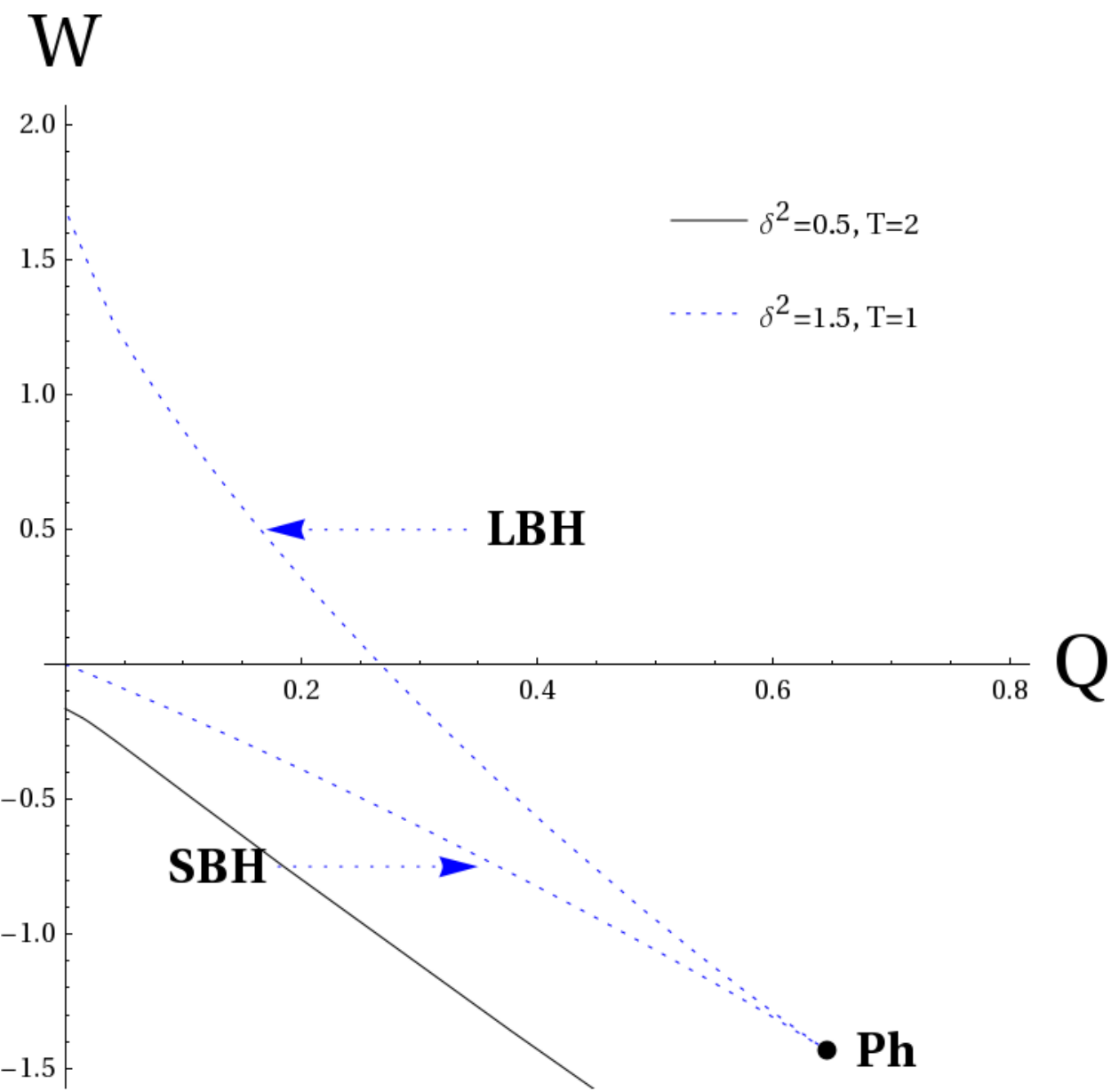}
 \end{tabular}
\caption[Free energy for the planar $\ga=\da$ EMD solution in the canonical ensemble]{Helmholtz potential versus the temperature at fixed charge (left), or versus charge at fixed temperature (right), for the planar $\ga=\da$ EMD solution \eqref{Sol2}, in the cases $\da^2=0.5$ (solid line), $\da^2=1$ (dotted line)  and $\da^2=1.5$ (dashed line).}
\label{Fig:FreeEnergyGa=DaCanonical}
\end{center}
\end{figure}

We now calculate the value of the Euclidean action \eqref{ActionCanonicalEMD}, subtracting the appropriate extremal black hole background,
\be
	I^c - I^c_e = -\beta \ell^{\da^2-2}\l[(1-\da^2)r_+^{3-\da^2}-\frac{4(3+\da^2)\ell^2Q^2}{(1+\da^2)r_+^{1+\da^2}} +\frac{8}{(1+\da^2)}r_e^{3-\da^2}\r],
\label{136}\ee
from which we can deduce the Helmholtz potential in terms of the thermodynamic variables,
\be
	W[T,Q] = \ell^{\da^2-2}\l[(\da^2-1)r_+^{3-\da^2}+\frac{4(3+\da^2)\ell^2Q^2}{(1+\da^2)r_+^{1+\da^2}} -\frac{8}{(1+\da^2)}\l(\frac{4\ell^2Q^2}{3-\da^2}\r)^{\frac{3-\da^2}4}\r].
	\label{Helmholtz2}
\ee
Slices at fixed charge density  or at fixed temperature are plotted in \Figref{Fig:FreeEnergyGa=DaCanonical}.

 We can check that the first law is satisfied using \eqref{MassGrandCanonical}, \eqref{EntropyGrandCanonical} and \eqref{EntropyGrandCanonical}:
\bsea
	E_W&=& E - E_e = 4(m-m_e)\,,\\
	\Phi_W&=& \Phi-\Phi_e = \frac{16Q\ell^{\da^2}}{(1+\da^2)}\l(r_+^{-1-\da^2}-r_e^{-1-\da^2}\r),\\
	S_W&=& S-S_e= r_+^2 = \frac{A_h}{4} \,.
\esea
The entropy, similarly to regular AdS-RN black holes, is zero at extremality, as is expected from the Euclidean path integral calculation.

\begin{figure}[t]
\begin{center}
\begin{tabular}{cc}
	 \includegraphics[width=0.45\textwidth]{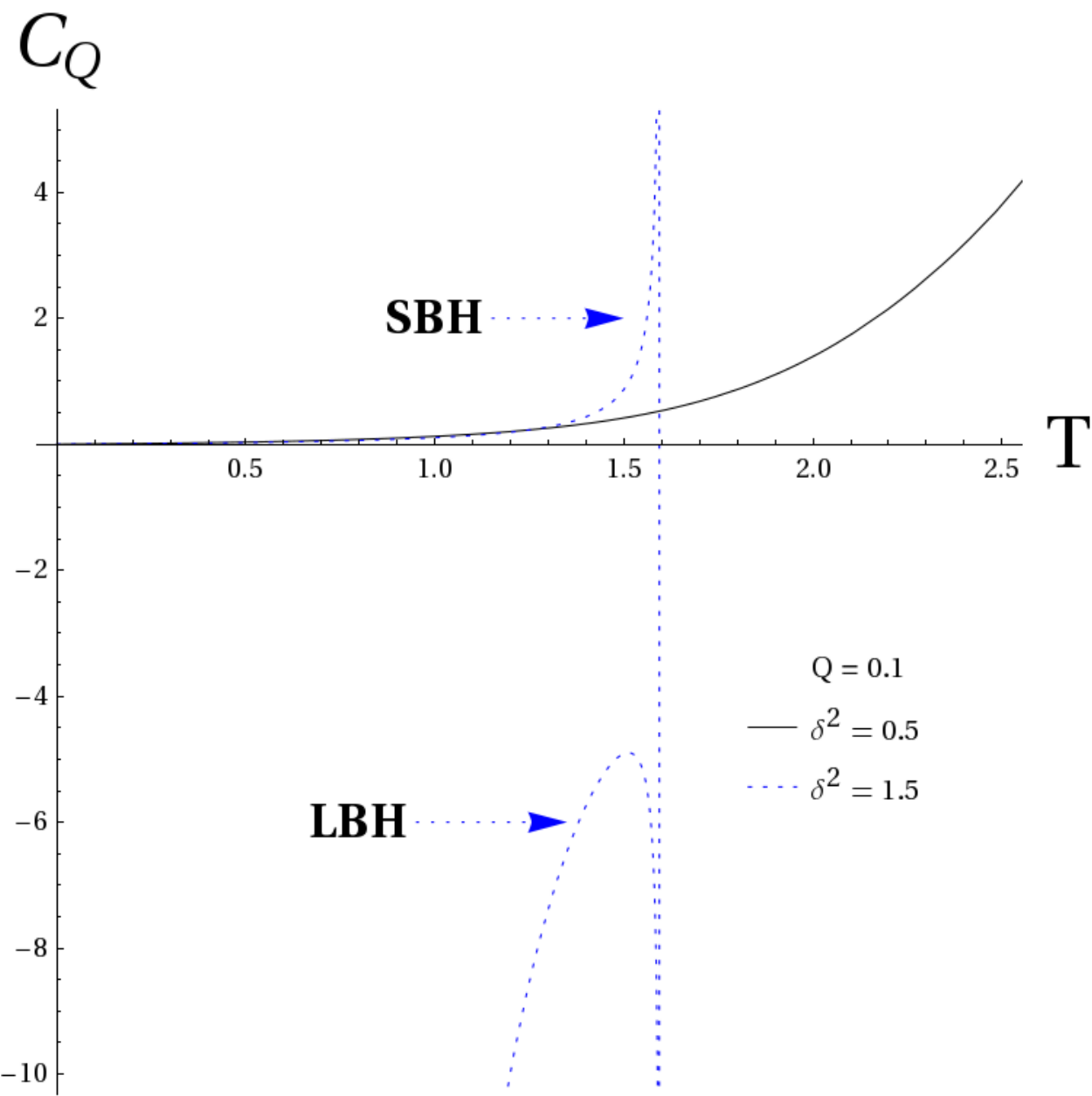}&
	 \includegraphics[width=0.45\textwidth]{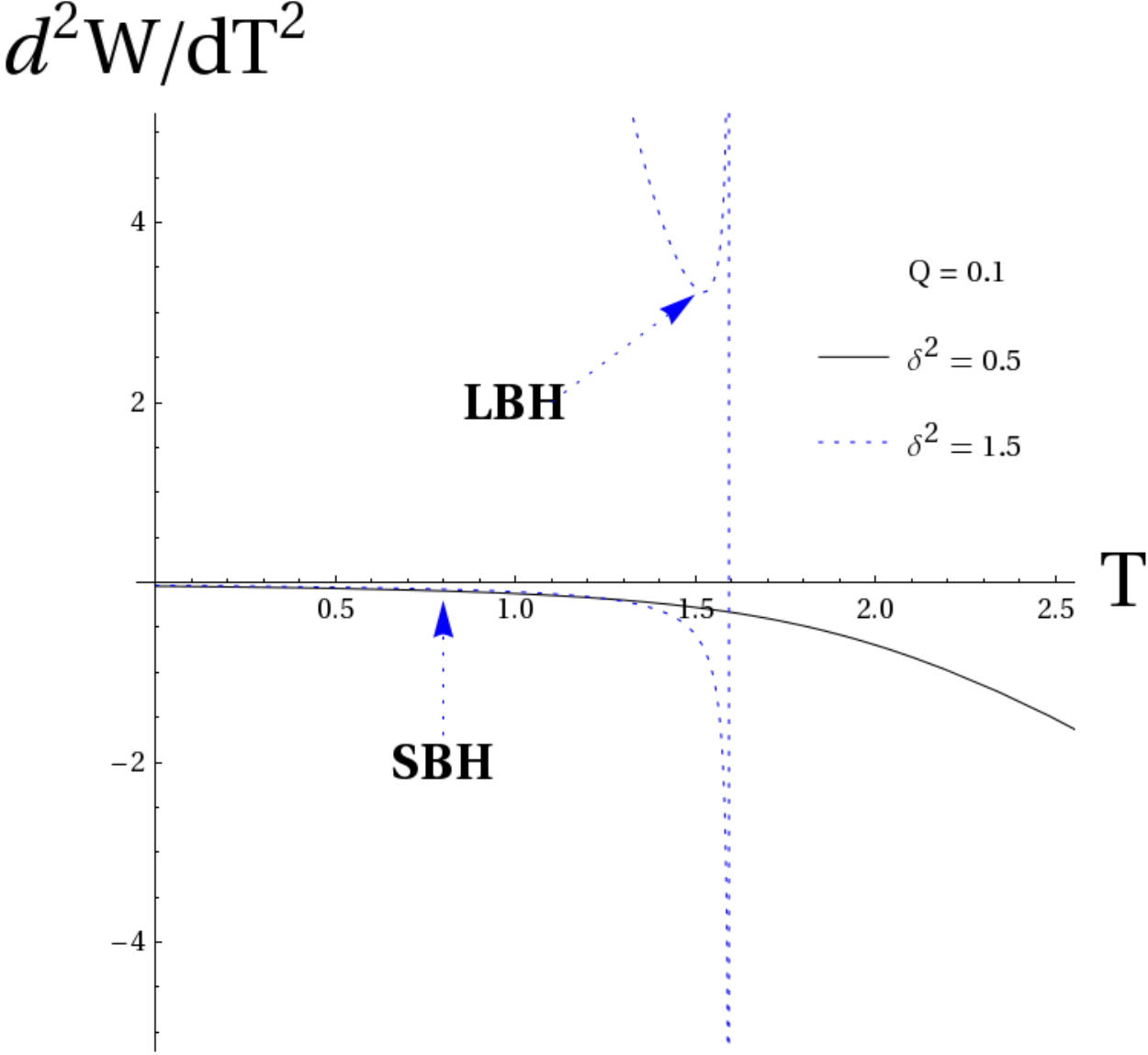}
 \end{tabular}
\caption[Local thermal stability for the planar $\ga=\da$ EMD solution in the canonical ensemble]{These plots show the heat capacity (left) and the second derivative of the Helmholtz potential with respect to the temperature (right) versus the temperature for the planar $\ga=\da$ EMD solution \eqref{Sol2}, in the cases $\da^2=0.5$ (solid) and $\da^2=1.5$ (dot).}
\label{Fig:ThermalStabilityGa=DaCanonical}
\end{center}
\end{figure}

\begin{figure}[t]
\begin{center}
\begin{tabular}{cc}
	 \includegraphics[width=0.45\textwidth]{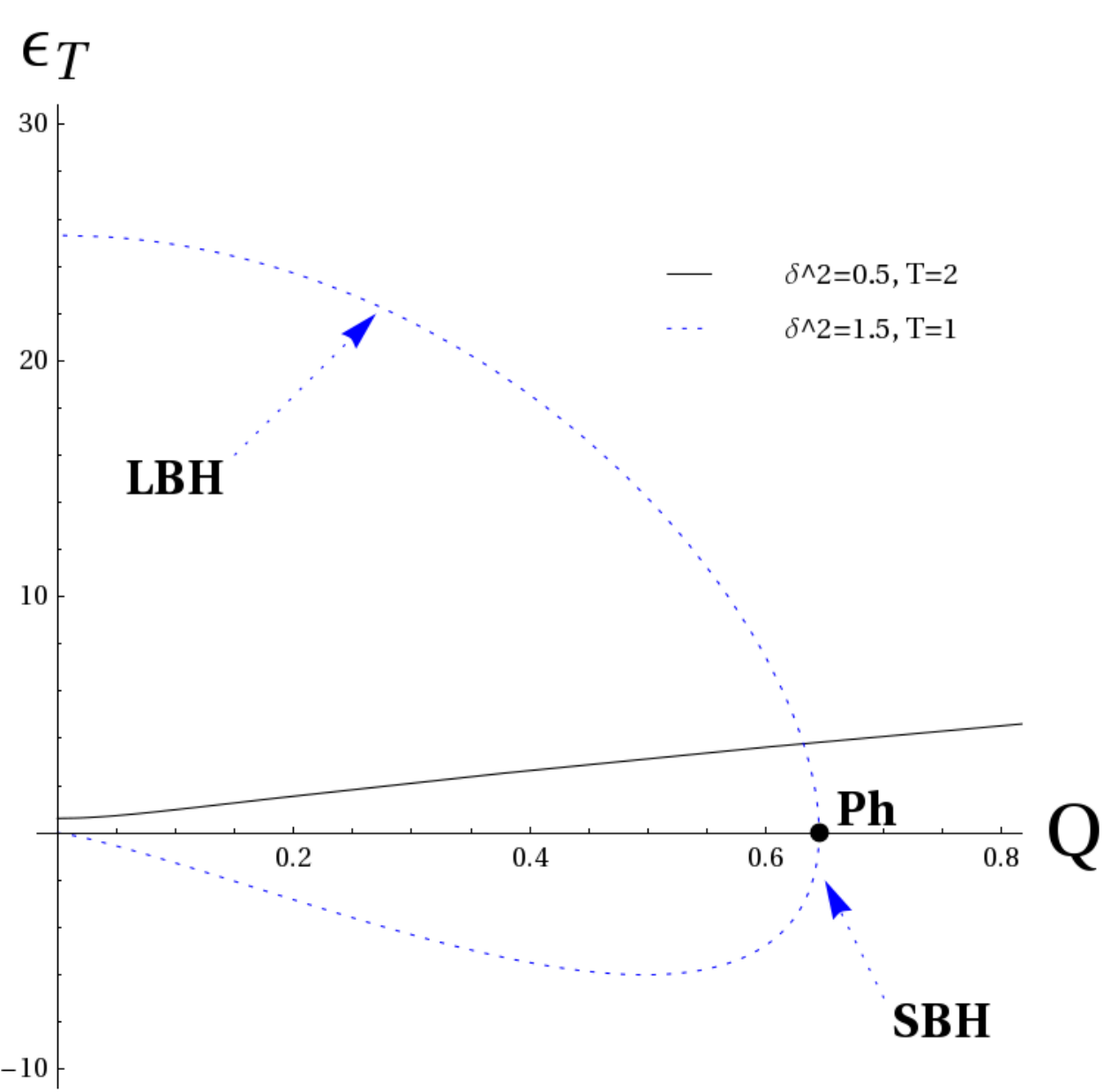}&
	 \includegraphics[width=0.45\textwidth]{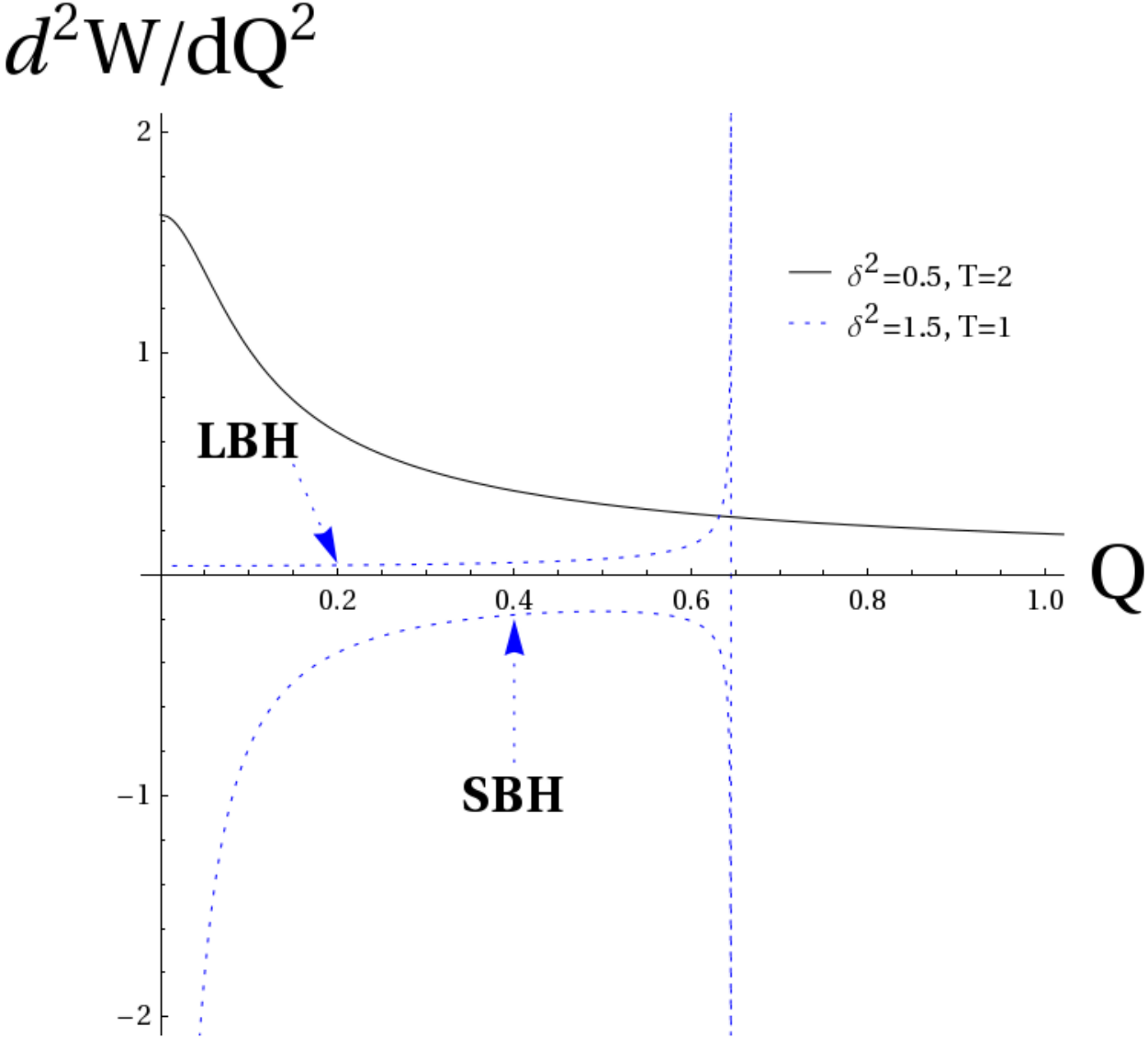}
 \end{tabular}
\caption[Local electric stability for the planar $\ga=\da$ EMD solution in the canonical ensemble]{Electric permittivity (left) and second derivative of the Helmholtz potential (right) versus charge, for $\da^2=0.5$ (solid line) and $\da^2=1.5$ (dotted line), for the planar $\ga=\da$ EMD solution \eqref{Sol2}.}
\label{Fig:ElectricPermittivityGa=DaCanonical}
\end{center}
\end{figure}

The heat capacity is explicitly:
\be
	C_{Q} = \frac{2r_+^2}{1-\da^2}\frac{r_+^4-r_e^4}{r_+^4-r_{Ph}^4}\,,
	\label{HeatCapacity2}
\ee
and is displayed in \Figref{Fig:ThermalStabilityGa=DaCanonical}. 
The electric permittivity is
\be
	 \epsilon_T=2\frac{1+\da^2}{3-\da^2}r_+^{1+\da^2}\frac{1+\frac{4(3+\da^2)Q^2}{(1-\da^2)(3-\da^2)r_+^4}}{1+\frac{4Q^2}{(3-\da^2)r_+^4}}\,,
	\label{ElectricPermittivityCanonical2}
\ee
see \Figref{Fig:ElectricPermittivityGa=DaCanonical}

We will now study the energetic competition between the black holes and the thermal background, e.g. the extremal limit.
\begin{itemize}
 \item Lower range: The single black hole branch is energetically favoured over the thermal background. The $\da^2=1$ displays an interesting maximal temperature, even though the separation into two branches has not yet occurred. The black holes are stable both against thermal and charge fluctuations, as seen in  \Figref{Fig:ThermalStabilityGa=DaCanonical} and  \Figref{Fig:ElectricPermittivityGa=DaCanonical}.
 \item Upper range: The small black holes are energetically favoured both with respect to the background and to the large black holes, up until the critical point where the two branches merge and cease to exist. When they are small enough, the large black holes become energetically favoured compared to the background, but still have a greater free energy than the small black holes. The small black holes are stable against thermal fluctuations, see  \Figref{Fig:ThermalStabilityGa=DaCanonical}, but not against electric fluctuations, see  \Figref{Fig:ElectricPermittivityGa=DaCanonical}. The large black holes have the opposite properties.
\end{itemize}

\begin{figure}[t]
\begin{center}
\begin{tabular}{cc}
	 \includegraphics[width=0.45\textwidth]{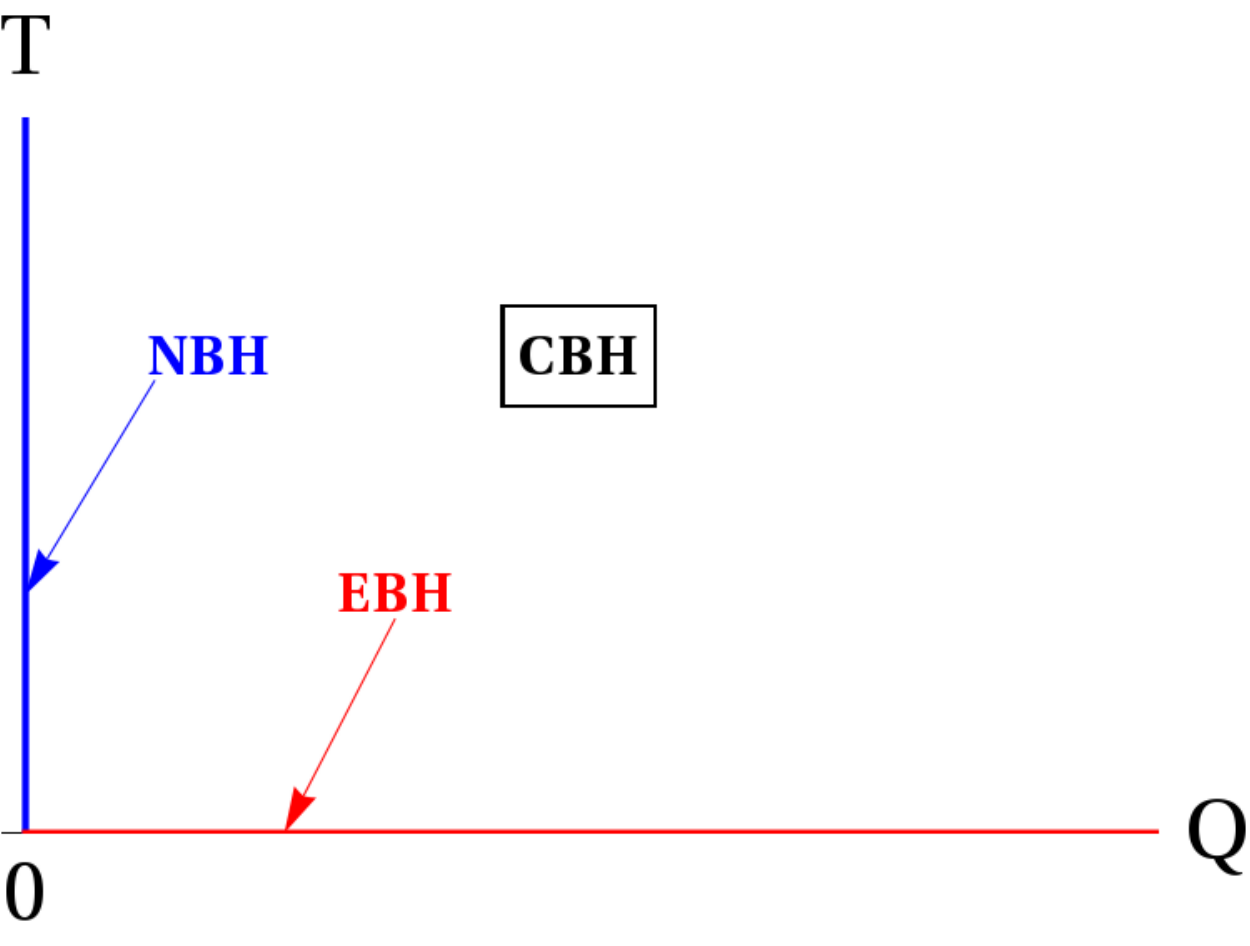}&
	 \includegraphics[width=0.45\textwidth]{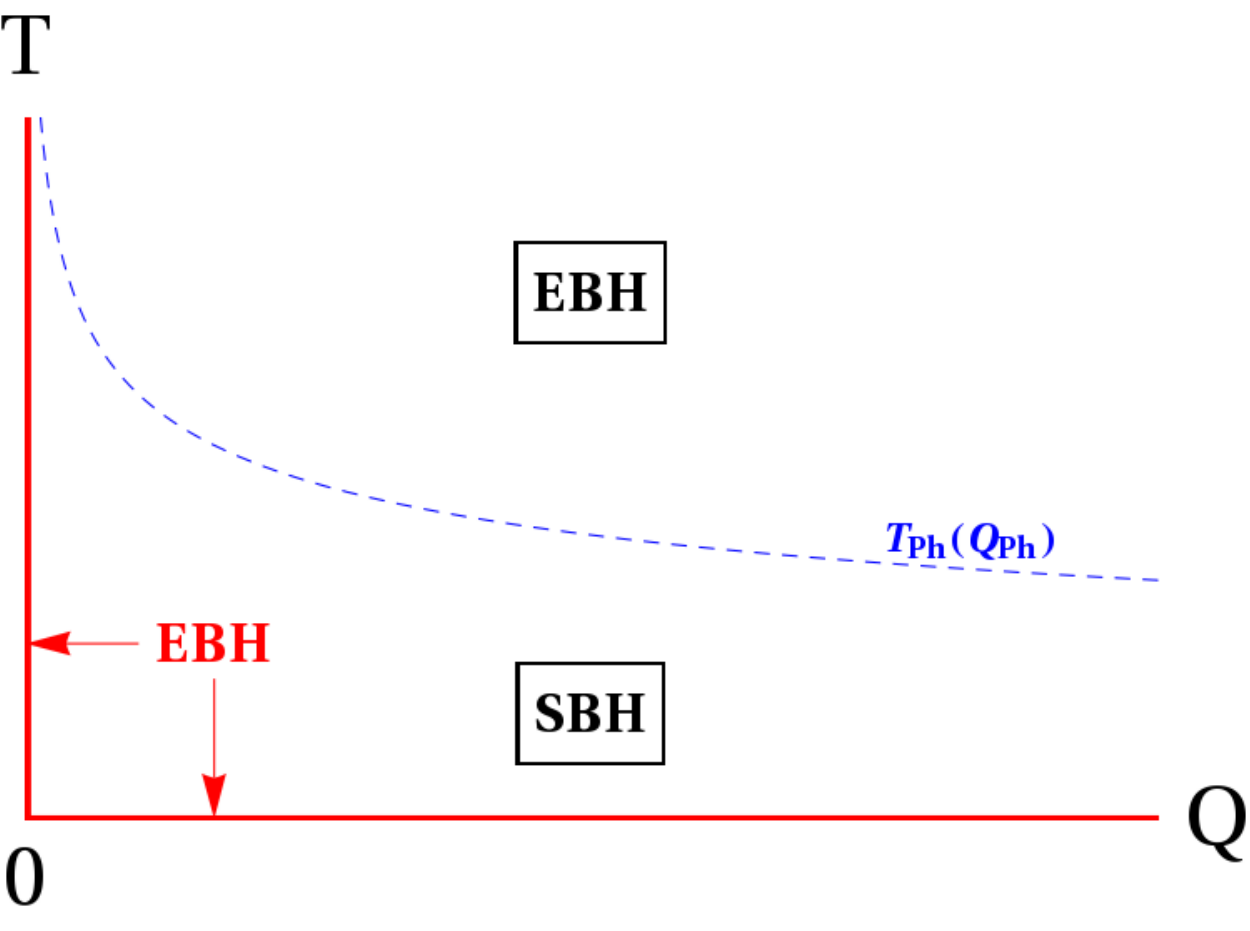}
\end{tabular}
\caption[Phase diagram for the planar $\ga=\da$ EMD solution in the canonical ensemble]{$(T,Q)$ phase diagram in the lower range (left) and the intermediate range (right) for the planar $\ga=\da$ EMD solution \eqref{Sol2}. EBH = Extremal Black Holes, CBH = Charged Black Holes, NBH=Neutral Black Holes, SBH = Small Black Holes.}
\label{Fig:PhaseDiagramGa=DaCanonical}
\end{center}
\end{figure}

The same remarks as for the $\ga\da=1$ black holes apply, regarding the effect of the scalar field on the (dis)appearance of various branches. The order of phase transitions is as follows, inspecting the free energy eq.~\eqref{Helmholtz2}
\begin{itemize}
 \item {Lower range: There is no phase transition to the extremal background at zero temperature for non-zero charge of the black hole. There appear however diverging mixed derivatives of the free energy as one approaches $Q=T=$ simultaneously, which are of at least third-order.

In the zero-charge limit at finite temperature, the neutral black holes dominate and there is a second-order phase transition from the charged black holes in this parameter range.
Finally, higher-order phase transitions appear as $T\rightarrow 0_+$ at exactly zero charge, as discussed around eq.~\eqref{PTuncharged}.

 }
 \item Intermediate range: There is a zeroth-order phase transition to the thermal background at the point $(T_\mathrm{Ph},Q_\mathrm{Ph})$ since the Helmholtz potential is discontinuous there and jumps to zero. {This is again to be remedied by an AdS-completion.

 At zero-temperature and finite charge, the small black holes again settle continuously in their extremal state, as seen by the continuity of the Helmholtz potential and its second derivative.

 Aproaching zero charge at finite temperature, the first derivative of the Helmholtz potential w.r.t. $Q$ diverges, so there is a first-order phase transition to the thermal background (which is the endpoint of the stable small black holes branch). In particular, on the vertical axis $Q=0$, the neutral black holes do not dominate (they are the endpoints of the unstable large black holes branch).
}
\end{itemize}

We have plotted the phase diagram ($T,Q$) for the lower and intermediate ranges in \Figref{Fig:PhaseDiagramGa=DaCanonical}.
\vfill
\pagebreak
	
\subsection{Near-extremal charged planar black holes}

We now examine the thermodynamic properties of the near-extremal solutions \eqref{Sol3} reported in Section \ref{section:NearExtremalPlanarEMD}.

The temperature is,
\be
	 T=\frac1{4\pi}\sqrt{-w\Lambda}\e^{-\frac\da2\phi_0}(2m)^{1-2\frac{(\ga-\da)^2}{wu}}\,,
	\label{Temperature3}
\ee
and we observe that it vanishes in the $m=0$ (extremal) limit if and only if the exponent is positive, that is
\be
wu-2(\ga-\da)^2=(3-\da^2)(1+\ga^2)+(1-\ga\da)^2-2(\ga-\da)^2>0.
\label{159} \ee
 This precisely reduces in the limit we already observed in the $\ga\da=1$ solution, $\da^2<1+\frac2{\sqrt3}$.  We recover the two standard dilatonic cases,
\begin{itemize}
 \item If  $1-2\frac{(\ga-\da)^2}{wu}>0$ (lower range), the extremal temperature is zero.
 \item If  $1-2\frac{(\ga-\da)^2}{wu}<0$ (upper range), the temperature diverges in the extremal limit.
\end{itemize}
The gravitational mass is
\be
	M_g = \frac{\omega_2}{4\pi}\sqrt{\frac{-\Lambda}{wu^2}}(\ga-\da)^2m\,,
	\label{Mass3}
\ee
and reduces to the corresponding near-extremal limits of the $\ga\da=1$ and $\ga=\da$ solutions. Notice that if $\ga=\da$ the mass is identically zero, which is expected since in that case the geometry is not a black hole but simply the direct product $\mathbf{AdS}_2\times \mathbf S_2$.
\noindent
The electric charge
\be
	Q = \frac{\omega_2}{8\pi}\sqrt{\frac{v\Lambda}{u}}\e^{\frac{\ga-\da}2\phi_0}\,,
	\label{ElectricCharge3}
\ee
is universal and does not contain an independent integration constant. Finally, the entropy is
\be
	S = \frac{\omega_2}{4}(2m)^{2\frac{(\ga-\da)^2}{wu}}\,,
	\label{Entropy3}
\ee
and we observe that it vanishes  at extremality provided that $\ga\neq\da$ (otherwise it is finite) and $1-2\frac{(\ga-\da)^2}{wu}>0$. This is consistent with the behaviour of the two previous solutions for $\ga\da=1$ and $\ga=\da$.

\begin{figure}[t]
\begin{center}
\begin{tabular}{cc}
	 \includegraphics[width=0.4\textwidth]{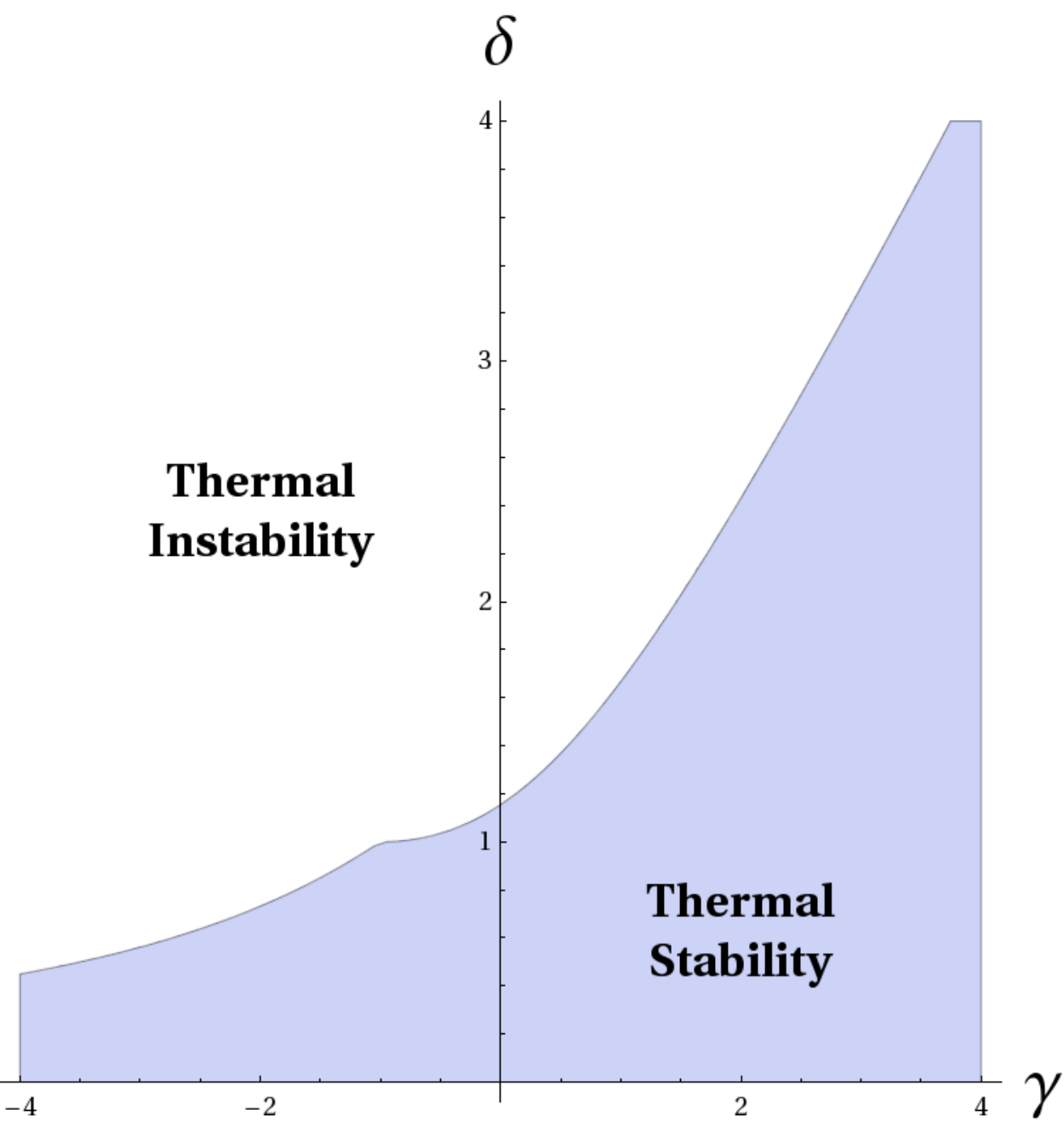} &
	 \includegraphics[width=0.4\textwidth]{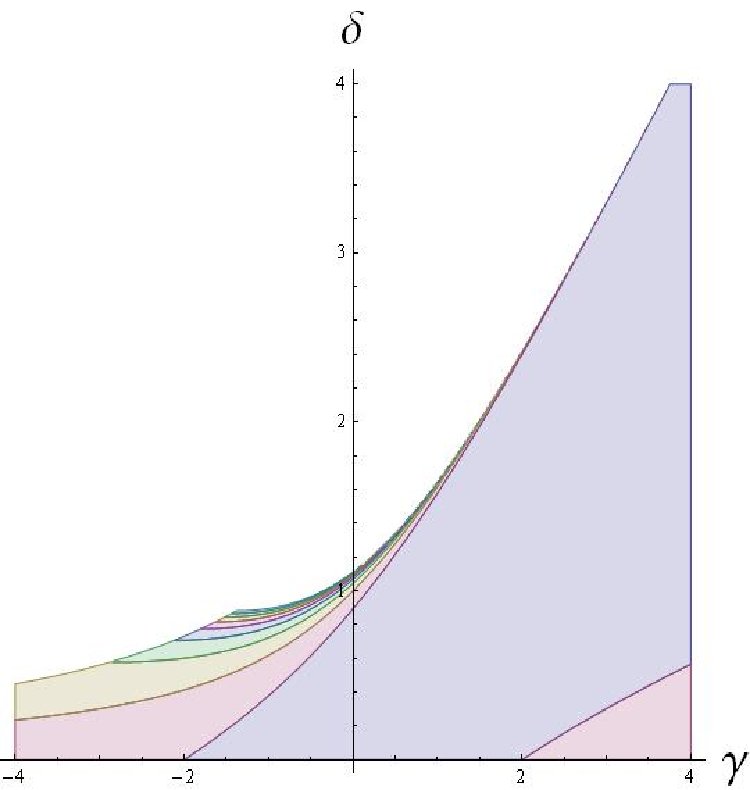}
\end{tabular}
\caption[Global and local thermal stability of the planar near-extremal EMD solutions]{{On the left panel, we plot the region of local stability of the near extremal black hole. The right panel shows a variety of phase transitions of the near extremal black hole to the background at zero temperature. In the blue region second-order transitions occur, in the purple region adjacent to the blue one the transitions are of third-order. The stripes starting with yellow to the left of the blue and purple regions depicts transitions of fourth-(yellow) up to tenth-order. Above them all higher-order transitions also occur.}}
\label{Fig:PhaseTransitionsNearExtremal}
\end{center}
\end{figure}

Given all of this, we work in the canonical ensemble, where only the temperature is allowed to vary (and the  charge density  is fixed to its value \eqref{ElectricCharge3}. Then the Helmholtz potential is
\be
	W = -\frac{\omega_2}{8\pi}\sqrt{-w\Lambda}\e^{-\frac\da2\phi_0}\l[1-2\frac{(\ga-\da)^2}{wu} \r]m\,,
	\label{Helmholtz3}
\ee
and again reduces both to \eqref{Helmholtz1} and \eqref{Helmholtz2} in the appropriate limits. Unsurprisingly, inspection of its sign reveals that the black holes are globally stable in the lower range but not in the upper range, similarly to the $\ga\da=1$ solution. Moreover, computing the heat capacity, we find
\be
	C_Q = \frac{\omega_2}{4}\frac{(\ga-\da)^2}{wu}\l[1-2\frac{(\ga-\da)^2}{wu}\r]^{-1}\l(\e^{\frac\da2\phi_0}\frac{4\pi T}{\sqrt{w\Lambda}} \r)^{\frac{2(\ga-\da)^2}{wu-2(\ga-\da)^2}}\,,
	\label{HeatCapacity3}
\ee
which will be positive in a range a little bit smaller than the lower range and negative in the upper range, thus rendering the black holes locally stable in the former and locally unstable in the latter, as displayed in the left panel of \Figref{Fig:PhaseTransitionsNearExtremal}.

Finally, second- or higher-order phase transitions will occur in the zero temperature limit if the appropriate derivative of the Helmholtz potential \eqref{Helmholtz3} diverges. In particular an $n^\textrm{th}$-order transition occurs if
\be
	-1 < \frac{wu}{wu-2(\gamma-\delta)^2} - n < 0\,.
\label{160}\ee
This is depicted in the right panel of \Figref{Fig:PhaseTransitionsNearExtremal}, showing the regions in which transitions of second-  to tenth-order occur. Two remarkable features are that, in most of the parameter space that is thermodynamically stable (which amounts to requiring a well-defined extremal limit and $vu<0$), second-order transitions appear, and there are no first-order transitions. We note that it does not make sense to analyse the thermodynamically unstable region in this way as long as the AdS-completion has not ben taken into account, since our experience with the solutions for $\gamma \delta=1$ show us that this is crucial for the correct structure of the phase transitions.

We stress again that all these considerations should be taken as valid in the near-zero temperature regime: as the temperature grows, one will need the (yet unknown) full solution and not its near-extremal approximation in order to study finite temperature thermodynamics.

\clearpage

\vfill

\part{Perspectives}
\label{part:four}
\section{The AdS/CFT correspondence and holography}

\label{section:Holography}

\subsection{The AdS/CFT correspondence}

At the end of the last decade, Maldacena put forward a surprising proposal, \cite{Maldacena:1997re} (see \cite{Aharony:1999ti} for a review). Maldacena's original idea is developed in the setting of String Theory, and in particular in its low energy (weak curvature) limit, supergravity (see \cite{Wess:1992cp} for a textbook introduction). Supergravity lives naturally in a ten-dimensional spacetime, and it is possible to find solutions to the equations of motion which describe an arbitrary number $N$ of parallel and coincident $D3$-branes\footnote{A $Dp$-brane is a surface with $p$ spatial and one time dimensions, moving around in a $D=p+1$-dimensional spacetime. They have Dirichlet boundary conditions, that is the coordinates of the endpoints of open strings are fixed on the brane for their normal components, so that the only allowed motion of the endpoints is tangential.}. The geometry of the near-horizon region reduces to a direct product of $AdS_{p+2}\times\mathbf S^{D-2-p}$. Originally, Maldacena concentrated on $D3$-branes, so that the reduced geometry was $AdS_5\times \mathbf S^5$ for critical superstring theories. It can be verified that the curvature scale of both spaces in the product is of the order of an inverse positive power in $N$, so indeed the approximation is valid at large $N$. A decoupling limit is then taken where the energy is bounded by an inverse power of the string loop-expansion parameter $\al'$. Then, taking $\al'\to0$ so that supergravity describes the low energy tree action of Type II String Theory, the bulk and boundary massless modes decouple.

On the other hand, it was known that the gauge theory living on the (four-dimensional) worldvolume of a $D3$-brane is $\mathcal N=4$ (number of supersymmetries) $SU(N)$ super Yang Mills (sYM) gauge theory, at low energies in the same decoupling limit as above. Since these two alternatives describe the same set of objects (the stack of coincident $D3$-branes), they should be equivalent somehow. This is made possible by the fact that $AdS_5$ has a boundary with geometry $\M_4$ (four-dimensional Minkowski space). Thus one postulates that type IIb String Theory on $AdS_5\times \mathbf S^5$ is equivalent to  $\mathcal N=4$  $SU(N)$ sYM: this is the essence of the AdS/CFT correspondence.

Gubser, Klebanov and Polyakov, \cite{Gubser:1998bc}, and independently Witten, \cite{Witten:1998qj}, showed that the correlators for massless fields were the same, computed on the (super)conformal gauge theory side and on the gravitational side. This went a long way to assert the validity of the conjecture. But perhaps the most interesting aspect is the regime in which the correspondence is defined. Maldacena postulated that the following parameters should be identified:
\be
	g_s=g_{YM}^2\,, \qquad \frac{R^4}{\al'^2}=g_{YM}^2N=\la\,.
	\label{AdSCFTCorrespondence}
\ee
$g_s$ is the string coupling constant, $g_{YM}$ is the (dimensionless) Yang Mills coupling, $N$ is both the number of $D3$-branes and the size of the gauge group $SU(N)$, $R$ is the curvature radius of both the $AdS_5$ and $\mathbf S^5$ spaces, while we remind that $\al'$ is the string loop-expansion parameter, and is proportional both to the square root of the string length and the inverse square root of the string tension.  $\la$ is the 't Hooft coupling. When are the previous considerations justified? Supergravity is a good approximation to String Theory at low curvatures, that is large 't Hooft coupling, $\la>>1$. On the other hand, perturbative field theory tools can be used in Yang Mills theory when the interaction constant is small. The effective interaction constant is precisely the 't Hooft coupling, so the regime of validity is $\la<<1$. Thus, the AdS/CFT correspondence is a duality between the weak coupling regime of the boundary gauge theory and the strong coupling regime of the bulk gravitational theory, or vice-versa.

Although the correspondence does not allow to describe the same regime of physics in both dual descriptions, and so does not provide a test of one with the other, it embodies a very interesting concept, advocated by 't Hooft, \cite{'tHooft:1993gx}, and then by Susskind, \cite{Susskind:1994vu}: the holographic principle. In essence, it states that the boundary degrees of freedom suffice to describe physics in the bulk, and that this is the only way to reconcile quantum mechanics and black hole physics. One of the main arguments in favour of this is that the entropy of the region of spacetime cloaked by the horizon is proportional to the area of the horizon (that, is the area of the boundary of this region), and not to its volume. This suggests that the degrees of freedom are stored on the boundary and not behind it.

Before closing off this section, we wish to mention one last (but not the least) piece of the puzzle. What happens once we put a black hole in the bulk? Indeed, it is possible to emblacken the supergravity solution, so that it could admit both regular AdS space and a Schwarszschild-AdS black hole. Then, Witten noted, \cite{Witten:1998qj,Witten:1998zw}, along the lines of Hawking and Page, \cite{Hawking:1982dh}, that two branches of black holes (small and large) could exist only above some minimal temperature, and that above a critical temperature (greater than the minimum temperature), the large black holes were both global and local extrema of the partition function. So, AdS space dominates the bulk physics below the critical temperature, while above it the large black holes do.

How does this translate on the gauge boundary theory? These thermodynamic considerations are valid in the large $N$, large fixed $\la$ limit, for which the behaviour of $\mathcal N=4$ sYM is not known (it lies in the non-perturbative region). But comparing to the expectations for pure (non-conformal) $SU(N)$ gauge theory allows to see that  Hawking-Page thermal phase transition is the equivalent of a transition from a confining phase at low temperatures to a deconfining phase at high temperatures. This is valid in the case where the volume of the boundary is finite, that is when the boundary of $AdS_5$ space is $\mathbf S^1\times \mathbf S^3$. However, one would ideally like the boundary gauge theory to live on Minkowski space (or equivalently on Euclidean $\mathbf S^1\times \mathbf R^3$). This implies taking the large volume limit for the sphere $\mathbf S^3$, which then turns into a planar subspace. The unfortunate consequence is that the Hawking-Page phase transition at finite temperature is destroyed, and the large black hole phase is stable everywhere (see Section \ref{section:ThermoRNAdS}). 

Thus, in order to recover a finite temperature phase transition, one has to turn to different bulk theories, where another scale is generated which may result in the desired properties: for instance, one can include bulk scalars, which are able to simulate positive horizon curvature in the thermodynamics, as we have seen in the previous part. This is the topic of the next section, where we will describe the so-called Effective Holographic Theories.

\subsection{Effective Holographic Theories}
\label{section:EHT}

The spirit of Effective Holographic Theories (EHT) is a bit different from the original spirit of the AdS/CFT correspondence. Indeed, as explained above, the firmest footing on which the latter was established was in settings which could be embedded in low energy limits of String Theory, such as the paradigmatic example of supergravity on $AdS_5\times\mathbf{S}^5$. However, there has been very early on suggestions that the correspondence could be generalised to other bulk geometries, and in particular ones which would not induce a CFT on the boundary, but simply a QFT. This is the case for the Domain Wall/QFT (DW/QFT) correspondence advocated in \cite{Boonstra:1998mp}. A main interest for this approach is that many physical systems possess not conformal, but simply Poincar\'e invariance. So, if one's intent were to reproduce some (or all) characteristics of a gauge theory like QCD (with gauge group $SU(3)$, generalised to an arbitrary number of colors $SU(N)$ so as to be able to work in the large $N$ limit), then one would clearly need to go beyond Maldacena's original idea and find bulk settings that break the conformal invariance on the boundary. Another way to phrase this is that one must introduce relevant deformations on the CFT. 

In our case, such a leading relevant operator will be a scalar field. On the boundary, its value will drive the dynamics from the Ultra-Violet to the Infra-Red, and provide the theory with a dynamical energy scale (effectively breaking conformal invariance, such as is needed in QCD for instance). In the bulk, the scalar field will have a non-trivial profile in the radial coordinate normal to the brane, and it is one of AdS/CFT prescriptions to identify this with the running of the energy scale, \cite{Susskind:1998dq,Peet:1998wn}. However, pure AdS/CFT dictionary would require that the scalar field reaches a constant value on the boundary, identified with the expectation value of the relevant operator. Breaking of conformal invariance is also naturally associated with non-critical string theories with a central charge deficit, as briefly discussed in Section \ref{section:StringEffectiveActions}. Mixing all of these ingredients, a worthy bottom-up approach for EHTs appears to be actions of the kind of \eqref{EMDaction}, which we studied in some detail in sections \ref{section:EMDBH} and \ref{section:ThermoEMD} and that we recall here:
\be
	\label{EMDactionHol}
	S = \int \ud^dx\sqrt{-g}\l[R-\frac12(\partial \phi)^2-\frac14\e^{\ga\phi}F^2-V(\phi)\r].
\ee
The exact shape of the potential may be kept undefined, and indeed general asymptotic studies of the solutions of such actions were carried out, where holographic properties are related only to the IR and UV behaviour of the potential, see for instance \cite{Gursoy:2007cb,Gursoy:2007er,gkmn2} (with a mind towards setups allowing to recover QCD-like properties). If the potential is specified to be a Liouville exponential potential, then its connection with a central charge deficit in a non-critical theory can be made clearer, again see Section \ref{section:StringEffectiveActions}. However, it turns out to be very difficult to find solutions with AdS asymptotics as Wiltshire \emph{et al.} proved, \cite{Poletti:1994ff,Poletti:1994ww}, except in the case of a constant potential. Then, one may find perturbative or numerical solutions that interpolate between the non-conventional IR geometry and the AdS UV boundary, see \cite{Poletti:1994ww} and more recently \cite{Goldstein:2009cv,Cadoni:2009xm,Bertoldi:2010ca}. The addition of a gauge field will be commented upon shortly, but may serve to introduce a conserved number of particles in the boundary theory. Let us note that these actions in the case of a Liouville potential do contain a string-embeddable setup, for the particular choice of coupling constants $\ga=\da=1$.

EMD actions \eqref{EMDactionHol} may then be used and trusted in the IR, at low energy, and expected to be modified appropriately in the UV. This is possible if for instance the potential has the shape
\be
	V(\phi) = 2\La_{IR}\e^{-\da\phi}+2\La_{UV}\,,
\ee
where one assumes that the exponential part decreases rapidly enough in the UV so that the constant part of the potential dominates the dynamics. The IR scale $\La_{IR}$ can then be completely decorrelated from the UV scale, which is the expectation value of the scalar operator in the boundary CFT. This is the view advocated in our work \cite{Charmousis:2010zz}\footnote{Another point of view, this time put forward by \cite{Perlmutter:2010qu}, relies fully on the DW/QFT correspondence and gives up entirely the goal of recovering AdS asymptotics. It is simply required that the IR geometry interpolates to a domain-wall solution. Such solutions may still be found from actions \eqref{EMDactionHol} with a Liouville potential, and constraints are placed on the value of $\da$ so that indeed the solutions possess a boundary. Note that once a finite temperature solution is found in the IR, it can perfectly well be matched in the UV to another domain-wall solution of the same action.}, where we aim to describe the IR region and argue that valid information can be extracted which are relevant in the UV. Thus we do not assume that the potential in \eqref{EMDactionHol} should be a constant, and use an exponential Liouville potential in order to obtain exact analytical solutions valid in the IR:
\be
	V(\phi) = 2\La\e^{-\da\phi}\,\qquad \textrm{in the IR.}
	\label{IRpotential}
\ee
If one insists to have analytic solutions that are asymptotically AdS, then this can be accomodated with potentials having a finite minimum at some $\phi_0$, where the scalar field rolls down as it approaches the boundary. Typical shapes for these potentials are hyperbolic sines or cosines, see \cite{Martinez:2004nb,Martinez:2006an}. In this case, the backreaction of the scalar field on the geometry is enough so that the boundary conditions present a slower fall-off towards AdS than usual and one has to be careful as to the definition of the energy, \cite{Henneaux:2004zi,Henneaux:2006hk}. However, we point out that the leading behaviour of such potentials in the IR will still be an exponential of the scalar field, so this does not contradict our previous considerations.

Specialising to the four-dimensional case, are the exact solutions presented in Section \ref{section:PlanarEMDBH} and whose thermodynamics were analysed in Section \ref{section:ThermoEMD} good IR  solutions regarding the UV behaviour of the potential? The chargeless (zero temperature) background is
\bsea
	\ud \bar s^2 &=& r^2\l(-\ud t^2 + \ud x^2 + \ud y^2\r) +r^{2\da^2-2}\ud r^2\,,\\
	\bar\phi &=& 2\da\log r\,,
	\label{DomainWallBackground}
\esea
and is associated with a rolling dilaton with a runaway minimum at the value $\bar\phi\underset{r\to+\infty}{\longrightarrow}+\infty$. The Liouville potential \eqref{IRpotential} behaves in the UV as
\be
	V(\phi)= 2\La\e^{-\da\phi} = r^{-2\da^2}\underset{r\to+\infty}{\to} 0\,,
\ee
and so it is AdS-completable.
Moreover, it is pointed out in \cite{Perlmutter:2010qu} that for $\da^2<1$, this background is indeed a domain wall, in the sense that it has a boundary at $r\to+\infty$: the coordinate time to $r\to+\infty$ along a null geodesic is finite, although the proper distance is infinite. This is crucial in order for the DW/QFT interpretation to hold. In our case, this is of less import, since we assume an AdS completion. 

The exact solutions for $\ga\da=1$ \eqref{Sol1} and $\ga=\da$ \eqref{Sol2} describe the whole IR geometry and do approach the backgound value $\bar\phi$ in the UV. So the Liouville potential also vanishes in the UV and can be AdS-completed, although we have not explicitly done the numerical interpolation. We now turn to the arbitrary $\ga$, $\da$ near-extremal solutions. Here, a little care is required in the analysis. Indeed, these solutions \eqref{Sol3} do not admit the above domain-wall background. As we have pointed out in Section \ref{section:NearExtremalPlanarEMD}, they are merely the near-extremal geometry of the full IR solutions. They only describe the latter in the deep IR, and this means that we are lacking some information and should not trust these solutions in the UV. The background they live on is the extremal charged black hole, as also recognised in the $\da=0$ case in \cite{Goldstein:2009cv}. So, relying on the actual proof provided in Section \ref{section:NearExtremalPlanarEMD} that the solution \eqref{Sol3} is indeed the near-extremal limit of the full solutions in the cases $\ga\da=1$, $\ga=\da$, we argue that we should still use the asymptotic value of the domain-wall solution \eqref{DomainWallBackground} to assess the AdS-completability of solution \eqref{Sol3}, and not the value exhibited in the near-extremal limit \eqref{Phi3bis}. To put this claim on a firmer ground, we have done an analysis of the phase space of the dynamical system of equations of motion, see Appendix E of \cite{Charmousis:2010zz}, using a coordinate-independent approach\footnote{We use the scalar field as the phase space coordinate. Thus, in the planar case, only three first-order equations for three independent variables remain. We recover results in qualitative agreement with those of Wiltshire et al., \cite{Poletti:1994ff},where the phase space was analysed without restricting to a planar subspace.}. Besides generically singular solutions, two extremal fixed points of interest are found, one neutral and one charged. The former is the dilatonic background \eqref{DomainWallBackground}, while the other is the extremal black hole. Study of the linear perturbations around these shows that, for a holographically meaningful behaviour\footnote{More precisely, we ask that the Liouville potential evaluated on-shell should increase toward the IR. So should the scalar field, since we use it as a coordinate. If it should decrease in that direction, then this means that our axis of coordinate is reversed, and we should change the sign of the scalar field, which changes the sign of the perturbations eigenvalues.}, the neutral fixed point is unstable in the IR\footnote{All eigenvalues are positive in cases of interest.} but stable in the UV, while the charged fixed is stable in the IR and unstable in the UV, provided one initial condition is tuned\footnote{In cases of interest, two eigenvalues have negative sign, the third is positive.}. This is in agreement with AdS/CFT lore, which usually requires the fixation of one initial condition. The dilatonic solutions we have shown will then become subdominant in the UV, which opens the way for them to be generically completed to AdS.

Up until now, we have only addressed the issue of AdS-completability to determine the validity of our solutions. Other criterions need to be examined. The first of these was established by Gubser, \cite{Gubser:2000nd}, and has to do with the nature of the singularity in the IR. Indeed, holographic backgrounds generically display such reprehensible behaviour. In a General Relativity context, one would argue against such occurrences and cloak the singularity with an event horizon. In holography however, such solutions may be acceptable on the condition that they derive from finite temperature solutions. Let us elaborate a little on this point. Turning on a finite temperature in the boundary field theory means that a black hole lives in the bulk, and the field theory temperature is the Hawking temperature on the horizon of the black hole. From the field theory point of view, there should be nothing wrong with taking the limit of zero temperature. In the bulk, having an event horizon is thus equivalent with putting a lower cut-off on the temperature, hiding the low temperature region and thus the deep IR physics. Such a region may be accessed by taking the zero temperature limit, which means either taking an extremal limit for a black hole with two horizons, or switching off the black hole entirely. We have seen that in both cases this can result in the appearance of a naked singularity. These bulk singular spacetimes should be construed as acceptable so long as they are the zero-temperature limit of a regular finite temperature solution. A necessary condition for the finite temperature solution to exist at all is then, \cite{Gubser:2000nd}, that 
\begin{quote}
\emph{The scalar potential evaluated on-shell should be bounded from above}\footnote{Our sign convention for the potential is the same as in \cite{Gubser:2000nd} but opposite to \cite{Charmousis:2010zz}.}. 
\end{quote}
This is always the case for the full solutions $\ga\da=1$ \eqref{Sol1} and $\ga=\da$ \eqref{Sol2}, both in the UV (as we saw) but also in the IR, once black-hole solutions with a negative $\La$ have been selected. For the near-extremal solutions, though one is not forced to choose $\La<0$ in order to get black-hole solutions, this will be necessary to satisfy Gubser criterion.

\begin{figure}[t]
\begin{center}
\begin{tabular}{c}
	 \includegraphics[width=0.45\textwidth]{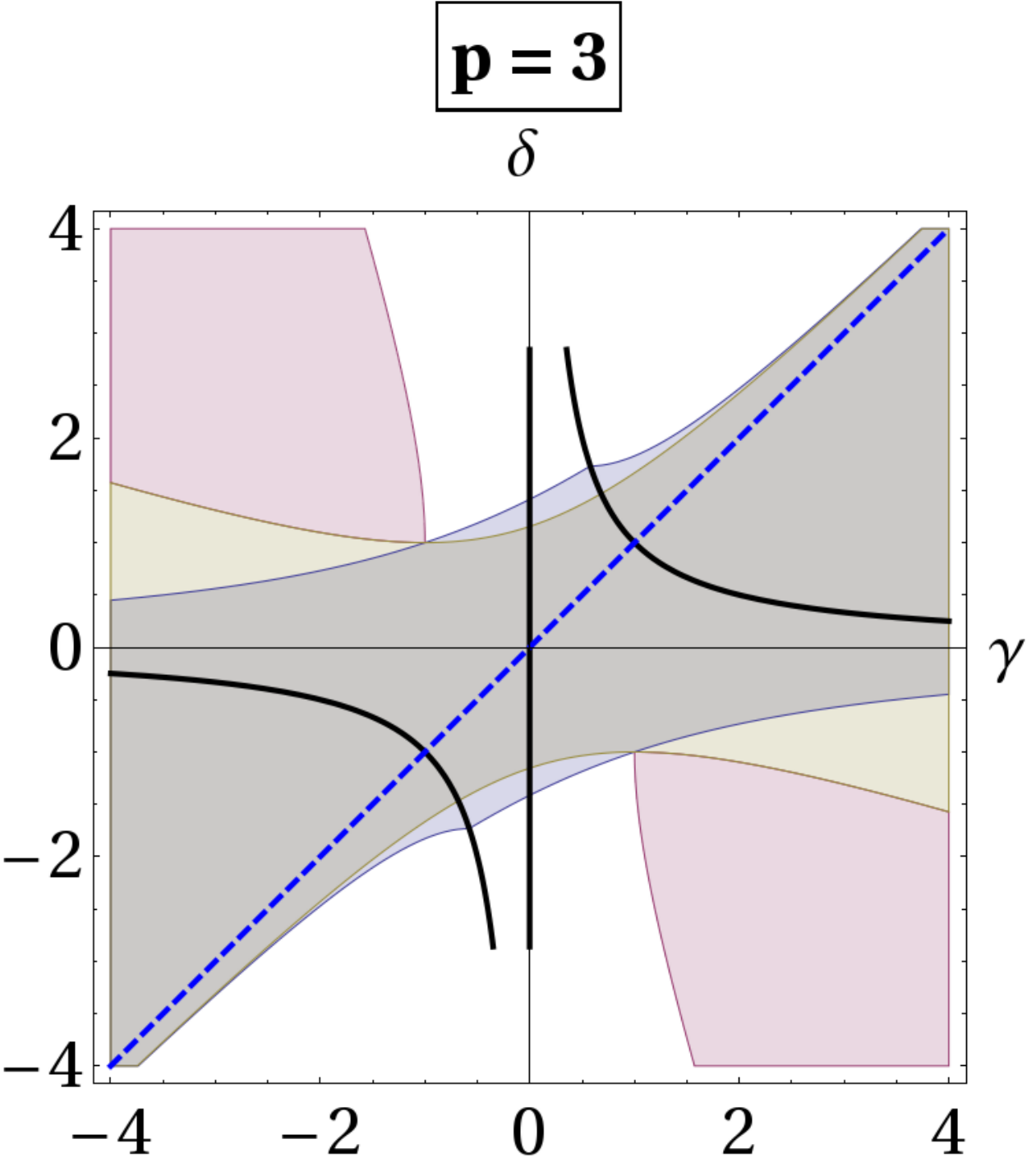}
\end{tabular}
\caption[Gubser and spin-2 fluctuations holographic constraints]{
The Gubser constraint coincides with the constraint to have a black hole, depicted as the blue region. The region where both the spin-2 fluctuations are reliable and the black holes thermodynamically stable is in yellow, while the region where the spin-2 fluctuations are reliable and the black holes thermodynamically unstable is in red. The dashed blue line is the $\gamma=\delta$ solutions while the solid black line corresponds to the $\gamma\delta=1$ solutions.
}
\label{Fig:GubserSpin2constraints}
\end{center}
\end{figure}

For the near-extremal solution \eqref{Sol3}, Gubser's criterion coincides exactly with the requirement to have a black-hole like solution, see \Figref{Fig:GubserSpin2constraints}.

\begin{figure}[t]
\begin{center}
\begin{tabular}{c}
	 \includegraphics[width=0.45\textwidth]{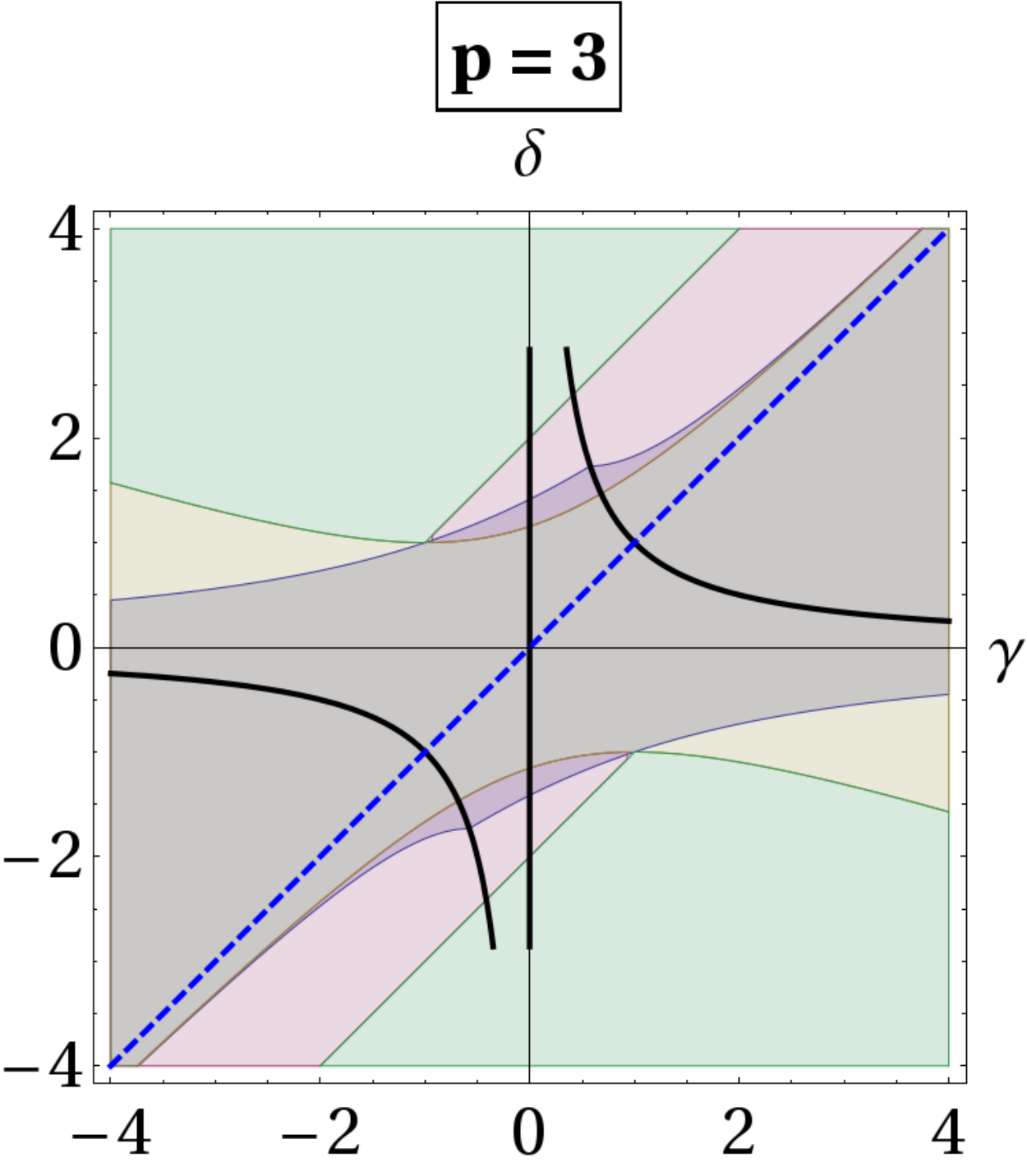}
\end{tabular}
\caption[Gubser and spin-1 fluctuations holographic constraints]{This plot summarises the reliability constraints on the spin-1 fluctuations.  The blue region depicts the part of the $(\gamma,\d)$ plane which satisfies the Gubser bound.
The yellow region is allowed and is identical to both spin-2 and thermal stability constraints.
The purple region is thermodynamically unstable and has $4\geq(\ga-\da)^2$, with a lower bound on charge density, while the green region is thermodynamically unstable and has $4\leq(\ga-\da)^2$. The overlap of the latter region with the Gubser one is trivial in four dimensions.
The solid line corresponds to $\gamma\delta=1$ and the dashed blue line to $\gamma=\delta$.  }
\label{Fig:GubserSpin1constraints}
\end{center}
\end{figure}

Other criterions were put forward by Kiritsis \emph{et al.} in \cite{Gursoy:2007cb,Gursoy:2007er} on the spectra of spin-1 and spin-2 fluctuations:
\begin{quote}
 \emph{A good singularity should be repulsive, e.g. screened to physical modes.}
\end{quote}
Basically, this amounts to checking that the perturbations in the spectra of spin-1 and spin-2 fluctuations in the metric and gauge field do not give rise to two normalisable solutions in the IR. Otherwise, this means that an extra boundary condition should be specified. This is not physical as we expect that only the boundary conditions on the UV should determine the physics, and no boundary conditions should come from the IR. We give them here in the four-dimensional case, but one can refer to Appendix D of \cite{Charmousis:2010zz} for generic dimensions.
\begin{itemize}
 \item Spin-2: the fluctuation problem is reliable in the thermodynamically stable region, $w_pu>2(\ga-\da)^2$, and also in the thermodynamically unstable region if $8<w_pu<2(\ga-\da)^2$. In four dimensions this second region is trivial as it only touches the Gubser-allowed region in two points, see \Figref{Fig:GubserSpin2constraints}.
 \item Spin-1: the fluctuations are always reliable in the thermodynamically stable region. Moreover, in the thermodynamically unstable region, they are reliable if $4\leq(\ga-\da)^2$, which also touches the Gubser-allowed region only in two points in four dimensions. Finally, in the remaining region, it is reliable if the charge is large enough, see Appendix D of \cite{Charmousis:2010zz}. These regions are depicted in \Figref{Fig:GubserSpin2constraints}.
\end{itemize}

All in all, it seems reassuring that both the near-extremal black holes should be thermodynamically stable and have well-defined IR physics from the point of view of holography in the same region. The other regions where the fluctuation problem is well-defined either do no meet the Gubser criterion (in four dimensions, but not necessarily so in higher dimensions) or impose some sort of lower bound on the charge density.

\subsection{Effective Holographic Theories for low-temperature Condensed Matter systems}

We now briefly turn to recent developments in the holography community, where the usual AdS/CFT paradigm between gravitational bulk theories and boundary gauge theories has been carried over to boundary field theories where the accent is placed on the recovery of behaviours more typical of strongly-coupled electrons Condensed Matter Systems. The philosophy of such an AdS/CMS correspondence is reviewed in the following references, \cite{Sachdev:2008ba,Hartnoll:2009sz,Herzog:2009xv,Sachdev:2010ch,Horowitz:2010gk}, with an eye towards quantum criticality and superconductors. Quantum criticality designates the occurrence of a phase transition at zero temperature, which is usually continuous (that is, higher than first order) and driven by quantum fluctuations. Superconductors are materials where below a certain critical temperature the electrons become strongly-correlated and the usual field theory techniques fail. A more phenomenological approach has to be followed, pairing the electrons in a boson-like state as proposed by Bardeen, Cooper and Schrieffer in 1957, \cite{Cooper:1956zz,Bardeen:1957kj,Bardeen:1957mv}. Here, we will not concern ourselves with these topics but rather turn to the properties of strange metals.

The so-called strange metals are phases of heavy fermion compounds, \cite{Stewart:2001zz}, and high critical temperature superconductors, \cite{Hussey:2008,Cooper:2009}. They display behaviours which are at odds with the usual Fermi liquid description in Condensed Matter physics. Fermi liquid theory is a phenomenological theory used to describe quantum mechanical liquid systems of fermions at low temperature and share many properties with Fermi gases. This is all the more interesting since fermions may be interacting non-weakly in the former while they are non-interacting in the latter. This approach has been quite successful in describing the behaviour of electrons, grouped as quasiparticles (which interact similarly as in a Fermi gas). For instance, their heat capacities (resistivities) show the same linear (quadratic) dependence on temperature as for Fermi gases.

Examples of strange metallic behaviours are a resistivity linear in the temperature for high chemical potential, \cite{Martin:1990}, or again an AC conductivity which scales like a negative power of the frequency different from unity, \cite{vandeMarel:2003wn}. These are the characteristics that we aim to find in examining our charged dilatonic setups. Numerous other articles tackling this topic with other setups exist (for instance and without trying to be exhaustive \cite{Karch:2007pd,Lee:2008xf,Hartnoll:2009ns}), which all find strange metallic behaviour to some measure, with more or less effort to provide String Theory embeddings.

\subsubsection{AC conductivity}

In the remainder of this section, we will thus sum up various results on Condensed Matter observables, obtained by a holographic computation in \cite{Charmousis:2010zz} (though note that the $\da=0$ was already thoroughly analysed in \cite{Goldstein:2009cv}), and we refer to this work for technical details. Indeed, holography allows to compute both the AC and DC conductivity. In the low charge density approximation where the charge carriers do not backreact on the bulk geometry, one may use a probe approximation. One solves first for the appropriate chargeless background, and then treats the gauge field as a perturbation. This is clearly not the case for the charged planar dilatonic solutions we have exhibited previously. In the backreacted case, the prescription is to introduce time-dependent perturbations around the solution in the transverse spatial components of both the gauge field $A_x$ and the metric $g_{tx}$. By an appropriate change of coordinates and redefinition of variables, one can represent the variation of the gauge field fluctuation by a second-order Schr\"odinger-like equation:
\be
	-\frac{\ud^2\Psi}{\ud z^2}+V\Psi=\omega^2\Psi\,,
	\label{SchrodingerEquation}
\ee
where $z\sim-1/r$ is a new radial coordinate which varies from $z=0$ on the UV boundary to $z\to-\infty$ in the IR and $\Psi=Z(\phi)A_x$. The study of the conductivity of the system thus reduces to the study of incoming plane waves on a potential. Of course, we will only know analytically the IR and UV parts of the potential, and one has to match numerically over the region in-between. The behaviour of the Schr\"odinger potential may then be derived in the near-horizon region from the IR solution, and in the UV by the AdS boundary conditions. At this point, Kachru \emph{et al.}, \cite{Goldstein:2009cv}, argue that if it is frequency-dominated both in the UV and in the IR, and potential-dominated in-between, one may deduce the IR scaling from the UV scaling by imposing conservation of the energy flux of $\Psi$ from one to the other. This will give a relation between the IR and the UV, where the conductivity is calculated.The AC conductivity was given in \cite{Goldstein:2009cv}, building upon previous work by \cite{Hartnoll:2008kx,Horowitz:2009ij},
\be
	\sig(\omega) = \frac{1-\mathcal R(\omega)}{1+\mathcal R(\omega)}-\frac{i}{2\omega}\l.\frac{\dot Z}{Z}\r|_b\,,
	\label{ACConductivity}
\ee
where $Z(\phi)=\e^{\ga\phi}$ is the gauge coupling function and dots denote derivation with respect to the radial coordinate $z$. $\mathcal R(\omega)$ is the energy-dependent reflection coefficient of the Schr\"odinger potential. One wants to solve the problem with the following boundary conditions: in the IR ($z\to-\infty$), the wave should be purely ingoing, so no information comes out from the horizon; in the UV ($z=0$), after extending the potential to positive $z$ by setting $V(z>0)=0$, there should be an ingoing wave and a reflected outgoing wave. This is summarised by:
\be
	A_x \underset{z>0}{\sim} \e^{-i\omega t} + \mathcal R\e^{+i\omega t}\,,\qquad
	A_x \underset{z\to-\infty}{\sim} \mathcal T\e^{-i\omega t}\,,
\ee
where $\mathcal R$ and $\mathcal T$ are the reflection and transmission coefficients in amplitude for the waves. On the UV boundary,
\bsea
	\Psi(0)&=&Z(0)A_x (0) = 1+\mathcal  R\,, \\
	\Psi_{,z}(0)&=&Z(0)A_{x,z}(0)+Z_{,z}(0)A_x(0)=-i\omega(1-\mathcal R)\,.
	\label{BoundaryConditionsPlaneWaves}
\esea
It was proven in \cite{Hartnoll:2008kx} that, with AdS boundary conditions, $A_x=A_x^{(0)}+r^{-1}A_x^{(1)}$, where $r$ is the usual black-hole coordinate. Using AdS/CFT, $A^{(0)}_x$ may be interpreted as a source on the boundary, while $A^{(1)}$ is a current generated by the linear response of the system. The conductivity is then inferred by Ohm's law:
\be
	\sigma = \frac{J_x}{E_x} = -\frac{iA^{(1)}_x}{A^{(0)}}\,,
	\label{ACConductivityGauge}
\ee
which reduces to \eqref{ACConductivity} using \eqref{BoundaryConditionsPlaneWaves}. These arguments rely crucially on our assumption of AdS asymptotics. Thus, one may deduce the scaling of the conductivity in the UV.

The results for our solutions are as follows, \cite{Charmousis:2010zz}:
\begin{figure}[!th]
\begin{center}
\begin{tabular}{ccc}
	\includegraphics[width=0.3\textwidth]{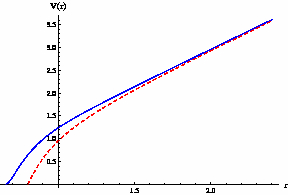}&
	\includegraphics[width=0.3\textwidth]{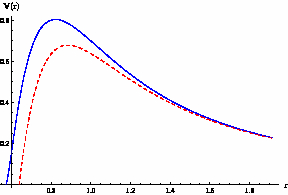}&	\includegraphics[width=0.3\textwidth]{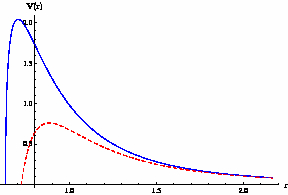}
	\end{tabular}
\caption[Schr\"odinger potential for gauge fluctuations around the $\gamma\delta=1$ planar EMD solution
 at and near extremality in the lower and middle range]{The Schr\"odinger potential $V(r)$
for the $A_x$ fluctuation around the $\gamma\delta=1$ solution
 at extremality (blue solid curves) and slightly away from extremality (red dashed curves) in the range $\da^2<1+2/\sqrt{3}$. The non-extremality is parametrized by the non-extremality parameter $\epsilon$, $\eps=0$ being extremality. From left to right, the plots show the respective potentials for  $(\da,\epsilon)=(0.7,0.3)$, $(1.3,0.1)$ and $(\sqrt{1+2/\sqrt{3}}-0.03,0.1)$.}
\label{Fig:gd1VSlowmiddle}
\end{center}
\end{figure}

\begin{itemize}
	\item In the $\ga\da=1$ case, the Schr\"odinger potential has the correct behaviour when $\da^2<1+2/\sqrt 3$, which is also the thermodynamically stable region. The AC conductivity scales like
	\be\label{gd1ACscaling}
\sigma(\omega) \simeq \omega^n \,,\qquad n = \sqrt{4c+1}-1 = \frac{(3-\da^2)(5\da^2+1)}{|3\da^4-6\da^2-1|} - 1\,.
	\ee
	There is a lower bound on the exponent $n\geq\frac53$, so that it never becomes negative. The system then behaves like a conductor, both at and near extremality since the potential is qualitatively unchanged\footnote{In the case $\da^2<1$, one should keep in mind that the Schr\"odinger potential potential is expected to be completed by an AdS contribution which vanishes on the UV boundary, see \cite{Goldstein:2009cv,Charmousis:2010zz}.}, see \Figref{Fig:gd1VSlowmiddle}.

\begin{figure}[ht]
\begin{center}
\begin{tabular}{cc}
	 \includegraphics[width=0.45\textwidth]{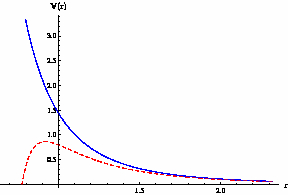}&
	 \includegraphics[width=0.45\textwidth]{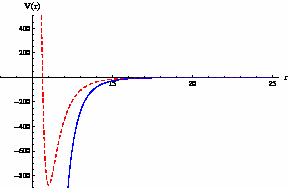}
	 \end{tabular}
\caption[Schr\"odinger potential for gauge fluctuations around the $\gamma\delta=1$ planar EMD solution
 at and near extremality in the upper range]{The Schr\"odinger potential $V(r)$  for the $A_x$ fluctuation around the $\gamma\delta=1$ solution  at extremality (blue solid curves) and slightly away from extremality (red dashed curves) in the range $\da^2\ge1+2/\sqrt{3}$. The non-extremality is parametrized by the non-extremality parameter $\epsilon$, with $\eps=0$ being extremality. From left to right, the plots show the respective potentials for  $(\da,\epsilon)=(\sqrt{1+2/\sqrt{3}},0.1)$ and $\left(\frac{1}{2}\sqrt{5+\sqrt{33}}+0.002,0.01\right)$.}
\label{Fig:gd1VSupper}
\end{center}
\end{figure}

	In the upper range $\da^2>1+2/\sqrt 3$, two ranges must be distinguished. If $1+2/\sqrt 3<\da^2<5/4+\sqrt{33}/4$, the potential diverges near the horizon at extremality, and so is insulating; away from extremality, this behaviour is immediately regularised and the system become conducting with a finite potential wall, see the left pannel of \Figref{Fig:gd1VSupper}. When $5/4+\sqrt{33}/4<\da^2<3$, the potential has an infinite well at extremality and so displays a continuum of bound states and can be conductive;  on the other hand, away from extremality, the potential displays a finite negative minimum and so is an insulator, see the right pannel of \Figref{Fig:gd1VSupper}.
	
	\item For the $\ga=\da$ solutions, the scaling of the potential and the conductivity is universal:
	\be
		\sig(\omega) = \omega^2\,,
		\label{ACConductivityG=D}
	\ee
	which can be checked independently by setting $\da=1$ in \eqref{gd1ACscaling}. The Schr\"odinger potential diverges at the UV, but we expect this to be corrected once it has been AdS-completed, so that the system is conducting.

\begin{figure}[!ht]
\begin{center}
\begin{tabular}{c}
	 \includegraphics[width=0.4\textwidth]{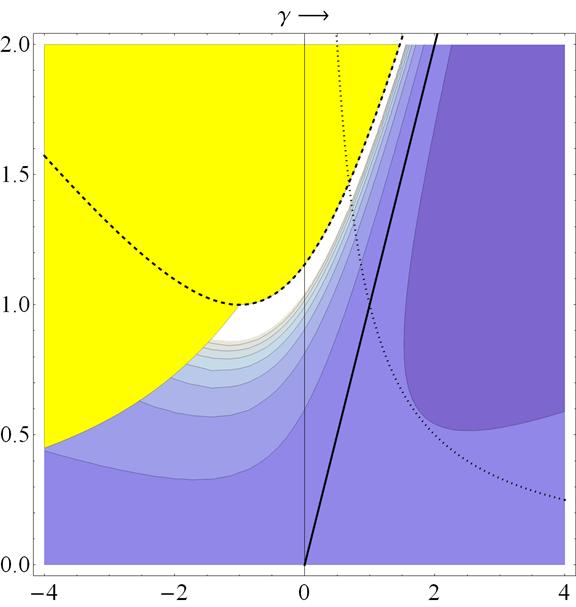}
\end{tabular}
\caption[Frequency scaling exponent of the AC conductivity for the $\ga\da=1$ planar EMD solution]{Contour plot of the scaling exponent $n$ in the $(\gamma,\delta)$ upper half plane, $0\leq \da \leq \sqrt{5/3}$. Contours correspond to $n=1.52,\dots,8.36$, starting with $n=1.52$ in the upper right corner and increasing in steps of $0.76$. The black solid line $\ga=\d$ is $n=2$, and brighter colors correspond to larger $n$.  The black dotted line is $\ga\da=1$. The yellow regions are forbidden by several constraints (see text). The scaling exponent diverges to $+\infty$ along the dashed black line in both cases.}
\label{Fig:ACNegExpSol3}
\end{center}
\end{figure}

	\item Let us finally turn to the near-extremal arbitrary $\ga$, $\da$ geometry. The AC conductivity is found to scale like
	\be
		\label{Sol3ACsigma}
		\sigma \sim \omega^n\,,\qquad n = \left|\frac{-12+(\da-\gamma)(3\gamma + 5\da)}{-4 + (\da - \gamma) (\gamma + 3\da )}\right| - 1\,.
	\ee
	This exponent is positive in the whole thermodynamically and holographically valid region, see \Figref{Fig:ACNegExpSol3}. There, the system is a conductor, once the proper AdS-completion has been taken into account.
\end{itemize}

\subsubsection{DC conductivity and resistivity}

In a system with charge carriers, the resistivity is the inverse of the DC conductivity, which itself is the zero-frequency limit of the AC conductivity. From \eqref{ACConductivityGauge}, one may expect some trouble in this limit, which will yield an infinite value. Another approach was pointed out in \cite{Karch:2007pd}: one may adapt a drag force calculation, felt by massive charged carriers in the charged dilatonic black hole background. A classic result from the Drude theory of electrons moving around in a metal is that a steady state may be reached, where the acceleration due to the background (bulk) electric field is completely compensated by the effective drag force felt by the electrons due to collisions.and is proportional to their momentum. This can in turn be expressed in terms of the current, and so one obtains a relation between the electric field and the current which gives the DC conductivity, by Ohm's law. We shall not enter into the details of how this carries over in String Theory, and rather refer the interested reader to \cite{Karch:2007pd} and \cite{Charmousis:2010zz}.

The results are summarised as follows, \cite{Charmousis:2010zz}:
\begin{itemize}
 \item In the $\ga\da=1$ case, one finds two occurrences of linear scaling of the resistivity at low temperature, one in the lower range $\da^2<1$ and one in the intermediate range $1<\da^2<1+2/\sqrt3$.
 \item In the $\ga=\da$ case, the resistivity has a regular Taylor expansion in integer powers of the temperature. The first term is a constant, but the second term is linear in the temperature.

In both cases, the validity of the low-temperature expansion is controlled by the value of the IR scale.

 \item In the generic, near-extremal case, the low-temperature approximation is built in. One finds that the scaling is linear when $\ga_\pm = 3\da+2\pm\sqrt{\da^2+2\da+2}$.
\end{itemize}

To conclude this section on applications of our IR geometries to Condensed Matter systems, we may say they allow to recover some characteristics of strange metallic behaviour, that is linear scaling of the dual resisitivity over a broad enough range of parameters. However, though the AC conductivity does scale with the frequency, the exponent is positive and not negative as is expected for strange metals.

\section*{Conclusion\markboth{Conclusion}{Conclusion}}
\addcontentsline{toc}{section}{\protect\numberline{}Conclusion}

Throughout this thesis, the red line we have been guided by has been the search for possible new black hole solutions in various gravity theories, departing from General Relativity in a number of ways: either by the inclusion of matter fields with specific couplings, or by the addition of extra spatial dimensions and higher powers of the Riemann tensor in the gravity Lagrangian. The former was the subject of the study in Section \ref{section:EMDBH}, while the latter was examined in Section \ref{section:EGBBH}.

The Einstein-Maxwell-Dilaton theories with a Liouville scalar potential we have studied in Section \ref{section:EMDBH} find their motivation from several sources: they are another example of the interplay between gravity and matter fields, and so it is interesting to replace them in the context of no-hair theorems; they can also be seen to descend from higher-dimensional theories for specific relations between the couplings, either by Kaluza-Klein reduction of one of the extra dimensions or by compactification of six of the ten dimensions of supergravity, the low-energy, classical approximation to String Theory.

Several remarks are in order. The study of the integrability of the equations of motion constitutes an essential part of the work presented in \cite{Charmousis:2009xr}: in the case of two-dimensional planar maximally symmetric subspaces, the system of equations boils down to a single second-order non-linear ordinary differential equation; in the case of two-dimensional non-planar maximally symmetric subspaces, the system of equations boils down to two coupled second-order non-linear ordinary differential equations. It is quite interesting that both cases which we can link to higher-dimensional theories benefit from extended integrability properties. They belong to a more general class for which the single remaining equation can be completely integrated in the case of solutions with a maximally symmetric planar subspace; where the two remaining equations can be decoupled in the case of  solutions with a maximally symmetric non-planar subspace. This hints at the possibility that some kind of hidden symmetry is at work here, inherited from the higher-dimensional theory\footnote{We are grateful to Pr. Marc Henneaux at ULB, Brussels, for suggesting this idea.}. This matter certainly invites deeper investigation. We then proceeded to classify and analyse the properties of the solutions of the theory.

The scalar field has a dramatic impact on spacetime. Since the potential is an exponential, it has a runaway minimum at which the scalar field takes an infinite value. Thus, even the background (no black hole) spacetime is endowed with a non-trivial scalar field which diverges asymptotically. It can be seen to depart from regular asymptotics, neither flat nor (Anti-)de Sitter. The black holes living in this unconventional background will thus suffer from the same irregular behaviour at spatial infinity. Concentrating on planar black holes, which made for the largest part of the novel material presented in \cite{Charmousis:2009xr}, we exhibited a large class of exact and general new solutions when the couplings are interrelated in specific ways. In the case of arbitrary couplings, we exhibited a class of solutions which we interpreted as near-extremal, near-horizons limits of the full solution. The latter remains elusive, and though there is no guarantee a closed form exists, it would still be worthwhile to write down its perturbative expansion at infinity. One class of solutions has truly dilatonic properties: it has a singularity at finite distance, cloaked by a regular event horizon, and an irregular extremal limit (the extremal radius is the locus of a naked singularity); moreover the horizon degenerates at this same finite radius. Another class of solutions retains some properties which are Reissner-Nordstr\"om-like: they have an outer (event) and inner (Cauchy) horizon cloaking a singularity, with a regular extremal limit; the horizon space degenerates at zero radius. The near-extremal solutions share properties with both classes in the appropriate limits.

We closed off this section by suggesting directions for future work. Dyonic solutions are still quite scarce, though promising, fragmentary work already exists. This would allow to go beyond the electromagnetic duality exposed in \cite{Charmousis:2009xr} (though it had already appeared in previous work by Charmousis \emph{et al.} in \cite{Charmousis:2006fx}), which maps the electric solutions we found to magnetic ones. Another direction involves the inclusion of a specific combination of quadratic powers of the Riemann tensor in the action: the so-called Gauss-Bonnet term, for which the theory remains ghostless (no negative energy degree of freedom). This term is expected in some String Theory low-energy effective actions and comes naturally in higher-dimensional theories.

These Gauss-Bonnet theories were the topic of Section \ref{section:EGBBH}. They are a natural generalisation of General Relativity, in that they conserve all of its properties (second-order equations, Bianchi identity, no ghost degrees of freedom), and can be shown to be the unique set of theories for which it is so in five and six dimensions (just as it was the case for General Relativity in four dimensions). This line of reasoning carries over in any dimension, at the price of including generalised, dimensionally-extended Euler characteristics in the action: this theory of gravity in generic dimension is known as Lovelock theory.

Coming back to Gauss-Bonnet theory, the main surprise is that they generically contain two vacua, as well as two branches of spherically-symmetric and static black holes, a fact which seems to circumvent Birkhoff's theorem. When the bare cosmological constant in the action is zero, one of the branches (the Einstein branch) is asymptotically flat, while the other (the Gauss-Bonnet branch) displays an effective cosmological constant and (Anti)-de Sitter asymptotics. The latter branch is a distinctive feature of Gauss-Bonnet gravity. A measure of solace comes from its instability, which helps to minimise the loss of unicity.

Spurred on by results obtained by the authors of \cite{Dotti:2005rc}, we made a full classification of the six-dimensional theory in \cite{Bogdanos:2009pc} (the five-dimensional classification had already appeared some time ago in \cite{Charmousis:2002rc}). The main novelty of the analysis was a full accounting of the exposure of the dynamics to the Weyl tensor of the four-dimensional horizon space, which is made possible only in Gauss-Bonnet theory and not in General Relativity. A non-zero Weyl tensor induces non-trivial, non-maximally symmetric topology on the horizon. On top of modifying the asymptotics of the blackness potential and generating event horizons, this allows to progress on a more fundamental problem with General Relativity in higher dimensions. It is a well-known fact that unicity properties of Einstein theory are generically lost upon going to higher dimensions: numerous new solutions coexist with Myers-Perry black holes, such as black strings, black rings, etc. Moreover, even simple black holes such as Schwarzschild suffer from degeneracy: the horizon topology is not limited to maximally symmetric spaces anylonger, but can simply be Einstein spaces. The former have zero Weyl tensor and so the Riemann tensor is proportional to the metric; the latter have possibly non-zero Weyl tensor and only the Ricci tensor is proportional to the metric. There exist many Einstein spaces, and the horizon topology is greatly degenerated. This state of affairs is quite different once the Gauss-Bonnet term is included: the square of the Weyl tensor of the horizon needs to be proportional to the metric, too. This imposes a severe trimming down of the plethora of Einstein spaces admissible, and non-trivial (e.g., non-maximally symmetric) topologies are quite hard to find: we exhibited two such instances, one of them the product of two two-spheres, the second a specific limit of Taub-NUT space called the Bergman space.

Lifting the degeneracy on the horizon's topology in higher-dimensional gravity seems to be a generic feature of Lovelock theory. These preliminary results need to be confirmed, since then this would figure in good place in Lovelock theory's list of achievements. The completion of this work is left to the near-future.

After an introductory section recounting the analogy between black-hole mechanics and thermodynamics, as well as classic results on the thermodynamics of black holes in General Relativity, we moved on to the study of the thermodynamics of black holes in Einstein-Maxwell-Dilaton theories in Section \ref{section:ThermoEMD}. These results were presented in \cite{Charmousis:2010zz}. They show that in the canonical ensemble, where the temperature and electric charge are kept fixed, one witnesses the appearance of two branches of small and large black holes, similarly as for \emph{spherical} Reissner-Nordstr\"om, for some intermediary range in the couplings. Only the former are stable, globally and locally, up to a maximum temperature at which they cease to exist. Beyond, a zeroth-order phase transition to the extremal background occurs. In a lower range of the couplings, there is also a single stable branch. This range is purely dilatonic in that the situation clearly does not resemble General Relativity. In the grand-canonical ensemble, a single stable branch exists in a lower range of the couplings, while the upper ranges are unstable. Further, the study of phase transitions at zero temperature shows that they can generically be of any order, depending on the values of the couplings. No Hawking-Page phase transition is present: we have focused on planar horizons, so there is no curvature scalar on the horizon which the cosmological constant can couple to in order to generate an extra critical temperature, as is the case for the spherical Reissner-Nordstr\"om-AdS black holes. It is quite interesting to notice that the effect of the scalar field varies depending on the values of the couplings: it can destroy the effect of the negative cosmological constant or the electric charge, but it is also able to simulate a positive horizon curvature in some cases. It would certainly be interesting to investigate further the thermodynamics of the non-planar solutions to check that the Hawking-Page transition can indeed be recovered in cases where Anit-de Sitter asymptotics are restored. 

Another way to consider these solutions is provided by the AdS/CFT correspondence and the principles guiding the building of Effective Holographic Theories, examined in Section \ref{section:Holography}. One may think of the EMD solutions as valid in the Infra-Red, where the scalar potential is generically expected to be dominated by exponentials (see \cite{Gursoy:2007cb,Gursoy:2007er,gkmn2} for a generic classification of the asymptotics of the potential and the thermodynamics of the neutral black holes). On the other hand, a suitable AdS-completion is assumed in the UV. This is embodied in our solutions by the vanishing of the scalar potential in the UV when evaluated on the solutions (except for a subtlety in the near-extremal case due to the lack of information: the near-extremal geometry should not be extrapolated in the UV). This is confirmed by a phase space analysis, revealing that the stable UV fixed point is the neutral extremal background, while the stable IR fixed point is the charged extremal background.  The two-dimensional phase space in the couplings can then be constrained through imposition of holographic bounds controlling that the IR singularity is well-behaved, and encompasses the region of thermodynamic stability.

Finally, we briefly exposed how one could compute holographic transport coefficients for our EMD solutions, and interpret them in the dual boundary field theory as conductivities (both AC and DC). The interest for such a procedure arose at first in relation with superfluid phase transitions. These setups contain a charged scalar field, which upon condensation yields the superfluid phase. In our case, the scalar field is real and one rather looks for classes of universality in the region of strange metal behaviours, the hallmarks of which are non-conventional scaling of the AC conductivity with the frequency and linear scaling of the resistivity at low temperatures. The latter can certainly be accommodated in some lines of the phase space, but the latter display positive exponents while strange metals show negative ones. There is thus room for improvement, especially in the addition of new ingredients: superfluid transitions are generated by complex charged scalars, and assume a magnetic field is added to the setup. This underlines the importance of looking for dyonic solutions, which are for the moment rather scarce.

Let us bring this work to a conclusion on a more general note. It is our hope that in some small way we have contributed to emphasise the central role black holes play in the physics of gravitation. They are as fundamental to General Relativity as group representations are to particle physics and the Standard Model. As higher dimensions entered the scene, the beautiful picture made by unicity and existence theorems seemed to fissure: numerous solutions were added and topological restrictions went down. On the classical gravity side, Lovelock theory seems to bring back some of the order that was lost. But on the quantum gravity side, plethora of solutions compete, in particular due to the difficulty of agreeing on a particular truncation of String Theory low-energy effective actions. One then has to wonder if the point is not being missed, and if black holes are sufficient to the task. It may very well be that String Theory requires another conceptual upheaval of the magnitude caused by the advent of General Relativity, and which is maybe yet to come. Or that in the end String Theory may mutate into some other theory of quantum gravity. All in all, whether black holes as we know them will survive the transition to fully quantum gravity remains quite an open question.

\vfill

\appendix

\selectlanguage{french}
\section{Appendix: Synopsis}

\subsection{Introduction}

\subsubsection{La th\'eorie de la Relativit\'e G\'en\'erale}

Au cours de cette th\`ese, nous allons passer en revue les solutions de trou noir de certaines th\'eories de gravitation, pr\'esentant des modifications vis-\`a-vis du cas paradigmatique de la Relativit\'e G\'en\'erale d'Einstein. Ces modifications rev\^etront la forme soit de champs de mati\`ere additionnels, soit de dimensions spatiales suppl\'ementaires.

En guise d'introduction, commen\c cons par quelques mots sur la Relativit\'e G\'en\'erale elle-m\^eme. Cette th\'eorie fut propos\'ee par Einstein en 1915, apr\`es de longues ann\'ees de labeur sur le lien entre la th\'eorie newtonienne de la gravitation et la Relativit\'e Restreinte\footnote{Introduite par Einstein lui-m\^eme en 1905 sous sa forme la plus aboutie, cette th\'eorie du mouvement inertiel \'etait construite autour de l'invariance de Lorentz du mouvement d'observateurs inertiels (c\`ad non-acc\'el\'er\'es). Elle postulait notamment que le temps devait \^etre trait\'e sur un pied d'\'egalit\'e avec les coordonn\'ees spatiales, et introduisait une vitesse maximale pour le mouvement, celle de la lumi\`ere, not\'ee $c$. Ainsi, cela permettait de construire une coordonn\'ee $x^0$ avec la dimension d'une longueur \`a partir de la coordonn\'ee habituelle de temps $t$ : $x^0=ct$. De nombreux effets contre-intuitifs \`a l'\'epoque furent d\'eduits de ces deux postulats, comme la contraction des longueurs et la dilatation des temps. Dans la suite, nous adopterons un syst\`eme d'unit\'es o\`u $\hbar=c=8\pi G_N=1$}. Afin de r\'econcilier  ces deux formalismes, Einstein eut l'id\'ee de ``g\'eom\'etriser'' la force de gravitation. Loin de  sortir du n\'eant, cette intuition provenait de l'analogie qu'Einstein \'etablit entre mouvement uniform\'ement acc\'el\'er\'e et chute libre sous l'action d'un champ gravitationnel. La gravit\'e pouvait \^etre ``compens\'ee'' par une acc\'el\'eration ad\'equate.

Les \'equations d'Einstein furent publi\'ees sous la forme suivante :
\be
	R_{\mu\nu}-\half R g_{\mu\nu}= T_{\mu\nu}\,, \label{EinsteinEqSV}
\ee
o\`u l'on d\'efinit $G_N$ comme la constante de gravitation de Newton, tandis que $g_{\mu\nu}$ est la m\'etrique (lorentzienne, c\`ad de signature $(-,+,+,+)$ pour trois dimensions d'espace et une de temps) permettant de mesurer les intervalles d'espace-temps\footnote{$\ud s^2=g_{\mu\nu}\ud x^\mu\ud x^\nu$; lorsque $\ud s^2$ est n\'egatif, positif ou nul, l'intervalle est dit respectivement de genre temps, espace ou lumi\`ere, en r\'ef\'erence \`a la nature du quadri-vecteur permettant de joindre les extr\^emit\'es de l'intervalle.}, et que le tenseur de Riemann $R_{\la\mu\nu\rho}$, construit \`a partir des d\'eriv\'ees secondes de la m\'etrique, mesure la courbure de l'espace-temps. Au vu de ses propri\'t\'es d'antisym\'etrie, on ne peut en former qu'une seule trace non-triviale, ce qui donne le tenseur de Ricci : $R_{\mu\nu}\doteq g^{\la\rho}R_{\la\mu\rho\nu}$. La trace de ce dernier constitue un invariant scalaire de l'espace-temps, le scalaire de Ricci : $R\doteq g^{\mu\nu}R_{\mu\nu}$, dont les propri\'et\'es physiques ne d\'ependent pas du syst\`eme de coordonn\'ees choisi, contrairement \`a un tenseur de rang non-nul comme le tenseur de Riemann. Par exemple, une singularit\'e de courbure de l'espace-temps sera signal\'ee par une divergence d'au moins un des invariants form\'es \`a partir du tenseur de Riemann : le scalaire de Ricci, mais aussi le carr\'e du tenseur de Riemann ou encore celui de Ricci. Enfin, le tenseur pr\'esent dans le membre de droite de \eqref{EinsteinEqSV} est le tenseur \'energie-impulsion qui encode les propri\'et\'es de la mati\`ere pr\'esente dans l'espace-temps.

Afin de formuler ces \'equations, Einstein se laissa guider par un certain nombre de principes qu'il para\^it utile de rappeler :
\begin{itemize}
 \item \emph{Le principe de Mach} cherche \`a d\'epasser le concept habitule d'espace et de temps absolus newtoniens. En effet, un mouvement est dit inertiel lorsqu'il est uniform\'ement acc\'el\'er\'e compar\'e \`a un r\'ef\'erentiel \emph{absolu}, postul\'e \emph{a priori} de toute distribution de mati\`ere. Dans les pas de Mach, Einstein refusait par essence un tel postulat, et tenta de construire une th\'eorie o\`u g\'eom\'etrie et mati\`ere seraient intimement li\'es : la mati\`ere d\'eterminerait la g\'eom\'etrie, et vice-versa. Ce principe est clairement incarn\'e par la forme des \'equations \eqref{EinsteinEqSV}, o\`u \`a gauche est pr\'esente la g\'eom\'etrie, \`a droite la mati\`ere\footnote{La forme initiale de la th\'eorie de la Relativit\'e G\'en\'erale n'impl\'ementait d\'ej\`a qu'une version faible de ce principe, puique l'espace-temps plat de Minkowski est une solution sans mati\`ere des \'equations \eqref{EinsteinEqSV}. Faible au sens o\`u effectivement une corr\'elation semblait survenir entre absence de courbure et absence de mati\`ere. Bien s\^ur, le lecteur averti sait qu'un d\'ementi cinglant fut apport\'e d\`es 1916 avec la publication de la solution de Schwarzschild, qui d\'ecrit un espace-temps courbe sans mati\`ere.}.
 \item \emph{Le principe d'\'equivalence} postule l'\'egalit\'e entre la masse inertielle $m_i$, qui intervient dans le Principe Fondamental de la Dynamique de Newton $m_i{\bf\overrightarrow{a}}={\bf\overrightarrow{F_{ext}}}$, la masse grave passive $m_p$, qui intervient dans l'expression de la force gravitationnelle ressentie par une particule ponctuelle massive dans un potentiel gravitationnel ${\bf \overrightarrow{F}}=m_p{\bf\overrightarrow{\nabla}}\Phi$, et enfin la masse grave active $m_a$, qui intervient dans le potentiel gravitationnel cr\'e\'e par une masse ponctuelle $\Phi=-\frac{G_Nm_a}r$. Cette \'egalit\'e, fortuite et non-n\'ecessaire dans la th\'eorie newtonienne, devient essentielle \`a la th\'eorie einsteinienne, sans laquelle ni r\'ef\'erentiels localement inertiels ni description coh\'erente de l'\'interaction g\'eom\'etrie-mati\`ere.
 \item \emph{Le principe de la covariance g\'en\'erale} postule que les lois de la physique sont invariantes sous un changement de coordonn\'ees. Tout observateur doit pouvoir localement d\'eterminer les lois de la physique. En termes plus math\'ematiques, la structure tensorielle de la th\'eorie est compl\'et\'ee par une invariance sous les diff\'eomorphismes.
\end{itemize}

Enfin, terminons cette courte introduction \`a la th\'eorie de la Relativit\'e G\'en\'erale par la mention de deux derni\`eres propri\'et\'es des \'equations d'Einstein. Au-del\`a des trois principes que nous venons de d\'ecrire, ce qui fixa la forme d\'efinitive des \'equations d'Einstein fut la conservation de l'\'energie. En effet, pour autant que les \'equations \eqref{EinsteinEqSV} d\'ecrivent un syt\`eme isol\'e, l'\'energie totale de ce syst\`eme se doit d'\^etre conserv\'ee, ce qui se traduit par l'identit\'e suivante :
\be \nabla_\mu T^{\mu}_\nu=0\,. \ee
Or la seule combinaison \`a deux indices faisant intervenir la m\'etrique et le tenseur de Riemann admissible pour le membre de droite des \'equations d'Einstein est tr\`es pr\'ecis\'ement le tenseur d'Einstein
\be G_{\mu\nu}\doteq R_{\mu\nu}-\half Rg_{\mu\nu}\,, \ee
de divergence nulle gr\^ace aux identit\'es de Bianchi du tenseur de Riemann. La beaut\'e des \'equations d'Einstein r\'eside donc aussi dans la coincidence de la propri\'et\'e de divergence nulle des deux membres, soit juste ce qu'il faut pour retrouver les lois habituelles de la physique.

Concluons en pr\'esentant l'action dite d'Einstein-Hilbert, qui permet de retrouver les \'equations d'Einstein par un principe variationnel de moindre action et de faire le lien avec une th\'eorie du champ gravitationnel :
\be
	S_{EH} = \frac1{2}\int \ud^4x\sqrt{-g}\l(R + \mathcal{L}_m\l[\Psi\r]\r)\,,
	\label{EinsteinHilbertActionSV}
\ee
o\`u apparaissent la racine carr\'ee du d\'eterminant de la m\'etrique, le scalaire de Ricci et le Lagrangian d\'ecrivant les champs de mati\`ere $\Psi$. En variant cette action par rapport \`a la m\'etrique $g_{\mu\nu}$ et en imposant que les trajectoires classiques correspondent \`a celles qui extr\'emisent l'action, on retrouve les \'equations \eqref{EinsteinEqSV} \`a condition d'identifier le tenseur \'energie-impulsion et la variation du Lagrangien de la mati\`ere
\be
	T_{\mu\nu}=\frac{\da\mathcal L_m}{\da g^{\mu\nu}}\,.
	\label{StressEnergyGeneralSV}
\ee
Cette formulation de la Relativit\'e G\'en\'erale ouvre la porte, en anticipant sur ce qui va suivre, \`a un traitement effectif de la gravitation, dans lequel le terme de Ricci ne constitue qu'une approximation valide \`a certaines \'echelles de la ``bonne'' th\'eorie de la gravitation.

\subsubsection{Trous noirs en Relativit\'e G\'en\'erale et extensions}

\paragraph{La solution de Schwarzschild}

(1916) fut l'une des toutes premi\`eres solutions exactes connue en Relativit\'e G\'en\'erale, quoiqu'il fall\^ut attendre quasiment cinquante ans pour qu'une interpr\'etation coh\'erente en soit donn\'ee. La voici :
\be
	\ud s^2 = -\l(1-\frac{2m}{r}\r)\ud t^2 +\frac{\ud r^2}{\l(1-\frac{2m}{r}\r)}+r^2\l(\ud\theta^2+\sin^2\theta\ud\varphi^2\r).
	\label{SchwarzschildSV}
\ee
Sous sa forme initiale, elle se pr\'esente comme d\'ecrivant la r\'egion (vide) ext\'erieure d'un objet \`a sym\'etrie sph\'erique (compos\'e de mati\`ere, comme une \'etoile), et effectivement il est possible d'en faire la jonction avec une solution d\'ecrivant une \'etoile \`a densit\'e d'\'energie constante. On remarque que sa signature change lorsque la surface $r=2m$ est franchie, de $(-,+,+,+)$ \`a $(+,-,+,+)$. Source de nombreuses confusions pendant de longues ann\'ees, ce lieu ``magique'' o\`u temps et espace s'interchangent sera finalement interpr\'et\'e comme un horizon des \'ev\'enements, c\`ad une hypersurface de genre lumi\`ere\footnote{Le vecteur de Killing $\xi^\mu=\delta^\mu_0\frac{\partial}{\partial x^0}$ y est de norme nulle.} o\`u tout signal est infiniment d\'ecal\'e vers le rouge pour un observateur \`a l'infini. Il lui est donc impossible de voir quoi que ce soit franchir cet horizon, d'o\`u le qualificatif de ``noir'' attribu\'e \`a cette solution. En revanche, et en tirant les le\c cons de la covariance g\'en\'erale, il est ais\'e de voir (a posteriori!) que des coordonn\'ees d'Eddington-Finkelstein permettent de suivre la chute libre d'un observateur dans le trou noir et se comportent tout \`a fait r\'eguli\`erement \`a l'horizon. La solution de Schwarzschild \'etant une solution du vide, son scalaire de Ricci est nul, mais le carr\'e du tenseur de Riemann pr\'esente une divergence au centre de la coordonn\'ee $r$ :
\be R_{\la\mu\nu\rho}R^{\la\mu\nu\rho}\sim\frac1{r^6}\,, \ee
ce qui signale une singularit\'e de courbure en cet endroit, les forces de mar\'ee devenant infiniment intenses. La singularit\'e de courbure en $r=0$ est une caract\'eristique intrins\`eque de cette solution, tandis que l'horizon des \'ev\`enements en $r^+=2m$ est un art\'efact du syst\`eme de coordonn\'ees utilis\'e par un observateur ext\'erieur au trou noir.

\begin{figure}[t]
\begin{center}
	 \includegraphics[width=0.45\textwidth]{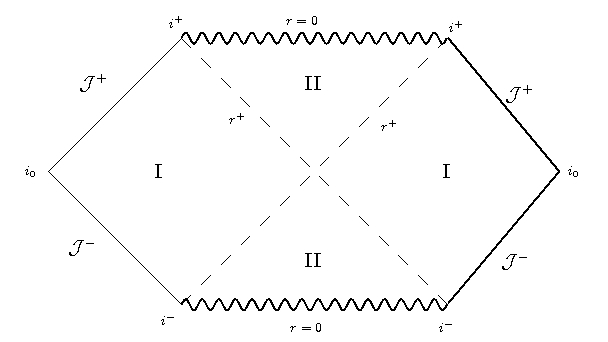}
	 \caption{Diagramme conformal de Penrose-Carter pour l'espace-temps de Schwarzschild.}
\label{Fig:SchwMaximalSV}
\end{center}
\end{figure}

Ces propri\'et\'es sont bien r\'esum\'ees par un diagramme d'espace-temps \`a la mani\`ere de Penrose et Carter, \Figref{Fig:SchwMaximalSV}, o\`u l'infini asymptotique du trou noir (de genre lumi\`ere) est ramen\'e \`a une distance finie par une transformation conforme\footnote{Du type $\tilde g_{\mu\nu}=\Omega^2(x^\mu)g_{\mu\nu}$.}. Celle-ci conserve les angles tout en dilatant ou en contractant les intervalles d'espace-temps, cela permet donc de d\'ecrire la r\'egion asymptotique sans modifier la structure causale de l'espace-temps. Le trou noir peut alors \^etre visualis\'e comme la r\'egion $II$, s\'epar\'ee causalement du futur de la r\'egion ext\'erieure $I$ par l'horizon, qui est le bord de genre lumi\`ere de la r\'egion $II$. La singularit\'e est de genre espace et est donc in\'evitable par tout observateur qui suivrait une g\'eod\'esique (de genre temps).

Cette m\'ethode a \'egalement pour cons\'equence de r\'ev\'eler la topologie non-triviale de l'extension des coordonn\'ees de la solution de Schwarzschild au-del\`a de l'horizon, puisqu'il faut y adjoindre un ``trou blanc'' en $II'$, dont on peut s'\'echapper mais non y p\'en\'etrer, ainsi qu'une seconde r\'egion asymptotique $I'$ causalement d\'econnect\'ee de la premi\`ere.

\paragraph{La solution de Reissner-Nordstr\"om}est une extension imm\'ediate aux trous noirs charg\'es :
\bsea
	\ud s^2 &=& -\l(1-\frac{2m}{r}+\frac{q^2}{r^2}\r)\ud t^2 +\frac{\ud r^2}{\l(1-\frac{2m}{r}+\frac{q^2}{r^2}\r)}+r^2\l(\ud\theta^2+\sin^2\theta\ud\varphi^2\r), \slabel{RNmetricSV} \\
	A&=&\l(\Phi-\frac{2q}r\r)\ud t\,. \slabel{RNASV}
	\label{RNSV}
\esea
Ce sont des extrema de l'action d'Einstein-Hilbert agr\'ement\'ee d'un terme de Maxwell :
\be
	S_{EM} = \frac1{2}\int \ud^4x\sqrt{-g}\l(R -\frac14 F^2\r)\,,\qquad F_{\mu\nu}=\partial_\mu A_\nu-\partial_\nu A_\mu\,.
	\label{EinsteinMaxwellActionSV}
\ee
Ils ont la particularit\'e de pr\'esenter deux horizons lorsque $|m|>|q|$, un seul lorsque $|m|=|q|$ et aucun lorsque $|m|<|q|$. Dans tous les cas, la singularit\'e en $r=0$ est de genre temps, et donc peut \^etre \'evit\'ee par un observateur suivant une g\'eod\'esique; dans le premier cas,  seul l'horizon ext\'erieur est un horizon des \'ev\'enements; dans le second cas, on a tout de m\^eme un trou noir, dit extr\'emal, o\`u l'horizon est d\'eg\'en\'er\'e; dans le dernier cas, il s'agit d'une singularit\'e nue. 

Nous avons jusqu'ici examin\'e des solutions de trou noir construits sur un espace-temps plat (on retrouve l'espace-temps de Minkowski en posant $m,q=0$ dans \eqref{SchwarzschildSV} et \eqref{RNSV}), que se passe-t-il dans le cas o\`u la courbure de l'\'espace-temps est globale et non locale?

\paragraph{L'espace-temps (Anti-)de Sitter}est une solution du vide lorsque l'on rajoute une constante positive (n\'egative) aux \'equations d'Einstein :
\be
	G_{\mu\nu}+\Lambda g_{\mu\nu}= T_{\mu\nu}\,, \label{EinsteinLambdaEqSV}
\ee
\be
	S_{E\Lambda} = \frac1{2}\int \ud^4x\sqrt{-g}\l(R -2\Lambda+ \mathcal{L}_m\l[\Psi\r]\r)\,.
	\label{EinsteinLambdaActionSV}
\ee
Dans ce cas, le vide de la th\'eorie s'\'ecrit:
\be
	\ud s^2 = -\l(-\frac{\Lambda}{3}r^2+1\r)\ud t^2 + \frac{\ud r^2}{\l(-\frac{\Lambda}{3}r^2+1\r)} +r^2\l(\ud\theta^2+\sin^2\theta\ud\varphi^2\r),
	\label{AdSSV}
\ee
et se ram\`ene comme attendu \`a Minkowski si l'on y pose $\Lambda=0$. Ils ont tous deux partout une courbure constante $R=4\Lambda$ comme on peut le voir \`a partir des \'equations \eqref{EinsteinLambdaEqSV}. Il est possible de montrer l\`a aussi que l'horizon, dit cosmologique, en $r_c=\sqrt{3/\Lambda}$ dans le cas o\`u $\Lambda>0$, est bien une singularit\'e de coordonn\'ees et non de courbure, par exemple en incorporant de Sitter comme une hypersurface de courbure positive dans un espace-temps de Minkowski \`a une dimension suppl\'ementaire. Similairement, quoique d\'epourvu d'horizon, Anti-de Sitter peut \^etre r\'e\'ecrit comme une hyperbolo\"ide dans une dimension suppl\'ementaire. 

\paragraph{Schwarzschild (Anti-)de Sitter}est la g\'en\'eralisation du trou noir de Schwarzschild sur un espace-temps \`a courbure constante positive (n\'egative) :
\be
	\ud s^2 = -\l(-\frac{\Lambda}{3}r^2+1-\frac{2m}r\r)\ud t^2 + \frac{\ud r^2}{\l(-\frac{\Lambda}{3}r^2+1-\frac{2m}r\r)} +r^2\l(\ud\theta^2+\sin^2\theta\ud\varphi^2\r),
	\label{KottlerSV}
\ee
Dans le cas de Schwarzschild-de Sitter, il est possible d'avoir \`a la fois un horizon des \'ev\'enements et un horizon cosmologique tels que $r_h<r_c$ si $9\Lambda m^2<1$, un horizon d\'eg\'en\'er\'e (extr\'emal) si $9\Lambda m^2=1$, ou bien une singularit\'e future/pass\'ee nue (c\`ad un Big Crunch ou un Big Bang) si $9\Lambda m^2>1$. Le fait que de Sitter soit asymptotiquement de genre espace (et non lumi\`ere comme Schwarzschild) ouvre la possibilit\'e d'avoir des horizons de particules. 

Le cas d'Anti-de Sitter est un peu plus int\'eressant, au sens o\`u la topologie de l'horizon peut devenir non-triviale. Nous avions eu affaire jusqu'ici \`a des horizons de topologie sph\'erique, tandis que Schwarzschild Anti-de Sitter peut \^etre g\'en\'eralis\'e \`a des topologies planaires ou hyperboliques :
\be
	\ud s^2 = -\l(-\frac{\Lambda}{3}r^2+\ka-\frac{2m}r+\frac{q^2}{r^2}\r)\ud t^2 + \frac{\ud r^2}{\l(-\frac{\Lambda}{3}r^2+\ka-\frac{2m}r+\frac{q^2}{r^2}\r)} +r^2\l(\ud\theta^2+\si^2\theta\ud\varphi^2\r),
	\label{TopoKottlerSV}
\ee
en d\'efinissant
\be
	 \si\theta = \begin{cases}
\sin\theta \, ,&\ka=1\,;\\
\theta\ ,&\ka=0\,;\\
\sinh\theta\ ,&\ka=-1\,.\\
\end{cases}
\label{HorizonKappaSV}
\ee
Dans le cas asymptotiquement plat, des th\'eor\`emes existent pour montrer que des horizons planaires ou hyperboliques ne sont pas possibles. Dans le cas d'Anti-de Sitter, cela n'est plus vrai et l'on peut avoir un horizon des \'ev\'enements topologique. Cela se comprend en imaginant qu'intuitivement, une surface plane (l'horizon) dans un espace-temps \`a courbure n\'egative (Anti-de Sitter) peut avoir les m\^emes effets qu'une surface de courbure positive dans un espace-temps plat.

\subsubsection{La Relativit\'e G\'en\'erale, une th\'eorie effective de la gravitation}

Dans un large r\'egime d'\'echelles, les pr\'edictions de la Relativit\'e G\'en\'erale se sont trouv\'ees extr\^emement bien v\'erifi\'ees. Ainsi, les tests dans le Syst\`eme solaire (avance du p\'erih\'elion de Mercure, d\'eflection de la lumi\`ere par le champ gravitationnel du Soleil ou encore l'effet Sapiro) atteignent aujourd'hui une excellente pr\'ecision de $10^{-4}-10^{-5}$. Il en est de m\^eme pour les tests en champ gravitationnel fort, comme la mesure de la p\'eriode de r\'evolution des pulsars. 

Ces \'echelles peuvent \^etre qualifi\'ees d'interm\'ediaires. En revanche, des probl\`emes surviennent aux \'echelles extr\^emes. Dans l'Ultra-Violet, c\`ad aux petites \'echelles/grandes \'energies, les diagrammes d'\'echange de gravitons divergent, ce qui est reli\'e au fait que la constante de couplage du processus, $G_N^{-1}$, a la dimension d'une masse au carr\'e. La th\'eorie n'est donc pas renormalisable, et l'on s'attend \`a ce que d'importantes corrections quantiques doivent \^etre prises en compte aux \'echelles de Planck ($\sim10^{19}GeV$). La Relativit\'e G\'en\'erale est donc vue comme une th\'eorie effective classique de la gravitation. Typiquement, les corrections quantiques pourraient prendre la forme de puissances suppl\'ementaires du tenseur de Riemann. Il existe plusieurs propositions de th\'eories quantiques de la gravitation, comme la th\'eorie des cordes ou encore la gravitation quantique \`a boucles. L'une est bas\'ee sur l'introduction d'une coupure ultra-violette naturelle, la longueur de Planck ($\sim10^{-33}cm$), \`a laquelle les champs sont r\'esolus comme \'etant des \'etats de vibrations de cordes. L'autre repose sur une discr\'etisation de l'espace-temps qui permet \'egalement d'\'eviter les divergences ultra-violettes.

\`A l'inverse, un autre probl\`eme se pose aux tr\`es grandes \'echelles (cosmologiques). Il a \'et\'e mesur\'e il y a un peu plus de dix ans que l'expansion de l'Univers acc\'el\'erait, au lieu de ralentir comme l'attractivit\'e de la gravitation newtonienne le laissait supposer. Cela signifie qu'aux grandes \'echelles la gravitation se comporte comme s'il existait une constante cosmologique $\Lambda>0$, dont l'effet ne dominerait pas auparavant. Il est bien s\^ur possible de se contenter d'ajouter une telle constante dans la th\'eorie et de ne plus s'en pr\'eoccuper, mais cette solution n'est gu\`ere satisfaisante : premi\`erement pour des raisons esth\'etiques, les physiciens rechignent \`a introduire de nouvelles constantes fondamentales, sauf si cela est rigoureusement n\'ecessaire et surtout si ces constantes sont extr\^emement faibles, comme dans le cas pr\'esent ($\sim10^{-29}g/cm^3$); deuxi\`emement, cette constante cosmologique devrait alors avoir un rapport avec l'\'energie des fluctuations quantiques dans le vide, mais tous les calculs pr\'edisent des dizaines d'ordres de grandeurs de diff\'erence; enfin, la th\'eorie des cordes, qui est la th\'eorie de gravitation quantique la plus aboutie aujourd'hui (quoiqu'encore largement imparfaite) peine \`a ``trouver'' de mani\`ere naturelle une constante cosmologique positive. Les solutions propos\'ees \`a ce probl\`eme sont multiples, mais font intervenir trois grandes cat\'egories de mod\`eles : ceux qui utilisent des champs de mati\`ere additionnels pour simuler cette acc\'el\'eration (mod\`eles de quintessence avec champs scalaires, entre autres); ceux qui modifient la gravitation aux grandes \'echelles (th\'eories $f(R)$) et enfin ceux qui introduisent des dimensions suppl\'ementaires (mondes branaires). Sans parler des coktails de plusieurs ou de tous ces ingr\'edients. 

Dans cette th\`ese, nous ne nous int\'eresserons pas directement \`a ces deux probl\`emes, mais nous examinerons plut\^ot les effets de ces ajouts ou de ces modifications de la gravitation d'Einstein sur les solutions de trou noir de la th\'eorie. Nous examinerons deux cat\'egories de mod\`eles : les th\'eories Einstein-Maxwell-Dilaton, qui consistent en la Relativit\'e G\'en\'erale coupl\'ee \`a des champs de mati\`ere scalaire et electro-magn\'etique; et les th\'eories Einstein-Gauss-Bonnet en six dimensions, o\`u la gravitation d'Einstein est intrins\`equement modifi\'ee.

\subsection{Solutions de trou noir des th\'eories Einstein-Maxwell-Dilaton}

Par la suite, nous examinerons ces th\'eories \`a quatre dimensions d'espace-temps. Toutefois, la structure de la gravitation n'y \'etant pas modifi\'ee, on peut s'attendre \`a ce que les propri\'et\'es qualitatives des solutions que nous exhiberons soient inchang\'ees avec des dimensions suppl\'ementaires\footnote{Ceci n'exclut \'evidemment en rien l'apparition de nouvelles solutions, comme c'est le cas en Relativit\'e G\'en\'erale avec les $p$-branes et les anneaux noirs.}.

\subsubsection{Les th\'eories Einstein-Maxwell-Dilaton}

L'action des th\'eories Einstein-Maxwell-Dilaton (EMD) s'\'ecrit de la mani\`ere suivante :
\be
	\label{EMDactionSV}
	S = \int \ud^dx\sqrt{-g}\l[R-\frac12(\partial \phi)^2-\frac14\e^{\ga\phi}F^2-2\Lambda\e^{-\delta\phi}\r],
\ee
tandis que les \'equations du mouvement de la th\'eorie, obtenues en variant la m\'etrique, le champ scalaire (dilaton) et le champ de Maxwell sont :
\bsea
	G_{\mu\nu}&=& \half\partial_\mu\phi\partial_\nu\phi-\frac{g_{\mu\nu}}4\l(\partial\phi\r)^2+\half \e^{\ga\phi}F^{\;\rho}_\mu F_{\nu\rho}-\frac{g_{\mu\nu}}8\e^{\ga\phi}F^2-\La\e^{-\da\phi}g_{\mu\nu}\,, \slabel{EinsteinEqEMDSV} \\
	\square{\phi}&=&\frac{\ga}4\e^{\ga\phi}F^2-2\da\La\e^{-\da\phi}\,, \slabel{DilatonEqEMDSV} \\
	0&=&\partial_\mu\l(\sqrt{-g}\e^{\ga\phi}F^{\mu\nu}\r). \slabel{MaxwellEqEMDSV}
	\label{EOMEMDSV}
\esea
On y voit appara\^itre la gravitation habituelle avec le scalaire de Ricci dans l'action ou encore le tenseur d'Einstein dans les \'equations du mouvement. \`A cela, on a rajout\'e un champ de Maxwell et un champ scalaire, minimalement coupl\'es \`a la gravit\'e mais non-minimalement coupl\'es entre eux. Enfin, le dilaton a un potentiel scalaire en exponentielle. Les th\'eories EMD peuvent \^etre classifi\'ees par deux param\`etres r\'eels $\gamma$ et $\delta$, qui sont en fait les pentes des exponentielles pr\'esentes dans le couplage de jauge et le potentiel scalaire.

Les motivations pour l'examen de ces th\'eories sont multiples: elles viennent aussi bien de l'\'etude de la gravit\'e en dimensions suppl\'ementaires, puisque l'on peut les obtenir en faisant une r\'eduction de Kaluza-Klein sur la Relativit\'e G\'en\'erale munie d'une constante cosmologique \`a cinq dimensions; que d'actions effectives de th\'eorie des cordes, valides \`a basse \'energie, dans lesquelles on a int\'egr\'e les modes massifs des cordes pour ne garder que les modes z\'ero et fait un d\'eveloppement au premier ordre dans la constante de couplage de la th\'eorie (\`a basse \'energie, on se trouve dans un r\'egime de couplage faible). Alternativement, ces th\'eories ont \'egalement \'et\'e consid\'er\'ees en cosmologie dans les premiers temps de l'\'etude des th\'eories de quintessence. Toutefois, au moins dans cette version ``simple'', elles ont d\^u \^etre mises de c\^ot\'e car en d\'esaccord avec des contraintes exp\'erimentales venant de la nucl\'eosynth\`ese.

D'un point de vue purement gravitationnel, on peut rattacher l'examen de ces th\'eories \`a celui de la Relativit\'e G\'en\'erale avec des champs de mati\`ere, et des c\'el\`ebres th\'eor\`emes de ``calvitie''. Cette formulation est due \`a Wheeler dans les ann\'ees 70. Il conjectura \`a cette \'epoque qu'un trou noir \'etait sp\'ecifi\'e de mani\`ere unique par un triplet de quantit\'es g\'eom\'etriques, sa masse, son moment angulaire et sa charge, et que tout autre nombre (quantique) caract\'erisant un champ existant lors de l'effondrement gravitationnel ne survirait pas \`a la formation du trou noir, mais tomberait tel un cheveu.

Ces conjectures ont \'et\'e reformul\'ees d'un point de vue plus moderne. En effet, il faut distinguer d\'esormais entre un cheveu primaire et un cheveu secondaire. Un cheveu primaire consiste en un champ induisant une constante d'int\'egration suppl\'ementaire et ind\'ependante des constantes usuelles (masse, moment angulaire, charge), qui caract\'eriserait le profil du cheveu. Par exemple, le champ électrique g\'en\`ere une constante d'int\'egration ind\'ependante, la charge. En ce sens, c'est un cheveu primaire. Au contraire, un cheveu secondaire consiste en un champ non-trivial dans la configuration de trou noir, mais qui ne g\'en\`ere pas de constante d'int\'egration ind\'ependante. C'est g\'en\'eralement le cas des champs scalaires.

Ainsi, la conjecture de calvitie prend la forme suivante : \'etant donn\'e certaines conditions asymptotiques (plates, Anti-de Sitter...) et un ensemble de charges conserv\'ees, calcul\'ees asymptotiquement par des int\'egrales de Gauss, peut-on avoir plusieurs solutions de trou noir? Alternativement, peut-on avoir des cheveux non-triviaux?

Bekenstein et d'autres ont tent\'e de r\'epondre \`a cette question d\`es les ann\'ees 70, dans le cas asymptotiquement plat. Tr\`es t\^ot, il fut \'etabli que pour un champ scalaire avec un potentiel convexe, le champ scalaire devait \^etre constant partout \`a l'ext\'erieur du trou noir (et m\^eme en pr\'esence d'un champ de Maxwell minimalement coupl\'e). Ce r\'esultat fut \'etendu vingt ans plus tard au cas d'un potentiel positif quelconcque.

Dans le cas de conditions asymptotiques (Anti-)de Sitter, il a \'et\'e prouv\'e au d\'ebut des ann\'ees 2000 que, lorsque le potentiel avait un minimum global, le champ scalaire tendait asymptotiquement vers le minimum effectif, c\`ad $V_{eff}(\phi_\infty)=V(\phi_\infty)-2\Lambda$, et le champ \'etait alors trivial et ne constituait pas un cheveu. Au contraire, s'il y a un maximum global n\'egatif, dans ce cas des cheveux non-triviaux sont autoris\'es pour le cas Anti-de Sitter car la constante cosmologique n\'egative peut g\'en\`erer un extremum positif en rendant positive la valeur du maximum global.

Dans le cas qui nous int\'eresse, celui des th\'eories EMD, le potentiel peut \^etre positif (n\'egatif) si $\Lambda>0$ ($\Lambda<0$). Dans le premier cas, on sait qu'il est impossible d'obtenir des cheveux non-triviaux dans le cas asymptotiquement plat. Toutefois, m\^eme dans le cas contraire, le fait que le potentiel admette un extremum global \`a l'infini a des cons\'equences dramatiques sur les conditions asymptotiques autoris\'ees.

En effet, une \'etude de Wiltshire montre que des asymptotes plates ne sont autoris\'ees que dans le cas sans potentiel ($\Lambda=0$), tandis que des asymptotes (Anti-)de Sitter ne le sont que pour un exposant $\delta=0$, c\`ad pour un potentiel plat. D'ailleurs, un trou noir charg\'e asymptotiquement plat avec champ scalaire non-trivial a \'et\'e d\'ecouvert d\`es la fin des ann\'ees 80 par Gibbons et Maeda. Cependant, pour $\gamma=0$, ce trou noir co\"incide avec le trou noir de Reissner-Nordstr\"om, et dans la limite de charge nulle se r\'eduit au trou noir de Schwarzschild. Cela ne contredit pas les pr\'ec\'edents th\'eor\`emes car le couplage non-minimal entre champs scalaire et \'electrique les invalident.

Par la suite, nous allons nous concentrer sur le cas des trous noirs \`a topologie planaire, et nous renvoyons vers le texte principal de ce manuscrit pour plus de d\'etails dans les autres cas. Nous allons maintenant d\'ecrire une mani\`ere tr\`es efficace de r\'esoudre les \'equations du mouvement.

\subsubsection{R\'esolution analytique dans le cas planaire}

On utilise un Ansatz quadri-dimensionnel avec une sym\'etrie cylindrique :
\be
\label{cylindrical_metricSV}
	\ud s^2 = \e^{2\chi}\al^{-\frac{1}{2}}(\ud \rho^2+ \ud \theta^2)+\al\l(-\e^{2 U}dt^2+\e^{-2 U}\ud \varphi^2\r),
\ee
o\`u toutes les fonctions $\al$, $\chi$ et $U$ ne d\'ependent que de la coordonn\'ee radiale $\rho$. La condition pour obtenir un horizon planaire homog\`ene sera d'imposer que
\be \e^{2\chi+2U}=\al^{-\frac{3}{2}}. \ee
\`A nouveau, nous renvoyons vers le texte principal pour le cas o\`u cette condition n'est pas impos\'ee. On prend un Ansatz \'electrique pour le champ de jauge :
\be A^\mu=A(\rho)\delta^\mu_0\ud x^0\,, \ee
car il existe une dualit\'e \'electromagn\'etique permettant d'obtenir une solution duale magn\'etique \`a partir de toute solution \'electrique, \`a condition de renverser le signe du param\`etre $\gamma$. Cela nous donne donc une solution d'une \emph{autre} th\'eorie $(-\gamma,\delta)$.

Un certain nombre d'\'equations du mouvement peuvent s'int\'egrer, mais pour pouvoir r\'eduire autant que possible le syst\`eme, on introduit une coordonn\'ee radiale 
\be p=\frac{\ud \alpha}{\ud \rho}\,,\ee
ce qui va nous permettre de r\'e\'ecrire le syst\`eme pour une seule fonction inconnue
\be B(p)=\int A\ud p\,,\ee
qui est simplement l'int\'egrale du champ de jauge dans la nouvelle coordonn\'ee radiale. On obtient alors une \'equation ma\^itresse
\be
	k+\frac{sq}4\dot B^2 -a\dot B + \l(1-\ga\da \r)(p\dot B -B) = \ddot B\l[X(p)-\frac{q}2\l(1-\ga\da \r)B\r].
	\label{radial_eqSV}
\ee
o\`u
\be 
	X(p)\doteq \frac{3-\da^2}{2}p^2 +\l[\l(1-\ga\da \r)\frac as - h\r]p +\frac{q}2 k-\frac{sh^2}{2(\gamma+\delta)^2} -\frac{a^2}{2s} \,,\quad s\doteq\gamma^2+1\,.
\ee
L'\'equation \eqref{radial_eqSV} est du second ordre, non autonome (elle d\'epend explicitement de l'inconnue $p$), non-lin\'eaire, et les points d\'enotent des d\'eriv\'ees par rapport \`a $p$. Une fois qu'une solution de cette \'equation est connue, il est possible d'obtenir tous les autres champs : m\'etrique, champ scalaire et de jauge. Le probl\`eme initial, tr\`es compliqu\'e et multi-dimensionnel, a \'et\'e r\'eduit \`a une dimension.

Des constantes d'int\'egration apparaissent dans \eqref{radial_eqSV} : $h$ est li\'ee \`a la masse de la solution, $q$ est la charge \'electrique obtenue par int\'egration directe de l'\'equation de Maxwell, et $k$ et $a$ sont des constantes d'int\'egration reli\'ees \`a la sym\'etrie de jauge du champ \'electrique. De m\^eme que le potentiel \'electrique peut \^etre translat\'e d'une constante, sa primitive peut \^etre translat\'ee d'un polyn\^ome du premier degr\'e, ce qui correspond \`a deux constantes : $a$ et $k$.

Pour r\'esoudre \eqref{radial_eqSV}, on remarque imm\'ediatement qu'un cas particulier risque fort d'\^etre int\'eressant : lorsque $\ga\da=1$, on peut effectivement int\'egrer compl\`etement l'\'equation \eqref{radial_eqSV}, ce qui donne \emph{la solution g\'en\'erale} de la th\'eorie dans ce cas. Passons en revue les solutions obtenues.

\paragraph{La solution pour $\ga\da=1$}est, comme annonc\'ee, la solution g\'en\'erale de la th\'eorie dans ce cas. Apr\`es un changement de coordonn\'ee radiale vers une coordonn\'ee plus habituelle $r$ et une red\'efinition des constantes d'int\'egration, on peut \'ecrire cette solution sous la forme suivante :
\bsea
	\ud s^2 &=& - \frac{V(r)\ud t^2 }{\l[1-\l(\frac{r_-}{r}\r)^{3-\da^2}\r]^{\frac{4(1-\da^2)}{(3-\da^2)(1+\da^2)}}}+ e^{\da\phi}\frac{\ud r^2}{V(r)}+ \nn\\
			&&\qquad\qquad\qquad+ r^2\l[1-\l(\frac{r_-}{r}\r)^{3-\da^2}\r]^{\frac{2(\da^2-1)^2}{(3-\da^2)(1+\da^2)}}\l(\ud x^2+\ud y^2\r), \slabel{Metric1SV} \\
	V(r) &=& \l(\frac r\ell\r)^2-2\frac{m\ell^{-\da^2}}{r^{1-\da^2}} +\frac{(1+\da^2)q^2\ell^{2-2\da^2}}{4\da^2(3-\da^2)^2r^{4-2\da^2}}\,, \slabel{Pot1SV} \\
	\l(r_{\pm}\r)^{3-\da^2} &=& \ell^{2-\da^2}\l[m\pm\sqrt{m^2-\frac{(1+\da^2)q^2}{4\da^2(3-\da^2)^2}}\r], \slabel{Horizon1SV}\\
	\e^{\phi}&=& \l(\frac r\ell\r)^{2\da}\l[1-\l(\frac{r_-}{r}\r)^{3-\da^2}\r]^{\frac{4\da(\da^2-1)}{(3-\da^2)(1+\da^2)}}\,, \slabel{Phi1SV}\\
	A &=&\frac{q\ell^{2-\da^2}}{(3-\da^2)} \l[1 -\l(\frac{r_+}{r}\r)^{3-\da^2}\r]\ud t\,, \slabel{A1SV}
	\label{Sol1SV}
\esea
o\`u les constantes d'int\'egration $m$ et $q$ sont reli\'ees \`a la masse et \`a la charge du trou noir. Nous avons d\'efini l'\'equivalent du rayon Anti-de Sitter de la th\'eorie
\be
	\ell^2=\frac{3-\da^2}{-\Lambda}\,,
\label{107SV}\ee
et il est \`a noter que des solutions de type de Sitter, mais non statique (type cosmologique), peuvent \^etre obtenues de la fa\c con habituelle en changeant $\ell\rightarrow ia$. Des solutions de type trou noir sont obtenues uniquement lorsque $\Lambda<0$ et $\da^2<3$. L'inspection du scalaire de Ricci r\'ev\`ele que, en plus de la divergence en $r=0$, il en existe une autre au niveau de ce qui serait habituellement l'horizon de Cauchy interne $r_-$, qui ici est singulier. On a donc un trou noir avec un horizon des \'ev\'enements $r_+$ qui cache une singularit\'e de taille finie en $r_-$, l'espace-temps ne s'\'etend donc pas jusqu'en $r=0$. De m\^eme, la limite extr\'emale $r_+=r_-$ est singuli\`ere.

\paragraph{La solution pour $\ga=\da$}peut \^etre obtenue par int\'egration de \eqref{radial_eqSV}, \`a condition de poser $h=0$. Malgr\'e cette restriction, elle comporte deux constantes d'int\'egration ind\'ependantes (masse $m$ et charge $q$) et s'\'ecrit dans le m\^eme syst\`eme de coordonn\'ees que \eqref{Sol1SV} comme :
\bsea
	\ud s^2 &=& - V(r)\ud t^2 + e^{\da\phi}\frac{\ud r^2}{V(r)} + r^2\l(\ud x^2+\ud y^2\r)\,, \slabel{Metric2SV} \\
	V(r) &=& \l(\frac r \ell\r)^{2}-2m\ell^{-\da^2}r^{\da^2-1} +\frac{q^2}{4(1+\da^2)r^2}\,, \slabel{Pot2SV} \\
	\e^{\phi}&=& \l(\frac r\ell\r)^{2\da} \slabel{Phi2SV}\,, \slabel{ElectricPotential2SV}\\
	A &=&\frac{\ell^{\da^2}q}{(1+\da^2)r_+^{1+\da^2}} \l[1 - \l(\frac{r_+}{r}\r)^{1+\da^2}\r]\ud t\,. \slabel{A2SV}
	\label{Sol2SV}
\esea
Les m\^emes remarques que pr\'ec\'edemment s'appliquent, en notant que cette fois l'horizon interne $r_-$ est bien r\'egulier et qu'il n'y a qu'une singularit\'e de courbure de taille nulle en $r=0$. La limite extr\'emale $r_+=r_-$ est donc r\'eguli\`ere.

Remarquons qu'asymptotiquement ces deux solutions tendent vers une m\'etrique de fond commune
\be
\ud s^2 = -r^2(\ud t^2+\ud x^2+\ud y^2) + r^{2\da^2-2}\ud r^2\,,\qquad \e^\phi=r^{2\da}\,.
\label{DilatonBackgroundSV}
\ee
Cette m\'etrique de fond partage avec Anti-de Sitter la particularit\'e d'avoir une sym\'etrie conforme sur son bord, mais brise la sym\'etrie d'Anti-de Sitter pour $\da\neq0$. De plus, elle poss\`ede une singularit\'e nue en $r=0$. Dans ce cas, le champ scalaire roule \`a l'infini vers le minimum global du potentiel scalaire exponentiel. En posant $\da=0$, nous retrouverions un espace localement Anti-de Sitter, mais il facile de voir dans les expressions \eqref{Sol1SV} et \eqref{Sol2SV} que ces deux solutions se r\'eduisent alors \`a Schwarzschild Anti-de Sitter et Reissner-Nordstr\"om Anti-de Sitter avec un horizon planaire et un champ scalaire constant. Ce ne sont donc pas des solutions avec un cheveu scalaire, m\^eme secondaire.

\paragraph{La solution pour $\ga,\da$ arbitraires}est obtenue au moyen d'un Ansatz polynomial du second degr\'e, et s'\'ecrit :
\bsea
	\ud s^2 &=& -V(p)p^{-4\frac{\ga(\ga-\da)}{wu}}\ud t^2 +\frac{\e^{\da\phi}\ud p^2}{-w\Lambda V(p)} + p^{2\frac{(\ga-\da)^2}{wu}}\l(\ud x^2+\ud y^2\r), \slabel{Metric3bisSV}\\
	\e^{\phi} &=& \e^{\phi_0}p^{-4\frac{(\ga-\da)}{wu}}\,, \slabel{Phi3bisSV}\\	
	A&=&2\sqrt{\frac{-v}{wu}}\e^{-\frac\ga2\phi_0} \l(p-2m\r)\ud t\,, \slabel{A3bisSV}\\
	V(r)&=& p(p-2m)\,, \slabel{Pot3bisSV} \\
	wu&=&3\ga^2-\da^2-2\ga\da+4\,, \qquad u=\ga^2-\ga\da+2\,, \qquad v=\da^2-\ga\da-2\,.\nn
	\label{Sol3SV}
\esea
Elle poss\`ede une singularit\'e en $p=0$, et un horizon en $p=2m$. La constante $\phi_0$ est simplement une \'echelle qu'il est utile de garder mais qui ne contribue pas \`a former de nouvel horizon. On peut \'etablir un lien entre cette solution et les deux pr\'ec\'edentes en en prenant la limite proche de l'horizon et presqu'extr\'emale dans \eqref{Sol1SV} et \eqref{Sol2SV} :
\be
	r_+\sim r_-\,,\qquad r=r_-+p
\ee
qui permet de retrouver la solution \eqref{Sol3SV} dans les cas $\ga\da=1$ et $\ga=\da$. 

En r\'esum\'e, nous avons \`a notre disposition des solutions d\'ecrivant la g\'eom\'etrie compl\`ete du trou noir dans les cas $\ga\da=1$ et $\ga=\da$, et une solution d\'ecrivant la g\'eom\'etrie proche de l'horizon dans la limite extr\'emale lorsque $\ga$ et $\da$ sont arbitraires.

\subsubsection{Thermodynamique des solutions de trou noir des th\'eories Einstein-Maxwell-Dilaton}

Dans les ann\'ees 70, il a \'et\'e \'etabli une correspondance entre les lois de la m\'ecanique des trous noirs et les lois de la thermodynamique.
\begin{itemize}
 \item  Loi z\'ero : \`a l'\'equilibre, il est possible d'associer une temp\'erature uniforme \`a l'horizon du trou noir. Elle est calcul\'ee \`a partir de la gravit\'e de surface \'evalu\'ee sur l'horizon, et dans un calcul semi-classique c\'el\`ebre, Hawking a d\'emontr\'e qu'un trou noir rayonnait avec un spectre de corps noir \`a cette m\^eme temp\'erature.
  \item Premi\`ere loi : il est possible de relier les variations de la masse, de l'entropie et de la charge \'electrique d'un trou noir par la formule
  \be
  	\ud M= T\ud S + \Phi\ud Q\,,
  \ee
  o\`u $M,T,S,\Phi,Q$ sont respectivement la masse, la temp\'erature et l'entropie de l'horizon, le potentiel chimique et la charge \'electriques du trou noir.
  \item Seconde loi : l'entropie associ\'ee \`a l'horizon d'un trou noir est proportionnelle \`a son aire et ne peut que cro\^itre (classiquement).
  \item Troisi\`eme loi : par un processus physique, on ne peut pas faire baisser arbitrairement la temp\'erature du trou noir en un temps fini.
\end{itemize}

On peut d\`es lors appliquer ces principes aux diverses solutions de trou noir dont nous disposons, calculer leur potentiel thermodynamique et autres quantit\'es int\'eressantes, d\'eterminer s'il existe des transitions de phase entre deux solutions en comp\'etition pour une temp\'erature donn\'ee, etc. Dans ce synopsis, nous allons nous limiter \`a exposer les r\'esultats obtenus pour la solution g\'en\'erale $\ga\da=1$ dans l'ensemble canonique (c\`ad \`a temp\'erature $T$ et charge \'electrique $Q$ fix\'es). Le potentiel thermodynamique correspondant est l'\'energie libre dite de Helmholtz
\be
 W(T,Q)=M-TS(T,Q)\,.
\ee

\begin{figure}[t]
\begin{center}
\begin{tabular}{cc}
	 \includegraphics[width=0.45\textwidth]{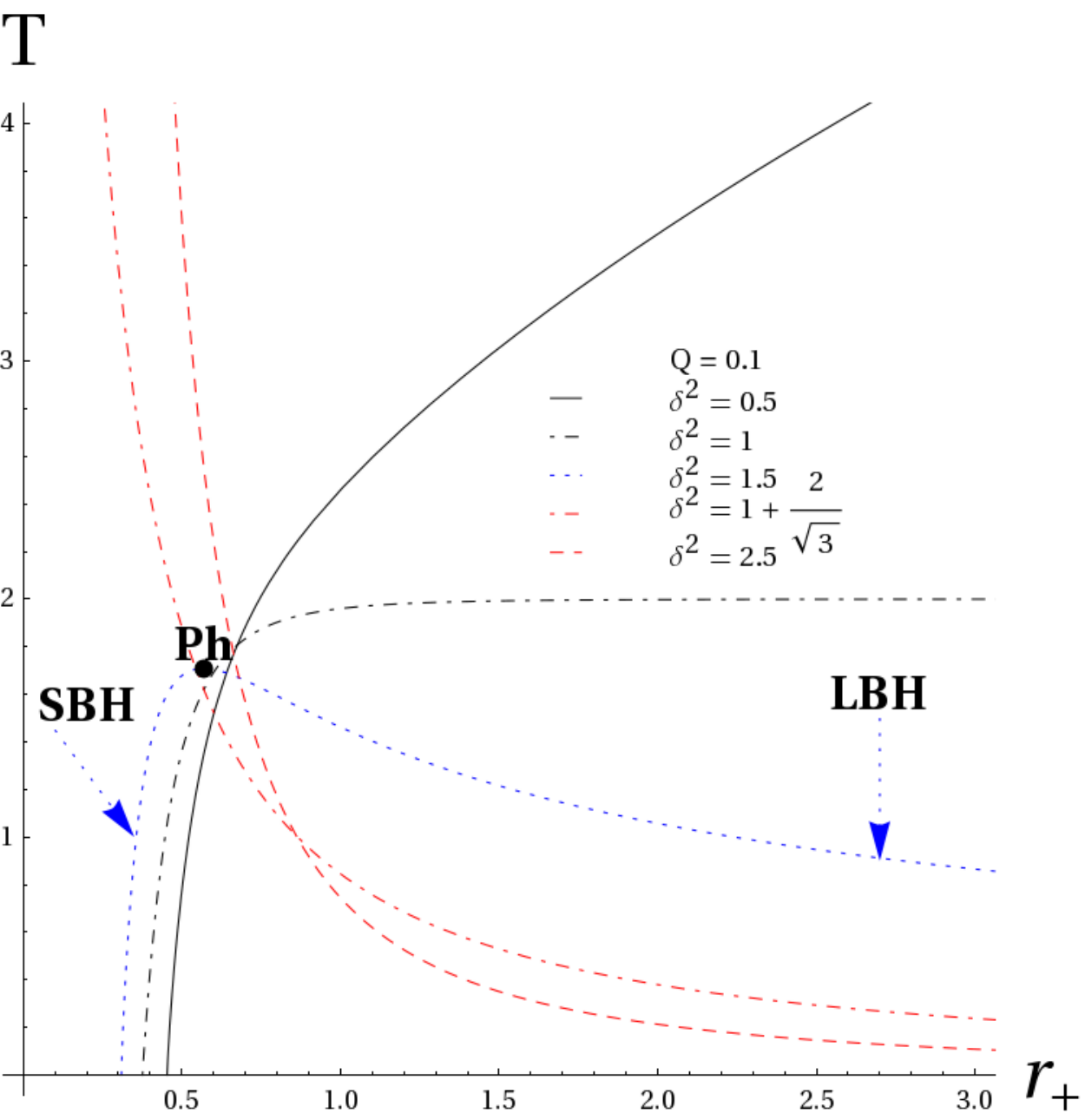}&	 
	 \includegraphics[height=0.3\textheight]{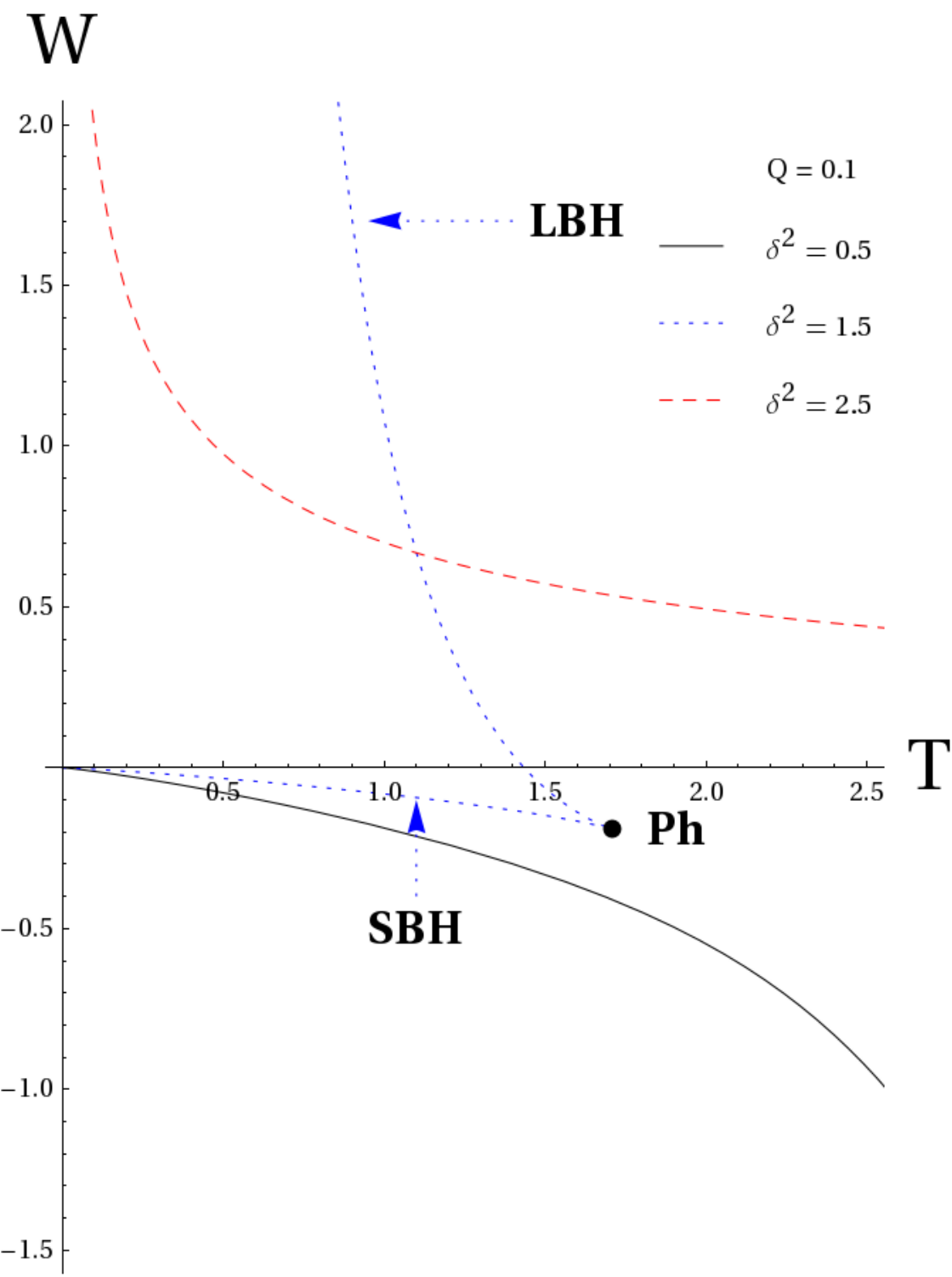}
 \end{tabular}
\caption{\'Equation d'\'etat $T(r_+)$ et \'energie libre $W(T)$ \`a charge fix\'ee pour la solution EMD \eqref{Sol1SV}.}
\label{Fig:ThermoGaDa1CanonicalSV}
\end{center}
\end{figure}

Une belle figure valant mieux qu'une \'equation compliqu\'ee, nous avons repr\'esent\'e dans la \Figref{Fig:ThermoGaDa1CanonicalSV} \`a la fois l'\'equation d'\'etat $T(r_+)$ et l'\'energie libre $W(T)$ \`a charge fix\'ee pour la solution EMD \eqref{Sol1SV}. On y constate que trois intervalles doivent \^etre distingu\'es. 
\begin{itemize}
 \item Lorsque $\da^2<1$, il existe une unique branche de trou noir, thermodynamiquement stable \`a la fois globalement (son \'energie libre est n\'egative par rapport \`a la m\'etrique de fond pour laquelle $W(T)=0\,\forall T$) et localement (l'\'energie libre est concave). Il n'y a donc aucune transition de phase \`a temp\'erature finie, et le trou noir EMD domine tout l'espace des phases.
  \item Lorsque $1<\da^2<1+2/\sqrt{3}$, il existe deux branches de trous noirs, des petits et des grands. Les grands sont instables localement, et aussi globalement par rapport aux petits trous noirs. Ces derniers sont stables thermodynamiquement dans tous les sens du terme, et dominent donc l'espace des phases. Ceci est vrai jusqu'\`a une temp\'erature maximale $T_{Ph}$ (voir \Figref{Fig:ThermoGaDa1CanonicalSV}), \`a laquelle les deux branches se rejoignent et au-del\`a de laquelle elles cessent d'exister. Il n'existe plus alors que la m\'etrique de fond, qui domine pour toutes les temp\'eratures sup\'erieures. L'\'energie libre \'etant discontinue \`a $T_{Ph}$, la transition est d'ordre z\'ero.
   \item Dans l'intervalle sup\'erieur $1+2/\sqrt{3}<\da^2<3$, une unique branche de trous noirs instable existe, donc l'espace des phases est domin\'e partout par la m\'etrique de fond.
\end{itemize}

\begin{figure}[t]
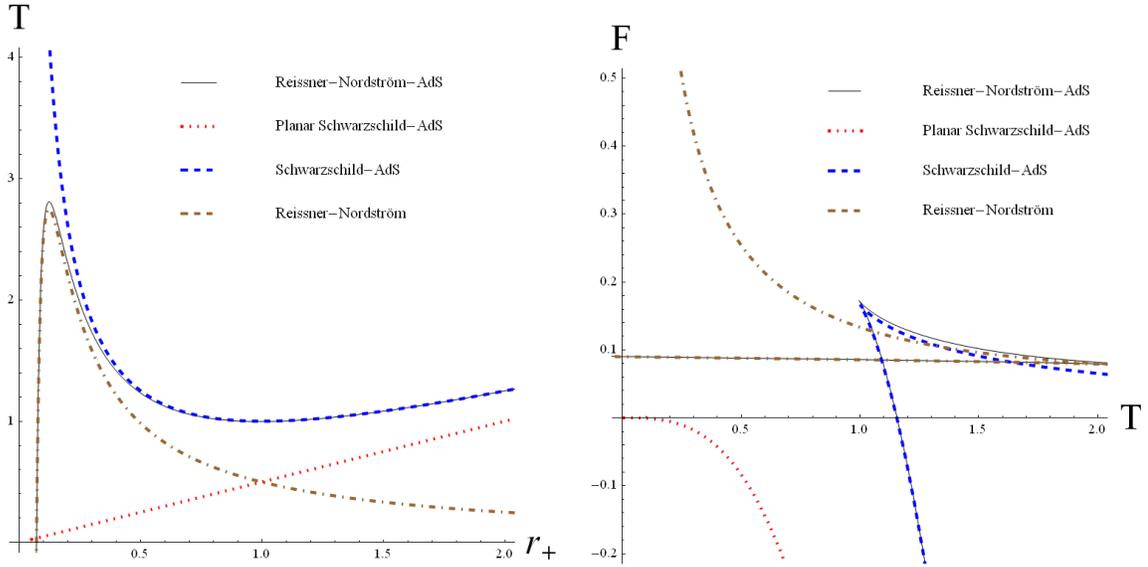

\begin{center}
\begin{tabular}{cc}
	 \includegraphics[width=0.45\textwidth]{Figs/ThirdPart/ReissnerNordstrom/TemperaturevsRadiusCanonicalRNAdS-150dpi}&	 
	 \includegraphics[width=0.45\textwidth]{Figs/ThirdPart/ReissnerNordstrom/FreeEnergyvsTemperatureCanonicalRNAdS-150dpi}
 \end{tabular}
\caption{\'Equation d'\'etat $T(r_+)$ et \'energie libre $W(T)$ \`a charge fix\'ee pour la solution EMD \eqref{Sol1SV}.}
\label{Fig:ThermoRNAdSCanonicalSV}
\end{center}
\end{figure}

Ces r\'esultats prennent tout leur int\'er\^et lorsqu'on les compare avec ceux obtenus pour les trous noirs Reissner-Nordstr\"om Anti-de Sitter, toujours dans l'ensemble canonique, cf. \Figref{Fig:ThermoRNAdSCanonicalSV}. Dans ce cas, on obtient trois branches. Les petits trous noirs (stables) sont g\'en\'er\'es par la charge \'electrique $Q\neq0$, et correspondent aux branches stables des trous noirs EMD dans les deux premiers intervalles. Les trous noirs interm\'ediaires (instables) sont g\'en\'er\'es par la courbure sph\'erique $\kappa=1$ de l'horizon et les branches instables des trous noirs EMD dans les deux derniers intervalles leurs correspondent. Les grands trous noirs (stables) sont g\'en\'er\'es par la constante cosmologique $\Lambda<0$ et sont absents dans les trous noirs EMD. Ainsi, on peut interpr\'eter l'effet du champ scalaire. Tout d'abord, en emp\^echant d'avoir un comportement asymptotique Anti-de Sitter, le champ scalaire annule l'effet de la constante cosmologique et les grands trous noirs ``Anti-de Sitter'' n'apparaissent jamais. Ainsi, la c\'el\`ebre transition de phase du premier ordre d'Hawking et Page entre les grands trous noirs et Anti-de Sitter est d\'etruite. On peut esp\'erer la recouvrer en r\'etablissant un comportement asymptotique r\'egulier, nous y reviendrons. Dans le premier intervalle, on retrouve un comportement typique des trous noirs Schwarzschild/Reissner-Nordstr\"om avec un horizon planaire dans Anti-de Sitter. En revanche, on constate que dans le second intervalle, le champ scalaire permet de simuler une courbure de l'horizon sph\'erique, tandis que dans le dernier intervalle, l'effet de la charge \'electrique est d\'etruit. On obtient donc une vari\'et\'e tr\`es int\'eressante de comportements.

Pour conclure, des r\'esultats similaires sont obtenus pour les trous noirs $\ga=\da$ (avec disparition du troisi\`eme intervalle), tandis que pour les trous noirs avec $\ga,\da$ arbitraires, on reproduit le comportements des petits trous noirs dans les deux premiers intervalles et de la branche instable dans le dernier.

\subsubsection{Applications holographiques des solutions de trou noir des th\'eories Einstein-Maxwell-Dilaton}

La correspondance AdS/CFT est une c\'el\`ebre conjecture \emph{holographique} formul\'ee par Maldacena en 1998, et \'etablit une dualit\'e entre le r\'egime fortement coupl\'e d'une th\'eorie de jauge (super Yang-Mills $\mathcal N=4$) et le r\'egime faible d'une th\'eorie gravitationnelle (la supergravit\'e de la th\'eorie des cordes de type $IIb$). Cette derni\`ere vit dans un espace-temps \`a dix dimensions, $AdS_5\times S^5$, qui est donc le produit d'Anti-de Sitter \`a cinq dimensions avec une sph\`ere \`a cinq dimensions \'egalement. Anti-de Sitter poss\`ede un bord quadri-dimensionnel, de topologie $\mathbf S^1\times\mathbf S^3$ qui a une sym\'etrie conforme, on peut donc y d\'efinir une th\'eorie des champs conforme \`a quatre dimensions, ce qui est le cas de super Yang-Mills $\mathcal N=4$. Cette conjecture est extr\^emement int\'eressante, car il est tr\`es difficile d'obtenir un contr\^ole analytique de la th\'eorie super Yang-Mills $\mathcal N=4$ fortement coupl\'ee, puisque dans ce r\'egime les techniques perturbatives de th\'eorie des champs ne sont pas d\'efinies.

Peu de temps apr\`es l'article fondateur de Maldacena, Witten a d\'emontr\'e qu'il y avait une correspondance exacte entre la transition de phase entre confinement et d\'econfinement dans la th\'eorie super Yang-Mills $\mathcal N=4$ sur le bord d'Anti-de Sitter et celle de Hawking-Page dans son volume, toutes deux du premier ordre. En effet, la correspondance stipule que le dual gravitationnel d'une th\'eorie surfacique \`a temp\'erature finie est un trou noir non-extr\'emal. Or, il est facile de constater sur la \Figref{Fig:ThermoRNAdSCanonicalSV} que cette transition n'a lieu que pour une courbure de l'horizon positive et non planaire. Dans l'hypoth\`ese o\`u nous souhaiterions appliquer ce principe holographique \`a une th\'eorie de jauge du mod\`ele standard comme la ChromoDynamique Quantique\footnote{Th\'eorie qui d\'ecrit les interactions fortes entre quarks, et pour laquelle on pr\'edit une transition vers une phase d\'econfin\'ee \`a haute temp\'erature, le plasma quark-gluon. Les scientifiques du Collisionneur d'Ions Lourds et Relativistes (RHIC) au Laboratoire National de Brookhaven, \'Etats-Unis, pr\'etendent l'avoir observ\'e \`a $4.10^{12}$ Kelvins. Trois exp\'eriences du Grand Collisionneur de Hadrons (LHC) au CERN, ALICE, ATLAS et CMS, sont \'egalement charg\'ees de son observation.}, il nous faudrait d\'efinir la th\'eorie surfacique sur un bord avec trois dimensions spatiales plates, ce qui correspondrait dans la th\'eorie volumique \`a un trou noir avec horizon planaire. La transition de Hawking-Page n'y survivrait pas. Il faut donc compliquer la th\'eorie volumique, en introduisant un champ scalaire, qui va permettre de simuler une nouvelle \'echelle de longueur et donc de former \`a nouveau une temp\'erature critique, tout en brisant la sym\'etrie conforme du bord. En dessous de cette temp\'erature critique o\`u a lieu la transition, l'op\'erateur scalaire sur le bord qui brise la sym\'etrie conforme condense, et la valeur attendue dans le vide de cet op\'erateur correspond \`a la valeur asymptotique du champ scalaire dans la th\'eorie volumique.

Examinons si nous pouvons trouver une application holographique \`a nos solutions EMD. Le probl\`eme le plus \'evident qui se pose \`a nous est qu'ils ne sont pas asymptotiquement Anti-de Sitter et que le champ scalaire ne prend pas une valeur constante \`a l'infini (au contraire, il diverge). Pour recouvrer un comportement asymptotique correct, nous allons donc limiter leur port\'ee \`a la region Infra-Rouge, loin du bord. En effet, on peut constater que dans le cas des g\'eom\'etries compl\`etes \eqref{Sol1SV} et \eqref{Sol2SV}, le potentiel scalaire exponentiel s'annule \`a l'infini et diverge au centre. Il est donc possible d'argumenter que le potentiel exponentiel constitue la partie dominante \`a l'Infra-Rouge ($r<<\infty$) du ``vrai'' potentiel scalaire, et que loin de la singularit\'e ($r>>0$), on peut rajouter une partie constante au potentiel :
\be
	\tilde V(\phi) = 2\Lambda_{UV} + 2\Lambda_{IR}\e^{-\da\phi} \underset{r<<\infty}{\sim} + 2\Lambda_{IR}\e^{-\da\phi},.
\ee
Cette derni\`ere va donc dominer la dynamique de la th\'eorie \`a l'Infra-Rouge, et autoriser des asymptotes Anti-de Sitter, ce qui aura pour effet de r\'etablir la transition de phase de Hawking-Page.

Les espaces-temps sur lesquels sont construits ces trous noirs sont, nous l'avons vu, g\'en\'eriquement singuliers. Les singularit\'es nues sont d'habitude un crit\`ere discriminatoire pour les th\'eories de gravitation comme la Relativit\'e G\'en\'erale, mais une th\'eorie holographique peut s'en accommoder, \`a certaines conditions. Une premi\`ere condition est qu'une connaissance de la physique \`a l'Infra-Rouge (c\`ad aux basses temp\'eratures) ne soit pas n\'ecessaire \`a la physique ultra-violette. On ne s'attend effectivement pas \`a ce que la physique des hautes \'energies soit conditionn\'ee par celle des basses \'energies, mais bien au contraire qu'elle en soit s\'epar\'ee par des transitions de phase brisant certaines sym\'etries. Traduit pour l'espace-temps gravitationnel, cela implique que la singularit\'e nue est acceptable si elle peut \^etre retrouv\'ee comme la limite \`a temp\'erature nulle d'un trou noir. Gubser a conjectur\'e que cela \'etait \'equivalent \`a ce que le potentiel scalaire soit born\'e par au-dessus, ce qui est le cas pour toutes nos solutions de trou noir EMD.

Un second crit\`ere, formul\'e par Kiritsis et al., est que les donn\'ees physiques doivent venir de la th\'eorie sur le bord, c\`ad des conditions asymptotiques. Il faut donc \'etudier les fluctuations de la m\'etrique et du champ de jauge, et imposer que sur les deux solutions ind\'ependantes que l'on obtient, une seule soit normalisable et n\'ecessite donc une condition initiale dans l'UV. Sinon, une autre condition initiale doit \^etre impos\'ee, mais cette fois dans l'IR, ce qui contrevient aux principes pr\'ec\'edents. En examinant nos solutions, nous trouvons que les fluctuations sont ad\'equates notamment pour la r\'egion o\`u elles sont thermodynamiquement stables, qui a une intersection non-nulle avec le crit\`ere de Gubser.

\paragraph{Applications holographiques aux syst\`emes de Mati\`ere Condens\'ee : }Une tendance r\'ecente de la communaut\'e holographique est l'application de ces id\'ees aux syst\`emes de Mati\`ere Condens\'ee en couplage fort, comme les supraconducteurs \`a haute temp\'erature critique ou encore les m\'etaux \'etranges. Le couplage fort se manifeste par la formation d'un condensat, qui l\`a aussi peut \^etre bien mod\'elis\'e par la valeur attendue dans le vide d'un op\'erateur scalaire de la th\'eorie surfacique, correspondant \`a la valeur asymptotique d'un champ scalaire dans la th\'eorie volumique. Pour les supraconducteurs, il semble qu'il faille examiner un champ scalaire complexe, aussi nous ne nous y int\'eresserons pas, en revanche les th\'eories EMD peuvent servir \`a tenter de mod\'eliser le comportement \`a basse temp\'erature des m\'etaux \'etranges. Leurs caract\'eristiques principales sont que dans ce r\'egime de temp\'eratures, la conductivit\'e en courant alternatif se comporte comme une puissance n\'egative de la fr\'equence, tandis que la r\'esistivit\'e (l'inverse de la conductivit\'e en courant continu) est lin\'eaire avec la temp\'erature. En calculant les coefficients de transport de la th\'eorie duale pour l'op\'erateur dual du champ de jauge volumique, nous trouvons que la conductivit\'e AC est bien proportionnelle \`a une puissance positive (et non n\'egative) de la fr\'equence, et que la conductivit\'e DC est bien lin\'eaire en la temp\'erature sur un certain intervalle.

\subsection{Trous noirs dans les th\'eories Einstein-Gauss-Bonnet}

\subsubsection{De la Relativit\'e G\'en\'erale aux th\'eories Gauss-Bonnet}

\`A quatre dimensions d'espace-temps, la Relativit\'e G\'en\'erale est l'unique th\'eorie v\'erifiant les principes suivants :
\begin{itemize}
 \item elle a une structure tensorielle, est invariante par diff\'eomorphime et la m\'etrique est un tenseur sym\'etrique de rang deux; 
 \item les \'equations du mouvement sont d'ordre deux dans la m\'etrique; 
 \item elles ob\'eissent aux identit\'es de Bianchi (pour un syst\`eme isol\'e, l'\'energie est conserv\'ee). 
\end{itemize}
Cette propri\'et\'e d'unicit\'e cesse d'\^etre vraie pour une th\'eorie d\'efinie dans plus de dimensions. De plus, en Relativit\'e G\'en\'erale, le th\'eor\`eme de Birkhoff-Jensen stipule que l'unique solution \`a sym\'etrie sph\'erique est statique et qu'il s'agit de la solution de Schwarzschild avec un horizon sph\'erique. Lorsque l'on rajoute \`a la th\'eorie une constante cosmologique (ce que rien n'interdit), l'unicit\'e de la topologie est perdue, des horizons planaires ou hyperboliques sont possibles. Lorsque l'on passe \`a cinq dimensions, apparaissent de nouveaux objets noirs statiques, comme la corde noire ou encore les $p$-branes noires\footnote{Ces solutions sont form\'ees d'un produit direct entre une ou $p$ directions plates et la m\'etrique de Schwarzschild \`a quatre dimensions}. Les anneaux noirs sont un exemple de nouvel objet noir stationnaire n'existant pas \`a quatre dimensions. Enfin, m\^eme pour la g\'en\'eralisation \`a plus grande dimension du trou noir de Schwarzschild (dite de Tangherlini), on constate que l'horizon n'est plus restreint \`a \^etre un espace \`a courbure constante, mais plus largement un espace d'Einstein.

Rappelons ici ces notions :
\begin{itemize}
 \item pour un espace \`a courbure constante, le tenseur de Riemann de l'horizon est proportionnel \`a la m\'etrique et son scalaire de Ricci est constant; s'il est nul (respectivement positif, n\'egatif), la topologie est planaire (respectivement sph\'erique, hyperbolique).
 \item pour un espace d'Einstein, le tenseur de Ricci seul est proportionnel \`a la m\'etrique sur l'horizon; le tenseur de Riemann conserve une partie de trace nulle, que l'on appelle le tenseur de Weyl\footnote{Il poss\`ede les m\^emes propri\'et\'es d'antisym\'etrie que le tenseur de Riemann et est invariant sous les transformations conformes de la m\'etrique. Tout espace conform\'ement plat a un tenseur de Weyl nul. R\'eciproquement, pour un espace de dimension plus grande que trois, un espace dont le tenseur de Weyl est nul est conform\'ement plat (c'est le cas pour Anti-de Sitter, par exemple).};
\end{itemize}

Ainsi, non contents d'avoir perdu l'unicit\'e des solutions de trou noir, nous subissons \'egalement une \'enorme d\'eg\'enerescence de la topologie admissible pour l'horizon.

Ces manques peuvent \^etre reli\'es au fait d'avoir augment\'e le nombre de dimensions sans modifier la th\'eorie de gravitation utilis\'ee. Ainsi,  Lovelock a d\'emontr\'e en 1971 que si l'on rajoutait dans l'action de dimension $D$ la densit\'e d'Euler correspondant \`a la dimension $d=[(D-1)/2]$, l'unicit\'e de la th\'eorie de gravitation v\'erifiant les propri\'et\'es pr\'ec\'edemment cit\'ees \'etait recouvr\'ee. En dimension $D=4$, la densit\'e d'Euler pour un espace de dimension $d=4-2=2$ est justement le scalaire de Ricci, celle de dimension $0$ une constante (cosmologique). En revanche, en dimension cinq ou six, la densit\'e d'Euler de dimension $d=2$ est le terme dit de Gauss-Bonnet, qui est la combinaison suivante de termes quadratiques en la courbure :
\be
	\hat G = R_{\la\mu\nu\rho}R^{\la\mu\nu\rho}-4R_{\mu\nu}R^{\mu\nu}+R^2\,. \label{GhatSV}
\ee
L'action totale \`a six dimensions devient alors
\be
	S_{EGB} = \half\int \ud^6x \sqrt{-g}\l(-2\La+R+\al\hat G\r)\,,
	\label{EGBactionSV}
\ee
o\`u $\al$ est la constante de couplage du terme de Gauss-Bonnet. Les \'equations du mouvement de cette th\'eorie sont
\be
	\E_{\mu\nu} = G_{\mu\nu} + \La g_{\mu\nu} -\al H_{\mu\nu} = T_{\mu\nu}\,,
	\label{EOMGBSV}
\ee
o\`u $H_{\mu\nu}$ est le tenseur de Lanczos :
\be
	H_{\mu\nu}=\half\hat Gg_{\mu\nu}-2RR_{\mu\nu}+4R_{\mu\rho}R^{\rho}_{\phantom{1}\nu}+4R_{\rho\sigma}R^{\rho\phantom{1}\sigma}_{\phantom{1}\mu\phantom{1}\nu}-2R_{\mu}^{\phantom{1}\rho\sigma\tau}R_{\nu\rho\sigma\tau}\,.
	\label{LanczosTensorSV}
\ee
Ces \'equations sont d'ordre deux dans la m\'etrique, et donc ne contiennent pas de degr\'e de libert\'e d'\'energie n\'egative (fant\^omes).

Les vides de la th\'eorie (espace-temps maximalement sym\'etriques) \`a six dimensions s'\'ecrivent 
\bsea
	\ud s^2&=&-V(r)\ud t^2 + \frac{\ud r^2}{V(r)} + r^2\ud\Omega^2_{4}\,,\\
	V_{\pm}(r)&=& 1-\frac{\La_e^{\pm}r^2}{10}\,,
\esea
avec une constante cosmologique effective $\La_e$
\bsea
	\La_{e}^{\pm} &=& 2\La_{CS}\l[1\mp\sqrt{1-\frac{\La}{\La_{CS}}}\r],\\ 
\Leftrightarrow	\La &=& \La_e\l(1-\frac{\La_e}{4\La_{CS}}\r)\,,\\
	\La_{CS}&=&-\frac{5}{12\al}\,.
	\label{GBEffectiveCCSV}
\esea	
Il y a donc deux branches, que l'on peut distinguer en prenant la limite o\`u le couplage de Gauss-Bonnet $\al$ tend vers z\'ero :
\bsea
	V_+(r)&\underset{\al\to0}{\sim}&1-\frac{\La r^2}{10} + O(\al)\,,\slabel{PlusVacSmallAlphaSV}\\
	V_-(r)&\underset{\al\to0}{\sim}&1+\frac{r^2}{6\al}\l[1+\frac{3\al\La}{5}\r] + O(\al)\,.\slabel{MinusVacSmallAlphaSV}
	\label{GBVacSmallAlphaSV}
\esea
La branche $(+)$ donne donc une limite o\`u l'on retrouve (Anti-)de Sitter comme attendu, nous l'appellerons la branche Einstein. La branche $(-)$ n'a au contraire pas de limite coh\'erente lorsque $\al$ tend vers z\'ero, elle est intrins\`eque \`a ces th\'eories et nous l'appellerons la branche Gauss-Bonnet. De plus, il est possible de montrer que cette derni\`ere est instable.

\subsection{Les trous noirs des th\'eories de Gauss-Bonnet}

Afin d'int\'egrer au mieux les \'equations du mouvement, nous allons choisir un Ansatz qui s'inspire de celui utilis\'e pour prouver le th\'eor\`eme de Birkhoff-Jensen en Relativit\'e G\'en\'erale :
\be
\label{GBmetricSV}
	\ud s^2  = e^{2\nu \left( {t,z} \right)} B\left( {t,z} \right)^{ - 3/4} \left( { - \ud t^2  + \ud z^2 } \right) + B\left({t,z}\right)^{1/2} h^{\left( 4 \right)} _{\mu \nu } \left( x \right)\ud x^\mu  \ud x^\nu  \,.
\ee
Il comporte une partie transverse bi-dimensionnelle, d\'ependant du temps et d'une coordonn\'ee radiale, ainsi qu'un produit d\'eform\'e par le facteur $B$ avec une m\'etrique interne quadri-dimensionnelle $h^{\left( 4 \right)} _{\mu \nu }(x^\mu)$, qui tiendra lieu d'horizon au final. En utilisant des coordonn\'ees du c\^one de lumi\`ere, 
\be
	u = \frac{{t - z}}{{\sqrt 2 }}\,,\qquad v = \frac{{t + z}}{{\sqrt 2 }}\,,
\ee
les \'equations du mouvement se factorisent :
\be
	\label{GBequuSV}
	{\cal E}_{uu}=\frac{2 \nu_{,u} B_{,u}- B_{,uu}}{B} \left[ 1+\alpha \left( B^{-1/2} R^{(4)}+\frac{3}{2} e^{-2\nu} B^{-5/4} B_{,u} B_{,v}  \right) \right]\,,
\ee
\be
	\label{GBeqvvSV}
	{\cal E}_{vv}=\frac{2 \nu_{,v} B_{,v}- B_{,vv}}{B} \left[ 1+\alpha \left( B^{-1/2} R^{(4)}+\frac{3}{2} e^{-2\nu} B^{-5/4} B_{,u} B_{,v}  \right) \right].
\ee
Nous n'avons reproduit que les deux premi\`eres par souci de concision. Le premier facteur est identique \`a celui en Relativit\'e G\'en\'erale et impose que les solutions consid\'er\'ees sont statiques : c'est celui-ci qui engendre le th\'eor\`eme de Birkhoff-Jensen. Le second est sp\'ecifique aux th\'eories Gauss-Bonnet, et impose notamment que $\Lambda=\Lambda_{CS}$. Nous ne nous pr\'eoccuperons pas de cette classe de solutions, non-statiques, dans le synopsis\footnote{Le lecteur int\'eress\'e peut se r\'ef\'erer au texte principal.}.

Lorsque le premier facteur s'annule, les autres \'equations du mouvement fixent la forme du potentiel du trou noir :
\be
	V(r) = \frac{{R^{\left( 4 \right)} }}{{12}} + \frac{{r^2 }}{{12\alpha }}\left[ {1 \pm \sqrt {1 +\frac{12\alpha \Lambda}{5}  +\frac{{\alpha ^2 \left({R^{(4)}}^2-6\hat G^{(4)} \right)}}{{r^4 }} + 24\frac{{\alpha M}}{{r^5 }}} } \right]\,,
	\label{potentialclassIISV}
\ee
ainsi que (pour une sous-classe) la m\'etrique de l'horizon soit un espace d'Einstein\footnote{L'autre sous-classe requiert une nouvelle fois d'imposer $\Lambda=\Lambda_{CS}$}. Les propri\'et\'es du tenseur de Riemann r\'ev\`elent que le tenseur de Weyl doit \'egalement \^etre proportionnel \`a la m\'etrique :
\be
	C^{\alpha \beta \gamma \mu} C_{\alpha \beta \gamma \nu}=\Theta \delta^\mu_\nu\,,
	\label{ConditionSV}
\ee
ce qui, report\'e dans le potentiel du trou noir, donne :
\be
	V(r)=\kappa+\frac{r^2}{12 \alpha} \left(1\pm \sqrt{1+\frac{12}5\alpha\Lambda - 24 \frac{\alpha^2 \Theta}{r^4} + 24 \frac{\alpha M}{r^5}} \right)\,,
	\label{potentialclassIIEinsteinSV}
\ee
o\`u $M$ est une constante d'int\'egration reli\'ee \`a la masse du trou noir. Ce r\'esultat est remarquable pour plusieurs raisons :
\begin{itemize}
 \item Le nombre de g\'eom\'etries admissibles pour l'horizon sont grandement diminu\'ees, car peu d'espaces d'Einstein v\'erifient la condition \eqref{ConditionSV}. Par exemple, on peut avoir \'evidemment les espaces \`a courbure constante, mais aussi le produit direct de deux $\mathbf S^2$. Une grande partie de la d\'eg\'en\'erescence sur la g\'eom\'etrie de l'horizon est donc lev\'ee.
  \item La topologie de l'horizon intervient directement dans le potentiel du trou noir. \`A titre de comparaison, pour le trou noir de Schwarzschild, on peut simplement distinguer le signe de la courbure de l'horizon. Dans le cas de \eqref{potentialclassIIEinsteinSV}, le carr\'e du tenseur de Weyl intervient \'egalement.
  \item Sous certaines conditions, le terme en $\Theta$ permet de g\'en\'erer un horizon des \'ev\'enements, m\^eme en l'absence d'un terme de masse.
\end{itemize}

On peut donc formuler le th\'eor\`eme de staticit\'e suivant : \emph{Si les param\`etres de la th\'eorie sont libres ($\La\neq\La_{CS}$), les solutions \`a sym\'etrie sph\'erique des th\'eories Gauss-Bonnet \`a six dimensions sont statiques localement et d\'ecrivent des trous noirs avec le potentiel \eqref{potentialclassIIEinsteinSV} et un horizon qui est un espace d'Einstein v\'erifiant \eqref{ConditionSV}}.

\subsection{Perspectives}

Passons en revue rapidement quelques perspectives pour un travail futur :
\begin{itemize}
 \item L'\'etude des trous noirs EMD pourrait \^etre poursuivie, notamment pour essayer de d\'eterminer la solution compl\`ete pour $\ga$ et $\da$ arbitraires. Il est n\'eanmoins fort possible que cette solution n'existe pas sous forme analytique.
 \item Modifier le potentiel en lui rajoutant une partie constante, qui dominerait asymptotiquement, permettrait peut-\^etre de retrouver des solutions asymptotiquement Anti-de Sitter.
 \item D'un point de vue holographique, l'inclusion d'un champ de Maxwell dans la th\'eorie volumique implique la conservation du nombre de particules dans la th\'eorie surfacique; \'etudier plus avant les solutions EMD aurait certainement un int\'er\^et du point de vue de la th\'eorie de jauge duale.
 \item L'\'etape suivante en complexit\'e est le rajout d'un couplage dilaton-terme de Gauss-Bonnet, ce qui permettrait de g\'en\'erer de nouvelles \'echelles.
 \item En ce qui concerne les applications au syst\`emes de Mati\`ere Condens\'ee, trouver des solutions dyoniques (avec \`a la fois un champ \'electrique et un champ magn\'etique) permettrait peut-\^etre de mod\'eliser des effets comme l'effet Nernst.
 \item La thermodynamique des solutions Gauss-Bonnet \`a six dimensions pourrait bien \^etre modifi\'ee par le terme en carr\'e du tenseur de Weyl de l'horizon.
 \item Cette lev\'ee de la d\'eg\'en\'erescence de la g\'eom\'etrie de l'horizon persiste-t-elle en th\'eorie de Lovelock, pour un nombre de dimensions quelconque?
\end{itemize}
\selectlanguage{english}


\thispagestyle{empty}
\strut\newpage

\addcontentsline{toc}{part}{References}

\bibliography{these}

\providecommand{\href}[2]{#2}\begingroup\raggedright\begin{thebibliography}{10%
0}

\bibitem{deruelle2006}
N.~Deruelle, ``{\em General Relativity: A Primer}.''
\newblock General Relativity Trimester, Institut Henri Poincar\'e, Paris, 2006.
  \url{http://luth2.obspm.fr/IHP06/}.

\bibitem{deruelle2009}
N.~Deruelle, ``{\em Black Holes in General Relativity}.''
\newblock Institut de Physique Th\'eorique, CEA, Saclay, 2009.
  \url{http://ipht.cea.fr/Phocea-SPhT/ast_visu_spht.php?id_ast=188}.

\bibitem{Eisenstaedt:2007}
J.~Eisenstaedt, {\em {Einstein et la Relativit\'e G\'en\'erale}}.
\newblock CNRS \'Editions, 2007.
\newblock In French.

\bibitem{Inverno:1992}
R.~d'Inverno, {\em {Introducing Einstein's Relativity}}.
\newblock Oxford University Press, 1992.

\bibitem{Hawking:1973uf}
S.~W. Hawking and G.~F.~R. Ellis, {\em {The Large Scale Structure of
  Space-time}}.
\newblock Cambridge University Press, Cambridge, 1973.

\bibitem{Charmousis:2009}
C.~Charmousis, ``{\em Anti de Sitter black holes}.''
\newblock Lecture notes from the fifth Aegean Summer School, \emph{From Gravity
  to Thermal Gauge Theories: The AdS/CFT Correspodences},
  \url{http://www.physics.ntua.gr/cosmo09/Milos2009/}.

\bibitem{Einstein:1915by}
A.~Einstein, {\it {On the General Theory of Relativity}},  Sitzungsber. Preuss.
  Akad. Wiss. Berlin (Math. Phys. ) {\bf 1915} (1915) 778--786.

\bibitem{Einstein:1915ca}
A.~Einstein, {\it {The Field Equations of Gravitation}},  Sitzungsber. Preuss.
  Akad. Wiss. Berlin (Math. Phys. ) {\bf 1915} (1915) 844--847.

\bibitem{Einstein:1916vd}
A.~Einstein, {\it {The Foundation of the General Theory of Relativity}},
  Annalen Phys. {\bf 49} (1916) 769--822.

\bibitem{Schwarzschild:1916uq}
K.~Schwarzschild, {\it {On the Gravitational Field of a Mass Point according to
  Einstein's Theory}},  Sitzungsber. Preuss. Akad. Wiss. Berlin (Math. Phys. )
  {\bf 1916} (1916) 189--196,
  [\href{http://xxx.lanl.gov/abs/physics/9905030}{{\tt physics/9905030}}].

\bibitem{Schwarzschild:1916ae}
K.~Schwarzschild, {\it {On the Gravitational Field of a Sphere of
  Incompressible Fluid according to Einstein's Theory}},  Sitzungsber. Preuss.
  Akad. Wiss. Berlin (Math. Phys. ) {\bf 1916} (1916) 424--434,
  [\href{http://xxx.lanl.gov/abs/physics/9912033}{{\tt physics/9912033}}].

\bibitem{Droste:1916}
J.~Droste, {\em {The Field of a Single Centre in Einstein's Theory of
  Gravitation, and the Motion of a Particle in that Field}}.
\newblock PhD thesis, Leiden University, 1918.
\newblock In Dutch.

\bibitem{Eddington:1987tk}
A.~S. Eddington, {\it {Space, Time and Gravitation. An Outline of the General
  Relativity Theory}}, . Cambridge, Uk: Univ. Pr. (1920; reissued, 1987) 218p.

\bibitem{Hilbert:1917}
D.~Hilbert, {\it {Die Grundlagen der Physik. 2. Mitteilungen}},  Gott. Nachr.
  (1917) 53--76.

\bibitem{Painleve:1921}
P.~Painlev\'e, {\it {La M\'ecanique Classique et la Th\'eorie de la
  Relativit\'e}},  C. R. Acad. Sci. (Paris) {\bf 173} (1921) 677--680.

\bibitem{Gullstrand:1922}
A.~Gullstrand, {\it {Allgemeine L\"osung des statischen Eink\"orperproblems in
  der Einsteinschen Gravitationstheorie}},  Arkiv. Mat. Astron. Fys. {\bf
  16(8)} (1922) 1--15.

\bibitem{Eddington:1924}
A.~S. Eddington, {\it {A Comparison of Whitehead's and Einstein's Formul\ae}},
  Nature {\bf 2832} (1924) 192.

\bibitem{Flamm:1917}
L.~Flamm, {\it {Beitr\"age zur Einsteinschen Gravitationstheorie}},  Physik Z.
  {\bf 17} (1916) 448--454.

\bibitem{Weyl:1919a}
H.~Weyl, {\em {Raum. Zeit. Materie. Vorlesungen \"{u}ber allgemeine
  Relativit\"{a}tstheorie}}.
\newblock J. Springer, 1919.

\bibitem{Einstein:1917ce}
A.~Einstein, {\it {Cosmological Considerations in the General Theory of
  Relativity}},  Sitzungsber. Preuss. Akad. Wiss. Berlin (Math. Phys. ) {\bf
  1917} (1917) 142--152.

\bibitem{deSitter:1917a}
W.~de~Sitter, {\it {On the Relativity of Inertia: Remarks Concerning Einstein's
  Latest Hypothesis}},  Proc. Kon. Ned. Akad. Wet. {\bf 19} (1917) 1217--1225.

\bibitem{deSitter:1917b}
W.~de~Sitter, {\it {The Curvature of Space}},  Proc. Kon. Ned. Akad. Wet. {\bf
  20} (1917) 229--243.

\bibitem{deSitter:1917c}
W.~de~Sitter, {\it On einstein's theory of gravitation and its astronomical
  consequences. iii},  Mon. Not. Roy. Astron. Soc. {\bf 78} (1917) 3.

\bibitem{deSitter:1918}
W.~de~Sitter, {\it Further remarks on the solutions of the field equations of
  einstein's theory of gravitation},  Proc. Kon. Ned. Akad. Wet. {\bf 20}
  (1918) 1309.

\bibitem{Birkhoff:1923}
G.~D. Birkhoff, {\em {Relativity and Modern Physics}}.
\newblock Harvard University Press, Cambridge, MA, 1923.

\bibitem{Jebsen:1921}
J.~T. Jebsen, {\it {}},  Ark. Mat. Ast. Fys. {\bf 15} (1921).

\bibitem{Lemaitre:1933}
G.~Lema\^itre, {\it {L'Univers en Expansion}},  Annales de la Soci\'et\'e des
  Sciences de Bruxelles {\bf 53A} (1933) 51--83. In French.

\bibitem{Chandrasekhar:1930}
S.~Chandrasekhar, {\it {The Maximum Mass of Ideal White Dwarfs}},  Astrophys.
  J. {\bf 74} (1931) 81.

\bibitem{Chadwick:1932ma}
J.~Chadwick, {\it {Possible Existence of a Neutron}},  Nature {\bf 129} (1932)
  312.

\bibitem{Oppenheimer:1939}
J.~R. Oppenheimer and G.~Volkov, {\it {On Massive Neutron Cores}},  Phys. Rev.
  {\bf 55} (1939) 374--381.

\bibitem{Oppenheimer:1939ue}
J.~R. Oppenheimer and H.~Snyder, {\it {On Continued Gravitational
  Contraction}},  Phys. Rev. {\bf 56} (1939) 455--459.

\bibitem{Finkelstein:1958zz}
D.~Finkelstein, {\it {Past-Future Asymmetry of the Gravitational Field of a
  Point Particle}},  Phys. Rev. {\bf 110} (1958) 965--967.

\bibitem{Synge:1950gk}
J.~L. Synge, {\it {The Gravitational Field of a Particle}},  Proc. Roy. Irish
  Acad. (Sect. A) {\bf 53} (1950) 83.

\bibitem{Kruskal:1959vx}
M.~D. Kruskal, {\it {Maximal Extension of Schwarzschild Metric}},  Phys. Rev.
  {\bf 119} (1960) 1743--1745.

\bibitem{Szekeres:1960gm}
G.~Szekeres, {\it {On the Singularities of a Riemannian Manifold}},  Publ.
  Math. Debrecen {\bf 7} (1960) 285--301.

\bibitem{Penrose:1962ij}
R.~Penrose, {\it {Asymptotic Properties of Fields and Space-times}},  Phys.
  Rev. Lett. {\bf 10} (1963) 66--68.

\bibitem{Penrose:1964ge}
R.~Penrose, {\it {Conformal Treatment of Infinity}},  in {\em {Relativity,
  Groups and Topology}} ({de Witt, C.M. and de Witt, B.}, ed.), Les Houches
  Summer School, {Gordon and Breach (New York}), 1963.

\bibitem{Carter:1966zz}
B.~Carter, {\it {Complete Analytic Extension of the Symmetry Axis of Kerr's
  Solution of Einstein's Equations}},  Phys. Rev. {\bf 141} (1966) 1242--1247.

\bibitem{Reissner:1916}
H.~Reissner, {\it {\"Uber die Eigengravitation des Elektrischen Feldes nach
  Einsteinschen Theorie}},  Ann. Phys. {\bf 59} (1916) 106--120.

\bibitem{Nordstrom:1916}
{G. Nordstr\"{o}m}, {\it {On the Energy of the Gravitational Field in
  Einstein's Theory}},  Proc. Kon. Ned. Akad. Wet. {\bf 20} (1918) 1238--1245.

\bibitem{Graves:1960zz}
J.~C. Graves and D.~R. Brill, {\it {Oscillatory Character of Reissner-Nordstrom
  Metric for an Ideal Charged Wormhole}},  Phys. Rev. {\bf 120} (1960)
  1507--1513.

\bibitem{Carter1966423}
B.~Carter, {\it {The Complete Analytic Extension of the Reissner-Nordstr\"{o}m
  Metric in the Special Case $e^{2} = m^{2}$}},  Physics Letters {\bf 21}
  (1966) 423 -- 424.

\bibitem{Gibbons:1976ue}
G.~W. Gibbons and S.~W. Hawking, {\it {Action Integrals and Partition Functions
  in Quantum Gravity}},  Phys. Rev. {\bf D15} (1977) 2752--2756.

\bibitem{Kottler:1918}
{Kottler, F.}, {\it {\"Uber die physikalischen Grundlagen der Einsteinschen
  Gravitationstheorie}},  Ann. Phys. (Berlin) {\bf 56} (1918) 401.

\bibitem{Gibbons:1977mu}
G.~W. Gibbons and S.~W. Hawking, {\it {Cosmological Event Horizons,
  Thermodynamics, and Particle Creation}},  Phys. Rev. {\bf D15} (1977)
  2738--2751.

\bibitem{Maldacena:1997re}
J.~M. Maldacena, {\it {The Large N Limit of Superconformal Field Theories and
  Supergravity}},  Adv. Theor. Math. Phys. {\bf 2} (1998) 231--252,
  [\href{http://xxx.lanl.gov/abs/hep-th/9711200}{{\tt hep-th/9711200}}].

\bibitem{Witten:1998qj}
E.~Witten, {\it {Anti-de Sitter Space and Holography}},  Adv. Theor. Math.
  Phys. {\bf 2} (1998) 253--291,
  [\href{http://xxx.lanl.gov/abs/hep-th/9802150}{{\tt hep-th/9802150}}].

\bibitem{Witten:1998zw}
E.~Witten, {\it {Anti-de Sitter Space, Thermal Phase Transition, and
  Confinement in Gauge Theories}},  Adv. Theor. Math. Phys. {\bf 2} (1998)
  505--532, [\href{http://xxx.lanl.gov/abs/hep-th/9803131}{{\tt
  hep-th/9803131}}].

\bibitem{Brill:1997mf}
D.~R. Brill, J.~Louko, and P.~{Peld\`an}, {\it {Thermodynamics of
  (3+1)-Dimensional Black Holes with Toroidal or Higher Genus Horizons}},
  Phys. Rev. {\bf D56} (1997) 3600--3610,
  [\href{http://xxx.lanl.gov/abs/gr-qc/9705012}{{\tt gr-qc/9705012}}].

\bibitem{Hawking:1971vc}
S.~W. Hawking, {\it {Black Holes in General Relativity}},  Commun. Math. Phys.
  {\bf 25} (1972) 152--166.

\bibitem{Israel:1967wq}
W.~Israel, {\it {Event Horizons in Static Vacuum Space-times}},  Phys. Rev.
  {\bf 164} (1967) 1776--1779.

\bibitem{Israel:1967za}
W.~Israel, {\it {Event Horizons in Static Electrovac Space-times}},  Commun.
  Math. Phys. {\bf 8} (1968) 245--260.

\bibitem{Gannon:1976}
D.~Gannon, {\it {On the Topology of Spacelike Hypersurfaces, Singularities and
  Black Holes}},  Gen. Rel. Grav. {\bf 7} (1976) 219.

\bibitem{Friedman:1993ty}
J.~L. Friedman, K.~Schleich, and D.~M. Witt, {\it {Topological Censorship}},
  Phys. Rev. Lett. {\bf 71} (1993) 1486--1489,
  [\href{http://xxx.lanl.gov/abs/gr-qc/9305017}{{\tt gr-qc/9305017}}].

\bibitem{Jacobson:1994hs}
T.~Jacobson and S.~Venkataramani, {\it {Topology of Event Horizons and
  Topological Censorship}},  Class. Quant. Grav. {\bf 12} (1995) 1055--1062,
  [\href{http://xxx.lanl.gov/abs/gr-qc/9410023}{{\tt gr-qc/9410023}}].

\bibitem{Hughes:1994ea}
S.~A. Hughes {\em et.~al.}, {\it {Finding Black Holes in Numerical
  Space-times}},  Phys. Rev. {\bf D49} (1994) 4004--4015.

\bibitem{Shapiro:1995rr}
S.~L. Shapiro, S.~A. Teukolsky, and J.~Winicour, {\it {Toroidal Black Holes and
  Topological Censorship}},  Phys. Rev. {\bf D52} (1995) 6982--6987.

\bibitem{Skenderis:2002wp}
K.~Skenderis, {\it {Lecture Notes on Holographic Renormalization}},  Class.
  Quant. Grav. {\bf 19} (2002) 5849--5876,
  [\href{http://xxx.lanl.gov/abs/hep-th/0209067}{{\tt hep-th/0209067}}].

\bibitem{Banados:1992wn}
M.~Banados, C.~Teitelboim, and J.~Zanelli, {\it {The Black Hole in
  Three-dimensional Space-time}},  Phys. Rev. Lett. {\bf 69} (1992) 1849--1851,
  [\href{http://xxx.lanl.gov/abs/hep-th/9204099}{{\tt hep-th/9204099}}].

\bibitem{Banados:1992gq}
M.~Banados, M.~Henneaux, C.~Teitelboim, and J.~Zanelli, {\it {Geometry of the
  (2+1) Black Hole}},  Phys. Rev. {\bf D48} (1993) 1506--1525,
  [\href{http://xxx.lanl.gov/abs/gr-qc/9302012}{{\tt gr-qc/9302012}}].

\bibitem{Mann:1997iz}
R.~B. Mann, {\it {Topological Black Holes: Outside Looking In}},
  \href{http://xxx.lanl.gov/abs/gr-qc/9709039}{{\tt gr-qc/9709039}}.

\bibitem{Vanzo:1997gw}
L.~Vanzo, {\it {Black holes with Unusual Topology}},  Phys. Rev. {\bf D56}
  (1997) 6475--6483, [\href{http://xxx.lanl.gov/abs/gr-qc/9705004}{{\tt
  gr-qc/9705004}}].

\bibitem{Weinberg:1980gg}
S.~Weinberg, {\it {Utraviolet Divergences in Quantum Theories of Gravitation}},
  . In *Hawking, S.W., Israel, W.: General Relativity*, 790- 831.

\bibitem{Polchinski:1998rq}
J.~Polchinski, {\it {String Theory. Vol. 1: An Introduction to the Bosonic
  String}}, . Cambridge, UK: Univ. Pr. (1998) 402 p.

\bibitem{Polchinski:1998rr}
J.~Polchinski, {\it {String Theory. Vol. 2: Superstring Theory and Beyond}}, .
  Cambridge, UK: Univ. Pr. (1998) 531 p.

\bibitem{Riess:1998cb}
{\bf Supernova Search Team} Collaboration, A.~G. Riess {\em et.~al.}, {\it
  {Observational Evidence from Supernovae for an Accelerating Universe and a
  Cosmological Constant}},  Astron. J. {\bf 116} (1998) 1009--1038,
  [\href{http://xxx.lanl.gov/abs/astro-ph/9805201}{{\tt astro-ph/9805201}}].

\bibitem{Perlmutter:1998np}
{\bf Supernova Cosmology Project} Collaboration, S.~Perlmutter {\em et.~al.},
  {\it {Measurements of Omega and Lambda from 42 High-Redshift Supernovae}},
  Astrophys. J. {\bf 517} (1999) 565--586,
  [\href{http://xxx.lanl.gov/abs/astro-ph/9812133}{{\tt astro-ph/9812133}}].

\bibitem{Eisenstein:2005su}
{\bf SDSS} Collaboration, D.~J. Eisenstein {\em et.~al.}, {\it {Detection of
  the Baryon Acoustic Peak in the Large-Scale Correlation Function of SDSS
  Luminous Red Galaxies}},  Astrophys. J. {\bf 633} (2005) 560--574,
  [\href{http://xxx.lanl.gov/abs/astro-ph/0501171}{{\tt astro-ph/0501171}}].

\bibitem{Spergel:2003cb}
{\bf WMAP} Collaboration, D.~N. Spergel {\em et.~al.}, {\it {First Year
  Wilkinson Microwave Anisotropy Probe (WMAP) Observations: Determination of
  Cosmological Parameters}},  Astrophys. J. Suppl. {\bf 148} (2003) 175--194,
  [\href{http://xxx.lanl.gov/abs/astro-ph/0302209}{{\tt astro-ph/0302209}}].

\bibitem{Weinberg:1988cp}
S.~Weinberg, {\it {The Cosmological Constant Problem}},  Rev. Mod. Phys. {\bf
  61} (1989) 1--23.

\bibitem{Carroll:2000fy}
S.~M. Carroll, {\it {The Cosmological Constant}},  Living Rev. Rel. {\bf 4}
  (2001) 1, [\href{http://xxx.lanl.gov/abs/astro-ph/0004075}{{\tt
  astro-ph/0004075}}].

\bibitem{Caldwell:1997ii}
R.~R. Caldwell, R.~Dave, and P.~J. Steinhardt, {\it {Cosmological Imprint of an
  Energy Component with General Equation-of-State}},  Phys. Rev. Lett. {\bf 80}
  (1998) 1582--1585, [\href{http://xxx.lanl.gov/abs/astro-ph/9708069}{{\tt
  astro-ph/9708069}}].

\bibitem{Randall:1999vf}
L.~Randall and R.~Sundrum, {\it {An Alternative to Compactification}},  Phys.
  Rev. Lett. {\bf 83} (1999) 4690--4693,
  [\href{http://xxx.lanl.gov/abs/hep-th/9906064}{{\tt hep-th/9906064}}].

\bibitem{Dvali:2000hr}
G.~R. Dvali, G.~Gabadadze, and M.~Porrati, {\it {4D Gravity on a Brane in 5D
  Minkowski Space}},  Phys. Lett. {\bf B485} (2000) 208--214,
  [\href{http://xxx.lanl.gov/abs/hep-th/0005016}{{\tt hep-th/0005016}}].

\bibitem{Dvali:2000rv}
G.~R. Dvali, G.~Gabadadze, and M.~Porrati, {\it {Metastable Gravitons and
  Infinite Volume Extra Dimensions}},  Phys. Lett. {\bf B484} (2000) 112--118,
  [\href{http://xxx.lanl.gov/abs/hep-th/0002190}{{\tt hep-th/0002190}}].

\bibitem{Dvali:2000xg}
G.~R. Dvali and G.~Gabadadze, {\it {Gravity on a Brane in Infinite-Volume Extra
  Space}},  Phys. Rev. {\bf D63} (2001) 065007,
  [\href{http://xxx.lanl.gov/abs/hep-th/0008054}{{\tt hep-th/0008054}}].

\bibitem{Lovelock:1971yv}
D.~Lovelock, {\it {The Einstein Tensor and its Generalizations}},  J. Math.
  Phys. {\bf 12} (1971) 498--501.

\bibitem{Lovelock:1972vz}
D.~Lovelock, {\it {The Four-Dimensionality of Space and the Einstein Tensor}},
  J. Math. Phys. {\bf 13} (1972) 874--876.

\bibitem{Regge:1957td}
T.~Regge and J.~A. Wheeler, {\it {Stability of a Schwarzschild Singularity}},
  Phys. Rev. {\bf 108} (1957) 1063--1069.

\bibitem{Kerr:1963ud}
R.~P. Kerr, {\it {Gravitational Field of a Spinning Mass as an Example of
  Algebraically Special Metrics}},  Phys. Rev. Lett. {\bf 11} (1963) 237--238.

\bibitem{Newman:1965my}
E.~T. Newman {\em et.~al.}, {\it {Metric of a Rotating, Charged Mass}},  J.
  Math. Phys. {\bf 6} (1965) 918--919.

\bibitem{Carter:1971zc}
B.~Carter, {\it {Axisymmetric Black Hole Has Only Two Degrees of Freedom}},
  Phys. Rev. Lett. {\bf 26} (1971) 331--333.

\bibitem{Wald:1971iw}
R.~M. Wald, {\it {Final States of Gravitational Collapse}},  Phys. Rev. Lett.
  {\bf 26} (1971) 1653--1655.

\bibitem{Robinson:1975bv}
D.~C. Robinson, {\it {Uniqueness of the Kerr Black Hole}},  Phys. Rev. Lett.
  {\bf 34} (1975) 905--906.

\bibitem{Mazur:1982db}
P.~O. Mazur, {\it {Proof of Uniqueness of the Kerr-Newman Black Hole
  Solution}},  J. Phys. {\bf A15} (1982) 3173--3180.

\bibitem{Mazur:1984wz}
P.~O. Mazur, {\it {A Global Identity for Nonlinear Sigma Models}},  Phys. Lett.
  {\bf A100} (1984) 341.

\bibitem{Ruffini:1973}
R.~Ruffini and J.~Wheeler, {\it {Introducing the Black Hole}},  Physics Today
  {\bf B24} (1971) 30.

\bibitem{Bekenstein:1973ur}
J.~D. Bekenstein, {\it {Black Holes and Entropy}},  Phys. Rev. {\bf D7} (1973)
  2333--2346.

\bibitem{Bekenstein:1980}
J.~D. Bekenstein, {\it {Black-Hole Thermodynamics}},  Phys. Today {\bf 33}
  (1980) 24--31.

\bibitem{Chase:1970}
J.~Chase, {\it {Final States of Gravitational Collapse}},  Commun. Math. Phys.
  {\bf 19} (1970) 276.

\bibitem{Bekenstein:1972ny}
J.~D. Bekenstein, {\it {Transcendence of the Law of Baryon-Number Conservation
  in Black Hole Physics}},  Phys. Rev. Lett. {\bf 28} (1972) 452--455.

\bibitem{Bekenstein:1972ky}
J.~D. Bekenstein, {\it {Nonexistence of Baryon Number for Black Holes. ii}},
  Phys. Rev. {\bf D5} (1972) 2403--2412.

\bibitem{Hartle:1971qq}
J.~B. Hartle, {\it {Long-Range Neutrino Forces Exerted by Kerr Black Holes}},
  Phys. Rev. {\bf D3} (1971) 2938--2940.

\bibitem{Teitelboim:1972qx}
C.~Teitelboim, {\it {Nonmeasurability of the Quantum Numbers of a Black Hole}},
   Phys. Rev. {\bf D5} (1972) 2941--2954.

\bibitem{Heusler:1992ss}
M.~Heusler, {\it {A No Hair Theorem for Selfgravitating Nonlinear Sigma
  Models}},  J. Math. Phys. {\bf 33} (1992) 3497--3502.

\bibitem{Bekenstein:1995un}
J.~D. Bekenstein, {\it {Novel 'No Scalar Hair' Theorem for Black Holes}},
  Phys. Rev. {\bf D51} (1995) 6608--6611.

\bibitem{Sudarsky:1995zg}
D.~Sudarsky, {\it {A Simple Proof of a No Hair Theorem in Einstein Higgs
  Theory,}},  Class. Quant. Grav. {\bf 12} (1995) 579--584.

\bibitem{Mayo:1996mv}
A.~E. Mayo and J.~D. Bekenstein, {\it {No Hair for Spherical Black Holes:
  Charged and Nonminimally Coupled Scalar Field with Self-Interaction}},  Phys.
  Rev. {\bf D54} (1996) 5059--5069,
  [\href{http://xxx.lanl.gov/abs/gr-qc/9602057}{{\tt gr-qc/9602057}}].

\bibitem{Greene:1992fw}
B.~R. Greene, S.~D. Mathur, and C.~M. O'Neill, {\it {Eluding the No Hair
  Conjecture: Black Holes in Spontaneously Broken Gauge Theories}},  Phys. Rev.
  {\bf D47} (1993) 2242--2259,
  [\href{http://xxx.lanl.gov/abs/hep-th/9211007}{{\tt hep-th/9211007}}].

\bibitem{Bekenstein:1971hc}
J.~D. Bekenstein, {\it {Nonexistence of Baryon Number for Static Black Holes}},
   Phys. Rev. {\bf D5} (1972) 1239--1246.

\bibitem{Cai:1997ij}
R.-G. Cai, J.-Y. Ji, and K.-S. Soh, {\it {Hairs on the Cosmological Horizon}},
  Phys. Rev. {\bf D58} (1998) 024002,
  [\href{http://xxx.lanl.gov/abs/gr-qc/9708064}{{\tt gr-qc/9708064}}].

\bibitem{Torii:1998ir}
T.~Torii, K.~Maeda, and M.~Narita, {\it {No-Scalar Hair Conjecture in
  Asymptotic de Sitter Spacetime}},  Phys. Rev. {\bf D59} (1999) 064027,
  [\href{http://xxx.lanl.gov/abs/gr-qc/9809036}{{\tt gr-qc/9809036}}].

\bibitem{Torii:2001pg}
T.~Torii, K.~Maeda, and M.~Narita, {\it {Scalar Hair on the Black Hole in
  Asymptotically Anti-de Sitter Spacetime}},  Phys. Rev. {\bf D64} (2001)
  044007.

\bibitem{Sudarsky:2002mk}
D.~Sudarsky and J.~A. Gonzalez, {\it {On Black Hole Scalar Hair in
  Asymptotically Anti-de Sitter Spacetimes}},  Phys. Rev. {\bf D67} (2003)
  024038, [\href{http://xxx.lanl.gov/abs/gr-qc/0207069}{{\tt gr-qc/0207069}}].

\bibitem{Martinez:2004nb}
C.~Martinez, R.~Troncoso, and J.~Zanelli, {\it {Exact Black Hole Solution with
  a Minimally Coupled Scalar Field}},  Phys. Rev. {\bf D70} (2004) 084035,
  [\href{http://xxx.lanl.gov/abs/hep-th/0406111}{{\tt hep-th/0406111}}].

\bibitem{Henneaux:2004zi}
M.~Henneaux, C.~Martinez, R.~Troncoso, and J.~Zanelli, {\it {Asymptotically
  Anti-de Sitter Spacetimes and Scalar Fields with a Logarithmic Branch}},
  Phys. Rev. {\bf D70} (2004) 044034,
  [\href{http://xxx.lanl.gov/abs/hep-th/0404236}{{\tt hep-th/0404236}}].

\bibitem{Henneaux:2006hk}
M.~Henneaux, C.~Martinez, R.~Troncoso, and J.~Zanelli, {\it {Asymptotic
  Behavior and Hamiltonian Analysis of Anti-de Sitter Gravity Coupled to Scalar
  Fields}},  Annals Phys. {\bf 322} (2007) 824--848,
  [\href{http://xxx.lanl.gov/abs/hep-th/0603185}{{\tt hep-th/0603185}}].

\bibitem{Hertog:2004ns}
T.~Hertog and G.~T. Horowitz, {\it {Designer Gravity and Field Theory Effective
  Potentials}},  Phys. Rev. Lett. {\bf 94} (2005) 221301,
  [\href{http://xxx.lanl.gov/abs/hep-th/0412169}{{\tt hep-th/0412169}}].

\bibitem{Hertog:2004bb}
T.~Hertog and K.~Maeda, {\it {Stability and Thermodynamics of AdS Black Holes
  with Scalar Hair}},  Phys. Rev. {\bf D71} (2005) 024001,
  [\href{http://xxx.lanl.gov/abs/hep-th/0409314}{{\tt hep-th/0409314}}].

\bibitem{Hertog:2004dr}
T.~Hertog and K.~Maeda, {\it {Black Holes with Scalar Hair and Asymptotics in N
  = 8 Supergravity}},  JHEP {\bf 07} (2004) 051,
  [\href{http://xxx.lanl.gov/abs/hep-th/0404261}{{\tt hep-th/0404261}}].

\bibitem{Hertog:2005hm}
T.~Hertog and S.~Hollands, {\it {Stability in Designer Gravity}},  Class.
  Quant. Grav. {\bf 22} (2005) 5323--5342,
  [\href{http://xxx.lanl.gov/abs/hep-th/0508181}{{\tt hep-th/0508181}}].

\bibitem{Martinez:2006an}
C.~Martinez and R.~Troncoso, {\it {Electrically Charged Black Hole with Scalar
  Hair}},  Phys. Rev. {\bf D74} (2006) 064007,
  [\href{http://xxx.lanl.gov/abs/hep-th/0606130}{{\tt hep-th/0606130}}].

\bibitem{Hertog:2006rr}
T.~Hertog, {\it {Towards a Novel No-Hair Theorem for Black Holes}},  Phys. Rev.
  {\bf D74} (2006) 084008, [\href{http://xxx.lanl.gov/abs/gr-qc/0608075}{{\tt
  gr-qc/0608075}}].

\bibitem{Hertog:2006wj}
T.~Hertog, {\it {Violation of Energy Bounds in Designer Gravity}},  Class.
  Quant. Grav. {\bf 24} (2007) 141--154,
  [\href{http://xxx.lanl.gov/abs/hep-th/0607171}{{\tt hep-th/0607171}}].

\bibitem{Amsel:2007im}
A.~J. Amsel, T.~Hertog, S.~Hollands, and D.~Marolf, {\it {A Tale of Two
  Superpotentials: Stability and Instability in Designer Gravity}},  Phys. Rev.
  {\bf D75} (2007) 084008, [\href{http://xxx.lanl.gov/abs/hep-th/0701038}{{\tt
  hep-th/0701038}}].

\bibitem{Bocharova:1970}
N.~M. Bocharova, K.~A. Bronnikov, and V.~N. Mel'nikov, {\it {An Exact Solution
  of the System of Einstein Equations and Mass-Free Scalar Field}},  Vestnik
  Moskov. Univ. Fizika {\bf 25} (1970) 706--709.

\bibitem{Bekenstein:1974sf}
J.~D. Bekenstein, {\it {Exact Solutions of Einstein Conformal Scalar
  Equations}},  Ann. Phys. {\bf 82} (1974) 535--547.

\bibitem{Bekenstein:1975ts}
J.~D. Bekenstein, {\it {Black Holes with Scalar Charge}},  Annals Phys. {\bf
  91} (1975) 75--82.

\bibitem{Martinez:2002ru}
C.~Martinez, R.~Troncoso, and J.~Zanelli, {\it {De Sitter Black Hole with a
  Conformally Coupled Scalar Field in Four Dimensions}},  Phys. Rev. {\bf D67}
  (2003) 024008, [\href{http://xxx.lanl.gov/abs/hep-th/0205319}{{\tt
  hep-th/0205319}}].

\bibitem{Winstanley:2002jt}
E.~Winstanley, {\it {On the Existence of Conformally Coupled Scalar Field Hair
  for Black Holes in (Anti-)de Sitter Space}},  Found. Phys. {\bf 33} (2003)
  111--143, [\href{http://xxx.lanl.gov/abs/gr-qc/0205092}{{\tt
  gr-qc/0205092}}].

\bibitem{Harper:2003wt}
T.~J.~T. Harper, P.~A. Thomas, E.~Winstanley, and P.~M. Young, {\it
  {Instability of a Four-Dimensional de Sitter Black Hole with a Conformally
  Coupled Scalar Field}},  Phys. Rev. {\bf D70} (2004) 064023,
  [\href{http://xxx.lanl.gov/abs/gr-qc/0312104}{{\tt gr-qc/0312104}}].

\bibitem{Winstanley:2005fu}
E.~Winstanley, {\it {Dressing a Black Hole with Non-Minimally Coupled Scalar
  Field Hair}},  Class. Quant. Grav. {\bf 22} (2005) 2233--2248,
  [\href{http://xxx.lanl.gov/abs/gr-qc/0501096}{{\tt gr-qc/0501096}}].

\bibitem{Martinez:2005di}
C.~Martinez, J.~P. Staforelli, and R.~Troncoso, {\it {Charged Topological Black
  Hole with a Conformally Coupled Scalar Field}},  Phys. Rev. {\bf D74} (2006)
  044028, [\href{http://xxx.lanl.gov/abs/hep-th/0512022}{{\tt
  hep-th/0512022}}].

\bibitem{Dotti:2007cp}
G.~Dotti, R.~J. Gleiser, and C.~Martinez, {\it {Static Black Hole Solutions
  with a Self Interacting Conformally Coupled Scalar Field}},  Phys. Rev. {\bf
  D77} (2008) 104035, [\href{http://xxx.lanl.gov/abs/0710.1735}{{\tt
  arXiv:0710.1735}}].

\bibitem{Gibbons:1987ps}
G.~W. Gibbons and K.-i. Maeda, {\it {Black Holes and Membranes in Higher
  Dimensional Theories with Dilaton Fields}},  Nucl. Phys. {\bf B298} (1988)
  741.

\bibitem{Garfinkle:1990qj}
D.~Garfinkle, G.~T. Horowitz, and A.~Strominger, {\it {Charged Black Holes in
  String Theory}},  Phys. Rev. {\bf D43} (1991) 3140.

\bibitem{Holzhey:1991bx}
C.~F.~E. Holzhey and F.~Wilczek, {\it {Black Holes as Elementary Particles}},
  Nucl. Phys. {\bf B380} (1992) 447--477,
  [\href{http://xxx.lanl.gov/abs/hep-th/9202014}{{\tt hep-th/9202014}}].

\bibitem{Poletti:1994ww}
S.~J. Poletti, J.~Twamley, and D.~L. Wiltshire, {\it {Charged Dilaton Black
  Holes with a Cosmological Constant}},  Phys. Rev. {\bf D51} (1995)
  5720--5724, [\href{http://xxx.lanl.gov/abs/hep-th/9412076}{{\tt
  hep-th/9412076}}].

\bibitem{Wiltshire:1994de}
D.~L. Wiltshire, {\it {Dilaton Black Holes with a Cosmological Term}},  J.
  Austral. Math. Soc. {\bf B41} (1999) 198--216,
  [\href{http://xxx.lanl.gov/abs/gr-qc/9502038}{{\tt gr-qc/9502038}}].

\bibitem{Poletti:1994ff}
S.~J. Poletti and D.~L. Wiltshire, {\it {The Global Properties of Static
  Spherically Symmetric Charged Dilaton Space-times with a Liouville
  Potential}},  Phys. Rev. {\bf D50} (1994) 7260--7270,
  [\href{http://xxx.lanl.gov/abs/gr-qc/9407021}{{\tt gr-qc/9407021}}].

\bibitem{Dudas:2000ff}
E.~Dudas and J.~Mourad, {\it {Brane Solutions in Strings with Broken
  Supersymmetry and Dilaton Tadpoles}},  Phys. Lett. {\bf B486} (2000)
  172--178, [\href{http://xxx.lanl.gov/abs/hep-th/0004165}{{\tt
  hep-th/0004165}}].

\bibitem{Callan:1985ia}
C.~G. Callan, Jr., E.~J. Martinec, M.~J. Perry, and D.~Friedan, {\it {Strings
  in Background Fields}},  Nucl. Phys. {\bf B262} (1985) 593.

\bibitem{Callan:1986jb}
C.~G. Callan, Jr., I.~R. Klebanov, and M.~J. Perry, {\it {String Theory
  Effective Actions}},  Nucl. Phys. {\bf B278} (1986) 78.

\bibitem{Polyakov:1981rd}
A.~M. Polyakov, {\it {Quantum Geometry of Bosonic Strings}},  Phys. Lett. {\bf
  B103} (1981) 207--210.

\bibitem{Polyakov:1981re}
A.~M. Polyakov, {\it {Quantum Geometry of Fermionic Strings}},  Phys. Lett.
  {\bf B103} (1981) 211--213.

\bibitem{Dixon:1986iz}
L.~J. Dixon and J.~A. Harvey, {\it {String Theories in Ten-Dimensions Without
  Space-Time Supersymmetry}},  Nucl. Phys. {\bf B274} (1986) 93--105.

\bibitem{AlvarezGaume:1986jb}
L.~Alvarez-Gaume, P.~H. Ginsparg, G.~W. Moore, and C.~Vafa, {\it {An O(16) x
  O(16) Heterotic String}},  Phys. Lett. {\bf B171} (1986) 155.

\bibitem{Sagnotti:1995ga}
A.~Sagnotti, {\it {Some Properties of Open String Theories}},
  \href{http://xxx.lanl.gov/abs/hep-th/9509080}{{\tt hep-th/9509080}}.

\bibitem{Sagnotti:1996qj}
A.~Sagnotti, {\it {Surprises in Open-String Perturbation Theory}},  Nucl. Phys.
  Proc. Suppl. {\bf 56B} (1997) 332--343,
  [\href{http://xxx.lanl.gov/abs/hep-th/9702093}{{\tt hep-th/9702093}}].

\bibitem{Sugimoto:1999tx}
S.~Sugimoto, {\it {Anomaly Cancellations in Type I D9-D9-Bar System and the
  USp(32) String Theory}},  Prog. Theor. Phys. {\bf 102} (1999) 685--699,
  [\href{http://xxx.lanl.gov/abs/hep-th/9905159}{{\tt hep-th/9905159}}].

\bibitem{Charmousis:2003wm}
C.~Charmousis and R.~Gregory, {\it {Axisymmetric Metrics in Arbitrary
  Dimensions}},  Class. Quant. Grav. {\bf 21} (2004) 527--554,
  [\href{http://xxx.lanl.gov/abs/gr-qc/0306069}{{\tt gr-qc/0306069}}].

\bibitem{Charmousis:2006fx}
C.~Charmousis, D.~Langlois, D.~Steer, and R.~Zegers, {\it {Rotating Spacetimes
  with a Cosmological Constant}},  JHEP {\bf 02} (2007) 064,
  [\href{http://xxx.lanl.gov/abs/gr-qc/0610091}{{\tt gr-qc/0610091}}].

\bibitem{Shapere:1991ta}
A.~D. Shapere, S.~Trivedi, and F.~Wilczek, {\it {Dual Dilaton Dyons}},  Mod.
  Phys. Lett. {\bf A6} (1991) 2677--2686.

\bibitem{Kallosh:1992ii}
R.~Kallosh, A.~D. Linde, T.~Ortin, A.~W. Peet, and A.~Van~Proeyen, {\it
  {Supersymmetry as a Cosmic Censor}},  Phys. Rev. {\bf D46} (1992) 5278--5302,
  [\href{http://xxx.lanl.gov/abs/hep-th/9205027}{{\tt hep-th/9205027}}].

\bibitem{Clement:2005vn}
G.~Clement, D.~Gal'tsov, C.~Leygnac, and D.~Orlov, {\it {Dyonic Branes and
  Linear Dilaton Background}},  Phys. Rev. {\bf D73} (2006) 045018,
  [\href{http://xxx.lanl.gov/abs/hep-th/0512013}{{\tt hep-th/0512013}}].

\bibitem{Poletti:1995yq}
S.~J. Poletti, J.~Twamley, and D.~L. Wiltshire, {\it {Dyonic Dilaton Black
  Holes}},  Class. Quant. Grav. {\bf 12} (1995) 1753--1770,
  [\href{http://xxx.lanl.gov/abs/hep-th/9502054}{{\tt hep-th/9502054}}].

\bibitem{Yazadjiev:2005du}
S.~S. Yazadjiev, {\it {Non-Asymptotically Flat, Non-dS/AdS Dyonic Black Holes
  in Dilaton Gravity}},  Class. Quant. Grav. {\bf 22} (2005) 3875--3890,
  [\href{http://xxx.lanl.gov/abs/gr-qc/0502024}{{\tt gr-qc/0502024}}].

\bibitem{Yazadjiev:2005pf}
S.~S. Yazadjiev, {\it {Generating Dyonic Solutions in 5D Low-Energy String
  Theory and Dyonic Black Rings}},  Phys. Rev. {\bf D73} (2006) 124032,
  [\href{http://xxx.lanl.gov/abs/hep-th/0512229}{{\tt hep-th/0512229}}].

\bibitem{Chan:1995fr}
K.~C.~K. Chan, J.~H. Horne, and R.~B. Mann, {\it {Charged Dilaton Black Holes
  with Unusual Asymptotics}},  Nucl. Phys. {\bf B447} (1995) 441--464,
  [\href{http://xxx.lanl.gov/abs/gr-qc/9502042}{{\tt gr-qc/9502042}}].

\bibitem{Cai:1997ii}
R.-G. Cai, J.-Y. Ji, and K.-S. Soh, {\it {Topological Dilaton Black Holes}},
  Phys. Rev. {\bf D57} (1998) 6547--6550,
  [\href{http://xxx.lanl.gov/abs/gr-qc/9708063}{{\tt gr-qc/9708063}}].

\bibitem{Charmousis:2009xr}
C.~Charmousis, B.~{Gout\'eraux}, and J.~Soda, {\it {Einstein-Maxwell-Dilaton
  Theories with a Liouville Potential}},  Phys. Rev. {\bf D80} (2009) 024028,
  [\href{http://xxx.lanl.gov/abs/0905.3337}{{\tt arXiv:0905.3337}}].

\bibitem{Charmousis:2001nq}
C.~Charmousis, {\it {Dilaton Spacetimes with a Liouville Potential}},  Class.
  Quant. Grav. {\bf 19} (2002) 83--114,
  [\href{http://xxx.lanl.gov/abs/hep-th/0107126}{{\tt hep-th/0107126}}].

\bibitem{Binetruy:2002ck}
P.~Binetruy, C.~Charmousis, S.~C. Davis, and J.-F. Dufaux, {\it {Avoidance of
  Naked Singularities in Dilatonic Brane World Scenarios with a Gauss-Bonnet
  Term}},  Phys. Lett. {\bf B544} (2002) 183--191,
  [\href{http://xxx.lanl.gov/abs/hep-th/0206089}{{\tt hep-th/0206089}}].

\bibitem{Charmousis:2003ke}
C.~Charmousis, S.~C. Davis, and J.-F. Dufaux, {\it {Scalar Brane Backgrounds in
  Higher Order Curvature Gravity}},  JHEP {\bf 12} (2003) 029,
  [\href{http://xxx.lanl.gov/abs/hep-th/0309083}{{\tt hep-th/0309083}}].

\bibitem{Amendola:2005cr}
L.~Amendola, C.~Charmousis, and S.~C. Davis, {\it {Constraints on Gauss-Bonnet
  Gravity in Dark Energy Cosmologies}},  JCAP {\bf 0612} (2006) 020,
  [\href{http://xxx.lanl.gov/abs/hep-th/0506137}{{\tt hep-th/0506137}}].

\bibitem{Amendola:2007ni}
L.~Amendola, C.~Charmousis, and S.~C. Davis, {\it {Solar System Constraints on
  Gauss-Bonnet Mediated Dark Energy}},  JCAP {\bf 0710} (2007) 004,
  [\href{http://xxx.lanl.gov/abs/0704.0175}{{\tt arXiv:0704.0175}}].

\bibitem{Amendola:2008vd}
L.~Amendola, C.~Charmousis, and S.~C. Davis, {\it {Mimicking General Relativity
  in the Solar System}},  Phys. Rev. {\bf D78} (2008) 084009,
  [\href{http://xxx.lanl.gov/abs/0801.4339}{{\tt arXiv:0801.4339}}].

\bibitem{Nojiri:2005vv}
S.~Nojiri, S.~D. Odintsov, and M.~Sasaki, {\it {Gauss-Bonnet Dark Energy}},
  Phys. Rev. {\bf D71} (2005) 123509,
  [\href{http://xxx.lanl.gov/abs/hep-th/0504052}{{\tt hep-th/0504052}}].

\bibitem{Carter:2005fu}
B.~M.~N. Carter and I.~P. Neupane, {\it {Towards Inflation and Dark Energy
  Cosmologies From Modified Gauss-Bonnet Theory}},  JCAP {\bf 0606} (2006) 004,
  [\href{http://xxx.lanl.gov/abs/hep-th/0512262}{{\tt hep-th/0512262}}].

\bibitem{Koivisto:2006xf}
T.~Koivisto and D.~F. Mota, {\it {Cosmology and Astrophysical Constraints of
  Gauss-Bonnet Dark Energy}},  Phys. Lett. {\bf B644} (2007) 104--108,
  [\href{http://xxx.lanl.gov/abs/astro-ph/0606078}{{\tt astro-ph/0606078}}].

\bibitem{Koivisto:2006ai}
T.~Koivisto and D.~F. Mota, {\it {Gauss-Bonnet Quintessence: Background
  Evolution, Large Scale Structure and Cosmological Constraints}},  Phys. Rev.
  {\bf D75} (2007) 023518, [\href{http://xxx.lanl.gov/abs/hep-th/0609155}{{\tt
  hep-th/0609155}}].

\bibitem{Mignemi:1991wa}
S.~Mignemi and D.~L. Wiltshire, {\it {Black Holes in Higher Derivative Gravity
  Theories}},  Phys. Rev. {\bf D46} (1992) 1475--1506,
  [\href{http://xxx.lanl.gov/abs/hep-th/9202031}{{\tt hep-th/9202031}}].

\bibitem{Mignemi:1992nt}
S.~Mignemi and N.~R. Stewart, {\it {Charged Black Holes in Effective String
  Theory}},  Phys. Rev. {\bf D47} (1993) 5259--5269,
  [\href{http://xxx.lanl.gov/abs/hep-th/9212146}{{\tt hep-th/9212146}}].

\bibitem{Kanti:1995vq}
P.~Kanti, N.~E. Mavromatos, J.~Rizos, K.~Tamvakis, and E.~Winstanley, {\it
  {Dilatonic Black Holes in Higher Curvature String Gravity}},  Phys. Rev. {\bf
  D54} (1996) 5049--5058, [\href{http://xxx.lanl.gov/abs/hep-th/9511071}{{\tt
  hep-th/9511071}}].

\bibitem{Torii:1996yi}
T.~Torii, H.~Yajima, and K.-i. Maeda, {\it {Dilatonic Black Holes with
  Gauss-Bonnet Term}},  Phys. Rev. {\bf D55} (1997) 739--753,
  [\href{http://xxx.lanl.gov/abs/gr-qc/9606034}{{\tt gr-qc/9606034}}].

\bibitem{Kanti:1997br}
P.~Kanti, N.~E. Mavromatos, J.~Rizos, K.~Tamvakis, and E.~Winstanley, {\it
  {Dilatonic Black Holes in Higher-Curvature String Gravity. II: Linear
  Stability}},  Phys. Rev. {\bf D57} (1998) 6255--6264,
  [\href{http://xxx.lanl.gov/abs/hep-th/9703192}{{\tt hep-th/9703192}}].

\bibitem{Melis:2005xt}
M.~Melis and S.~Mignemi, {\it {Global Properties of Dilatonic Gauss-Bonnet
  Black Holes}},  Class. Quant. Grav. {\bf 22} (2005) 3169--3180,
  [\href{http://xxx.lanl.gov/abs/gr-qc/0501087}{{\tt gr-qc/0501087}}].

\bibitem{Melis:2005ji}
M.~Melis and S.~Mignemi, {\it {Global Properties of Charged Dilatonic
  Gauss-Bonnet Black Holes}},  Phys. Rev. {\bf D73} (2006) 083010,
  [\href{http://xxx.lanl.gov/abs/gr-qc/0512132}{{\tt gr-qc/0512132}}].

\bibitem{Guo:2008hf}
Z.-K. Guo, N.~Ohta, and T.~Torii, {\it {Black Holes in the Dilatonic
  Einstein-Gauss-Bonnet Theory in Various Dimensions I -- Asymptotically Flat
  Black Holes --}},  Prog. Theor. Phys. {\bf 120} (2008) 581--607,
  [\href{http://xxx.lanl.gov/abs/0806.2481}{{\tt arXiv:0806.2481}}].

\bibitem{Guo:2008eq}
Z.-K. Guo, N.~Ohta, and T.~Torii, {\it {Black Holes in the Dilatonic
  Einstein-Gauss-Bonnet Theory in Various Dimensions II -- Asymptotically AdS
  Topological Black Holes --}},  Prog. Theor. Phys. {\bf 121} (2009) 253--273,
  [\href{http://xxx.lanl.gov/abs/0811.3068}{{\tt arXiv:0811.3068}}].

\bibitem{Ohta:2009tb}
N.~Ohta and T.~Torii, {\it {Black Holes in the Dilatonic Einstein-Gauss-Bonnet
  Theory in Various Dimensions III -- Asymptotically AdS Black Holes with
  $k=\pm 1$ --}},  Prog. Theor. Phys. {\bf 121} (2009) 959--981,
  [\href{http://xxx.lanl.gov/abs/0902.4072}{{\tt arXiv:0902.4072}}].

\bibitem{Ohta:2009pe}
N.~Ohta and T.~Torii, {\it {Black Holes in the Dilatonic Einstein-Gauss-Bonnet
  Theory in Various Dimensions IV - Topological Black Holes with and without
  Cosmological Term}},  Prog. Theor. Phys. {\bf 122} (2009) 1477--1500,
  [\href{http://xxx.lanl.gov/abs/0908.3918}{{\tt arXiv:0908.3918}}].

\bibitem{Maeda:2009uy}
K.-i. Maeda, N.~Ohta, and Y.~Sasagawa, {\it {Black Hole Solutions in String
  Theory with Gauss-Bonnet Curvature Correction}},  Phys. Rev. {\bf D80} (2009)
  104032, [\href{http://xxx.lanl.gov/abs/0908.4151}{{\tt arXiv:0908.4151}}].

\bibitem{Stelle:1977ry}
K.~S. Stelle, {\it {Classical Gravity with Higher Derivatives}},  Gen. Rel.
  Grav. {\bf 9} (1978) 353--371.

\bibitem{Gross:1986mw}
D.~J. Gross and J.~H. Sloan, {\it {The Quartic Effective Action for the
  Heterotic String}},  Nucl. Phys. {\bf B291} (1987) 41.

\bibitem{Brustein:1987qw}
R.~Brustein, D.~Nemeschansky, and S.~Yankielowicz, {\it {Beta Functions and S
  Matrix in String Theory}},  Nucl. Phys. {\bf B301} (1988) 224.

\bibitem{Callan:1988hs}
C.~G. Callan, Jr., R.~C. Myers, and M.~J. Perry, {\it {Black Holes in String
  Theory}},  Nucl. Phys. {\bf B311} (1989) 673.

\bibitem{Myers:1987qx}
R.~C. Myers, {\it {Superstring Gravity and Black Holes}},  Nucl. Phys. {\bf
  B289} (1987) 701--716.

\bibitem{Zwiebach:1985uq}
B.~Zwiebach, {\it {Curvature Squared Terms and String Theories}},  Phys. Lett.
  {\bf B156} (1985) 315.

\bibitem{Lanczos:1938sf}
C.~Lanczos, {\it {A Remarkable Property of the Riemann-Christoffel Tensor in
  Four Dimensions}},  Annals Math. {\bf 39} (1938) 842--850.

\bibitem{Charmousis:2008kc}
C.~Charmousis, {\it {Higher Order Gravity Theories and their Black Hole
  Solutions}},  Lect. Notes Phys. {\bf 769} (2009) 299--346,
  [\href{http://xxx.lanl.gov/abs/0805.0568}{{\tt arXiv:0805.0568}}].

\bibitem{Garraffo:2008hu}
C.~Garraffo and G.~Giribet, {\it {The Lovelock Black Holes}},  Mod. Phys. Lett.
  {\bf A23} (2008) 1801--1818, [\href{http://xxx.lanl.gov/abs/0805.3575}{{\tt
  arXiv:0805.3575}}].

\bibitem{Wheeler:1985qd}
J.~T. Wheeler, {\it {Symmetric Solutions to the Maximally Gauss-Bonnet Extended
  Einstein Equations}},  Nucl. Phys. {\bf B273} (1986) 732.

\bibitem{Myers:1988ze}
R.~C. Myers and J.~Z. Simon, {\it {Black Hole Thermodynamics in Lovelock
  Gravity}},  Phys. Rev. {\bf D38} (1988) 2434--2444.

\bibitem{Crisostomo:2000bb}
J.~Crisostomo, R.~Troncoso, and J.~Zanelli, {\it {Black hole scan}},  Phys.
  Rev. {\bf D62} (2000) 084013,
  [\href{http://xxx.lanl.gov/abs/hep-th/0003271}{{\tt hep-th/0003271}}].

\bibitem{Zegers:2005vx}
R.~Zegers, {\it {Birkhoff's theorem in Lovelock gravity}},  J. Math. Phys. {\bf
  46} (2005) 072502, [\href{http://xxx.lanl.gov/abs/gr-qc/0505016}{{\tt
  gr-qc/0505016}}].

\bibitem{Edgar:2001vv}
S.~B. Edgar and A.~Hoglund, {\it {Dimensionally Dependent Tensor Identities by
  Double Antisymmetrisation}},  J. Math. Phys. {\bf 43} (2002) 659--677,
  [\href{http://xxx.lanl.gov/abs/gr-qc/0105066}{{\tt gr-qc/0105066}}].

\bibitem{Banados:1993ur}
M.~Banados, C.~Teitelboim, and J.~Zanelli, {\it {Dimensionally Continued Black
  Holes}},  Phys. Rev. {\bf D49} (1994) 975--986,
  [\href{http://xxx.lanl.gov/abs/gr-qc/9307033}{{\tt gr-qc/9307033}}].

\bibitem{Charmousis:2008ce}
C.~Charmousis and A.~Padilla, {\it {The Instability of Vacua in Gauss-Bonnet
  Gravity}},  JHEP {\bf 12} (2008) 038,
  [\href{http://xxx.lanl.gov/abs/0807.2864}{{\tt arXiv:0807.2864}}].

\bibitem{Deser:2002jk}
S.~Deser and B.~Tekin, {\it {Energy in Generic Higher Curvature Gravity
  Theories}},  Phys. Rev. {\bf D67} (2003) 084009,
  [\href{http://xxx.lanl.gov/abs/hep-th/0212292}{{\tt hep-th/0212292}}].

\bibitem{Boulware:1985wk}
D.~G. Boulware and S.~Deser, {\it {String Generated Gravity Models}},  Phys.
  Rev. Lett. {\bf 55} (1985) 2656.

\bibitem{Wheeler:1985nh}
J.~T. Wheeler, {\it {Symmetric Solutions to the Gauss-Bonnet Extended Einstein
  Equations}},  Nucl. Phys. {\bf B268} (1986) 737.

\bibitem{Wiltshire:1985us}
D.~L. Wiltshire, {\it {Spherically Symmetric Solutions of Einstein-Maxwell
  Theory with a Gauss-Bonnet Term}},  Phys. Lett. {\bf B169} (1986) 36.

\bibitem{Wiltshire:1988uq}
D.~L. Wiltshire, {\it {Black Holes in String Generated Gravity Models}},  Phys.
  Rev. {\bf D38} (1988) 2445.

\bibitem{Charmousis:2002rc}
C.~Charmousis and J.-F. Dufaux, {\it {General Gauss-Bonnet Brane Cosmology}},
  Class. Quant. Grav. {\bf 19} (2002) 4671--4682,
  [\href{http://xxx.lanl.gov/abs/hep-th/0202107}{{\tt hep-th/0202107}}].

\bibitem{Bogdanos:2009pc}
C.~Bogdanos, C.~Charmousis, B.~{Gout\'eraux}, and R.~Zegers, {\it
  {Einstein-Gauss-Bonnet Metrics: Black Holes, Black Strings and a Staticity
  Theorem}},  JHEP {\bf 10} (2009) 037,
  [\href{http://xxx.lanl.gov/abs/0906.4953}{{\tt arXiv:0906.4953}}].

\bibitem{Bowcock:2000cq}
P.~Bowcock, C.~Charmousis, and R.~Gregory, {\it {General Brane Cosmologies and
  their Global Spacetime Structure}},  Class. Quant. Grav. {\bf 17} (2000)
  4745--4764, [\href{http://xxx.lanl.gov/abs/hep-th/0007177}{{\tt
  hep-th/0007177}}].

\bibitem{Zanelli:2005sa}
J.~Zanelli, {\it {Lecture Notes on Chern-Simons (Super-)Gravities. Second
  Edition (February 2008)}},
  \href{http://xxx.lanl.gov/abs/hep-th/0502193}{{\tt hep-th/0502193}}.

\bibitem{Dotti:2005rc}
G.~Dotti and R.~J. Gleiser, {\it {Obstructions on the Horizon Geometry from
  String Theory Corrections to Einstein Gravity}},  Phys. Lett. {\bf B627}
  (2005) 174--179, [\href{http://xxx.lanl.gov/abs/hep-th/0508118}{{\tt
  hep-th/0508118}}].

\bibitem{Maeda:2006hj}
H.~Maeda and N.~Dadhich, {\it {Matter Without Matter: Novel Kaluza-Klein
  Spacetime in Einstein-Gauss-Bonnet Gravity}},  Phys. Rev. {\bf D75} (2007)
  044007, [\href{http://xxx.lanl.gov/abs/hep-th/0611188}{{\tt
  hep-th/0611188}}].

\bibitem{Maeda:2006iw}
H.~Maeda and N.~Dadhich, {\it {Kaluza-Klein Black Hole with Negatively Curved
  Extra Dimensions in String Generated Gravity Models}},  Phys. Rev. {\bf D74}
  (2006) 021501, [\href{http://xxx.lanl.gov/abs/hep-th/0605031}{{\tt
  hep-th/0605031}}].

\bibitem{Molina:2008kh}
A.~Molina and N.~Dadhich, {\it {On Kaluza-Klein Spacetime in
  Einstein-Gauss-Bonnet Gravity}},  Int. J. Mod. Phys. {\bf D18} (2009)
  599--611, [\href{http://xxx.lanl.gov/abs/0804.1194}{{\tt arXiv:0804.1194}}].

\bibitem{Dotti:2010bw}
G.~Dotti, J.~Oliva, and R.~Troncoso, {\it {Static Solutions with Nontrivial
  Boundaries for the Einstein-Gauss-Bonnet Theory in Vacuum}},
  \href{http://xxx.lanl.gov/abs/1004.5287}{{\tt arXiv:1004.5287}}.

\bibitem{Dotti:2005sq}
G.~Dotti and R.~J. Gleiser, {\it {Linear Stability of Einstein-Gauss-Bonnet
  Static Spacetimes. Part. I: Tensor Perturbations}},  Phys. Rev. {\bf D72}
  (2005) 044018, [\href{http://xxx.lanl.gov/abs/gr-qc/0503117}{{\tt
  gr-qc/0503117}}].

\bibitem{Gleiser:2005ra}
R.~J. Gleiser and G.~Dotti, {\it {Linear Stability of Einstein-Gauss-Bonnet
  Static Spacetimes. II: Vector and Scalar Perturbations}},  Phys. Rev. {\bf
  D72} (2005) 124002, [\href{http://xxx.lanl.gov/abs/gr-qc/0510069}{{\tt
  gr-qc/0510069}}].

\bibitem{Beroiz:2007gp}
M.~Beroiz, G.~Dotti, and R.~J. Gleiser, {\it {Gravitational Instability of
  Static Spherically Symmetric Einstein-Gauss-Bonnet Black Holes in Five and
  Six Dimensions}},  Phys. Rev. {\bf D76} (2007) 024012,
  [\href{http://xxx.lanl.gov/abs/hep-th/0703074}{{\tt hep-th/0703074}}].

\bibitem{Deser:2002rt}
S.~Deser and B.~Tekin, {\it {Gravitational Energy in Quadratic Curvature
  Gravities}},  Phys. Rev. Lett. {\bf 89} (2002) 101101,
  [\href{http://xxx.lanl.gov/abs/hep-th/0205318}{{\tt hep-th/0205318}}].

\bibitem{Taub:1950ez}
A.~H. Taub, {\it {Empty Space-times Admitting a Three Parameter Group of
  Motions}},  Annals Math. {\bf 53} (1951) 472--490.

\bibitem{Newman:1963yy}
E.~Newman, L.~Tamubrino, and T.~Unti, {\it {Empty Space Generalization of the
  Schwarzschild Metric}},  J. Math. Phys. {\bf 4} (1963) 915.

\bibitem{Hunter:1998qe}
C.~J. Hunter, {\it {The Action of Instantons with Nut Charge}},  Phys. Rev.
  {\bf D59} (1999) 024009, [\href{http://xxx.lanl.gov/abs/gr-qc/9807010}{{\tt
  gr-qc/9807010}}].

\bibitem{Hawking:1998jf}
S.~W. Hawking and C.~J. Hunter, {\it {Gravitational Entropy and Global
  Structure}},  Phys. Rev. {\bf D59} (1999) 044025,
  [\href{http://xxx.lanl.gov/abs/hep-th/9808085}{{\tt hep-th/9808085}}].

\bibitem{Chamblin:1998pz}
A.~Chamblin, R.~Emparan, C.~V. Johnson, and R.~C. Myers, {\it {Large N Phases,
  Gravitational Instantons and the Nuts and Bolts of AdS Holography}},  Phys.
  Rev. {\bf D59} (1999) 064010,
  [\href{http://xxx.lanl.gov/abs/hep-th/9808177}{{\tt hep-th/9808177}}].

\bibitem{Hawking:1998ct}
S.~W. Hawking, C.~J. Hunter, and D.~N. Page, {\it {Nut Charge, Anti-de Sitter
  Space and Entropy}},  Phys. Rev. {\bf D59} (1999) 044033,
  [\href{http://xxx.lanl.gov/abs/hep-th/9809035}{{\tt hep-th/9809035}}].

\bibitem{Zoubos:2002cw}
K.~Zoubos, {\it {Holography and Quaternionic Taub-NUT}},  JHEP {\bf 12} (2002)
  037, [\href{http://xxx.lanl.gov/abs/hep-th/0209235}{{\tt hep-th/0209235}}].

\bibitem{Kleban:2004bv}
M.~Kleban, M.~Porrati, and R.~Rabadan, {\it {Stability in Asymptotically AdS
  Spaces}},  JHEP {\bf 08} (2005) 016,
  [\href{http://xxx.lanl.gov/abs/hep-th/0409242}{{\tt hep-th/0409242}}].

\bibitem{Charmousis:2008bt}
C.~Charmousis and A.~Papazoglou, {\it {Self-Properties of Codimension-2
  Braneworlds}},  JHEP {\bf 07} (2008) 062,
  [\href{http://xxx.lanl.gov/abs/0804.2121}{{\tt arXiv:0804.2121}}].

\bibitem{Charmousis:2009uk}
C.~Charmousis, G.~Kofinas, and A.~Papazoglou, {\it {The Consistency of
  Codimension-2 Braneworlds and their Cosmology}},  JCAP {\bf 1001} (2010) 022,
  [\href{http://xxx.lanl.gov/abs/0907.1640}{{\tt arXiv:0907.1640}}].

\bibitem{CuadrosMelgar:2008kn}
B.~Cuadros-Melgar, E.~Papantonopoulos, M.~Tsoukalas, and V.~Zamarias, {\it
  {Black Holes on Thin 3-branes of Codimension-2 and their Extension into the
  Bulk}},  Nucl. Phys. {\bf B810} (2009) 246--265,
  [\href{http://xxx.lanl.gov/abs/0804.4459}{{\tt arXiv:0804.4459}}].

\bibitem{Dotti:2008pp}
G.~Dotti, J.~Oliva, and R.~Troncoso, {\it {Vacuum Solutions with Nontrivial
  Boundaries for the Einstein-Gauss-Bonnet Theory}},  Int. J. Mod. Phys. {\bf
  A24} (2009) 1690--1694, [\href{http://xxx.lanl.gov/abs/0809.4378}{{\tt
  arXiv:0809.4378}}].

\bibitem{Tangherlini:1963bw}
F.~R. Tangherlini, {\it {Schwarzschild Field in n Dimensions and the
  Dimensionality of Space Problem}},  Nuovo Cim. {\bf 27} (1963) 636--651.

\bibitem{Emparan:2008eg}
R.~Emparan and H.~S. Reall, {\it {Black Holes in Higher Dimensions}},  Living
  Rev. Rel. {\bf 11} (2008) 6, [\href{http://xxx.lanl.gov/abs/0801.3471}{{\tt
  arXiv:0801.3471}}].

\bibitem{Gregory:1993vy}
R.~Gregory and R.~Laflamme, {\it {Black Strings and p-Branes Are Unstable}},
  Phys. Rev. Lett. {\bf 70} (1993) 2837--2840,
  [\href{http://xxx.lanl.gov/abs/hep-th/9301052}{{\tt hep-th/9301052}}].

\bibitem{Emparan:2001wn}
R.~Emparan and H.~S. Reall, {\it {A Rotating Black Ring in Five Dimensions}},
  Phys. Rev. Lett. {\bf 88} (2002) 101101,
  [\href{http://xxx.lanl.gov/abs/hep-th/0110260}{{\tt hep-th/0110260}}].

\bibitem{Wald:1999vt}
R.~M. Wald, {\it {The Thermodynamics of Black Holes}},  Living Rev. Rel. {\bf
  4} (2001) 6, [\href{http://xxx.lanl.gov/abs/gr-qc/9912119}{{\tt
  gr-qc/9912119}}].

\bibitem{Ross:2005sc}
S.~F. Ross, {\it {Black Hole Thermodynamics}},
  \href{http://xxx.lanl.gov/abs/hep-th/0502195}{{\tt hep-th/0502195}}.

\bibitem{Bardeen:1973gs}
J.~M. Bardeen, B.~Carter, and S.~W. Hawking, {\it {The Four Laws of Black Hole
  Mechanics}},  Commun. Math. Phys. {\bf 31} (1973) 161--170.

\bibitem{Smarr:1972kt}
L.~Smarr, {\it {Mass Formula for Kerr Black Holes}},  Phys. Rev. Lett. {\bf 30}
  (1973) 71--73.

\bibitem{Bekenstein:1973mi}
J.~D. Bekenstein, {\it {Extraction of Energy and Charge from a Black Hole}},
  Phys. Rev. {\bf D7} (1973) 949--953.

\bibitem{Jacobson:2003wv}
T.~Jacobson and R.~Parentani, {\it {Horizon Entropy}},  Found. Phys. {\bf 33}
  (2003) 323--348, [\href{http://xxx.lanl.gov/abs/gr-qc/0302099}{{\tt
  gr-qc/0302099}}].

\bibitem{Penrose:1969pc}
R.~Penrose, {\it {Gravitational Collapse: The Role of General Relativity}},
  Riv. Nuovo Cim. {\bf 1} (1969) 252--276.

\bibitem{Penrose:1971uk}
R.~Penrose and R.~M. Floyd, {\it {Extraction of Rotational Energy from a Black
  Hole}},  Nature {\bf 229} (1971) 177--179.

\bibitem{Christodoulou:1970wf}
D.~Christodoulou, {\it {Reversible and Irreversible Transforations in Black
  Hole Physics}},  Phys. Rev. Lett. {\bf 25} (1970) 1596--1597.

\bibitem{Christodoulou:1972kt}
D.~Christodoulou and R.~Ruffini, {\it {Reversible Transformations of a Charged
  Black Hole}},  Phys. Rev. {\bf D4} (1971) 3552--3555.

\bibitem{Hawking:1971tu}
S.~W. Hawking, {\it {Gravitational Radiation from Colliding Black Holes}},
  Phys. Rev. Lett. {\bf 26} (1971) 1344--1346.

\bibitem{Israel:1986}
W.~Israel, {\it Third law of black-hole dynamics: A formulation and proof},
  Phys. Rev. Lett. {\bf 57} (Jul, 1986) 397--399.

\bibitem{Wald:1997qp}
R.~M. Wald, {\it {The *Nernst Theorem* and Black Hole Thermodynamics}},  Phys.
  Rev. {\bf D56} (1997) 6467--6474,
  [\href{http://xxx.lanl.gov/abs/gr-qc/9704008}{{\tt gr-qc/9704008}}].

\bibitem{Bekenstein:1972tm}
J.~D. Bekenstein, {\it {Black Holes and the Second Law}},  Nuovo Cim. Lett.
  {\bf 4} (1972) 737--740.

\bibitem{Bekenstein:1974ax}
J.~D. Bekenstein, {\it {Generalized Second Law of Thermodynamics in Black Hole
  Physics}},  Phys. Rev. {\bf D9} (1974) 3292--3300.

\bibitem{Hawking:1976de}
S.~W. Hawking, {\it {Black Holes and Thermodynamics}},  Phys. Rev. {\bf D13}
  (1976) 191--197.

\bibitem{Hawking:1974rv}
S.~W. Hawking, {\it {Black Hole Explosions}},  Nature {\bf 248} (1974) 30--31.

\bibitem{Hawking:1974sw}
S.~W. Hawking, {\it {Particle Creation by Black Holes}},  Commun. Math. Phys.
  {\bf 43} (1975) 199--220.

\bibitem{Unruh:1976db}
W.~G. Unruh, {\it {Notes on Black Hole Evaporation}},  Phys. Rev. {\bf D14}
  (1976) 870.

\bibitem{Wald:1993nt}
R.~M. Wald, {\it {Black hole Entropy is the Noether Charge}},  Phys. Rev. {\bf
  D48} (1993) 3427--3431, [\href{http://xxx.lanl.gov/abs/gr-qc/9307038}{{\tt
  gr-qc/9307038}}].

\bibitem{Jacobson:1993xs}
T.~Jacobson and R.~C. Myers, {\it {Black Hole Entropy and Higher Curvature
  Interactions}},  Phys. Rev. Lett. {\bf 70} (1993) 3684--3687,
  [\href{http://xxx.lanl.gov/abs/hep-th/9305016}{{\tt hep-th/9305016}}].

\bibitem{Banados:1993qp}
M.~Banados, C.~Teitelboim, and J.~Zanelli, {\it {Black Hole Entropy and the
  Dimensional Continuation of the Gauss-Bonnet Theorem}},  Phys. Rev. Lett.
  {\bf 72} (1994) 957--960, [\href{http://xxx.lanl.gov/abs/gr-qc/9309026}{{\tt
  gr-qc/9309026}}].

\bibitem{York:1972sj}
J.~W. York, Jr., {\it {Role of Conformal Three Geometry in the Dynamics of
  Gravitation}},  Phys. Rev. Lett. {\bf 28} (1972) 1082--1085.

\bibitem{Gibbons:1978ac}
G.~W. Gibbons, S.~W. Hawking, and M.~J. Perry, {\it {Path Integrals and the
  Indefiniteness of the Gravitational Action}},  Nucl. Phys. {\bf B138} (1978)
  141.

\bibitem{Monteiro:2010cq}
R.~Monteiro, {\em {Classical and Thermodynamic Stability of Black Holes}}.
\newblock PhD thesis, {University of Cambridge}, 2010.
\newblock \href{http://xxx.lanl.gov/abs/1006.5358}{{\tt arXiv:1006.5358}}.

\bibitem{Hawking:1982dh}
S.~W. Hawking and D.~N. Page, {\it {Thermodynamics of Black Holes in Anti-de
  Sitter Space}},  Commun. Math. Phys. {\bf 87} (1983) 577.

\bibitem{Henneaux:1985tv}
M.~Henneaux and C.~Teitelboim, {\it {Asymptotically Anti-De Sitter Spaces}},
  Commun. Math. Phys. {\bf 98} (1985) 391--424.

\bibitem{Balasubramanian:1999re}
V.~Balasubramanian and P.~Kraus, {\it {A Stress Tensor for Anti-de Sitter
  Gravity}},  Commun. Math. Phys. {\bf 208} (1999) 413--428,
  [\href{http://xxx.lanl.gov/abs/hep-th/9902121}{{\tt hep-th/9902121}}].

\bibitem{Emparan:1999pm}
R.~Emparan, C.~V. Johnson, and R.~C. Myers, {\it {Surface Terms as Counterterms
  in the AdS/CFT Correspondence}},  Phys. Rev. {\bf D60} (1999) 104001,
  [\href{http://xxx.lanl.gov/abs/hep-th/9903238}{{\tt hep-th/9903238}}].

\bibitem{Cai:1999xg}
R.-G. ~ and N.~Ohta, {\it {Surface Counterterms and Boundary Stress-Energy
  Tensors for Asymptotically Non-Anti-de Sitter Spaces}},  Phys. Rev. {\bf D62}
  (2000) 024006, [\href{http://xxx.lanl.gov/abs/hep-th/9912013}{{\tt
  hep-th/9912013}}].

\bibitem{Regge:1974zd}
T.~Regge and C.~Teitelboim, {\it {Role of Surface Integrals in the Hamiltonian
  Formulation of General Relativity}},  Ann. Phys. {\bf 88} (1974) 286.

\bibitem{Brown:1992br}
J.~D. Brown and J.~W. York, Jr., {\it {Quasilocal Energy and Conserved Charges
  Derived from the Gravitational Action}},  Phys. Rev. {\bf D47} (1993)
  1407--1419, [\href{http://xxx.lanl.gov/abs/gr-qc/9209012}{{\tt
  gr-qc/9209012}}].

\bibitem{Hawking:1995fd}
S.~W. Hawking and G.~T. Horowitz, {\it {The Gravitational Hamiltonian, Action,
  Entropy and Surface Terms}},  Class. Quant. Grav. {\bf 13} (1996) 1487--1498,
  [\href{http://xxx.lanl.gov/abs/gr-qc/9501014}{{\tt gr-qc/9501014}}].

\bibitem{Hawking:1996ww}
S.~W. Hawking and C.~J. Hunter, {\it {The Gravitational Hamiltonian in the
  Presence of Non- Orthogonal Boundaries}},  Class. Quant. Grav. {\bf 13}
  (1996) 2735--2752, [\href{http://xxx.lanl.gov/abs/gr-qc/9603050}{{\tt
  gr-qc/9603050}}].

\bibitem{Booth:1998eh}
I.~S. Booth and R.~B. Mann, {\it {Moving Observers, Non-Orthogonal Boundaries,
  and Quasilocal Energy}},  Phys. Rev. {\bf D59} (1999) 064021,
  [\href{http://xxx.lanl.gov/abs/gr-qc/9810009}{{\tt gr-qc/9810009}}].

\bibitem{Booth:2000iq}
I.~S.~N. Booth, {\em {A Quasilocal Hamiltonian for Gravity with Classical and
  Quantum Applications}}.
\newblock PhD thesis, University of Waterloo, 2000.
\newblock \href{http://xxx.lanl.gov/abs/gr-qc/0008030}{{\tt gr-qc/0008030}}.

\bibitem{Brown:2000dz}
J.~D. Brown, S.~R. Lau, and J.~W. York, Jr., {\it {Action and Energy of the
  Gravitational Field}},  \href{http://xxx.lanl.gov/abs/gr-qc/0010024}{{\tt
  gr-qc/0010024}}.

\bibitem{Brown:1992bq}
J.~D. Brown and J.~W. York, Jr., {\it {The Microcanonical Functional Integral.
  1. The Gravitational Field}},  Phys. Rev. {\bf D47} (1993) 1420--1431,
  [\href{http://xxx.lanl.gov/abs/gr-qc/9209014}{{\tt gr-qc/9209014}}].

\bibitem{Arnowitt:1962hi}
R.~L. Arnowitt, S.~Deser, and C.~W. Misner, {\it {The Dynamics of General
  Relativity}},  \href{http://xxx.lanl.gov/abs/gr-qc/0405109}{{\tt
  gr-qc/0405109}}.

\bibitem{Strominger:1996sh}
A.~Strominger and C.~Vafa, {\it {Microscopic Origin of the Bekenstein-Hawking
  Entropy}},  Phys. Lett. {\bf B379} (1996) 99--104,
  [\href{http://xxx.lanl.gov/abs/hep-th/9601029}{{\tt hep-th/9601029}}].

\bibitem{Sen:2007qy}
A.~Sen, {\it {Black Hole Entropy Function, Attractors and Precision Counting of
  Microstates}},  Gen. Rel. Grav. {\bf 40} (2008) 2249--2431,
  [\href{http://xxx.lanl.gov/abs/0708.1270}{{\tt arXiv:0708.1270}}].

\bibitem{Hawking:1994ii}
S.~W. Hawking, G.~T. Horowitz, and S.~F. Ross, {\it {Entropy, Area, and Black
  Hole Pairs}},  Phys. Rev. {\bf D51} (1995) 4302--4314,
  [\href{http://xxx.lanl.gov/abs/gr-qc/9409013}{{\tt gr-qc/9409013}}].

\bibitem{Teitelboim:1994az}
C.~Teitelboim, {\it {Action and Entropy of Extreme and Nonextreme Nlack
  Holes}},  Phys. Rev. {\bf D51} (1995) 4315,
  [\href{http://xxx.lanl.gov/abs/hep-th/9410103}{{\tt hep-th/9410103}}].

\bibitem{Ghosh:1996gp}
A.~Ghosh and P.~Mitra, {\it {Understanding the Area Proposal for Extremal Black
  Hole Entropy}},  Phys. Rev. Lett. {\bf 78} (1997) 1858--1860,
  [\href{http://xxx.lanl.gov/abs/hep-th/9609006}{{\tt hep-th/9609006}}].

\bibitem{Kiefer:1998rr}
C.~Kiefer and J.~Louko, {\it {Hamiltonian Evolution and Quantization for
  Extremal Black Holes}},  Annalen Phys. {\bf 8} (1999) 67--81,
  [\href{http://xxx.lanl.gov/abs/gr-qc/9809005}{{\tt gr-qc/9809005}}].

\bibitem{Carroll:2009maa}
S.~M. Carroll, M.~C. Johnson, and L.~Randall, {\it {Extremal Limits and Black
  Hole Entropy}},  JHEP {\bf 11} (2009) 109,
  [\href{http://xxx.lanl.gov/abs/0901.0931}{{\tt arXiv:0901.0931}}].

\bibitem{Balbinot:2007kr}
R.~Balbinot, A.~Fabbri, S.~Farese, and R.~Parentani, {\it {Hawking Radiation
  from Extremal and Non-Extremal Black Holes}},  Phys. Rev. {\bf D76} (2007)
  124010, [\href{http://xxx.lanl.gov/abs/0710.0388}{{\tt arXiv:0710.0388}}].

\bibitem{Louko:1994tv}
J.~Louko and B.~F. Whiting, {\it {Hamiltonian Thermodynamics of the
  Schwarzschild Black Hole}},  Phys. Rev. {\bf D51} (1995) 5583--5599,
  [\href{http://xxx.lanl.gov/abs/gr-qc/9411017}{{\tt gr-qc/9411017}}].

\bibitem{Braden:1990hw}
H.~W. Braden, J.~D. Brown, B.~F. Whiting, and J.~W. York, Jr., {\it {Charged
  Black Hole in a Grand Canonical Ensemble}},  Phys. Rev. {\bf D42} (1990)
  3376--3385.

\bibitem{Davies:1978mf}
P.~C.~W. Davies, {\it {Thermodynamics of Black Holes}},  Proc. Roy. Soc. Lond.
  {\bf A353} (1977) 499--521.

\bibitem{Chamblin:1999tk}
A.~Chamblin, R.~Emparan, C.~V. Johnson, and R.~C. Myers, {\it {Charged AdS
  Black Holes and Catastrophic Holography}},  Phys. Rev. {\bf D60} (1999)
  064018, [\href{http://xxx.lanl.gov/abs/hep-th/9902170}{{\tt
  hep-th/9902170}}].

\bibitem{Chamblin:1999hg}
A.~Chamblin, R.~Emparan, C.~V. Johnson, and R.~C. Myers, {\it {Holography,
  Thermodynamics and Fluctuations of Charged AdS Black Holes}},  Phys. Rev.
  {\bf D60} (1999) 104026, [\href{http://xxx.lanl.gov/abs/hep-th/9904197}{{\tt
  hep-th/9904197}}].

\bibitem{Louko:1996dw}
J.~Louko and S.~N. Winters-Hilt, {\it {Hamiltonian Thermodynamics of the
  Reissner-Nordstr\"om- Anti-de Sitter Black Hole}},  Phys. Rev. {\bf D54}
  (1996) 2647--2663, [\href{http://xxx.lanl.gov/abs/gr-qc/9602003}{{\tt
  gr-qc/9602003}}].

\bibitem{York:1986it}
J.~W. York, Jr., {\it {Black Hole Thermodynamics and the Euclidean Einstein
  Action}},  Phys. Rev. {\bf D33} (1986) 2092--2099.

\bibitem{Brown:1989fa}
J.~D. Brown {\em et.~al.}, {\it {Thermodynamic Ensembles and Gravitation}},
  Class. Quant. Grav. {\bf 7} (1990) 1433--1444.

\bibitem{Cvetic:1999ne}
M.~Cvetic and S.~S. Gubser, {\it {Phases of R-Charged Black Holes, Spinning
  Branes and Strongly Coupled Gauge Theories}},  JHEP {\bf 04} (1999) 024,
  [\href{http://xxx.lanl.gov/abs/hep-th/9902195}{{\tt hep-th/9902195}}].

\bibitem{Preskill:1991tb}
J.~Preskill, P.~Schwarz, A.~D. Shapere, S.~Trivedi, and F.~Wilczek, {\it
  {Limitations on the Statistical Description of Black Holes}},  Mod. Phys.
  Lett. {\bf A6} (1991) 2353--2362.

\bibitem{Charmousis:2010zz}
C.~Charmousis, B.~{Gout\'eraux}, B.~S. Kim, E.~Kiritsis, and R.~Meyer, {\it
  {Effective Holographic Theories for Low-Temperature Condensed Matter
  Systems}},  \href{http://xxx.lanl.gov/abs/1005.4690}{{\tt arXiv:1005.4690}}.

\bibitem{Creighton:1995au}
J.~D.~E. Creighton and R.~B. Mann, {\it {Quasilocal Thermodynamics of Dilaton
  Gravity Coupled to Gauge Fields}},  Phys. Rev. {\bf D52} (1995) 4569--4587,
  [\href{http://xxx.lanl.gov/abs/gr-qc/9505007}{{\tt gr-qc/9505007}}].

\bibitem{Creighton:1996st}
J.~D.~E. Creighton, {\em {Gravitational Calorimetry}}.
\newblock PhD thesis, {University of Waterloo}, 1996.
\newblock \href{http://xxx.lanl.gov/abs/gr-qc/9610038}{{\tt gr-qc/9610038}}.

\bibitem{Leygnac:2004bb}
C.~Leygnac, {\em {Non-Asymptotically Flat Black Holes / Branes}}.
\newblock PhD thesis, {University of Claude Bernard - Lyon 1}, 2004.
\newblock \href{http://xxx.lanl.gov/abs/gr-qc/0409040}{{\tt gr-qc/0409040}}.
\newblock In French.

\bibitem{gkmn2}
U.~Gursoy, E.~Kiritsis, L.~Mazzanti, and F.~Nitti, {\it {Holography and
  Thermodynamics of 5D Dilaton-Gravity}},  JHEP {\bf 05} (2009) 033,
  [\href{http://xxx.lanl.gov/abs/0812.0792}{{\tt arXiv:0812.0792}}].

\bibitem{Aharony:1999ti}
O.~Aharony, S.~S. Gubser, J.~M. Maldacena, H.~Ooguri, and Y.~Oz, {\it {Large N
  Field Theories, String Theory and Gravity}},  Phys. Rept. {\bf 323} (2000)
  183--386, [\href{http://xxx.lanl.gov/abs/hep-th/9905111}{{\tt
  hep-th/9905111}}].

\bibitem{Wess:1992cp}
J.~Wess and J.~Bagger, {\em {Supersymmetry and Supergravity}}.
\newblock Princeton Univ. Pr., Princeton, USA, 1992.
\newblock 259 p.

\bibitem{Gubser:1998bc}
S.~S. Gubser, I.~R. Klebanov, and A.~M. Polyakov, {\it {Gauge Theory
  Correlators from Non-critical String Theory}},  Phys. Lett. {\bf B428} (1998)
  105--114, [\href{http://xxx.lanl.gov/abs/hep-th/9802109}{{\tt
  hep-th/9802109}}].

\bibitem{'tHooft:1993gx}
G.~'t~Hooft, {\it {Dimensional Reduction in Quantum Gravity}},
  \href{http://xxx.lanl.gov/abs/gr-qc/9310026}{{\tt gr-qc/9310026}}.

\bibitem{Susskind:1994vu}
L.~Susskind, {\it {The World as a Hologram}},  J. Math. Phys. {\bf 36} (1995)
  6377--6396, [\href{http://xxx.lanl.gov/abs/hep-th/9409089}{{\tt
  hep-th/9409089}}].

\bibitem{Boonstra:1998mp}
H.~J. Boonstra, K.~Skenderis, and P.~K. Townsend, {\it {The Domain Wall/QFT
  Correspondence}},  JHEP {\bf 01} (1999) 003,
  [\href{http://xxx.lanl.gov/abs/hep-th/9807137}{{\tt hep-th/9807137}}].

\bibitem{Susskind:1998dq}
L.~Susskind and E.~Witten, {\it {The Holographic Bound in Anti-de Sitter
  Space}},  \href{http://xxx.lanl.gov/abs/hep-th/9805114}{{\tt
  hep-th/9805114}}.

\bibitem{Peet:1998wn}
A.~W. Peet and J.~Polchinski, {\it {UV/IR Relations in AdS Dynamics}},  Phys.
  Rev. {\bf D59} (1999) 065011,
  [\href{http://xxx.lanl.gov/abs/hep-th/9809022}{{\tt hep-th/9809022}}].

\bibitem{Gursoy:2007cb}
U.~Gursoy and E.~Kiritsis, {\it {Exploring Improved Holographic Theories for
  QCD: Part I}},  JHEP {\bf 02} (2008) 032,
  [\href{http://xxx.lanl.gov/abs/0707.1324}{{\tt arXiv:0707.1324}}].

\bibitem{Gursoy:2007er}
U.~Gursoy, E.~Kiritsis, and F.~Nitti, {\it {Exploring Improved Holographic
  Theories for QCD: Part II}},  JHEP {\bf 02} (2008) 019,
  [\href{http://xxx.lanl.gov/abs/0707.1349}{{\tt arXiv:0707.1349}}].

\bibitem{Goldstein:2009cv}
K.~Goldstein, S.~Kachru, S.~Prakash, and S.~P. Trivedi, {\it {Holography of
  Charged Dilaton Black Holes}},  \href{http://xxx.lanl.gov/abs/0911.3586}{{\tt
  arXiv:0911.3586}}.

\bibitem{Cadoni:2009xm}
M.~Cadoni, G.~D'Appollonio, and P.~Pani, {\it {Phase Transitions between
  Reissner-Nordstr\"om and Dilatonic Black Holes in 4D AdS Spacetime}},  JHEP
  {\bf 03} (2010) 100, [\href{http://xxx.lanl.gov/abs/0912.3520}{{\tt
  arXiv:0912.3520}}].

\bibitem{Bertoldi:2010ca}
G.~Bertoldi, B.~A. Burrington, and A.~W. Peet, {\it {Thermal Behavior of
  Charged Dilatonic Black Branes in AdS and UV Completions of Lifshitz-like
  Geometries}},  \href{http://xxx.lanl.gov/abs/1007.1464}{{\tt
  arXiv:1007.1464}}.

\bibitem{Perlmutter:2010qu}
E.~Perlmutter, {\it {Domain Wall Holography for Finite Temperature Scaling
  Solutions}},  \href{http://xxx.lanl.gov/abs/1006.2124}{{\tt
  arXiv:1006.2124}}.

\bibitem{Gubser:2000nd}
S.~S. Gubser, {\it {Curvature Singularities: The Good, the Bad, and the
  Naked}},  Adv. Theor. Math. Phys. {\bf 4} (2000) 679--745,
  [\href{http://xxx.lanl.gov/abs/hep-th/0002160}{{\tt hep-th/0002160}}].

\bibitem{Sachdev:2008ba}
S.~Sachdev and M.~Mueller, {\it {Quantum Criticality and Black Holes}},
  \href{http://xxx.lanl.gov/abs/0810.3005}{{\tt arXiv:0810.3005}}.

\bibitem{Hartnoll:2009sz}
S.~A. Hartnoll, {\it {Lectures on Holographic Methods for Condensed Matter
  Physics}},  Class. Quant. Grav. {\bf 26} (2009) 224002,
  [\href{http://xxx.lanl.gov/abs/0903.3246}{{\tt arXiv:0903.3246}}].

\bibitem{Herzog:2009xv}
C.~P. Herzog, {\it {Lectures on Holographic Superfluidity and
  Superconductivity}},  J. Phys. {\bf A42} (2009) 343001,
  [\href{http://xxx.lanl.gov/abs/0904.1975}{{\tt arXiv:0904.1975}}].

\bibitem{Sachdev:2010ch}
S.~Sachdev, {\it {Condensed Matter and AdS/CFT}},
  \href{http://xxx.lanl.gov/abs/1002.2947}{{\tt arXiv:1002.2947}}.

\bibitem{Horowitz:2010gk}
G.~T. Horowitz, {\it {Introduction to Holographic Superconductors}},
  \href{http://xxx.lanl.gov/abs/1002.1722}{{\tt arXiv:1002.1722}}.

\bibitem{Cooper:1956zz}
L.~N. Cooper, {\it {Bound Electron Pairs in a Degenerate Fermi Gas}},  Phys.
  Rev. {\bf 104} (1956) 1189--1190.

\bibitem{Bardeen:1957kj}
J.~Bardeen, L.~N. Cooper, and J.~R. Schrieffer, {\it {Microscopic Theory of
  Superconductivity}},  Phys. Rev. {\bf 106} (1957) 162.

\bibitem{Bardeen:1957mv}
J.~Bardeen, L.~N. Cooper, and J.~R. Schrieffer, {\it {Theory of
  Superconductivity}},  Phys. Rev. {\bf 108} (1957) 1175--1204.

\bibitem{Stewart:2001zz}
G.~R. Stewart, {\it {Non-Fermi-Liquid Behavior in d- and f-Electron Netals}},
  Rev. Mod. Phys. {\bf 73} (2001) 797--855.

\bibitem{Hussey:2008}
N.~E. Hussey, {\it Phenomenology of the normal state in plane-transport
  properties of high-tc cuprates},  Journal of Physics: Condensed Matter {\bf
  20} (2008), no.~12 123201.

\bibitem{Cooper:2009}
R.~A. {Cooper}, Y.~{Wang}, B.~{Vignolle}, O.~J. {Lipscombe}, S.~M. {Hayden},
  Y.~{Tanabe}, T.~{Adachi}, Y.~{Koike}, M.~{Nohara}, H.~{Takagi}, C.~{Proust},
  and N.~E. {Hussey}, {\it {Anomalous Criticality in the Electrical Resistivity
  of La$_{2-x}$Sr$_{x}$CuO$_{4}$}},  Science {\bf 323} (Jan., 2009) 603--.

\bibitem{Martin:1990}
S.~Martin, A.~T. Fiory, R.~M. Fleming, L.~F. Schneemeyer, and J.~V. Waszczak,
  {\it Normal-state transport properties of $bi2+x$$sr2-y$$cuo6+\delta{}$
  crystals},  Phys. Rev. B {\bf 41} (Jan, 1990) 846--849.

\bibitem{vandeMarel:2003wn}
D.~van~de Marel {\em et.~al.}, {\it {Quantum Critical Behaviour in a High-Tc
  Superconductor}},  Nature {\bf 425} (2003) 271.

\bibitem{Karch:2007pd}
A.~Karch and A.~O'Bannon, {\it {Metallic AdS/CFT}},  JHEP {\bf 09} (2007) 024,
  [\href{http://xxx.lanl.gov/abs/0705.3870}{{\tt arXiv:0705.3870}}].

\bibitem{Lee:2008xf}
S.-S. Lee, {\it {A Non-Fermi Liquid from a Charged Black Hole: A Critical Fermi
  Ball}},  Phys. Rev. {\bf D79} (2009) 086006,
  [\href{http://xxx.lanl.gov/abs/0809.3402}{{\tt arXiv:0809.3402}}].

\bibitem{Hartnoll:2009ns}
S.~A. Hartnoll, J.~Polchinski, E.~Silverstein, and D.~Tong, {\it {Towards
  Strange Metallic Holography}},  JHEP {\bf 04} (2010) 120,
  [\href{http://xxx.lanl.gov/abs/0912.1061}{{\tt arXiv:0912.1061}}].

\bibitem{Hartnoll:2008kx}
S.~A. Hartnoll, C.~P. Herzog, and G.~T. Horowitz, {\it {Holographic
  Superconductors}},  JHEP {\bf 12} (2008) 015,
  [\href{http://xxx.lanl.gov/abs/0810.1563}{{\tt arXiv:0810.1563}}].

\bibitem{Horowitz:2009ij}
G.~T. Horowitz and M.~M. Roberts, {\it {Zero Temperature Limit of Holographic
  Superconductors}},  JHEP {\bf 11} (2009) 015,
  [\href{http://xxx.lanl.gov/abs/0908.3677}{{\tt arXiv:0908.3677}}].

\end{thebibliography}\endgroup

\bibliographystyle{these}

\end{document}